\documentclass[longauth]{aa}  
\usepackage{graphicx}
\usepackage{txfonts}
\bibpunct{(}{)}{;}{a}{}{,} % to follow the A&A style
\usepackage[normal]{threeparttable}
\usepackage[nottoc, notlof, notlot]{tocbibind}   %biblio

\usepackage{amsmath} 
\usepackage{float}
\usepackage{mathabx}
\usepackage{tabularx}
\usepackage{siunitx}
\usepackage{colortbl}
\usepackage{soul}
\usepackage[caption = false]{subfig}
\usepackage[inkscapeformat=png]{svg}
\usepackage{hyperref}
\hypersetup{colorlinks,linkcolor={blue},citecolor={blue},urlcolor={red}} 
\usepackage{multirow}
\newcommand{\vlsr}{\mbox{V$_{\rm LSR}$\;}}
\newcommand{\Msun}{\mbox{\,M$_{\odot}$\;}}

\newcommand{\kms}{\mbox{km~s$^{-1}$\,}}
\newcommand{\mJybeamkms}{\mbox{mJy~beam$^{-1}$~km~s$^{-1}$\,}}
\usepackage{bm}

\newcommand{\col}[1]{\multicolumn{1}{c}{#1}}
\newcolumntype{N}{S[table-format=2.1]}

\defcitealias{Fred2022}{Paper~I}
\defcitealias{Ginsburg2022}{Paper~II}
\defcitealias{Pouteau2022}{Paper~III}
\defcitealias{Brouillet2022}{Paper~IV}
\defcitealias{Nony2023}{Paper~V}
\defcitealias{Pouteau2023}{Paper~VI}
\defcitealias{Cunningham2023}{Paper~VII}
\defcitealias{louvet22}{Paper~VIII}
\defcitealias{towner2023}{Paper~IX}
\defcitealias{diaz2023}{Paper~X}
\defcitealias{armante2023}{Paper~XI}

\begin{document} 

   \title{ALMA-IMF XVII - Census and lifetime of high-mass prestellar cores in 14 massive protoclusters}
%   \title{ALMA-IMF XX - A statistical study of the population of massive prestellar core candidates in 15 massive protoclusters}

   \author{M. Valeille-Manet\inst{1} \and %\orcidlink{0009-0005-5343-1888},
          S. Bontemps\inst{1} \and %\orcidlink{0000-0002-4093-7178},
          T. Csengeri\inst{1} \and %\orcidlink{0000-0002-6018-1371}, 
          T. Nony \inst{2, 3} \and %\orcidlink{0000-0003-3246-0821},
          F. Motte \inst{4} \and %\orcidlink{0000-0003-1649-8002},
          A. M. Stutz \inst{5, 17} \and %\orcidlink{0000-0003-2300-8200},
          A. Gusdorf \inst{6,7} \and %\orcidlink{0000-0002-0354-1684},
          A. Ginsburg \inst{8} \and %\orcidlink{0000-0001-6431-9633},
          R. Galv\'an-Madrid \inst{2} \and %\orcidlink{0000-0003-1480-4643}, 
          P. Sanhueza \inst{12,13} \and %\orcidlink{0000-0002-7125-7685},
          M. Bonfand \inst{9} \and %\orcidlink{0000-0001-6551-6444},  
          N. Brouillet \inst{1} \and %\orcidlink{0000-0002-3274-7024},
          P. Dell'Ova \inst{6,7} \and %,
          F. Louvet \inst{4} \and %\orcidlink{0000-0003-3814-4424},
          N. Cunningham \inst{4, 10} \and %\orcidlink{0000-0003-3152-8564},
          M. Fern\'andez-L\'opez \inst{11} \and %\orcidlink{0000-0001-5811-0454},
          F. Herpin \inst{1} \and %\orcidlink{0000-0003-2977-5072}, 
          F. Wyrowski \inst{15} \and %\orcidlink{0000-0003-4516-3981},
          R.\ H.\ Álvarez-Gutiérrez \inst{5} \and %\orcidlink{0000-0002-9386-8612},
          M. Armante \inst{6, 7} \and %,
          A. E. Guzm\'an \inst{14} \and %\orcidlink{0000-0003-0990-8990},
          N. Kessler \inst{1} \and %,
          A. Koley \inst{5} \and %\orcidlink{0000-0003-2713-0211},
          J. Salinas \inst{5} \and %\orcidlink{0009-0009-4976-4320},
          T. Yoo \inst{8} \and %\orcidlink{0000-0003-2968-5333},
          L. Bronfman \inst{16} \and %\orcidlink{0000-0002-9574-8454},
          N. Le Nestour \inst{4}
          }
   \institute{Laboratoire d’astrophysique de Bordeaux, Univ. Bordeaux, CNRS, B18N, all\'ee Geoffroy Saint-Hilaire, 33615 Pessac, France
    \and Instituto de Radioastronom\'ia y Astrof\'isica, Universidad Nacional Aut\'onoma de M\'exico, Morelia, Michoac\'an 58089, M\'exico      %2
    \and INAF-Osservatorio Astrofisico di Arcetri, Largo E. Fermi 5, I-50125 Firenze, Italy        %3
    \and Univ. Grenoble Alpes, CNRS, IPAG, 38000 Grenoble, France             %4
    \and Departamento de Astronom\'{i}a, Universidad de Concepci\'{o}n, Casilla 160-C, 4030000 Concepci\'{o}n, Chile    %14 --> 5
    \and Laboratoire de Physique de l'\'Ecole Normale Sup\'erieure, ENS, Universit\'e PSL, CNRS, Sorbonne Universit\'e, Universit\'e de Paris, Paris, France  %6
    \and Observatoire de Paris, PSL University, Sorbonne Universit\'e, LERMA, 75014, Paris, France %7
    \and Department of Astronomy, University of Florida, PO Box 112055, USA %11 --> 8
    \and Departments of Astronomy and Chemistry, University of Virginia, Charlottesville, VA 22904, USA %8 --> 9
    \and SKA Observatory, Jodrell Bank, Lower Withington, Macclesfield SK11 9FT, United Kingdom %9 --> 10
    \and Instituto Argentino de Radioastronom\'\i a (CCT-La Plata, CONICET; CICPBA), C.C. No. 5, 1894, Villa Elisa, Buenos Aires, Argentina %10 --> 11
    \and Department of Astronomical Science, SOKENDAI (The Graduate University for Advanced Studies), 2-21-1 Osawa, Mitaka, Tokyo 181-8588, Japan   %12 --> 12
    \and The Graduate University for Advanced Studies (SOKENDAI), 2-21-1 Osawa, Mitaka, Tokyo 181-0015, Japan %13 --> 13
    \and Joint Alma Observatory (JAO), Alonso de Córdova 3107, Vitacura, Santiago, Chile %14
    \and Max Planck Institute for Radio Astronomy, Auf dem Hügel 69, 53121 Bonn, Germany %15
    \and Department of Astronomy, Universidad de Chile, Camino El Observatorio 1515, Santiago, Chile. %16
    \and Franco-Chilean Laboratory for Astronomy, IRL 3386, CNRS and Universidad de Chile, Santiago, Chile.%17
}

   \date{Received ...; accepted ...}

% \abstract{}{}{}{}{} 
% 5 {} token are mandatory
 
  \abstract
    {High-mass prestellar cores are extremely rare.  Indeed, the search for such objects has until recently been hampered by small sample sizes, leading to large ambiguities in their lifetimes and hence the conditions in the cores in which high mass stars ($\rm \gtrsim 8\, M_{\odot}$) form.}
    {Here we leverage the large sample ($\sim$\,580 cores) detected in the ALMA-IMF survey to identify both protostellar and prestellar cores to estimate their relative lifetimes.}
   % Methods
    {We use CO and SiO outflows to identify protostellar cores. We present a new automated method based on aperture line emission and background subtraction to systematically detect outflows associated with each of the 141 most massive cores. Massive cores which are not driving an outflow in either tracer are identified as prestellar. After careful scrutiny of the sample, we derive statistical lifetime estimates for the prestellar phase.}
   % Results
    {Our automated method allows for efficient detection of CO and SiO outflows and has a similar performance efficiency as more cumbersome classical techniques.  We identify 30 likely prestellar cores with M\,$\gtrsim8$\,$\rm M_{\odot}$, of which 12 have core masses M\,$\gtrsim16$\,$\rm M_{\odot}$. The latter contains the best candidates for high-mass star precursors. Moreover, most of these 12 are located inside the crowded central regions of the protoclusters, where most high-mass stars are expected to form. Using the relative ratios of prestellar to protostellar cores, and assuming a high-mass protostellar lifetime of 300~kyr, we derive a prestellar core lifetime of 120~kyr to 240~kyr for cores with masses $\rm 8\,M_{\odot} < M < 16\,M_{\odot}$.  For $\rm 30\,M_{\odot} < M < 55\,M_{\odot}$, lifetimes range from 50~kyr to 100~kyr. The spread in timescales reflects different assumptions for scenarios for the mass reservoir evolution. These timescales are remarkably long compared to the 4~kyr to 15~kyr free-fall time of the cores. Hence, we suggest that high mass cores live $\sim$\,10 to 30 free-fall times, with a tentative trend of a slight decrease with core mass. Such large ratios suggest that the collapse of massive cores is slowed down by non-thermal support of turbulent, magnetic or rotational origin at or below the observed scale.}
    % Conclusions (optional) {} leave it empty if necessary
    {}

    \titlerunning{ALMA-IMF XVI - High-mass prestellar cores}
    \authorrunning{M. Valeille-Manet et al.}
    \maketitle

%-------------------------------------------------------------------
\section{Introduction}
Understanding star formation strongly relies on unveiling the youngest stages of the evolution from molecular gas to young stars, i.e. the stages of the so-called protostars and pre-collapse or prestellar cores (e.g. \citealp{Lada1992, Andre2000} and references therein). These stages still carry the imprint of the physical properties at the onset of collapse and of the prevailing physical processes leading to the birth of a new star. For low-mass star formation, a fairly good census of all protostars and prestellar cores has been obtained with a series of infrared (IR) space telescopes, such as IRAS, ISO, Spitzer and Herschel leading to a view with a sequence of different observational stages led by clear physical processes. 

Low-mass ($\rm M<8\,M_{\odot}$) prestellar cores (hereafter PSCs) are believed to originate from quasi-static contractions of dense cores and clumps, well described by Bonnort-Ebert spheres, and leading to a gravitational collapse when gravity cannot be contained by the other supports, i.e. thermal, magnetic and turbulent pressures.  In nearby molecular clouds, the properties of low-mass PSCs have been widely studied leading to a picture, namely the gravo-turbulent scenario \citep{Padoan2002}, of a probable slow contraction of low density clump to high density cores over $\sim 1$~Myr (e.g. \citealt{Jessop2000}; \citealt{Konyves2015}). The typical thermal Jeans masses are also found to correspond well to the stellar masses suggesting no additional support to thermal pressure is required to account for low-mass star masses.

In contrast the steps to form high-mass stars ($\rm M > 8 \,M_{\odot}$) are much less understood. From the observational point of view this is due to the rarity of high-mass stars and the need to search for high-mass protostellar objects at large galactic distances (typically at a distance d > 2 kpc; \citealp{Motte_review_2018}). From the theoretical point of view, the modeling and simulations have difficulties in accounting for the full complexity of the required physics to explain the highest stellar masses, which certainly have to involve magnetic fields, turbulence and radiative energy at the same time and over decades of different scales in molecular clouds \citep[e.g.][]{Tan2014, Stutz2016, Commercon2022}.
%\timea{Before talking about statistics, one needs to have first at least a few reliable detections.} 
Low-mass PSCs are easy to study and can be convincingly seen as the reservoir of mass to form stars \citep{Motte1998, Launhardt2013, Konyves2015, Furlan2016} possibly formed by thermal Jeans instability in dense filaments and clumps, following the so-called core-fed scenario. In contrast, the reservoirs to form high-mass stars are much elusive and seem too massive to be explained by simple thermal Jeans fragmentation, certainly requiring magnetic and turbulent supports to keep possible the core-fed scenario. Continuously driven turbulence could increase the local Jeans masses enough to explain high-mass reservoirs \citep{McKee&Tan2003}. The global gravitational potential well of dense clumps (> 0.1 pc scale) could also lead to a competition for accretion from the large infall rates  \citep{Bonnell&Bate2001}. The reservoir of mass to form high-mass stars could therefore mostly reside at clump scales with powerful convergence of gas down to protostellar scales in a so-called clump-fed scenario (\citealp{Peretto2013,Csengeri2011,Duarte-Cabral2013,Peretto2020}). %\timea{I think this is not clear. A thourough referee would point out the lack of discussion on the role of B fields.}
%It is unclear which precise physics is regulating and at which scale the fragmentation process may form reservoirs of mass (cores) massive enough to account for the formation of high-mass stars (high-mass PSCs).
It is unclear which exact physical processes are regulating or dominating the evolution of these massive cradles, and at which scale(s) a fragmentation process might yield small scale concentrations of mass that we would call "cores". Hence it remains unclear if high mass reservoirs can be pinpointed observationally as well-identified objects distinct from their surroundings \citep{Motte2007,Tige2017,Sanhueza2019,Morri2023}. 

%The quasi-static monolitic turbulent scenario by \cite{McKee&Tan2003} proposed that continuously driven turbulence in dense cores could increase the Jeans mass enough to explain a monolitic collapse of high-mass turbulent cores. Alternatively the competitive accretion scenario \citep{Bonnell&Bate2001} proposed that only the global gravitational dwell of a protocluster can drive accretion streams massive enough to build up high-mass stars from low-mass seeds at the center of protoclusters. More recently the inertial-inflow model by \cite{Padoan2020} proposes that low-mass protostellar seeds could have their mass increasing at relatively low paces (a few $10^{-5}\,$M$_\odot$/yr) during more than a million of years to reach high masses thanks to inertial large scale converging flows present in the turbulent cascade.

Observationally the protoclusters are very dynamical in nature \citep{Schneider2010,Peretto2013,Avison2021,Alvarez-Gutierrez2024} and the dense high-mass clumps do not have the required level of turbulence to explain high-mass Jeans fragmentation (e.g. \citealp{Csengeri2011}). In order to explain the very rare occurence of high-mass PSCs, \cite{Motte_review_2018} proposed a scenario in which protostellar cores and their parental clumps simultaneously grow in mass and accrete gas from their surroundings without any high-mass reservoir formed at early stage. %This scenario, in line with the global hierarchical collapse model \citep{Vazquez-Semadeni2009}, does not require the existence of massive starless cores, but proposes that the earliest phase of high-mass star formation corresponds to massive clumps hosting low-mass protostars.  
The top-heavy core mass functions recently observed in the ALMA-IMF protoclusters (\citealp{Motte2018, Pouteau2022, louvet22}) and its proposed evolution (\citealp{Pouteau2023, Nony2023, Armante2024}) could reflect this effect of progressive increase %and then decrease of mass in ridges/hubs before and after a star-formation burst, as well as the increase 
of core mass from the prestellar to the protostellar phase.

To understand the precise origin of high-mass stars, it is most vital to find very good examples of high-mass PSCs, and to study their physical properties. 
Such high-mass PSCs are, however, extremely rare as recognised early on by \cite{Motte2007} (see also  \citealp{Motte_review_2018} for a review).
%If they exist, the consequently short formation time-scales of the order of 10$^4\,$yr (10 kyr\footnote{We will adopt in the following the practical unit of kyr (1000 years) for the prestellar and free-fall time scales.}) invoked the role of supersonic flows which are indeed observed in high-mass clumps \citep{Schneider2010,Csengeri2011,Peretto2013,Avison2021}. To collect mass in only 10~kyr along the typical size of low-mass PSCs of 4000 au implies a velocity of 2 km/s (10 times the typical sound speed) while for low-mass PSCs living typically 10$^6\,$yr, the speed is 0.02 km/s (10 times smaller than the sound speed). If they exist, the estimate of the lifetime of high-mass PSCs is however very uncertain so far due to the lack of statistically large enough samples of well observed (at high resolution enough) high-mass protostellar cores to fully establish their existence and to estimate their number. 
So far only a few candidates for high mass prestellar cores have been found as compact cores that are both massive enough (at least above 16\,M$_{\odot}$) and without any outflow detection (i.e., likely tracing no protostellar activity, see below): 

\begin{itemize}
    \item CygX-N53-MM2 discovered in \cite{Duarte-Cabral2013} among nine other high-mass protostellar cores which does not show any strong sign of CO outflow but which is also situated on the side of a strong protostellar outflow from the nearby CygX-N53-MM1 protostellar core. 
    \item G11.11-P6-SMA1 found in \cite{Wang2014} which is reprieved of any outflow emission.
    \item W43-MM1~\#6 discovered in \cite{Nony2018} which does not drive any strong outflow (it is close to a strong outflow driving high-mass protostar W43-MM1~\#3) but has an interesting and unusual weak emission of complex organic molecules pointing to the presence of some warm gas inside it (between 20 and 90K; \citealp{Molet2019}). 
    \item C2c1a in the Dragon cloud from \cite{Barnes2023} which is a core not driving any CO outflow but which might not be massive enough to be in the high mass regime. The relatively high-mass in \cite{Barnes2023} is due to an adopted very low temperature close to 9\,K (from NH$_3$ gas temperature estimate) and a favorable dust opacity (see Sect.~\ref{subsection:unique_sample}).
\end{itemize}
Some additional candidates in the past have been discarded since then: cores G11.92-0.61-MM2 of \cite{Cyganowski2014} and C1S of \cite{Tan2013} have been found to be protostellar in nature after follow up observations (\citealp{Cyganowski2022} and \citealp{Tan2016} respectively). Some recent studies have searched for high-mass prestellar cores but without success most probably due to low statistics in the sample of high-mass cores (e.g. \citealt{Louvet2019} and \citealt{Morri2023} for the ALMA program ASHES). 
It is thus vital to observe some more significant samples of high-mass PSCs. %, and to study the physical properties of a representative sample of such true PSCs cores to constrain the physics of the formation and evolution of these PSCs which should correspond to the initial conditions to form high-mass stars, and test the scenarii in competition. 

ALMA-IMF is the largest ALMA survey for high-mass protostellar objects in the relatively nearby high-mass protoclusters of the Galaxy (see \citealp{Fred2022} for an overview). As suggested by \cite{Sanhueza2019}, if high-mass prestellar cores exist they are most likely to be found in more massive and possibly more evolved environments than clumps in so-called IR Dark Clouds (IRDC). \cite{Xu2024} also observed that starless cores become more massive in evolved clumps such as protoclusters. ALMA-IMF survey allows us thus for the first time to search for high-mass prestellar cores in such environments with unprecedented statistics. 

At disk and protostar scale (few hundreds of au, see e.g. \citealp{Commercon2022}), there is evidence that low and high-mass star accretion and ejection rely on the same mechanism of the magneto-centrifugal accretion-ejection process \citep{Blandford&Payne1982,Ferreira2006} with a magnetic regulation of the angular momentum in the inner parts of the accretion disk (see \citealp{Matsushita2017, Kolligan&kuiper2018,Csengeri2018,Olguin2023} for high-mass protostars). 
Accretion onto protostars is possible only if angular momentum in excess is driven away by ejecting part of the accretion flow through a magnetised jet and wind in the polar directions. Ejection of angular momentum allows the main accretion flows to spiral down onto the inner protostar in the equatorial plane. %Therefore both ejection and accretion must take place at the formation of a newborn star.
Since protostars are embedded inside their collapsing envelope and in surrounding dense clumps and filaments, jets and winds have to strongly interact with the surroundings. These interactions lead to a large amount of entrained gas in a roughly momentum conserved manner easily seen in CO line wings since CO is the most abundant observable species in cold dense gas. CO outflows are thus the best tracers of accretion for protostellar objects of all masses. This strong link between CO outflows and accretion onto protostars was first recognized observationally by \cite{Bontemps1996} for low-mass protostars. 
%These outflows are particularly unavoidable to be observed as CO outflows for the youngest protostars since they are expected to have still dense enough envelopes to reveal any jet escaping from the central regions where accretion is occurring. 
In addition to CO, SiO is a great tracer of shocks (see e.g. \citealp{Gusdorf2008}) and can help to identify outflows especially in very crowded regions where the confusion of CO can be important. These outflows can therefore be used to recognize accretion in cold cores to differentiate between prestellar cores (potential infall or slow contraction but no central accretion on a protostar) and protostellar objects (accretion on the central protostar). Here we use the CO and SiO outflow diagnostics as a signpost of protostellar accretion in the ALMA-IMF cores. 
Despite the above described arguments pointing to the expected and observed strong link between the protostellar status defined as actively accreting stellar embryos and the detection of CO/SiO outflows, we may consider possibilities that some truely protostellar cores may lack detectable such outflows. We discuss these possibilities in Sect.~\ref{subsection:alt-scenarii}.

After presenting the dataset in Sect.~\ref{sec:dataset}, we describe the CO and SiO outflow automated method to detect in a homogeneous way the outflows associated with the detected protostars of the ALMA-IMF fields in Sect. \ref{Method}. We then show that, as expected, most of the detected cores are driving outflows (Sect.~\ref{ResultsandAnalysis}). However, we identify a number of clearly young high-mass cores without outflow, making them excellent candidates to be high-mass PSCs. These sources are described in Sect.~\ref{ResultsandAnalysis}. We define high-mass PSCs and present the robust sample of high-mass PSC candidates in Sect.~\ref{sec:discussion}. Their rarity is discussed in Sect.~\ref{section:HMPSC_rare}. % In Sect.~\ref{subsection:fraction_with_regions} we present the statistics of PSCs in the different protoclusters of ALMA-IMF and discuss the rarity of the high-mass PSCs in Sect.~\ref{section:HMPSC_rare}. 
In Sect.~\ref{sec:lifetimes}, we then discuss this result in the context of the expected evolution of a high-mass protostar, deriving a strongly improved lifetime for PSCs for the high-mass regime which can be used to discuss first implications on the way high-mass stars may form.

%-----------------------------------------------------------------
%-----------------------------------------------------------------
%-----------------------------------------------------------------

\section{Observations and data reduction} \label{sec:dataset}
%\pagestyle{empty}
%\begin{landscape}
\begin{table*}[htbp!]
\centering
%\small
%    \setlength{\tabcolsep}{2pt}
\begin{threeparttable}[c]
\caption{ Overview of the ALMA-IMF protocluster clouds, their evolutionary stage and angular resolution.}
%(using global measurements)}
\label{tab:overview}
 \begin{tabular}{lcccccccccc}
\hline \noalign {\smallskip}
Protocluster  & RA\tnote{1}    & Dec\tnote{1}    & \vlsr\tnote{2} & $d$\tnote{1}   &  $\rm M_{cloud}\tnote{1}$ & Evolutionary  &  $\nu_{1.3mm}$\tnote{4} & $\theta$\tnote{5} & $\theta^{ave}_{spw1}$\tnote{6} & $\theta^{ave}_{spw5}$\tnote{7}\\  
cloud name\tnote{1}    & \multicolumn{2}{c}{[ICRS (J2000)]}   & [$\kms$]   & [kpc]  &  $\rm \times 10^3 M_{\odot}$ & stage\tnote{3}  & [GHz] &  ["] & ["] & ["]\\

\hline \noalign {\smallskip}

G008.67     
    & 18:06:21.12 & $-$21:37:16.7 & $+35.0$   & 3.4   &  3.1 %& 
    & I & 228.732 & 0.79 & 0.80 & 0.76 \\ 

G010.62     %(W31C), 010.624 -00.384   
    & 18:10:28.84 & $-$19:55:48.3 & $-2.7$     & 4.95   &  6.7 %& 
    & E & 229.268 & 0.55 & 0.60 & 0.54\\

G012.80     %(W33-IRS2), 012.888 +00.489  
    & 18:14:13.37 & $-$17:55:45.2 & $+36.1$   & 2.4   &  4.6 %& 
    & E & 229.080 & 1.13 & 1.07 & 1.00 \\
    
G327.29     %327.293 -00.579    %OLD: 15:53:08.62 & $-$54:37:06.3
    & 15:53:08.13 & $-$54:37:08.6 & $-45.0$   & 2.5   &  5.1 %&
    & Y & 229.507 & 1.08 & 0.78 & 0.73 \\ 

G328.25     %328.254 -00.532
    & 15:57:59.68 & $-$53:58:00.2 & $-43.4$   & 2.5   &  2.5 %& 
    & Y & 227.575 & 1.08 & 0.66 & 0.61 \\

G333.60	    %353.409 -00.361
    & 16:22:09.36 & $-$50:05:58.9 & $-47.8$   & 4.2  &  12.0 %& 
    &  E & 229.062 & 0.64 & 0.71 & 0.65 \\

G337.92 	%337.916 -00.477
    & 16:41:10.62 & $-$47:08:02.9 & $-39.6$   & 2.7   &  2.5 %&
    & Y & 227.503 & 1.00 & 0.73 & 0.68 \\ %& $1.5\times 10^5$ & 34
    % $d$ changed 4.2 --> 2.7       %3  %0.841
    % size: 24.3" ->

G338.93     %338.926 +00.554
    & 16:40:34.42 & $-$45:41:40.6 & $-61.1$   & 3.9  &  7.1 %& 
    & Y & 229.226 & 0.69 & 0.72 & 0.67 \\ 	%& $9.4\times 10^4$ & 24    %Flux30K=5.1
    % size: 35.3" ->

G351.77    
    & 17:26:42.62 & $-$36:09:20.5 & $-3.9$   & 2.0   &  2.5 %&
    &  I & 227.991 & 1.35 & 0.95 & 0.88 \\

G353.41     %353.409 -00.361
    & 17:30:26.28 & $-$34:41:49.7 & $-17.6$   & 2.0   &  2.5 %&
    &  I & 229.431 & 1.35 & 0.97 & 0.90 \\
    
W43-MM1     %(G030.82), 030.818 -00.05	
    & 18:47:47.00 &  $-$01:54:26.0 & $+97.4$    & 5.5  &  13.4 %& 2
    & Y  & 229.680 & 0.49 & 0.55 & 0.50 \\
    
W43-MM2     %(G030.70), 030.703 -00.067	
    & 18:47:36.61 & $-$02:00:51.1 & $+91.0$   & 5.5   &  11.6 %& 2 
    & Y & 227.597 & 0.49 & 0.56 & 0.56 \\

W43-MM3     %(G030.72), 030.718 -00.082	
    & 18:47:41.46 & $-$02:00:27.6 & $+93.0$   & 5.5  &  5.2 %& 2, 3 
    & I & 228.931 & 0.49 & 0.61 & 0.58 \\
    
W51-E       %(G049.49Main), 049.489 -00.369
    & 19:23:44.18 & $+$14:30:29.5 & $+55.2$   & 5.4   &  32.7 %& 1 
    & I & 228.918 & 0.50 & 0.4 & 0.46 \\ 

W51-IRS2    %(G049.49), 049.489 -00.389 
    & 19:23:39.81 & $+$14:31:03.5 & $+61.4$   & 5.4   &  20.6 %& 1 
    & E & 228.530 &  0.50 & 0.62 & 0.55 \\

\hline \noalign {\smallskip}
\end{tabular}
\begin{tablenotes}
\item[1] Protocluster name, central position used for the ALMA-IMF observations. Distances and masses are taken from \cite{Fred2022} and references therein.
\item[2] Velocity relative to the local standard of rest extracted from \cite{Cunningham2023}. 
\item[3] Classification of the ALMA-IMF protocluster clouds: Young (Y), Intermediate (I), and Evolved (E) (see Section 4.1 by \cite{Fred2022}). 
\item[4] Central frequencies of observations in Band 6 extracted from \cite{louvet22}
\item[5] Angular resolution in Band 6 extracted from \cite{louvet22}
\item[6] Angular resolution in spw1 Band 6 (217.150\,GHz, which includes the SiO\,(5--4) and DCN\,(3--2) lines) define as $\theta_{ave} = \sqrt{\theta_{maj} \times \theta_{min}}$ and extracted from \cite{Cunningham2023}
\item[7] Angular resolution in spw5 Band 6 (230.530\,GHz, which includes the CO(2--1) line) extracted from \cite{Cunningham2023}

\end{tablenotes}
\end{threeparttable}
\end{table*}
\begin{table*}[htbp!]
\label{tab:spws}
    \centering
    \begin{threeparttable}[c]
    \caption{Spectral lines used and their spectral resolution.}
    \begin{tabular}{ccccccc}
    \hline \noalign {\smallskip}
    Line & Frequency & ALMA Band & Spectral window & Cube bandwidth & \multicolumn{2}{c}{Resolution}\\
        &  [GHz] &   &  & [MHz] & [kHz] & [$\kms$] \\
    \hline \noalign {\smallskip}
    SiO\,(5-4) & 217.105 & B6 & SPW 1 & 234 & 244 & 0.34 \\
    DCN\,(3-2) & 217.238 & B6 & SPW 1 & 234 & 244 & 0.34 \\
    CO\,(2-1) & 230.538 & B6 & SPW 5 & 974 & 468 & 1.27 \\
    \hline
    \end{tabular}
    \end{threeparttable}
\end{table*}

The data used for this paper are part of the large program ALMA-IMF\footnote{\url{https://www.almaimf.com} : ALMA transforms our view of the origin of stellar masses} (\#2017.1.01355.L, PIs: Motte, Ginsburg, Louvet, Sanhueza), which targets 15 massive protoclusters located at distances from 2 to 5.5~kpc. \cite{Fred2022} (ALMA-IMF Paper I) describe in detail the ALMA-IMF large program, its objectives and first results. In short, 
the targets were selected from the ATLASGAL survey carried out with the APEX telescope \citep{Csengeri2014}, from which \cite{Csengeri2017} identified the 200 sub-millimeter brightest star forming clumps covering different evolutionary stages. We used the Atacama Large Millimeter/submillimeter Array (ALMA) interferometer to image these regions in  two frequency bands : Band~3 (B3; $\sim$91-106~GHz) and Band~6 (B6; $\sim$216-234~GHz) as described in \cite{Fred2022}. Table~\ref{tab:overview} lists the main characteristics of each protocluster, such as their name and central positions, \vlsr, distance from the Sun, evolutionary stage from \cite{Fred2022}, the synthesized beam of the continuum maps in arc seconds, and the corresponding physical scale in astronomical unit (au), as well as the synthesized beam in arc seconds of the line datacubes used.

Details of the data reduction process is described in \cite{Ginsburg2022} (ALMA-IMF Paper II) that presents the continuum maps at 1.3~mm and 3~mm for the 15 protoclusters. We use here the catalogs of continuum sources  from \cite{louvet22} that were obtained with the source extraction algorithm getsf \citep{Men'shchikov2021}. We use the source catalogs extracted from the cleanest continuum maps smoothed to a common physical resolution of 2700~au that corresponds to the poorest resolution of the sample. Sources potentially contaminated by free-free emission were removed from this catalog based on spectral index estimations (see \citealp{Galvan-Madrid2024} and \citealp{louvet22} for details). In short, the cleanest continuum maps are obtained by using only line free channels to estimate the continuum. %, and the sources removed are $100 \%$ free-free sources or sources on the edges of the free-free bubbles. 
The smoothing was done in order to obtain the same spatial resolution of 2700~au in Band 6 for all the targeted regions. This provides us a sample with a total of 580 gravitationally bound cores on the same linear scale (cores are considered as bounded for $\rm M_{BE}/M_{core}<2$ in \citealp{louvet22}).  %The catalogs can be found on the ALMA-IMF website. 
%For the rest of the paper the fluxes and sizes of the cores (non deconvolved from the beam) are directly extracted from these catalogs, and the sources will be called by their number in the catalog, with a numbering in the decreasing order of brightness.
We identify the sources based on the core name from \citet{louvet22}, and the core numbering per each protocluster.

We also make use of the spectral line datacubes from ALMA-IMF, in particular the CO (J=2--1), the SiO (J=5--4), and the DCN (J=3--2) transitions. While the CO and SiO lines are used to trace protostellar outflows, DCN is used to estimate a \vlsr for each source individually, similarly as done by \cite{Cunningham2023}. % in which the kinematics of each protocluster is studied. 
The data reduction of the line cubes is presented in detail in \cite{Cunningham2023}. We use the spectral line datacubes with the "JvM" correction \citep{JvM1995} of the residual flux scale. The continuum is subtracted from the datacubes using the  STATCONT procedure \citep{Sanchez-Monge2018}. The SiO (J=5--4) line is covered in our Band~6 setup and spectral window 1 together with the DCN (J=3--2) line. The CO (J=2--1) transition is also in our Band~6 setup and the spectral window 5. We use data only from the ALMA 12~m array configurations. The beam sizes of the linecubes used here are comparable to the continuum ones with a difference up to $35 \%$ depending on the region. The synthesized beam sizes and sensitivity of the linecubes is described in \cite{Cunningham2023}. We summarize the observing parameters for the spectral line datacubes used here in Table \ref{tab:spws}.

As described in \citet{Cunningham2023}, due to bright and extended CO emission in some regions, we masked channels around their \vlsr to avoid cleaning divergences in this spectral window. The G351.77 protocluster still presents strong sidelobes at high velocities in the CO datacube, preventing us from studying the protostellar outflows in this region, overall leaving us with the remaining 14 protoclusters of ALMA-IMF.

\cite{Nony2020} and \cite{Nony2023} used the same CO transition as we use here to distinguish pre- and protostellar cores in the W43 protoclusters. \cite{towner2023} used the SiO\,(J=5-4) transition of ALMA-IMF to search for shocked gas potentially associated with outflows although with no attempt to associate them to their driving sources. \cite{Armante2024} studied the evolved protocluster, G012.80, from ALMA-IMF and they also classified pre- and protostellar cores using the same CO and SiO transitions. A comparison of these studies and our work is presented in Sect. \ref{subsection:comparison_Nony}.

%In addition to the continuum maps, the ALMA-IMF consortium released line cubes which allows to do a full chemical and kinematic study of each protocluster thanks to the lines wealth of the ALMA-IMF data (see Table 2 from \cite{Motte2022}). \cite{Cunningham2023} presents the full data reduction process and data release of the line cubes. The line cubes used are "JvM" corrected (\cite{JvM1995}) which is a correction of the residual flux scale, and their continuum is substracted using the procedure STATCONT (\cite{Sanchez-Monge2018}). Table \ref{tab:spws} presents the specific lines used for this study and their spectral resolution obtained in the data cubes. While CO and SiO lines are used to trace protostellar outflows, DCN is used to estimate a \vlsr for each source individually, as it is done in \cite{Cunningham2023} in which the kinematics of each protocluster is studied. From now on, the Band 6 spw 1 and Band 6 spw 5 datacubes will be refered as the SiO and CO datacubes respectively. 

%-----------------------------------------------------------------
%-----------------------------------------------------------------
%-----------------------------------------------------------------

\section{A systematic method to detect proto-stellar outflows} \label{Method}

Our goal here is to classify cores from ALMA-IMF into protostellar or prestellar cores. To do so, we make use of the CO\,(J=2--1) and SiO\,(J=5--4) lines to perform a systematic search for high-velocity emission. %associated to dense cores identified by ALMA-IMF collaboration in \citet{louvet22}. 
Such emission is likely to correspond to directly ejected and gas entrained by ejected material due to protostellar accretion, and can be thus efficiently used to identify protostellar sources. %High-mass dense cores without outflowing emission are excellent candidates for high-mass prestellar cores. 

\subsection{Detection of high velocity excess emission: On-Off spectra} \label{subsection:on-off_spectra}

\begin{figure*}
    \centering
    \includegraphics[width=0.49\textwidth]{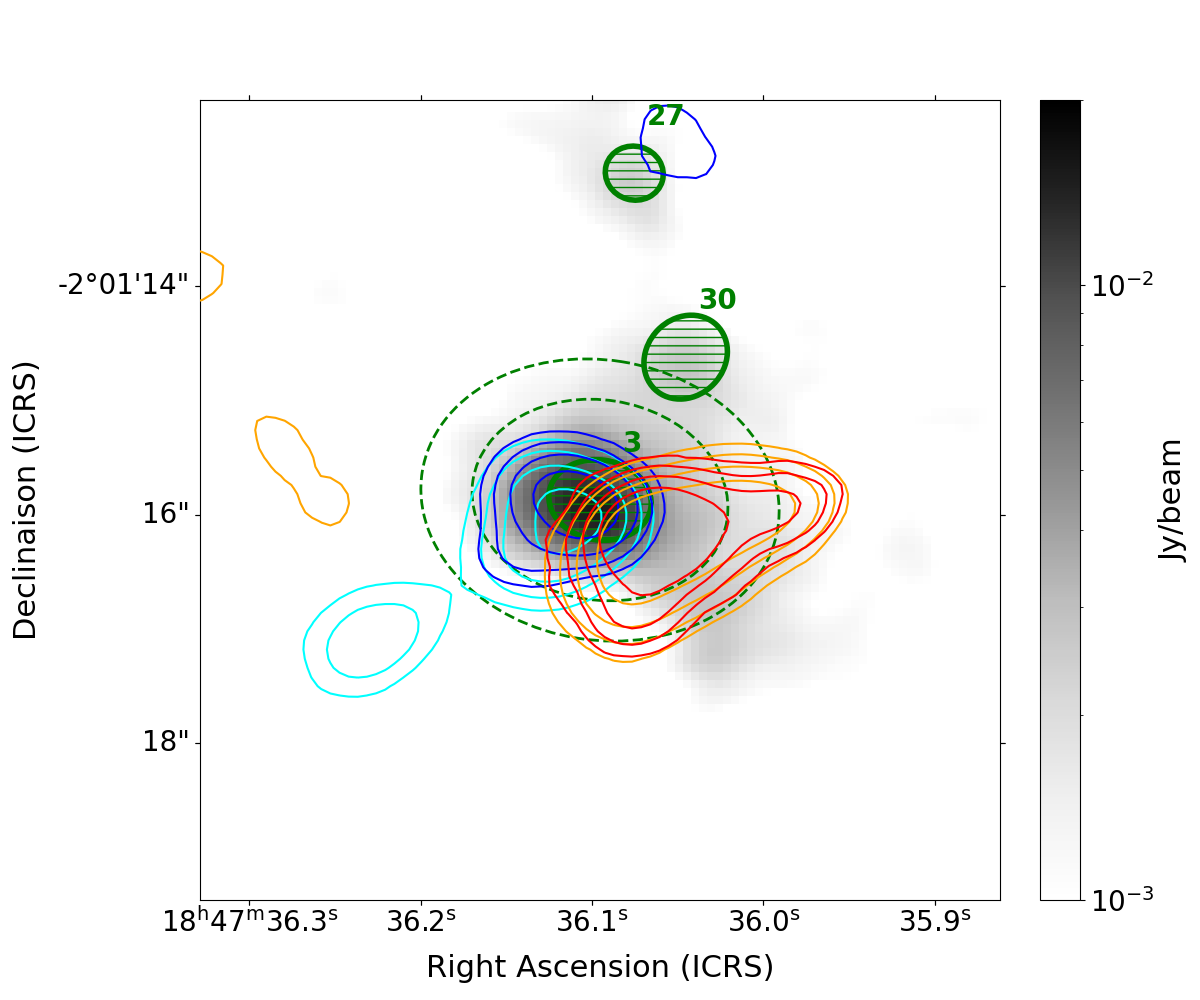}
    \includegraphics[width=0.45\textwidth]{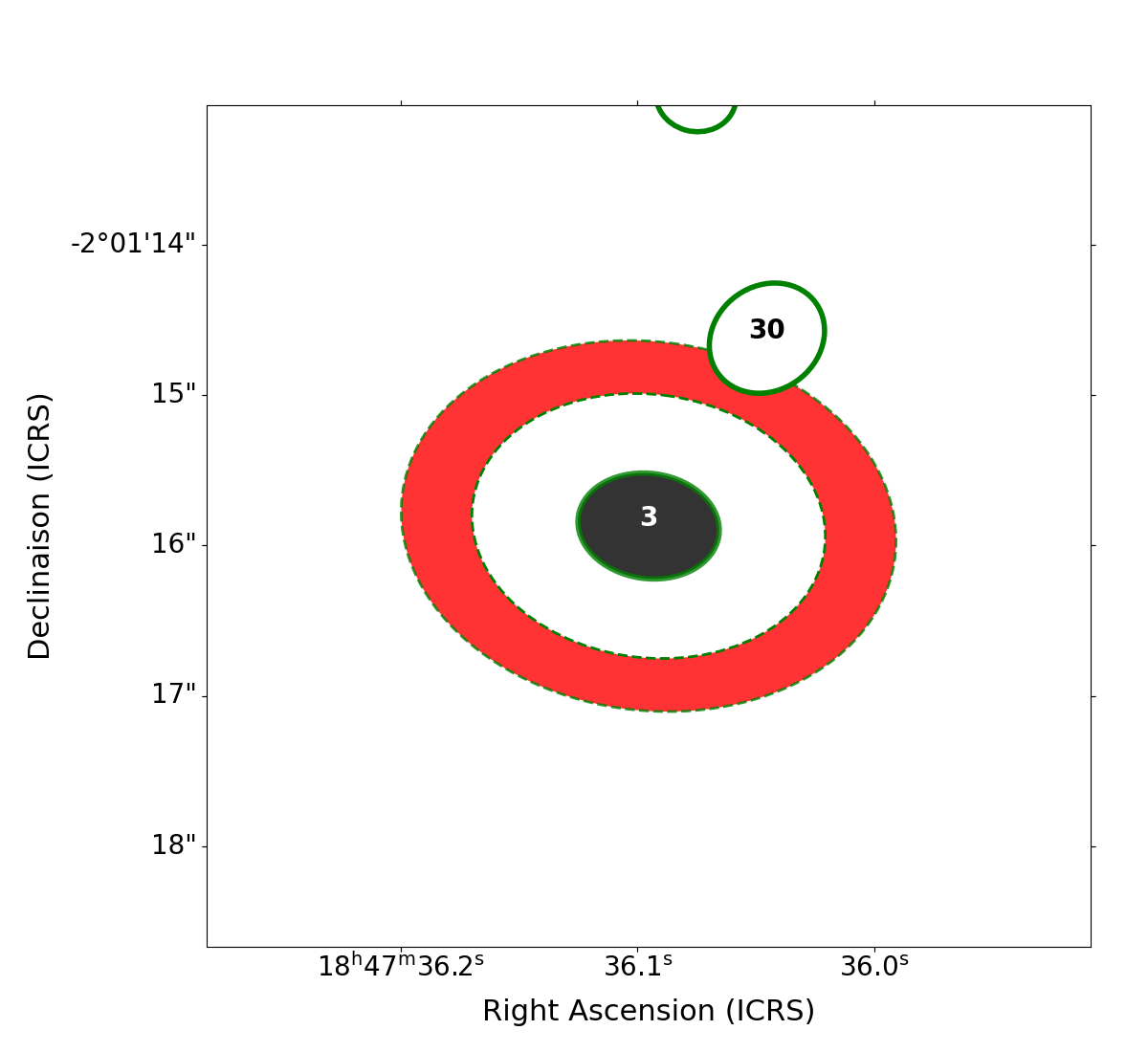}
    \includegraphics[width=0.95\textwidth]{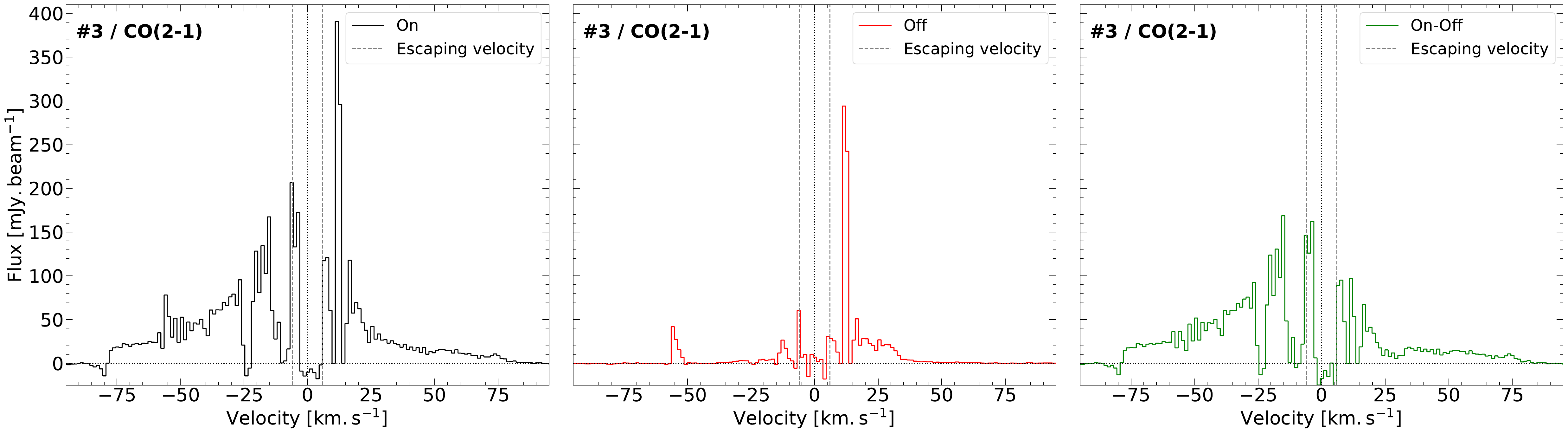}
    \includegraphics[width=0.95\textwidth]{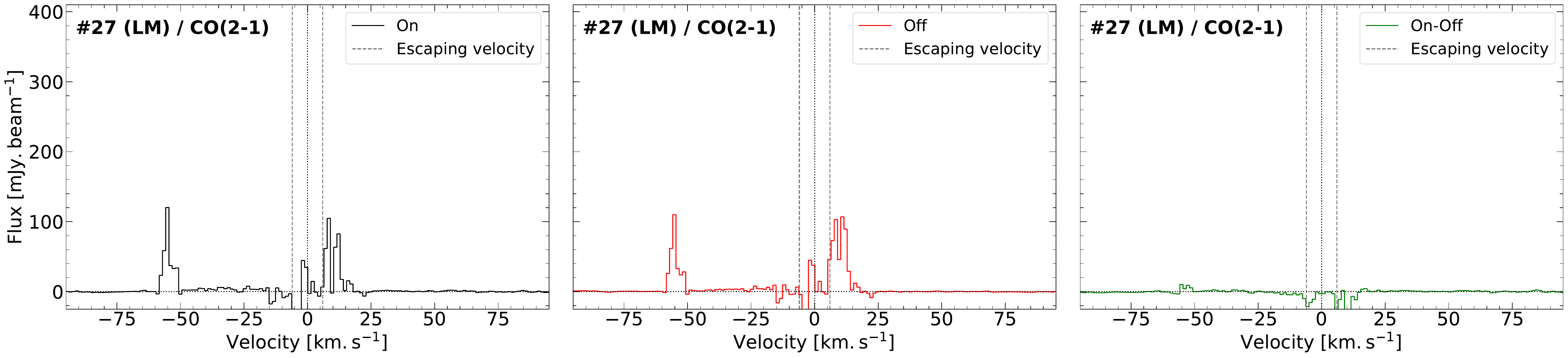}
    \includegraphics[width=0.95\textwidth]{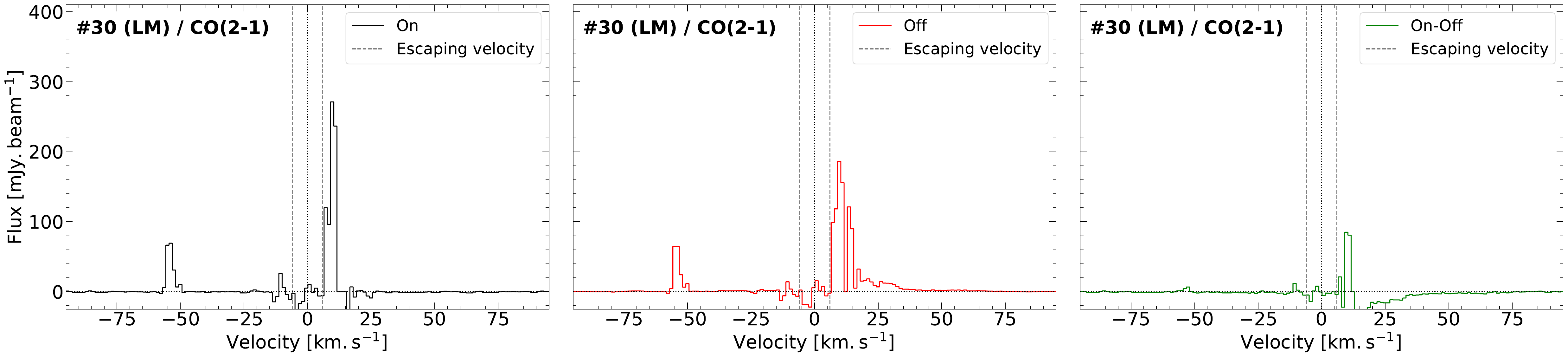}

   % \vskip -0.3cm
    \caption{Overview of the outflow detection procedure using the On-Off spectra presented in Sect.~\ref{subsection:on-off_spectra}. \textbf{Top Left:} Zoom on the continuum core \#3 overlaid on the 1.3mm dust continuum map of W43-MM2 (in grey scale). The green ellipses correspond to the FWHM of the extracted source sizes to the continuum emission by getsf. Colored CO\,(2--1) contours are 10, 20, 40 and 80 in units of $\sigma$, with $\sigma$ = $14.3$, $12.6$, $10.8$, $9.0$ \mJybeamkms for cyan, blue, orange and red contours respectively. The corresponding velocity ranges are $\pm$\,15--23\,\kms for cyan and orange, $-$\,31--42\,\kms for blue and $+$\,31--50\,\kms for red. These velocities have been centered the \vlsr of the core (see text). \textbf{Top Right:} Example of the On (black ellipse) and Off (red annulus) spectra computation used to estimate the core-averaged background-subtracted spectra, here for core \#3. \textbf{Bottom :} Resulting CO(2-1) On (black), Off (red), and On-Off (green) spectra for cores \#3, \#27 and \#30. Core \#3 shows clear line wings, representative of its bipolar outflow that can be analysed in its On-Off spectra. The black and red spectra are computed in the black ellipse and red annulus of the top right panel. The grey dashed lines correspond to the escaping velocity (i.e. the velocity to escape the gravitational orbit) presented in Sect. \ref{subsection:identification_outflow}. The spectra of cores \#27 and \#30 are extracted with the same method. These core do not show any sign of outflow, as expected for core \#30 when looking at the map in the top panel, while the On-Off spectrum of core \#27 do not present any blue line wing.}
    
    \label{On_Off_explanation}
   
\end{figure*}

 We develop an automatic method to systematically identify protostellar outflows by looking for high-velocity excess emission in the CO and SiO spectra. For this purpose, we use the ellipse corresponding to the extracted continuum source, where the major and minor axes of the ellipse are defined as the FWHM of the fitted 2D Gaussian major and minor axes. We then take the mean of the spectra within this region. We use this spectrum as an On source measurement, and compute an Off source measurement, using an annulus between $2.5$ and $3.5 \times$ the FWHM of the on-source ellipse. In this annulus we exclude pixels belonging to another continuum source. See example in Fig.\,\ref{On_Off_explanation} and in App.~\ref{appendix:On_Off_explanation}. With this method, our aim is to measure emission from the ejected and entrained gas from the relatively close vicinity of the source itself. By subtracting the Off measurement from the On, we can search for high-velocity residual emission in the immediate vicinity of the protostellar core, as first introduced by \citet{Bontemps1996} and explained in detail in \citet{Duarte-Cabral2013}. 
 This method relies on the principle that any ejection events should accelerate a significant amount of gas inside the core itself so that an excess of outflowing gas on the core (On spectrum) should be detected compared to the average surrounding (Off spectrum) . 
 %This method relies on the principle that if outflowing emission stopped due to episodic accretion for a time long enough to not present excess in the Off annulus, we should still detect momentum left in the envelope. 

%TCS: J'ai reorganise les prochains deux paragraphes
%In the case of a single, collimated outflow driven by the source, its emission is diluted in the area of the annulus and the excess emission from the source is detected in the On-Off spectrum. In the simplest cases where another outflow is contaminating the source, the On-Off spectrum allows us to detect the outflow from the source.

We compute this differential spectra for all sources with $\rm M > 8~M_{\odot}$ where we used a conservative temperature estimate of 20\,K (see Sect. \ref{subsection:4.1}) from the catalog of \citet{louvet22}. %\timea{This mass-based selection threshold gives us a sample size of ??? sources, where we use} \st{for} both the CO and the SiO lines.}

ALMA-IMF targets several of the most active star forming regions of the Galaxy, and therefore we have several complex areas where many cores and outflows overlap. Therefore, to complement our analysis based on differential spectra, we also produced maps of molecular outflows using moment zero maps to study their spatial distribution, as presented in Sect.~\ref{subsection:outflow_maps}.

 We show in Fig.~\ref{On_Off_explanation} the cores \#3, \#27 and \#30 of the W43-MM2 region as an example. The CO\,(2--1) On-Off spectrum of core \#3 shows an excess emission at velocities offset from the source \vlsr corresponding to line wings.
 This example demonstrates that core \#3 exhibits an excess in high-velocity emission up to $\pm 75$\kms, corresponding to outflowing and ejected gas, and thus can be classified as a protostellar core. On the opposite, the On-Off spectra of cores \#27 and \#30 do not show any sign of outflow. The SiO\,(5--4) spectra of these three cores are shown in App. \ref{appendix:On_Off_explanation}. From now on all the SiO spectra shown are smoothed by a factor of two in spectral resolution purely for visual clarity, but the analysis has been carried out using the full spectral resolution. The corresponding spectra of each PSC candidate identified in this paper are shown in App.~\ref{appendix:MPSC_fig_part}.

%-----------------------------------------------------------------
\subsection{Noise estimates} \label{Noise_estimation_section}
 
The noise distribution of the ALMA-IMF datacubes exhibits some spatial and spectral variations. Spatial variations of the noise distribution for other transitions in the ALMA-IMF has been discussed for example in \citet{Bonfand2024}. The CO (J=2--1) datacube is particular in this context, because channels close to the source \vlsr exhibit a higher noise mostly due to side-lobes from missing short spacings. 

\begin{figure}[htbp!]
    \centering
    \includegraphics[width=0.49\textwidth]{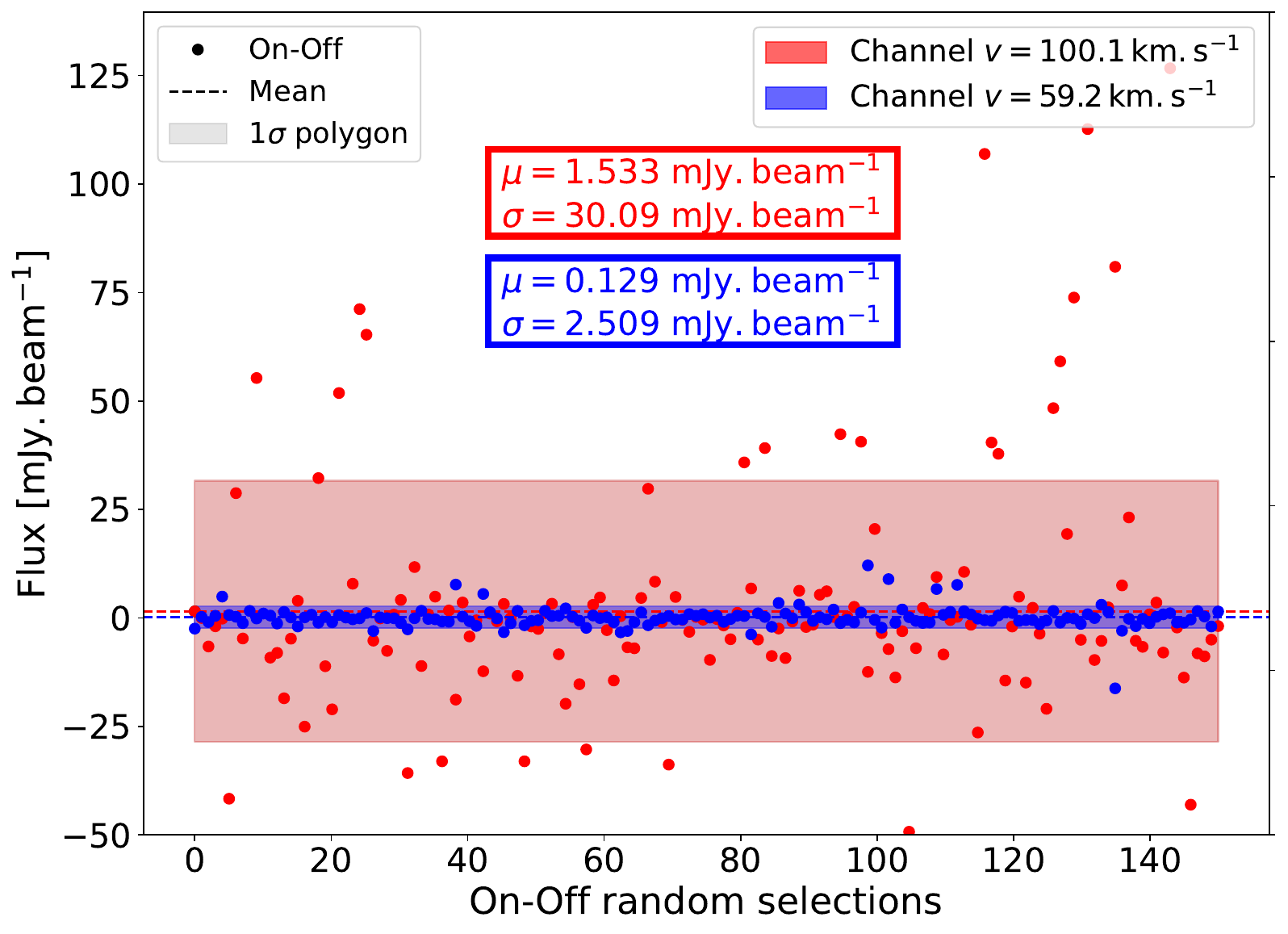}
    \includegraphics[width=0.49\textwidth]{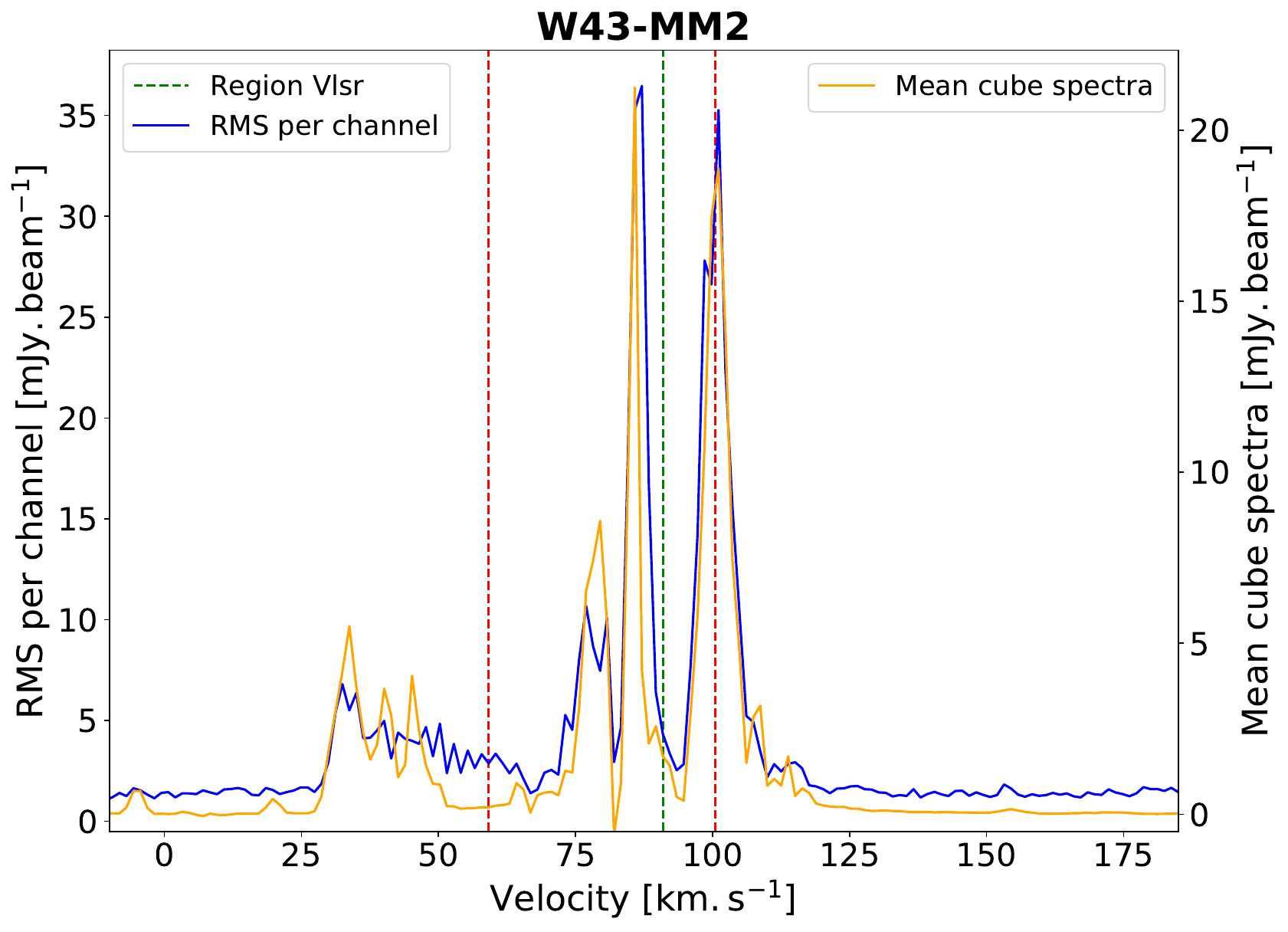}
    \includegraphics[width=0.49\textwidth]{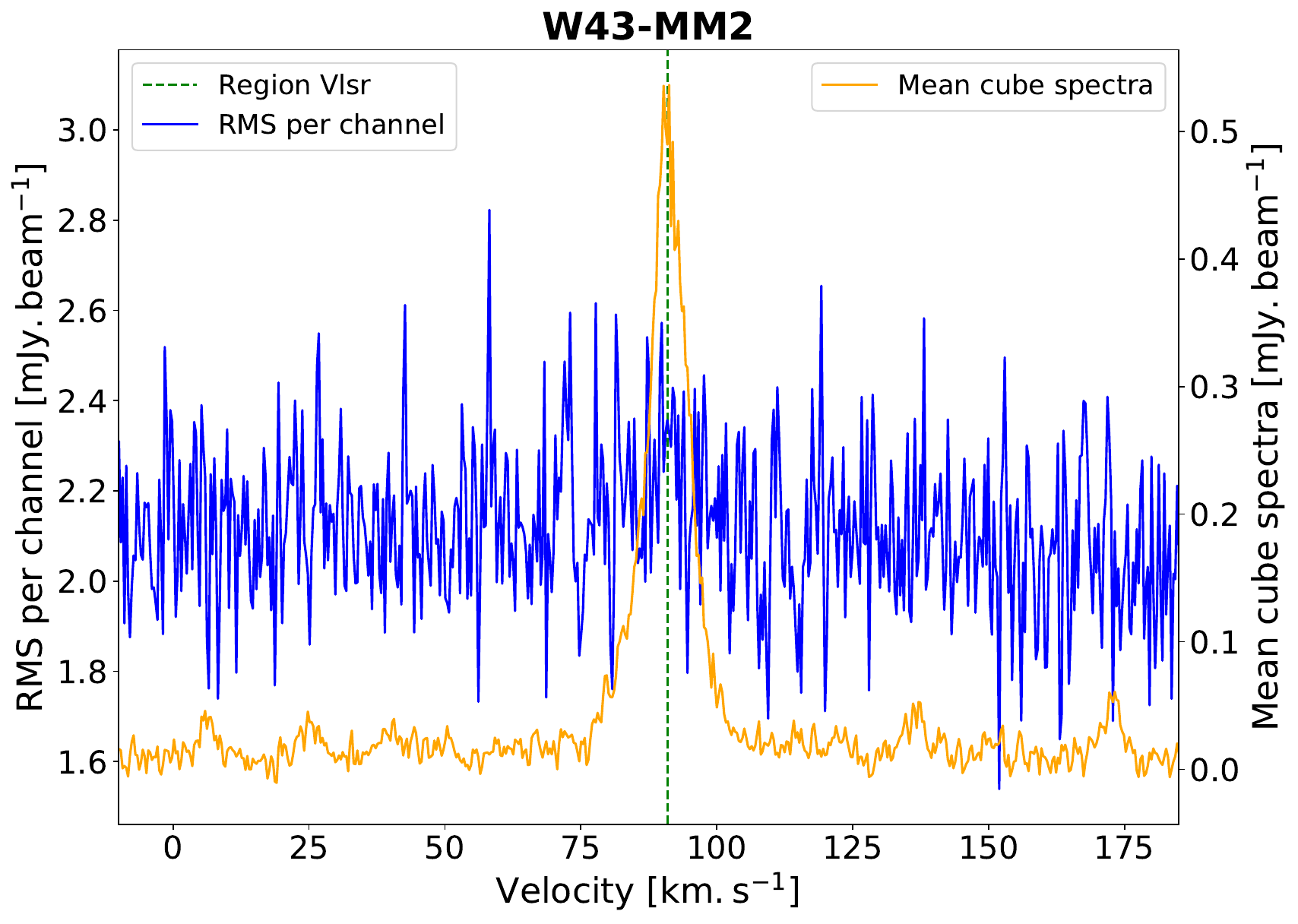}

   % \vskip -0.3cm
    \caption{ \textbf{Top :} Dispersion of the 150 On-Off random selections fluxes in the CO datacube of the region W43-MM2 for channels at velocities of $59.2\,\mathrm{km\,s^{-1}}$ and $100.1\,\mathrm{km\,s^{-1}}$. The mean ($\mu$) and standard deviation ($\sigma$), which is assumed for the noise in a respective channel, are indicated in the panel. \textbf{Middle :} Noise spectra of the CO datacube of the region W43-MM2 (in blue, left y axis) overlaid with its mean cube spectra (in orange, right y axis). The mean cube spectra is computed by averaging the fluxes of all the pixels (except the ones of the edges) at every channel. The green doted line is centered on the \vlsr of the region. The two channels represented in the top panel are spotted with the red dashed lines. \textbf{Bottom :} Same as above for the SiO datacube.
    \label{Noise estimation}
    }
   
\end{figure}

In order to be able to estimate the significance of CO and SiO emission individually in each velocity channels we have estimated the emission noise using the spatial variation of emission in both for the CO and the SiO lines.   %A noise estimation is needed to detect the line wings and characterize them as an excess in the On-Off spectra in an automatic and systematic way. 
For that, we measured the dispersion of the emission for each channel among randomly located On-Off positions. %For that we measured some randomly chosen On-Off spectra to estimate the emission dispersion of each channel inside the collection of spatially different On-Off spectra.
%Our aim is to obtain an estimate of the noise per channel that we use then to build a "noise spectrum". This is used then as a reference to judge the significance of detected emission in the On-Off spectra. 
To do so, we first mask all the pixels from the cores of the ALMA-IMF catalog and the noisy edges of the datacubes. We then compute the On-Off spectrum for randomly placed 150 sources with convolved FWHM sizes between 3000 and 5500 au, a random position angle and eccentricity between 0 and 0.8 for each (see Fig.~\ref{appendix:150_Noise_fig} in App.~\ref{appendix:Noise_estimation}). For each channel, we then extract the mean and standard deviation among the 150 On-Off positions. %We compute for each channel the mean and extract the standard deviation using the 150 On-Off fluxes together. 
We use this standard deviation as the noise in each channel (i.e. RMS per channel, see Fig.\,\ref{Noise estimation}). Placing 150 On-Off random locations allows to account for the spatial variations of the datacube, while doing it for every channel account for the spectral variations.
 
As an example, we show the noise estimation in Fig.\,\ref{Noise estimation} for the W43-MM2 region for two channels (top panel, one with bright and one with weak mean CO emission), and then the noise spectra extracted for both CO and SiO datacubes (middle and bottom panels). The noise spectrum recovers well the fluctuations of emission in the cube (shown on the right y-axis) when we compare the noise over the channels with the mean spectrum of the original datacube (i.e. without masked areas). In Fig.~\ref{Noise estimation} we show our velocity axis centered on the \vlsr of the region. We find that the noise is increasing where the CO emission is bright, i.e. close to the source $\rm V_{LSR}$, and between $25$ and $55\,\kms$ for W43-MM2 where there is a significant foreground CO emission (foreground cloud on the line of sight; see \citealp{Nony2023}, \citealp{Nguyen2017}). 
%As discussed in \citet{Nony2023} this can be due to a foreground CO cloud at these velocities (see also \citealp{Nguyen2017}). 
The noise is globally decreasing for velocities largely offsetted from CO emission at the \vlsr, confirming the interest to estimate such a precise noise variation along the velocity channels.  In Fig.~\ref{Noise estimation}, bottom panel, we see that the typical noise spectrum (for W43-MM2) for SiO is much more uniform than for CO. The mean and standard deviation of the derived noise levels in each of the 14 ALMA-IMF regions, for B6 spw1 and B6 spw5, are shown in Fig.~\ref{appendix:noise_levels_fig} in App.~\ref{appendix:Noise_estimation}.
%In contrast with the CO noise spectrum, the SiO one is stable as expected for this less bright and optically thin line. %The noise is also increasing at low velocities around the \vlsr, due to CO contamination again and the high quantity of low velocities outflows which can be taken into account in the random 150 On-Off. This allows the automated detection of significant excess of flux for each source spectra.
%allowing us to properly probe the significant part of each core spectra.

%-----------------------------------------------------------------

\subsection{\vlsr estimation of dust cores}

In order to define the velocity ranges corresponding to our definition of high-velocity emission, we 
%first 
need precise and systematic estimates of the \vlsr of each dust core. 
%We estimate individual core \vlsr instead of using one single \vlsr for the whole protocluster to be the more accurate possible to identify outflows. 
As shown in \cite{Cunningham2023}, the typical dispersion of the \vlsr for individual sources within our protoclusters can be as high as  12~\kms, meaning that the determination of the \vlsr of individual cores is important to well define the velocity range sufficiently offset from the source \vlsr\!. 

Using the same approach as in \cite{Cunningham2023}, we perform here an independent estimate of the \vlsr of each dense core using the DCN (J=3--2) transition. We extract the On-Off spectrum of the DCN line towards each core in the same fashion as discussed above, and as in \citet{Cunningham2023}, then fit a single component 1D Gaussian. We consider a line detected if the fit has a FWHM larger than the spectral resolution and the measured area is greater than $5 \sigma$. We estimate $\sigma$ by using the noise spectrum presented in Fig. \ref{Noise estimation} on the channels of the fit area. For a discussion on the noise estimation we refer to  Sect.\,\ref{Noise_estimation_section}. If these criteria are not met, we perform the same fitting using the On spectrum.
Finally, if none of the fits are good, we use the mean \vlsr of each region, as the value for the individual core, estimated with the DCN measurements on the cores where the fits can be performed. The DCN (J=3--2) spectra and the respective fit for every prestellar core candidate is provided in App.~\ref{appendix:MPSC_fig_part}. Our \vlsr estimates are in agreement with the estimates of \citet{Cunningham2023} for sources that are classified as exhibiting a single component DCN emission. Compared to the analysis shown in \citet{Cunningham2023}, here we go a step further by fitting also the On spectra.
%, therefore, we recover a systematically measured \vlsr for 20 more cores from the sample of 141 cores presented in Sect. \ref{subsection:complete_list}. 
From the sample of 141 cores presented in Sect. \ref{subsection:complete_list}, we get 136 individual \vlsr, among which 116 are also in \citet{Cunningham2023}, from the fit of 106 On-Off and 30 On spectra, while the region \vlsr is used for the 5 cores without a good DCN detection.

%-----------------------------------------------------------------

\subsection{Identification of outflow emission} \label{subsection:identification_outflow}

From our analysis above, we obtain a spectrum representing the CO and SiO emission for each dust core. As a next step we analyse these spectra to search for emission at high velocities compared to the source \vlsr\negthinspace. We define the velocity range to search for gas that can escape the core gravitational potential by computing the escape velocity for a core mass of $M=50\, M_{\odot}$ and a radius $r$ of 3000\,au as $V_{esc}=\sqrt{\frac{2GM}{r}}$, where $G$ is the gravitational constant. 
%\begin{equation}
%    \centering
%     V_{esc}=\sqrt{\frac{2GM}{r}}
%    \label{v_escape}
%\end{equation}
%where G is the gravitational constant, M is the mass of the core, and r its radius. With a typical radius size of around $3000 \, au$ in the ALMA-IMF catalogs and a typical massive core of $50\, M_{\odot}$, 
We obtain an escape velocity of $5.4 \, \kms$which is then used to define the velocity limits to identify outflowing gas. Since the CO emission from the protocluster is usually broader than this limit we actually adopt a slightly larger value of $>\pm{6}$\,\kms around the \vlsr to search for possible significant outflowing gas emission. This threshold in velocity offset is used systematically for all cores\footnote{The same threshold of 6\,\kms has been adopted in order to stay systematic and for simplicity. For cores with lower masses than 50\,$M_{\odot}$, this velocity threshold could be smaller.}. %(above $5 \sigma$ using the noise estimate discussed in the previous section) outflowing gas emission. 
We consider emission as significant in two steps: if at least two consecutive channels exceed 5$\sigma$ (sigma computed with the RMS per channel presented in Sect.\,\ref{Noise_estimation_section}), then the emission is directly considered as significant; if this first criteria is not reached but 
the sum of the emission in five consecutive channels or less exceed the 5$\sigma$ threshold (sigma still computed with the RMS per channel, taking into account the variations per channel), it is also considered as significant emission. These criteria are considered for both CO and SiO lines, however significant emission in the high-velocity part in one of the two lines is considered sufficient to be considered as an outflow.% over two consecutive channels. %The flux needed for the two channels is then $3 \sqrt{\sigma_i + \sigma_{i+1}}$. %This velocity range was chosen in order to recover both low and high velocity outflows without being limited by the bandwidth of the datacube. \timea{This is somewhat hard to follow, needs to be slightly rephrased.}

% To be considered as significant, we define the criteria that the flux in a channel must exceed $3 \sigma$.

The method described above
%search for significant outflowing gas emission in both CO and SiO in the On-Off spectra 
is the first step of our detection procedure. In the case of crowded regions, outflows from nearby protostars can deposit some momentum on the detected cores, therefore it is necessary to also visually check the spatial distribution of outflowing gas to support our systematic outflow detection method (see next section). For this we have also automatically produced moment zero maps of outflows in both CO and SiO.
%This provides us a first crude estimate on the presence of CO and SiO emission for each source. Due to the complication of the broad CO emission, and source confusion in the most active regions, outflow emission from the sources may overlap. Therefore, we follow-up these results by visually inspecting the distribution of the CO and SiO emission in the most complex regions. 

%-----------------------------------------------------------------

%TCS: not outflows but outflow
\subsection{Molecular outflow maps} \label{subsection:outflow_maps}

\begin{figure*} [htbp!]
    \centering
    \includegraphics[width=1\textwidth]{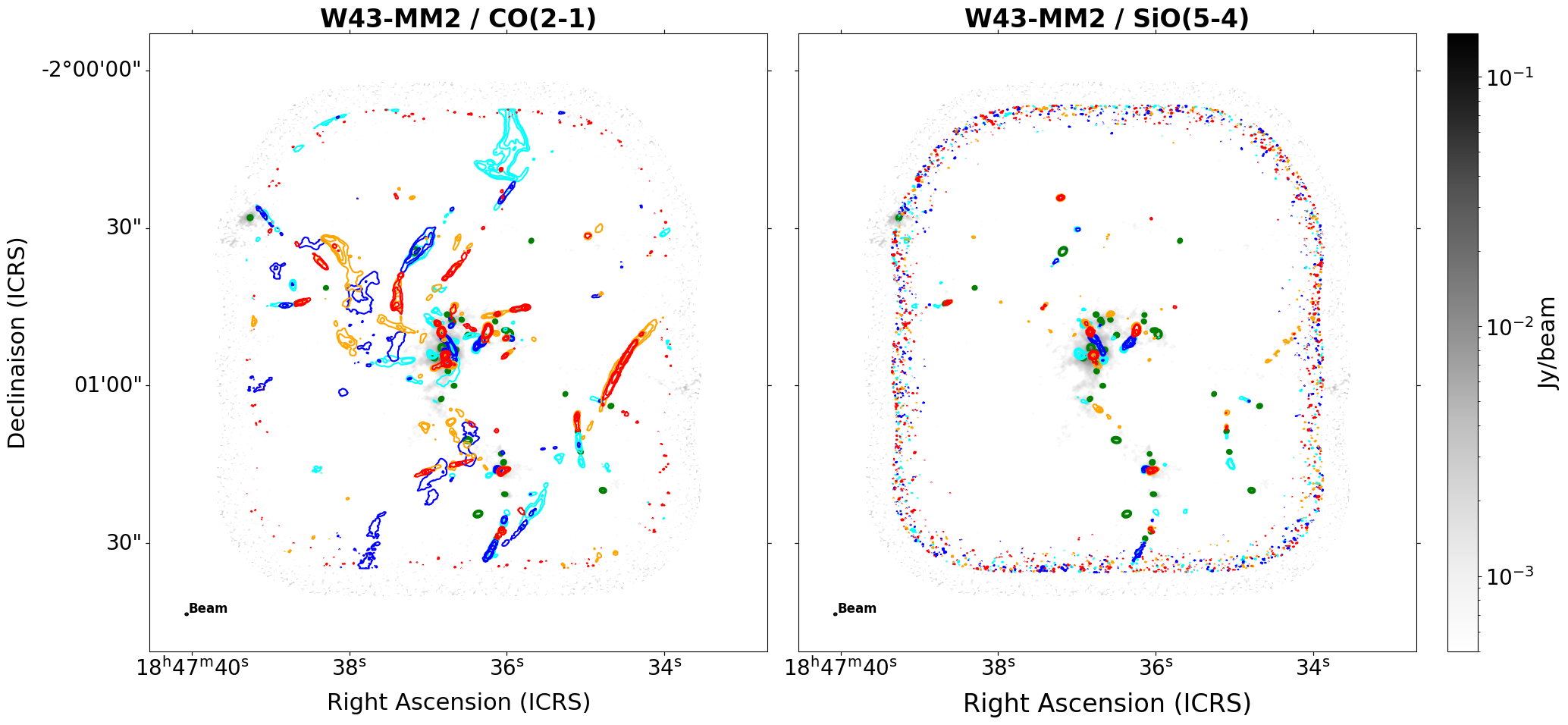}
   % \vskip -0.3cm
    \caption{CO (left) and SiO (right) molecular outflows map of the  W43-MM2 region overlaid on the 1.3mm dust continuum map (in grey scale). The green ellipses represent the FWHM of the continuum cores convolved by the beam size. Moment 0 contours of the blue-shifted wings are overlaid on the continuum map at low velocity in cyan and high velocity in blue. Moment 0 contours of the red-shifted wings are overlaid on the continuum map at low velocity in orange and high velocity in red. CO and SiO contours are 10, 20, 40, and 80 in
    units of $\sigma$, with $\sigma$ = $18.9$, $11.0$, $18.1$, $6.4$ \mJybeamkms for CO, and $\sigma$ = $4.4$, $5.7$, $4.5$, $5.7$ \mJybeamkms for SiO, for cyan, blue, orange and red contours respectively. The corresponding velocity ranges are $\pm$\,15--30\,\kms and $\pm$\,30--45\,\kms for CO and $\pm$\,10--25\,\kms and $\pm$\,25--50\,\kms for SiO.
    \label{W43-MM2_maps_example}
    }
   
\end{figure*}

To complement the On-Off method presented above, we compute moment zero maps of the blue- and red-shifted wings of the CO\,(2-1) and SiO\,(5-4) lines, that are then compared to the continuum map and the position of continuum cores. The visual inspection of maps of high velocity lobes is the usual method to recognise outflows and their driving sources which are placed at the center of the usual bipolar configuration of molecular outflows (see e.g. \citealp{Armante2024}, \citealp{Nony2023}, \citealp{Avison2021}, \citealp{Nony2020}, \citealp{Li2020}, \citealp{Cunningham2016}, \citealp{Duarte-Cabral2013}). Here the velocity ranges of integration are chosen as the best compromise to cover the full CO and SiO emission in the line wings. We try to avoid a too strong contamination from the broad CO emission of the region, while keeping the same velocity ranges (compared to $\rm V_{LSR}$) for the 14 ALMA-IMF protoclusters. We then choose as a reference to have low and high velocity contours at velocity offsets with respect to the \vlsr at $\pm$\,15--30\,\kms and $\pm$\,30--50\,\kms for CO, and $\pm$\,10--25\,\kms and $\pm$\,25--50\,\kms for SiO. These ranges can slightly change depending on the region to avoid some possible strong CO foreground or background cloud contamination. Figure \ref{W43-MM2_maps_example} shows the example for the W43-MM2 region with the CO and SiO molecular outflow maps on the left and right panel, respectively. Some CO contamination is, however, still present in the outflow map, shown by the north blue-shifted low-velocity emission (cyan contours) and the south-east blue-shifted high-velocity emission (blue contours). This emission is not collimated and not associated with a continuum core, therefore it is not considered as outflow emission.
The CO and SiO outflow maps of each high-mass PSC candidate identified from this work (see Sect.~\ref{subsection:4.1} below) are presented in  Appendix \ref{appendix:MPSC_fig_part} with the corresponding velocity range plotted on the spectra and the characteristics %parameters 
for each map in the caption.
The noise for each velocity range to represent the contours is obtained from the noise estimation presented in Sect. \ref{Noise_estimation_section}.

%With the noise estimation presented in Sect.\, \ref{Noise_estimation_section}, we can obtain automatically the sigma for each velocity range to represent the contours. 

%\input{tables/Contours_table.tex}

%-----------------------------------------------------------------
%-----------------------------------------------------------------
%-----------------------------------------------------------------

\section{Results and analysis} \label{ResultsandAnalysis}

\subsection{High-mass cores in ALMA-IMF}
%\subsection{The selection of high-mass cores in the ALMA IMF source list} 
\label{subsection:4.1}

%In order to search for high-mass PSC candidates, we use the core catalog derived by \cite{louvet22} that identified compact dust cores in the ALMA-IMF continuum maps and removed sources with potential free-free contamination. 
We estimate the core masses using the classical formula modified to correct from optically thick emission as in \cite{Pouteau2022} (see their appendix B), which converts the integrated flux ($\rm S_{1.3mm}^{int}$) and peak flux ($\rm S_{1.3mm}^{peak}$) at 1.3 mm  into a mass : 

\begin{equation}
    \centering
    \rm{M_{source} = -\Omega _{beam}^{1.3mm}\, \frac{d^2}{\kappa_{1.3mm} } \, \frac{S_{1.3mm}^{int}}{S_{1.3mm}^{peak}} \, ln\left ( 1-\frac{S_{1.3mm}^{peak}}{\Omega _{beam}^{1.3mm} B(T_d, \nu )} \right )}    
    \label{Mass_computation}
\end{equation}
where $\rm \Omega_{beam}^{1.3mm}$ is the solid angle of the beam at 1.3mm, $\rm d$ is the distance between the region and the Sun, $\rm \kappa_{1.3mm}$ is the dust opacity defined by  $\rm \kappa_{1.3mm} = 0.1(\nu/1000 \, GHz)^{\beta}$~cm$^2$g$^{-1}$ which encompasses a gas-to-dust ratio of 100, at the central frequency $\nu$ of the observations in the Band 6 of each region, with $\beta = 1.5$ chosen to be the same as in \cite{louvet22} and \cite{Pouteau2022}, which is the typical opacity index for cold and dense gas at the core scale (\citealp{Andre1993}, \citealp{Ossenkopf1994}). 
$\rm B(T_d, \nu)$ is the Planck function at the frequency $\nu$ and dust temperature $\rm T_{d}$. 

%TCS: I put a comment on this in the first paragraph with the aimt o limit the repetition.
%We note that the \cite{louvet22} catalog is obtained from the ALMA-IMF continuum "cleaned" images where all free-free contaminated sources have been removed to keep only cores with well determined dust emission to derive their masses. \timea{This is repetition, this has already been said in the beginning.}

To cover most of the potential precursors of high-mass stars, we perform the systematic outflow detection on all cores above 8\Msun using a dust temperature of 20~K.
%To cover all potential precursors of high-mass stars, we perform the systematic outflow detection on all cores above 8\Msun using a dust temperature of 20~K.
%In order to be conservative and to find all PSC candidates in the catalog, we have decided to analyse all cores above 8 $M_{\odot}$ using a single 20K dust temperature for all cores. 
This temperature could be in practice lower for PSCs without internal source as the cooling times are short at such high densities (shorter than free-fall times for an isothermal collapse; see for instance \citealp{Commercon2022} and references therein), and could be as low as  10-15~K as observed in recent claims for high-mass PSCs such as in \cite{Mai2024}, \cite{Barnes2023}, \cite{Morri2023}, or \cite{Sanhueza2017} in which they used NH$_3$ or rotational diagram of $\rm CH_3OH$ lines to derive gas temperatures. At a temperature of 15/10 K, our proposed threshold would then correspond to 12/16 M$_{\odot}$. A total of 141 cores were found above this threshold. The final adopted dust temperatures are further discussed object by object in Sect.~\ref{subsection:basic_properties}
below.

\subsection{Systematic detection of outflows}
\label{subsection:systematic_detection}
%The outflow systematic detection}
%\subsection{The outflow systematic detection}

We use both the On-Off spectra (step 1) and the outflow lobe maps in CO and in SiO (step 2) described above in Sect.\,\ref{Method} to search for outflows %driven by each 
%automated outflow detection tool described in Sect.\,\ref{Method} 
%to systematically identify outflow emission for all 
associated with the 141 cores from ALMA-IMF with M\,$>$\,8\,M$_{\odot}$.

Our primary and first criterium to decide whether a core is driving an outflow is based on the detection of a significant emission in the high velocity ranges in CO and/or in SiO  in the On-Off spectrum towards the core. As a second step then we use the outflow lobe maps to check whether a possible significant detection of an excess (compared to the surroundings as measured in the Off spectrum) of outflowing gas towards the core could not be due to a nearby outflow (from another source). This would deposit some detectable momentum onto the targeted core. This happens often in the crowded regions like the ALMA-IMF central clumps. When such a clear nearby interacting outflow is present we consider to revise the On-Off result leading to a possible non detection of an outflow despite a significant excess on the source. This second criterion is slightly more subjective since it is based on the visual inspection of the maps but is necessary to reject clear cases of a significant influence of nearby outflows.

%We first search for significant outflowing gas emission in the on-off spectra (both in CO and SiO), and analysed the outflow maps for each core to decide whether the core is driving an outflow or not.
%We confirm all the detections and non detections of outflows by inspecting the CO and SiO outflow lobe maps to sort out the possible mis-classifications when for instance a core lies in the middle a strong outflow lobe coming clearly from another core. %criterionusing of the spectra on the source, as well as information from the spatial distribution of the CO and SiO emission to decide whether or not each of the 145 cores drive an outflow. We consider that a core does not drive an outflow if: 
In practice to select potential prestellar core candidates not driving an outflow we adopt the following two steps:
\begin{itemize}
    \item The CO and SiO spectra show no excess above 5\,$\sigma$ in the blue- or red-shifted high-velocity ranges as defined in Sect.~\ref{subsection:outflow_maps} ; 
    \item If an excess is detected in the On-Off spectra onto the source, there is a nearby outflow driven by another source which convincingly explain this excess. %There is no nearby outflow driven by another source which would convincingly explain all the observed excess in the On-Off spectra onto the source. 
\end{itemize}

\begin{figure*} [htbp!]
    \centering
    \includegraphics[width=0.49\textwidth]{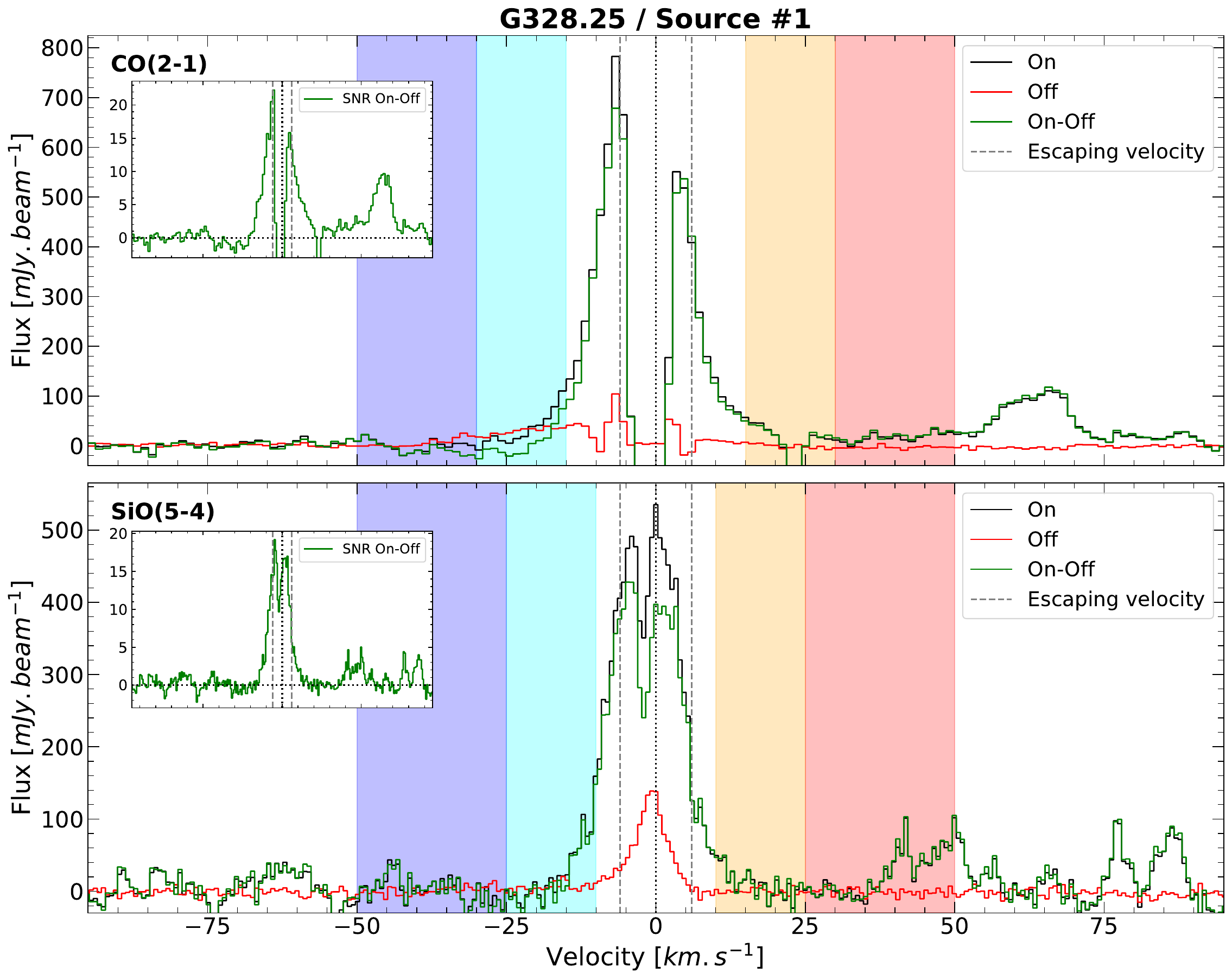}
    \includegraphics[width=0.49\textwidth]{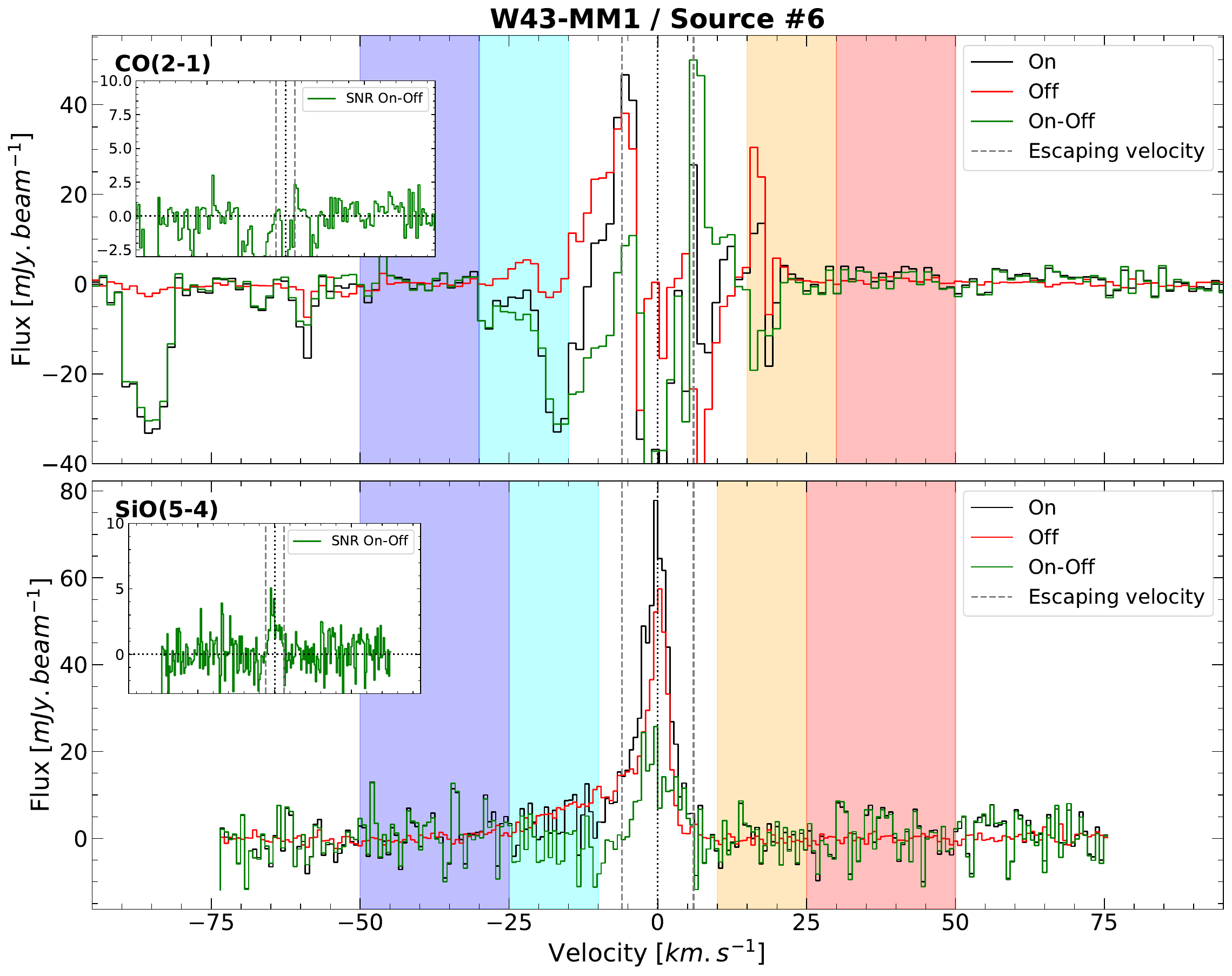}
    \includegraphics[width=0.49\textwidth]{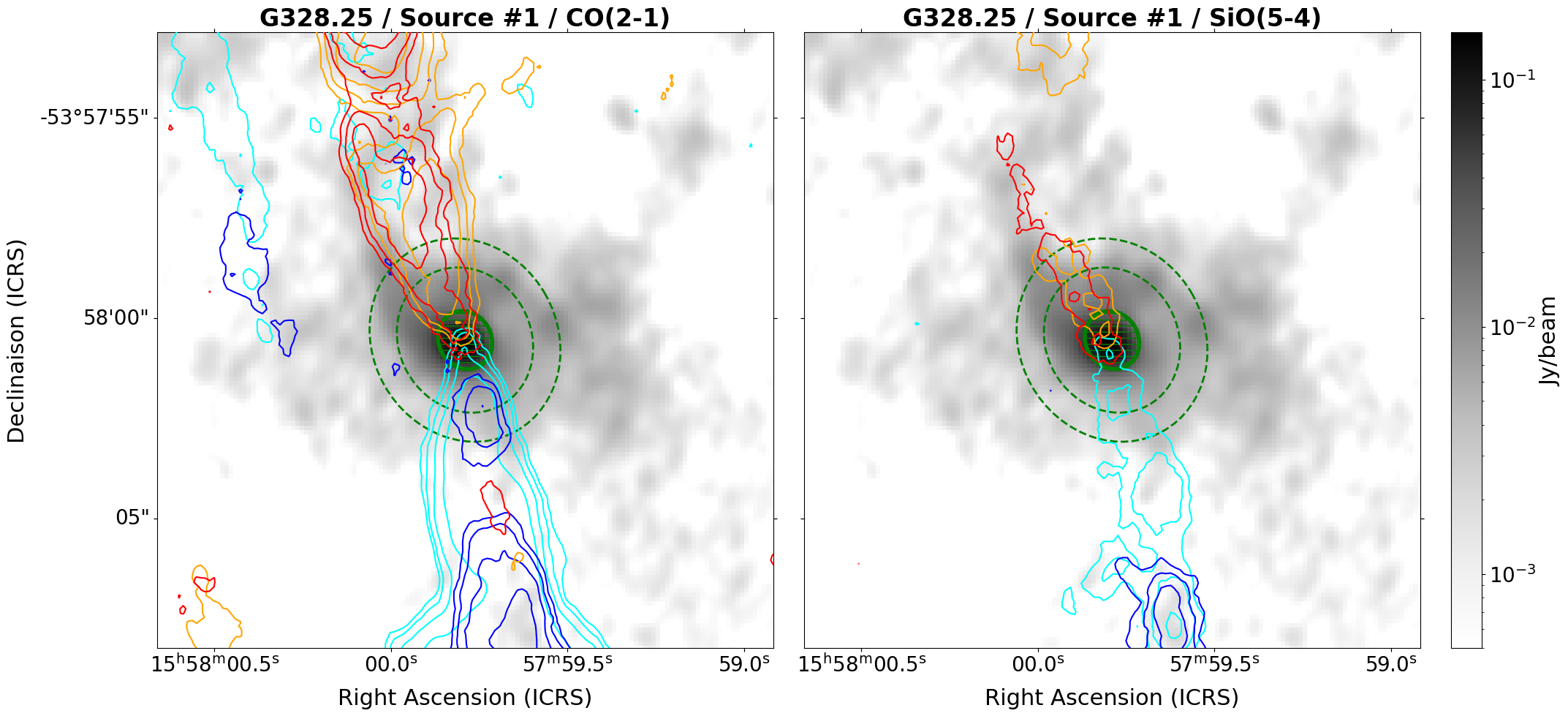}
    \includegraphics[width=0.49\textwidth]{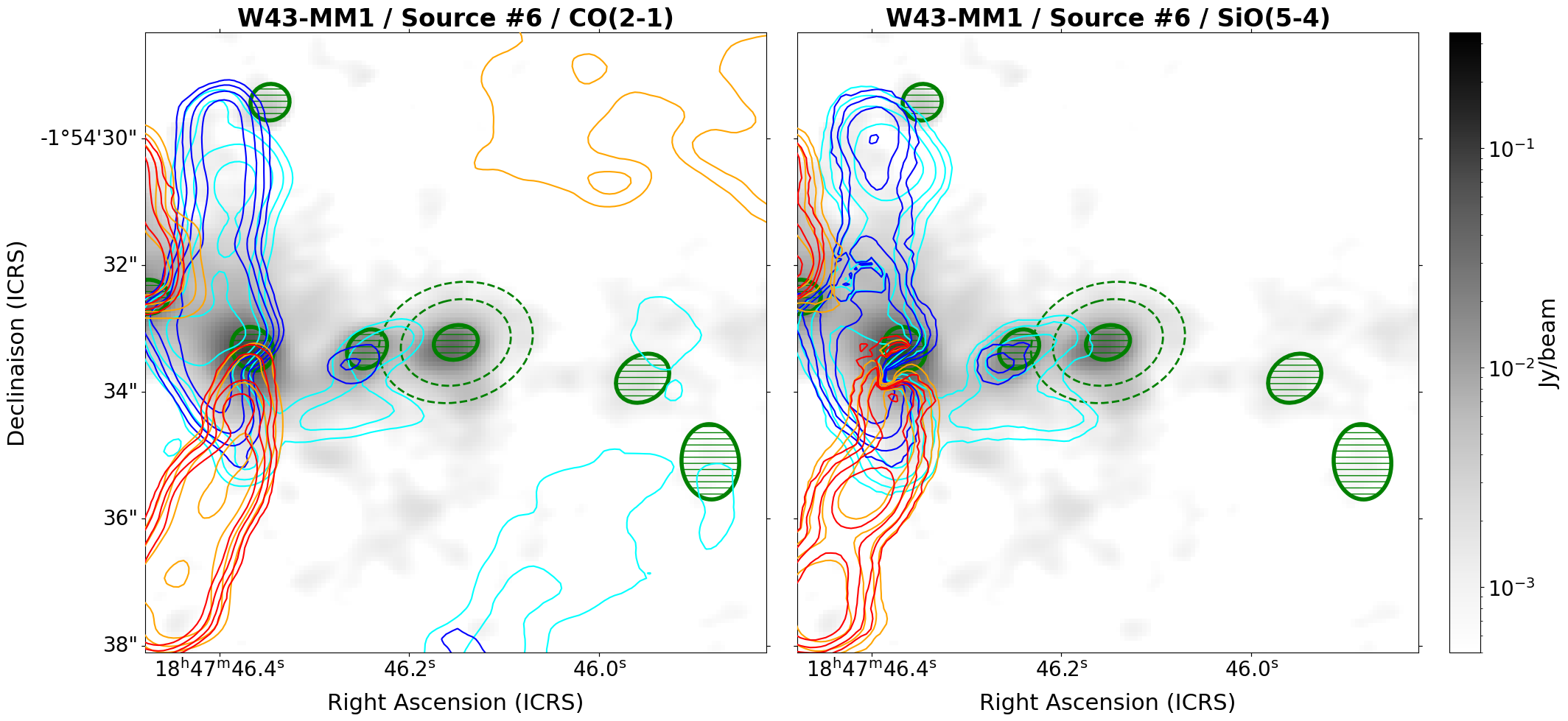}

   % \vskip -0.3cm
    \caption{ \textbf{Top Left:} CO and SiO spectra of the protostellar source \#1 of the G328.25 region. On, Off, and On-Off spectra are the black, red, and green spectra respectively. The small subplots on the top left of each spectra are the SNR for On and On-Off spectra, revealing the significant part of each one. The colored spaces on the spectra are the velocities ranges used to make molecular outflows maps. \textbf{Bottom Left:} Zoom on the \#1 of the G328.25 region driving a bipolar outflow. Contours colors correspond to the same as in figure \ref{W43-MM2_maps_example}. CO and SiO contours are 5, 10, 20, and 40 in units of $\sigma$, with $\sigma$ = $60.1$, $60.4$, $61.9$, $62.7$ \mJybeamkms for CO, and $\sigma$ = $38.9$, $49.0$, $38.1$, $49.6$ \mJybeamkms for SiO, for cyan, blue, orange and red contours respectively. \textbf{Top Right:} Same as top left panel but for the prestellar source \#6 of the W43-MM1 region. \textbf{Bottom Right:} Same as the bottom left panel but for the prestellar source \#6 of the W43-MM1 region. CO and SiO contours are 10, 20, 40, and 80 in units of $\sigma$, with $\sigma$ = $32.6$, $18.6$, $49.5$, $15.3$ \mJybeamkms for CO, and $\sigma$ = $6.4$, $7.4$, $7.3$, $7.2$ \mJybeamkms for SiO, for cyan, blue, orange and red contours respectively.}
    \label{Pro_vs_Pre_example}
\end{figure*}

In Fig. \ref{Pro_vs_Pre_example} we show an example of a protostellar core and a prestellar source
%make the difference between prestellar and protostellar cores 
using both CO and SiO spectra and maps. These sources are representative of the simplest cases where there is no outflow confusion. The left panel shows the protostellar core \#1 of the G328.25 region, known to be a hot core precursor \citep{Csengeri2019, Bouscasse2022}, where both the CO and SiO spectra show line wings representative of its bipolar outflow \citep[e.g.][]{Csengeri2018}, clearly driven by the central source. The right panel confirms the prestellar nature of the core \#6 of W43-MM1 as proposed by \cite{Nony2018}. Neither the CO, nor the SiO spectra exhibit emission in the line wings, and no outflow is driven from the core shown in the CO and SiO maps.

%Using the method presented in Sect.\,\ref{Method}, we searched for all outflows associated with the 145 selected cores. 
As expected, a large number of the 141 targeted cores are found to drive outflows. However, 42 of these cores are found with no significant sign of outflowing activity, which are then considered to be PSC candidates. The CO and SiO outflow maps of the 12 most massive of these 42 PSC candidates are displayed in Fig.~\ref{12_massive_cores_outflows_highlighted}. %(see Sect.~\ref{} for details)
In the different panels of Fig.~\ref{12_massive_cores_outflows_highlighted}, centered on these 12 high-mass PSCs, we see a number of cores associated with outflow lobes qualifying them as protostellar cores. On the other hand, at the center of these maps we see each PSC candidate that we propose showing no sign of outflow driven by these cores. %Most of these examples of dust cores lacking outflow detections are easily identified in the outskirts of the crowded central regions, however, they mostly coincide with cores with masses lower than 8\,M$_{\odot}$. Taking core \#6 of W43-MM1 in this figure as an example, we can clearly identify a nearby high mass protostellar core driving a bipolar outflow, and also continuum sources that present no outflow emission such as the central high-mass PSC candidate. These cores without an outflow detection are therefore good candidates to be prestellar in nature.

We note that outflows mostly in the plane of the sky could be missed when searching for outflowing gas at velocities projected on the line of sight larger than 6\,$\rm km\,s^{-1}$. For a typical maximum, de-projected velocity of the outflow of 50\,$\rm km\,s^{-1}$ (see typical outflows in Fig.\,\ref{12_massive_cores_outflows_highlighted}), and assuming uniform distributions of the opening angles of the outflow from 0 to 30$^\circ$ and of the inclination angles between 0 and 90$^\circ$, we obtain a fraction of non detectable (above 6 km/s) outflows of 0.0175 (1.75\,\%). This fraction represents less than one missed outflow among the 42 PSC candidates. Even with more extreme values with the maximum velocity at 30\,$\rm km\,s^{-1}$ and the maximum opening at 20$^\circ$ the fraction of missed outflows stays relatively low at 7.35\,\%, leading to typically 3 (1) missed outflows among the 42 PSC candidates (12 HM PSC candidates) respectively.

Finally we note that in crowded regions, where the confusion between outflow lobes is large, at least in projection, the adopted methodology may result in some PSCs being miss-classified as protostellar cores. It is difficult to evaluate to which extend this happens but we argue that the On-Off approach is the only systematic and homogeneous way to identifiy the most probable driving sources. 

%Applying this systematic classification, we identify outflow lobes, and 

%In addition to the large number of outflows associated with most of the 145 targted cores, we find several cores without an outflow detection which are therefore good candidates to be prestellar in nature. While a number of least massive cores are found without outflow are situated in the more isolated environment, we see that we also identify several high-mass cores without outflow emission which then tend to be more in the crowded central regions. %These are often situated in the crowded central regions of protoclusters close to well identified outflows, they are, however, unlikely to be the driving source to any of the outflow lobes.  
%We identify these sources as the best candidates to be high-mass prestellar cores. In Fig.~\ref{12_massive_cores_outflows_highlighted} we show an overview of 11 sub-regions from ALMA-IMF that contain the most massive PCS candidates.

\begin{figure*} [htbp!]
    \centering
    \includegraphics[width=0.49\textwidth]{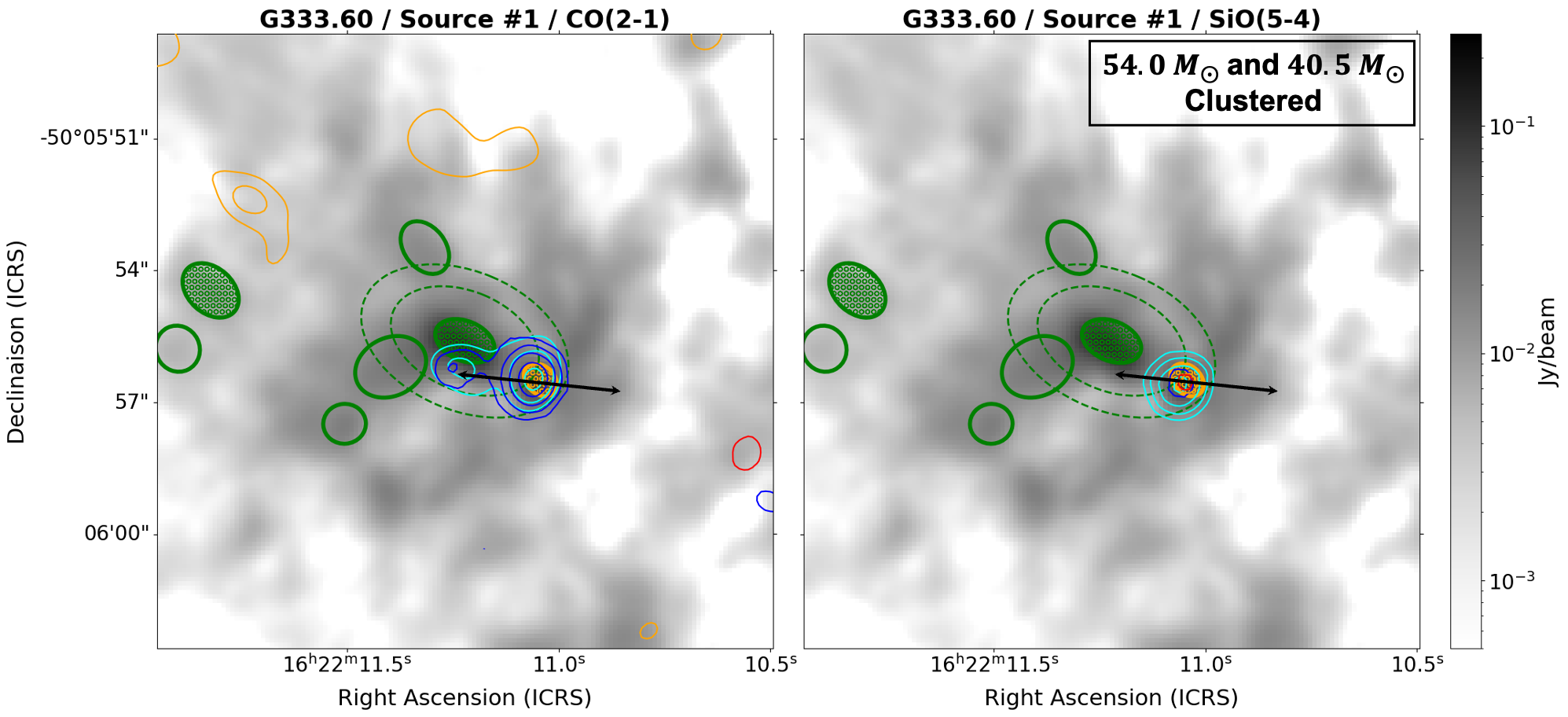}
    \includegraphics[width=0.49\textwidth]{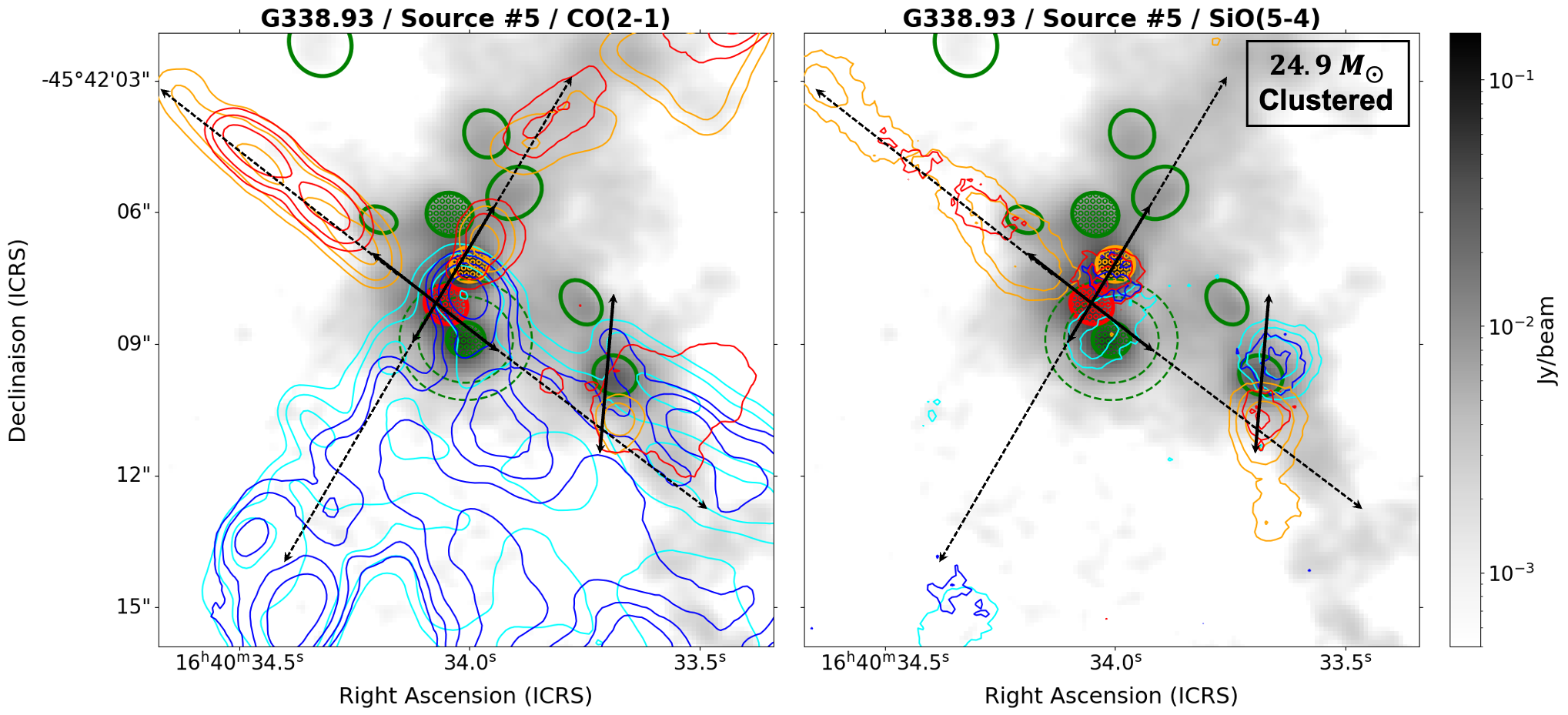}
    \includegraphics[width=0.49\textwidth]{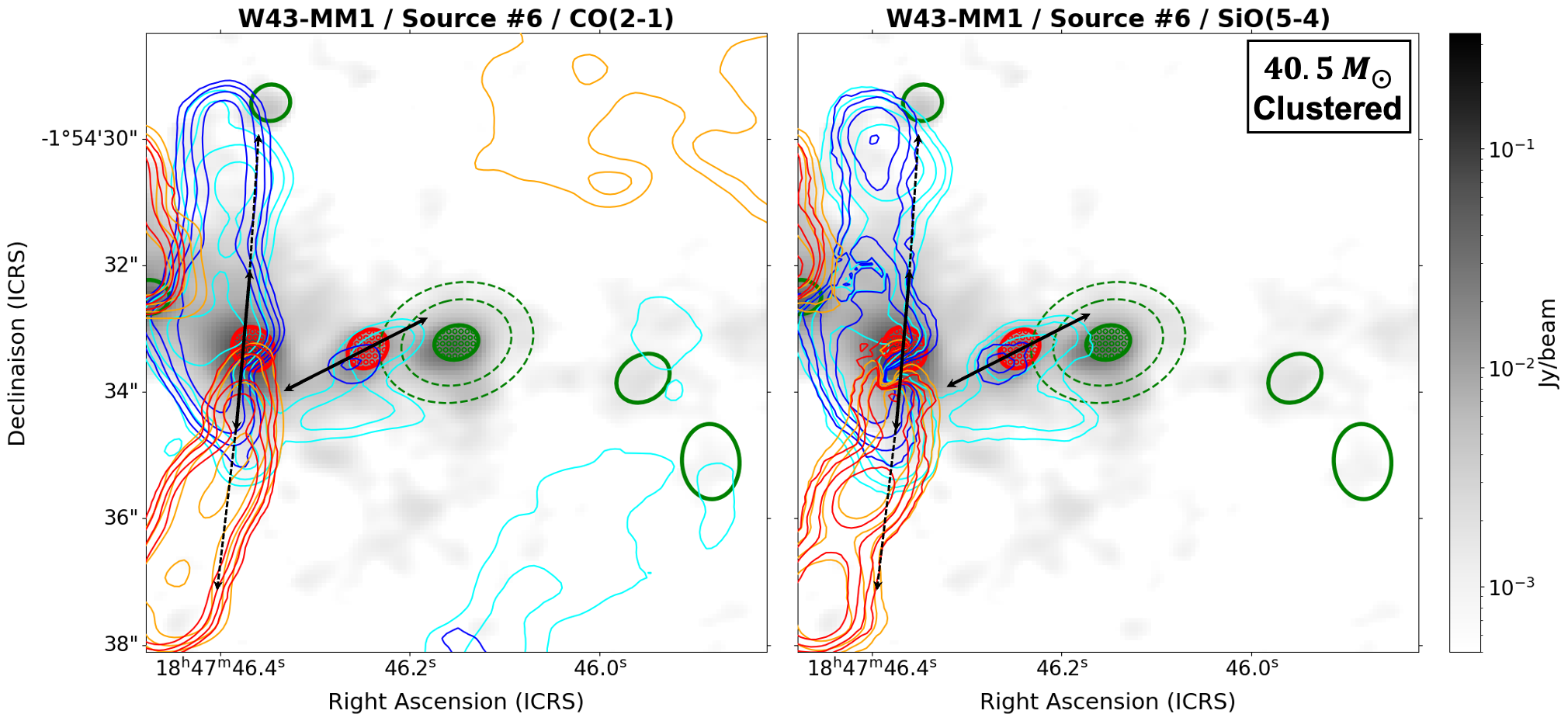}
    \includegraphics[width=0.49\textwidth]{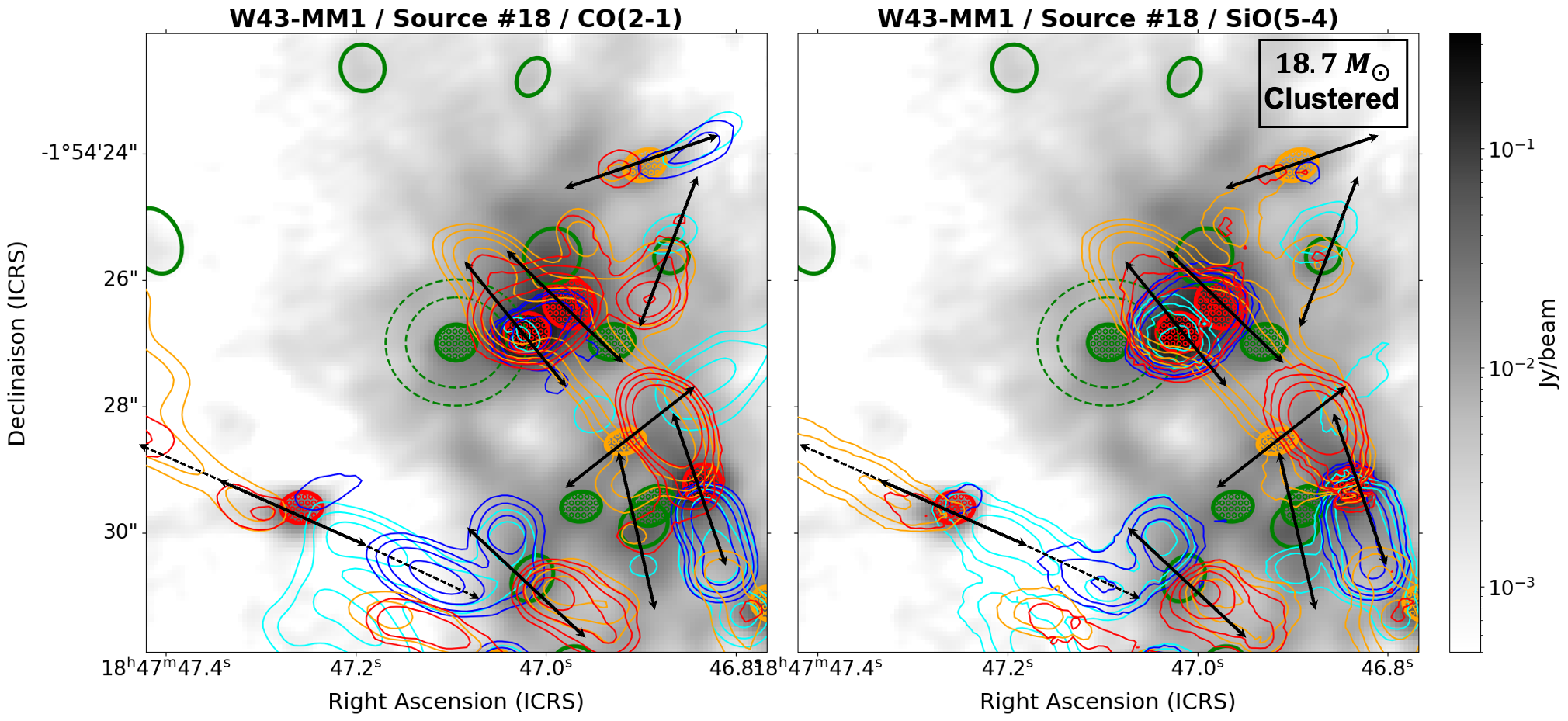}
    \includegraphics[width=0.49\textwidth]{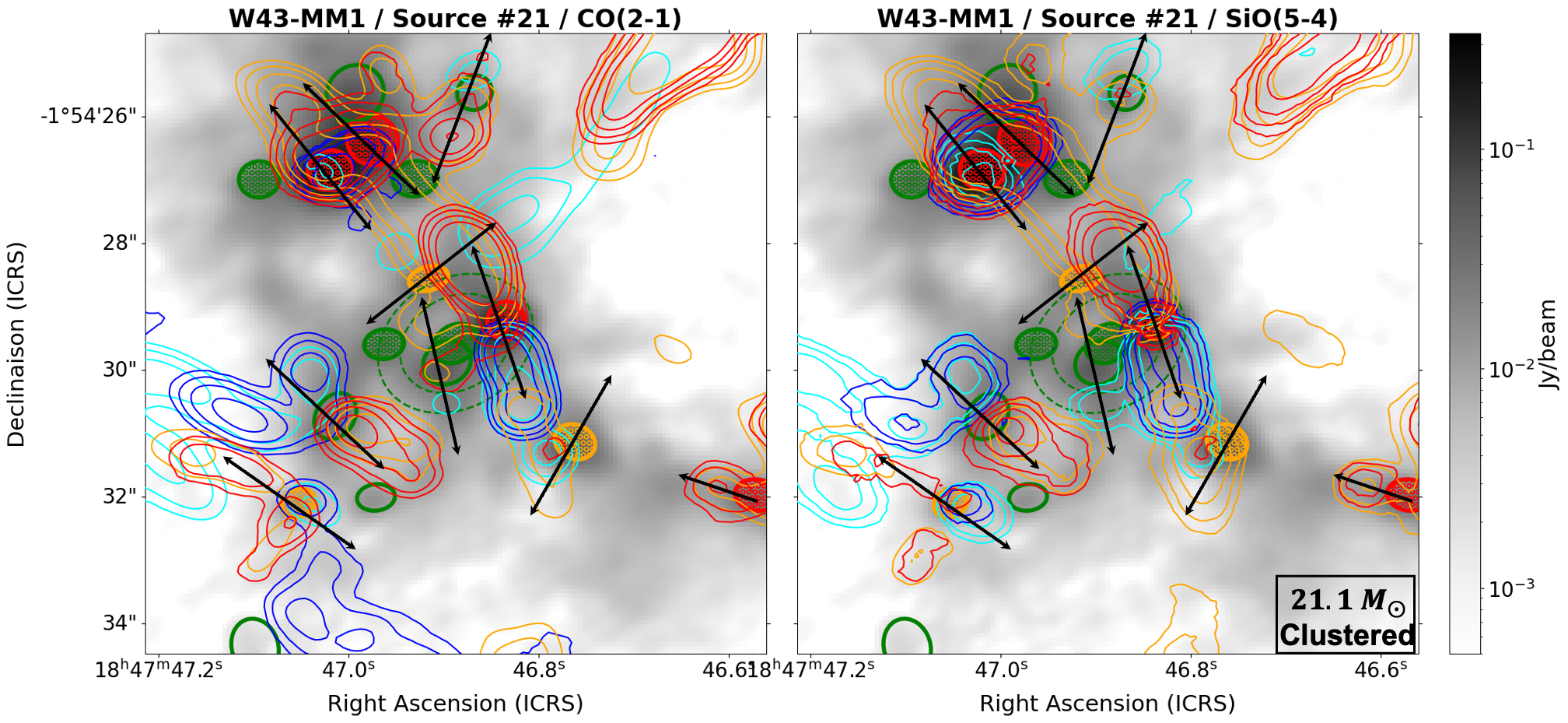}
    \includegraphics[width=0.49\textwidth]{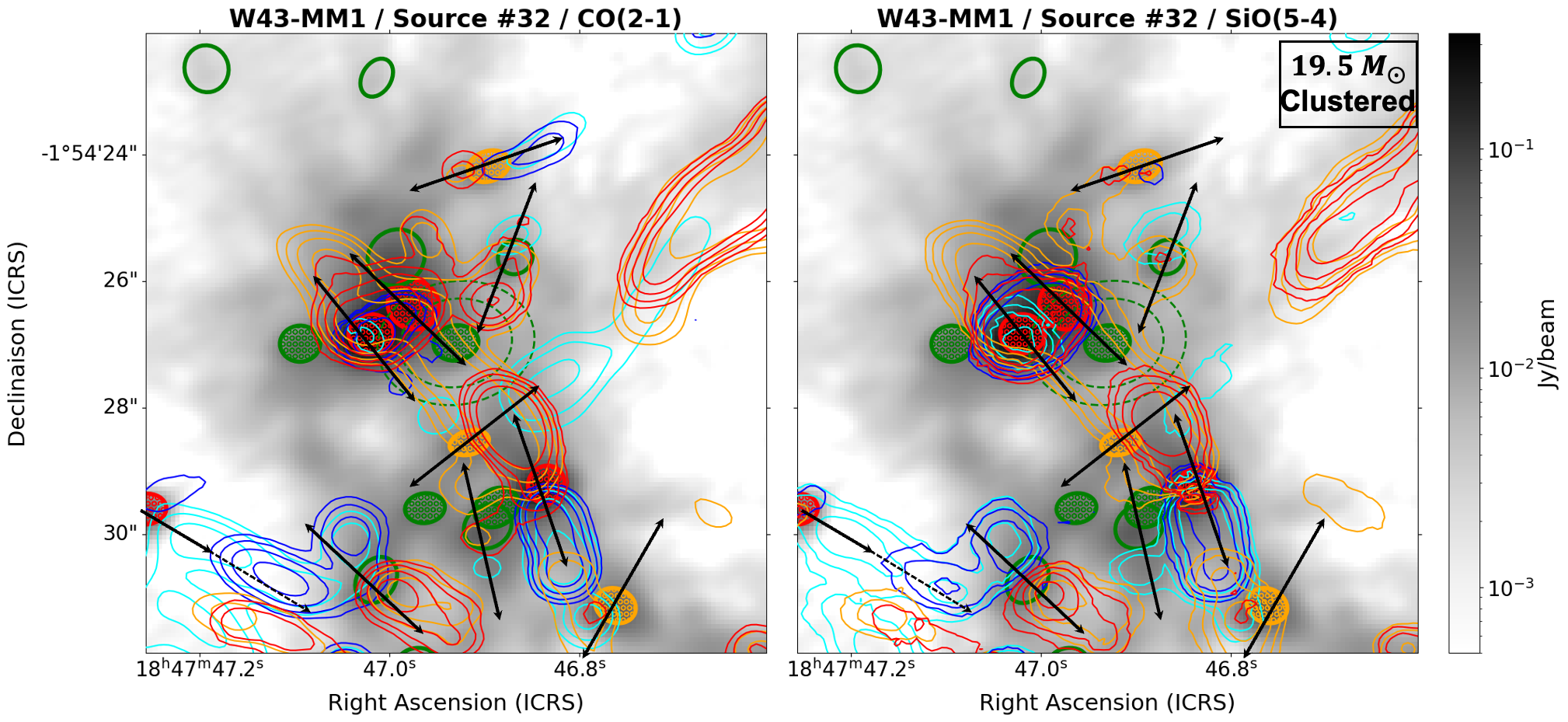}
    \includegraphics[width=0.49\textwidth]{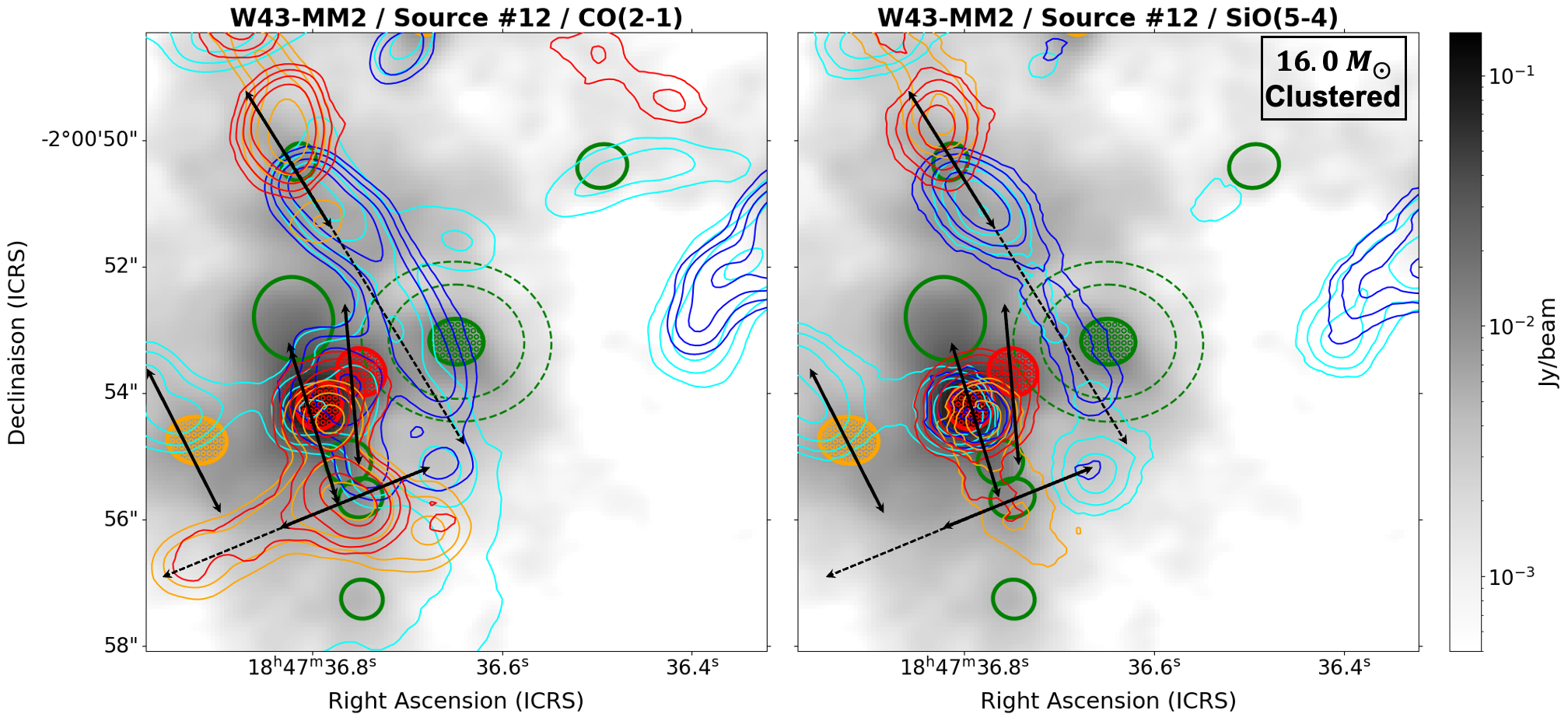}
    \includegraphics[width=0.49\textwidth]{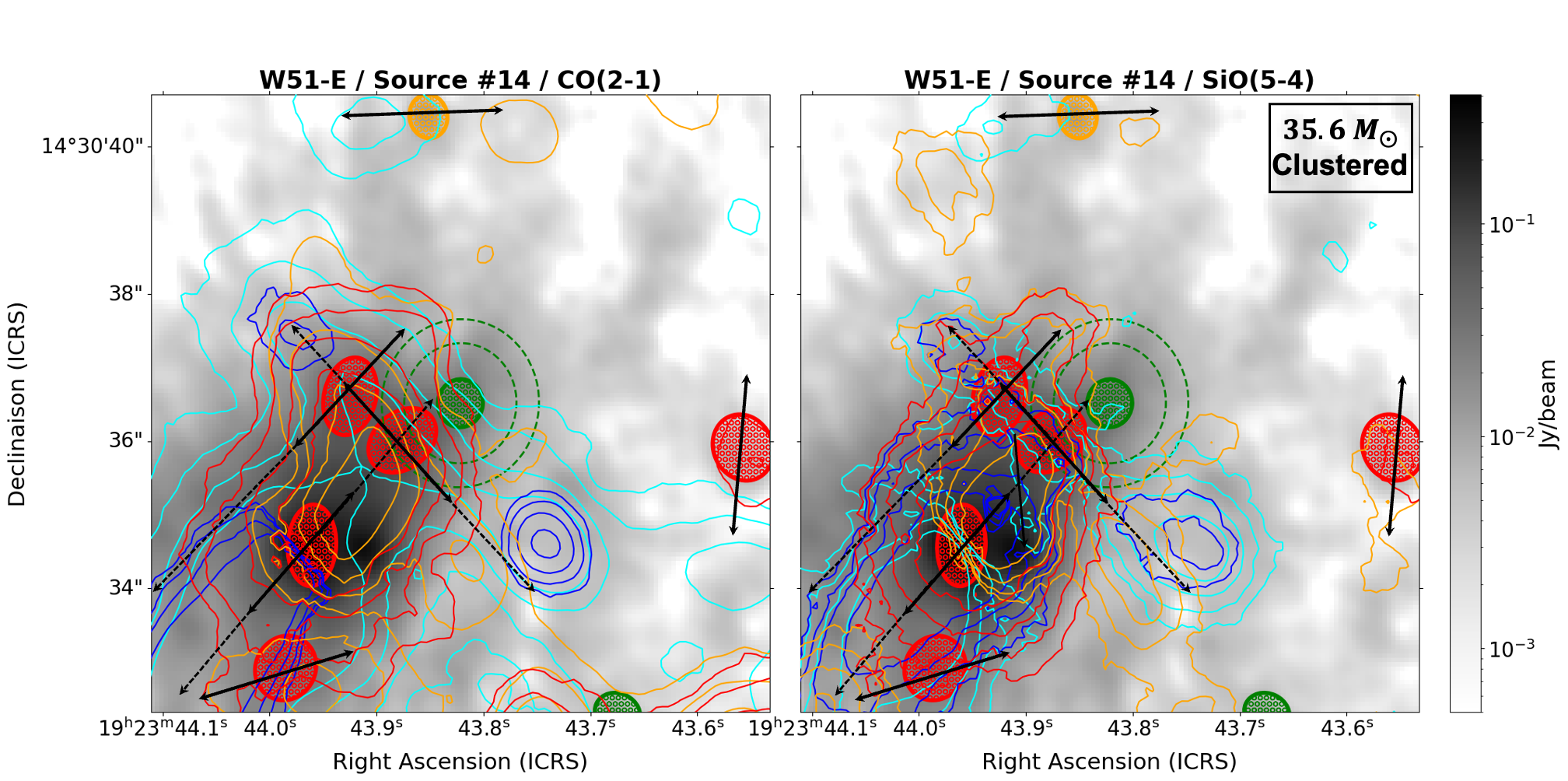}
    \includegraphics[width=0.49\textwidth]{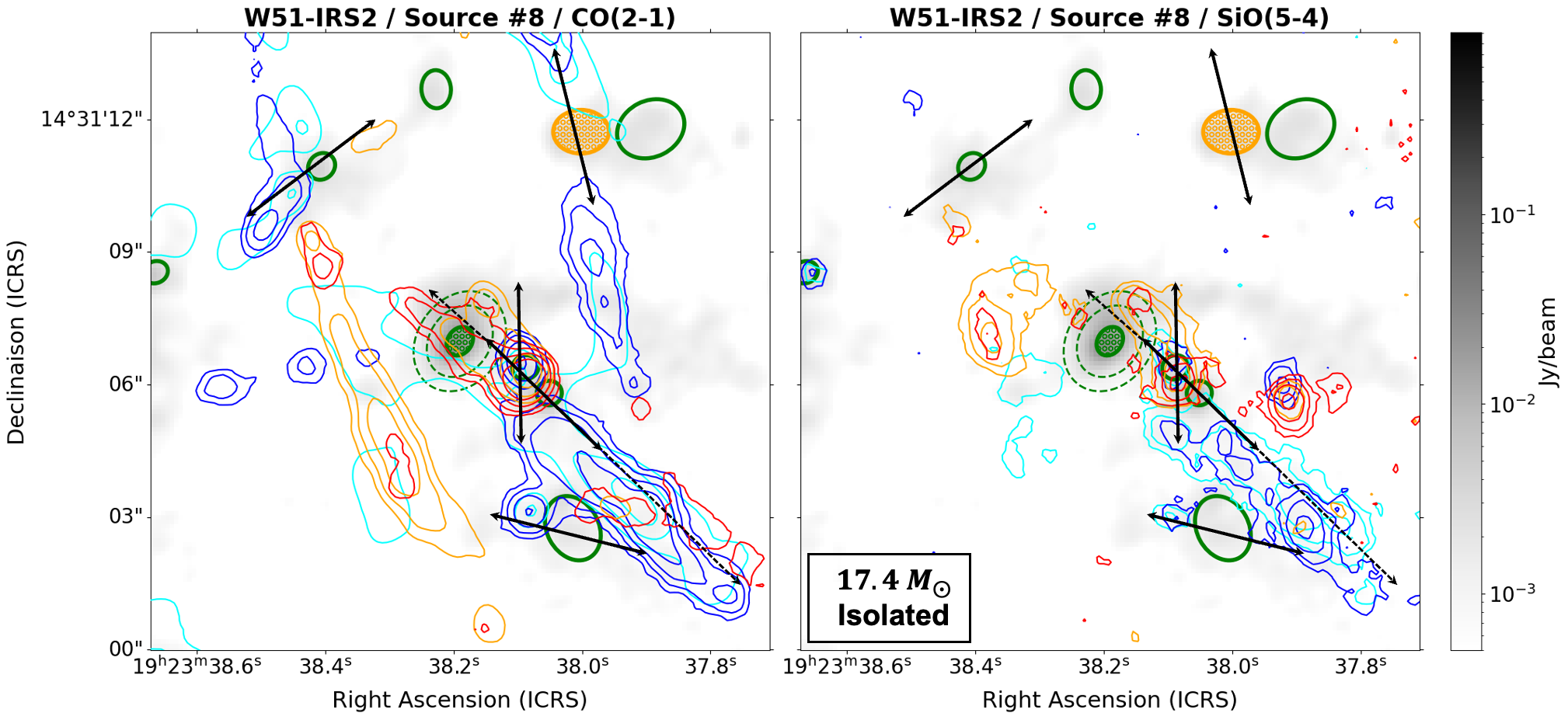}
    \includegraphics[width=0.49\textwidth]{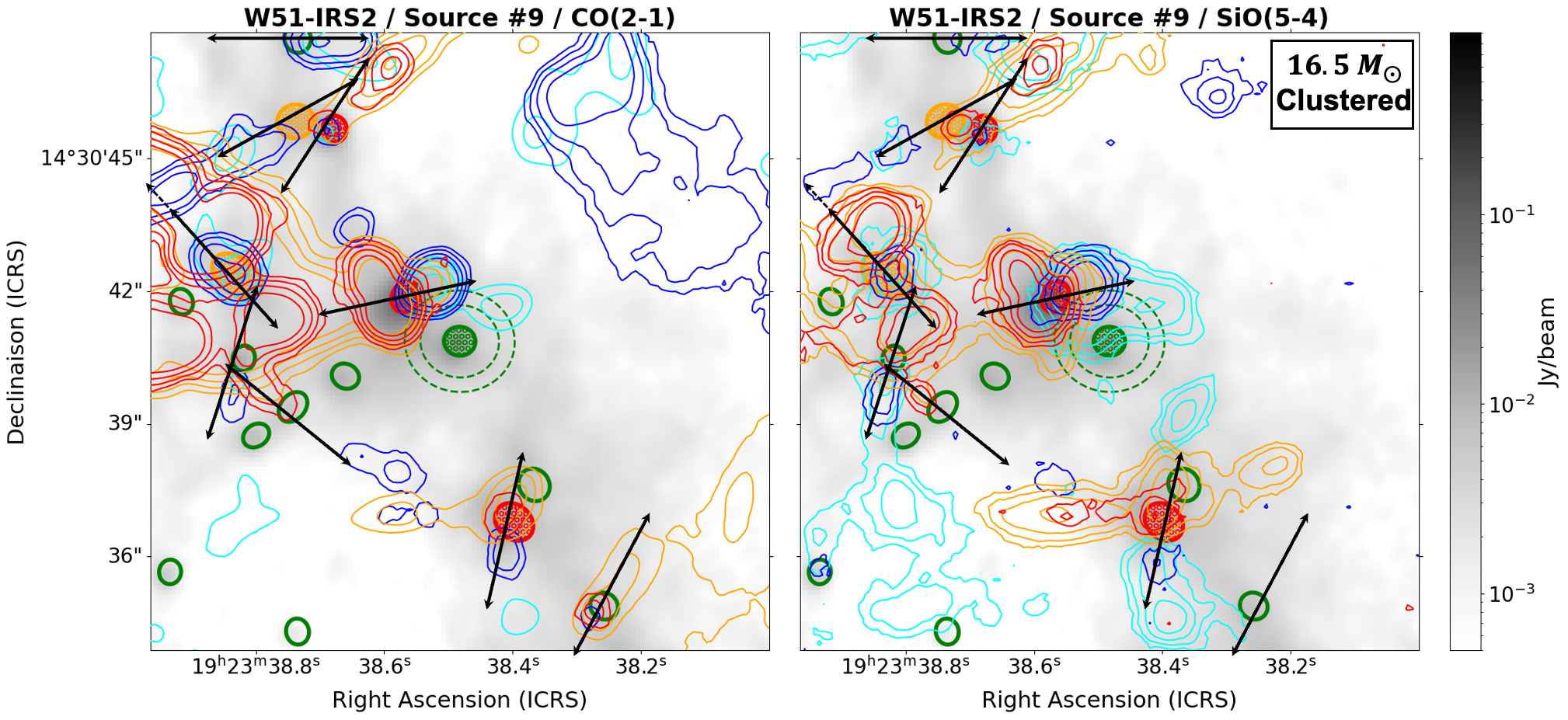}
    
   % \vskip -0.3cm
    \caption{CO(2-1) and SiO(5-4) molecular outflow maps centered on the 12 most massive PSC candidates, represented as the center green ellipse with an annulus in each panel (cores \#1A and \#1B of G333.60 are displayed as one core here, see Sect.\,\ref{subsection:complete_list}). Ellipses in red are cores classified as protostellar (due to an association with an outflow) with a mass greater than $8$ M$_{\odot}$, and ellipses in orange are protostellar cores with a mass between 4 and 8\,M$_{\odot}$. Filled green ellipses are other PSC candidates (i.e with M\,$ > 8$ M$_{\odot}$ at 20\,K) while the empty green ellipses are low mass cores which have not been analysed here. Moment 0 contours of the blue-shifted wings are overlaid on the continuum map at low velocity in cyan and high velocity in blue. Moment 0 contours of the red-shifted wings are overlaid on the continuum map at low velocity in orange and high velocity in red. Arrows represent the direction of the outflows driven by the protostars. The mass and the location of the candidate (i.e. classified as clustered or as isolated, see Sect.~\ref{subsection:unique_sample} for the details of the classification) are presented in each SiO map. Parameters of each map can be found in Appendix \ref{appendix:MPSC_fig_part}.}
    \label{12_massive_cores_outflows_highlighted}
\end{figure*}

\begin{figure}\ContinuedFloat
    \centering
    \includegraphics[width=0.49\textwidth]{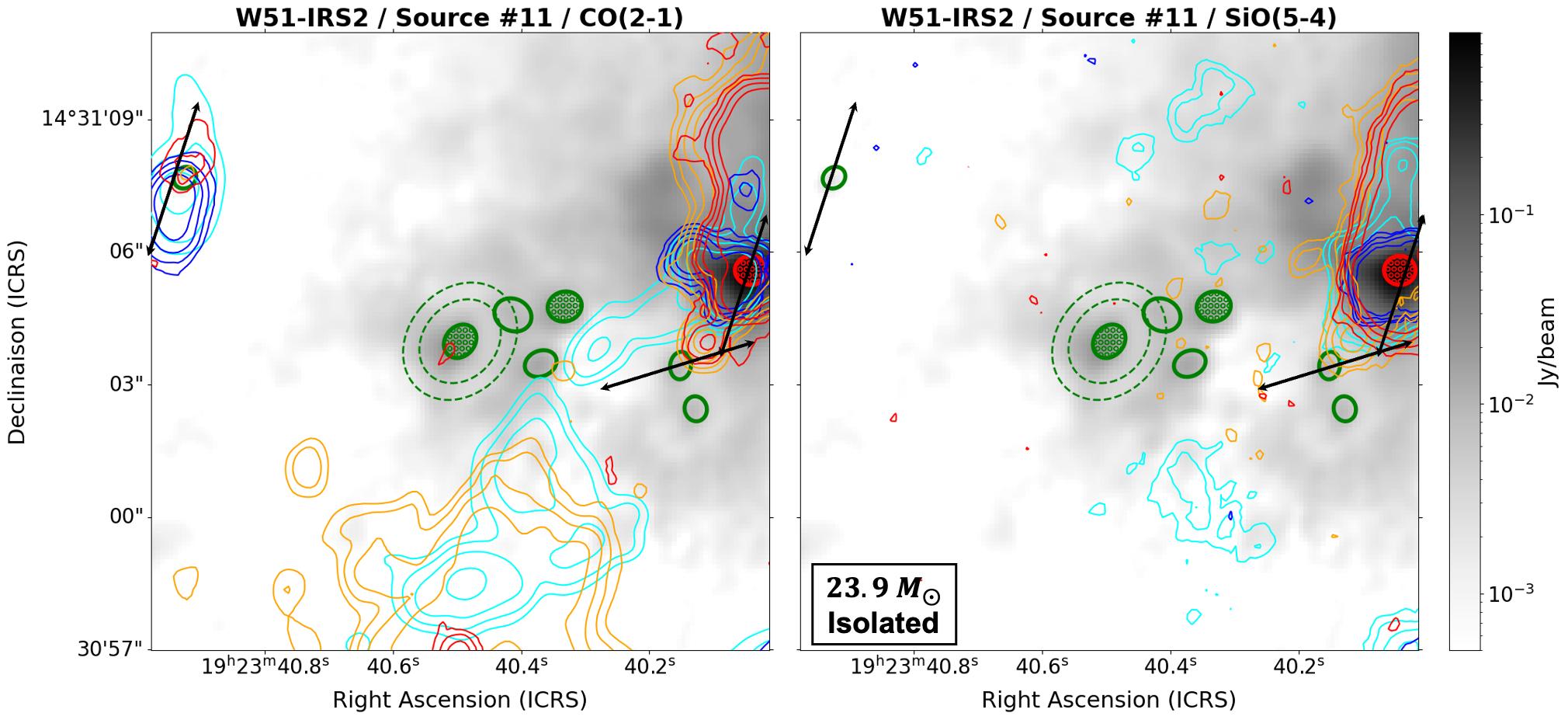}
    \caption{continued.}
\end{figure}

% {\color{red} Among the candidates, 8 are considered as uncertain due to their possible outflow association as they are located in complex areas. Those 8 cores are marked in table \ref{tab:MPSC_candidates} with a cross in column "uncertain candidates". Cores \#3 and \#5 from G337 are located in a big red outflow zone in which they could contribute. In addition, cores \#5 and \#13 from G338 and cores \#21 and \#32 from W43-MM1 are in complex area where outflows overlap. Finally core \#12 from W43-MM2 is located near a blue outflow that could be associated to the source. From now on we call robust list the list of 32 cores and complementary list the list of 8 uncertain cores. }

%-----------------------------------------------------------------

%\subsection{Comparison with \cite{Nony2023} and \cite{towner2023}} \label{subsection:comparison_Nony}
\subsection{Comparison with previous ALMA-IMF outflow results} \label{subsection:comparison_Nony}

\cite{Nony2020} and \cite{Nony2023} previously used the CO (J=2--1) outflow lobe maps to visually identify outflows in the W43-MM1, W43-MM2 and W43-MM3 regions. %This identification was based on visual inspection of the maps of the high velocity lobes. 
%To fully validate our automated method to derived the outflow status of each core, we have compared our results with previous works based on more classical methods (maps of the CO/SiO lobes to recognize all outflows in a region). 
%As we discriminate prestellar from protostellar cores using outflows identification, 
%We compare here our results with the studies of \cite{Nony2023} and \cite{towner2023}. \cite{Nony2023} is based on the CO outflows in the W43-MM2 and W43-MM3 regions, following \cite{Nony2020} in the W43-MM1 region. \cite{towner2023} is based on the SiO outflows among the 15 ALMA-IMF regions.
These works were based on careful and relatively time consuming investigation of the outflow maps and cubes.

The comparison with our present results is therefore extremely interesting to validate the method since the data are the same (ALMA-IMF dataset). %Comparing our outflow identification with the classification of \citet{Nony2020} and \citet{Nony2023} for the W43-MM1/MM2/MM3 regions, 
We see that our more automated method based on the On-Off spectra gives exactly the same identification of outflows for 26 % $100\%$ ($26/26$) on 
protostellar common sources with \citet{Nony2020} and \citet{Nony2023}. We see discrepancies in the identification of outflows for one core that we classify here as protostellar and for three out of 11 cores which we here consider as not driving any outflow. We classify W43-MM1 source \#20\footnote{This core is labeled as \#37 in \cite{Nony2020} and \cite{Nony2023}} as protostellar due to a detection in the red-shifted wing in the On-Off spectrum and its associated lobe, while \cite{Nony2020} classified it as prestellar. Among the three other sources, two (W43-MM1\,\#30 and W43-MM2\,\#9)\footnote{They correspond to sources \#14 in \citet{Nony2020} and \#15 in \citet{Nony2023} respectively.} do not have any significant emission at high velocities in their On-Off spectrum and their outflow maps show no sufficient evidence of association with the surrounding lobes to validate an outflow from these cores. W43-MM1\,\#30 was indeed labeled as tentative in \citet{Nony2020}. W43-MM2\#2\footnote{It corresponds to source \#6 in \citet{Nony2023}.} has no significant emission at high velocities in the On-Off spectrum in CO, but has a possible weak extend in high velocities (up to $\sim \pm$10 \kms) in SiO, at the limit of the significance (4.8 and 2.5~$\sigma$ in the blue and red parts respectively). In \citet{Nony2023} this source is an outflow source due to some weak red lobes in CO. But these lobes are weak and are situated at relatively large distance from the core making very uncertain the association of these lobes with the core. % which happens to be detected only at relatively large distance (5$^{\prime\prime}$ i.e. 0.13 pc) from this 0.37$^{\prime\prime}$ size source casting doubt on the outflow identification also using this method. 
Altogether, we therefore can conclude that our automated fast method is validated against the more classical approach of \citet{Nony2020,Nony2023} for 36 out 38 cores (95~$\%$) in W43 with a disagreement for only two sources, W43-MM1\,\#20 and W43-MM2\,\#2. %This last source is anyway intriguing since it is surprising not to detect any CO excess at high velocities on the source while there is a weak excess in SiO and possibly some far distance CO lobes. 
%We have thus decided to consider this source as uncertain in the following concerning its outflow status. 

%and see a few difference of identifiction for 4 out of the 13 cores we propose to be not driving a strong outflow. Concerning the prestellar cores, from the 13 prestellar cores identified in our study, we have a common agreement on 9 of them. Two of these 4 cores were indeed found to be uncertain outflows in Nony's works and using the on-off master criterium to decide whether a core is driving an outflow these 2 cores are not outflow drives anymore. 

%Including both pre and protostellar cores we have a $90\%$ total agreement with the study of \cite{Nony2023}. This slight difference on prestellar cores is due to the classification of four cores (\#2 and \#9 from W43-MM2 and cores \#30 and \#42 from W43-MM1) as PSC candidates that are classified as protostellar sources by \cite{Nony2020} and \cite{Nony2023}. A deeper analysis on these four cores is provided in Appendix~\ref{appendix:difference_4cores_Nony}. 

\cite{towner2023} used the SiO (J=5--4) line to search for emission that could be associated with jet and outflow activity, independent of the core positions towards all ALMA-IMF fields. In that work however there was no attempt to associate the detected high velocity lobes with the cores, making it difficult to compare with the results presented here. However, we perform a visual comparison between our SiO lobes with the catalog of SiO lobes in \cite{towner2023}. We recover $\sim 80\%$ of the SiO structures find in \cite{towner2023}. The remaining 20\% lobes do not affect our classification since none are directly associated with the cores in consideration here.

Finally, \cite{Armante2024} classified pre- and protostellar cores using the same CO and SiO lines as this study. Above our mass threshold, we have the same identification for four protostellar cores and one prestellar. For core \#6\footnote{This core is labeled as core \#5 in \cite{Armante2024}.} in G012.80 however, there is a difference in the identification of outflows. \cite{Armante2024} classified it as prestellar, while we classify it as protostellar due to a red excess detection in the On-Off spectrum, and blue and red lobes centered on the core in the outflow lobe maps. \cite{Armante2024} rejected it as protostellar due to the lack of SiO lobes centered on the core, leading to the difference of classification. Overall we have an agreement on five out of six cores ($\sim 80$\,\%).

%As \cite{towner2023} is not focusing on outflows driven by cores, we compare the SiO outflow lobes find in this study with the ones in our maps and have a final agreement of around $80\%$ among the ALMA-IMF fields. The remaining 20\% lobes does not affect the classification made in this paper as they are lobes not associated to a continuum core.
%-----------------------------------------------------------------

\subsection{The complete list of PSC candidates}\label{subsection:complete_list}

To arrive to the 141 cores above 8\Msun (using a temperature of 20\,K) presented in the previous sections, that we classified into pre and protostellar cores, we have performed an additional verification to scrutinize and hence verify the continuum emission of these cores and verify that they are indeed compact individual cores.
%above a mass threshold of M\,$>$\,8\Msun using a dust temperature of 20~K. %We verified whether all these cores are solid continuum source detection since the automated method to extract cores may sometime fail and report very uncertain or even fake sources. 

\begin{figure} [htbp!]
    \centering
    \includegraphics[width=0.49\textwidth]{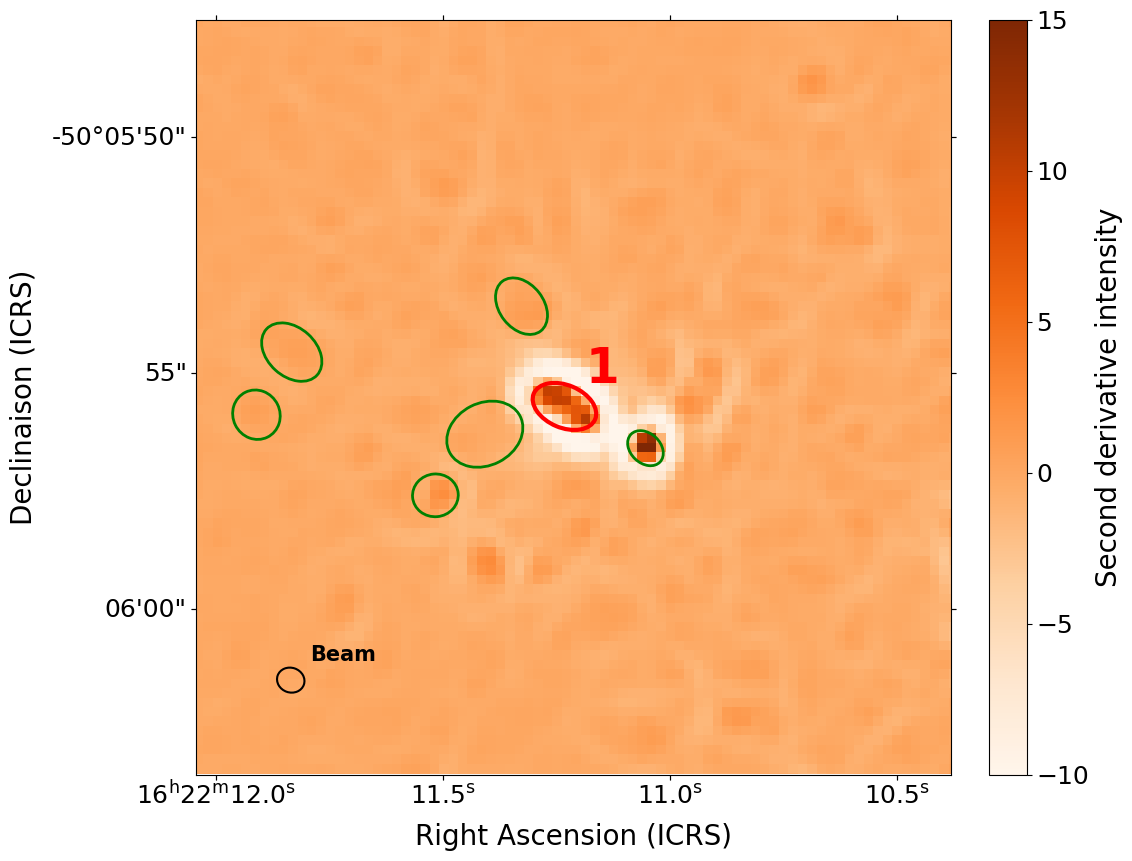}

    \caption{Second derivative map of G333.60 with a zoom on core \#1. The beam of the smoothed continuum map of G333.60 is shown in bottom left.}
   \label{fig:G333_core1_2ndder}
\end{figure}

%We further checked the continuum emission of these candidates to verify that these cores are indeed compact individual cores. %This is particularly important since at the end we select objects which do not show outflows, and may have therefore no strong confirmation that they are indeed compact cores. 
We use the results of the GExt2D algorithm that is based on the 2D second derivative of the continuum map to check the presence of strong local curvature peak for each core. 
%Among the 47 candidates, one core (G333 \#30) appears as doubtful based on local curvature in the second derivative map and has been then excluded. 
Also since the \cite{louvet22} catalog, used in this work, is actually based on images smoothed to a homogeneous physical scale of 2700~au, we may have signs of sub-fragmentation when investigating the original ALMA-IMF images and the 2D second derivative maps at the native resolution. We remove any cores that do not show either a strong local curvature or that have indications of multiple peaks in the second derivative map. In this later case, sources that lead to fragments below the 8\,$\rm M_{\odot}$ cutoff are removed.
%We therefore removed all the cores that were not presenting any sign of strong local curvature and the lowest mass cores that wear appearing as multiple peaks in the second derivative map, leading to fragments less massive than 8\,$\rm M_{\odot}$.
%At the native spatial resolution \st{of the maps}, some of the candidates might actually appear as multiple (double) so that they would be typically two times less massive, which could move them away from our prime list of PSC candidates which are truly high-mass. 
%Using the curvature maps from the 2D second derivatives at the native resolution we find that indeed a few cores have definite sign of being multiple. Three of them (G010\#22 being double, W51-E\#17 triple and W51-E\#39 double) are in the low mass range of our selection and the resulting individual cores at the native resolution are therefore clearly lower mass cores and have therefore been removed from our selection. %end up below our mass threshold of 8\Msun when considering the fragmentation, and therefore we omit them from our final selection. %driving them out of our selection. 
On the other hand, some sources could appear as multiple at the native resolution, but each fragment can still be massive enough to stay in our list of cores. This is the case for the G333 \#1 source, which corresponds to the most massive core of the complete list of PSCs with 148\,M$_{\odot}$ at 20\,K.
%and is thus most probably made of at least one core that is still in the high-mass regime. 
Figure~\ref{fig:G333_core1_2ndder} shows the second derivative map produced by GExt2D revealing that this core splits into two continuum peaks. From the ratio of flux density measurements towards both peaks at the native resolution, we estimate the fraction of the total mass distributed in both cores which are presented in Table \ref{tab:MPSC_candidates}. In the final complete list of PSC candidates, we have therefore split  G333 \#1 into G333 \#1A (North-East core) and \#1B (South-West core) adding a candidate in the list.\footnote{Their respective sizes are taken from the GExt2D catalog which will be published on CDS with the catalogs of \cite{louvet22}. Cores \#1A and \#1B correspond to cores \#9 and \#7 in the GExt2D catalog of the G333.60 protocluster.} 
%This additional prestellar core leads to a total of 581 ALMA-IMF dense cores.

%This makes the masses of these cores comparable to the overall mass range of our high-mass PSC candidates. We consider the North-East peak as a robust high-mass PSC candidate, while the classification of the South-West peak is uncertain (cross in the last column of Table \ref{tab:MPSC_candidates}) due to a blue outflow lobe close to the dust continuum peak. 
% From now on core \#1 will be considered as two separated cores \#1A (North-East core) and \#1B (South-West core) (see Table \ref{tab:MPSC_candidates}). 

%Finally, we also excluded cores that are very extended with sizes larger than 7000\,au which therefore rather consist of clumps than more compact cores.

%\#15 from W43-MM2 based on the fact that it is a very extended (and therefore more uncertain) core with a FWHM size of $\sim 7000$ au. Altogether we have therefore excluded five cores and perform the same cleaning on our protostellar cores sample leading to the removal of three sources.

We also verify that the PSC candidates we recognize are not associated with a hot core which would point to a strong central heating source. We checked into the 76 hot core catalog of \cite{Bonfand2024}, and found that only one of our PSC candidates (W43-MM1 \#30) seem to spatially coincide with a methyl formate peak. However in \cite{Bonfand2024} this peak is weak (weakest methyl formate peak of the W43-MM1 hot core catalog), and could actually correspond to a spatial contamination from the extended hot core/methyl formate emission which is observed in this part of the W43 region. We also note that none of our PSC candidates in the W43 regions have been classified as hot cores in the \cite{Brouillet2022} study.

Our final catalog contains 42 individual PSC candidates and 99 individual protostellar cores candidates, above the threshold of M $> 8$ M$_{\odot}$ at 20\,K. The PSC candidates are presented in Table \ref{tab:MPSC_candidates} along with their physical properties we discuss in the following section.

Except the eight PSC candidates in common with the W43 study of \cite{Nony2023} and the one in common with the G012.80 study of \cite{Armante2024}, the other PSC candidates from the sample of 42 cores presented here are newly found PSCs. Among these 8 PSCs in the W43 regions that are in common to our study, core \#6 from W43-MM1 is one of our most massive candidates and was discovered in \cite{Nony2018}. Core \#12 from W43-MM2 and core \#21 from W43-MM1 were also presented as possible high-mass prestellar cores in \cite{Nony2023} (labeled as core \#22 and \#134). The five other cores were under 16\Msun and not proposed to be high-mass PSC candidates in \cite{Nony2023}. %Except these three prestellar cores, all the cores presented here are new high-mass PSC candidates. 

%Among the sample of MPSC candidates, core \#1 ofthe G333.60 protocluster is by far the most massive candidate with more than three times the mass of core \#6 of W43-MM1, with a total mass about 150~M$_\odot$ using a dust temperature of 20\,K. %This outlier raises questions about its individuality at such mass. 
%By looking into the second derivative map \timea{produced by the} GExt2D \timea{algorithm}, \st{it can be seen that}  

\subsection{Basic properties of the PSC candidates}\label{subsection:basic_properties}

Table \ref{tab:MPSC_candidates} presents the list of 42 candidates above 8\Msun at 20~K, together with the name of the host region, their number, their right ascension and declination coordinates (ICRS), the range of mass and the size of the core as well as the corresponding average density and free-fall time.

The dust temperatures to adopt to estimate the masses from dust emission are difficult to establish for such distant regions for which the FIR data required to observe the peak of the spectral energy distribution of dust emission have too low spatial resolution to resolve individual cores. As argued in Sect.~\ref{subsection:4.1} in such dense cores the cooling times are low pointing to possible low temperatures most probably below 20~K (see also \citealp{Bhandare2018}). On the other hand, the observed regions are filled by warm gas and dust due to high-mass star feedbacks as shown in the dust temperatures obtained with PPMAP in \cite{Dell'Ova2023} using Herschel data and derived down to 2.5\arcsec\ scale (i.e. between 2 and 5 times larger than the typical core size). These PPMAP temperatures are obtained from the low resolution Herschel maps with beam sizes between 12\arcsec\ and 18\arcsec\ at 160 and 250~$\mu$m by a guided (bayesian) method using also the high resolution 1.3mm ALMA-IMF images. The PPMAP dust temperatures are estimated by averaging over the pixels of each source in the "non-corrected" dust temperature maps of \cite{Dell'Ova2023}. We use these 
2.5\arcsec\ scale temperatures to derive a maximum dust temperature for each core and adopt 20 K as a conservative lowest temperature. Using these as upper and lower limits in temperature, we obtain lower and upper limits in mass, giving a possible range of mass for each PSC candidate. %also added a 50\,\% additional uncertainty on the masses to account for uncertainty on dust emissivity.
%uncertain estimate of It may therefore tend to associate unresolved {\it Herschel} emission to the compact millimeter sources possibly increasing the obtained PPMAP temperatures for strong 1.3mm peaks. Altogether we can therefore consider that the PPMAP temperatures are upper limits for the cores and that 20 K can be a conservative lower limit. 
In Table \ref{tab:MPSC_candidates} we have listed the corresponding range of masses for each core and give an adopted central mass taken at the middle of this uncertainty range. 
%We note however that the PPMAP method  (i.e. in the surroundings of the cores), their deconvolved FWHM, volume density and corresponding free-fall time. 
%Additional uncertainties on the mass estimates are obtained from the uncertainties on the peak and integrated fluxes. 
%into account. 
%A temperature of 20~K is a good approximation for prestellar cores, as adopted in \cite{Bontemps2010}, and is close to the temperatures of massive dense cores found between 16 to 20K in \cite{Duarte-Cabral2013}.
%Masses computed at 20~K and PPMAP temperatures correspond to upper and lower limits in mass of the PSC candidates presented in Table \ref{tab:MPSC_candidates}. 
%These two estimates reflect the uncertainties of the mass due to the poorly constrained dust temperature. 
We use these lower and upper masses to extract a symmetric error on the density and free-fall time computations (see below).

Similarly we propose to adjust the dust temperatures to be adopted in the following to best estimate the masses of the protostellar cores. We propose to also adopt the PPMAP temperatures from \cite{Dell'Ova2023} but using the 70\,$\rm \mu m$ corrected ones such as in \citet{louvet22} and which are best suitable for internally heated sources such as high-mass protostellar cores.

Using these best estimate dust temperatures for both PSCs and protostellar cores we thus finally get a reduced number of cores above 8~\Msun with 30 PSCs and 52 protostellar cores\footnote{The full list of protostellar cores down to $\sim 4$\Msun using the PPMAP temperature is presented in appendix\,\ref{appendix:proto_candidates_list}.} from the full sample of 42 PSCs and 99 protostellar cores originally analyzed for outflows in the previous sections. These 82 cores (30 PSCs and 52 protostellar cores) are then going to be the ones further discussed in following sections. The spectra and molecular outflows maps for each 30 PSC candidates not shown in Fig.~\ref{12_massive_cores_outflows_highlighted} are shown in Appendix~\ref{appendix:MPSC_fig_part}. We also display in appendix \ref{appendix:MPSC_highlight_part} the location of the PSC candidates and protostellar cores identified in each protocluster.

%We find clear often bipolar CO and SiO outflows for a significant fraction of the high mass cores (see examples of outflows in Fig.~\ref{12_massive_cores_outflows_highlighted} and also in \citealp{Nony2023}) which were initially selected using a 20 K dust temperature. A total of 99 cores were found driving an outflow while 42 are not. 
%This is expected as these objects are the most massive cores in the survey and should thus correspond to the main accretion/ejection phase of the formation of the highest mass stars in the observed fields. %The protoclusters targeted by the ALMA-IMF survey have been selected to represent well the early evolutionary phase of rich clusters which should then contain precursors of highest mass stars formed in the relatively nearby neighborhood within Milky way (see discussion of the selection of observed fields of ALMA-IMF in \citealp{Fred2022}). 

%Accounting for improved dust temperatures to be used to select protostellar (PPMAP 70\,$\rm \mu m$ corrected temperature) and prestellar cores (average temperature between 20\,K and PPMAP non corrected temperature) it is finally 30 PSCs and 52 protostellar cores which are found above 8\,M$_{\odot}$. %The list of cores classified as protostellar, with a mass greater than 4\Msun using the PPMAP temperature is presented in appendix\,\ref{appendix:proto_candidates_list}.

To calculate the volume density $\rm n_{\rm H_2}$, we use as a radius
%derive a 1D size used as 3D radius and define from 
the 1.3mm continuum deconvolved size FWHM$^{\rm dec}_{\rm 1.3mm}$ as: FWHM$^{\rm dec}_{\rm 1.3mm} \simeq \sqrt{a_{\rm 1.3mm} \times b_{\rm 1.3mm} - \Theta_{\rm beam}^2} \times d$, with $a_{\rm 1.3mm}$ and $b_{\rm 1.3mm}$ the major and minor FWHM axes of the core , $\Theta_{\rm beam}$ the geometric mean of the beam FWHM of the smoothed continuum map, and $d$ the distance of the region. As in \cite{Pouteau2022}, we fix a minimum deconvolved size at half the beam FWHM. %of the beam in order to limit the deconvolution effects that can give too small sizes for the cores. 
%Using formula \ref{volume_density_computation}, assuming a spherical core, 
We compute the $\rm H_2$ volume density as:%densities of each candidate : 

\begin{equation}
    \centering
    n_{\rm H_2} = \frac{\rm M_{\rm source}}{\frac{4}{3} \pi \, \mu \, \rm m_{\rm H} \left( \rm FWHM^{\rm dec}_{\rm 1.3mm} \right)^3}    
    \label{volume_density_computation}
\end{equation}
with M$_{\rm source}$ the mass of the source, $\mu = 2.8$ the mean molecular weight of interstellar gas, and $\rm m_{\rm H}$ the mass of the hydrogen atom. Using the average volume density, we compute the free-fall time of each core as:

\begin{equation}
    \centering
    t_{\rm ff} = \left(\frac{3 \pi}{32G \rho_0} \right)^{1/2} \simeq 10^4 \, \rm yr \, \left( \frac{n_{\rm H_2}}{10^7 \rm cm^{-3}}\right)^{-1/2}   
    \label{freefall_computation}
\end{equation}
with $G$ the gravitational constant, $\rho_0$ and $\rm n_{\rm H_2}$ the volumetric mass and volume density of the core, respectively.

%Some slightly more uncertain cases due to the strong confusion in central regions are marked with a cross in the last column. These sources are located in overlapping outflows regions, causing a difficulty to be certain on their prestellar nature.

We also classify the PSC candidates according to their environmental conditions. We define an environment as clustered if the PSC has two or more high-mass (M$\,>\,$8\,M$_{\odot}$) cores within a radius of 0.15~pc, and as isolated if they have less than two neighbouring high-mass cores. This criterion was chosen to take into account both crowded environment and mass.
%The high-mass PSC candidates can also harbour low mass cores in their environment but they are not accounted for this classification.
We present in Table~\ref{tab:properties_PSC} the ranges in mass, size, volume density, and free-fall time of the 30 PSC candidates with $\rm M > 8\,M_{\odot}$, for the three different hypothesis (average mass, mass at 20\,K, and mass using the PPMAP dust temperature). The minimum, mean, and maximum of each value are presented in each column. The range of free-fall times, between $\sim 2 \times 10^3\,$yr and $\sim 2 \times 10^4\,$yr, is in agreement with previous studies on massive starless cores (e.g. \citealp{Kong2017}, \citealp{Duarte-Cabral2013}). We have roughly as many clustered PSC candidates as isolated ones when including all the cores (17 clustered and 13 isolated), but 10 of the 12 most massive ones are found in clustered regions (see Fig. \ref{12_massive_cores_outflows_highlighted}), which is expected as most of the mass is located in these crowded clumps of the protoclusters. %This highlights that the most massive cores are forming in the clumps of the protoclusters.

From all the 14 protoclusters studied, only G328.25 does not host any PSC candidate above 8\Msun using a dust temperature of 20\,K. Core \#22 in W51-IRS2 identified as PSC candidate is also covered by the W51-E field (identified as core \#20). We keep this object in W51-IRS2 as the signal to noise ratio in this region is higher. 
We have a large number of PSC candidates in the W51-IRS2 region with ten cores, followed by the W43-MM1 region with seven of them. In between, we have the G333.60 region with five candidates, G338.93 with four, G008.67, W43-MM2 and W51-E with three candidates each, and G337.92 with two candidates. Finally, the G010.62, G012.80, G327.29, G353.41, and W43-MM3 regions host only one candidate each.

%-----------------------------------------------------------------
 
\begin{table*} [htbp!]
    \centering     
    \begin{threeparttable}[c]
    \caption{List and properties of the sample of prestellar core (PSC) candidates.}
    
    \begin{tabular}{lllllclrc}
    \hline \noalign {\smallskip}
    Region & Source &  \col{RA}  &  \col{DEC}  & \col{Mass $\left[ \rm range \right]$\tnote{1}} & \col{FWHM$^{\rm dec}$} & \col{$n_{\rm H_2}\tnote{2}$} & \col{$t_{\rm ff}$\tnote{2}} & Location\tnote{3} \\ \noalign {\smallskip}
        &   &  \col{[deg]}  & \col{[deg]}  &  \col{$\left[\rm M_\odot \right]$} & \col{$\left[\rm au \right]$} & \col{$\left[\times 10^6 \rm \, cm^{-3} \right]$} & \col{$\left[\rm kyr \right]$}  &  {\smallskip}\\  
    \hline \noalign {\smallskip}
    
	 G008.67   &   \#4   &   271.580761   &   -21.6237093   &   7.8  $\left[ 6.6 - 8.9 \right]$   &   3870   &   4.1 $\pm$ 0.6   &   15.7 $\pm$ 1.2   &   C   \\
	 G008.67   &   \#7   &   271.5800126   &   -21.6251966   &   13.7  $\left[ 11.2 - 16.2 \right]$   &   2780   &   19.3 $\pm$ 3.5   &   7.2 $\pm$ 0.7   &   C   \\
	 G008.67   &   \#10   &   271.5774356   &   -21.6271193   &   7.4  $\left[ 5.9 - 8.9 \right]$   &   4120   &   3.2 $\pm$ 0.6   &   17.7 $\pm$ 1.8   &   I   \\

    \hline \noalign {\smallskip}

	 G010.62   &   \#4   &   272.6216933   &   -19.9291438   &   11.7  $\left[ 8.7 - 14.7 \right]$   &   1840   &   56.7 $\pm$ 14.6   &   4.2 $\pm$ 0.6   &   I   \\
 
    \hline \noalign {\smallskip}
    
	 G012.80   &   \#5   &   273.5482204   &   -17.9458195   &   7.4  $\left[ 6.8 - 8.1 \right]$   &   1450   &   73.6 $\pm$ 6.6   &   3.7 $\pm$ 0.2   &   I   \\

    \hline \noalign {\smallskip}
    
	 G327.29   &   \#11   &   238.2800319   &   -54.6194279   &   8.6  $\left[ 6.6 - 10.6 \right]$   &   3480   &   6.2 $\pm$ 1.4   &   12.7 $\pm$ 1.5   &   C   \\

    \hline \noalign {\smallskip}
    \textbf{G333.60}  &  \tnote{4}\;\,\textbf{\#1A}  & \textbf{245.5468839}  &  \textbf{-50.0987477}  &  \textbf{54.0 $\mathbf{\left[35.7-72.3\right]}$}  &  \textbf{2640}  &  $\bm{88.8\pm30.1}$  &  $\bm{3.4\pm0.6}$  &  \textbf{C}\\
    \textbf{G333.60} &  \tnote{4}\;\,\textbf{\#1B}  & \textbf{245.5466503}  &  \textbf{-50.0988518}  &  \textbf{40.5 $\mathbf{\left[26.9-54.2\right]}$}  &  \textbf{2020}  &  $\bm{148.8\pm50.2}$  &  $\bm{2.6\pm0.5}$  &  \textbf{C}\\
    G333.60   &   \#8   &   245.5282954   &   -50.1068134   &   6.7  $\left[ 5.4 - 8.0 \right]$   &   2510   &   12.8 $\pm$ 2.5   &   8.8 $\pm$ 0.9   &   C   \\
    G333.60   &   \#26   &   245.535353   &   -50.1041745   &   7.0  $\left[ 4.9 - 9.1 \right]$   &   2310   &   17.3 $\pm$ 5.1   &   7.6 $\pm$ 1.2   &   I   \\
    G333.60   &   \#45   &   245.549303   &   -50.0984873   &   6.4  $\left[ 4.5 - 8.2 \right]$   &   4370   &   2.3 $\pm$ 0.7   &   20.8 $\pm$ 3.2   &   C   \\

    \hline \noalign {\smallskip}
    
	 G337.92   &   \#3   &   250.2949021   &   -47.1343006   &   12.0  $\left[ 9.2 - 14.8 \right]$   &   2750   &   17.4 $\pm$ 4.1   &   7.6 $\pm$ 0.9   &   C   \\
	 G337.92   &   \#5   &   250.295556   &   -47.134715   &   11.2  $\left[ 8.7 - 13.7 \right]$   &   3450   &   8.3 $\pm$ 1.9   &   11.0 $\pm$ 1.3   &   C   \\
 
    \hline \noalign {\smallskip}
    
	\textbf{G338.93}  &  \textbf{\#5}  &   \textbf{250.1417392}  &  \textbf{-45.7025004}  &  \textbf{24.9 $\mathbf{\left[17.4-32.4\right]}$}  &  \textbf{1750}  &  $\bm{140.5\pm42.4}$  &  $\bm{2.7\pm0.4}$  &  \textbf{C}\\
	 G338.93   &   \#7   &   250.1378537   &   -45.7040233   &   10.0  $\left[ 8.2 - 11.8 \right]$   &   1470   &   95.4 $\pm$ 16.8   &   3.2 $\pm$ 0.3   &   C   \\
	 G338.93   &   \#13   &   250.1423231   &   -45.6931039   &   11.1  $\left[ 9.5 - 12.7 \right]$   &   3090   &   11.4 $\pm$ 1.7   &   9.4 $\pm$ 0.7   &   I   \\
	 G338.93   &   \#16   &   250.1418854   &   -45.701708   &   6.5  $\left[ 4.8 - 8.2 \right]$   &   2930   &   7.8 $\pm$ 2.1   &   11.3 $\pm$ 1.6   &   C   \\

    \hline \noalign {\smallskip}
    
	 G353.41   &   \#5   &   262.6101515   &   -34.6960014   &   12.7  $\left[ 9.9 - 15.6 \right]$   &   2630   &   21.2 $\pm$ 4.7   &   6.9 $\pm$ 0.8   &   C   \\

    \hline \noalign {\smallskip}
    
	\textbf{W43-MM1}  &  \textbf{\#6}  &   \textbf{281.9423152}  &  \textbf{-1.909246}  &  \textbf{40.5 $\mathbf{\left[37.3-43.6\right]}$}  &  \textbf{2030}  &  $\bm{146.5\pm11.4}$  &  $\bm{2.6\pm0.1}$  &  \textbf{I}\\
	 W43-MM1   &   \#8   &   281.938703   &   -1.9102889   &   8.3  $\left[ 7.5 - 9.0 \right]$   &   1350   &   101.6 $\pm$ 8.9   &   3.1 $\pm$ 0.1   &   C   \\
	 W43-MM1   &   \#14   &   281.944974   &   -1.9044595   &   9.1  $\left[ 8.4 - 9.7 \right]$   &   1350   &   111.6 $\pm$ 8.2   &   3.0 $\pm$ 0.1   &   I   \\
	\textbf{W43-MM1}  &  \textbf{\#18}  &   \textbf{281.9462451}  &  \textbf{-1.9075171}  &  \textbf{18.7 $\mathbf{\left[17.9-19.4\right]}$}  &  \textbf{1930}  &  $\bm{78.6\pm3.0}$  &  $\bm{3.6\pm0.1}$  &  \textbf{I}\\
	\textbf{W43-MM1}  &  \textbf{\#21}  &   \textbf{281.9453834}  &  \textbf{-1.9082357}  &  \textbf{21.1 $\mathbf{\left[20.3-21.9\right]}$}  &  \textbf{2390}  &  $\bm{46.8\pm1.7}$  &  $\bm{4.6\pm0.1}$  &  \textbf{C}\\
	 W43-MM1   &   \#30   &   281.9456971   &   -1.9082389   &   12.2  $\left[ 11.8 - 12.6 \right]$   &   1350   &   150.0 $\pm$ 5.4   &   2.6 $\pm$ 0.0   &   C   \\
	\textbf{W43-MM1}  &  \textbf{\#32}  &   \textbf{281.9455609}  &  \textbf{-1.9075139}  &  \textbf{19.5 $\mathbf{\left[18.8-20.1\right]}$}  &  \textbf{2180}  &  $\bm{56.9\pm1.9}$  &  $\bm{4.2\pm0.1}$  &  \textbf{C}\\

    \hline \noalign {\smallskip}
    
	 W43-MM2   &   \#2   &   281.9001273   &   -2.0224281   &   13.0  $\left[ 11.7 - 14.3 \right]$   &   2020   &   47.7 $\pm$ 4.8   &   4.6 $\pm$ 0.2   &   C   \\
	 W43-MM2   &   \#9   &   281.9034863   &   -2.0173945   &   7.8  $\left[ 6.7 - 8.8 \right]$   &   2160   &   23.3 $\pm$ 3.3   &   6.6 $\pm$ 0.5   &   I   \\
	\textbf{W43-MM2}  &  \textbf{\#12}  &   \textbf{281.9027193}  &  \textbf{-2.014794}  &  \textbf{16.0 $\mathbf{\left[14.0-18.1\right]}$}  &  \textbf{3390}  &  $\bm{12.4\pm1.6}$  &  $\bm{9.0\pm0.6}$  &  \textbf{I}\\

    \hline \noalign {\smallskip}
    
	 W43-MM3   &   \#3   &   281.9207197   &   -2.0057611   &   10.8  $\left[ 9.4 - 12.2 \right]$   &   2420   &   23.1 $\pm$ 3.0   &   6.6 $\pm$ 0.4   &   C   \\

    \hline \noalign {\smallskip}
    
	\textbf{W51-E}  &  \textbf{\#14}  &   \textbf{290.9326071}  &  \textbf{14.510126}  &  \textbf{35.6 $\mathbf{\left[28.8-42.3\right]}$}  &  \textbf{2080}  &  $\bm{119.7\pm22.7}$  &  $\bm{2.9\pm0.3}$  &  \textbf{I}\\
	 W51-E   &   \#28   &   290.9319946   &   14.5089561   &   7.5  $\left[ 6.7 - 8.2 \right]$   &   1600   &   55.1 $\pm$ 5.8   &   4.3 $\pm$ 0.2   &   C   \\
	 W51-E   &   \#31   &   290.9288221   &   14.5159753   &   9.5  $\left[ 9.3 - 9.7 \right]$   &   4170   &   3.9 $\pm$ 0.1   &   15.9 $\pm$ 0.2   &   C   \\

    \hline \noalign {\smallskip}
    
	\textbf{W51-IRS2}  &  \textbf{\#8}  &   \textbf{290.9091553}  &  \textbf{14.5185781}  &  \textbf{17.4 $\mathbf{\left[14.0-20.9\right]}$}  &  \textbf{1930}  &  $\bm{73.4\pm14.5}$  &  $\bm{3.7\pm0.4}$  &  \textbf{I}\\
	\textbf{W51-IRS2}  &  \textbf{\#9}  &   \textbf{290.9103723}  &  \textbf{14.5113251}  &  \textbf{16.5 $\mathbf{\left[13.0-19.9\right]}$}  &  \textbf{2510}  &  $\bm{31.5\pm6.7}$  &  $\bm{5.6\pm0.6}$  &  \textbf{I}\\
	\textbf{W51-IRS2}  &  \textbf{\#11}  &   \textbf{290.9187587}  &  \textbf{14.5177443}  &  \textbf{23.9 $\mathbf{\left[17.8-30.0\right]}$}  &  \textbf{2950}  &  $\bm{28.2\pm7.2}$  &  $\bm{6.0\pm0.8}$  &  \textbf{C}\\
	 W51-IRS2   &   \#17   &   290.9245968   &   14.5197752   &   11.5  $\left[ 9.4 - 13.5 \right]$   &   3950   &   5.6 $\pm$ 1.0   &   13.3 $\pm$ 1.2   &   I   \\
	 W51-IRS2   &   \#22   &   290.9255723   &   14.5112676   &   7.6  $\left[ 6.1 - 9.1 \right]$   &   1350   &   93.5 $\pm$ 18.6   &   3.3 $\pm$ 0.3   &   I   \\
	 W51-IRS2   &   \#24   &   290.9114738   &   14.5186878   &   7.7  $\left[ 6.4 - 9.1 \right]$   &   3120   &   7.7 $\pm$ 1.4   &   11.4 $\pm$ 1.0   &   I   \\
	 W51-IRS2   &   \#28   &   290.9052888   &   14.5184717   &   10.1  $\left[ 8.7 - 11.5 \right]$   &   3910   &   5.1 $\pm$ 0.7   &   14.0 $\pm$ 1.0   &   I   \\
	 W51-IRS2   &   \#32   &   290.9069467   &   14.5064897   &   9.7  $\left[ 8.3 - 11.2 \right]$   &   1900   &   43.0 $\pm$ 6.4   &   4.8 $\pm$ 0.4   &   I   \\
	 W51-IRS2   &   \#38   &   290.9262262   &   14.515357   &   7.7  $\left[ 6.1 - 9.4 \right]$   &   1800   &   40.1 $\pm$ 8.6   &   5.0 $\pm$ 0.6   &   I   \\
	 W51-IRS2   &   \#42   &   290.9180791   &   14.5179626   &   11.5  $\left[ 8.7 - 14.2 \right]$   &   2820   &   15.5 $\pm$ 3.7   &   8.0 $\pm$ 1.0   &   I   \\

    \hline
    \end{tabular}
    \label{tab:MPSC_candidates}
\begin{tablenotes}
\item Bold font is used to present the robust sample of PSC candidates with $\rm M> 16 M_{\odot}$.
\item[1] Mean masses of the candidates with their range using the non-corrected PPMAP dust temperature (see \citealp{Dell'Ova2023}) and 20\,K as high and lower limits in temperature.%, and applying a 50\% additional uncertainty on the mass to account for uncertainty on the dust emissivity.
\item[2] Volume density and free-fall time (see section \ref{subsection:basic_properties}).
\item[3] Location of the source with C meaning clustered and I meaning isolated, based on the criteria described in Sect. \ref{subsection:basic_properties}.
\item[4] This source contains two continuum peaks that can be seen in the GExt2D second derivative map. We then use the flux density and size of each GExt2D source to compute the masses and densities of the two peaks (see section \ref{subsection:complete_list}).
%
%\item[*] Sources classified as uncertain due to a very complex environment. 

\end{tablenotes}    
\end{threeparttable}
\end{table*}

%\pagestyle{empty}
%\begin{landscape}
\begin{table*}[htbp!]
\centering
%\small
%    \setlength{\tabcolsep}{2pt}
\begin{threeparttable}[c]
\caption{Range of the properties of the 30 PSC candidates with $\rm M>8\,M_{\odot}$.}
%(using global measurements)}
\label{tab:properties_PSC}
\begin{tabular}{ccccc}
\hline \noalign {\smallskip}
    & Mass\tnote{1}  & FWHM$^{\rm dec}$\tnote{1} & n$_{\rm H_2}$\tnote{1} & t$_{\rm ff}$\tnote{1}\\ \noalign {\smallskip}% & Clustered/Isolated\\   
    &  $\left[\rm M_{\odot} \right]$  &  $\left[\rm au \right]$ &   $\left[\times 10^6 \rm cm^{-3}\right]$    &  $\left[\rm kyr \right]$ \\

\hline \noalign {\smallskip}

Adopted mass &   8.3 - 17.5 - 54.0  & 1350 - 2509 - 4170  & 3.9 - 54.1 - 150.0  &  2.6 - 6.3 - 15.9 \\ %& 22/20\\
M$_{\rm 20K}$ &   9.0 - 20.8 - 72.3   &  1350 - 2509 - 4170 & 4.0 - 62.6 - 182.9  &  2.3 - 5.9 - 15.7 \\ %& 22/20 \\
M$_{\rm T_{\rm PPMAP}}$ &   6.6 - 14.2 - 37.3   &  1350 - 2509 - 4170 & 3.9 - 45.7 - 144.6  &  2.6 - 7.0 - 16.1 \\ %&  22/20 \\ 

\hline
\end{tabular}

\begin{tablenotes}
\item[1] Each range is presented as minimum - mean - maximum.
\end{tablenotes}    

\end{threeparttable}
\end{table*}

\section{A new sample of high-mass PSC candidates} \label{sec:discussion}

\subsection{High-Mass PSCs defined as massive cores not hosting yet any highly accreting protostars}
 \label{sec:discuss:highly-accreting-hosts}

%\subsection{Strong CO and SiO outflows from high-mass protostars}
% \label{sec:discuss:strongCOoutflow}
It is difficult and uncertain to decide which observed mass reservoirs are going to truly form high-mass stars (larger than 8\,M$_{\odot}$). Here we adopt a mass threshold of 16~M$_\odot$ corresponding to a core to star efficiency $\epsilon_{\rm cse}$ of 50~\% to form a final 8\,M$_{\odot}$ star. In Table~\ref{tab:MPSC_candidates} (in bold font), using our adopted mass estimate, we find that 12 candidates then qualify as high-mass and we propose that they represent the robust sample of high-mass PSC candidates to be considered below for discussion. The rest of the 30 candidates will be labeled as PSC candidates in the following.

The most probable driver for high-mass star formation is the high level of dynamical (convergent) motions at the scale of dense clumps (\citealp{Schneider2010,Galvan-Madrid2010,Peretto2013,Alvarez-Gutierrez2024}; see also references in \citealp{Motte_review_2018}). These convergent flows should lead to large infall rates at core scales and then to large final accretion rates of the newly formed high-mass protostars.
In order to form a 30\Msun protostar in 0.3\,Myr, the accretion rate has to be of the order of $\rm 10^{-4}\,M_{\odot} yr^{-1}$.
These large accretion rates are compatible with the global infall rates at the scales of clumps which are directly observed with for instance $\rm 2.5\times10^{-3}\,M_{\odot} yr^{-1}$ in the SDC335 clump which contains only a few high-mass protostars; (\citealp{Peretto2013}; see also \citealp{Contreras2018,Redaelli2022,Olguin2023}).
It is also compatible with accretion rates derived from CO outflows for high-mass/intermediate mass protostars \citep{Duarte-Cabral2013,Maud2015,Avison2021}.
A young high-mass protostar is therefore expected to be a strongly accreting object. High rate accretion has always to be accompanied by similarly high rate ejection (e.g. \citealp{Cabrit&Bertout1992,Bontemps1996,Beuther2002,Duarte-Cabral2013,Maud2015}) to allow for angular momentum removal as predicted in the magneto-centrifugal jet/wind processes at disk and protostellar scales \citep{Ferreira2006,Mignon-Risse2021,Commercon2022}. Young highly accreting high-mass protostars should thus all be powerful outflow drivers. In addition since the observed cores are massive and dense, any jets or winds escaping from high-mass protostars should always be revealed as CO outflows entrained inside the cores themselves. 

We therefore adopt here for the criterium to recognize an excellent candidate to be a HM PSC that it should be a compact core ($\rm FWHM^{dec} < 5000 \, au$) massive enough to form a high-mass star (16 $\rm M_{\odot}$) and not driving any CO/SiO outflow, indicating that the core does not host yet any high-mass protostar with a large accretion rate. This definition of an excellent candidate to be a HM PSC may allow for the existence of low-mass protostars in the core with a global (adding all accretion rates of protostars) low accretion rate. This could result in low-mass protostars within these cores being misclassified as prestellar cores. We note that in the view promoted by \cite{Motte_review_2018}, the earliest phases to form high-mass stars would indeed correspond to cores hosting few low-mass protostars competing for mass from the global infall rate. We note, however, that this stage with only low-mass protostars in a high-mass core cannot last for long if the infall is globally large. Indeed, material would accumulate very fast in the protostars, with a large sum of all accretion rates leading to a sum of low-mass ejections which would cumulate as a detectable global outflow from the core (see discussion for the possible effects of fragmentation, case 6 in Sect.~\ref{subsection:alt-scenarii}).

We see in Fig.~\ref{12_massive_cores_outflows_highlighted} that these 12 high-mass PSCs (cores \#1A and \#1B of G333.60 are displayed as one core in this figure) are mostly situated in the crowded, central regions of the protoclusters where the confusion between outflows from nearby (also massive) protostars is the highest. This is expected since high-mass stars are believed to form mostly at the center of the densest and most massive clumps where several high-mass stars may form roughly at the same time. Despite this unavoidable confusion, we are confident that we have a robust selection due to our systematic approach. The On-Off spectra are expected to reveal any excess of outflow emission in CO or SiO centered on the source, and thus are expected to provide a reliable mean of detecting outflow emission coincident with the position of the core. We can certainly exclude for all these 12 high-mass cores that they could drive a powerful outflow as expected if they would be strongly accreting objects. 
%(see Sect.~\ref{subsection:alt-scenarii} for more discussion).

\subsection{Other protostellar tracers towards the HM PSC candidates}
 \label{sec:discuss:other-tracers}

One of the traditional ways to recognize a protostar is to probe mid-IR emission (typically below 100$\,\mu$m) to detect warm dust emission indicative of a significant internal heating inside the collapsing core/envelope. We cannot use the mid-IR probe towards the ALMA-IMF compact cores, since they are not resolved individually in the available Herschel or Spitzer maps (see e.g. \citealp{Gutermuth2009, Motte2010}). The spatial resolution of Herschel at 70$\,\mu$m or of Spitzer at 24$\,\mu$m used for instance by \cite{Bontemps2010} in Aquila are of the order of 6 arcsec, i.e. between 5 and 15 times (depending on the distance of our regions) larger than the angular size of the here studied cores. %The JWST/MIRI could reach the required scales below 1 arcsec at 25$\,\mu$m but we could not get observing time yet to map the ALMA-IMF fields. 
Alternatively, molecular line emission, such as hot-core tracers, could also trace the presence of warm gas in the cores. We have verified that none of our HM PSC candidates are hot cores in \cite{Bonfand2024}. A forthcoming paper (Valeille-Manet et al. in prep) will further investigate the molecular emission of the HM PSC candidates. A preliminary analysis show that all cores have chemical properties compatible with cold gas typically between 20 and 25\,K with clear sign of CO depletion and detections of N$_2$H$^+$ towards the cores.

\subsection{Comparison with previous high-mass PSC surveys} \label{subsection:unique_sample}
%\subsection{Robust high-mass PSC candidates} \label{subsection:unique_sample}

We have therefore discovered 12 robust HM PSC candidates above 16\Msun and 18 lower mass PSCs between 8 and 16\Msun (see Table~\ref{tab:MPSC_candidates}) as well as 52 protostellar cores above 8\Msun in the ALMA-IMF fields that we can then discuss further below.

%We present here a unique sample of MPSC candidates at a resolution of $\sim 2700 \, au$. Those candidates are mass reservoirs that do no show any trace of outflows and could possibly form protostars from several solar masses to high mass protostars. 
It is important to have discovered and catalogued a significant number of PSCs with potential mass reservoirs massive enough to qualify as high-mass star precursors.
The number of candidates appears large compared to what is known in the literature (see below) but still low compared to the total number of cores in the ALMA-IMF sample. The 12 robust candidates of the sample above 16\Msun represent $\sim2\%$ (12/584) of the total number of cores in \cite{louvet22}. %Using the best temperatures for the protostars (cores with outflows), i.e. the PPMAP dust temperature used in \cite{louvet22}, among the 145 cores we investigated we actually end up with only 51 protostars (with outflows) above 8\,M$_{\odot}$. The raw proportion of high-mass PSC candidates is then of $\sim20\%$ (12/63) which is a very small number compared to what is observed in low-mass star-forming regions where PSCs are typically several times to 10 times more numerous than protostars (see \citealp{Konyves2015}).%compared to the total number of high-mass cores (M\,$>$\,8\,M$_{\odot}$) of the ALMA-IMF survey appears to be consequent with a fraction of $\sim46\%$ (43/93). %MPSC candidates are hard to find and to identify especially in the crowded regions where cores and outflows overlap.

To our knowledge, no work has yet reached large enough statistics in high-mass SFRs to investigate well high-mass PSC candidates. First systematic surveys such as \cite{Motte2007} and \cite{Bontemps2010} in Cygnus X could only find a couple of high-mass cores (9 objects at a $\sim$2000 au scale in \citealp{Bontemps2010}) and only one candidate to be prestellar (CygX-N53-MM2) could be identified among them using CO outflows in \cite{Duarte-Cabral2013}. This candidate still stands as a probable intermediate mass PSC with a here revised mass of 9.5\Msun in the average convolved size of $3800\,$au of the ALMA-IMF PSC candidates, using the same dust temperature of 20\,K as in their study and our emissivity (see Sect.~\ref{subsection:4.1} and \ref{subsection:basic_properties}). Similarly \cite{Tan2013} led a survey in the first galactic quadrant towards cold IRDCs detected in N$_2$D$^+$, but their best candidate happened to correspond to two early stage protostellar cores (see \citealp{Tan2016}). \cite{Louvet2019} did not find any HM PSC candidate in NGC2264 either.
%We compare the mass range of our objects to that of other studies by adapting their estimates to the parameters used here. The high-mass PSC candidate of \cite{Barnes2023} is $31 M_{\odot}$ ($23 M_{\odot}$ with background subtracted) in the dragon cloud. Using the dust opacity presented in \ref{subsection:4.1} and their temperature, we obtain a mass of  $17.6~M_{\odot}$ ($13.2~M_{\odot}$ with background subtracted). However, adopting a temperature of 20~K, the mass of this PSC candidate decreases further to $7.0 M_{\odot}$ and $5.2 M_{\odot}$ with background removed. As discussed in Sect.\,\ref{subsection:4.1}, this core would not make it in our initial list of massive PSC candidates as it mass would fall below our threshold. 
In \cite{Barnes2023} one PSC candidate of $\sim$2200\,au was identified in the dragon cloud (core C2c1a) using 1.3\,mm continuum, with a mass of 23\,M$_{\odot}$. For a direct comparison with the present work, using the dust opacity presented in Sect.~\ref{subsection:4.1} and a temperature of 20\,K, the mass of this PSC candidate would actually be 5.2\,M$_{\odot}$ (a dust temperature of 10.4\,K is used in their study). 
\cite{Wang2014} used the SMA and similarly found a PSC candidate (G11.11-P6-SMA1) of 28\,M$_{\odot}$ and $\sim$2600\,au, which, using our dust emission flux to mass conversion and temperature, would actually have a mass of 7.4\,M$_{\odot}$ (a dust temperature of 11\,K is used in their study). As discussed in Sect.\,\ref{subsection:4.1}, these two cores would not make it in our complete list of 42 PSC candidates above 8~M$_{\odot}$. Finally, the ALMA project ASHES which mapped massive IR dark clumps did not succeed in finding high-mass PSC candidates with $\rm M>16\,M_{\odot}$ (see \citealp{Morri2023}). Their most massive cores without outflow have lower masses than 16\Msun and have deconvolved FWHM larger than the core sizes in our sample with an average factor of $\sim 5$. %Their two most massive cores without outflow have masses of 7.6 and 8.6\Msun using our dust emission flux to mass conversion but for deconvolved FWHM of $\sim$12000 and $\sim$17000\,au respectively (\citealp{Morri2023}), which are significantly larger than our core sizes. %Their most massive core without outflow but with warm gas tracer is slightly more massive (8.6 M$_{\odot}$) but is also much larger than the PSC candidates of our sample with a deconvolved FWHM of $\sim17000$ au.
%The detection of 42 PSCs among which 12 are prime candidates with a mass larger than 16\Msun is therefore an important improvement over what was obtained so far. It is an unique and extremely valuable sample of cores to further confirm as PSCs (no high-mass protostar formed yet) and to be used to properly investigate the earliest phase and initial conditions to form high-mass stars. 

\begin{figure*} [htbp!]
    \centering
    \includegraphics[width=0.80\textwidth]{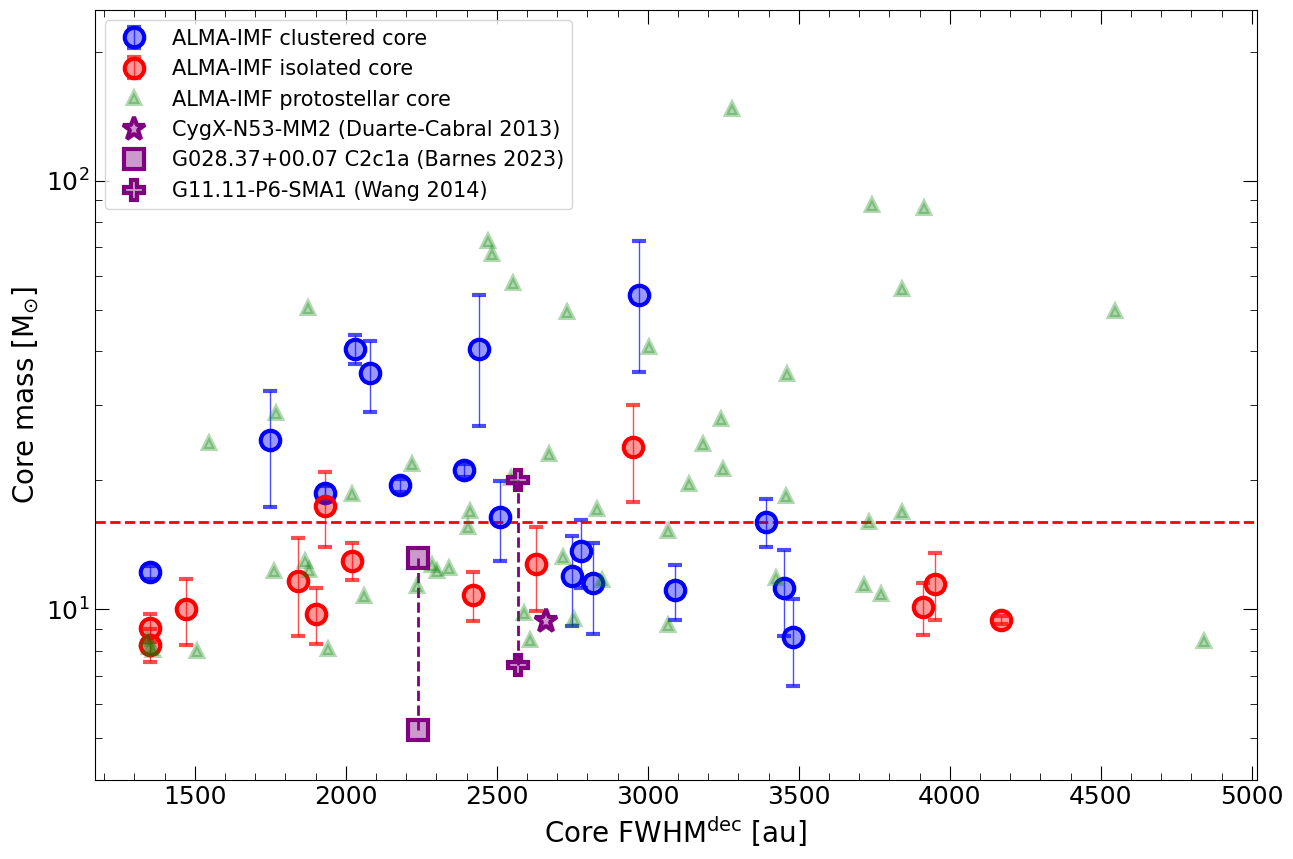}

    \caption{Mass distribution in function of the size of the total sample of 30 PSC candidates. Blue and red circles correspond to clustered and isolated candidates respectively. Errorbar symbols indicate the whole range of possible mass justified in Sect. \ref{subsection:basic_properties}, using 20\,K and PPMAP non corrected temperatures as lower and upper limits in temperature. The red dashed line corresponds to the threshold of 16\Msun. The markers in purple are the previous high-mass PSC candidates from \cite{Duarte-Cabral2013}, \cite{Barnes2023}, \cite{Wang2014} using the dust emissivity from the present work, a temperature of 20\,K and the temperature used in these studies. The prestellar core from \cite{Duarte-Cabral2013} has only one symbol because a temperature of 20\,K was already adopted in their work. Protostellar cores with M\,$>$\,8\,M$_{\odot}$ are shown in green triangles.
    }
   \label{fig:Mass_distrib}
\end{figure*}

In Fig. \ref{fig:Mass_distrib} we present the distribution of the masses and sizes of the 30 PSC candidates with M$\,>\,$8\Msun and added the candidates discussed above from the literature, where we recomputed their mass using our dust emissivity with a dust temperature of 20\,K and the temperature used in their studies (see the paragraph above). We clearly see that our new detections are probing for the first time the high mass regime to investigate the initial conditions prevailing to form high-mass stars (i.e. above 8\Msun). We note however that among the 12 robust high-mass PSC candidates, only 4 cores have mass reservoirs greater than 30\Msun most probably highlighting the difficulty to find cores in the highest stellar mass regime. 
% (only $\sim 3\%$ of the ALMA-IMF dust cores have M$\,>\,$30\,M$_{\odot}$).
Clustered and isolated candidates are displayed as blue and red circle points respectively (see Sect.~\ref{subsection:basic_properties} for details on this classification). %Robust and uncertain candidates are represented by filled and emptied circle points respectively.  
Protostellar cores with M$\,>\,$8\Msun are shown as green triangles. It appears that 10 of the 12 most massive candidates are found in a clustered environment as expected since these crowded regions are the most probable places to form highest mass stars. 
In contrast, a significant fraction of the cores below 16\Msun are found to be isolated with 11 of the 18 remaining cores having one or no high-mass core neighbor.

\section{Are high-mass PSCs really rare?}\label{section:HMPSC_rare}

So far an extremely small number of candidates to be HM PSCs have been recognized in the literature (Sect.~\ref{subsection:unique_sample}) suggesting that the compact high-mass cores in pre-collapse or at the very beginning of collapse/assembly without high-mass protostellar activity  yet are rare and may even not exist at all (see \citealp{Motte_review_2018}). 
We here actually detect some robust HM PSC candidates up to masses of $\sim50\,$M$_\odot$ (between 35 and 70\,M$_\odot$ accounting for the uncertain dust temperature). This indicates that high-mass PSCs exist up to relatively high masses. Since these masses are similar to those observed for the protostellar cores (see green triangles in Fig.~\ref{fig:Mass_distrib}), it suggests that our ability to probe even larger mass reservoirs might actually be limited by the statistics of cores we could detect in ALMA-IMF. 

%Using the large statistics of ALMA-IMF with 580 cores in the smoothed catalogs of \cite{louvet22}, we here find a number of good candidates to correspond to this so-called prestellar phase. %, and only in a couple of protoclusters (3/4 of the candidates are in 5 regions) with typically 10~\% of the detected cores being candidates to be HM PSCs (see previous section). This result therefore clearly revises our view of the pre-collapse phase of HM SF and allows us to envision detailed studies of the initial conditions for the collapse towards high-mass stars inside such detected objects. 
%We find no %There are no 
%HM PSC candidates above $\sim70\,$M$_{\odot}$, however such massive sources only represent $\sim0.7\%$ (4/583) of the whole ALMA-IMF sample. Despite being a representative sample of massive star forming protoclusters, neither protostellar (purely dust), nor prestellar cores probe well the extremely massive regime with only four sources of $\gtrsim$~70~M$_{\odot}$ within a 2700~au scale.
%leaving the question mark on the existence of extremely massive prestellar core candidates, but better statistics at such mass are needed to conclude about this type of cores. 
%TCS: I'm not sure that a larger statistics solves the problem. I think we have a problem of tracing the envelope mass.
Investigating the number statistics of prestellar versus protostellar cores, we find about half as many HM PSCs than high-mass protostars above 16\Msun with 12 [9-14]\footnote{range of number of PSCs above 16 \Msun from the range of uncertain mass for PSCs} HM PSCs and 27 high-mass protostars among a total of 39 high-mass cores; i.e. 69~\% [66-75~\%] of protostars among these 39 [36-41] cores. This fraction is statistically compatible with the value found by \cite{Nony2023} on the sub-sample obtained only in the three protoclusters of W43 with a value of $\sim 80\,\%$ of protostars. %These two values are statistically compatible with a lower value of the Poisson uncertainty range in \cite{Nony2023} of 68~\% and an upper value of our new estimate of 78~\%. \timea{IT IS COMPLETELY UNCLEAR WHAT YOU ARE TALKING ABOUT HERE. WE NEED TO TALK ABOUT THIS.}
We note also that the mass scale of the present work is slightly higher as we use a source catalog on smoothed images at a single physical size beam (for homogeneity reason over all regions).
%and the adopted temperatures of the PSCs in \cite{Nony2023} are also slightly larger with for instance an average value of 22 K for the three HM PSCs while we adopt 20 K. 
Since in \cite{Nony2023} the fraction of protostars is strongly dependent of the mass, the difference between these fractions is also explained by this shift toward slightly higher masses in the present work.

%With the number of HM PSCs being half the number of high-mass cores, it suggests that the corresponding lifetime of HM PSCs is half of that of HM protostars. \timea{I do not REALLY UNDERSTAND THE POINT OF THIS SENTENCE, YOU WILL DISCUSS THIS IN MORE DETAIL WITH THE FIGURE. } 

Typically low-mass star forming regions host significantly more numerous prestellar than protostellar cores.
%Typically in low-mass star formation, the prestellar cores are found significantly more numerous than the protostars. 
%In \cite{Konyves2015} 
In Aquila, \cite{Konyves2015} found a total of 446 prestellar cores (gravitationally bound cores without 70~$\mu$m emission) for only 58 protostars, i.e a factor of 8 times more prestellar cores to be compared to the factor $\sim$0.5 found here for HM PSCs.
%Using the ratio of prestellar cores (gravitationally bound cores without 70~$\mu$m emission) to the number of young stars (Class II objects) \cite{Konyves2015} estimate a prestellar lifetime of 1.2 Myr i.e. 6 times larger than what we obtain here in the high-mass regime. 
The high-mass PSCs are thus indeed significantly rarer than their low-mass counterparts, and are globally extremely rare due to the rarity of high-mass cores, protostars, and stars. 

The reason why previous surveys for HM PSCs could not detect almost any good candidates is probably two-fold. First they are indeed very rare, slightly rarer than high-mass protostars (driving the need for large samples of high-mass cores), and second they are mostly clustered with the high-mass protostellar objects in dense clumps making their identification extremely difficult (confusion with the strong outflows from the very nearby high-mass protostars). We see that most of the HM PSCs above 16\Msun are indeed found clustered in the central clumps of the regions (see Fig.~\ref{fig:Mass_distrib}). 
%\timea{IN THE NUMBER STATISTICS ONE SHOULD INCLUDE THE CERTAIN VERSUS INCERTAIN CANDIDATES.}
 
%They are also 2600 times rarer than low-mass prestellar cores.
%ans we look at the ratio of MPSC candidates to number of massive cores characterized as protostellar for different mass bins. 

%We can then use the below formula \ref{lifetime_formula} to estimate the lifetime of the HM PSC phase where $N_{pre}$ is the number of massive prestellar cores, $N_{proto}$ is the number of massive protostellar cores (including free-free sources), and $t_{proto}$ is the lifetime of the massive protostellar phase: 

%\begin{equation}
%    \centering
%    t_{pre} = \frac{N_{pre}}{N_{proto}} \times t_{proto}   
%    \label{lifetime_formula}
%\end{equation}

%Until this paper, the relatively low number of MPSC candidates discovered has been explained by a massive prestellar lifetime much shorter than the protostellar one. As discussed above, one of the main goal of this work is to estimate the lifetime of the massive prestellar phase thanks to our first statistical sample of MPSC candidates. To do so, we use formula \ref{lifetime_formula} where $N_{pre}$ is the number of massive prestellar cores, $N_{proto}$ is the number of massive protostellar cores, and $t_{proto}$ is the lifetime of the massive protostellar phase: 

%-----------------------------------------------------------------
\section{Statistical lifetimes of high-mass prestellar cores} \label{sec:lifetimes}

The number of PSC candidates can be used to estimate the statistical duration of the prestellar phase to form high-mass stars providing that we can choose a favored protostellar lifetime for high-mass stars.

\subsection{Accretion phase and protostellar lifetimes}
\label{subsection:protostellar-lifetime}

The lifetime of (high-mass) protostars which should correspond to the main accretion of the final star is not a well determined quantity due to the difficulty to estimate it observationally or theoretically. Statistical estimate of protostellar lifetimes was firstly introduced by \cite{Beichman1986} who studied the embedded young stellar objects (YSOs) in the sample of cores of \cite{Myers_Linke1983}. Only few papers focused on the lifetime of the massive protostellar phase. %(i.e. the timescale during which the protostar is accreting). 
\cite{Duarte-Cabral2013} adopted 300 kyr based on the observed trend that the protostellar lifetime for the intermediate/high mass regime is similar to the low-mass value which was debated to range from 200 to 400 kyr, and thanks to indirect estimates of accretion rates using CO outflows. 
% after observing a direct proportionality between momentum flux of CO outflows of massive objects and their mass reservoirs from low to intermediate/high masses suggesting a similar accretion timescale from low to high-mass regime. 
Independently and in the high-mass regime, \cite{Mottram2011} surveyed massive YSOs and compact HII regions at the scale of the galactic plane. Their obtained luminosity distribution were used to compare the number of massive YSOs (high-mass protostars) and of HII regions with the total number of massive stars. This allowed them to estimate statistical lifetimes as a function of the luminosity of the objects (i.e. equivalent to the final mass of the object). 
At low luminosities (B0.5 stars, lowest masses) the obtained lifetime of 500 kyr represents the total protostellar time. At higher luminosities a part of the compact HII regions can actually be still accreting protostars and should then be included in the protostellar lifetime (see below in Sect.~\ref{subsection:missing_UCHII}). Interestingly enough \cite{Nony2024} in the W49 region found that of the order of one third of the UC HII regions are still dust emission cores and are possibly still protostellar (accreting) objects. Adding massive YSO and one third of the compact HII region lifetimes from \cite{Mottram2011} we then get a lifetime of the order of 200 kyr for O7-O8 final stars. It also suggests that the protostellar lifetime may decrease with mass from 500-600 to 100-200 kyr. In W49, \cite{Nony2024} derived a protostellar lifetime of 140 kyr. %but which is actually seen as a lower limit. 

Altogether, accounting for above constrains and uncertainties, we adopt as a baseline a constant protostellar lifetime of $300\pm100$ kyr. At one point of the discussion we will also consider the possibility of a decrease of this time for highest masses that we will adopt from 300 to 100 kyr for final mass from 8 to $\sim 30$\Msun.  %Adding these two phases, they obtain a relatively constant lifetime of around 0.5 Myr over the whole range of stellar masses (luminosities) containing massive YSOs. On the A part of the 
%In the following we then adopt an intermediate value for the protostellar lifetime $t_{\rm proto}$ of 0.4 Myr. 

\subsection{Mass dependence of the statistical lifetime}
\label{subsection:lifetime-clump-fed}

Down to a core mass of 8\Msun using our adopted dust temperature (Sect.~\ref{subsection:basic_properties} and Tab.~\ref{tab:MPSC_candidates}), we detect a total of 30 PSC candidates. Compared to the total of 52 protostellar (with outflows) cores above 8\Msun which correspond to a statistical time of 0.3 Myr, we then get a global statistical time for PSCs above 8\Msun of $\sim 170\,$kyr. These 30 candidates also provides a statistical basis to investigate possible mass dependencies of the fractions and statistical lifetimes of prestellar cores in the observed mass regime. In \cite{Nony2023} with 8 PSCs (5 above 8\Msun and 3 above 16\,M$_{\odot}$) in the W43 regions, a trend of a decrease of the fraction of PSCs toward the high-mass regime was observed. We can here go one step further with a larger statistics.

In Fig. \ref{fig:histo_lifetime_no_correction} left panel we show the histogram of
%we have then built an histogram of 
the number of prestellar cores (blue histogram) compared to the number of protostellar cores (red histogram). % shown 
%in a cumulative way (the red histogram is put on the top of the blue histogram). 
To compute the masses, PPMAP temperatures are taken for protostellar cores, while we use our best temperature estimates (see  Sect.~\ref{subsection:basic_properties} and Tab.~\ref{tab:MPSC_candidates}) for the prestellar cores.
%The bins are constructed in order to have 10 cores per bin, except for the last one with MPSC candidates which contains the six most massive ones above $30~M_{\odot}$, leading to 6 cores in the second to last bin containing candidates. 
Assuming a 50~\% core to star efficiency (hereafter $\epsilon_{\rm cse}$), we have indicated the corresponding final mass of the stars on the upper x-axis. Between core masses of 8 and $\sim$16\Msun (two first bins) we obtain a relatively similar number of prestellar and of protostellar cores (fraction PSCs/proto close to 1) and with a trend of a decreasing fraction towards high masses. This fraction continues to decrease above 16\Msun to a fraction of PSCs more of the order of 0.5 of the number of protostars. In terms of lifetimes (green points in Fig.~\ref{fig:histo_lifetime_no_correction}), this suggests a decrease from $\sim 300$~kyr to $\sim$150~kyr in the highest mass regime which for $\epsilon_{\rm cse}$ of 50~\% would correspond to final masses of the order of 20-30\Msun (see upper x-axis of Fig.~\ref{fig:histo_lifetime_no_correction}). We note that we do not detect HM PSC candidates in the last bin of mass of Fig.~\ref{fig:histo_lifetime_no_correction} leading to a large uncertainty on the resulting upper limit of lifetime for these highest masses (last green point in the figure). 

\subsection{Correction for missing UCHII protostars} \label{subsection:missing_UCHII}

Some of the bright millimeter cores in the survey have a clear contribution at 3~mm from free-free emission, pointing to the existence of young ionising high-mass protostellar objects, i.e. of so-called Ultra compact HII (UCHII) regions inside the detected cores\footnote{Free-free contamination could also originate from other mechanisms such as radiojets \citep{Anglada1996}.}. Some of these free-free contaminated millimeter sources are expected to be still accreting. High-mass stars have indeed to still strongly accrete mass while being already an ionising protostar so that they can reach masses well above 10 M$_{\odot}$ \citep{Yorke2002,Krumholz2007,Keto2007,Tanaka2016}. For the most massive stars, it is even a significant fraction of the protostellar lifetime (accretion phase) which is expected to appear as an UCHII region \citep{Peters2010}.
In the \cite{louvet22} study, 68 such compact free-free sources are found, i.e. of the order of 10~\% of the whole sample of cores. 
%Some of these sources actually correspond to fluctuations of free-free emission on the border of the few developped HII regions seen in some of the ALMA-IMF fields (mostly in W51-IRS2, G012.80, W43-MM3 and G333.60). 
These free-free contaminated cores are either high-mass protostars that are still accreting, or slightly more evolved objects, i.e. main sequence stars with accretion finished with a remaining compact halo of ionised gas (see for instance \citealp{Nony2024} in W49). We note that with our outflow detection tool we indeed found some outflows around a few of these bright free-free contaminated compact objects in the survey confirming the protostellar nature of these objects. We even decided to add to our list of high-mass protostars the three least free-free contaminated objects initially excluded from the catalogs of \citet{louvet22}, those with a spectral index close to two (cores \#5 and \#9 of W51-E and core \#5 of W51-IRS2) and found them indeed associated with an outflow\footnote{Since they have some weak free-free emission, they cannot be prestellar cores and were then expected to be driving an outflow.}. %These 3 additional protostellar cores gives a total of ALMA-IMF dust cores of 584. % in our protostellar cores sample.  The presence of this significant population of free-free contaminated protostars validates the nature of the targeted protoclusters as nurseries for high-mass stars since only high-mass protostars can ionise UCHII regions.

\begin{figure*} [htbp!]
    \centering
    \includegraphics[width=0.49\textwidth]{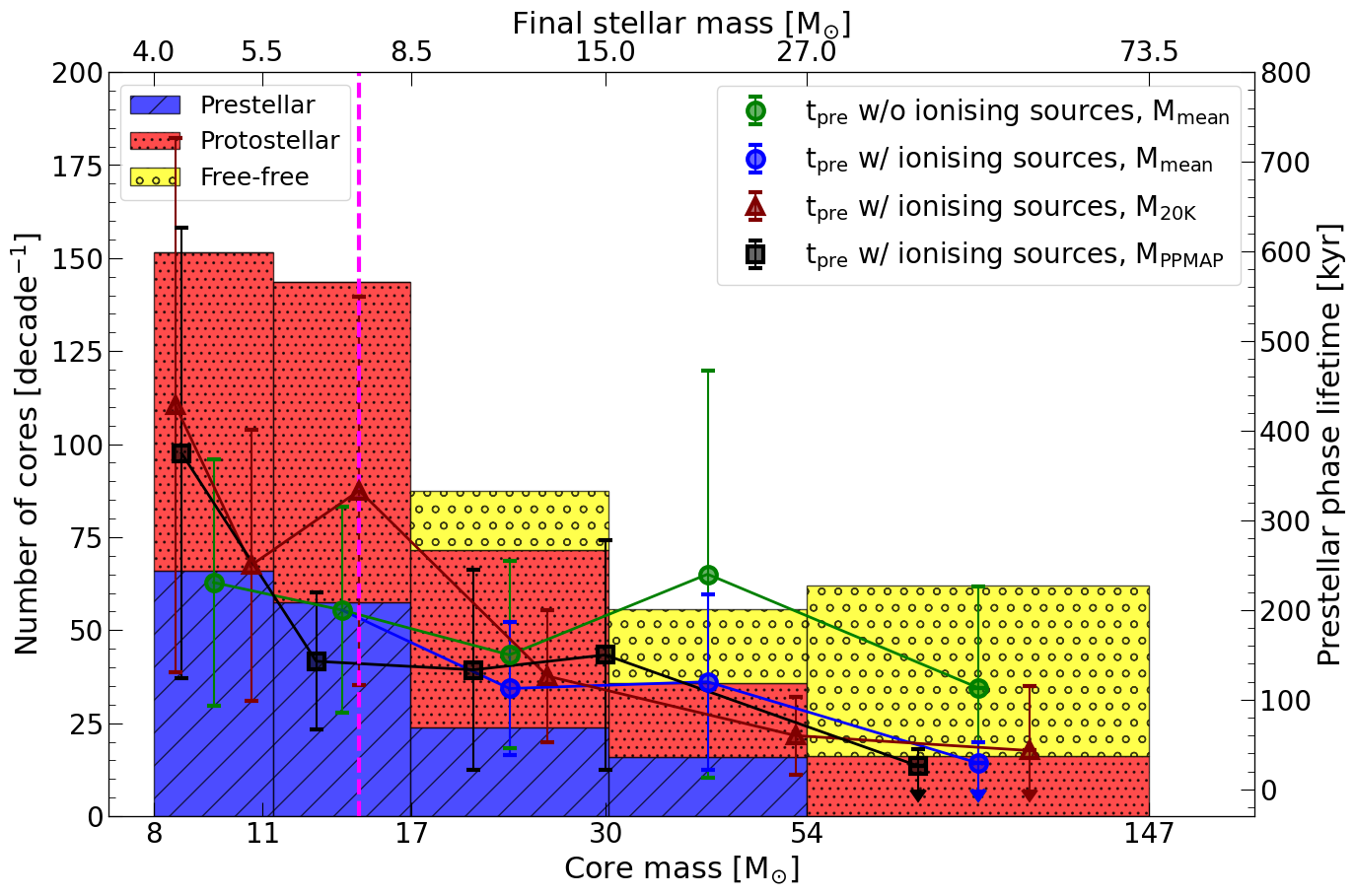}
    \includegraphics[width=0.49\textwidth]{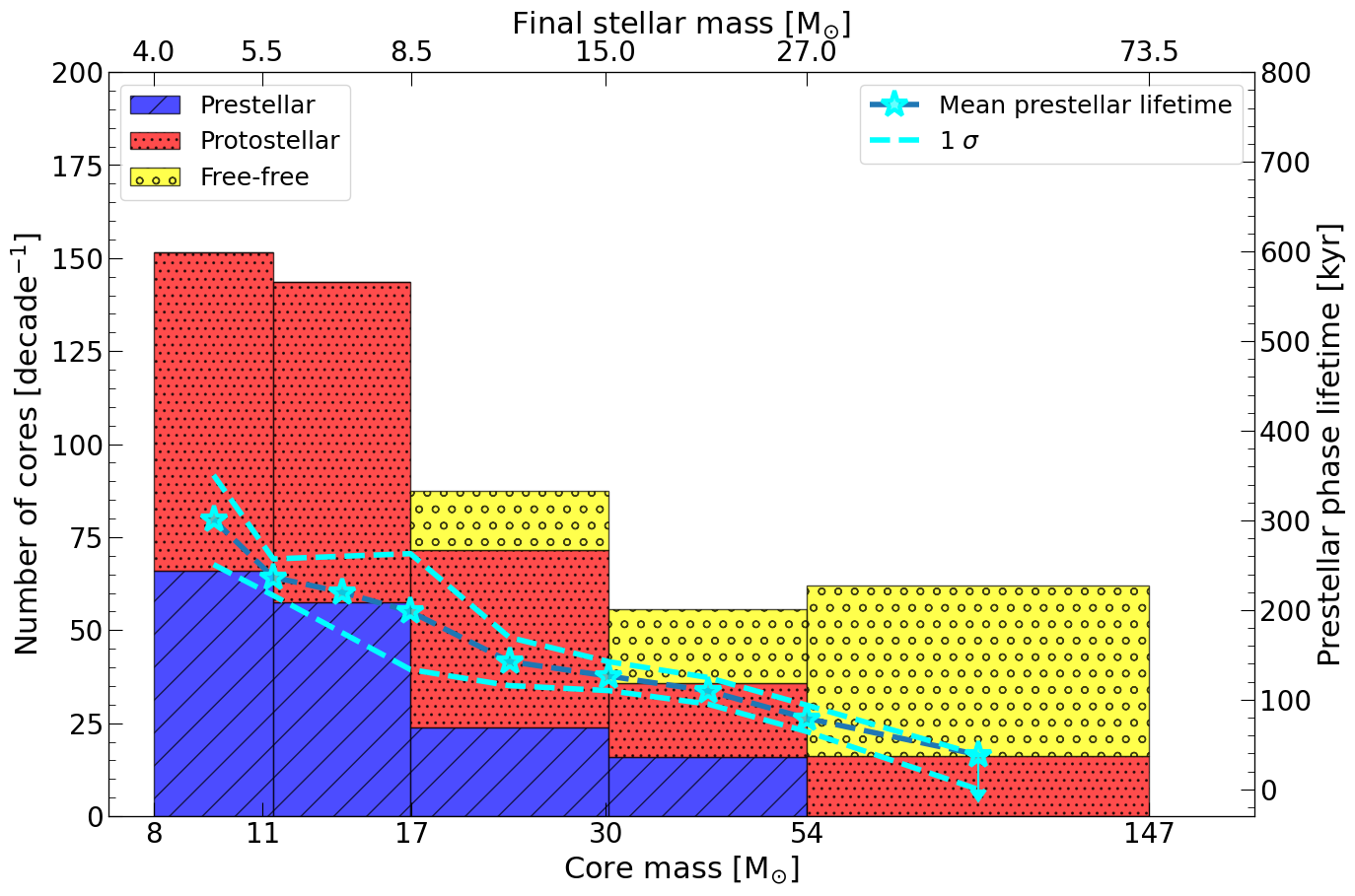}

    \caption{\textbf{Left:} Weighted histogram of the number of prestellar and protostellar cores per bin and lifetime of the (massive) prestellar phase for each bin (points). The left y-axis indicates the density of number of cores per decade. Prestellar cores are shown in blue bars, protostellar cores in red, and free-free sources in yellow. Bins are built using the average mass adopted in Table~\ref{tab:MPSC_candidates}. Prestellar lifetimes (right y-axis) are presented with green (without ionising sources) and blue (with ionising sources) points for the average mass adopted. We also display with brown and black points the histograms obtained using PSC masses for the lowest and highest dust temperatures respectively we considered in Table~\ref{tab:MPSC_candidates}. The magenta dashed line represents the threshold in mass of the protostar to start ionising its envelope. The upper x axis represent the final stellar mass assuming a core to star efficiency $\epsilon_{\rm cse}$ of 50\%. \textbf{Right:} Same histogram with the mean prestellar lifetime in cyan stars extracted along the range of PSC masses (blue, brown and black points in left panel). Cyan dashed lines represent the one sigma dispersion.}
   \label{fig:histo_lifetime_no_correction}
\end{figure*}

%The protostellar time (accretion time) should therefore also comprise a phase during which the protostar is already massive enough to burn hydrogen and to be therefore an ionising star while still growing by accretion. For final masses above $\sim8-10$\Msun (right of the pink vertical line) the protostar is expected to be ionising before the end of the accretion/protostellar phase. 
The more massive a star is going to be, the larger is the fraction of its accretion time as an ionising protostar (ultra or hyper compact HII regions, hereafter UCHII regions). These ionising protostars are not included in the sample of cores of \cite{louvet22} due to the fact that it is not trivial to evaluate a proper dust emission at 1.3~mm to estimate their envelope masses. They are therefore not accounted for in the red histogram of Fig.~\ref{fig:histo_lifetime_no_correction} and in the PSC lifetime estimate. In order to compensate for these missing high-mass protostars we have estimated for each bins of final stellar masses of Fig.~\ref{fig:histo_lifetime_no_correction} the number of expected UCHII protostars. For that we use the model of protostellar evolution of \cite{Hosokawa2009} as expressed in \cite{Duarte-Cabral2013}. Assuming a constant accretion rate and a core to star efficiency $\epsilon_{\rm cse}$ of 50\%, for each final stellar mass bin we can estimate the percentage of lifetime a protostar is going to be ionising as an UCHII region (see appendix \ref{appendix:ionising_models}). This leads to a number of free-free contaminated protostars in each bin, as an additional fraction of the number of observed non-ionising protostellar cores (red histogram). The number of these expected ionising protostars are shown as a yellow histogram in Fig.~\ref{fig:histo_lifetime_no_correction}. This larger number of protostars in the highest mass bins then decreases the fraction of PSCs in the highest mass regime and decreases the corresponding PSC lifetime.

%Using the same method we also compute the prestellar lifetime using the lower and upper limits in mass of the PSC candidates (presented in Sec. \ref{subsection:basic_properties}), that we show in brown and black points and curve in Fig. \ref{fig:histo_lifetime_no_correction}. This finally gives us a range of prestellar lifetimes evolving with the mass of the PSC candidates, showing a global decreasing trend as the mass of the PSC candidates increase.

Adopting this correction we have plotted in Fig.~\ref{fig:histo_lifetime_no_correction} the lifetime values for all mass bins (blue connected points). In addition we show the dispersion in the histograms resulting from the range of possible masses for PSCs (see Tab.~\ref{tab:MPSC_candidates}) as black and brown connected points. Together with the uncertainty bars for each points reflecting the Poisson noise, we then get a view of the evolution of the lifetimes with masses and of the dispersion of the obtained values due to PSC mass uncertainties. A decrease of the lifetime is now visible with the correction while without correction there was a tendency of a flattening (connected green points). In the right panel of Fig.\,\ref{fig:histo_lifetime_no_correction} we present the mean PSC lifetime extracted from the three hypotheses of the left panel with its one sigma dispersion. The values in ranges of mass of this mean PSC lifetime and its dispersion are given in Tab.\,\ref{tab:lifetime}. We see that the decrease now goes from approximately 300\,kyr in the lowest masses to 100\,kyr in the highest mass regime of detected PSC candidates of 40 to 50 \Msun (corresponding to final stellar masses from $\sim 5$ to $\sim$ 25\Msun for $\epsilon_{\rm cse}$ of 50~\%).

%\subsubsection{A short massive prestellar lifetime}
%\subsection{A corrected lifetime for the high-mass prestellar phase}
\subsection{Accounting for the evolution of protostellar core masses} \label{subsection:evolution_protostellar_masses}

%\textcolor{violet}{Explain the difference between the classical ratio between NPSC / Nproto to discuss the lifetime for PSC which works OK for a complete survey of proto and of PSC above a final mass threshold. Here in the the previous section, we use as proxy for the final mass the present-day mass of the envelope. This would correct of the mass of the envelope stays roughly constant as the protostar grows. Alternatively one could assume that the envelope mass decreases as the protostar evolved and grows in mass by accretion. In this case the infla mass proxy as the present-day mass of envelope could bias the final mass to larger masses for protostars compared to PSCs. This is particularly important here as we propose a discussion of the lifetime as a function of the final mass in relatively small bins of mass. } 

So far we have compared the number of PSCs to the number of protostellar cores above the same threshold in core masses and for the bins of Fig.~\ref{fig:histo_lifetime_no_correction}. 
However, in the simplest view of an initially set mass reservoir (without strong replenishment during protostellar times, i.e. a core-fed view), the protostellar core masses (also referred as protostellar envelope) are expected to decrease as the inner protostar accretes and ejects material. 

%The fraction of PSCs and the PSC lifetimes in Fig.~\ref{fig:histo_lifetime_no_correction} have been obtained assuming that the protostellar core/envelope masses stay constant over time which is possible if flows of gas from larger (clump) scales continuously refill the core while accreting and ejecting material.Alternatively in a more classical (usually assumed for low-mass star formation) view the core/envelope mass decreases due to accretion/ejection until it ends up empty at the end of the accretion (protostellar) phase.

%As we estimate the statistical lifetime, we compare the number of prestellar sources with the number of protostellar ones using the mass of their envelopes. 
%However, protostellar cores have already begun the accretion on the protostellar embryo, causing the loss of part of the envelope due to the accretion and the outflows of the protostar. 

Accounting for a constant rate for the decrease of core/envelope mass and constant accretion rate, a protostar is statistically observed at half of its protostellar time which would then correspond to be at half of its initial core mass (i.e. its original mass at the stage of PSC). We have revised Fig.~\ref{fig:histo_lifetime_no_correction} to account for such a possible decrease in protostellar core masses by doubling the present day mass of protostellar cores (as a proxy for their initial core mass). The result of this correction is displayed in Fig.~\ref{fig:histo_ratio}. In practice, however, this simple adjustment has to be slightly revised for the protostars which have a significant part of their accretion time appearing as ionising protostars. For all mass bins for which a part of the protostars are expected to be ionising (bins with a contribution of ionising objects displayed as a yellow histogram), we applied an adapted correction. It accounts for this shorter protostellar time as a non ionised protostar to evaluate the shift in average envelope mass to apply when counting only non ionised high-mass protostars. The detailed of this correction calculated using the model of protostellar evolution of \cite{Hosokawa2009} and expressed in \cite{Duarte-Cabral2013} are provided in Appendix \ref{appendix:ionising_models}. Also in order to properly sample the two first bins below 16\Msun (i.e. below 8\Msun before doubling mass) we had to consider cores down to 4\Msun for PSC/protostellar classification using CO/SiO outflow detections. We find a total of 48 additional protostellar cores associated with an outflow with masses between 4 and $\rm 8\,M_{\odot}$.
%To build this figure, we thus search for outflows for xx cores between 4 and 8\Msun among which xx were found to drive an outflow and then to be protostellar cores. 

In Fig.~\ref{fig:histo_ratio} we show that compared to the case of a constant core mass which can correspond to a continuous replenishment of the core mass in a clump-fed view, the number of prestellar cores is now always smaller than the number of protostellar cores. A fraction of 0.3 to 0.5 for mass bins without ionising sources is observed, and this fraction decreases to less than 0.25 above an initial core mass of 16\Msun in the more realistic case which includes the expected number of ionising protostars (yellow histogram and blue, black and brown points). We note that the total number of ionising protostars is of 49 objects, and appears as a reasonable value since the total number of possible ionising sources found in \cite{louvet22} was 68 objects. As for Fig.\,\ref{fig:histo_lifetime_no_correction} we present in the right panel of Fig.\,\ref{fig:histo_ratio} the mean PSC lifetime extracted from the left panel with values presented in Tab.\,\ref{tab:lifetime}. In terms of lifetime in this scenario of a progressive decrease of the core mass as the protostars evolve, we obtain a decrease from $\sim150$\,kyr to $\sim50$\,kyr from 4 to $\sim 25$\Msun final stellar masses.
%and accounting for the binning and PSC mass uncertainties (dispersion of blue, black and brown points and Poisson error bars), we then see that the PSC lifetime would progressively decreases from $170$ [130-220]~kyr to close to $50$ [30-70]~kyr from 4 to $\sim 25$\Msun final stellar masses. 

%\subsection{A corrected lifetime for the high-mass prestellar phase}
\subsection{Prestellar lifetimes and free fall times}
\label{subsection:lifetime_ff-times}

The obtained HM PSC lifetimes of the order of 100~kyr (see above) are therefore clearly lower than the values measured for low-mass star formation, with a best value of 1.2 Myr in Aquila by \cite{Konyves2015}. Such 10 times shorter times confirm that the growth of compact mass reservoirs to form high-mass stars has to be faster than that of low-mass stars while having to collect larger amount of material to form high-mass stars. This leads to the need for fast contraction and/or rapid converging flows of gas. Starting from typical densities in dense clumps\footnote{Values obtained for the central clumps of for instance G338 or W43-MM2 in Tab.1 of \citealp{Fred2022}} in the observed protoclusters of $\sim10^5\,$cm$^{-3}$, a 30\Msun sphere has a radius of 0.11 pc, and travelling over 0.11 pc during $10^5\,$yr leads to a velocity of $1.1\,$\kms which is clearly supersonic (at a temperature of 20\,K). %At the other extreme, interpreting the 0.1 Myr lifetime as a crossing time over the size of the PSCs leads to a velocity of $\sim 0.13\,$km/s, i.e. close to the sound speed in cold gas. 
This velocity of 1.1 \kms is of the order of magnitude of the observed gas dispersions in the clumps of ALMA-IMF \citep{Cunningham2023}. Moreover, in particular in G353, where we detect one PSC, the previous study by \cite{Alvarez-Gutierrez2024} shows that this core is located in the kinematically complex convergence point of 3 filaments and near a "V-shape" in position-velocity (PV) space (V-shape "F" in \cite{Alvarez-Gutierrez2024}, see their Table~D.1 and Fig.~4). These PV structures are interpreted in this study as cold $\rm N_2H^+$ inflows onto cores or dense regions in G353.  In particular, V-shape F has an associated inflow timescale based on the PV structure of about 63~kyr, broadly consistent with the timescales presented above, and that remains far longer than the $\rm t_{ff}$ estimate for this core (7\,kyr, see Tab.\,\ref{tab:MPSC_candidates}). This consistency in the contradiction between timescales albeit in one core in G353, derived from completely independent methods, appears remarkable and tends to confirm the results presented here and in \cite{Alvarez-Gutierrez2024}.  %We note that \cite{Alvarez-Gutierrez2024} also found, broadly over the G353 system, $\rm N_2H^+$ velocity structures that were consistent with slow collapse compared to $\rm t_{ff}$, indicating that the mechanisms slowing collapse in the PSCs may also be in operation on larger, protocluster, scales (see below). 

\begin{figure*} [htbp!]
    \centering
    \includegraphics[width=0.49\textwidth]{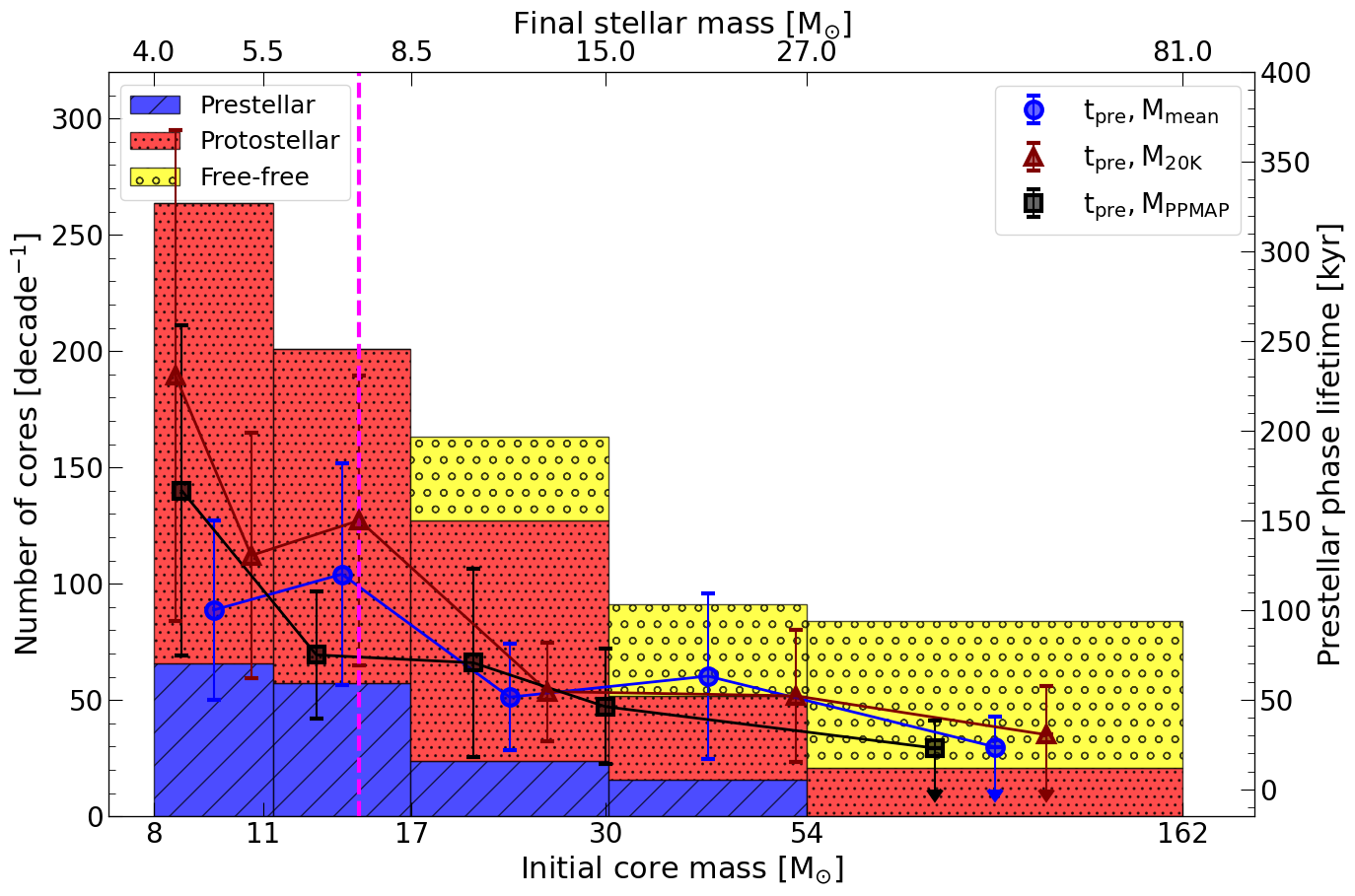}
    \includegraphics[width=0.49\textwidth]{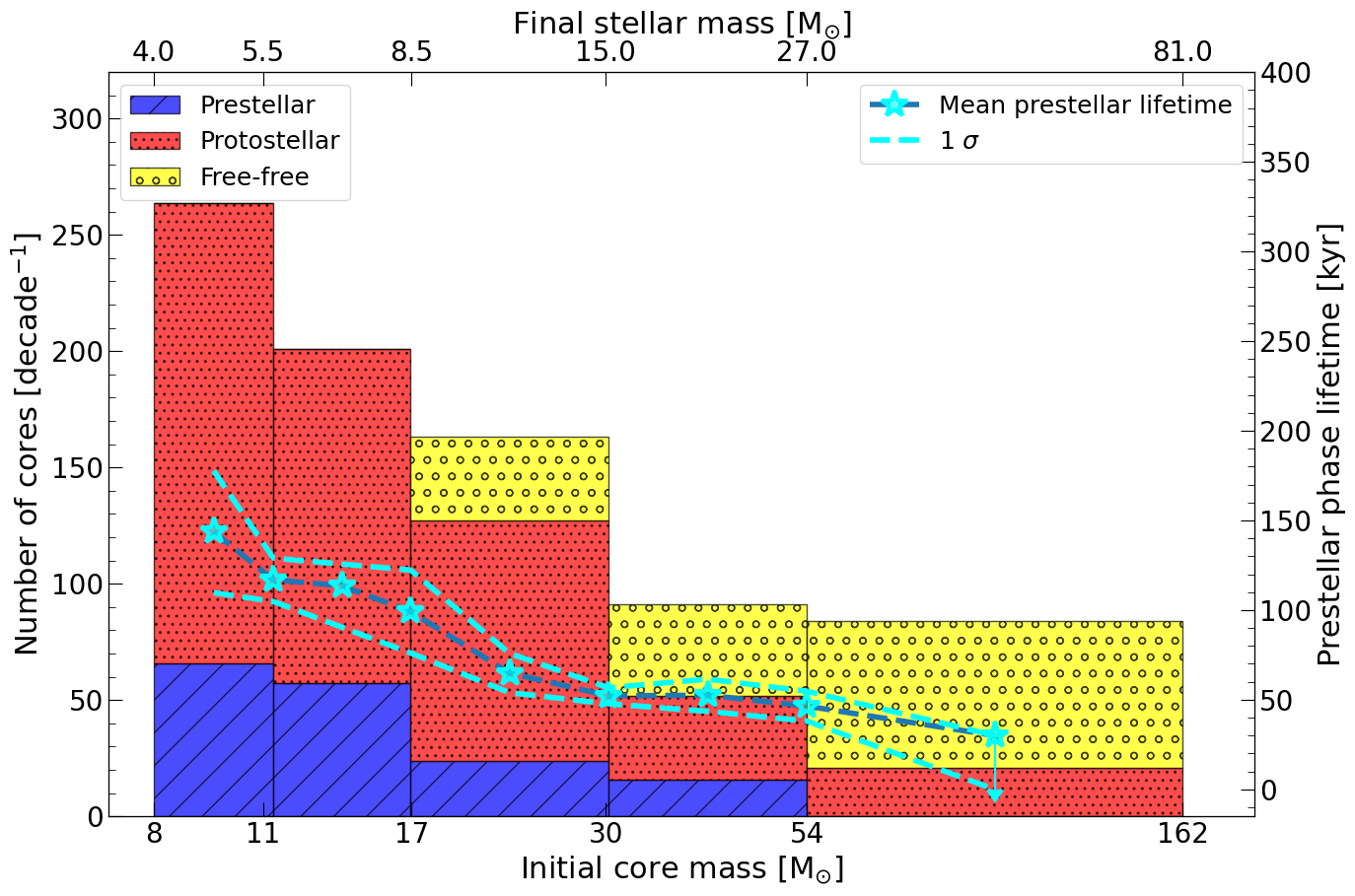}
    
    \caption{\textbf{Left:} Weighted histogram of the number of prestellar and protostellar cores per bin (left y-axis) and lifetime of the (massive) prestellar phase (right y-axis) for each bin (points). Prestellar cores are shown in blue bars, protostellar cores in red, and free-free sources in yellow. Bins are constructed using the average mass adopted in Table~\ref{tab:MPSC_candidates}. Protostellar cores masses are corrected assuming a decrease of the core mass envelope over time (see Sect. \ref{subsection:evolution_protostellar_masses}), with a constant accretion and a core to star efficiency $\epsilon_{\rm cse}$ of 50\%. Prestellar lifetimes are presented with blue points for the average mass adopted, black and brown points for the lower and upper mass limits. The magenta dashed line represents the threshold in mass of the protostar to start ionising its envelope. The upper x axis represent the final stellar mass assuming a core to star efficiency $\epsilon_{\rm cse}$ of 50\%. \textbf{Right:} Same histogram with the mean prestellar lifetime in cyan stars extracted from the three hypotheses of left panel. Cyan dashed lines represent the one sigma dispersion.}
   \label{fig:histo_ratio}
\end{figure*}

The above results are also compatible with the velocities of the so-called converging flows of \cite{Csengeri2011} in Cygnus X for instance (see also \citealp{Galvan-Madrid2010}).
%Such short crossing times at the clump scales in the high-mass regime points to dynamical processes to form the here observed compact HM PSCs. 
At a density of $10^5\,$cm$^{-3}$ in the clump, the free-fall time is actually $\sim100\,$kyr which is close to the observed HM PSCs lifetime. %It suggests that the probable clump scale dynamical process which may explain the formation of the HM PSCs can be driven mostly by the local clump scale gravity. 
In contrast the free-fall times at the scale of the PSCs are much lower than 100 kyr since the average densities have significantly increased. The obtained densities range from $\sim 2 \times 10^6\,$ to $1.5 \times 10^8\,$cm$^{-3}$ leading to free-fall times as low as $\sim 3$ to 20 kyr (see Table~\ref{tab:properties_PSC}). They are more than 10 times lower than the typical observed statistical lifetime we get for these HM PSCs. We have summarized the  lifetimes of Fig.~\ref{fig:histo_lifetime_no_correction} and Fig.~\ref{fig:histo_ratio} in several ranges of mass and the corresponding average free-fall times and densities in Table~\ref{tab:lifetime}.

%A contraction through magnetic field, i.e. driven by ambipolar diffusion, can probably not explain the short times at the scale of 0.1 pc in the clumps since the ambipolar timescale at 10 since the typical ambipolar timescale remains larger than 0.1 Myr, of the order of 0.3 yr, even at $\sim10^5\,$cm$^{-3}$ is of the order of 3 Myr (see Fig.~8 in \citealp{Bergin2007}; the ambipolar time scales as roughly 30 times the free-fall time), much larger than the observed crossing times which are more of the order of 0.1 Myr (0.11 pc at 1 km/s; see above). In contrast at the high densities of the observed cores above $\sim10^7\,$cm$^{-3}$ the ambipolar diffusion time reaches values of a few 0.1 Myr (0.3 Myr at $10^7\,$cm$^{-3}$ in \citealp{Bergin2007}) slightly larger but close to the observed lifetimes of the HM PSCs.

%This is in agreement with previous studies which had difficulties to find MPSC candidates, confirming that finding these cores is difficult and that large statistics are needed. In opposition to low mass objects, we also find less massive prestellar than protostellar cores, increasing the difficulty to find MPSC candidates as it is easier to detect protostellar cores.

%\pagestyle{empty}
%\begin{landscape}
\begin{table}[htbp!]
\centering
%\small
\setlength{\tabcolsep}{2pt}
\begin{threeparttable}[c]
\caption{Prestellar phase lifetime.}
%(using global measurements)}
\label{tab:lifetime}
%\begin{tabular}{ccccc}
%\hline \noalign {\smallskip}
%Mass range  & \# of  PSC\tnote{1}  & $\left<\rm n_{\rm H2}\right>$ \tnote{2} & $\rm t_{\rm PSC}$\tnote{3} &  $\left<\rm t_{\rm ff}\right>$ \tnote{4}\\   \noalign {\smallskip}
%$\left[\rm M_{\odot} \right]$ &      &  $\left[\times 10^7 \rm cm^{-3} \right]$ &   $\left[\rm kyr \right]$    &  $\left[\rm kyr \right]$ \\

\begin{tabular}{ccccc}
\hline \noalign {\smallskip}
  &  Mass range  & $\left<\rm n_{\rm H_2}\right>$ \tnote{1} &  $\left<\rm t_{\rm ff}\right>$ \tnote{2}  & $\rm t_{\rm PSC}$\tnote{3} \\   \noalign {\smallskip}
  &  $\left[\rm M_{\odot} \right]$  &  $\left[\times 10^7 \rm cm^{-3} \right]$    &  $\left[\rm kyr \right]$  &   $\left[\rm kyr \right]$  \\

\hline 

\multicolumn{1}{c|}{\multirow{3}{*}{\begin{tabular}[c]{@{}c@{}}Constant \\ protostellar \\ envelope\end{tabular}}} &  8 - 16 & 2.0  &  7.1 &  240 [190 - 290] \\
\multicolumn{1}{c|}{}  & 16 - 30  & 3.8  &  5.1 &  160 [120 - 200]\\
\multicolumn{1}{c|}{}  & 30 - 55  &  7.0  &  3.8  &  100 [80 - 120] \\

\hline 

\multicolumn{1}{c|}{\multirow{3}{*}{\begin{tabular}[c]{@{}c@{}}Decreasing \\ protostellar \\ envelope\end{tabular}}} & 8 - 16 & 2.0  &  7.1  &  120 [90 - 150]   \\
\multicolumn{1}{c|}{}  & 16 - 30 & 3.8  &  5.1  &  70 [50 - 90] \\
\multicolumn{1}{c|}{}  & 30 - 55 &  7.0  &  3.8  &  50 [40 - 60] \\
        
\hline
\end{tabular}

\begin{tablenotes}
\item[1] Mean volume density
\item[2] Mean free-fall time per bin
\item[3] Mean prestellar lifetime obtained from the right panels of Fig.\,\ref{fig:histo_lifetime_no_correction} and \ref{fig:histo_ratio}. The range correspond to the uncertainty obtained from the dispersion of the points.
\end{tablenotes}    
\end{threeparttable}
\end{table}

\begin{figure} [htbp!]
    \centering
    \includegraphics[width=0.49\textwidth]{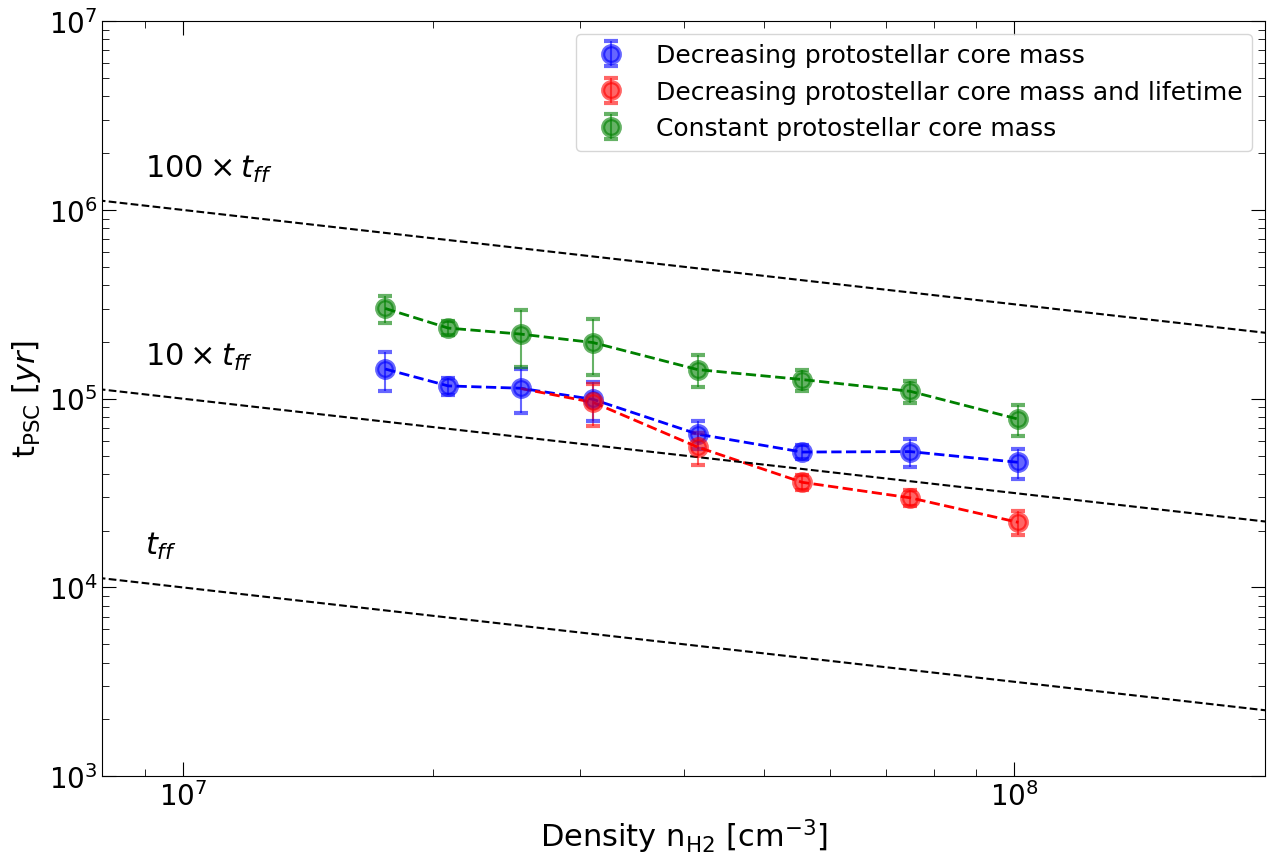}
    \caption{Statistical lifetime of the prestellar phase as a function of the density. Green points represent the lifetimes and densities extracted from the right panel of Fig. \ref{fig:histo_lifetime_no_correction}, corresponding to the case of a constant protostellar core mass over time (clump-fed). Blue and red points correspond to the case of a decreasing protostellar core mass over time (core-fed). Blue points are extracted from the right panel of Fig. \ref{fig:histo_ratio}, while red points are computed assuming a decreasing protostellar lifetime between 16\Msun and 60\,M$_{\odot}$, from 300\,kyr to 100\,kyr. The black dashed lines represent one, ten and a hundred free-fall times.}
   \label{fig:lifetime_vs_density}
\end{figure}

To further illustrate the relationship between free fall and statistical PSC lifetimes, we have plotted both of them as a function of the core densities in Fig.~\ref{fig:lifetime_vs_density}. 
%The blue, red, and black points correspond to all the points in Fig. \ref{fig:histo_ratio} representing the different bin masses as well as the effect of the range of masses for the HM PSC candidates.
Each of the green and blue points corresponds to the eight histogram points in the right panels of Fig.~\ref{fig:histo_lifetime_no_correction} and \ref{fig:histo_ratio}. The fourth blue and green points roughly correspond to core mass of 16\Msun. 
The green points are for the clump-fed scenario with a constant core mass over time (continuously refilled by converging flows; Fig.~\ref{fig:histo_lifetime_no_correction} and Sect.~\ref{subsection:lifetime-clump-fed}) and suggest very large values of the statistical lifetime over free-fall time ratio with values between 35 and 40 below 16\Msun and between 25 and 30 above 16\Msun. The blue points are for the core-fed scenario with a decrease of the core masses along protostellar times (Fig.~\ref{fig:histo_ratio} and Sect.~\ref{subsection:evolution_protostellar_masses}). They point to lower ratios ranging from 17 to 20 below 16\Msun and between 12 and 15 above 16\,$\rm M_{\odot}$. We also show with red points the effects of considering a possible decrease of the protostellar lifetime from 300 to 100 kyr for cores masses between 16 and 60\Msun (final stellar masses of 8 to 30\,M$_{\odot}$) (see Sect.~\ref{subsection:protostellar-lifetime}). This shows that even such a strong decrease of the protostellar time would lead to ratios with free-fall times which would still be of the order of 7 to 10. 

Altogether with values ranging from 50 to 160 kyr above 16\Msun (last 5 blue and green points) the prestellar lifetimes obtained for high-mass cores are clearly shorter in absolute value than for low mass cores with values of the order of 1 Myr (\citealp{Andre2000}). In \cite{Konyves2015} the obtained prestellar lifetime of 1.2 Myr at a density of $4\times10^4\,$cm$^{-3}$ corresponds to a ratio of lifetime over the free-fall time of 8. The prestellar cores in Aquila have however slightly larger sizes (average size of 0.03 pc) than our HM PSC candidates. Assuming a typical linear relationship between the size and the mass of cores, the Aquila value would then shift upward in density by a factor of $\sim 5$ at a size of 0.013~pc (density scales as size$^{-2}$ when mass scales as size) leading to a ratio of statistical lifetime to free-fall time of $\sim 18$ very similar to the values we obtain in the high-mass regime.
%This figure draws the picture from literature results in the low-mass regime \cite{Konyves2015} (green square) to the present work. 
%Despite having PSC lifetime clearly shorter than for low-mass cores in Aquila, 

The ratios of the obtained lifetimes with the free-fall times are therefore surprisingly large with values systematically larger than 10 and reaching values possibly as large as 30. They seem to be relatively similar to what is observed for low-mass PSCs.
%and may tend to decrease a bit above 16\Msun . Only a strong 
%This is actually more than what is obtained for Aquila's prestellar cores with a ratio of lifetime over the free-fall time a bit below 10 (of the order of 8). 
%On the other hand, the prestellar cores in Aquila from \cite{Konyves2015} have slightly larger sizes (average size of 0.03 pc) than our HM PSC candidates. Assuming a typical linear relationship between the size and the mass of cores, the Aquila's point could be shifted upward in density by a factor of $\sim 5$ (density scales as size$^{-2}$ when mass scales as size). Moving the Aquila's point to the expected value at a scale of 0.013~pc (see blue filled circles), the ratio value can actually be similar to the one we obtain in ALMA-IMF in the high-mass regime. 
%We may therefore wonder whether low and high-mass PSCs could have relatively large lifetimes compared to free-fall times with ratios between 10 and 30 suggesting gravity is not driving their lifetime at this scale all over the mass ranges. 
There is a tendency of a decrease of this ratio for the highest masses above 16\Msun (i.e. 8\Msun for the final stellar mass) but with values which stay above 10. It is only if we would consider that the protostellar lifetime could significantly decrease for the highest masses that we could have values below 10. In order to reach a plausible value of 1 to 2 for the ratio expected for a direct collapse without additional support (see Sect.~\ref{subsec:non-thermal} below), one would need to consider a decrease of the protostellar lifetime down to values as low as 20 kyr for the bins of mass we probe here, i.e. for typically  30\Msun final mass star (60\Msun core mass). This would lead to average accretion rates as large as $\rm 1.5\times 10^{-3}\,M_{\odot} yr^{-1}$ for a 30\Msun star.

\subsection{Non thermal supports in dynamical convergent flows}\label{subsec:non-thermal}

The ratios between the statistical lifetimes and the free-fall times of the cores are surprisingly large. If only thermal support is considered, such compact high-mass cores would collapse roughly in one or two free-fall times. How can they have a statistical lifetime at least 10 times larger? %This is particularly surprising in the above discussed framework of a probable dynamical formation of the cores starting at the scale of the clumps.  

If as expected the formation and evolution of the observed prestellar cores have a dynamical origin
with converging flows from large scales, one explanation is that the observed cores are dynamical features apparently stable for $\sim 50-100\,$kyr. They would be collapsing on a free-fall time (typically 5~kyr; see Tab.\ref{tab:lifetime}) but would be continuously replenished by the large scale flows of infall. In this case one has to understand how we can have a long continuous and strong collapse (typically 30\,\Msun over each free-fall time of $\sim 5\,$kyr, leads to infall rate of $6\times10^{-3}\,$\Msun$/$yr) on such a small scale without having formed yet a single high-mass protostar inside it (then driving a powerful outflow). %Despite the clear existence of dynamical flows associated with cores in ALMA-IMF fields as pointed out in \cite{Alvarez-Gutierrez2024} and observed in the literature, we thus conclude that it is improbable then that the observed HM PSCs are continuously collapsing dynamical features. 

Alternatively, the observed HM PSCs have to be significantly regulated and supported by other forces than thermal pressure to keep them that numerous before to further converge/contract or to collapse.
%If alternatively most of these HM PSCs are not collapsing yet while having such small free fall times, they have to exhibit some levels of additional support to keep them that numerous before to collapse. 
In this case, the additional support could also lead to effective Jeans masses large enough to favor high-mass star formation. For instance in the case of turbulent additional support, an effective Jeans mass of 10 \Msun would be obtained at all scales from 0.11 down to 0.013 pc (clump to PSC scale) for effective turbulent velocities ranging from 0.7 to 4.5 \kms respectively. 
This possible turbulent support can be partly discussed using the line width in DCN(3-2) for some of the detected HM PSCs (see also \citealp{Cunningham2023}). These line widths are typically ranging from 1 to 2 \kms. We see that most of these cores have then virial parameter close to 1 or below 1 suggesting turbulence could indeed play a role to support some of the cores (the ones with virial parameters close to or above 1). %Also in a very dynamical scenario to form the observed PSCs, the convergent flows could easily imprint a significant level of rotation in the cores leading to possible centrifugal braking of the convergence of matter. The keplerian velocities at the radius of the cores (taken here as half the sizes) typically range from \hl{Missing part here, do we put the keplerian velocities or we suppress this phrase ?}
The complete study of the line widths in different dense gas tracers goes out of the scope of the present work. A forthcoming paper will investigate carefully the complex spectra toward these sources and their surroundings, in different line tracers to draw a complete and reliable view of the possible turbulent support in the cores.
Concerning the potential magnetic support, the most recent measurements of B fields intensity in high-mass star forming region tend to point to relatively weak fields not strong enough to prevent collapse (e.g. \citealp{Sanhueza2021, Cortes2021}). On the other hand no measurement could have been done yet on HM PSCs due to the lack of good HM PSC candidates so far. We note however that the ambipolar diffusion time happens to be of the order of 30 times the free-fall time (see \citealp{Bergin2007}) i.e. similar to the largest observed lifetimes of the PSCs we discuss here (i.e. in the 8 to 16 \Msun mass regime, see Tab.\,\ref{tab:lifetime}). %These long lifetimes could then reflect the time of diffusion of magnetic flux concentrated by the fast concentration of matter in the supersonic inflows. 
B-field measurements toward our cores will be invaluable to verify this possibility. 
We conclude that the tendency to have large statistical lifetimes of HM PSCs suggest that turbulent and magnetic support may act together or alternatively in different cores to statistically indicate a slowing down of or a delay for an immediate collapse. %Since at the same time we observed in the same fields clear streams of gas converging towards cores \citep{Alvarez-Gutierrez2024}, we believe that the way to form these magnetic and/or turbulent cores is probably dynamical either driven by the clump-size gravity or by more complex inertial motions from large scales as witnessed in the velocity separations seen in molecular lines tracing dense gas at the scales of the ALMA-IMF protoclusters \citep{Cunningham2023}. This is inline with numerous observations of convergent flows in literature (e.g. \citealp{Schneider2010,Galvan-Madrid2010,Csengeri2011,Peretto2013,Avison2021,Schneider2023,Bonne2023}). 
The observed convergent dynamical supersonic flows (e.g. \citealp{Schneider2010,Galvan-Madrid2010,Csengeri2011,Peretto2013,Avison2021,Schneider2023,Bonne2023}) are expected to concentrate sources of support (magnetic field, rotation and turbulence) leading to some possible stop or braking of the collapse at smaller scales if dispersion/radiation of the related energies/pressures cannot dissipate fast enough.
The alternative view of continuously collapsing and replenished cores is more difficult to consider as we would expect a fast formation of at least one high-mass protostars inside the cores which should drive a powerful outflow. %Timescale for ambipolar diffusion seems indeed compatible with the observed lifetime of the lowest mass PSCs of our complete list of cores.

%We presented in this work a first sample of massive cores that are prestellar at a scale of $\sim 2700 au$. One question that still remains is : are these cores all monolithic ? Or are they in the process of subfragmentation into several lower mass cores or will be in the future ? This hypothesis does not question however the prestellar status of these candidates. Even if some cores are subfragmenting, no outflows or consequent trace of heating in the PPMAP temperatures are shown by those cores. High resolution data would be needed to settle the monolithic status of all the candidates. We know thanks to high resolution data of the W43-MM1 mini starbust that some cores are subfragmentating into several smaller ones, including core \#6 which is one of our candidate here (Brouillet N., private communication).
%\textcolor{orange}{The CMF found to be flatter (also called top heavy) than the traditional Salpeter slope in \cite{Nony2023} and \cite{Pouteau2022}, could be due to the fragmentation of the cores. The lack of high resolution data could lead to consider massive cores instead of several less massive. This consideration would explain that this type of CMF is likely to become a Salpeter-like IMF.}
%-----------------------------------------------------------------

\subsection{Effect of possible sub-fragmentation of the cores}
\label{subsection:sub-fragmentation}

So far in the discussion we assumed that each observed core is the precursor of a single (high-mass) protostar. We only accounted for a certain core to star efficiency $\epsilon_{\rm cse}$ (of 50~\%) to convert the mass reservoir into a final mass star. Confronted however with the complexity described above of dynamical collapses mediated by turbulence, magnetic field (and some possible angular momentum), one cannot be sure that the observed cores are not fragmenting at lower spatial scales (e.g. \citealp{Louvet2021}). The observations of outflows from the high-mass protostars in \cite{Nony2020,Nony2023} and in the present work mostly point to single outflows coming from the protostellar cores indicating that the single protostar hypothesis might be mostly correct. On the other hand very recent results of very high resolution ALMA observations of similar high-mass cores have revealed some fragmentation inside the cores with typically 2 to 3 very small scale (typically 500 au) sources which point to a multiplicity of such 2000-3000 au high mass cores \citep{Budaiev2024, Li2024, Olguin2022, goddi2020}. %One of our high-mass PSCs, in W43-MM1 has also recently been found to be multiple with three compact objects inside this $\sim40$~\Msun PSC core (Brouillet N., private communication).

This possible sub-fragmentation of the HM PSCs discussed here does not change our conclusion concerning the prestellar nature of the cores as we defined it in Sect.~\ref{sec:discuss:highly-accreting-hosts}. Only the mass will be sub divided into several cores, leading to low-mass prestellar fragments. If some sub-fragments are present inside the cores the non detection of outflows still indicates that there are not yet any strongly accreting objects inside them which would be revealed by a detectable outflow. Concerning the statistical lifetime of the cores the multiplicity would not affect the validity of the number ratios between pre and protostellar cores if the multiplicity level does not change between the two evolutionary stages. %Only the effect of some of the corrections we applied in Sect.~\ref{sec:lifetimes} which depends on the core to star efficiency $\epsilon_{\rm cse}$ would be affected. One needs to refine the effective core to star efficiency $\epsilon_{\rm cse}$ used to estimate the correction. With a core to star efficiency $\epsilon_{\rm cse}$ of 25~\% (2 sub-fragments and a global $\epsilon_{\rm cse}$ of 50\%) for instance the values obtained in Sect.~\ref{sec:lifetimes} after all corrections would give 27 estimated ionising sources instead of the previous 49 in the high-mass bins, leading to lifetimes up to $\sim20$~\% larger in the last mass bins which then also corresponds to final mass of stars formed 2 times lower (see Fig.~\ref{appendix:histo_25_fig} in appendix \ref{appendix:fragmentation lifetime}). These results point to actually even larger lifetimes compared to free-fall times (Sect.~\ref{sec:lifetimes}) therefore not changing our main conclusion on the need for non thermal support (see previous section) at the presently observed HM PSC scale. 
This would only move our conclusions lower in final stellar masses (typically from 25 to 12.5\,M$_{\odot}$) obtainable from the present day prestellar mass reservoirs. On the other hand the dynamical fast inflows have no reason to stop to flow, making possible to keep constant the mass reservoirs during the protostellar time (clump-fed scenario corresponding to Sect.~\ref{subsection:missing_UCHII} and Fig.~\ref{fig:histo_lifetime_no_correction}). This would allow for increasing the possible final stellar masses. This would also be in line with the protostellar mass evolution proposed to explain evolution of core mass functions in \cite{Nony2023} and \cite{Pouteau2023} (see also \citealp{Sanhueza2019}).

%and thus questioning to which extend the protoclusters in ALMA-IMF manage to form many high-mass stars, since the protostellar cores at the here observed scale would also be sub-fragmented presumably with the same statistics as PSCs.

% \subsection{Towards a new vision of high mass star formation ?}
% Larger scale mass reservoir\\
% This reservoir is accesible and near the cores\\
% Formation scenarios discussion ? \\

% This part is dedicated to investigate a new star formation scenario where monolithic cores at 2700 au do exist, but could sub fragmentate into lower mass and scale cores at higher resolution. However, massive prestellar phase do exist here in opposite to Bonnel and Bate etc...\\
% Those cores do not show any clear sign of outflow which should be huge at these masses, and should appear even if subfragments were protostellar.

%-----------------------------------------------------------------

\subsection{Alternative scenarios}
\label{subsection:alt-scenarii}

The coincidence of dynamical processes to form the compact cores and of relatively long statistical lifetimes compared to free-fall times appears difficult to explain without the above claim for additional support of magnetic, turbulent or rotational nature. On the other hand our approach relies on the expected strong link between a detectable CO/SiO outflow inside the cores themselves (On-Off criterium) and the expected strong accretion of any high-mass (i.e. highly accreting) protostar in the cores. We can question each of the steps of this link between high accretion and strong CO outflows in the cores. In other words, can our HM PSC candidates hide protostars with large accretion rates? 

We list below several alternative scenarios to non-thermal support at the scale of the observed cores:

   \begin{enumerate}
      \item One possibility is that the strong slowing down of the expected collapse does not occur at the presently observed scale of 2700 AU but in smaller fragments similar to those observed in \cite{Budaiev2024} for instance. If at such smaller scale, the fragments stop collapsing due to a strong support of magnetic, turbulent or rotational nature they would not have immediately a strong accretion activity, and would not drive any strong outflows such as the strong outflow observed for instance in a protostellar core of W51-IRS2 in \cite{goddi2020}.

      \item A fraction of our HM PSCs could be mis-classified due to a wrong non detection of CO/SiO outflow (false negative outflow detections). A significant fraction of the HM PSC candidates are indeed situated in crowded regions where it is difficult to firmly exclude any outflow driven by a core. This difficulty to firmly reject outflows in crowded regions may however not affect too much statistically our results since at the same time it is difficult to be sure that all protostellar cores we propose are indeed all the driving sources of the observed lobes in the same crowded regions  (false positive outflow detections) making them then candidates to be prestellar cores. These false positive outflow detections can compensate, at least partially, for the false negative outflow detections.

      \item The direct link between ejection and accretion could also be broken or altered for young high mass protostars. Contrary to what the basic physics of the magneto-centrifugal ejection dictate, this would allow for a high accretion without a strong ejection counterpart. The magneto-centrifugal processes explain the efficient removal of angular momentum, allowing accretion while ejecting material through magnetic mediation. No ejection would then point to no angular momentum to be removed for youngest high-mass protostars. In the here favored dynamical formation processes, the initial angular momentum of dynamically concentrated gas is however certainly larger than in more quasi-static processes which can best evacuate angular momentum during the slow gas concentration for low-mass star formation. Concerning the magnetic field, it is similarly difficult to expect no concentration at all of the magnetic field as the cores form and collapse. Altogether only the existence of so-called dead zones due to non ideal MHD processes or to very low levels of ionisation could be claimed to explain a suppressed magneto-centrifugal ejection/accretion process. But in this case the angular momentum could not be removed and compact rotating cores would then grow rapidly in mass without accretion onto an inner protostar and could then lead to non accreting compact rotating fragments. 

      \item The expected strong accretion/ejection activity could be variable enough to make that some cores have their CO outflows faded away between two events of accretion to become undetectable. A significant variability of the accretion/ejection is indeed expected. To let a CO outflow disappear between two accretion/ejection events, the time between events has to be large enough. The typical time for a CO outflow to fade away should roughly correspond to the crossing time of the CO gas. Here we typically search for outflows at 6 \kms for an averaged core radius of $\sim\,1350\,$au. This leads to crossing times of $\sim\,1100\,$yr. This is larger than the typical variability of 500\,yr recognized as typical for the CO outflows in W43 by \citep{Nony2020}. This would suggest that strong accretion/ejection has to stop for periods longer than $\sim$ 1100\,yr at the earliest stages of the high-mass star formation. This is a large value for a stop in accretion for cores which have free-fall times of only 3000 to 7000 yr.

      \item The expected strong ejection from the hidden high-mass protostars would not interact strongly enough with the compact cores we observed in dust continuum to be detected in CO or SiO at high velocities in the On-Off spectra. This is only possible if the cores are actually mostly empty with the detected dust emission being actually concentrated in even more compact cores. 
      Since most of the high-mass cores we studied are seen as resolved sources with the ALMA beam, this would indicate that they would be made of several such compact cores which would be distant enough from each other to mimic a single resolved core. In this case the strong ejected jets/winds could escape the cores without interacting much with the immediate surroundings.
%A fourth possibility is that the expected strong accretion/ejection activity is variable enough to make that some cores have their CO outflows faded away enough between two events of accretion to become undetectable. A significant variability of the accretion/ejection is expected. To let a CO outflow disappears between two events the time between events has however to be large enough. The typical time for a CO outflow to fade away should roughly correspond to crossing time of the CO gas. Here we typically search for outflows at 10 km/s for core sizes of $\sim\,2700\,$AU. This leads to crossing times of $\sim\,1300\,$yr. This is larger than the typical variability of 500 yr recognized as typical for the CO outflows in W43 by \citep{Nony2020}. This would suggest that strong accretion/ejection has to stop for periods longer than $\sim$ 1000 yr at the earliest stages of the high-mass star formation. This is a large value for cores which have free-fall times of only 3 to 7 kyr.

      \item In the case of sub-fragmentation of the cores, %A fifth possibility would be that 
      the expected large global accretion rate could be spread on several individual low-mass protostars with a reduced global effects of the different outflows driven by these protostars. Since the rate of momentum carried by outflows is proportional to the accretion rates, the global momentum of the sum of these outflows should be similar to that of a single outflow from a single high-mass protostar. The possible different directions of the outflows could limit the ability to well identify outflows in the maps, but some excess of high-velocity CO gas should be detectable towards the cores in the spectra.
  \end{enumerate}

Altogether we therefore see that possible alternative scenarios to a turbulent, magnetic or rotational support at the core scale exist and will need to be further investigated in the future. The main limitation of the present work is related to the still possible misclassification of cores as prestellar due to the difficulty to assure a non-detection of an outflow in the most crowded regions. It would be extremely valuable to further search for protostellar activity from the cores using mid-IR emission at high spatial resolution with the JWST/MIRI to validate their prestellar nature. We will also further study this unique sample of HM PSC candidates using detailed molecular line observations from ALMA-IMF (Valeille-Manet et al. in prep) and will gather new high spatial resolution ALMA studies both in line and continuum to investigate possible existence and effects of fragmentation of the cores. 
%-----------------------------------------------------------------
%-----------------------------------------------------------------
%-----------------------------------------------------------------

\section{Conclusions} \label{conclusions}
We used the core catalog of \cite{louvet22}, extracted from the continuum, combined with CO\,(J=2--1) and SiO\,(J=5--4) datacubes of the large program ALMA-IMF to identify a sample of high-mass prestellar cores in 14 protoclusters. Our main results are~:

   \begin{enumerate}
      \item We built a new automated method combining spectral information with the comparison of On and Off spectra and spatial information with molecular outflow maps to identify outflowing emission from compact cores. Our method is efficient at detecting outflows from protostellar cores and has proved essential for clarifying the classification of cores in the most clustered regions where outflows overlap, and is particularly adapted to systematic analysis of large CO/SiO data cubes.

      \item Using this method, we identified for the first time a significant sample of robust high-mass PSC candidates inside a single survey, with masses reaching up to 50\,M$_{\odot}$ as a mass reservoir and which may then form stars as massive as 25\,M$_{\odot}$ (for a core to star mass efficiency of 50 \%).
      
      \item From 82 cores above 8\,M$_{\odot}$, we identify 30 cores that do not have any detectable outflow with the current data and methodology used. Of these 30 cores, 12 are more massive than 16\,M$_{\odot}$, the most massive core being 54\,$\rm M_{\odot}$. These 30 PSC candidates range from 1350 to 4200 au in deconvolved FWHM size, and are all high density cores between $\sim4\times10^6$\,cm$^{-3}$ and $\sim1.5\times10^8$\,cm$^{-3}$.
      
      \item We found that most of the robust HM PSC candidates are located in the central clump regions where most of the mass is most probably dynamically gathered. We propose that it is mostly the difficulty to identify cores without outflows in such highly clustered environments which explains why so few high-mass PSCs could be identified so far.%but where it increases the difficulty to identify them as prestellar cores due to the overlapping of outflows of the neighbouring cores. In contrast, the other PSC candidates between 8 and 16\Msun are more located in more isolated areas with one or less high-mass cores.

      % \item Studying the number of PSC candidates in the 14 protoclusters we found that most of them are located in the young and the evolved regions rather than in the intermediate ones. We propose that this may be due to two bursts of (high-mass) star formation, the first occurring in the youngest regions and the second one later when these regions are evolved. 
      
      \item Adopting a high-mass protostellar lifetime of 300 kyr, we could estimate the lifetime of the high-mass prestellar phase in several mass bins. 
      %As high-mass protostars spend a significant part of their lifetime as already ionising (proto)stars, we have applied corrections for  missing protostellar cores which are UCHII regions in order to improve the precision of our lifetime estimates. Furthermore we also considered two scenarii for the possible evolution of envelope/core masses during the protostellar times providing us with a view of the spread of the plausible range of PSC lifetimes as a function of the mass. 
     % \item 
      We ended up with a range of most plausible statistical lifetimes between 8 and 16\Msun of 150 to 250 kyr with a trend of decrease to 50 to 100 kyr between 16 and 40-50\,M$_{\odot}$. %Above these masses the lifetime seems to continue to decrease according to the obtained upper limits expressing the non detection of HM PSC above 50\Msun but this remains uncertain and will need to be confirmed with even larger samples of rich protoclusters. 

      \item The surprisingly large ratios (10 to 30) of HM PSC lifetimes with their free-fall times strongly point to the need for non thermal support of the observed cores. Such support can either be turbulent, magnetic, or rotational in nature. Further studies are required to find out what is the best scenario to precisely account for these large timescale ratios.

      %\item Combining the masses of the observed reservoirs and the obtained short lifetimes we show that the formation of these cores has to be dynamical with supersonic gas inflows compatible with claims for such dynamical streams in the literature and with the observed velocity dispersion in ALMA-IMF fields \citep{Nony2018,Cunningham2023,Alvarez-Gutierrez2024}.

     % \item We therefore propose a formation/evolution scenario of high-mass PSCs which is dynamical (converging supersonic inflows) at least at scales from 0.1 pc down to 0.015 pc probably forcing sources of support, i.e. turbulence and magnetic field, to concentrate without time to disperse or diffuse so that it may at one point slow down or stop further fast concentration and/or collapse, expected (for pure thermal support) to occur close to the free-fall speed. Interestingly enough the typical timescale for magnetic ambipolar diffusion scales as free-fall times with a ratio to free-fall time of 30 which corresponds to the upper part of the range of PSC lifetimes we could derive.
   \end{enumerate}

These results confirm the extreme rarity of HM PSCs with only 12 candidates above 16\Msun found in 14 very active protoclusters of the Galaxy where hundreds of cores have been recognized, with only four of them more massive than 30\,M$_{\odot}$. 

\begin{acknowledgements}
We thank the referee for the careful reading of the manuscript and for providing useful comments.
This paper makes use of the following ALMA data: ADS/JAO.ALMA\#2017.1.01355.L. 
ALMA is a partnership of ESO (representing its member states), NSF (USA) and 
NINS (Japan), together with NRC (Canada), MOST and ASIAA (Taiwan), and KASI 
(Republic of Korea), in cooperation with the Republic of Chile. The Joint ALMA 
Observatory is operated by ESO, AUI/NRAO and NAOJ.

This work was supported by the Programme National “Physique et Chimie du Milieu Interstellaire” (PCMI) of CNRS/INSU with INC/INP co-funded by CEA and CNES.
This work was supported by the "Programme National de Physique Stellaire" (PNPS) of CNRS/INSU co-funded by CEA and CNES.
T. Cs. and M.B. acknowledge support from the French State in the framework of the IdEx Universit\'e de Bordeaux Investments for the future Program. 
AS gratefully acknowledges support by the Fondecyt Regular (project code 1220610), and ANID BASAL project FB210003.
AG acknowledges support from the NSF under grants AST 2008101 and CAREER 2142300.
R.G.M. acknowledges support from UNAM-PAPIIT project IN108822 and from CONACyT Ciencia de Frontera project ID 86372.
PS was partially supported by a Grant-in-Aid for Scientific Research (KAKENHI Number JP22H01271 and JP23H01221) of JSPS.
R.A. gratefully acknowledges support from ANID Beca Doctorado Nacional 21200897.
LB gratefully acknowledges support by ANID Basal project FB210003.
\end{acknowledgements}


%\bibliographystyle{aasjournal}
\bibliographystyle{aa}
\bibliography{biblio}

\newpage
\begin{appendix}

\section{On and Off spectra computation method} \label{appendix:On_Off_explanation}
\begin{figure*}[htbp!]
    \centering
    % Colonne gauche (image unique centrée verticalement)
    \begin{minipage}[c]{0.47\textwidth}
        \centering
        \includegraphics[width=\textwidth]{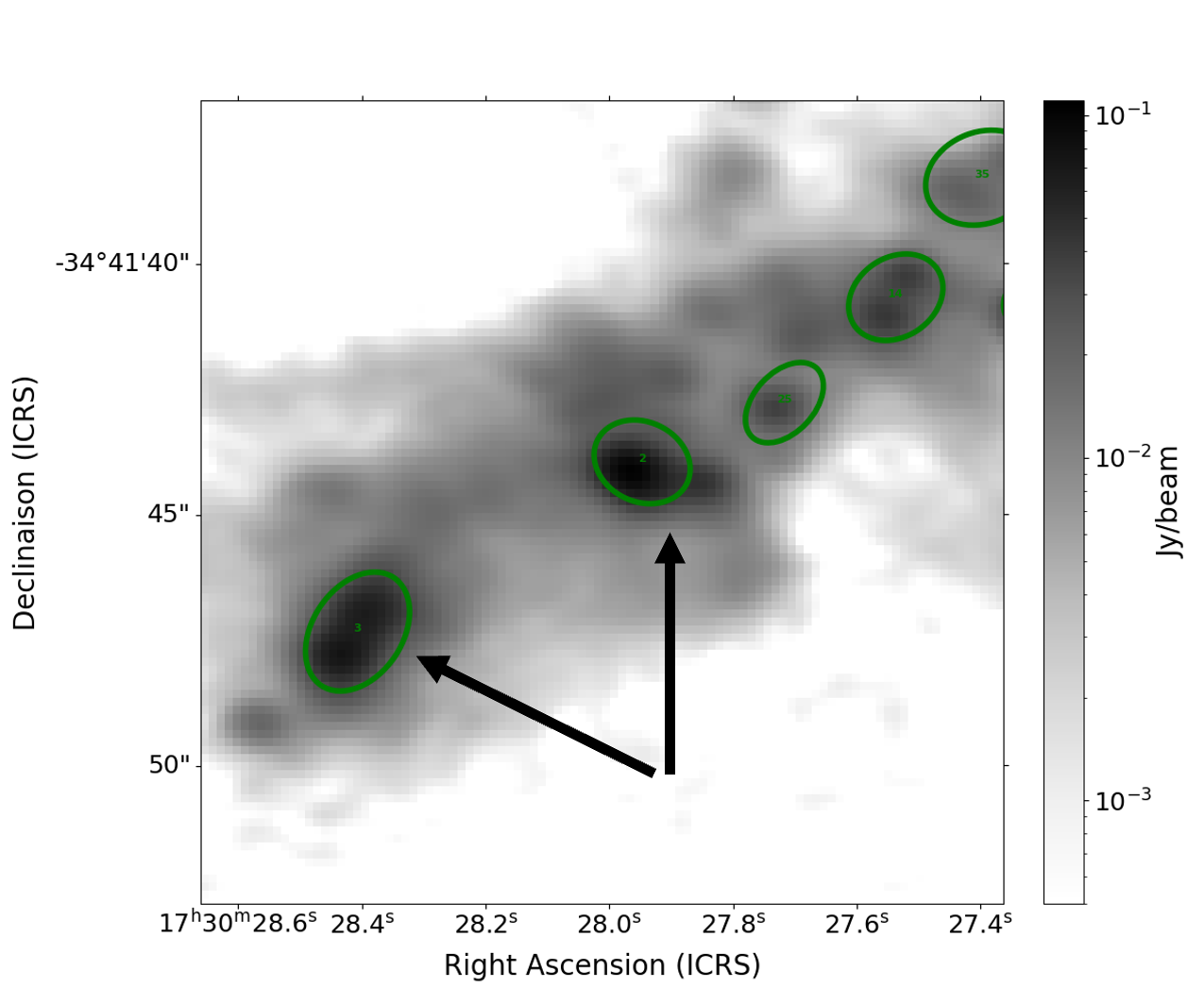}
    \end{minipage}
    \hfill
    % Colonne droite (deux images empilées et centrées)
    \begin{minipage}[c]{0.47\textwidth}
        \centering
        \raisebox{0.25\height}{ % Centrer verticalement
            \includegraphics[width=\textwidth]{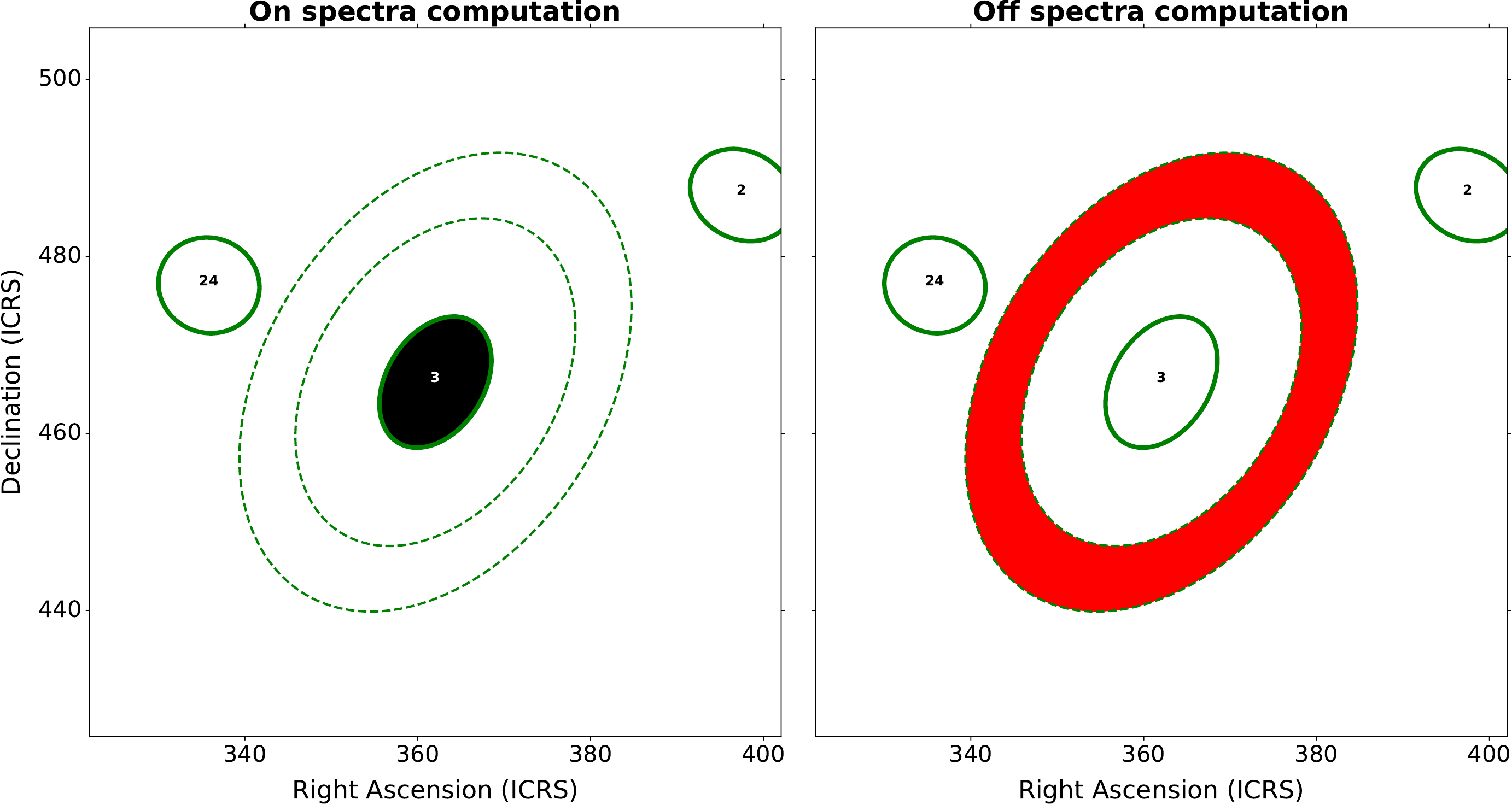}
        }
        \vspace{0.1cm}
        \includegraphics[width=\textwidth]{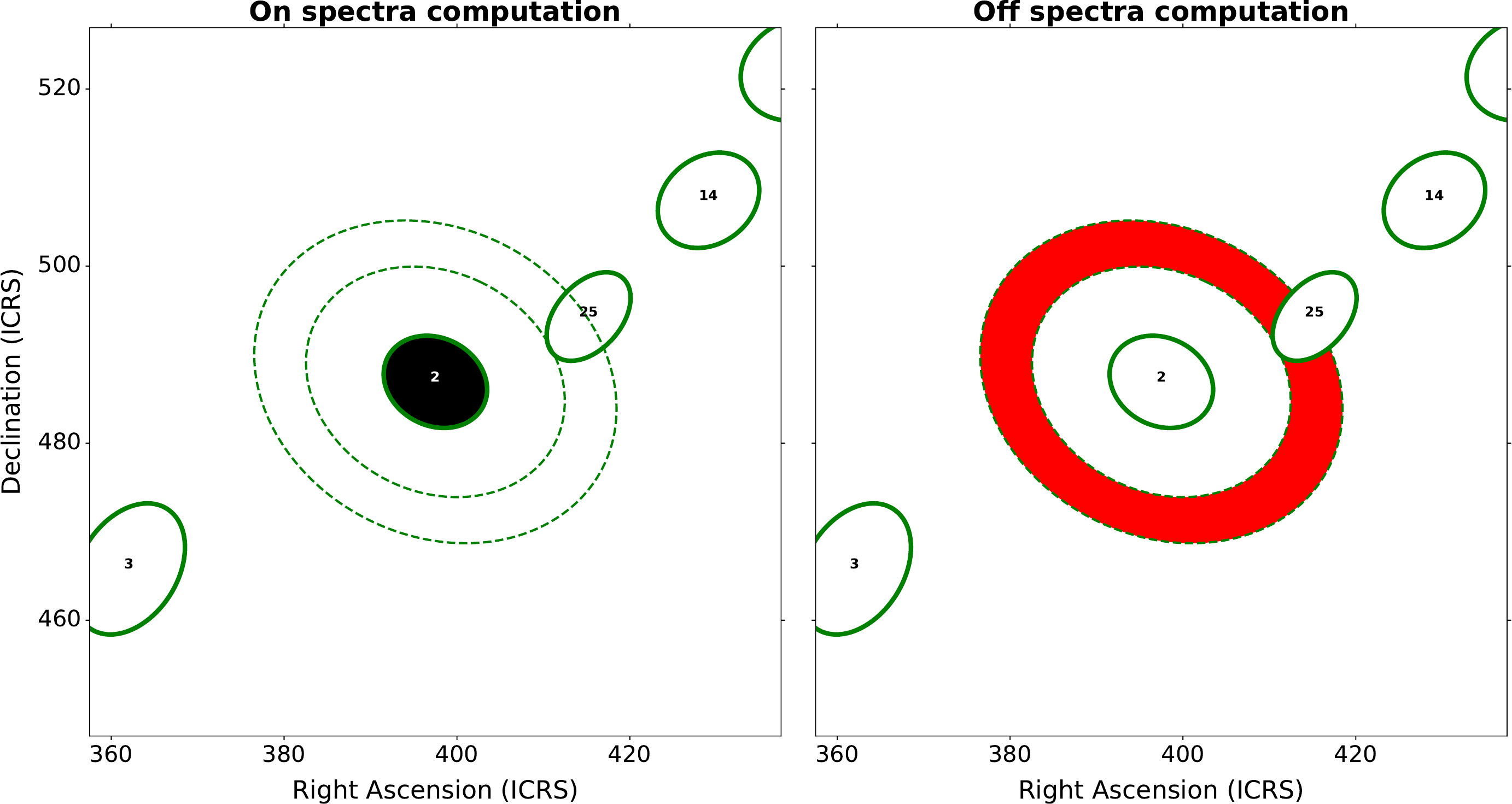}
    \end{minipage}
    \vskip -0.1cm
    \caption{
        \textbf{Left:} Zoom on the continuum core \#2 (top right) and \#3 (bottom left) overlaid on the 1.3mm continuum map of G353.41 (in gray scale). The green ellipses represent 1 time the FWHM of the continuum core. 
        \textbf{Right:} Example of the On (black ellipse) and Off (red annulus) spectra computation for each core, used to estimate the core-averaged background-subtracted spectra. Core \#3 (top) is the simplest case where no other core overlaps with its annulus (computed between 2.5 and 3.5 times the FWHM of the continuum core). Core \#2 (bottom) has another core overlapping its annulus. The overlapping pixels are subtracted from the background estimation.
    }
    \label{appendix:On_Off_overlapping_fig}
\end{figure*}

Figure\,\ref{appendix:On_Off_overlapping_fig} shows the two different cases for On and Off spectra computation. Top panel presents the continuum cores \#2 and \#3 of G353.41 overlaid on the 1.3mm continuum map in grey scale, and the computation of the On and Off spectra in the ellipses and annulus of the cores are presented in the middle and bottom panels. Core \#2 has another source crossing its annulus causing the pixels overlapping to be subtracted from the background annulus.

\begin{figure*}[htbp!]
    \centering
    \includegraphics[width=0.83\textwidth]{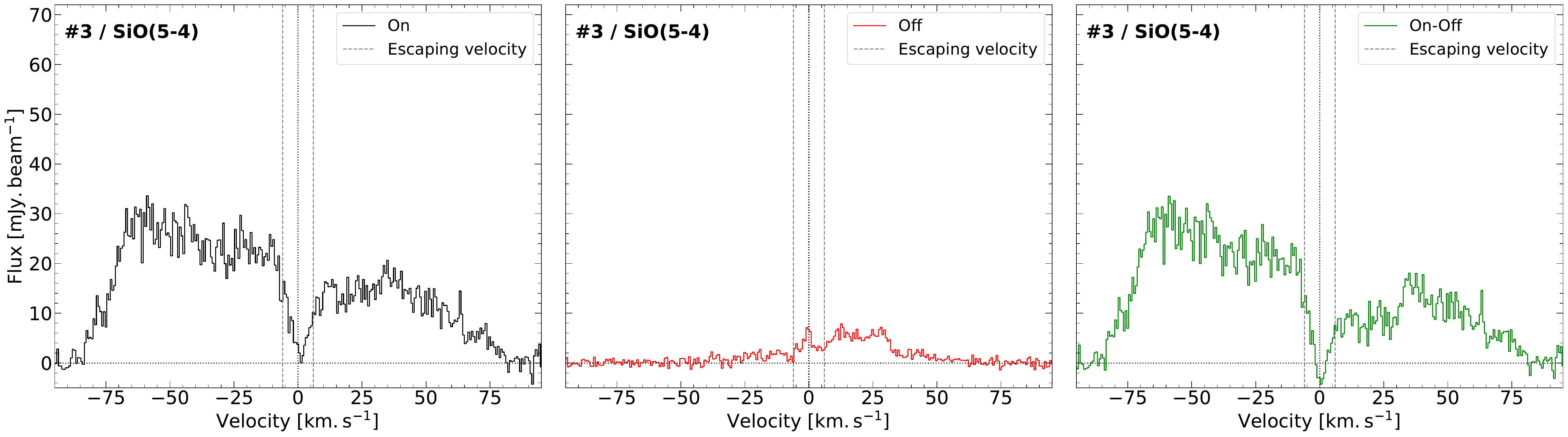}
    \includegraphics[width=0.83\textwidth]{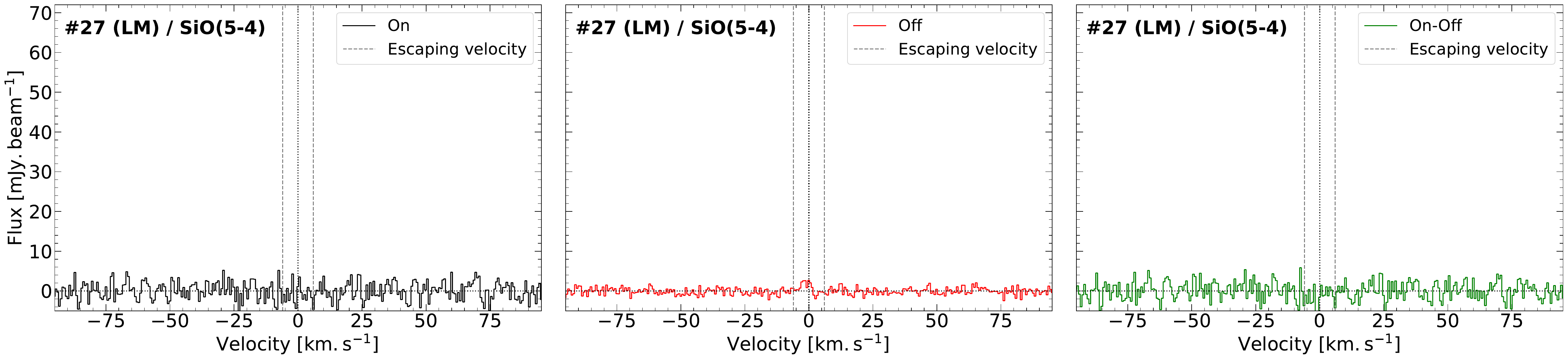}
    \includegraphics[width=0.83\textwidth]{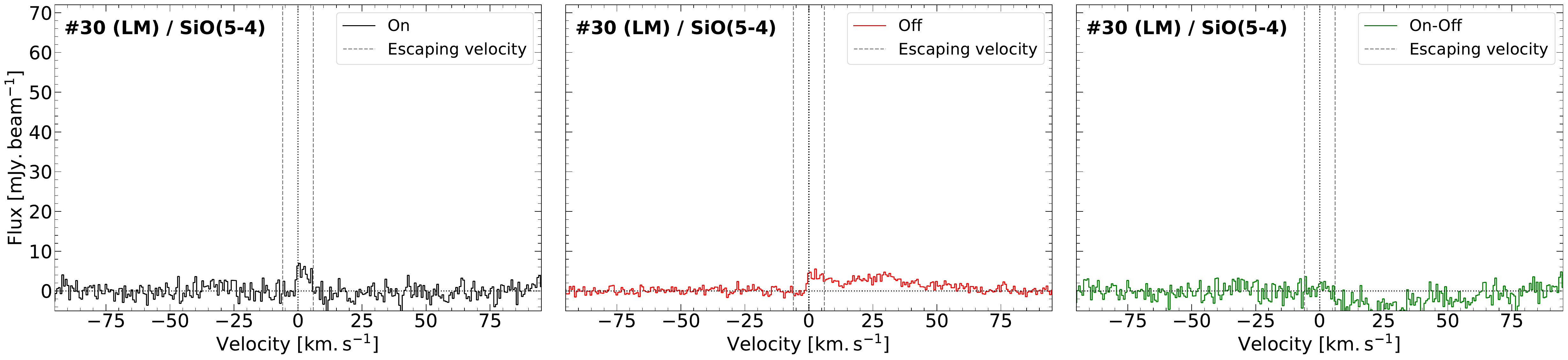}

   % \vskip -0.3cm
    \caption{Resulting SiO\,(5--4) spectra for cores \#3, \#27 and \#30 of the W43-MM2 region, presented in Fig.\,\ref{On_Off_explanation}.}

    \label{appendix:SiO_spectra_W43-MM2_fig}
   
\end{figure*}

Figure\,\ref{appendix:SiO_spectra_W43-MM2_fig} presents the obtained SiO\,(5--4) spectra for cores \#3 and \#30 of the W43-MM2 region, which are presented in Fig.\,\ref{On_Off_explanation}.

\vspace{-0.2cm}

\section{Noise estimation}\label{appendix:Noise_estimation}
\begin{figure}[h!]
    \centering
    \includegraphics[width=0.49\textwidth]{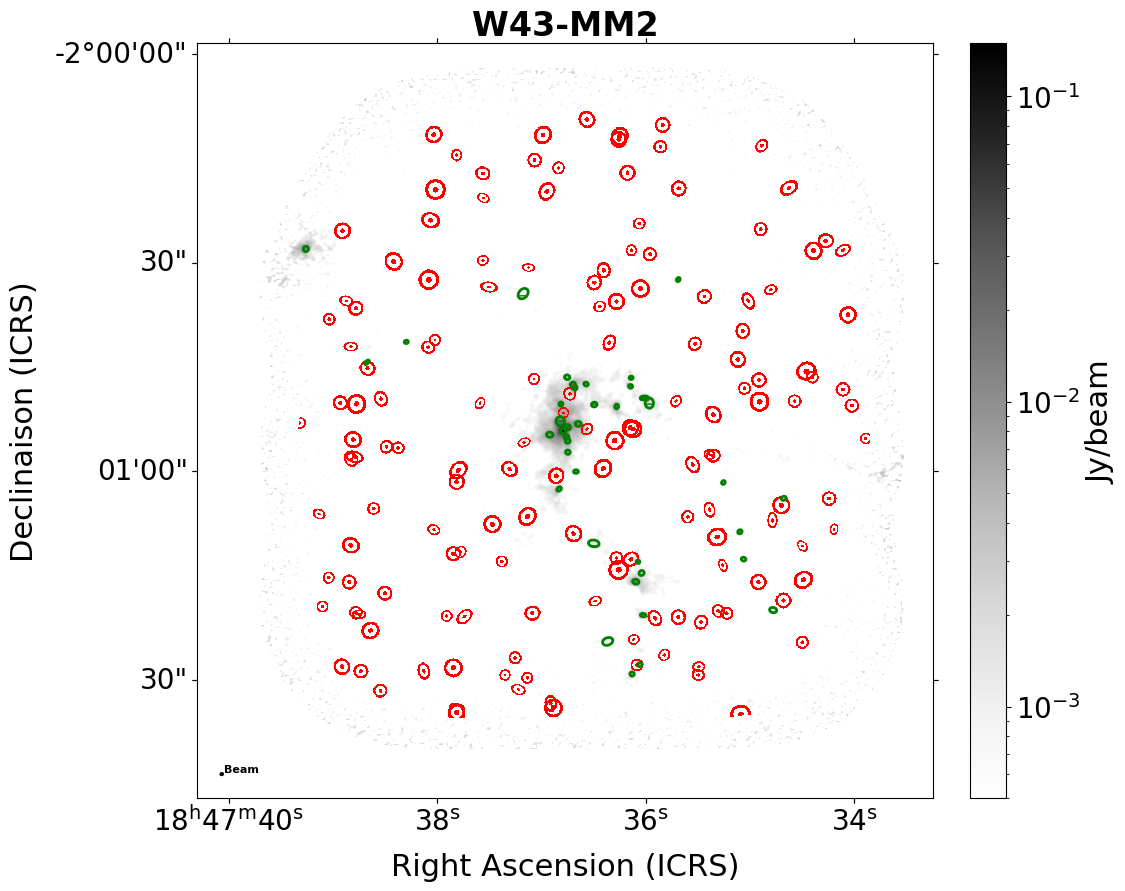}

   % \vskip -0.3cm
\caption{W43-MM2 field with in green ellipses the ALMA-IMF cores, and in red the 150 On and Off random selections. The grey background represents the 1.3mm dust continuum.}

\label{appendix:150_Noise_fig}
\end{figure}

\begin{figure}[h!]
    \centering
    \includegraphics[width=0.49\textwidth]{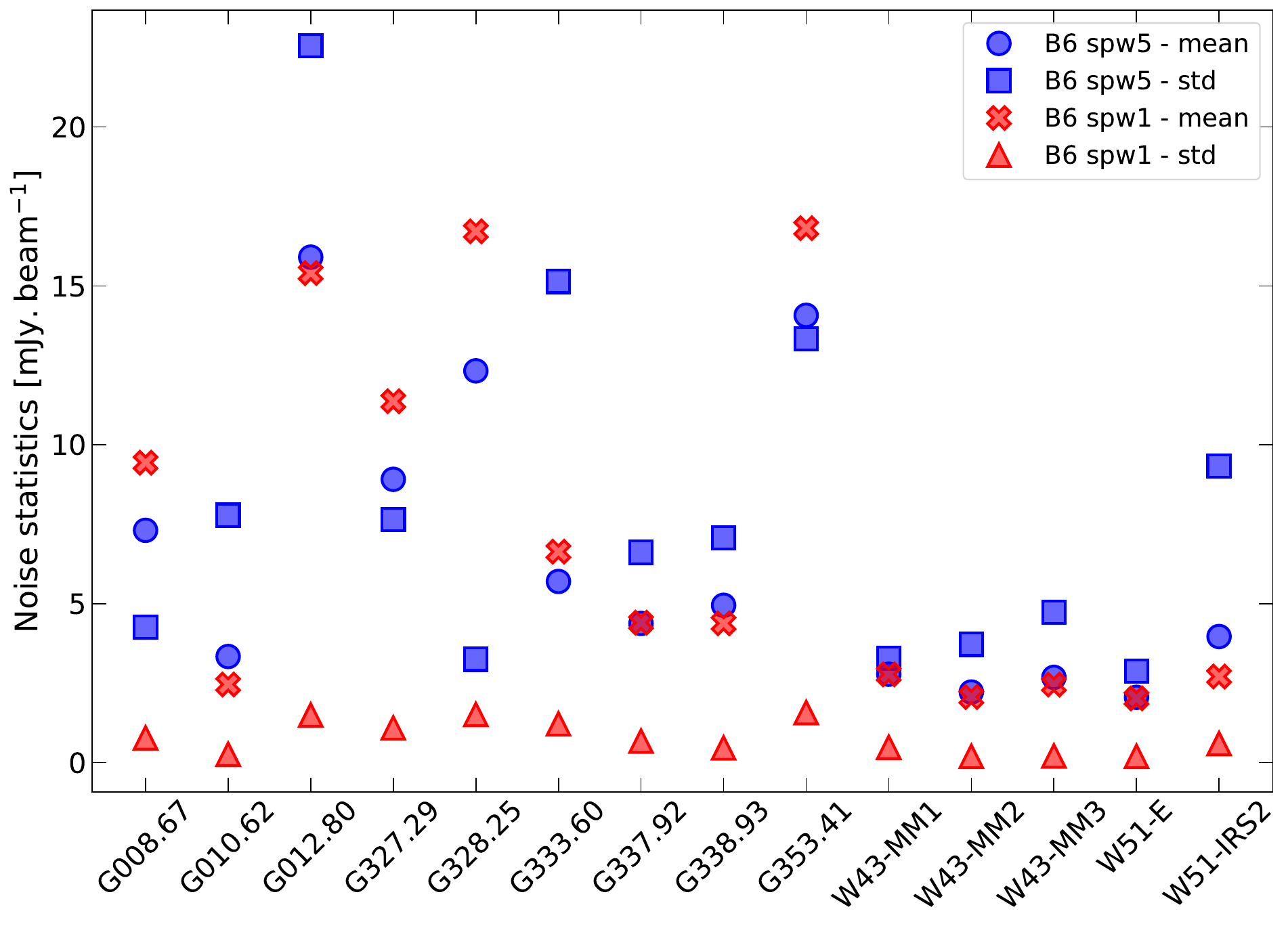}

   % \vskip -0.3cm
\caption{Mean and standard deviation of the derived rms noise levels in B6 spw1 and B6 spw5, for the 14 ALMA-IMF fields.}

\label{appendix:noise_levels_fig}
\end{figure}

Figure \ref{appendix:150_Noise_fig} presents an example of the method used to estimate the noise in the ALMA-IMF datacubes. Figure \ref{appendix:noise_levels_fig} presents the derived mean and standard deviation of the noise levels of the CO and SiO/DCN datacubes (i.e. over all velocities of the spectral windows), in the 14 ALMA-IMF regions.

\section{List of protostellar core candidates}\label{appendix:proto_candidates_list}
Here we present the 100 cores of \cite{louvet22} with $\rm M > 4 M_{\odot}$ classified as protostellar core candidates due to an association with an outflow. They are listed in Table\,\ref{tab:proto_candidates} with their host region, number, coordinates (ICRS), mass, temperature and size. To compute their masses the same PPMAP temperatures as in \cite{louvet22} were used. The error on the masses are computed by taking into account only the uncertainties on the integrated and peak fluxes.

\begin{table*} [h!]
    \centering     
    \begin{threeparttable}[c]
    \caption{List and properties of the sample of protostellar core candidates with $\rm M > 4 M_{\odot}$.}
    
    \begin{tabular}{lllllcc}
    \hline \noalign {\smallskip}
    Region & Source &  RA  &  DEC  & Mass\tnote{1} & Temperature\tnote{2} & FWHM$^{\rm dec}$  \\ \noalign {\smallskip} 
        &   &  [deg]  & [deg]  &  $\left[\rm M_\odot \right]$ & $\left[\rm K\right]$ & $\left[\rm au \right]$ {\smallskip}\\  
    \hline \noalign {\smallskip}
    
      G008.67   &   \#3   &   271.5796091   &   -21.6250348   &   8.5 $\pm$ 0.7   &   35   &   1340 \\
	 G008.67   &   \#5   &   271.5785345   &   -21.6222306   &   4.9 $\pm$ 0.2   &   25   &   2000 \\

    \hline \noalign {\smallskip}

	 G010.62   &   \#8   &   272.6185252   &   -19.9301787   &   8.0 $\pm$ 1.0   &   44   &   1360 \\
	 G010.62   &   \#14   &   272.6213499   &   -19.9296433   &   7.0 $\pm$ 0.6   &   40   &   2970 \\
	 G010.62   &   \#16   &   272.6207417   &   -19.929286   &   4.7 $\pm$ 0.6   &   41   &   1630 \\
	 G010.62   &   \#24   &   272.6222764   &   -19.930054   &   4.9 $\pm$ 0.5   &   36   &   3040 \\
 
    \hline \noalign {\smallskip}
    
	 G012.80   &   \#1   &   273.5493292   &   -17.9256817   &   6.1 $\pm$ 0.1   &   100   &   1850 \\
	 G012.80   &   \#3   &   273.5573504   &   -17.9225106   &   5.3 $\pm$ 0.1   &   100   &   2960 \\
	 G012.80   &   \#6   &   273.5531484   &   -17.9208183   &   4.4 $\pm$ 0.1   &   33   &   1680 \\
	 G012.80   &   \#8A\tnote{*}   &   273.5487793   &   -17.9261916   &   6.8 $\pm$ 0.3   &   31   &   2120 \\
	 G012.80   &   \#8B\tnote{*}   &   273.5484473   &   -17.9261587   &   6.4 $\pm$ 0.2   &   31   &   1910 \\
	 G012.80   &   \#11   &   273.5547682   &   -17.9278997   &   9.8 $\pm$ 1.4   &   42   &   2590 \\

    \hline \noalign {\smallskip}
    
      \textbf{G327.29}  &  \textbf{\#1}  &   \textbf{238.282335}  &  \textbf{-54.618403}  &  $\bm{49.6\pm0.2}$  &  \textbf{300}  &  \textbf{4550}\\
	 G327.29   &   \#2A\tnote{*}   &   238.2893669   &   -54.6167726   &   15.2 $\pm$ 0.3   &   32   &   2040 \\
	 G327.29   &   \#2B\tnote{*}   &   238.2895574   &   -54.616967   &   9.2 $\pm$ 0.2   &   32   &   1270 \\
	 G327.29   &   \#3   &   238.280656   &   -54.6186811   &   9.5 $\pm$ 0.3   &   42   &   2760 \\
	 G327.29   &   \#5   &   238.2955904   &   -54.6128047   &   11.7 $\pm$ 0.3   &   28   &   2850 \\
	 G327.29   &   \#6   &   238.2938644   &   -54.6110507   &   4.9 $\pm$ 0.2   &   28   &   1390 \\
	 G327.29   &   \#7   &   238.2961992   &   -54.6133521   &   8.5 $\pm$ 0.4   &   26   &   2610 \\
	 G327.29   &   \#8   &   238.2983074   &   -54.6069915   &   4.3 $\pm$ 0.3   &   27   &   1590 \\
	 G327.29   &   \#9   &   238.2858254   &   -54.6172552   &   4.0 $\pm$ 0.4   &   36   &   1590 \\

    \hline \noalign {\smallskip}

	 G328.25   &   \#1   &   239.4991778   &   -53.9668493   &   11.3 $\pm$ 0.1   &   100   &   2240 \\
    
    \hline \noalign {\smallskip}
      G333.60   &   \#2   &   245.5356169   &   -50.104813   &   12.7 $\pm$ 0.3   &   41   &   2290 \\
	 G333.60   &   \#3   &   245.528392   &   -50.1050657   &   5.8 $\pm$ 0.3   &   36   &   2080 \\
	 G333.60   &   \#4   &   245.5460571   &   -50.0990531   &   4.7 $\pm$ 0.3   &   100   &   1500 \\
	 G333.60   &   \#5   &   245.5285258   &   -50.1045541   &   4.7 $\pm$ 0.4   &   36   &   1420 \\
      \textbf{G333.60}  &  \textbf{\#7}  &   \textbf{245.5370048}  &  \textbf{-50.1027976}  &  $\bm{16.8\pm0.7}$  &  \textbf{46}  &  \textbf{3840}\\
	 G333.60   &   \#11   &   245.5357042   &   -50.1033887   &   5.2 $\pm$ 0.3   &   100   &   4390 \\
	 G333.60   &   \#62   &   245.5358049   &   -50.1009901   &   7.7 $\pm$ 1.4   &   45   &   3660 \\

    \hline \noalign {\smallskip}
    
      \textbf{G337.92}  &  \textbf{\#1}  &   \textbf{250.2935548}  &  \textbf{-47.1342727}  &  $\bm{23.0\pm0.4}$  &  \textbf{100}  &  \textbf{2670}\\
      \textbf{G337.92}  &  \textbf{\#2}  &   \textbf{250.2936172}  &  \textbf{-47.133748}  &  $\bm{19.6\pm0.7}$  &  \textbf{48}  &  \textbf{3140}\\
	 G337.92   &   \#8   &   250.2927529   &   -47.1352673   &   4.4 $\pm$ 0.6   &   39   &   2360 \\
	 G337.92   &   \#9   &   250.294383   &   -47.1348356   &   4.3 $\pm$ 0.8   &   40   &   2890 \\
 
    \hline \noalign {\smallskip}
    
	 G338.93   &   \#1   &   250.1422011   &   -45.6934028   &   12.9 $\pm$ 0.1   &   100   &   1870 \\
	 G338.93   &   \#2   &   250.1427268   &   -45.6936217   &   12.4 $\pm$ 0.1   &   100   &   1880 \\
	 G338.93   &   \#3   &   250.1417044   &   -45.7020244   &   7.6 $\pm$ 0.1   &   100   &   1750 \\
	 G338.93   &   \#4   &   250.1397543   &   -45.6936984   &   6.2 $\pm$ 0.1   &   100   &   1870 \\
	 G338.93   &   \#6   &   250.1419183   &   -45.7022799   &   15.5 $\pm$ 0.4   &   40   &   2400 \\
	 G338.93   &   \#8   &   250.1433977   &   -45.6938542   &   11.8 $\pm$ 0.3   &   27   &   3420 \\

    \hline \noalign {\smallskip}
    
	 G353.41   &   \#2   &   262.6165032   &   -34.6955865   &   12.3 $\pm$ 0.7   &   29   &   2300 \\
	 G353.41   &   \#8   &   262.6039531   &   -34.6936374   &   4.6 $\pm$ 0.2   &   26   &   2630 \\

    \hline \noalign {\smallskip}
    
      \textbf{W43-MM1}  &  \textbf{\#1}  &   \textbf{281.9459334}  &  \textbf{-1.9074736}  &  $\bm{72.4\pm0.6}$  &  \textbf{100}  &  \textbf{2470}\\
      \textbf{W43-MM1}  &  \textbf{\#2}  &   \textbf{281.9451648}  &  \textbf{-1.9081343}  &  $\bm{28.7\pm0.7}$  &  \textbf{100}  &  \textbf{1770}\\
      \textbf{W43-MM1}  &  \textbf{\#3}  &   \textbf{281.9432069}  &  \textbf{-1.9092736}  &  $\bm{20.3\pm0.3}$  &  \textbf{100}  &  \textbf{2550}\\
      \textbf{W43-MM1}  &  \textbf{\#4}  &   \textbf{281.945748}  &  \textbf{-1.9073463}  &  $\bm{35.4\pm0.5}$  &  \textbf{100}  &  \textbf{3460}\\
      \textbf{W43-MM1}  &  \textbf{\#5}  &   \textbf{281.9469228}  &  \textbf{-1.908243}  &  $\bm{24.4\pm0.3}$  &  \textbf{26}  &  \textbf{1550}\\
	 W43-MM1   &   \#7   &   281.9438998   &   -1.9064175   &   7.4 $\pm$ 0.2   &   100   &   1910 \\
	 W43-MM1   &   \#9   &   281.9448606   &   -1.9086666   &   6.3 $\pm$ 0.3   &   100   &   1860 \\

    \hline
    \end{tabular}
    \label{tab:proto_candidates}
\begin{tablenotes}
\item Bold font is used to present protostellar core candidates with $\rm M> 16 M_{\odot}$.
\item[1] Mass of the protostellar core candidates obtained with the PPMAP 70\,$\rm \mu m$ corrected dust temperatures (see \citealp{Dell'Ova2023}).%, and applying a 50\% additional uncertainty on the mass to account for uncertainty on the dust emissivity.
\item[2] PPMAP 70\,$\rm \mu m$ corrected dust temperatures (see \citealp{Dell'Ova2023}).

\item[*] These sources were found to be multiple using the GExt2D second derivative map. The sizes and masses are computed using the ellipses parameters and fluxes of the GExt2D catalogs.

\end{tablenotes}    
\end{threeparttable}
\end{table*}

\begin{table*}[h!]\ContinuedFloat

    \centering     
    \begin{threeparttable}[c]
    \caption[]{Continued.}
    
    \begin{tabular}{lllllcc}
    \hline \noalign {\smallskip}
    Region & Source &  RA  &  DEC  & Mass & Temperature & FWHM$^{\rm dec}$  \\ \noalign {\smallskip}
        &   &  [deg]  & [deg]  &  $\left[\rm M_\odot \right]$ & $\left[\rm K\right]$ & $\left[\rm au \right]$ {\smallskip}\\  
    \hline \noalign {\smallskip}

	 W43-MM1   &   \#13   &   281.9427046   &   -1.9092745   &   10.7 $\pm$ 0.6   &   26   &   2060 \\
	 W43-MM1   &   \#16   &   281.9372707   &   -1.9118969   &   5.0 $\pm$ 0.3   &   25   &   1890 \\
	 W43-MM1   &   \#20   &   281.9456983   &   -1.9036124   &   8.4 $\pm$ 0.4   &   24   &   4840 \\
	 W43-MM1   &   \#24   &   281.9440629   &   -1.9089   &   12.3 $\pm$ 1.4   &   26   &   1760 \\
	 W43-MM1   &   \#29   &   281.9460739   &   -1.9089351   &   4.6 $\pm$ 0.9   &   25   &   1350 \\
	 W43-MM1   &   \#37   &   281.9454183   &   -1.9067383   &   6.3 $\pm$ 1.1   &   28   &   1590 \\
	 W43-MM1   &   \#40   &   281.9440532   &   -1.9057986   &   4.5 $\pm$ 0.5   &   26   &   3060 \\
	 W43-MM1   &   \#42   &   281.9455026   &   -1.9079514   &   4.9 $\pm$ 1.3   &   31   &   1350 \\
  
    \hline \noalign {\smallskip}    
    
      \textbf{W43-MM2}  &  \textbf{\#1}  &   \textbf{281.9033208}  &  \textbf{-2.0150752}  &  $\bm{40.9\pm0.3}$  &  \textbf{100}  &  \textbf{3000}\\
      \textbf{W43-MM2}  &  \textbf{\#3}  &   \textbf{281.9004197}  &  \textbf{-2.0210949}  &  $\bm{18.4\pm0.3}$  &  \textbf{25}  &  \textbf{3460}\\
	 W43-MM2   &   \#5   &   281.9011837   &   -2.0141012   &   8.1 $\pm$ 0.2   &   27   &   1940 \\
      \textbf{W43-MM2}  &  \textbf{\#6}  &   \textbf{281.9031379}  &  \textbf{-2.0149264}  &  $\bm{24.3\pm0.5}$  &  \textbf{37}  &  \textbf{3180}\\
	 W43-MM2   &   \#8   &   281.8962619   &   -2.0191026   &   4.6 $\pm$ 0.1   &   23   &   1820 \\
      \textbf{W43-MM2}  &  \textbf{\#10}  &   \textbf{281.9135971}  &  \textbf{-2.0078124}  &  $\bm{21.3\pm0.6}$  &  \textbf{29}  &  \textbf{3250}\\
	 W43-MM2   &   \#13   &   281.9028491   &   -2.0133518   &   6.0 $\pm$ 0.2   &   28   &   2120 \\
	 W43-MM2   &   \#26   &   281.9038579   &   -2.0152273   &   4.1 $\pm$ 0.3   &   31   &   3720 \\

    \hline \noalign {\smallskip}
    
	 W43-MM3   &   \#1   &   281.9238036   &   -2.0079434   &   7.3 $\pm$ 0.2   &   100   &   1570 \\
	 W43-MM3   &   \#4   &   281.9238914   &   -2.0076105   &   8.2 $\pm$ 0.5   &   33   &   1350 \\
	 W43-MM3   &   \#6   &   281.9234345   &   -2.007048   &   5.4 $\pm$ 0.5   &   32   &   2130 \\
	 W43-MM3   &   \#10   &   281.924284   &   -2.008162   &   4.2 $\pm$ 0.5   &   31   &   2840 \\

    \hline \noalign {\smallskip}
    
      \textbf{W51-E}  &  \textbf{\#2}  &   \textbf{290.9331854}  &  \textbf{14.5095896}  &  $\bm{87.9\pm0.6}$  &  \textbf{300}  &  \textbf{3740}\\
      \textbf{W51-E}  &  \textbf{\#4}  &   \textbf{290.9329076}  &  \textbf{14.5078281}  &  $\bm{147.0\pm6.7}$  &  \textbf{100}  &  \textbf{3280}\\
      \textbf{W51-E}  &  \textbf{\#5}\tnote{$\dagger$}  &   \textbf{290.9328333}  &  \textbf{14.5099861}  &  $\bm{86.4\pm3.1}$  &  \textbf{59}  &  \textbf{3910}\\
      \textbf{W51-E}  &  \textbf{\#8}  &   \textbf{290.9330369}  &  \textbf{14.5101522}  &  $\bm{55.8\pm3.3}$  &  \textbf{54}  &  \textbf{3840}\\
      \textbf{W51-E}  &  \textbf{\#9}\tnote{$\dagger$}  &   \textbf{290.9426667}  &  \textbf{14.4964278}  &  $\bm{27.8\pm2.2}$  &  \textbf{24}  &  \textbf{3240}\\
      \textbf{W51-E}  &  \textbf{\#12}  &   \textbf{290.933288}  &  \textbf{14.5091267}  &  $\bm{16.0\pm1.5}$  &  \textbf{49}&  \textbf{3730}\\
      \textbf{W51-E}  &  \textbf{\#13}  &   \textbf{290.9324479}  &  \textbf{14.5054808}  &  $\bm{17.1\pm3.4}$  &  \textbf{32}  &  \textbf{2830}\\
	 W51-E   &   \#15   &   290.9315055   &   14.5099588   &   10.9 $\pm$ 0.8   &   29   &   3770 \\
	 W51-E   &   \#16   &   290.9327544   &   14.5073581   &   12.5 $\pm$ 5.5   &   53   &   2340 \\
	 W51-E   &   \#19   &   290.9327315   &   14.5112086   &   6.0 $\pm$ 1.0   &   27   &   1350 \\
	 W51-E   &   \#22   &   290.9277814   &   14.500986   &   4.1 $\pm$ 0.4   &   23   &   2970 \\
	 W51-E   &   \#29   &   290.9296394   &   14.5149377   &   7.5 $\pm$ 0.9   &   24   &   4520 \\
	 W51-E   &   \#34   &   290.9325349   &   14.5131101   &   6.6 $\pm$ 0.6   &   23   &   4730 \\
	 %W51-E   &   \#38   &   290.934952   &   14.5078333   &   5.3 $\pm$ 0.7   &   26   &   3180 \\
    
    \hline \noalign {\smallskip}
    
      \textbf{W51-IRS2}  &  \textbf{\#1}  &   \textbf{290.9168699}  &  \textbf{14.5181896}  &  $\bm{57.8\pm0.7}$  &  \textbf{300}  &  \textbf{2550}\\
      \textbf{W51-IRS2}  &  \textbf{\#2}  &   \textbf{290.9164704}  &  \textbf{14.5181596}  &  $\bm{50.4\pm1.4}$  &  \textbf{100}  &  \textbf{1870}\\
      \textbf{W51-IRS2}  &  \textbf{\#4}  &   \textbf{290.9107235}  &  \textbf{14.5116053}  &  $\bm{16.9\pm0.2}$  &  \textbf{100}  &  \textbf{2410}\\
      \textbf{W51-IRS2}  &  \textbf{\#5}\tnote{$\dagger$}  &   \textbf{290.9159167}  &  \textbf{14.5180806}  &  $\bm{67.6\pm2.9}$  &  \textbf{60}  &  \textbf{2480}\\
      \textbf{W51-IRS2}  &  \textbf{\#6}  &   \textbf{290.916647}  &  \textbf{14.5182923}  &  $\bm{49.5\pm1.4}$  &  \textbf{100}  &  \textbf{2730}\\
      \textbf{W51-IRS2}  &  \textbf{\#7}  &   \textbf{290.9156148}  &  \textbf{14.5181323}  &  $\bm{18.6\pm1.1}$  &  \textbf{100}  &  \textbf{2020}\\
	 W51-IRS2   &   \#10   &   290.9111836   &   14.5126596   &   8.0 $\pm$ 0.3   &   31   &   1510 \\
      \textbf{W51-IRS2}  &  \textbf{\#13}  &   \textbf{290.9059966}  &  \textbf{14.5052314}  &  $\bm{21.8\pm0.6}$  &  \textbf{20}  &  \textbf{2220}\\
	 W51-IRS2   &   \#14   &   290.9091924   &   14.5092115   &   6.1 $\pm$ 0.4   &   31   &   1770 \\
	 W51-IRS2   &   \#15   &   290.9100189   &   14.510189   &   11.4 $\pm$ 0.5   &   31   &   3720 \\
	 W51-IRS2   &   \#27   &   290.9083688   &   14.5198952   &   4.2 $\pm$ 0.1   &   28   &   5390 \\
	 W51-IRS2   &   \#33   &   290.9235087   &   14.5171682   &   4.7 $\pm$ 0.3   &   34   &   3150 \\
	 W51-IRS2   &   \#41   &   290.9161814   &   14.5181418   &   13.3 $\pm$ 1.9   &   62   &   2720 \\
	 W51-IRS2   &   \#49   &   290.9114383   &   14.5127072   &   4.3 $\pm$ 0.3   &   31   &   3370 \\
	 W51-IRS2   &   \#60   &   290.911816   &   14.5117357   &   5.9 $\pm$ 0.5   &   30   &   4860 \\
	 W51-IRS2   &   \#68   &   290.9149185   &   14.5178211   &   4.1 $\pm$ 0.7   &   45   &   2980 \\
	 W51-IRS2   &   \#76   &   290.9074011   &   14.506935   &   4.5 $\pm$ 0.5   &   27   &   2020 \\
	 W51-IRS2   &   \#81   &   290.9082835   &   14.5082754   &   5.4 $\pm$ 0.3   &   27   &   5180 \\

    \hline
    \end{tabular}
\begin{tablenotes}

\item[$\dagger$] These sources were initially considered as free-free sources by \cite{louvet22}, but were added here to the protostellar sample due to their spectral index close to two and their association to an outflow.

\end{tablenotes}    
\end{threeparttable}
\end{table*}
\section{Correction of the masses of the protostellar cores using ionising models} 
\label{appendix:ionising_models}

\begin{figure}[htbp!]
    \centering
    \includegraphics[width=0.49\textwidth]{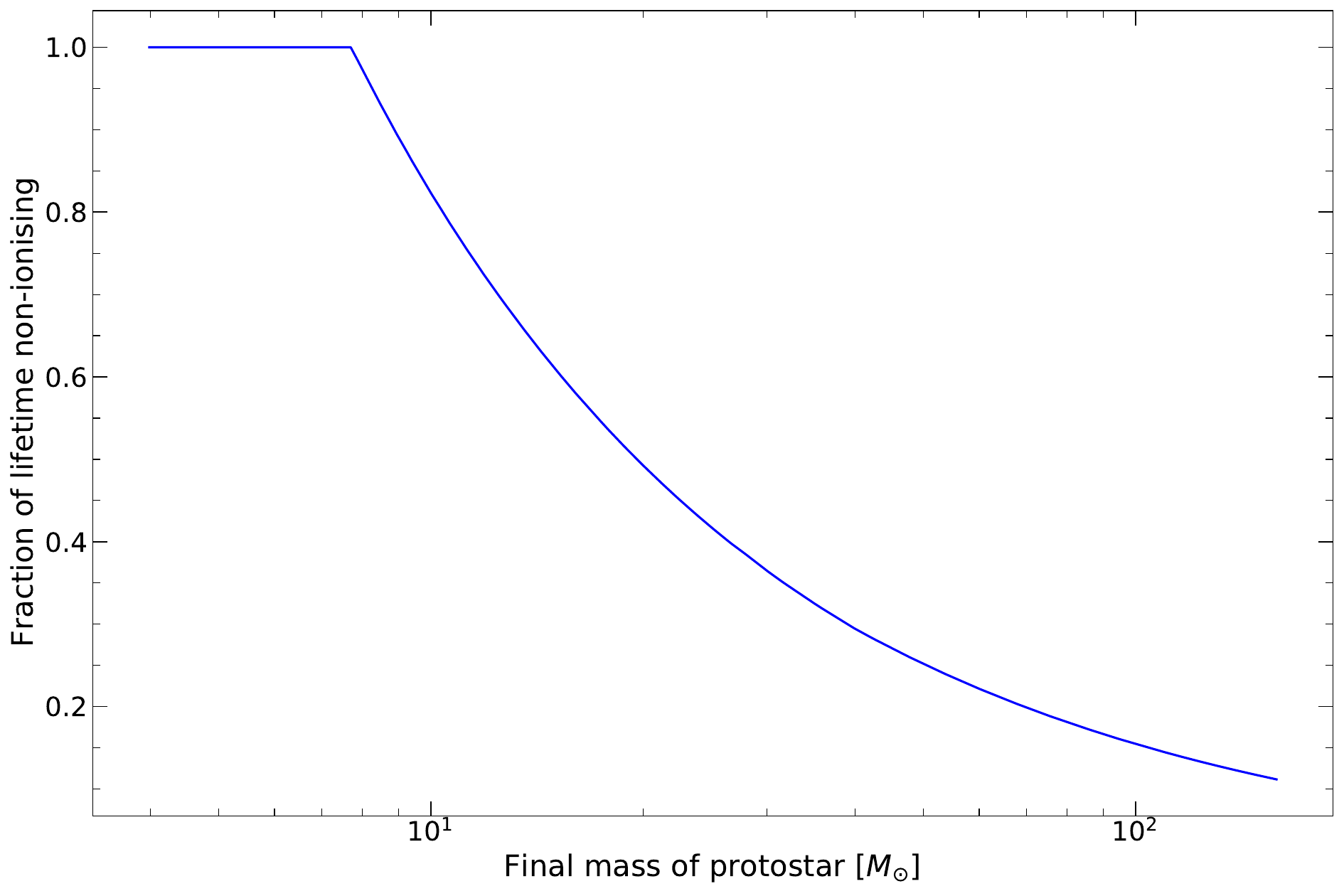}
    \includegraphics[width=0.49\textwidth]{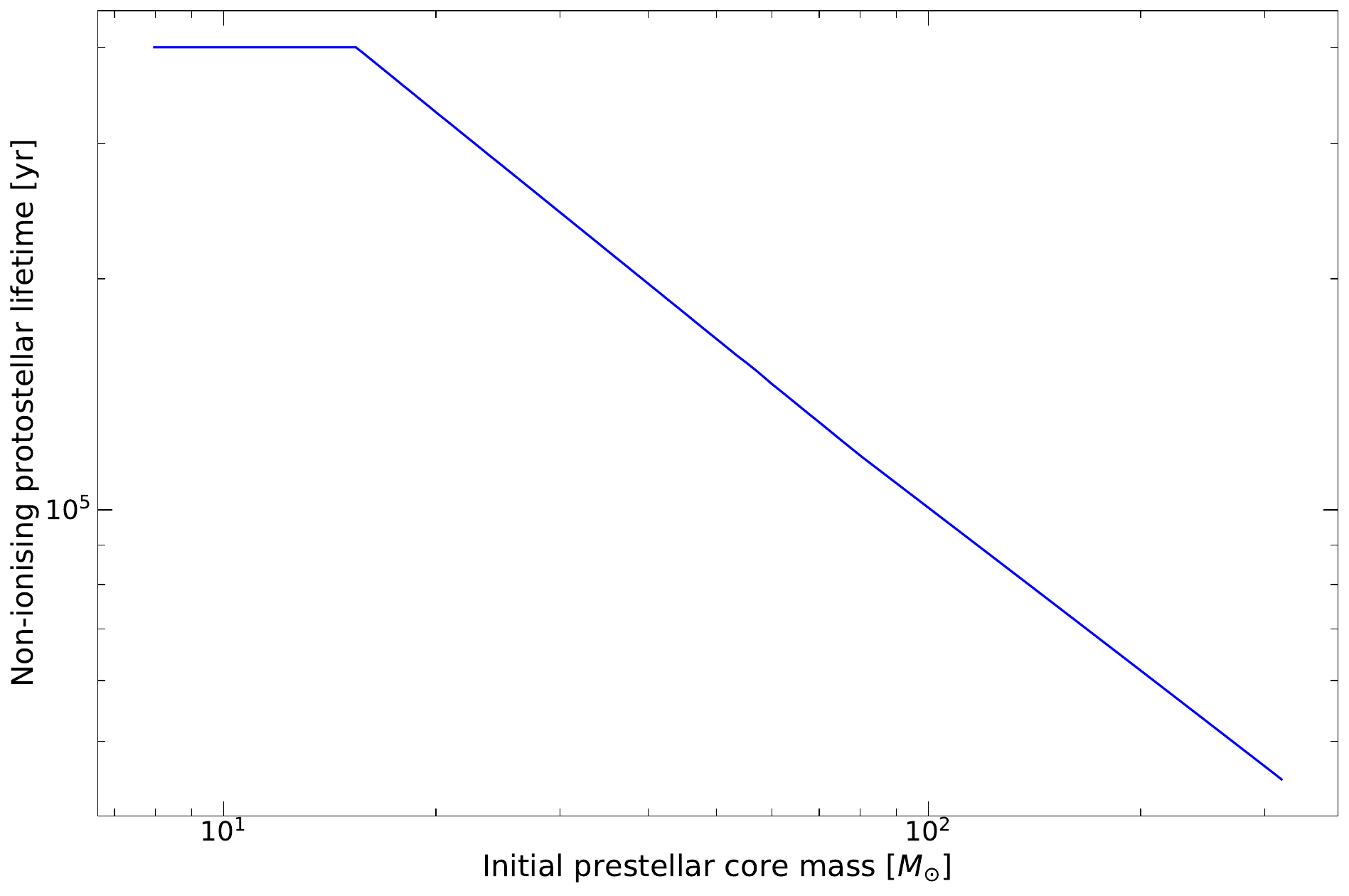}

   % \vskip -0.3cm
    \caption{ \textbf{Top: } Fraction of the non-ionising protostellar lifetime as function of the final stellar mass. \textbf{Bottom: } Non-Ionising protostellar lifetime as function of the initial prestellar mass reservoir.}

    \label{appendix:ionising_models_fig}
   
\end{figure}

Here we present the correction applied to the observed protostellar cores masses in order to estimate their initial prestellar mass reservoir. We assume a constant accretion and a core to star efficiency ($\epsilon_{\rm cse}$) of 50\%. As we work on core catalogs with ionising protostars removed, we can assume that on average we observe protostellar cores at half of their non-ionising part life (i.e. while the protostar is accreting to reach the mass ionising threshold). Using the models from \cite{Hosokawa2009} as in \cite{Duarte-Cabral2013}, we can infer the fraction of time each protostellar core is going to pass in the ionising phase in function of the final mass of the protostar. Then we correct from half of the mass lost during the non-ionising accretion time to obtain the initial mass reservoir of the protostellar core. In practice this correction doubles the mass of protostellar cores with protostars not massive enough to ionise their envelope, while the correction factor is lower than two for protostars that will be ionising for part of their protostellar lifetime. In Fig.\,\ref{appendix:ionising_models_fig} we present the fraction of the non-ionising protostellar lifetime as function of the final stellar mass, and the non-ionising protostellar lifetime as function of the initial prestellar mass reservoir.

\section{Massive prestellar core candidates figures} \label{appendix:MPSC_fig_part}
In this section we present the CO and SiO spectra and the individual molecular outflows maps of every candidate of the survey used to classify them as prestellar. The On and On-Off DCN spectra are also presented with the fit adopted to estimate the \vlsr of each candidate. 
 
%%%%%%%%%%%%%%%%%%%%%%% G008 %%%%%%%%%%%%%%%%%%%%%%%%%%%%%%%%%%%%%
\begin{figure*}
    \label{appendix:G008_MPSC_fig}
    % \centering
    % \raisebox{-0.5\height}{\includegraphics[width=0.49\textwidth]{Appendix/Appendix_figures/G008/G008_Source_4.pdf}}
    % \raisebox{-0.0\height}{\includegraphics[width=0.49\textwidth]{Appendix/Appendix_figures/G008/G008_Contours_Source_4.png}}
    % \raisebox{-0.9\height}{\includegraphics[width=0.30\textwidth]{Appendix/Appendix_figures/G008/G008_DCN_4.pdf}}

    \centering
    \begin{minipage}[c]{0.49\textwidth}
        \centering
        \includegraphics[width=\textwidth]{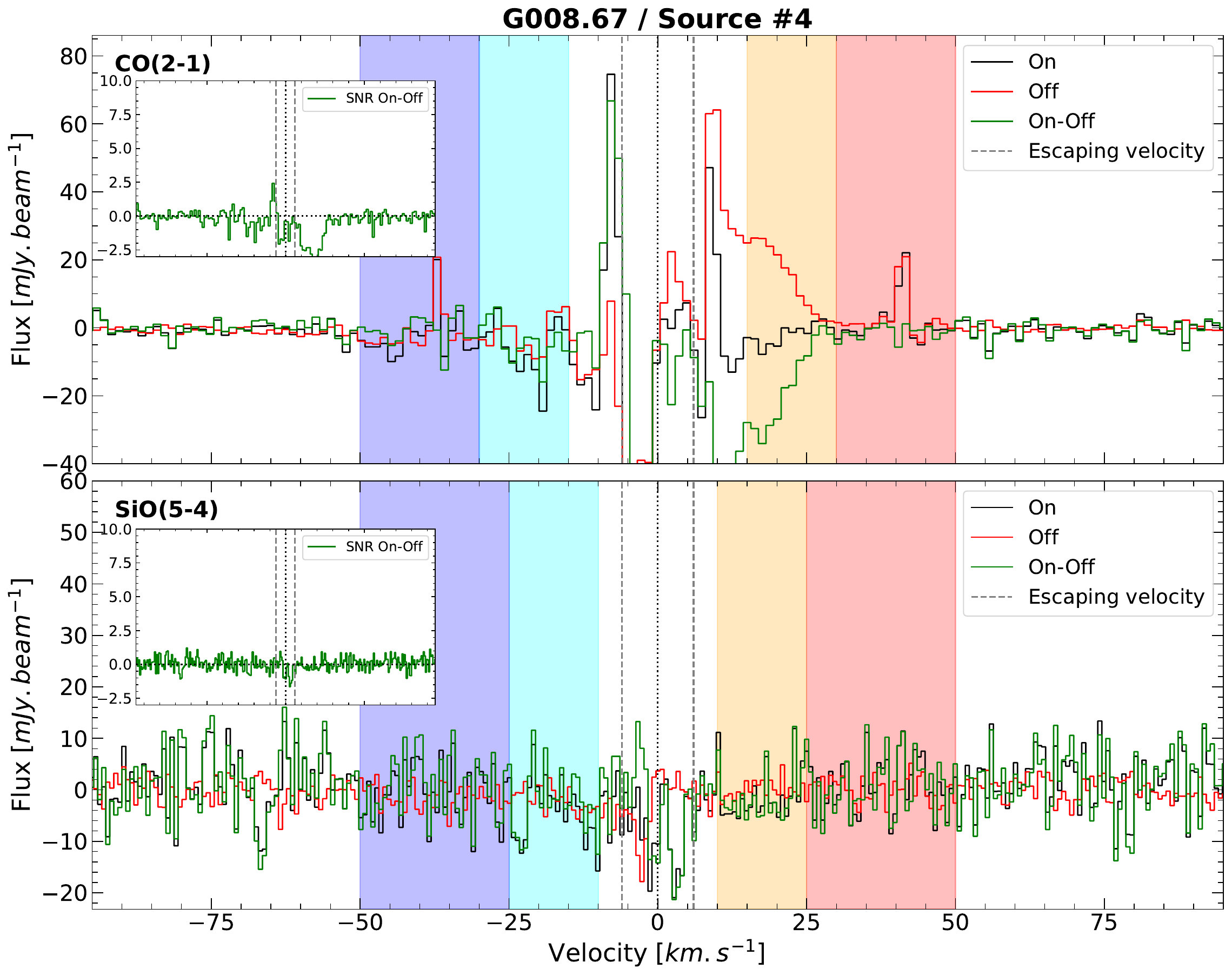}
    \end{minipage}
    \begin{minipage}[c]{0.49\textwidth}
        \centering
        \begin{minipage}[c]{\textwidth}
            \centering
            \includegraphics[width=0.9\textwidth]{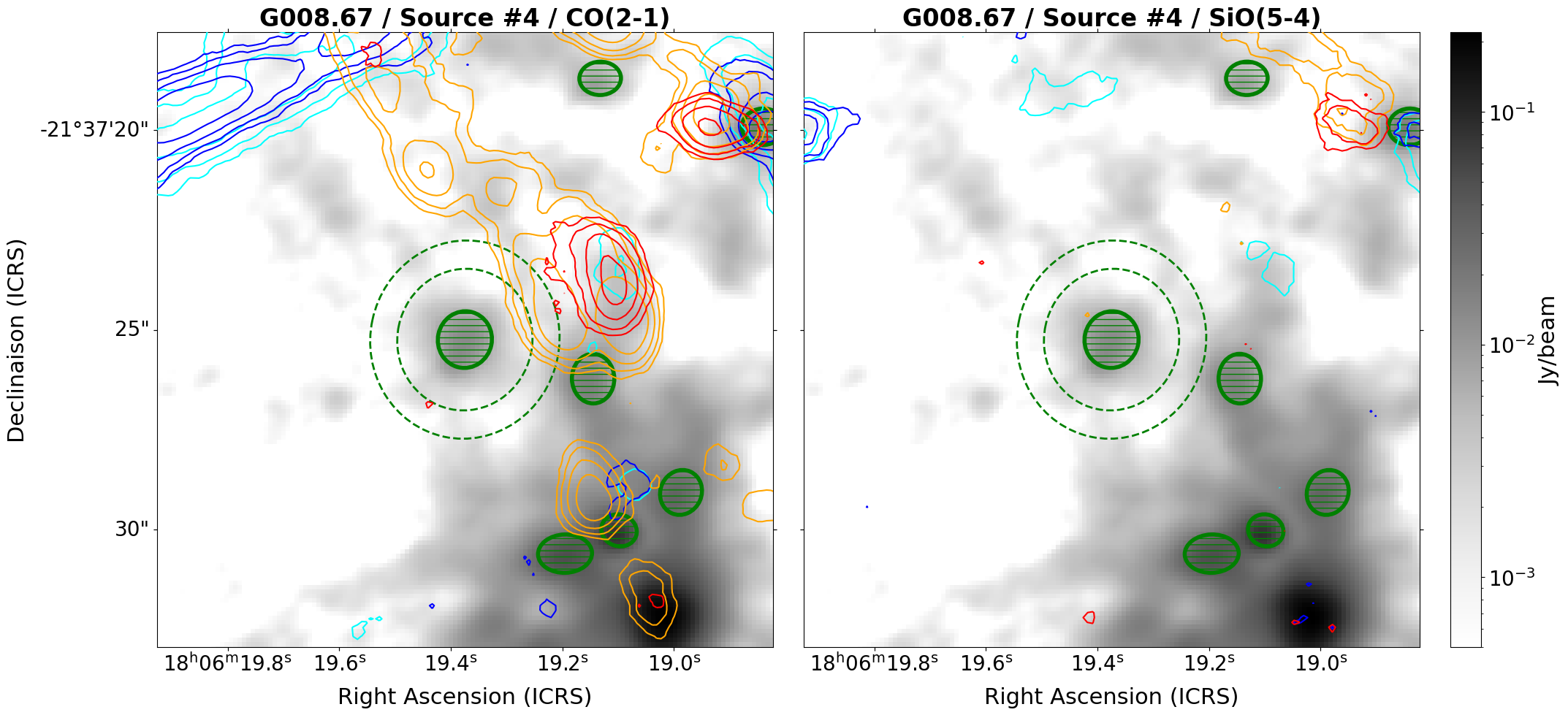}
        \end{minipage}
        \vfill
        \begin{minipage}[c]{\textwidth}
            \centering
            \includegraphics[width=0.7\textwidth]{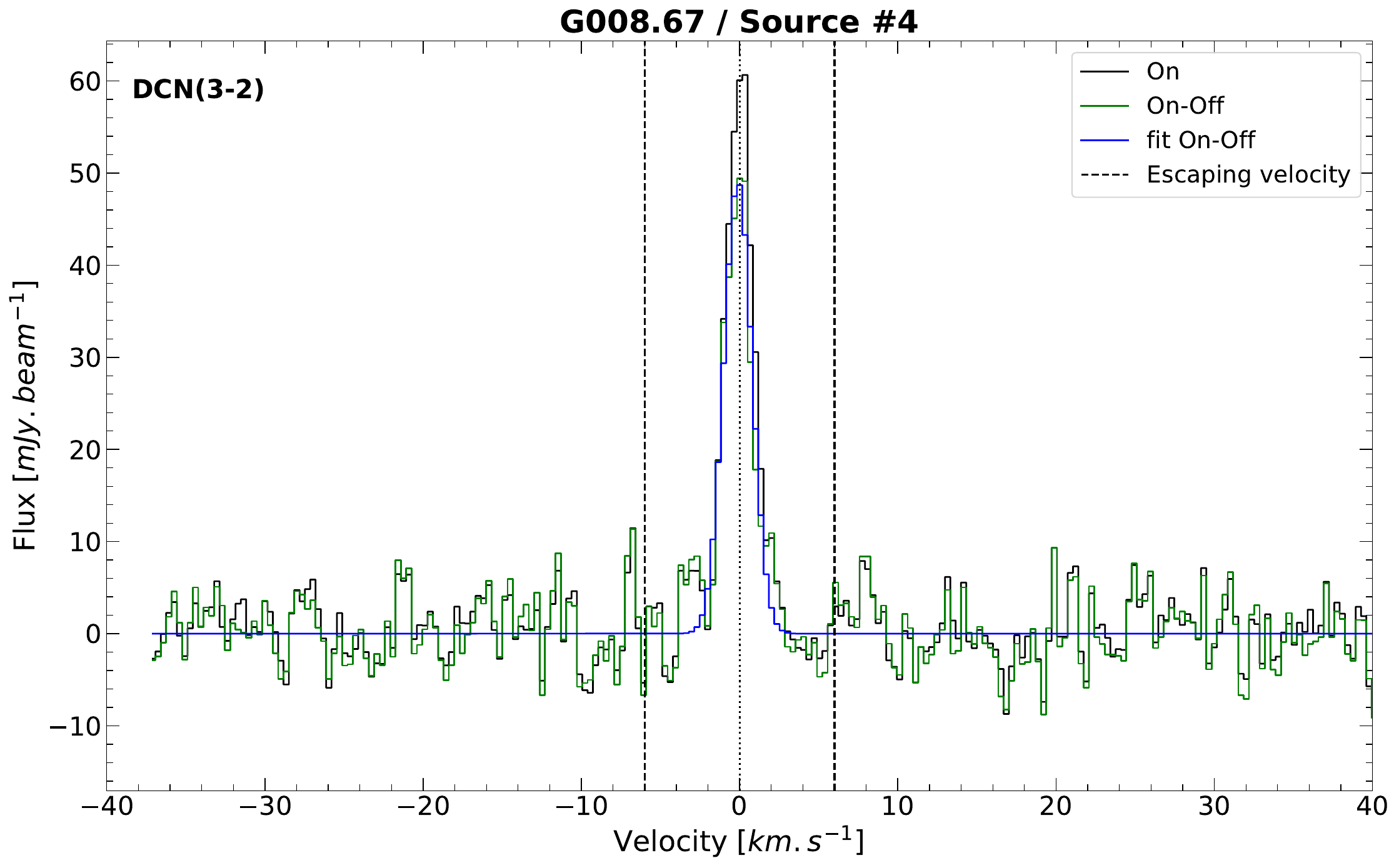}
        \end{minipage}
    \end{minipage}

\vspace{0.2cm}

    \centering
    \begin{minipage}[c]{0.49\textwidth}
        \centering
        \includegraphics[width=\textwidth]{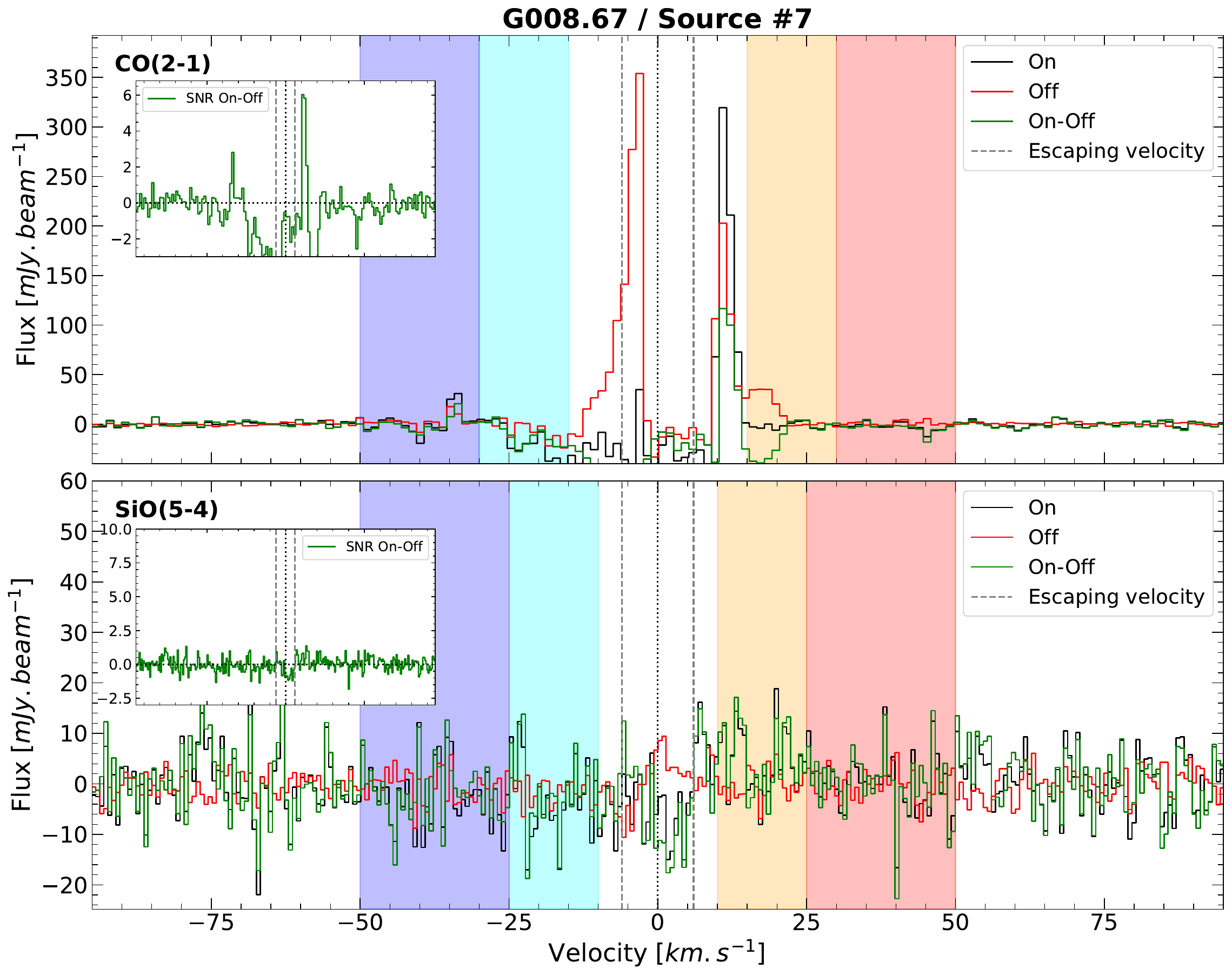}
    \end{minipage}
    \begin{minipage}[c]{0.49\textwidth}
        \centering
        \begin{minipage}[c]{\textwidth}
            \centering
            \includegraphics[width=0.9\textwidth]{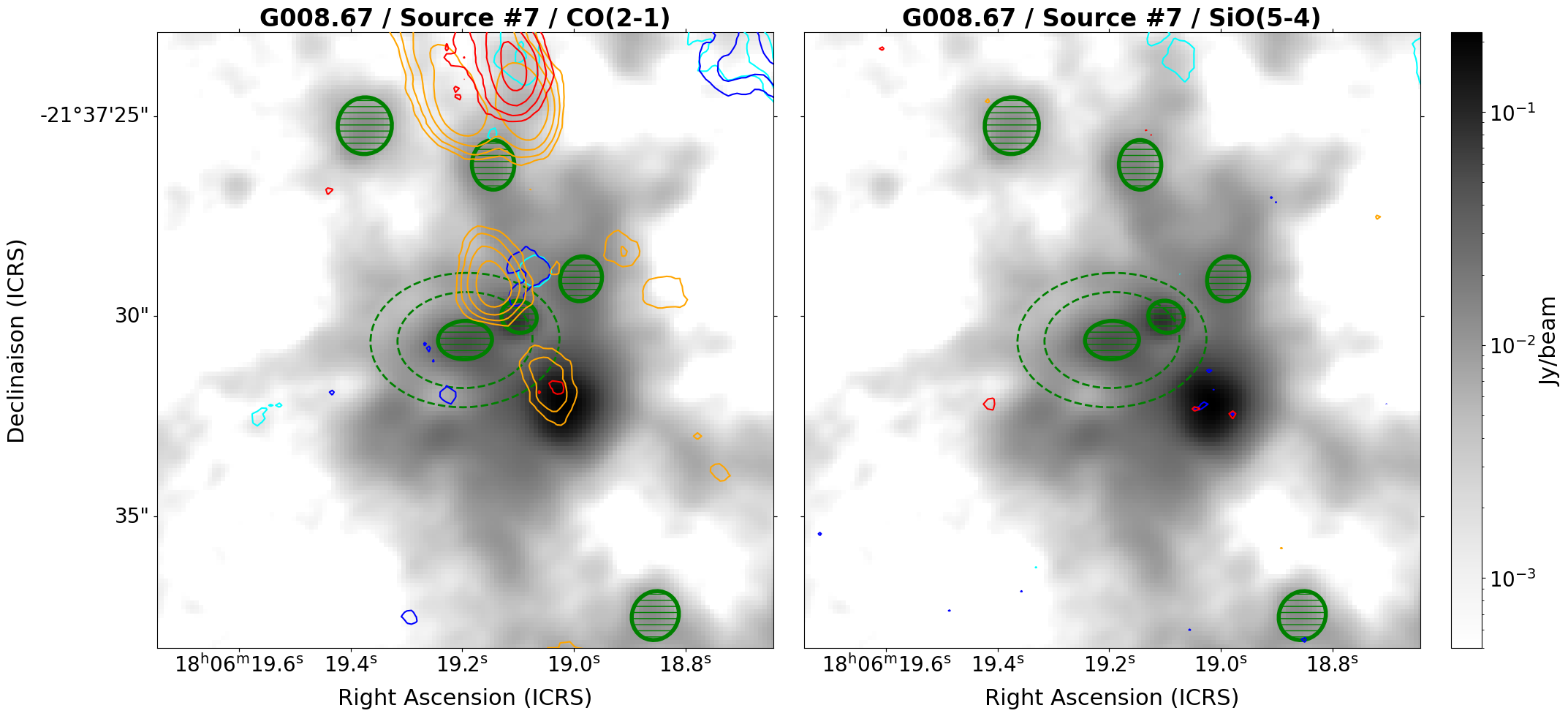}
        \end{minipage}
        \vfill
        \begin{minipage}[c]{\textwidth}
            \centering
            \includegraphics[width=0.7\textwidth]{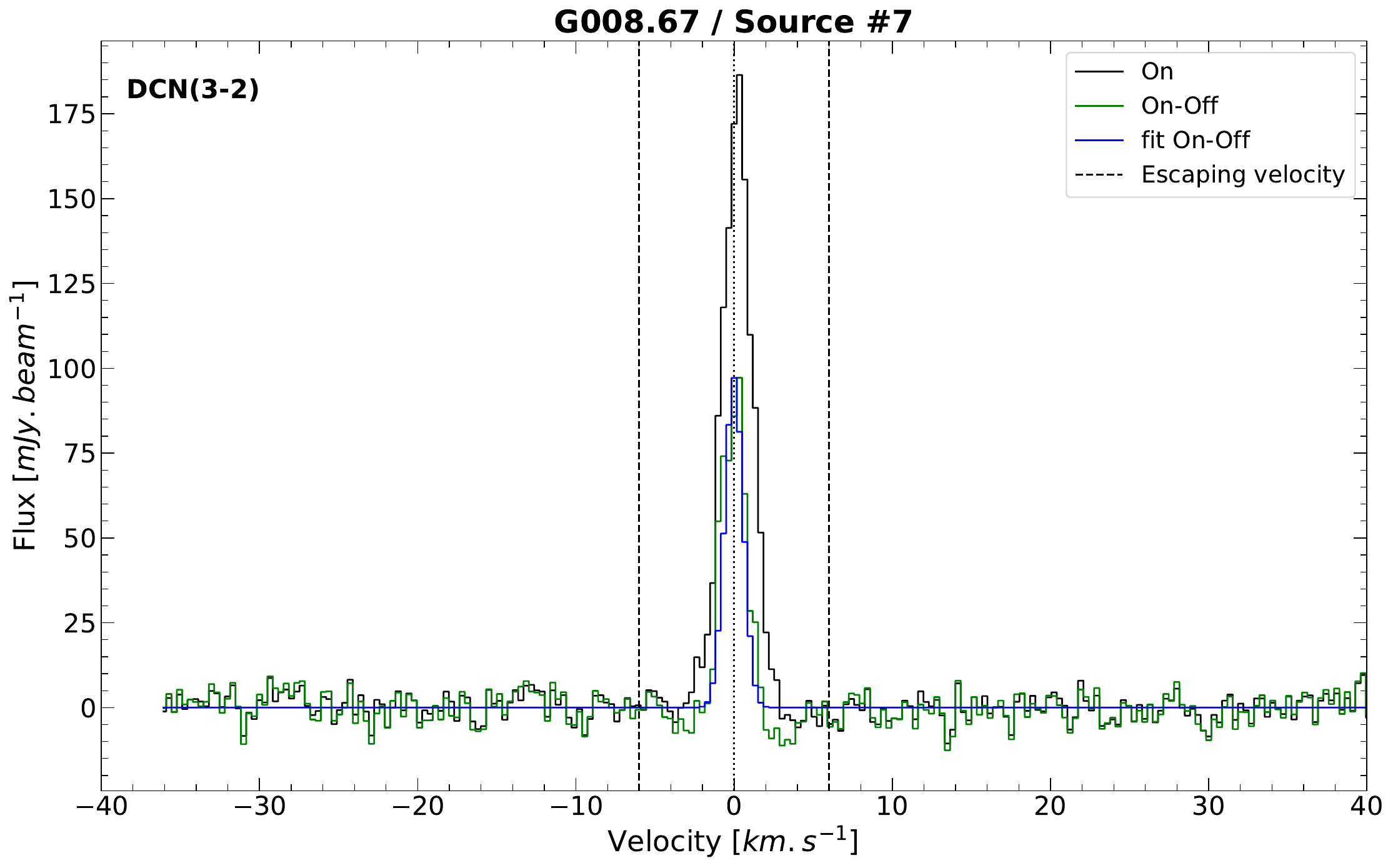}
        \end{minipage}
    \end{minipage}

\vspace{0.2cm}

    \centering
    \begin{minipage}[c]{0.49\textwidth}
        \centering
        \includegraphics[width=\textwidth]{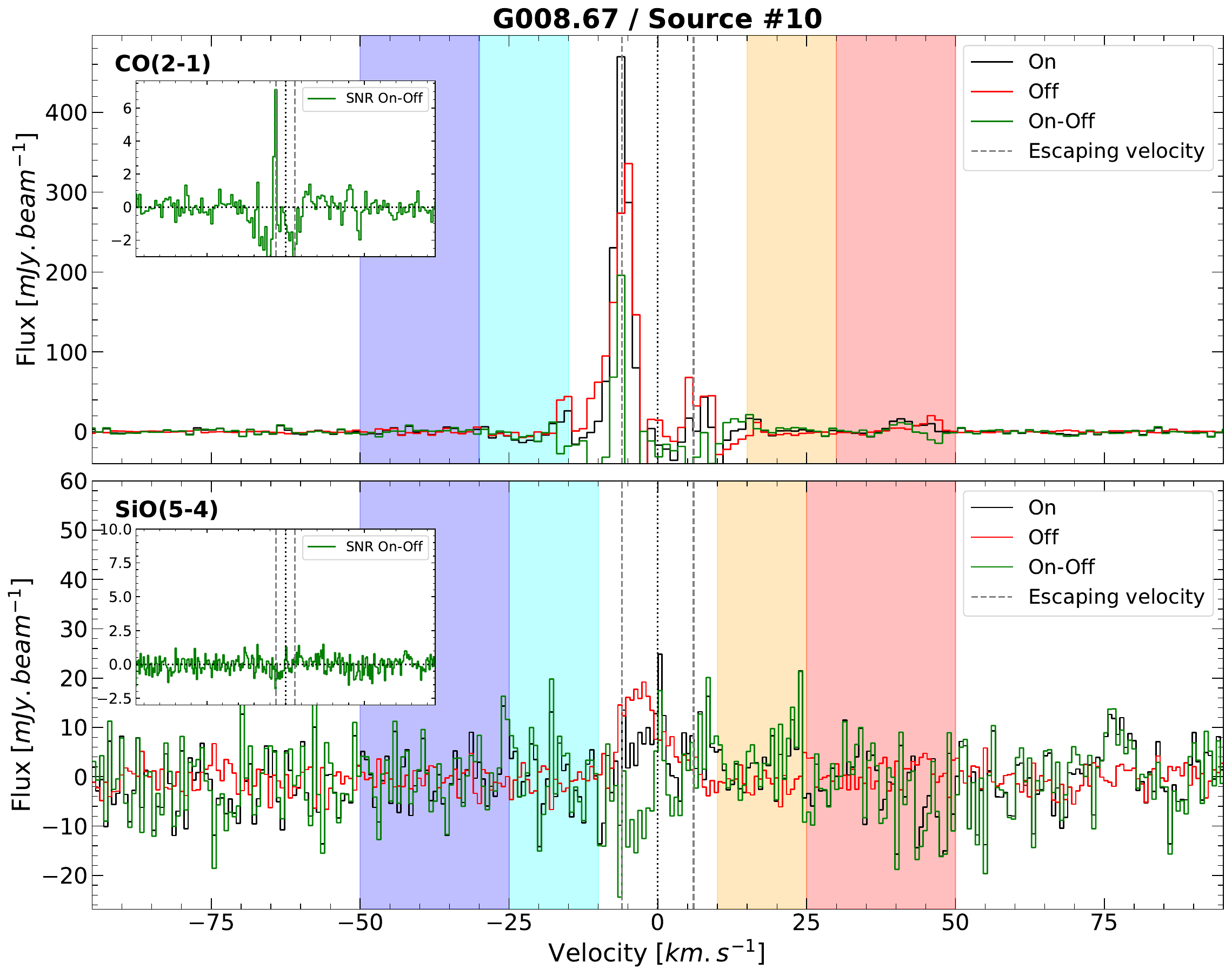}
    \end{minipage}
    \begin{minipage}[c]{0.49\textwidth}
        \centering
        \begin{minipage}[c]{\textwidth}
            \centering
            \includegraphics[width=0.9\textwidth]{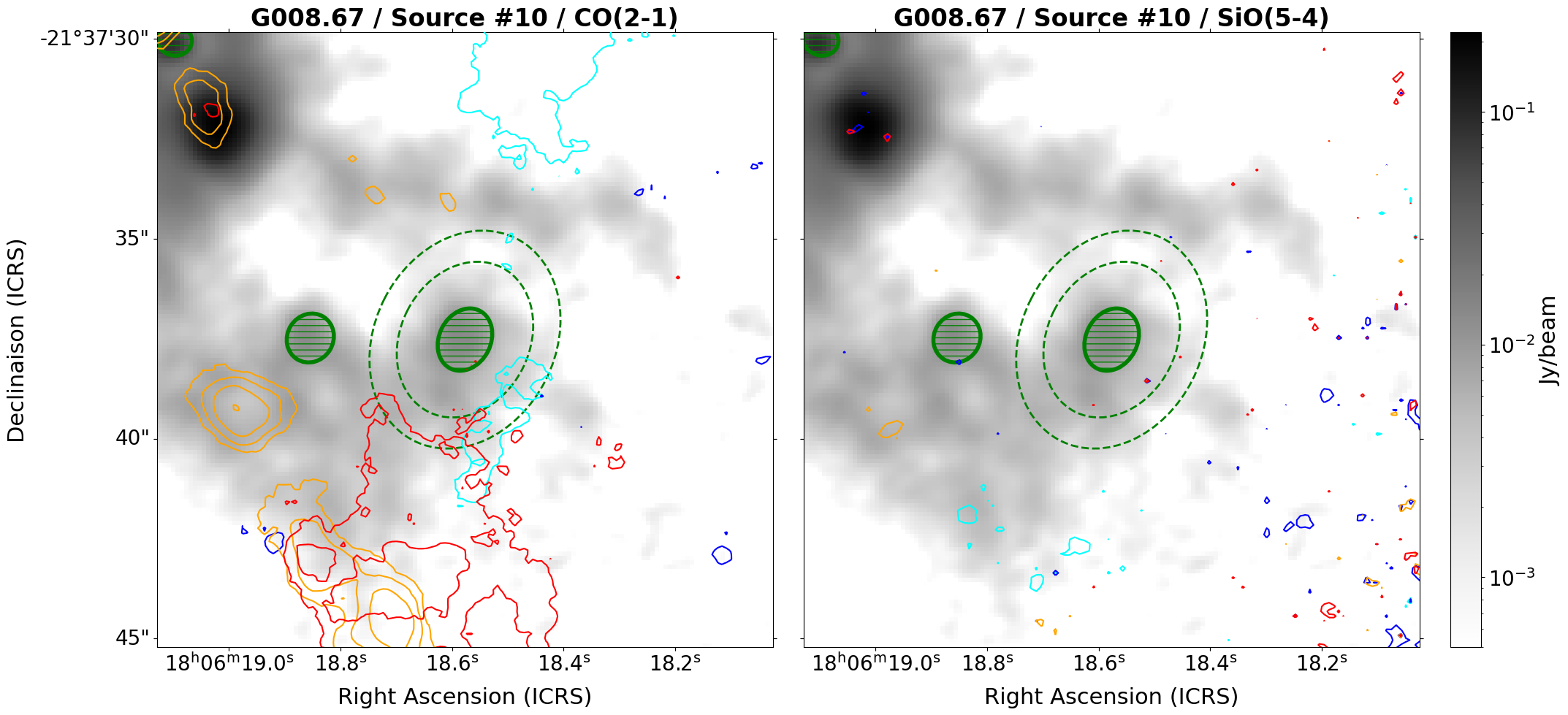}
        \end{minipage}
        \vfill
        \begin{minipage}[c]{\textwidth}
            \centering
            \includegraphics[width=0.7\textwidth]{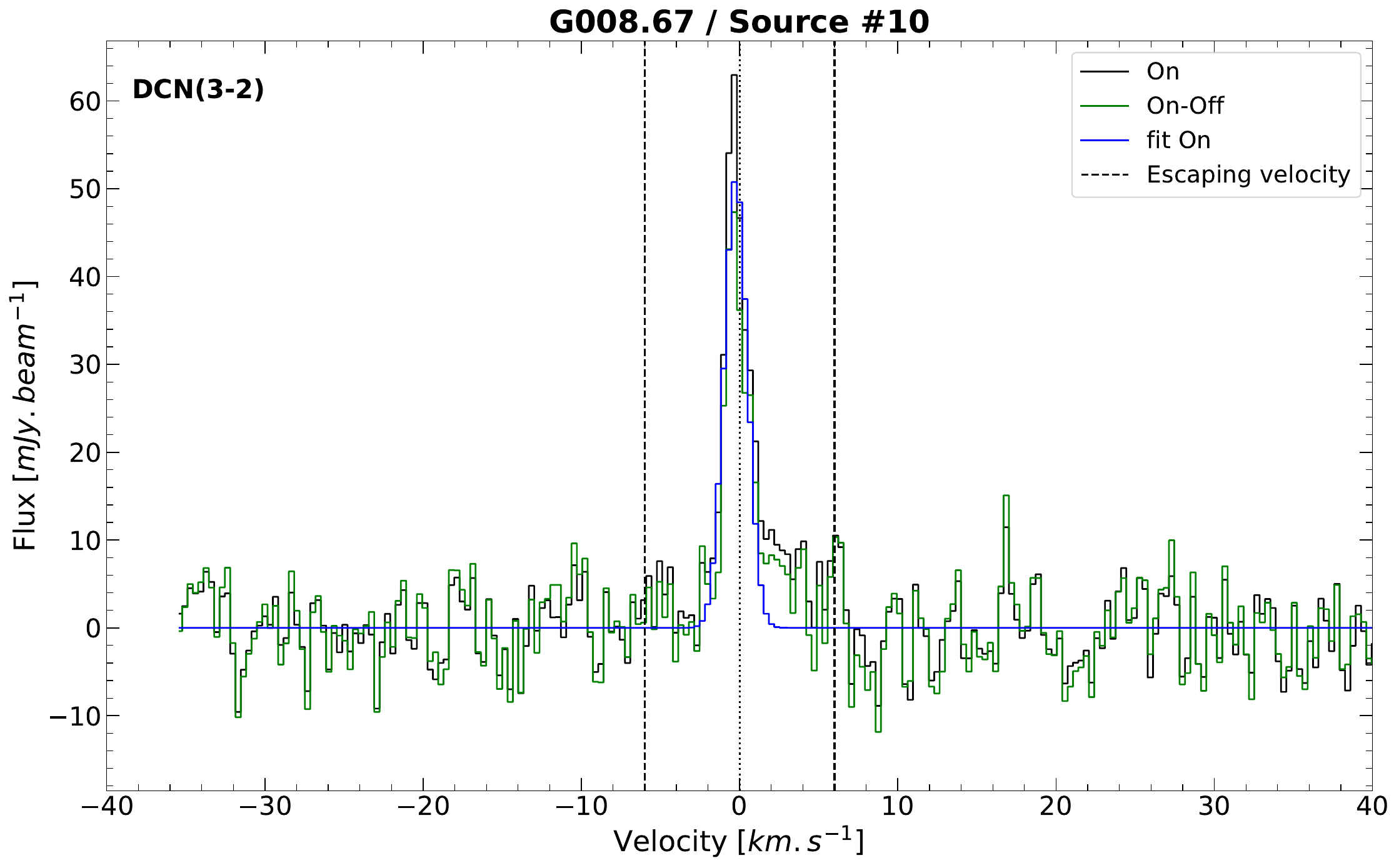}
        \end{minipage}
    \end{minipage}

    % \raisebox{-0.5\height}{\includegraphics[width=0.49\textwidth]{Appendix/Appendix_figures/G008/G008_Source_7.pdf}}
    % \raisebox{-0.5\height}{\includegraphics[width=0.49\textwidth]{Appendix/Appendix_figures/G008/G008_Contours_Source_7.png}}

    % \raisebox{-0.5\height}{\includegraphics[width=0.49\textwidth]{Appendix/Appendix_figures/G008/G008_Source_10.pdf}}
    % \raisebox{-0.5\height}{\includegraphics[width=0.49\textwidth]{Appendix/Appendix_figures/G008/G008_Contours_Source_10.png}}
   % \vskip -0.3cm
    \caption{CO and SiO spectra (left) and molecular outflow maps (top right) of the high-mass PSC candidates of the G008.67 region. CO contours are 5, 10, 20, and 40 in units of $\sigma$, with $\sigma$ = $39.4$, $37.5$, $35.0$, $38.0$ \mJybeamkms for cyan, blue, orange and red contours respectively. SiO contours are 5, 10, 20, and 40 in units of $\sigma$, with $\sigma$ = $21.7$, $27.7$, $21.5$, $27.3$ \mJybeamkms for cyan, blue, orange and red contours respectively. DCN spectra and fit (bottom right) of the high-mass PSC candidates of the G008.67 region.}

\end{figure*}

%%%%%%%%%%%%%%%%%%%%%%% G010 %%%%%%%%%%%%%%%%%%%%%%%%%%%%%%%%%%%%%
\begin{figure*}
    \label{appendix:G010_MPSC_fig}
    \centering
    \begin{minipage}[c]{0.49\textwidth}
        \centering
        \includegraphics[width=\textwidth]{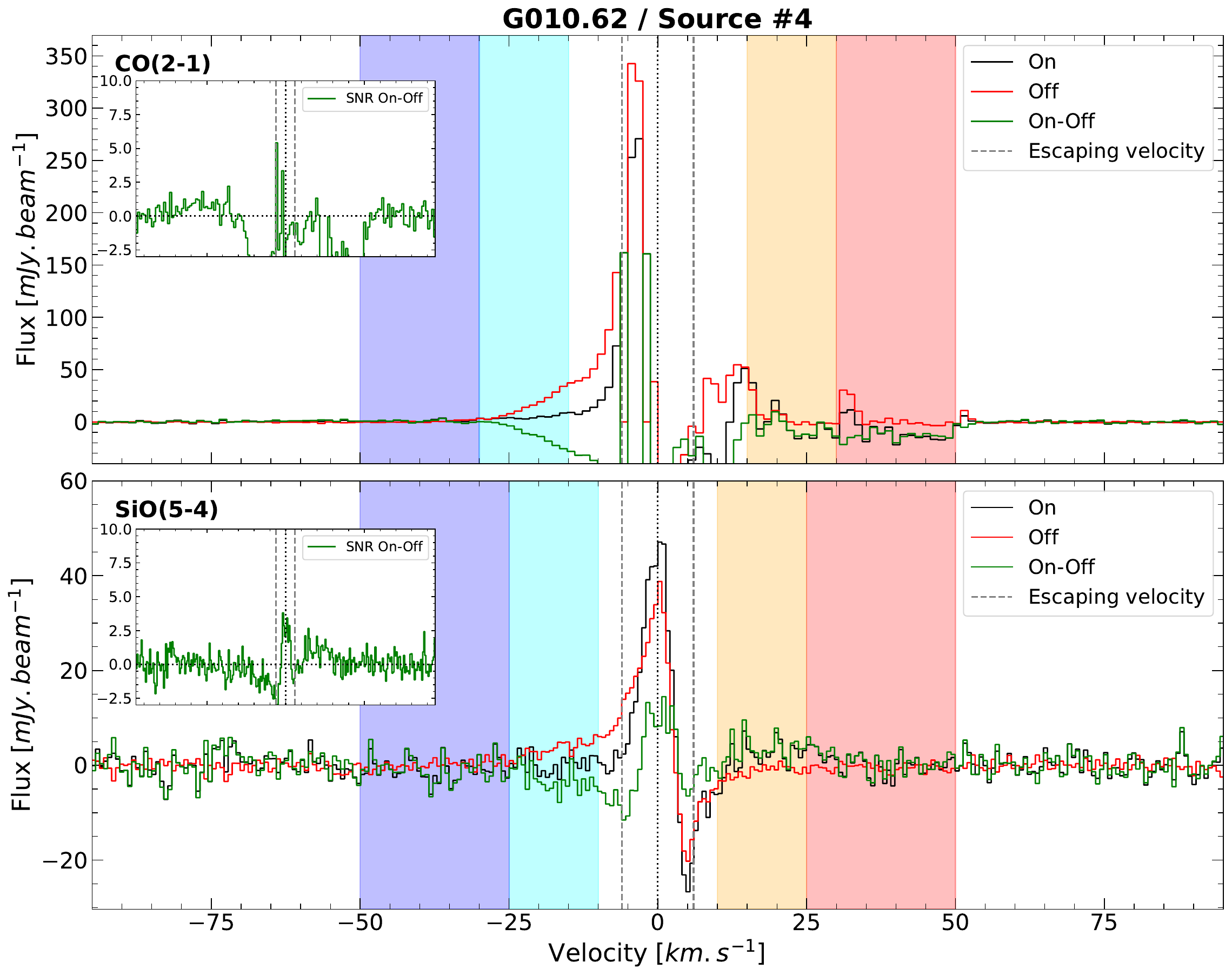}
    \end{minipage}
    \begin{minipage}[c]{0.49\textwidth}
        \centering
        \begin{minipage}[c]{\textwidth}
            \centering
            \includegraphics[width=0.9\textwidth]{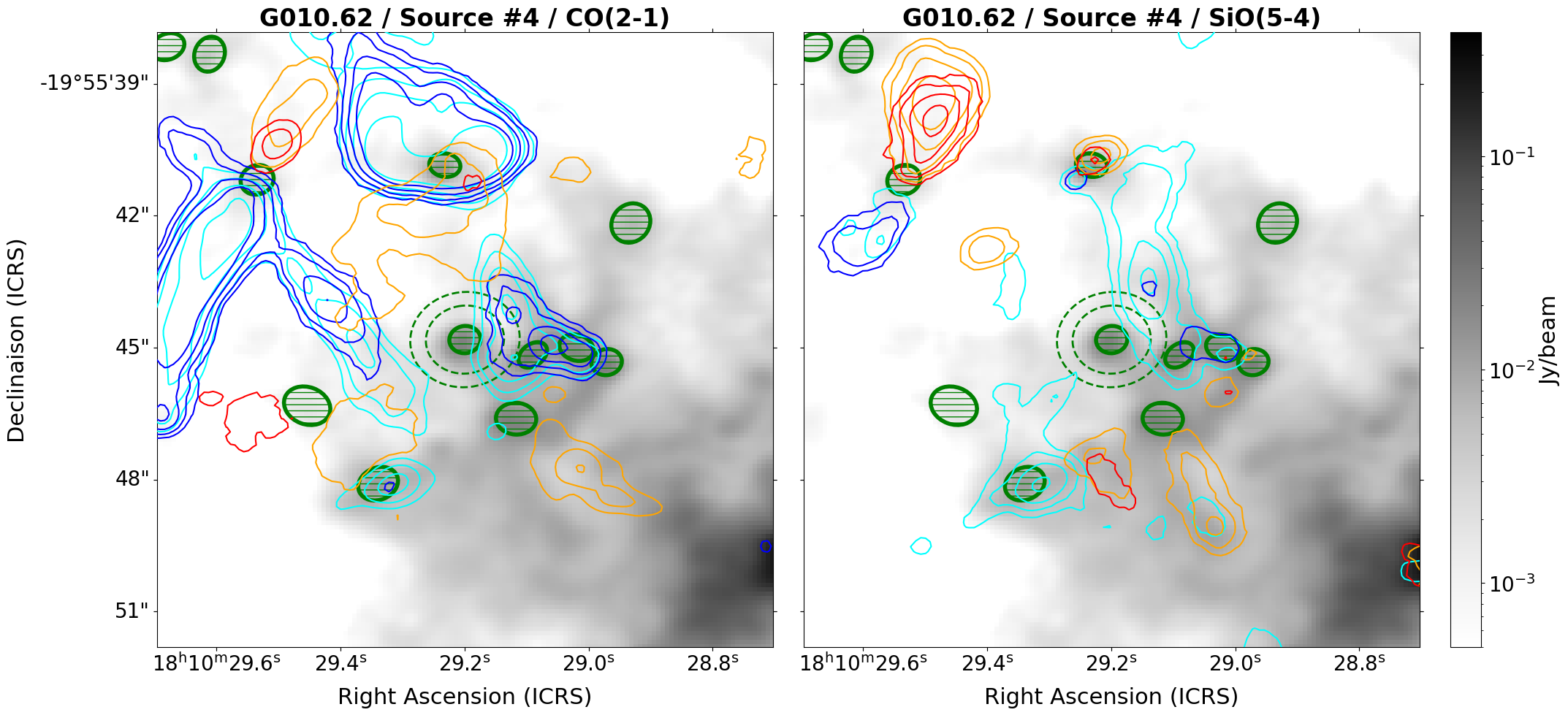}
        \end{minipage}
        \vfill
        \begin{minipage}[c]{\textwidth}
            \centering
            \includegraphics[width=0.7\textwidth]{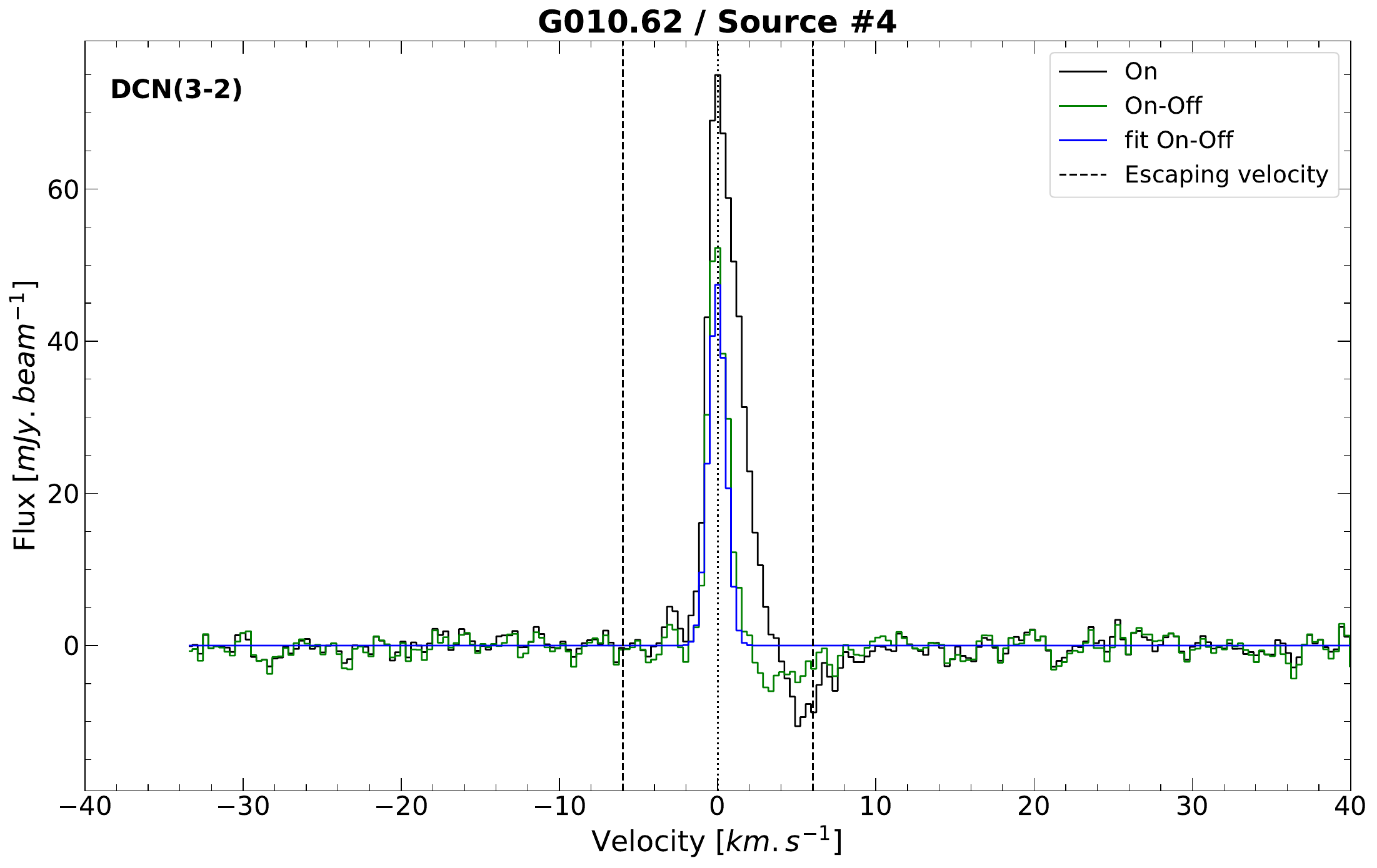}
        \end{minipage}
    \end{minipage}

   % \vskip -0.3cm
    \caption{CO and SiO spectra and molecular outflow maps of the high-mass PSC candidate of the G010.62 region. CO contours are 10, 20, 40, and 80 in units of $\sigma$, with $\sigma$ = $27.6$, $8.6$, $20.6$, $15.4$ \mJybeamkms for cyan, blue, orange and red contours respectively. SiO contours are 10, 20, 40, and 80 in units of $\sigma$, with $\sigma$ = $5.6$, $7.5$, $5.4$, $7.2$ \mJybeamkms for cyan, blue, orange and red contours respectively. DCN spectra (bottom right) of the high-mass PSC candidate of the G010.62 region.}

\end{figure*}

%%%%%%%%%%%%%%%%%%%%%%% G012 %%%%%%%%%%%%%%%%%%%%%%%%%%%%%%%%%%%%%
\begin{figure*}
    \label{appendix:G012_MPSC_fig}
    \centering
    \begin{minipage}[c]{0.49\textwidth}
        \centering
        \includegraphics[width=\textwidth]{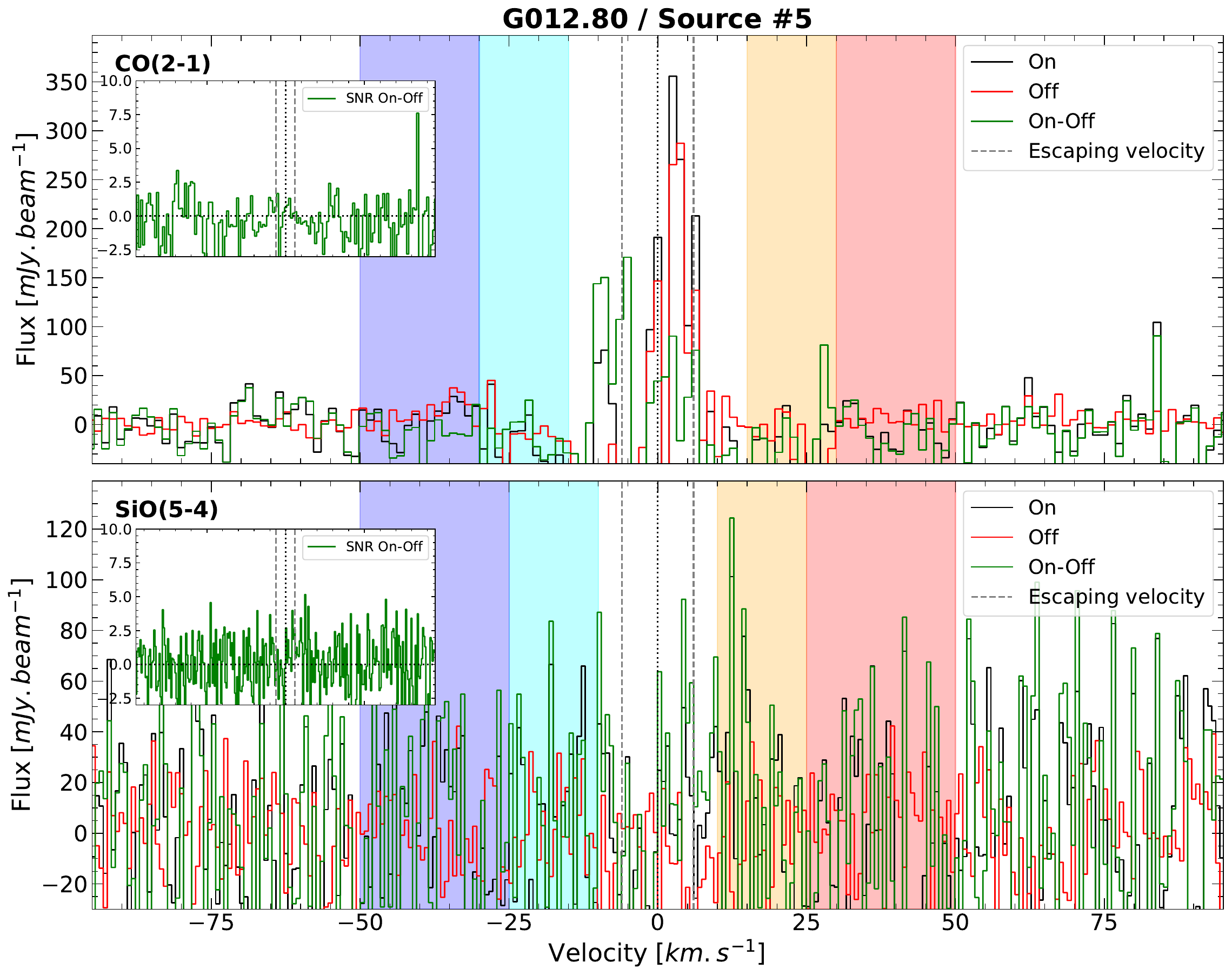}
    \end{minipage}
    \begin{minipage}[c]{0.49\textwidth}
        \centering
        \begin{minipage}[c]{\textwidth}
            \centering
            \includegraphics[width=0.9\textwidth]{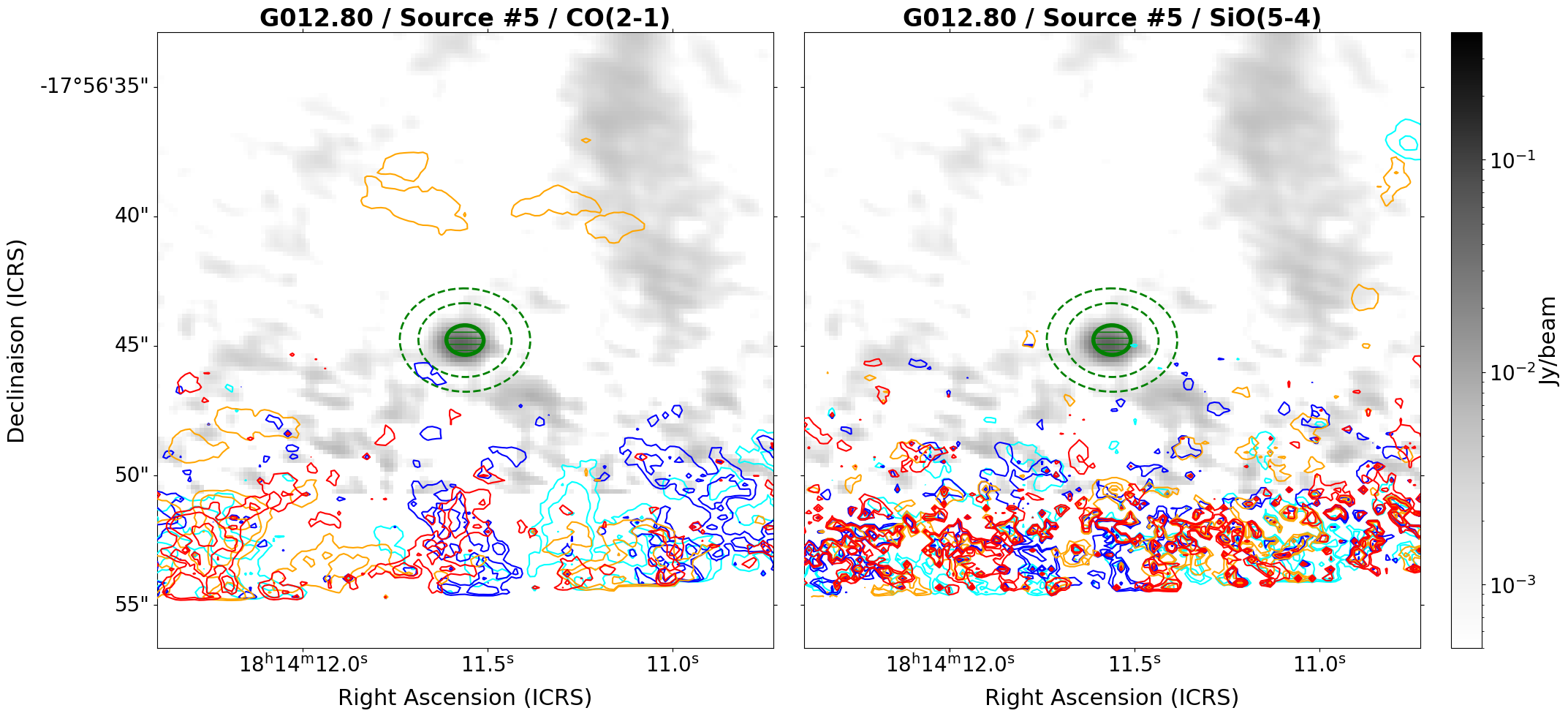}
        \end{minipage}
        \vfill
        \begin{minipage}[c]{\textwidth}
            \centering
            \includegraphics[width=0.7\textwidth]{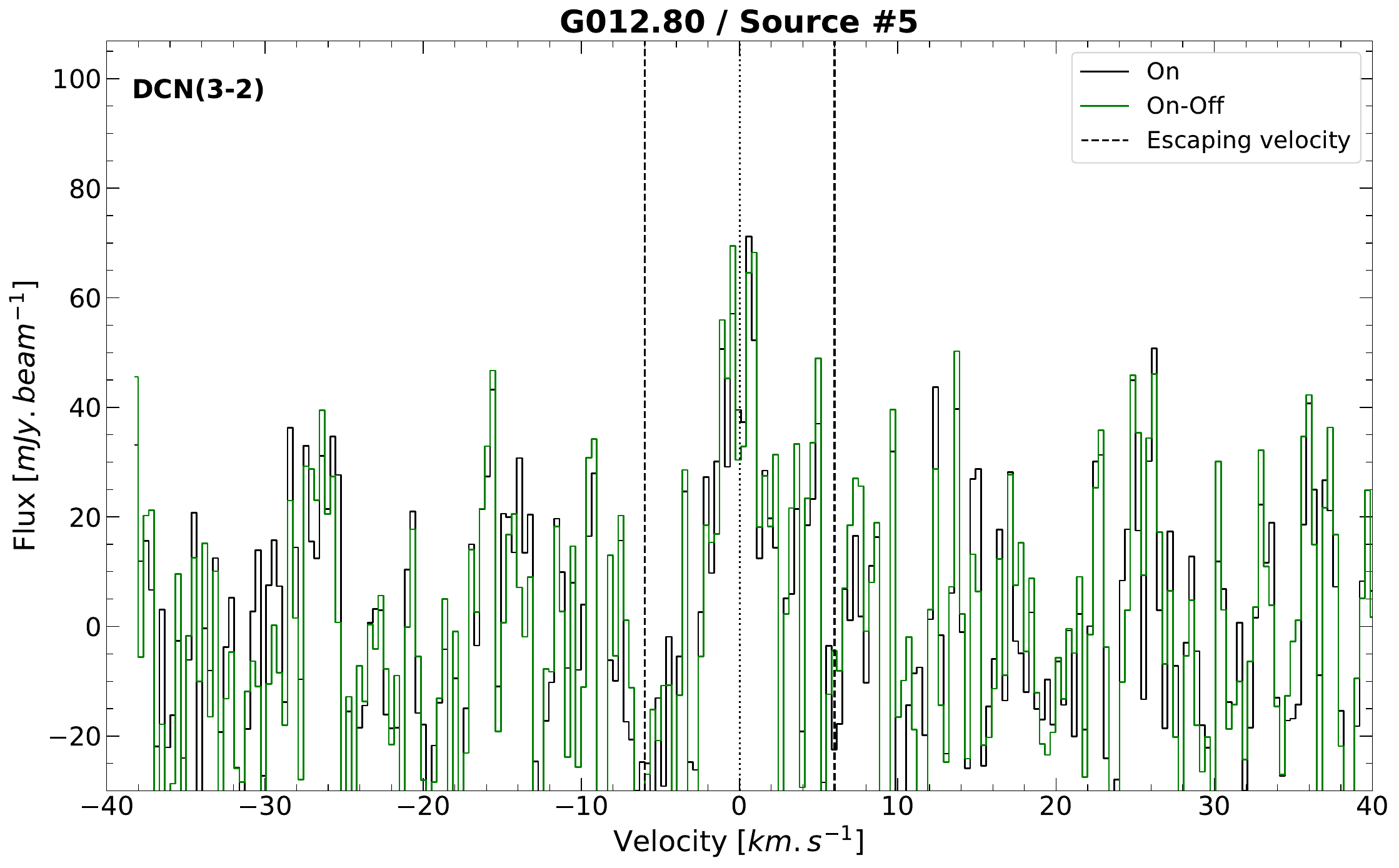}
        \end{minipage}
    \end{minipage}

   % \vskip -0.3cm
    \caption{CO and SiO spectra and molecular outflow maps of the high-mass PSC candidate of the G012.80 region. CO contours are 10, 20, 40, and 80 in units of $\sigma$, with $\sigma$ = $112.3$, $64.1$, $126.2$, $72.8$ \mJybeamkms for cyan, blue, orange and red contours respectively. SiO contours are 10, 20, 40, and 80 in units of $\sigma$, with $\sigma$ = $26.5$, $34.0$, $26.2$, $33.1$ \mJybeamkms for cyan, blue, orange and red contours respectively. DCN spectra (bottom right) of the high-mass PSC candidate of the G012.80 region.}

\end{figure*}

%%%%%%%%%%%%%%%%%%%%%%% G327 %%%%%%%%%%%%%%%%%%%%%%%%%%%%%%%%%%%%%
\begin{figure*}
    \label{appendix:G327_MPSC_fig}
    \centering
    \begin{minipage}[c]{0.49\textwidth}
        \centering
        \includegraphics[width=\textwidth]{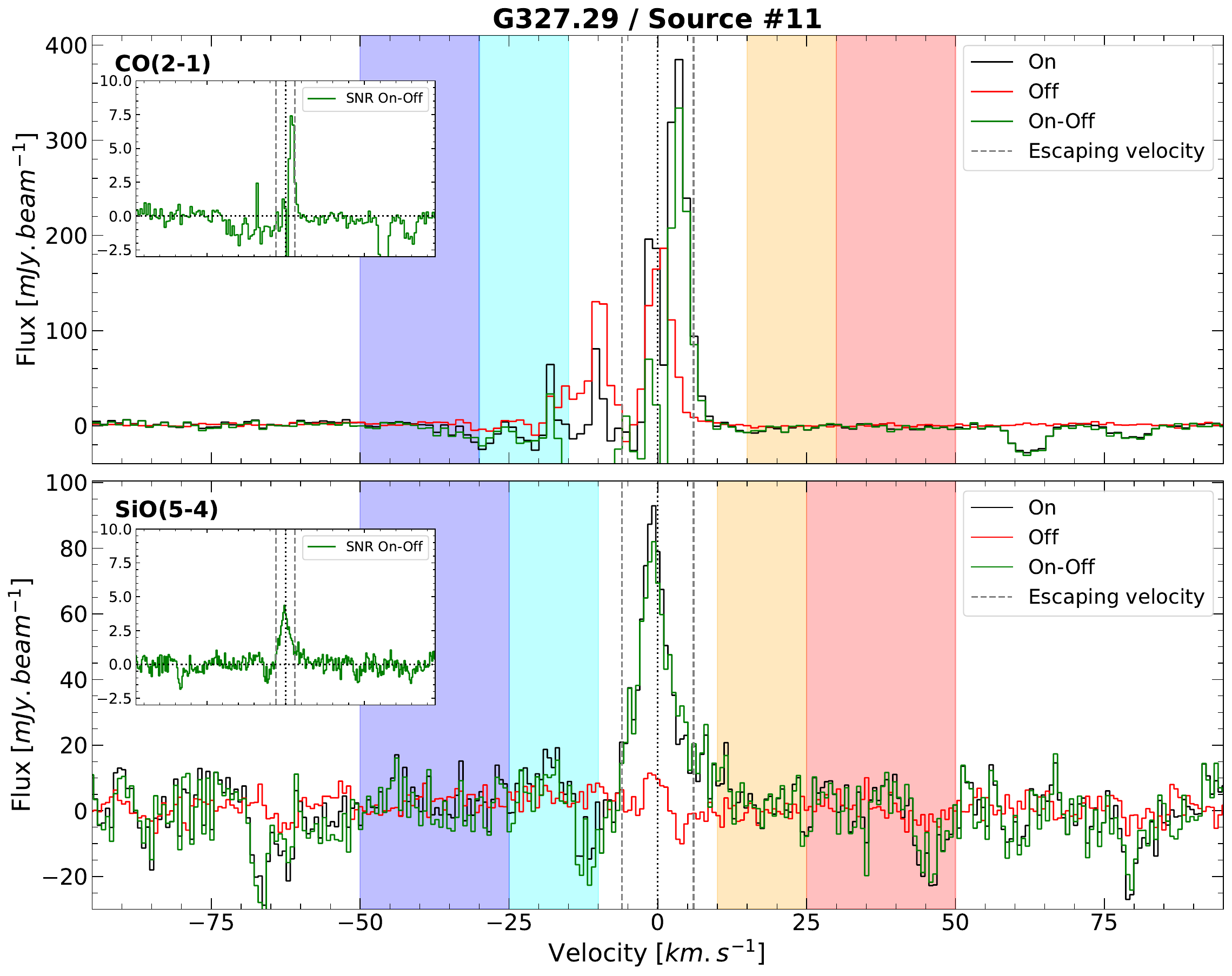}
    \end{minipage}
    \begin{minipage}[c]{0.49\textwidth}
        \centering
        \begin{minipage}[c]{\textwidth}
            \centering
            \includegraphics[width=0.9\textwidth]{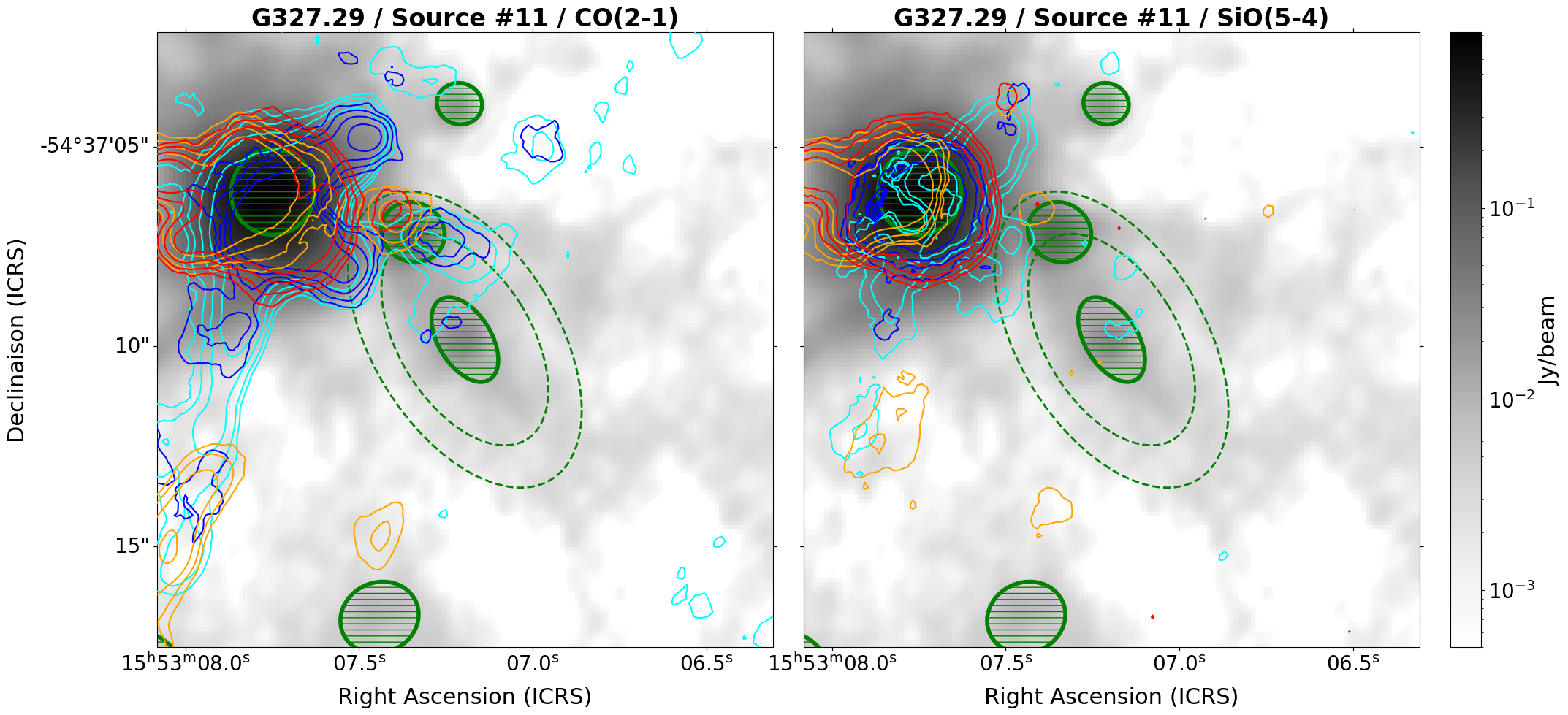}
        \end{minipage}
        \vfill
        \begin{minipage}[c]{\textwidth}
            \centering
            \includegraphics[width=0.7\textwidth]{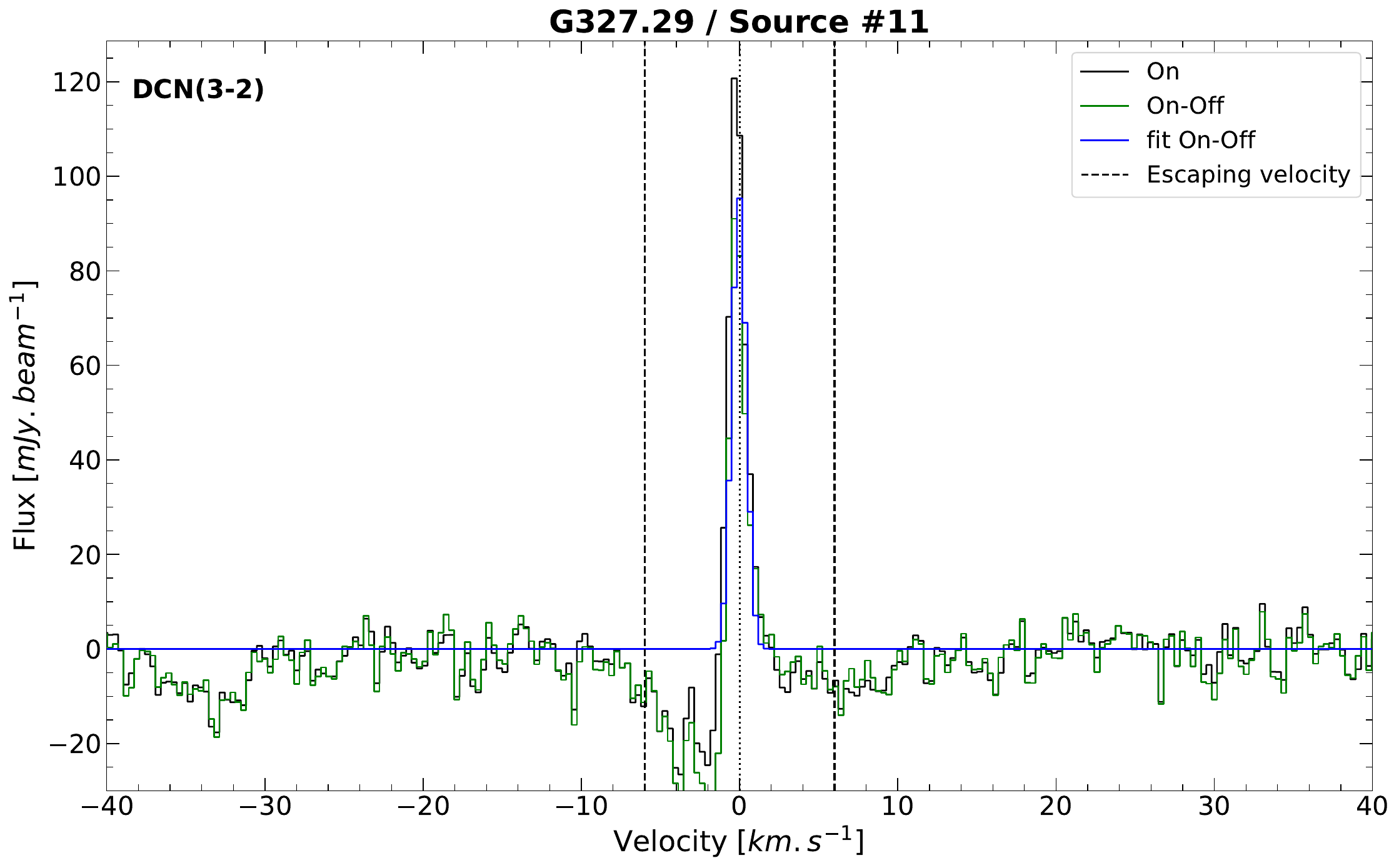}
        \end{minipage}
    \end{minipage}

   % \vskip -0.3cm
    \caption{CO and SiO spectra (left) and molecular outflow maps (top right) of the high-mass PSC candidates of the G327.29 region. CO contours are 5, 10, 20, and 40 in units of $\sigma$, with $\sigma$ = $48.2$, $39.3$, $56.1$, $37.8$ \mJybeamkms for cyan, blue, orange and red contours respectively. SiO contours are 5, 10, 20, and 40 in units of $\sigma$, with $\sigma$ = $35.5$, $44.7$, $35.7$, $45.5$ \mJybeamkms for cyan, blue, orange and red contours respectively. DCN spectra and fit (bottom right) of the high-mass PSC candidate of the G327.29 region.}

\end{figure*}

%%%%%%%%%%%%%%%%%%%%%%% G333 %%%%%%%%%%%%%%%%%%%%%%%%%%%%%%%%%%%%%
\begin{figure*}
    \label{appendix:G333_MPSC_fig}
    \centering
    \begin{minipage}[c]{0.49\textwidth}
        \centering
        \includegraphics[width=\textwidth]{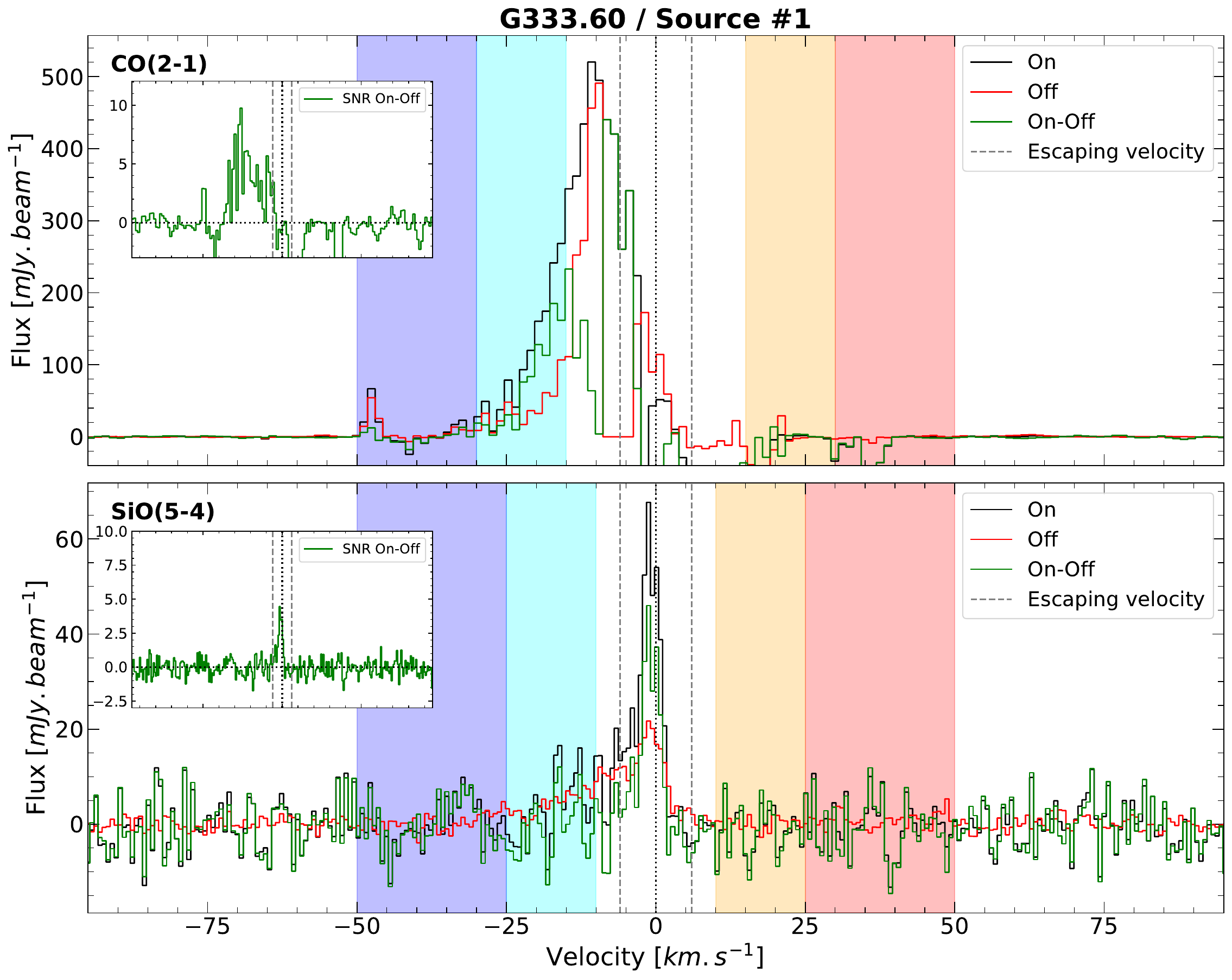}
    \end{minipage}
    \begin{minipage}[c]{0.49\textwidth}
        \centering
        \begin{minipage}[c]{\textwidth}
            \centering
            \includegraphics[width=0.9\textwidth]{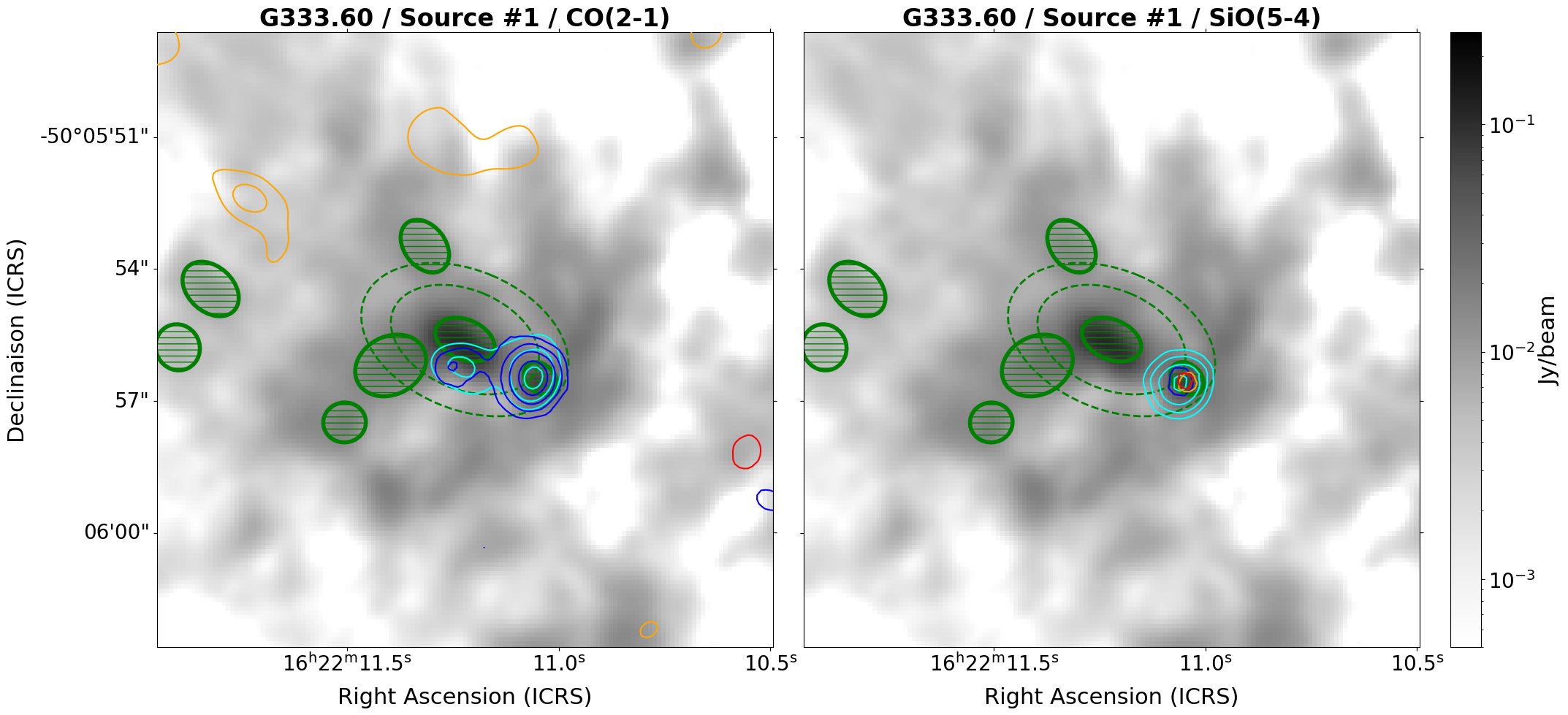}
        \end{minipage}
        \vfill
        \begin{minipage}[c]{\textwidth}
            \centering
            \includegraphics[width=0.7\textwidth]{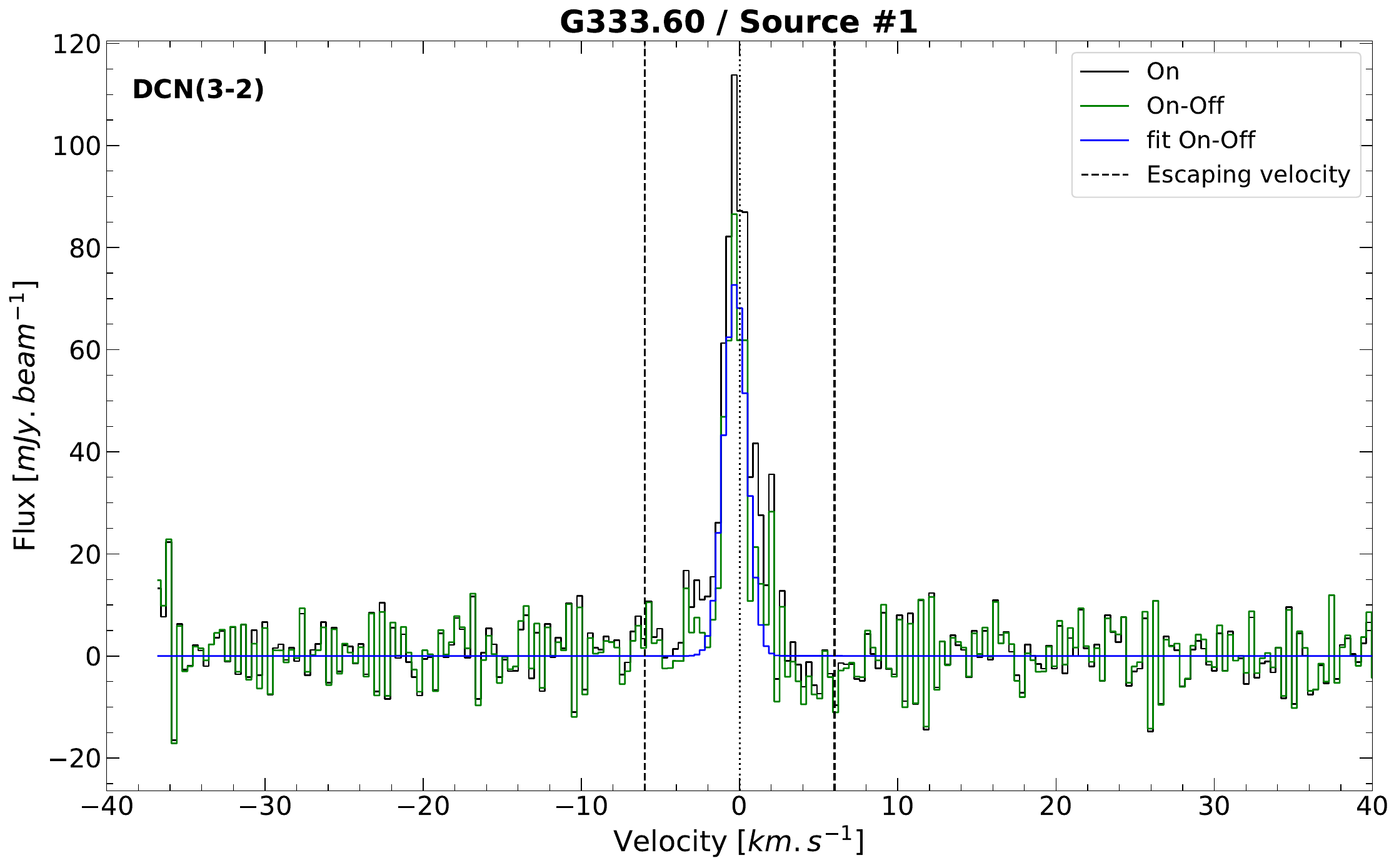}
        \end{minipage}
    \end{minipage}

\vspace{0.2cm}

    \centering
    \begin{minipage}[c]{0.49\textwidth}
        \centering
        \includegraphics[width=\textwidth]{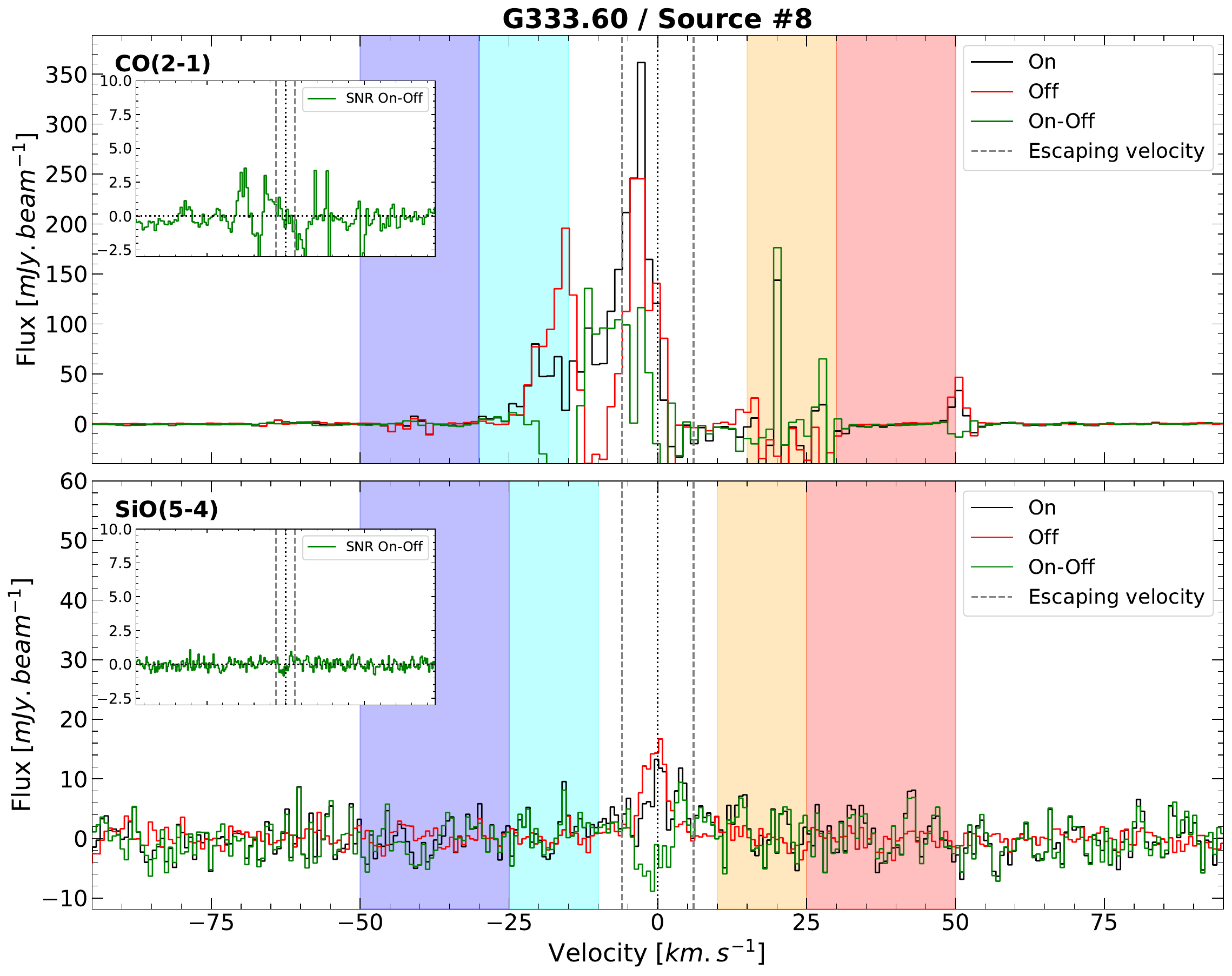}
    \end{minipage}
    \begin{minipage}[c]{0.49\textwidth}
        \centering
        \begin{minipage}[c]{\textwidth}
            \centering
            \includegraphics[width=0.9\textwidth]{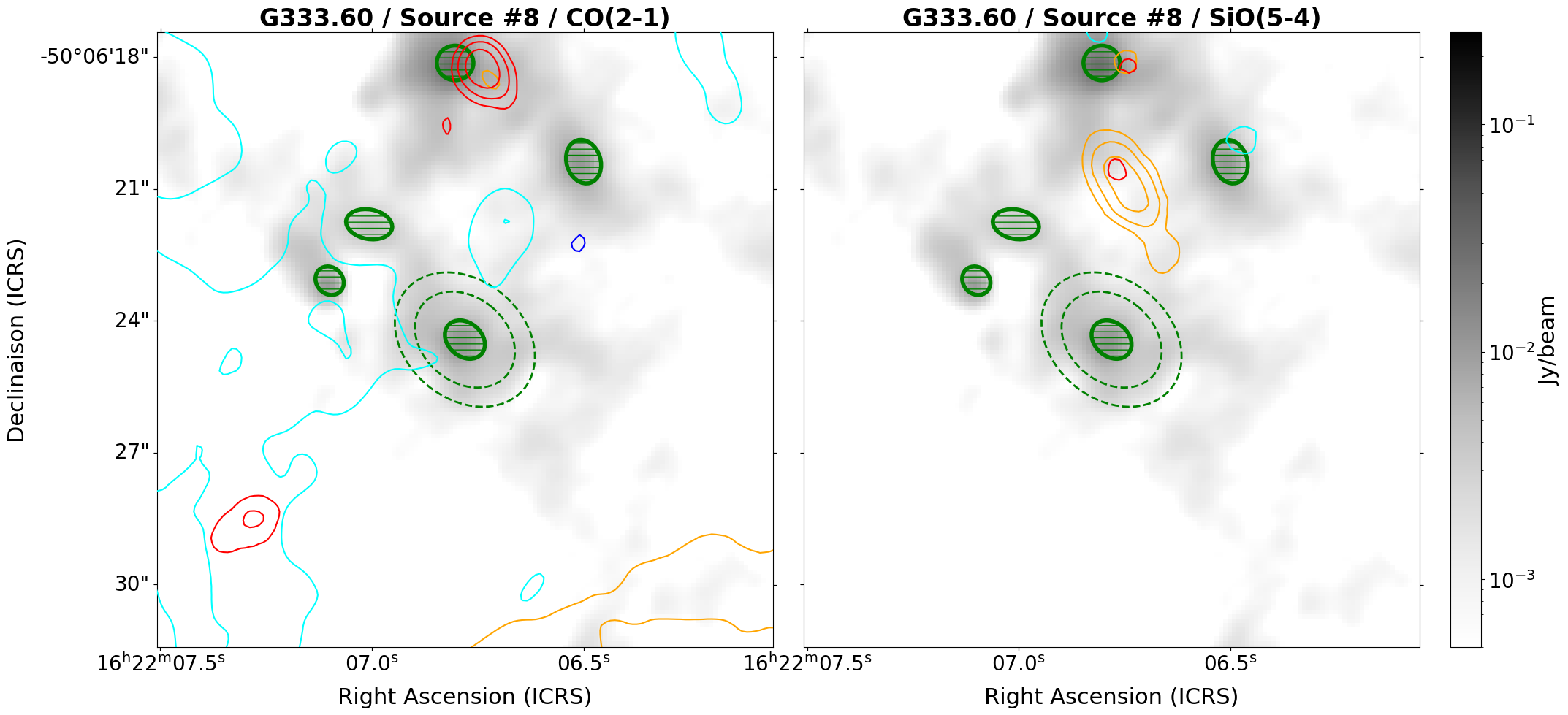}
        \end{minipage}
        \vfill
        \begin{minipage}[c]{\textwidth}
            \centering
            \includegraphics[width=0.7\textwidth]{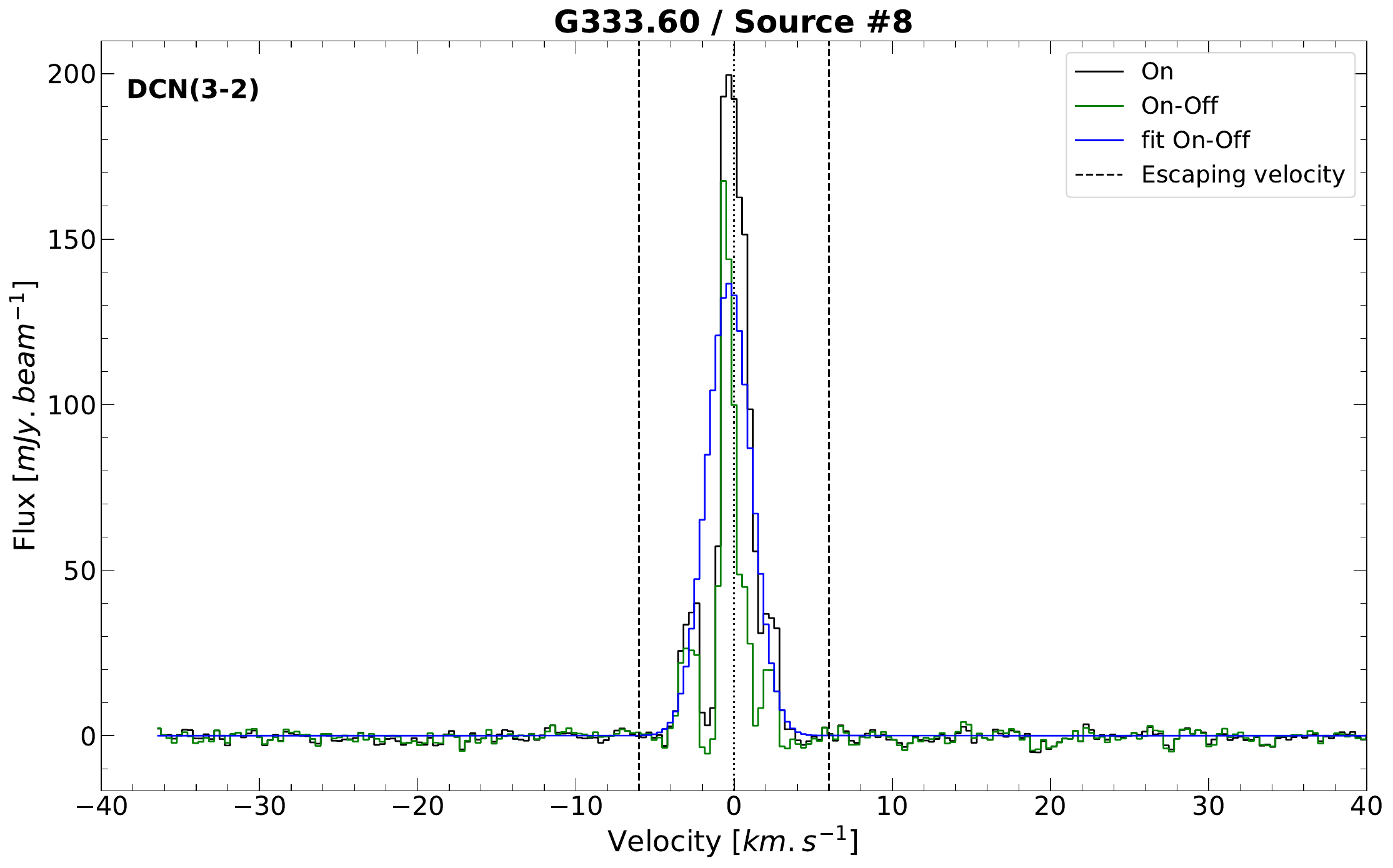}
        \end{minipage}
    \end{minipage}

\vspace{0.2cm}

    \centering
    \begin{minipage}[c]{0.49\textwidth}
        \centering
        \includegraphics[width=\textwidth]{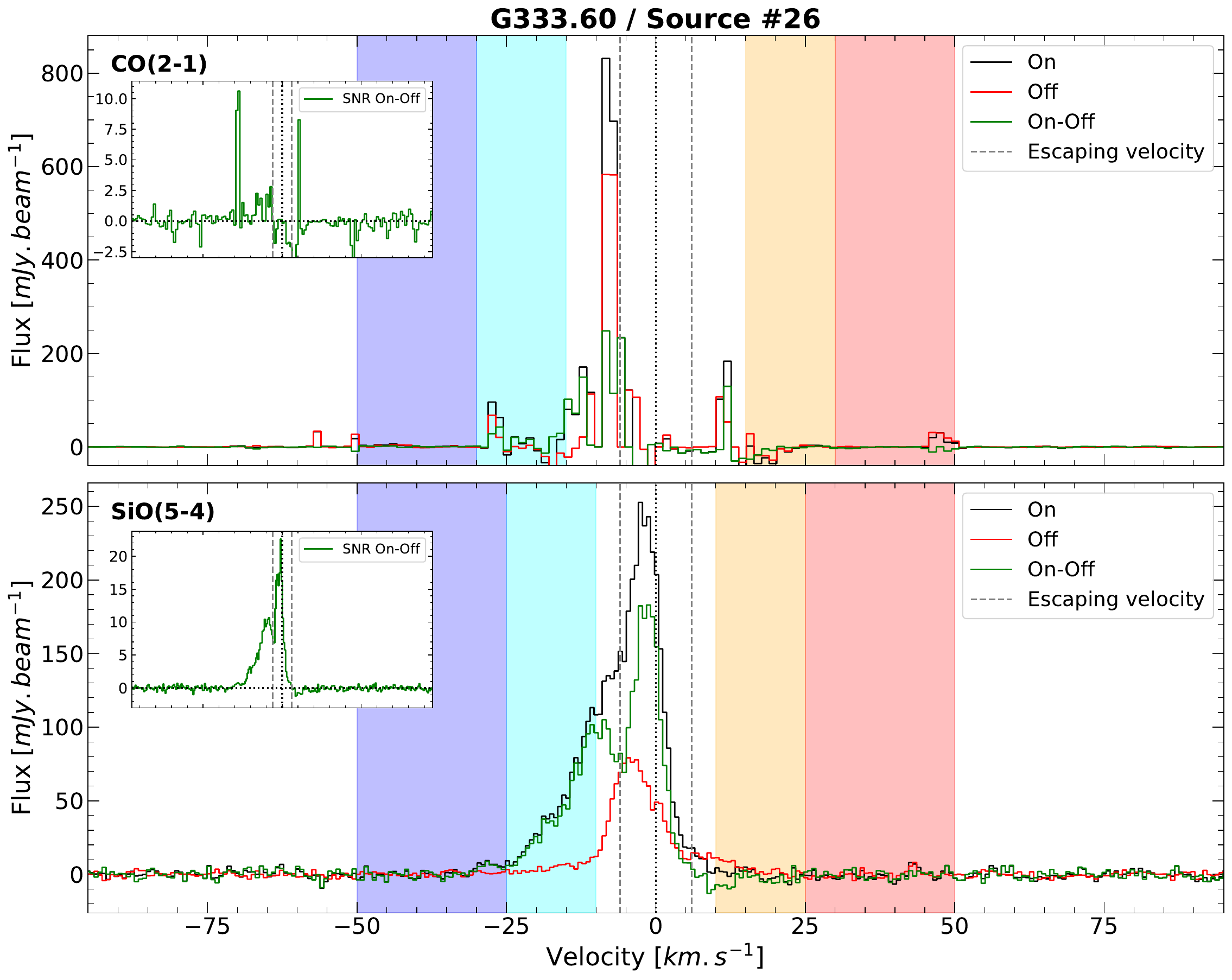}
    \end{minipage}
    \begin{minipage}[c]{0.49\textwidth}
        \centering
        \begin{minipage}[c]{\textwidth}
            \centering
            \includegraphics[width=0.9\textwidth]{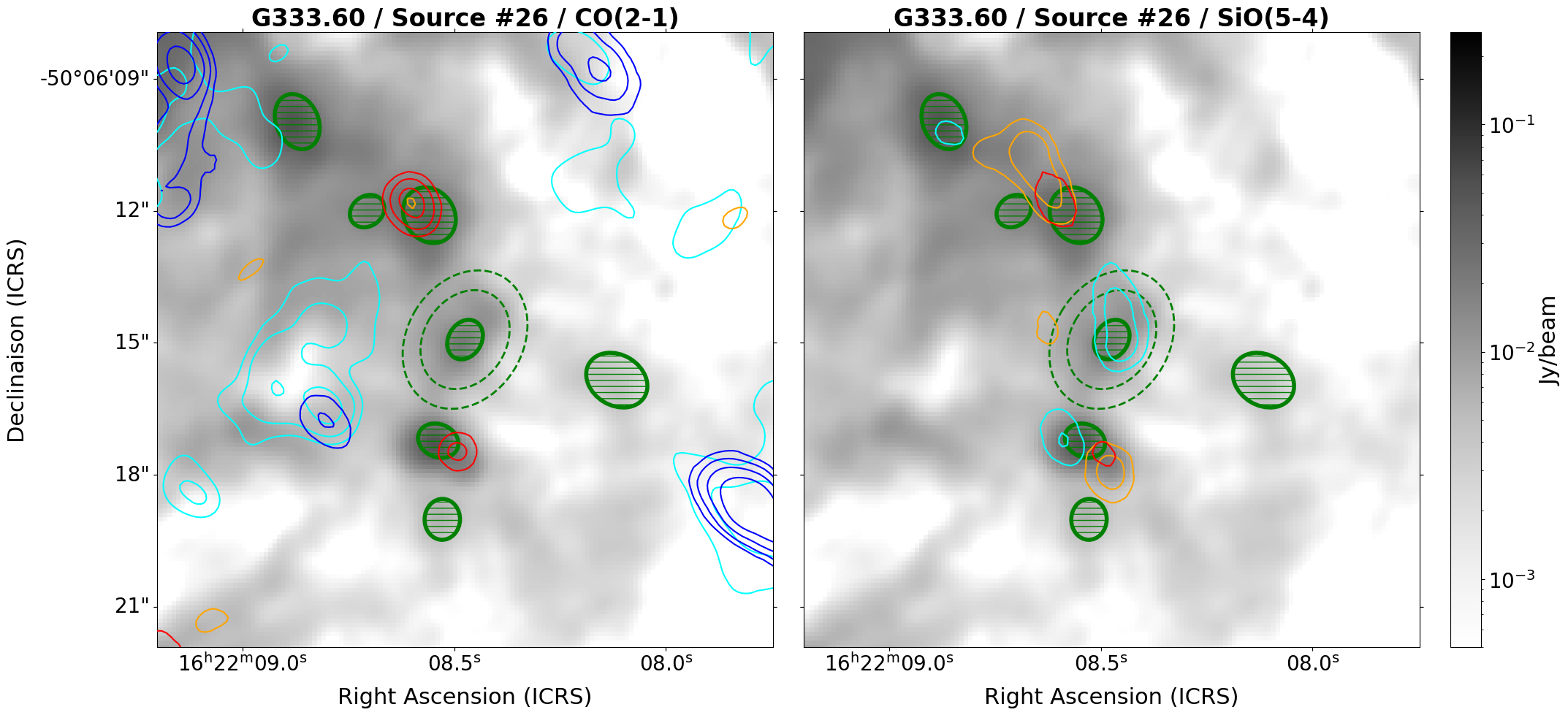}
        \end{minipage}
        \vfill
        \begin{minipage}[c]{\textwidth}
            \centering
            \includegraphics[width=0.7\textwidth]{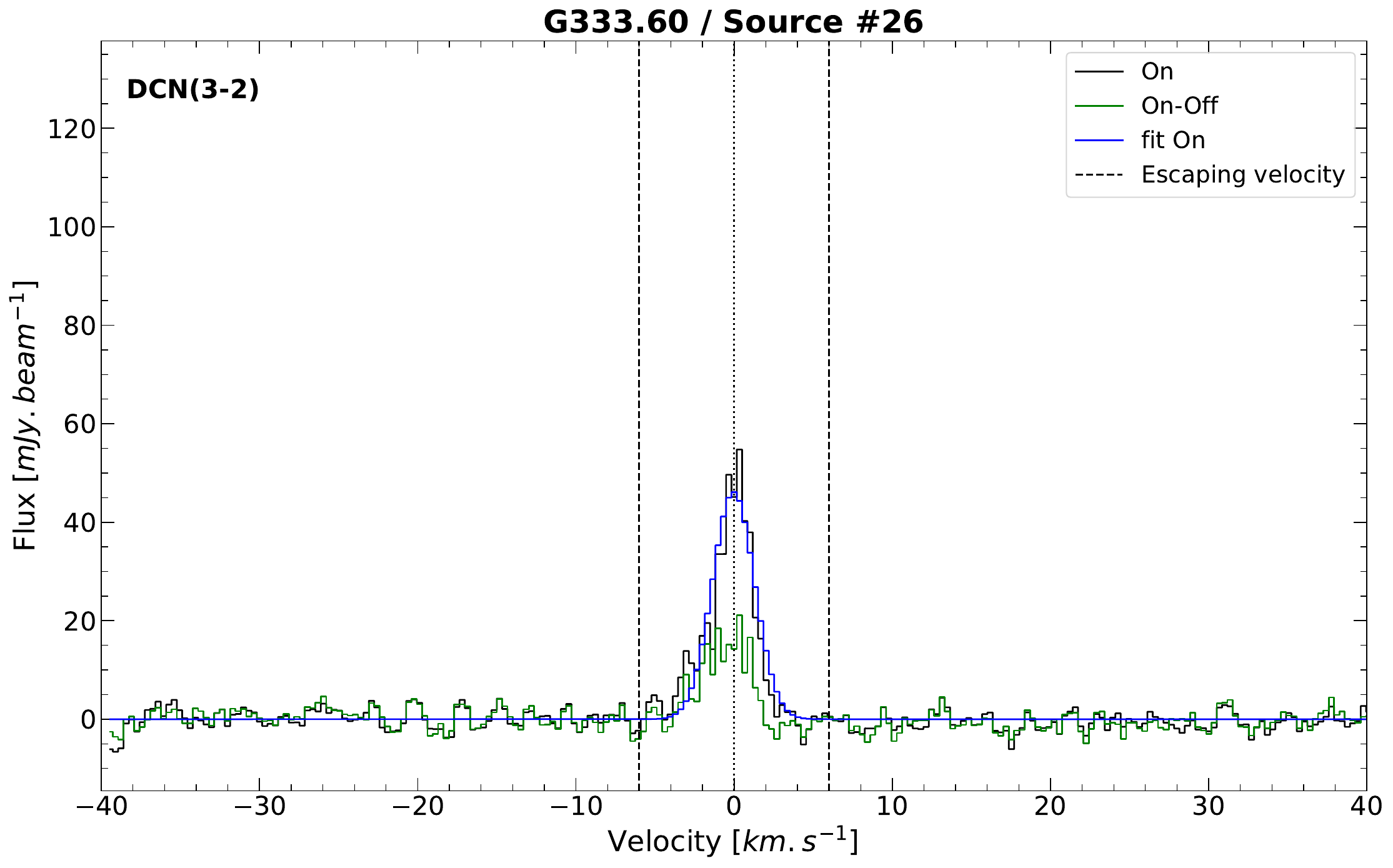}
        \end{minipage}
    \end{minipage}

   % \vskip -0.3cm
    \caption{CO and SiO spectra (left) and molecular outflow maps (top right) of the high-mass PSC candidates of the G333.60 region. CO contours are  10, 20, 40 and 80 in units of $\sigma$, with $\sigma$ = $110.8$, $17.0$, $98.6$, $20.3$ \mJybeamkms for cyan, blue, orange and red contours respectively. SiO contours are 10, 20, 40 and 80 in units of $\sigma$, with $\sigma$ = $15.8$, $20.1$, $16.1$, $19.8$ \mJybeamkms for cyan, blue, orange and red contours respectively. DCN spectra and fits (bottom right) of the high-mass PSC candidates of the G333.60 region.}

\end{figure*}

\begin{figure*}\ContinuedFloat
    
        \centering
        \begin{minipage}[c]{0.49\textwidth}
            \centering
            \includegraphics[width=\textwidth]{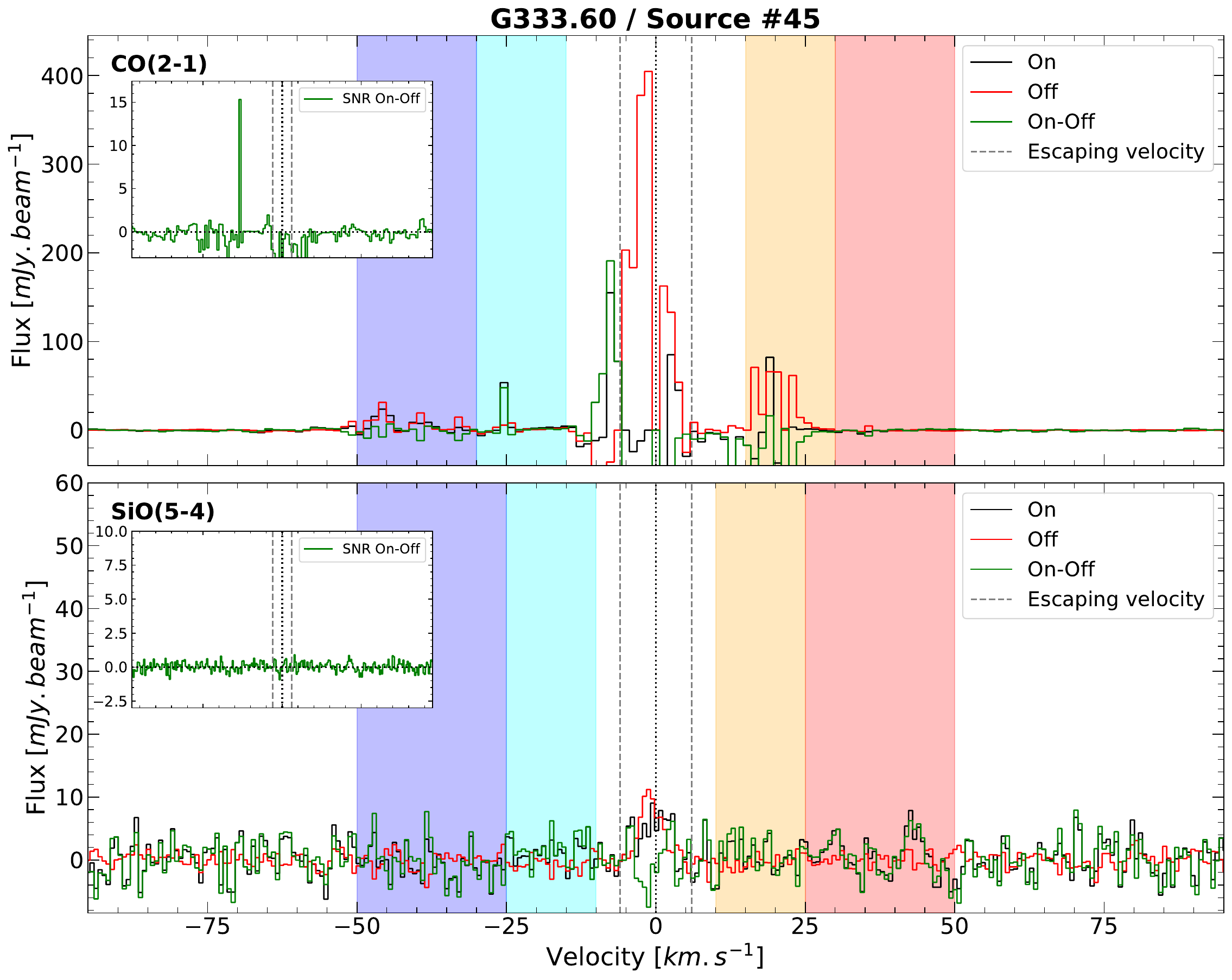}
        \end{minipage}
        \begin{minipage}[c]{0.49\textwidth}
            \centering
            \begin{minipage}[c]{\textwidth}
                \centering
                \includegraphics[width=0.9\textwidth]{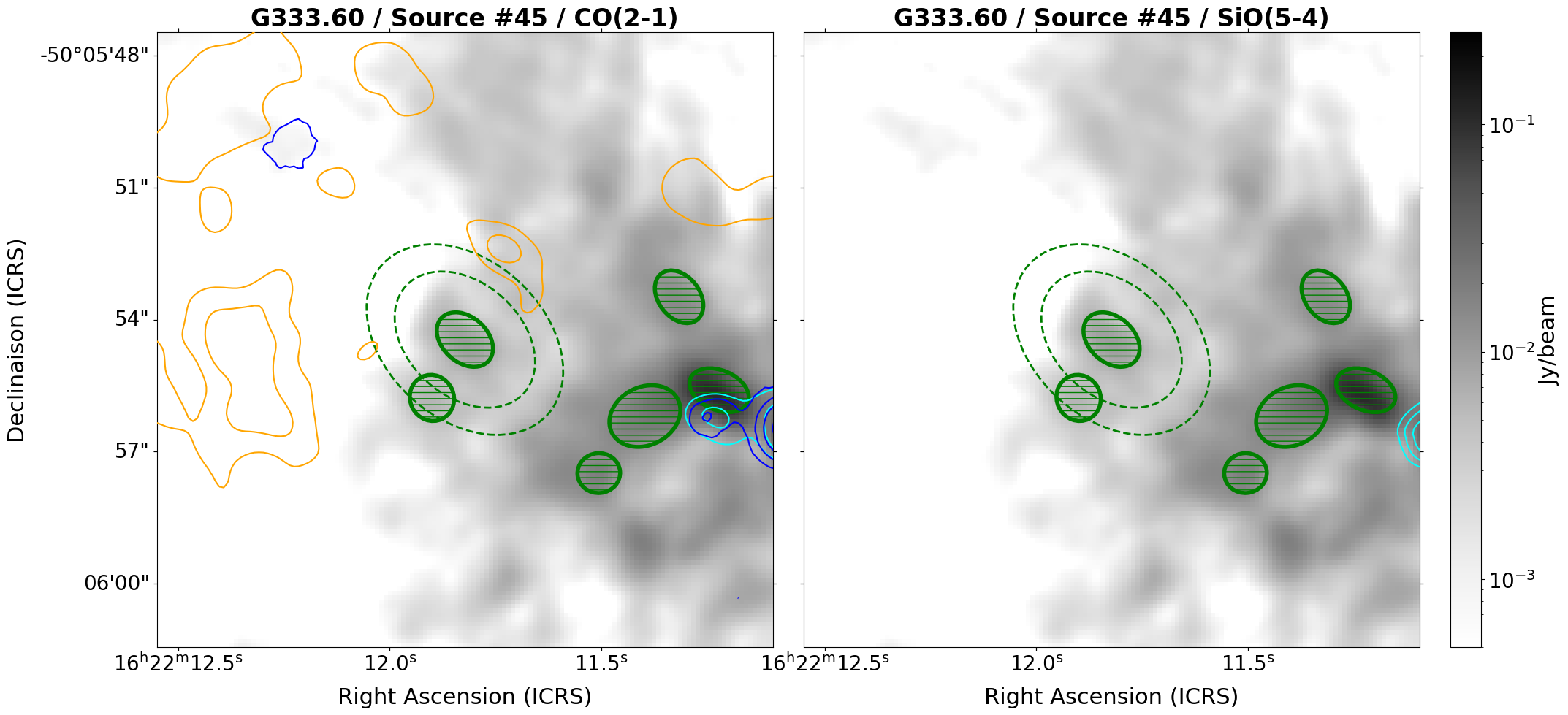}
            \end{minipage}
            \vfill
            \begin{minipage}[c]{\textwidth}
                \centering
                \includegraphics[width=0.7\textwidth]{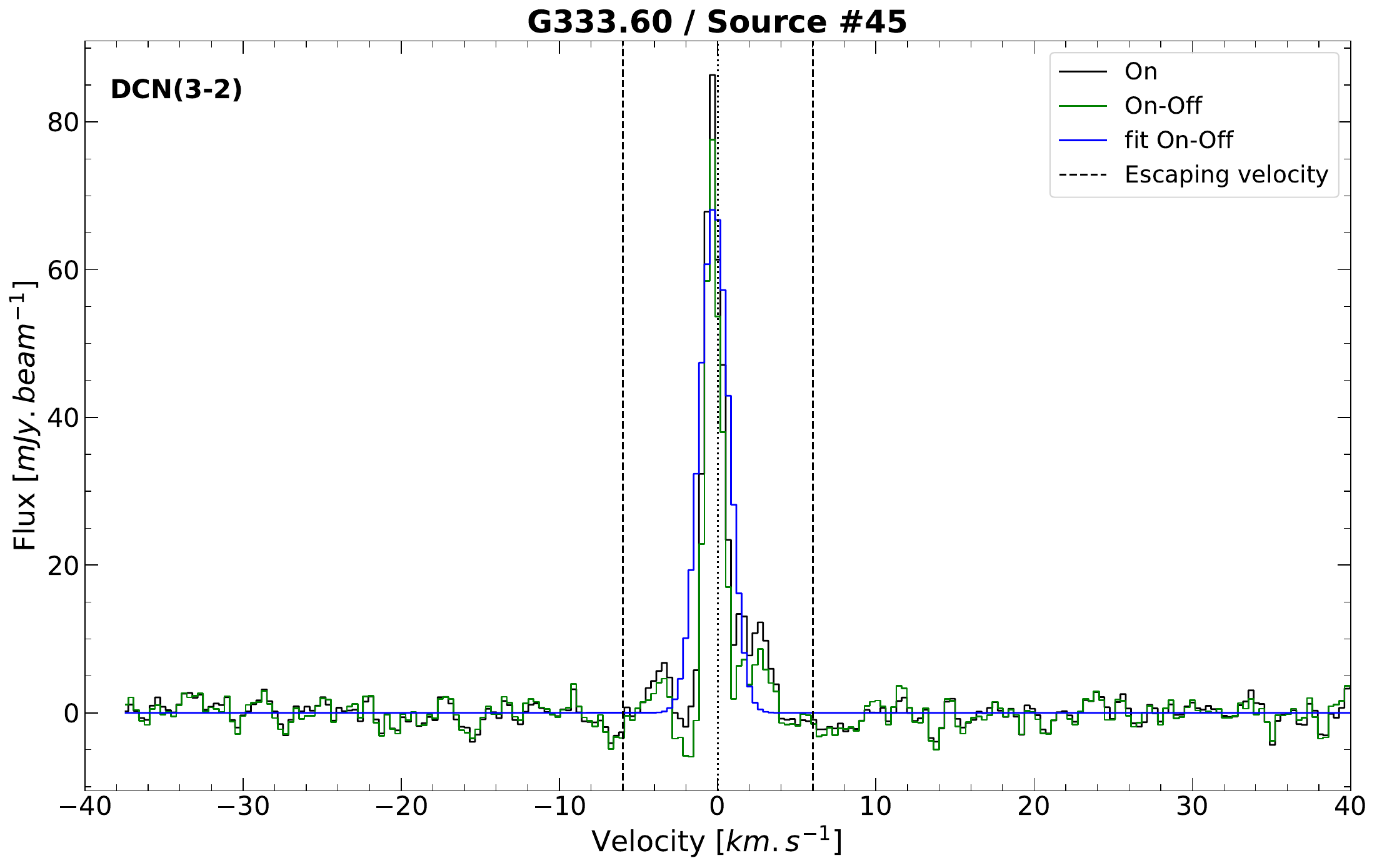}
            \end{minipage}
        \end{minipage}

   % \vskip -0.3cm
    \caption{continued.}

\end{figure*}

%%%%%%%%%%%%%%%%%%%%%%% G337 %%%%%%%%%%%%%%%%%%%%%%%%%%%%%%%%%%%%%
\begin{figure*}
    \label{appendix:G337_MPSC_fig}
    \centering
        \begin{minipage}[c]{0.49\textwidth}
            \centering
            \includegraphics[width=\textwidth]{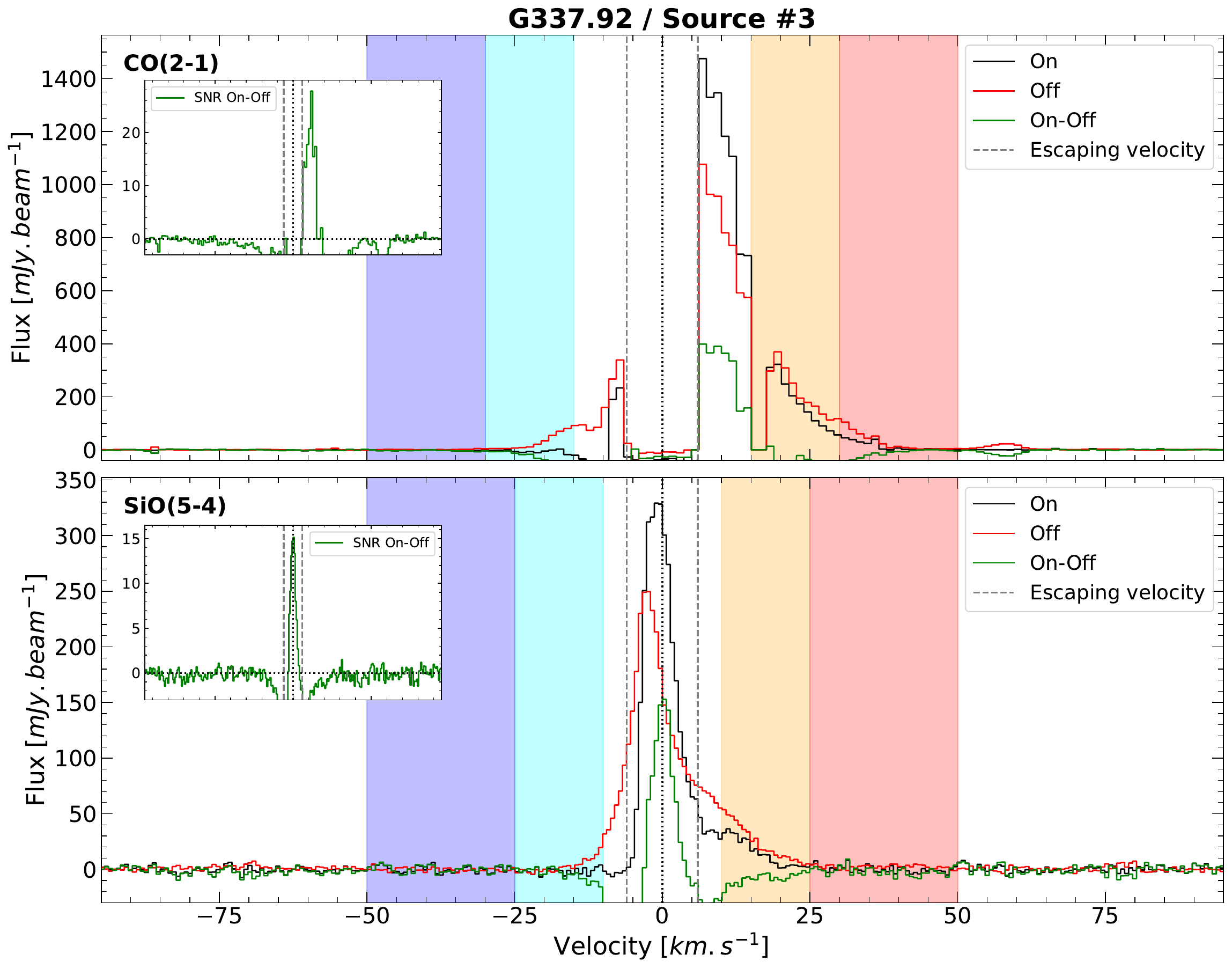}
        \end{minipage}
        \begin{minipage}[c]{0.49\textwidth}
            \centering
            \begin{minipage}[c]{\textwidth}
                \centering
                \includegraphics[width=0.9\textwidth]{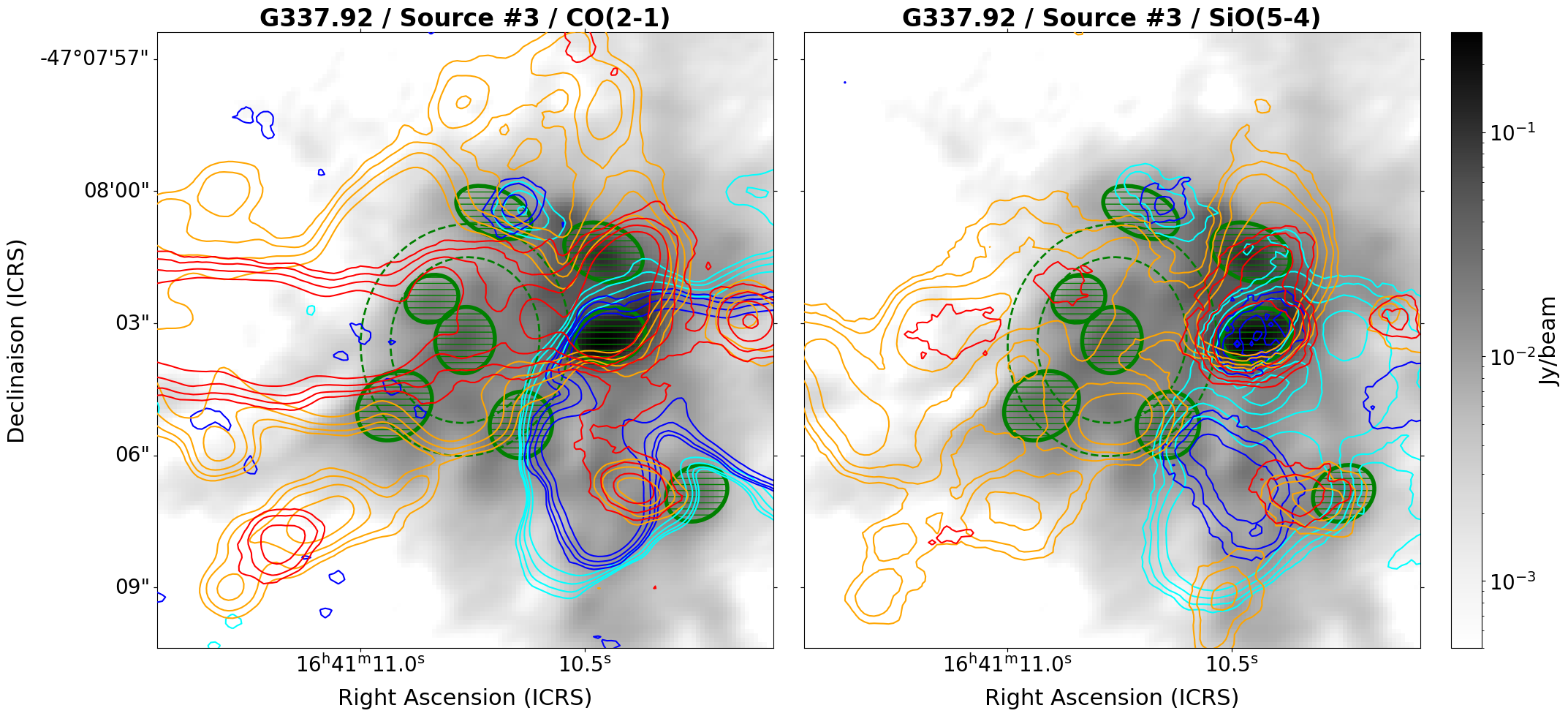}
            \end{minipage}
            \vfill
            \begin{minipage}[c]{\textwidth}
                \centering
                \includegraphics[width=0.7\textwidth]{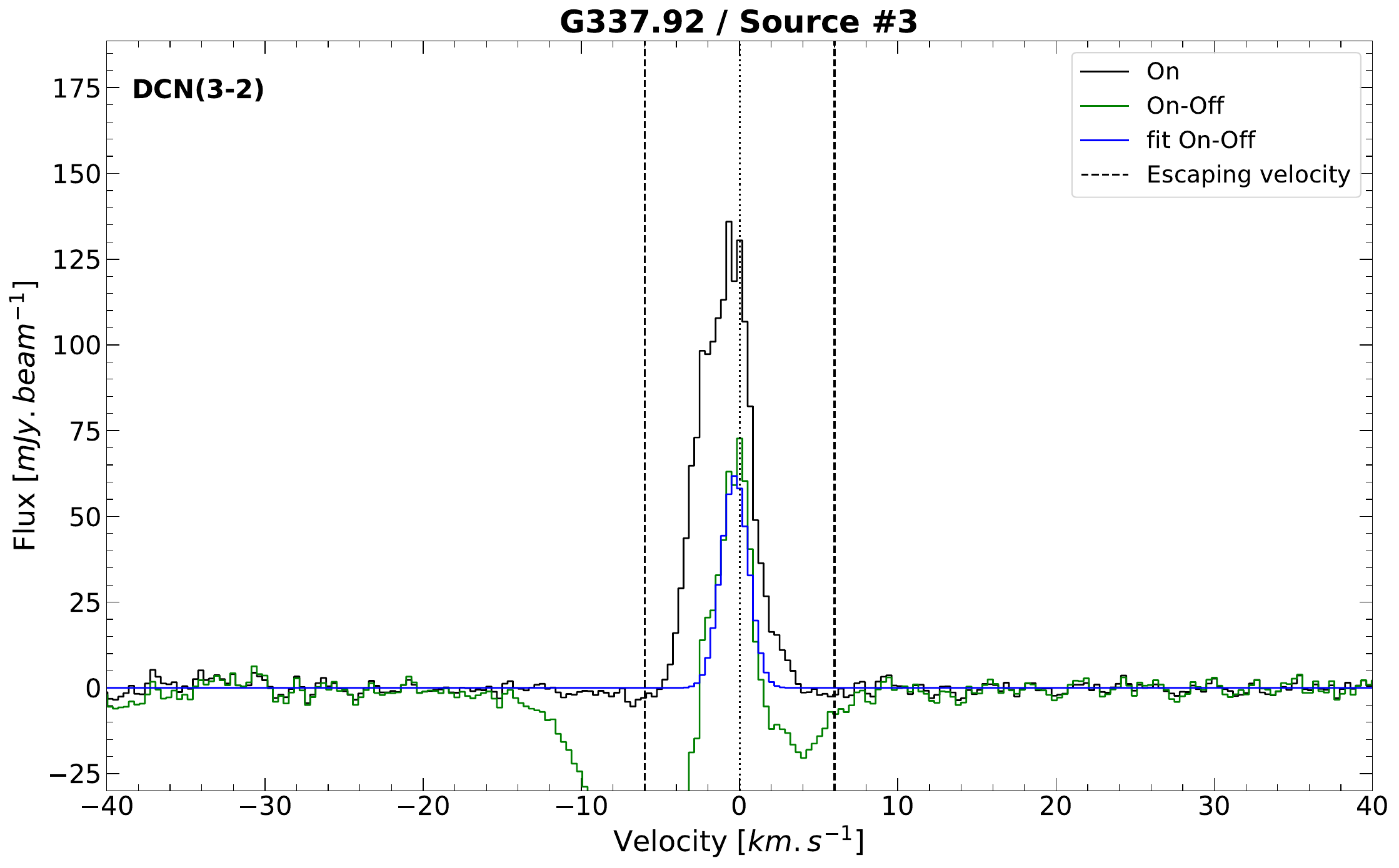}
            \end{minipage}
        \end{minipage}
    
    \vspace{0.2cm}
    
        \centering
        \begin{minipage}[c]{0.49\textwidth}
            \centering
            \includegraphics[width=\textwidth]{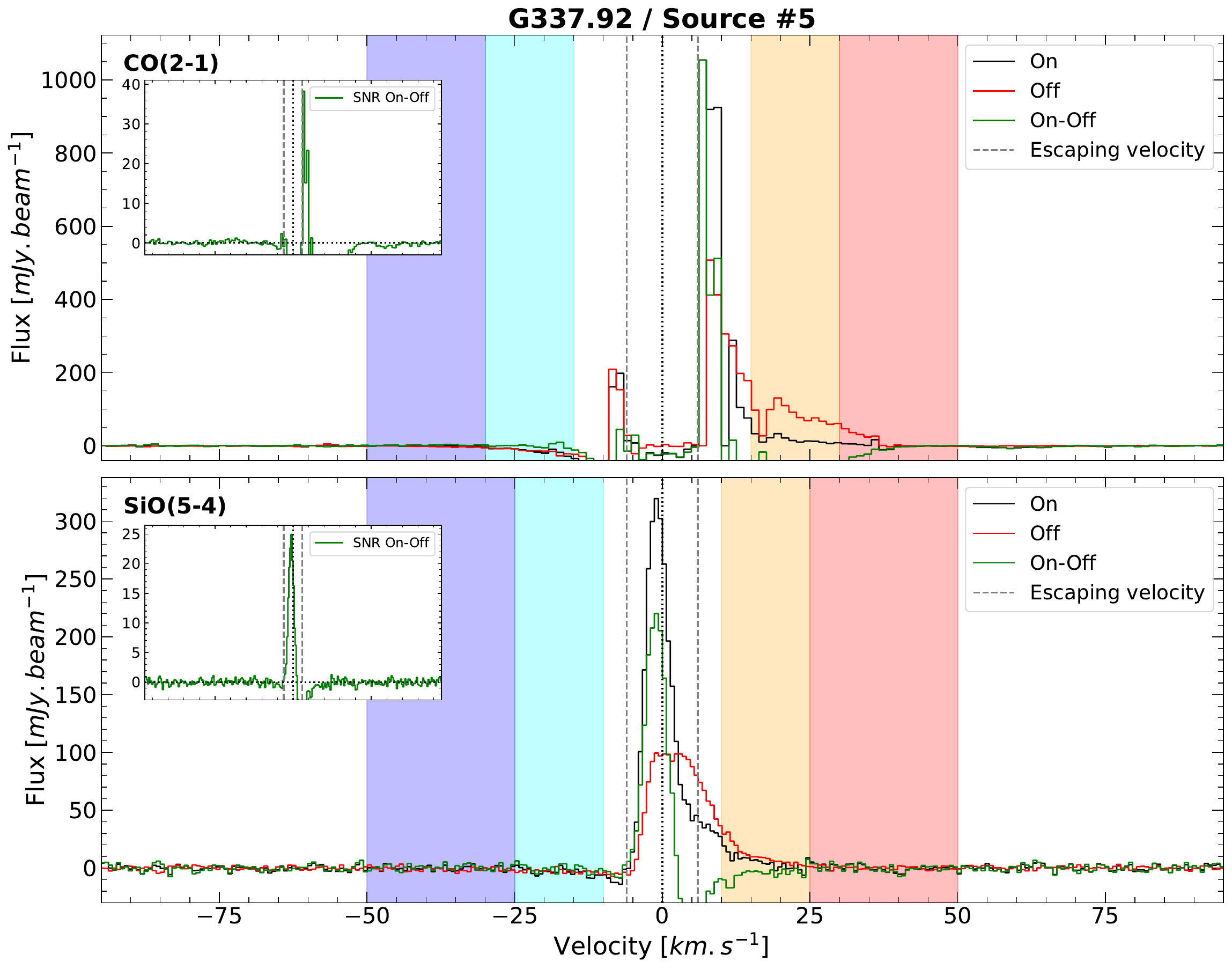}
        \end{minipage}
        \begin{minipage}[c]{0.49\textwidth}
            \centering
            \begin{minipage}[c]{\textwidth}
                \centering
                \includegraphics[width=0.9\textwidth]{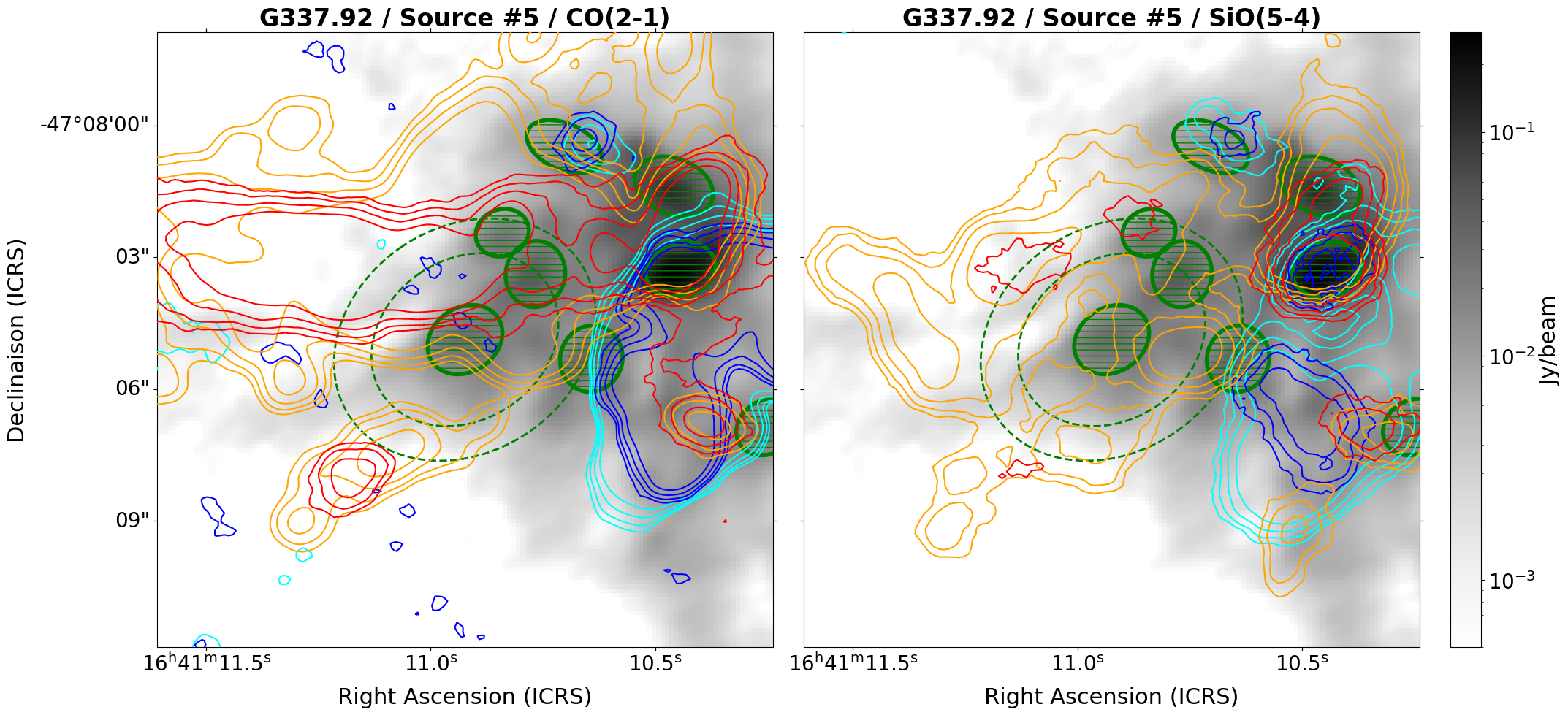}
            \end{minipage}
            \vfill
            \begin{minipage}[c]{\textwidth}
                \centering
                \includegraphics[width=0.7\textwidth]{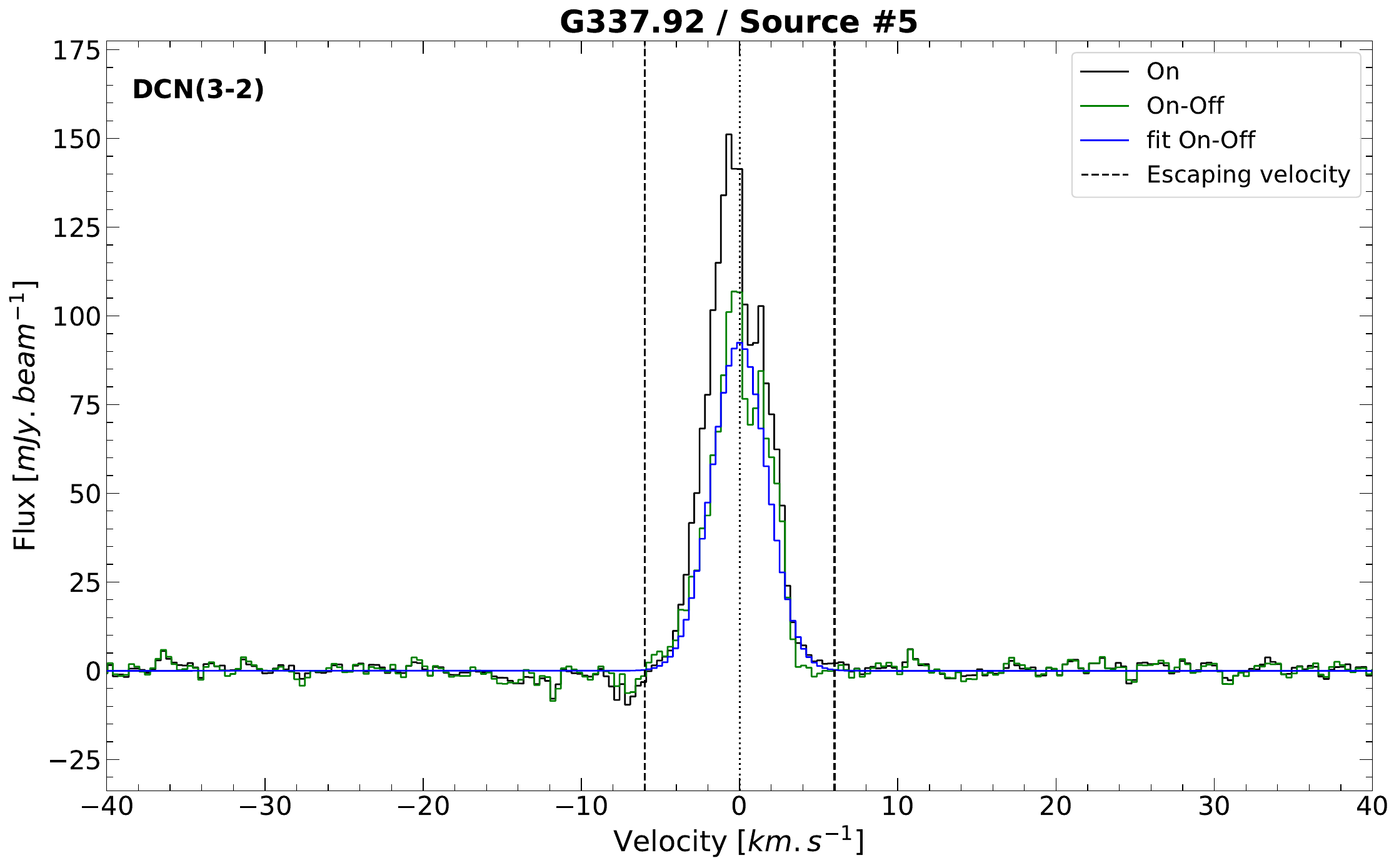}
            \end{minipage}
        \end{minipage}

   % \vskip -0.3cm
    \caption{CO and SiO spectra (left) and molecular outflow maps (top right) of the high-mass PSC candidates of the G337.92 region. CO contours are 5, 10, 20, and 40 in units of $\sigma$, with $\sigma$ = $38.8$, $16.6$, $40.3$, $24.4$ \mJybeamkms for cyan, blue, orange and red contours respectively. SiO contours are 5, 10, 20, and 40 in units of $\sigma$, with $\sigma$ = $9.4$, $12.5$, $10.2$, $12.2$ \mJybeamkms for cyan, blue, orange and red contours respectively.  DCN spectra and fits (bottom right) of the high-mass PSC candidates of the G337.92 region.}

\end{figure*}

%%%%%%%%%%%%%%%%%%%%%%% G338 %%%%%%%%%%%%%%%%%%%%%%%%%%%%%%%%%%%%%
\begin{figure*}
    \label{appendix:G338_MPSC_fig}
    \centering
        \begin{minipage}[c]{0.49\textwidth}
            \centering
            \includegraphics[width=\textwidth]{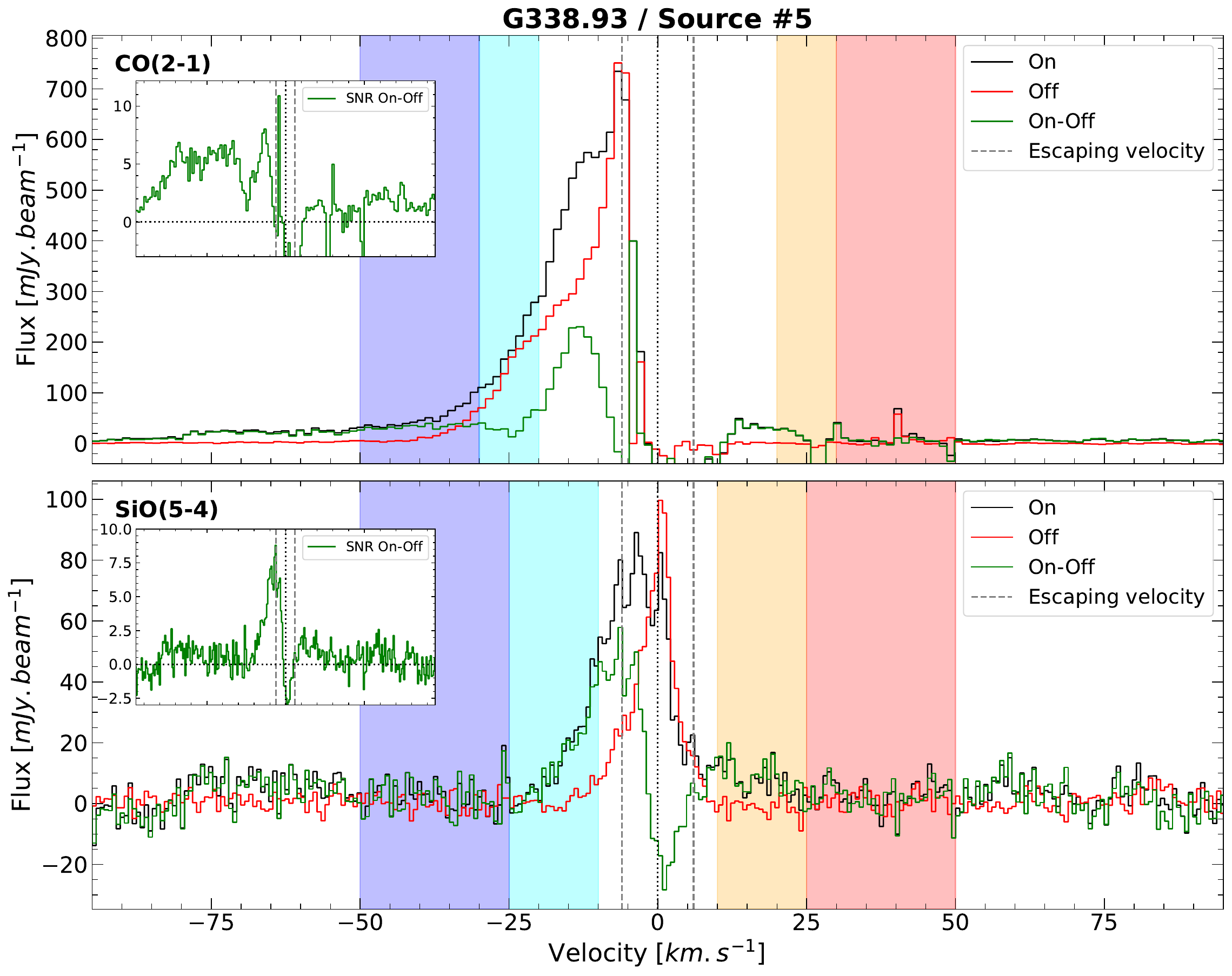}
        \end{minipage}
        \begin{minipage}[c]{0.49\textwidth}
            \centering
            \begin{minipage}[c]{\textwidth}
                \centering
                \includegraphics[width=0.9\textwidth]{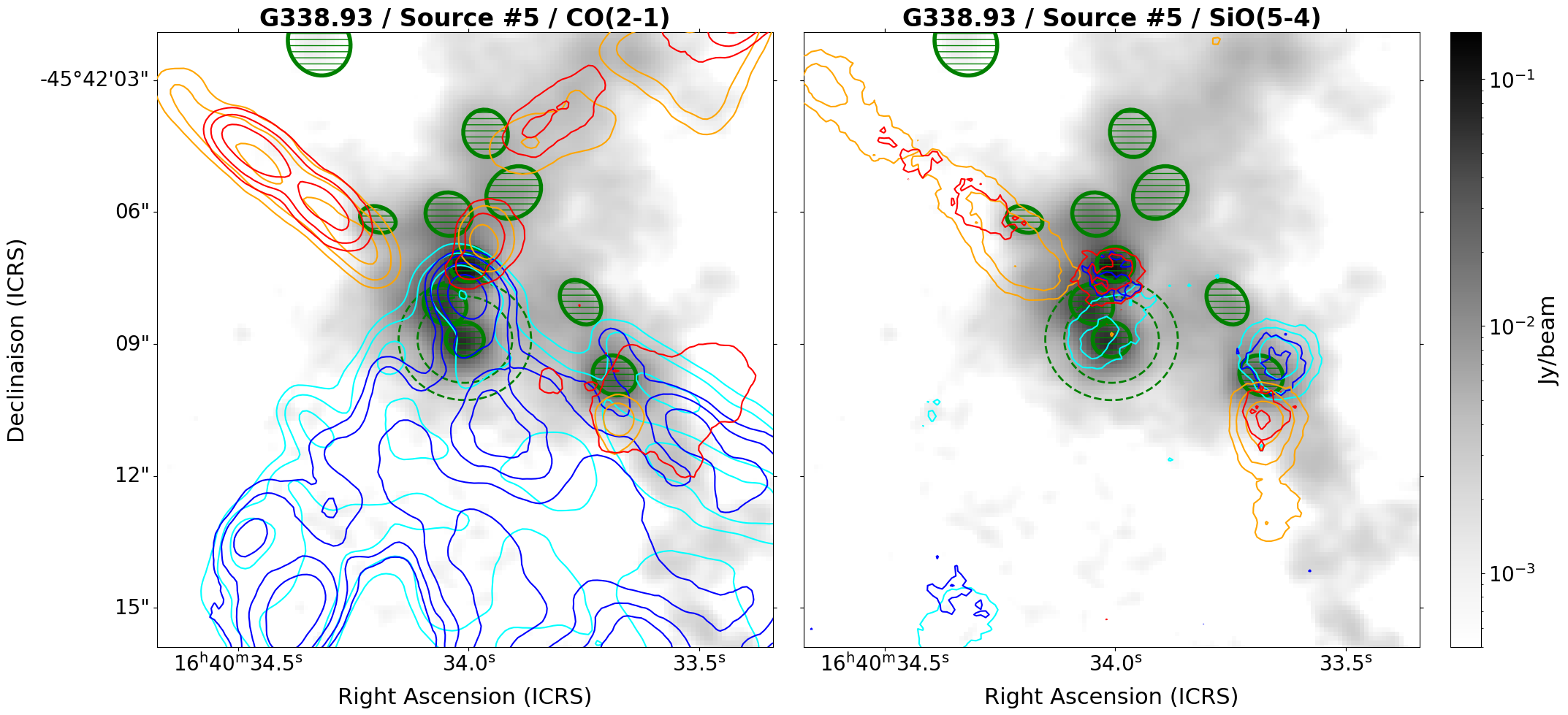}
            \end{minipage}
            \vfill
            \begin{minipage}[c]{\textwidth}
                \centering
                \includegraphics[width=0.7\textwidth]{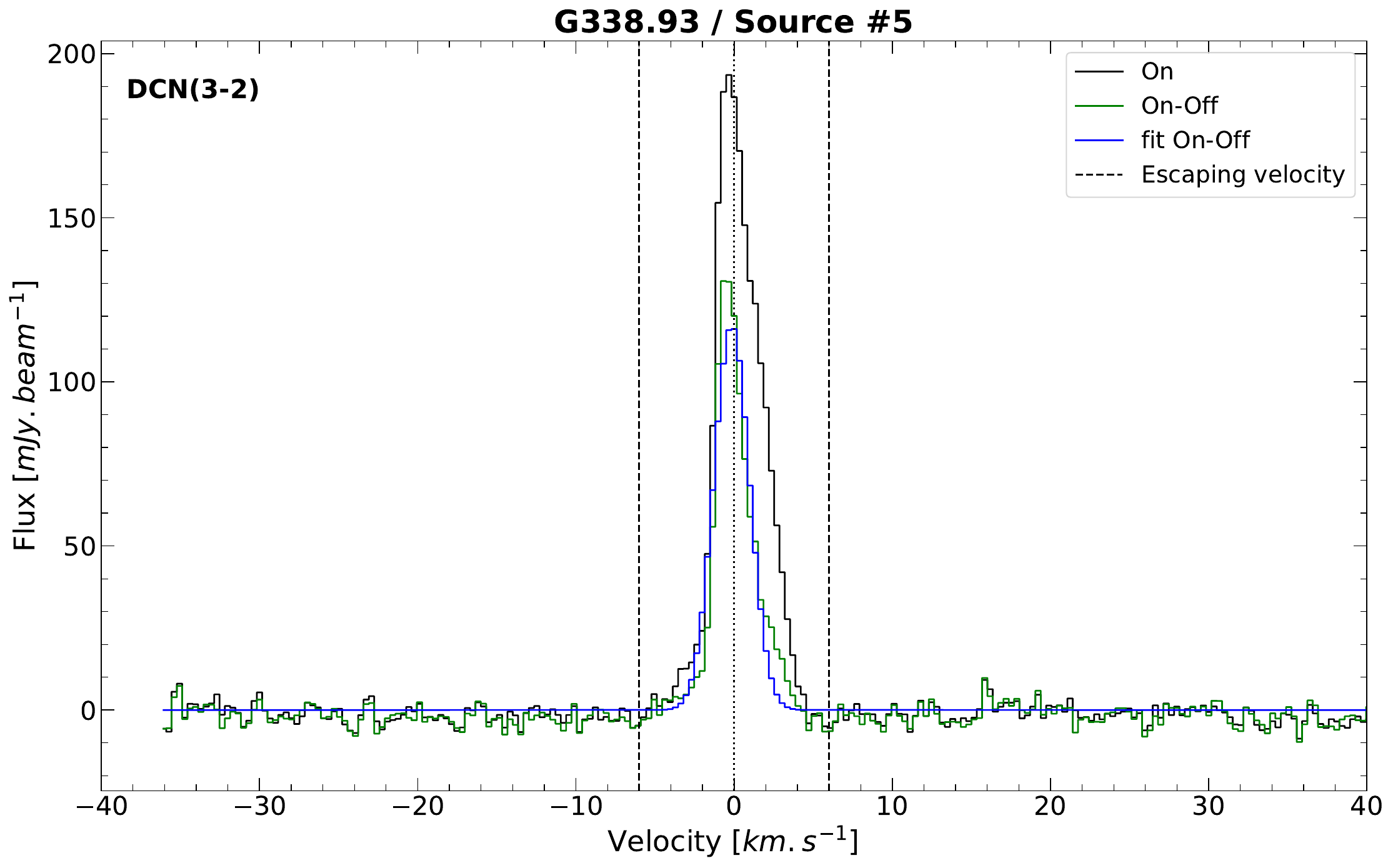}
            \end{minipage}
        \end{minipage}
    
    \vspace{0.2cm}
    
        \centering
        \begin{minipage}[c]{0.49\textwidth}
            \centering
            \includegraphics[width=\textwidth]{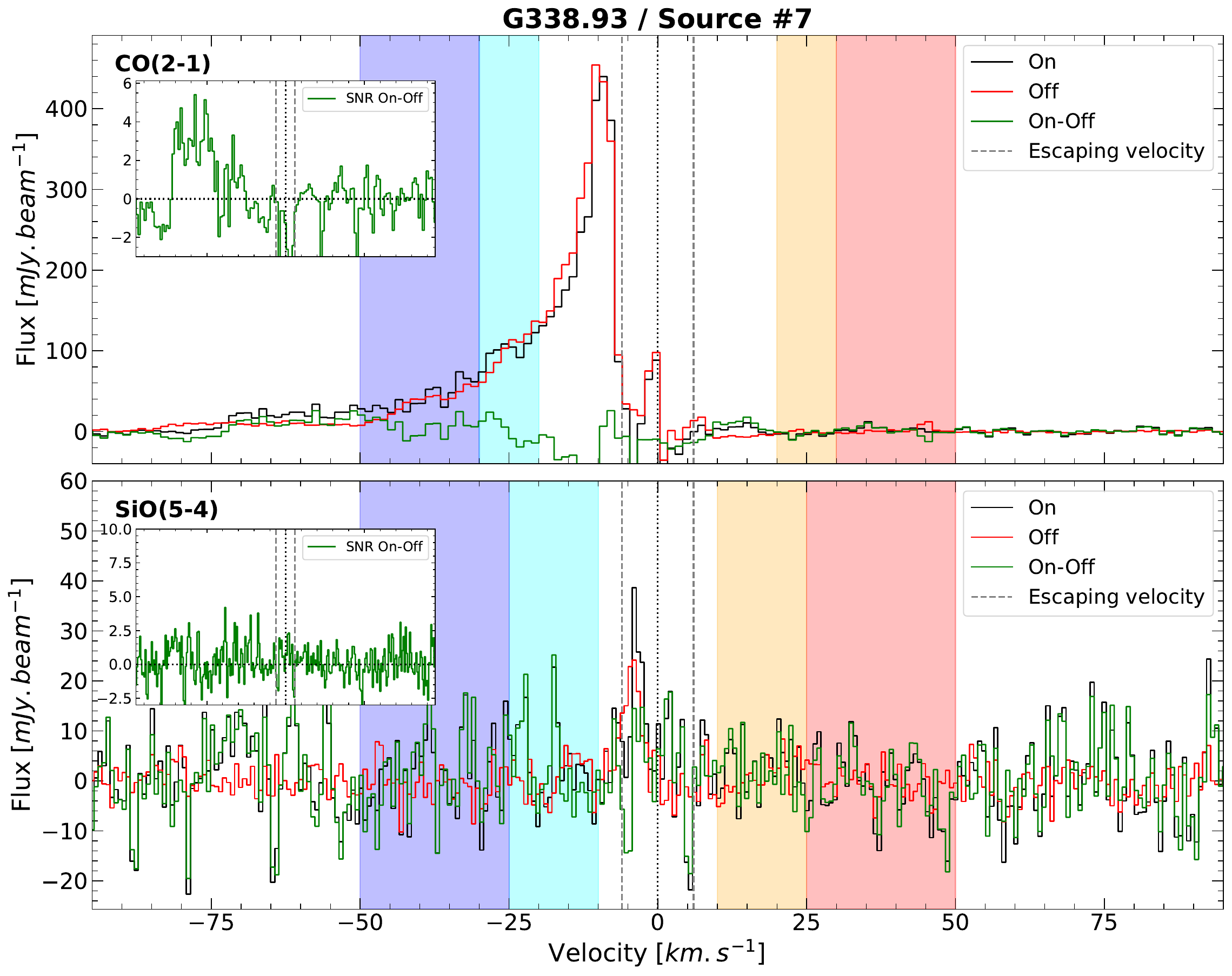}
        \end{minipage}
        \begin{minipage}[c]{0.49\textwidth}
            \centering
            \begin{minipage}[c]{\textwidth}
                \centering
                \includegraphics[width=0.9\textwidth]{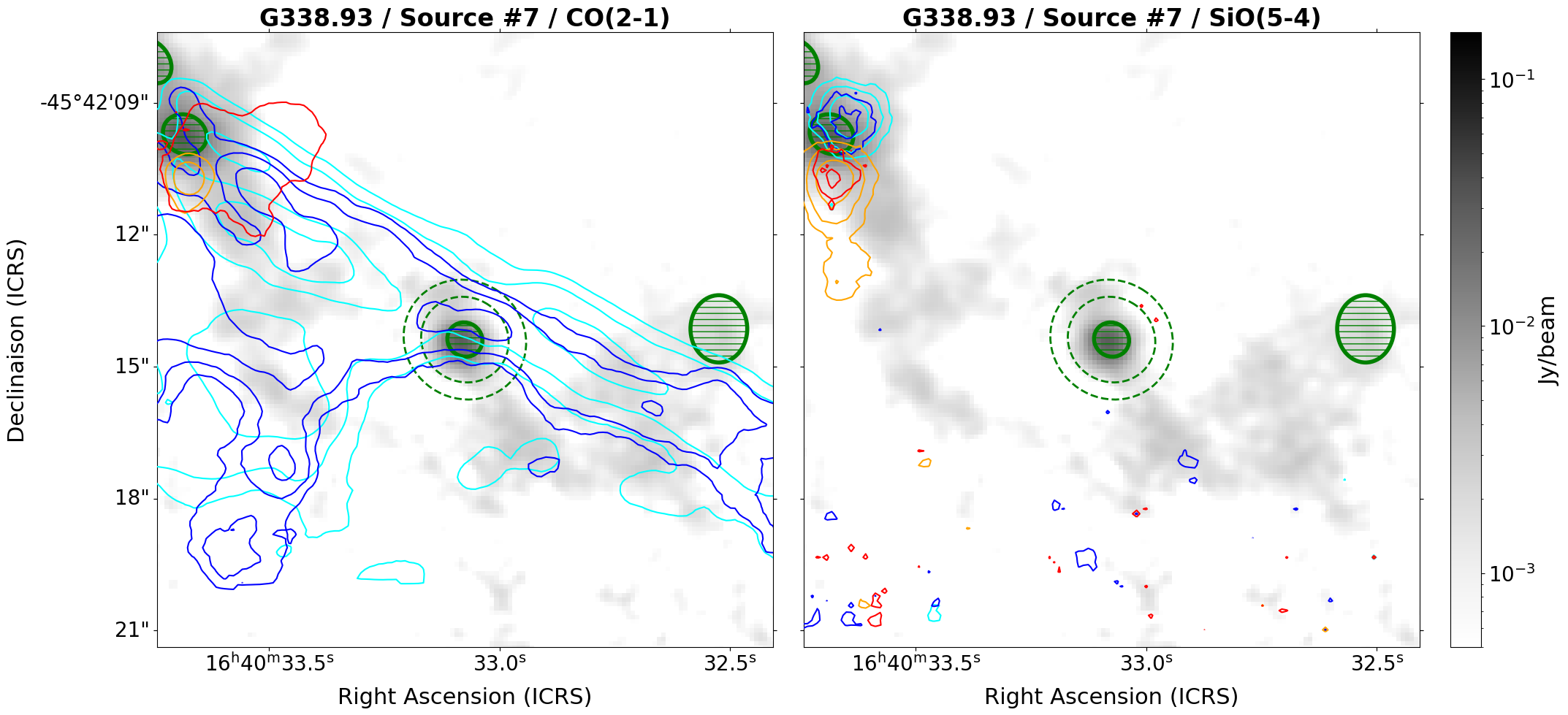}
            \end{minipage}
            \vfill
            \begin{minipage}[c]{\textwidth}
                \centering
                \includegraphics[width=0.7\textwidth]{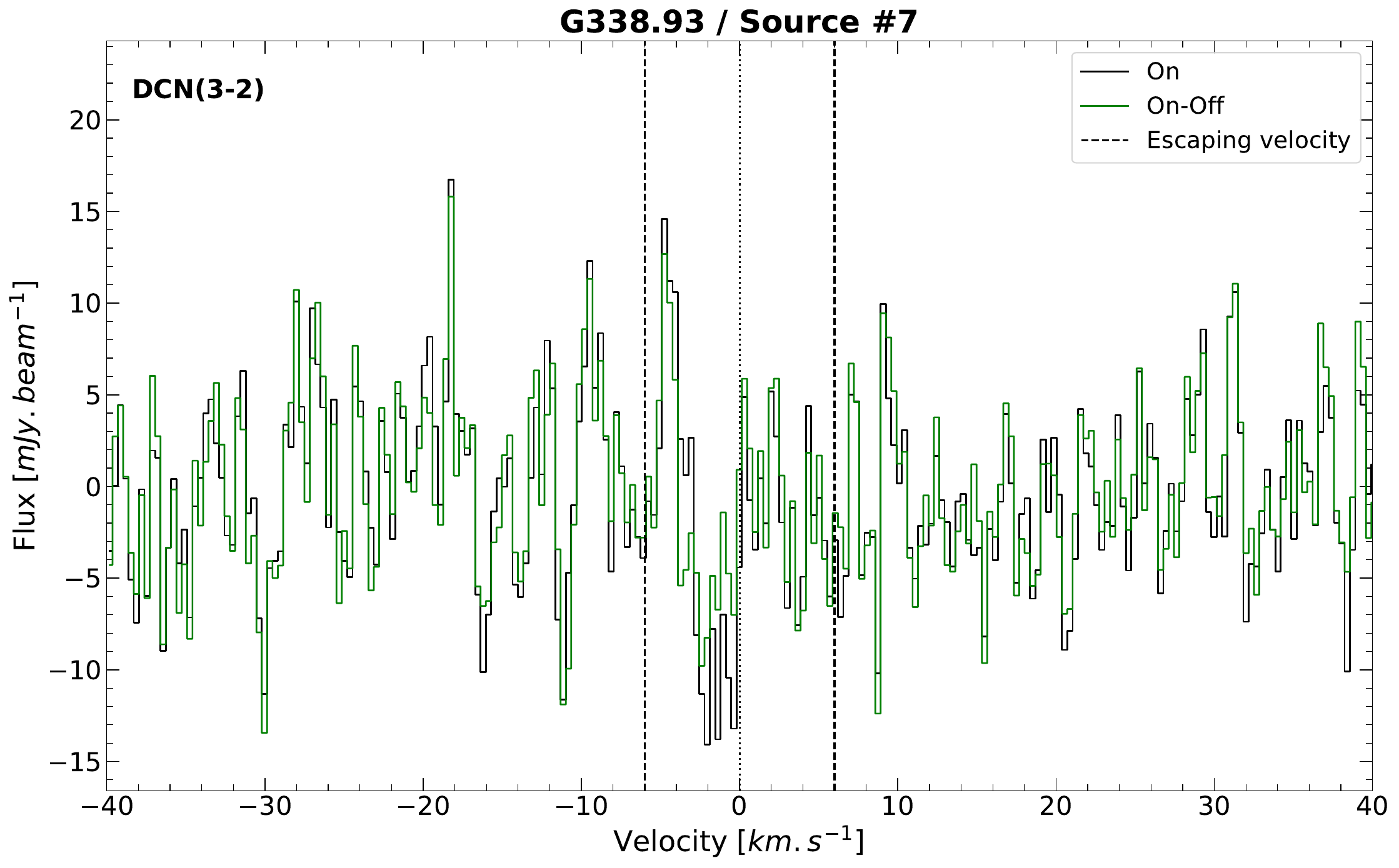}
            \end{minipage}
        \end{minipage}
    
    \vspace{0.2cm}
    
        \centering
        \begin{minipage}[c]{0.49\textwidth}
            \centering
            \includegraphics[width=\textwidth]{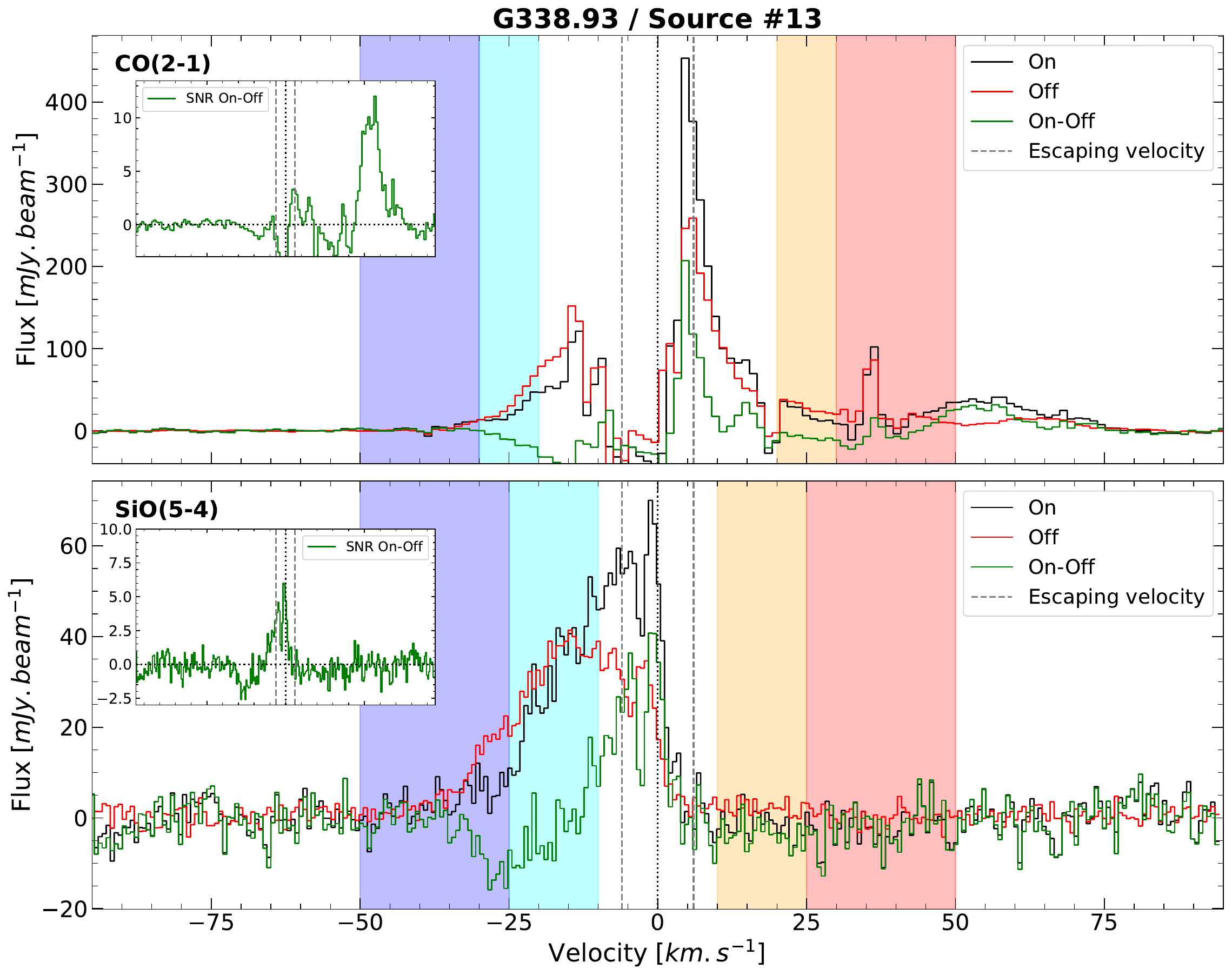}
        \end{minipage}
        \begin{minipage}[c]{0.49\textwidth}
            \centering
            \begin{minipage}[c]{\textwidth}
                \centering
                \includegraphics[width=0.9\textwidth]{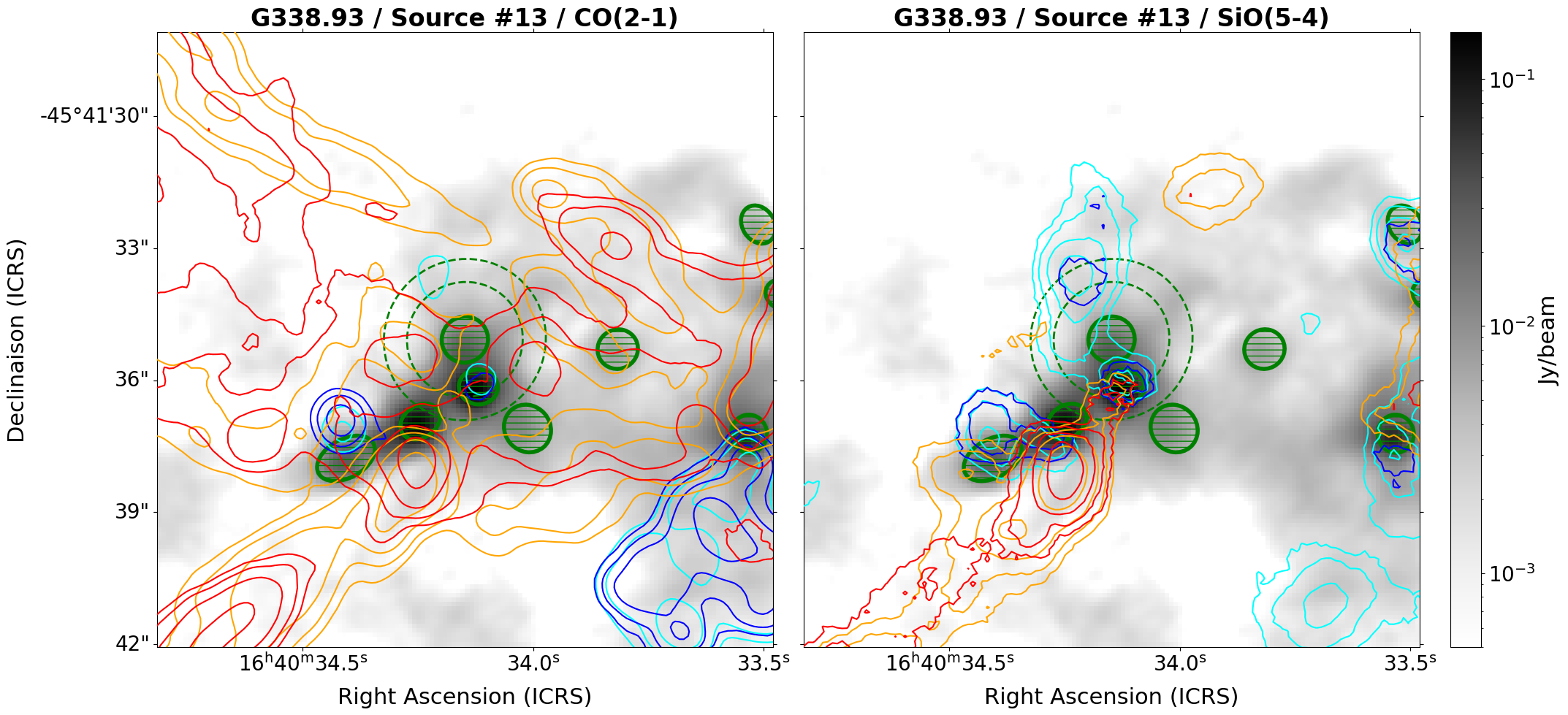}
            \end{minipage}
            \vfill
            \begin{minipage}[c]{\textwidth}
                \centering
                \includegraphics[width=0.7\textwidth]{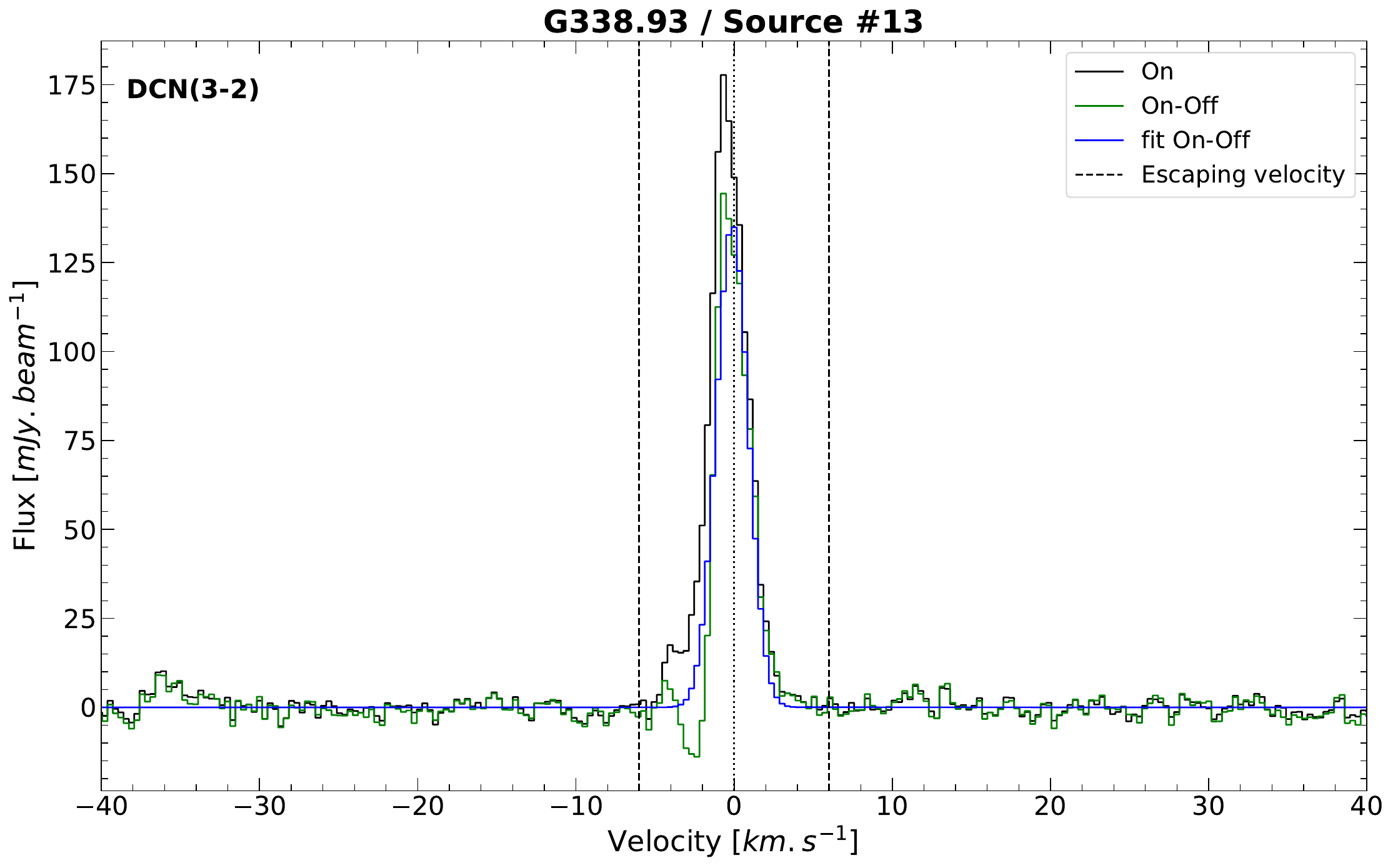}
            \end{minipage}
        \end{minipage}

   % \vskip -0.3cm
    \caption{CO and SiO spectra (left) and molecular outflow maps (top right) of the high-mass PSC candidates of the G338.93 region. CO contours are 10, 20, 40, and 80 in units of $\sigma$, with $\sigma$ = $69.1$, $27.1$, $70.0$, $27.3$ \mJybeamkms for cyan, blue, orange and red contours respectively. SiO contours are 10, 20, 40, and 80 in units of $\sigma$, with $\sigma$ = $10.3$, $12.7$, $11.0$, $12.5$ \mJybeamkms for cyan, blue, orange and red contours respectively. DCN spectra and fits (bottom right) of the high-mass PSC candidates of the G338.93 region.}

\end{figure*}

\begin{figure*}\ContinuedFloat
    \centering
        \begin{minipage}[c]{0.49\textwidth}
            \centering
            \includegraphics[width=\textwidth]{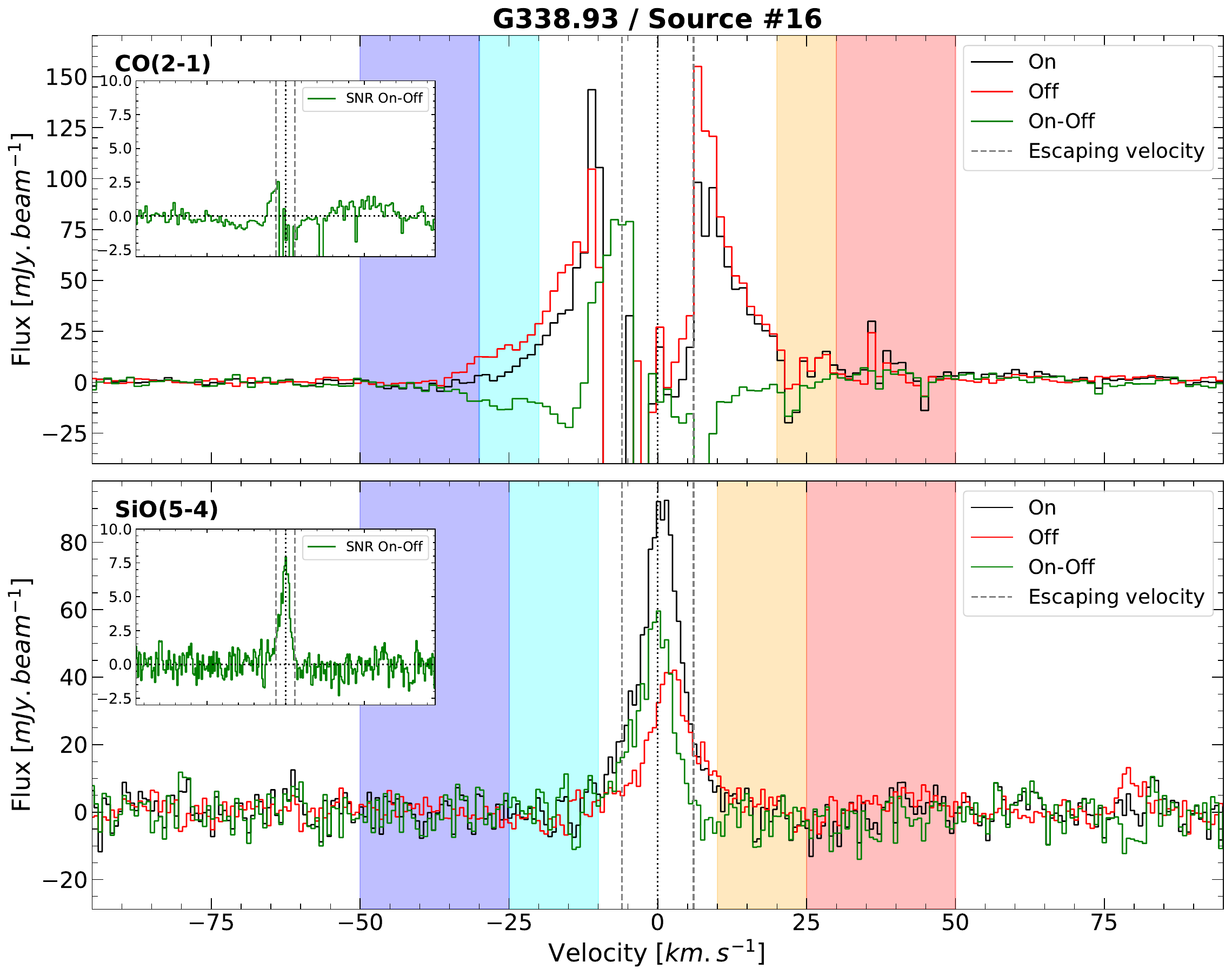}
        \end{minipage}
        \begin{minipage}[c]{0.49\textwidth}
            \centering
            \begin{minipage}[c]{\textwidth}
                \centering
                \includegraphics[width=0.9\textwidth]{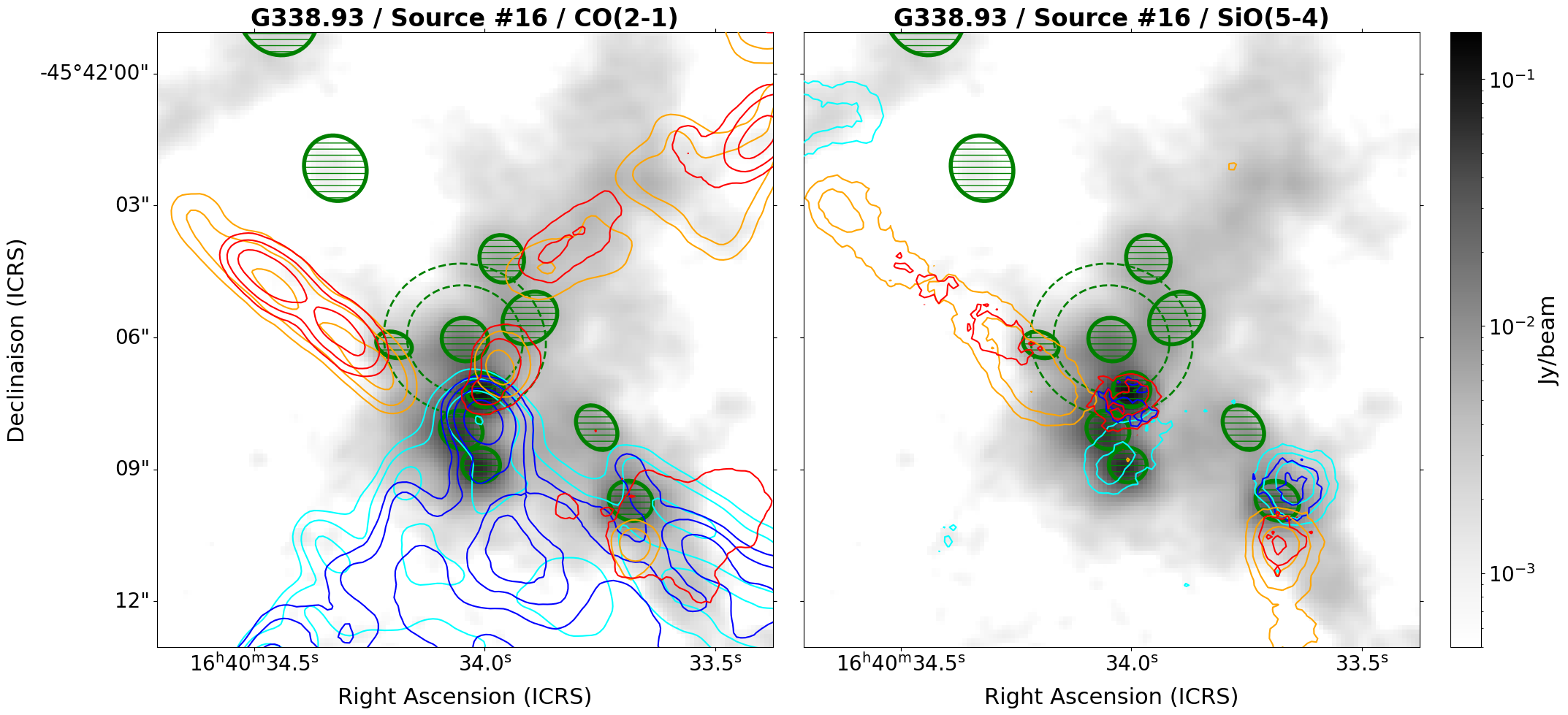}
            \end{minipage}
            \vfill
            \begin{minipage}[c]{\textwidth}
                \centering
                \includegraphics[width=0.7\textwidth]{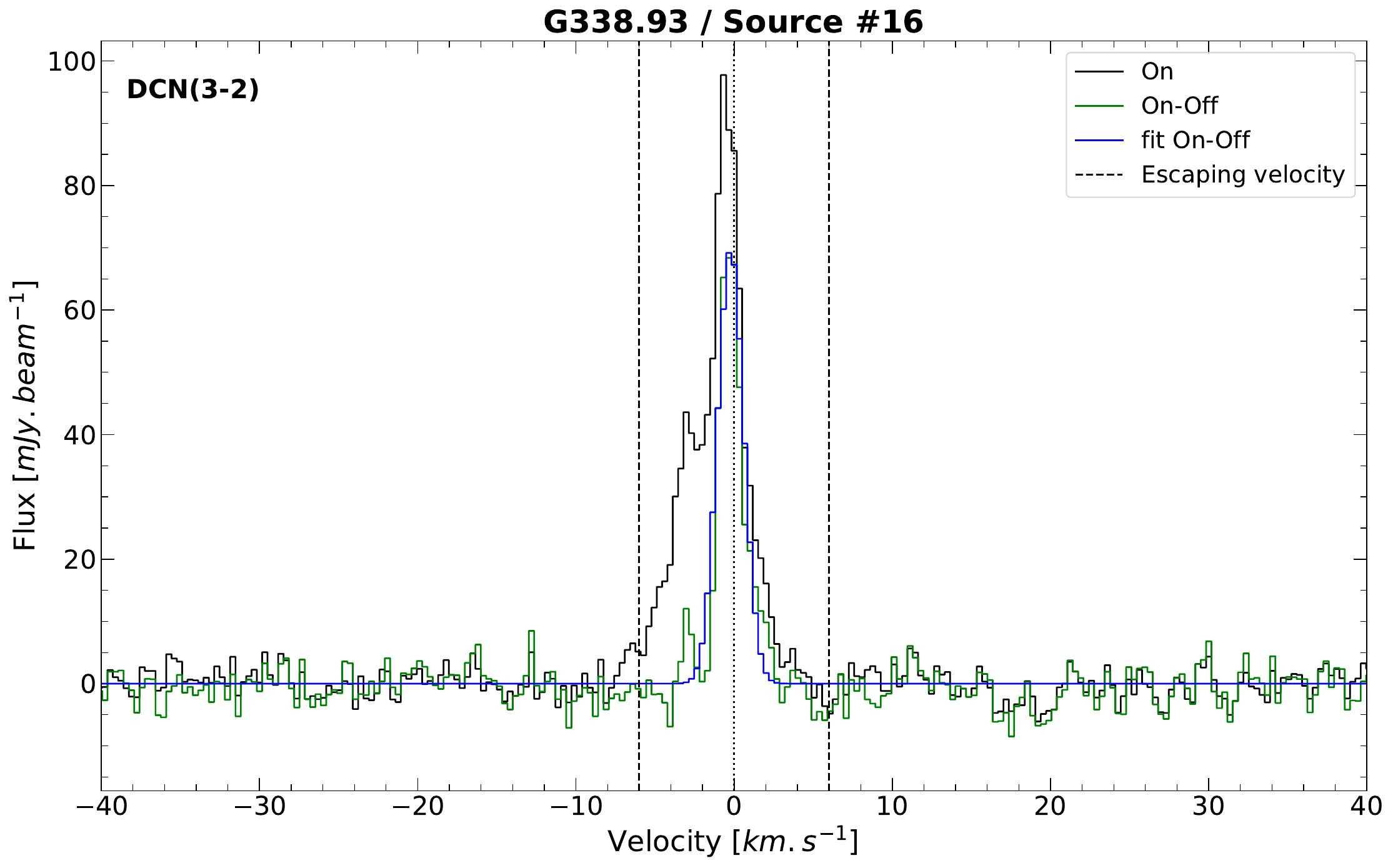}
            \end{minipage}
        \end{minipage}
    \caption{continued.}

\end{figure*}

%%%%%%%%%%%%%%%%%%%%%%% G353 %%%%%%%%%%%%%%%%%%%%%%%%%%%%%%%%%%%%%
\begin{figure*}
    \label{appendix:G353_MPSC_fig}
    \centering
        \begin{minipage}[c]{0.49\textwidth}
            \centering
            \includegraphics[width=\textwidth]{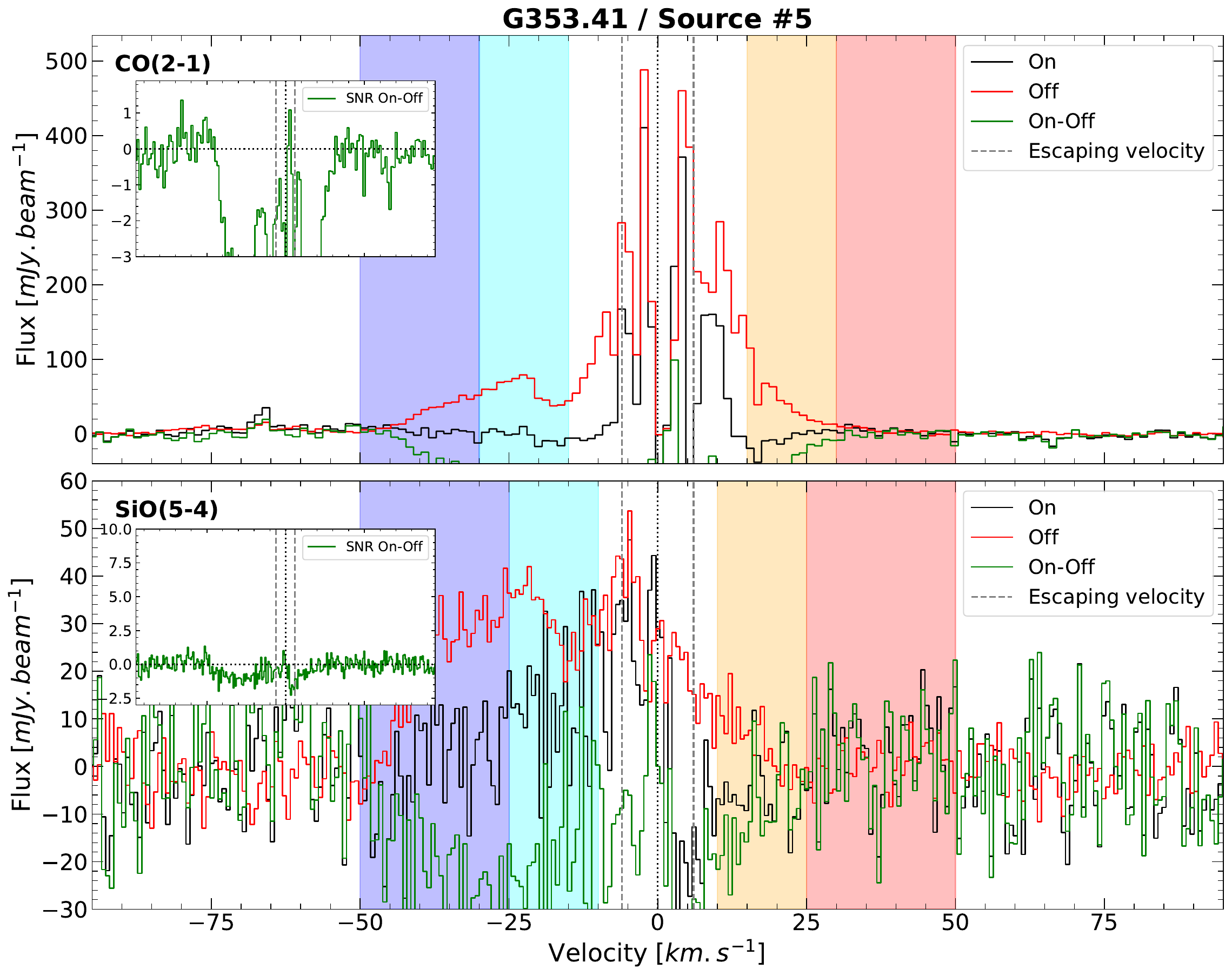}
        \end{minipage}
        \begin{minipage}[c]{0.49\textwidth}
            \centering
            \begin{minipage}[c]{\textwidth}
                \centering
                \includegraphics[width=0.9\textwidth]{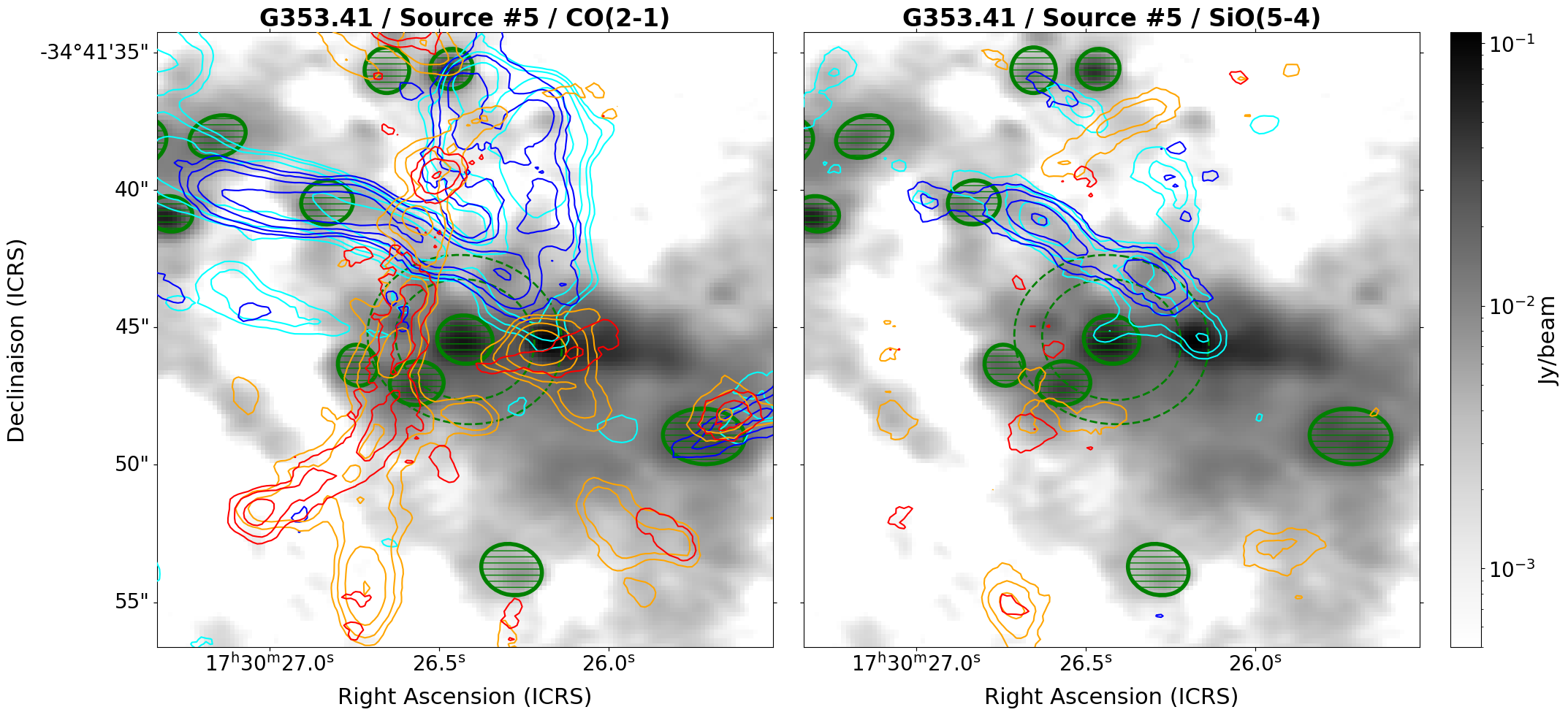}
            \end{minipage}
            \vfill
            \begin{minipage}[c]{\textwidth}
                \centering
                \includegraphics[width=0.7\textwidth]{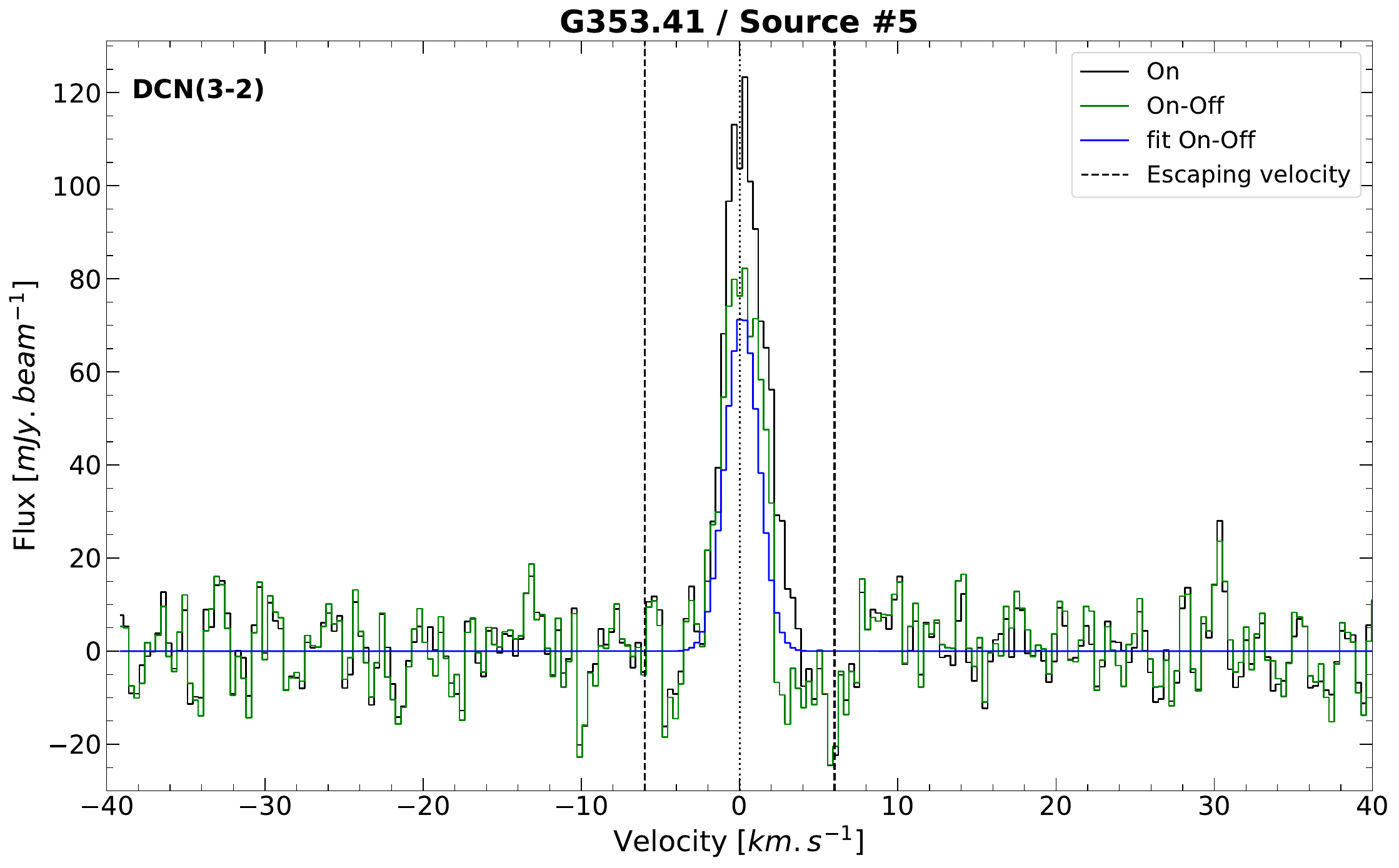}
            \end{minipage}
        \end{minipage}

   % \vskip -0.3cm
    \caption{CO and SiO spectra (left) and molecular outflow maps (top right) of the high-mass PSC candidate of the G353.41 region. CO contours are 5, 10, 20, and 40 in units of $\sigma$, with $\sigma$ = $95.3$, $66.0$, $62.9$, $64.1$ \mJybeamkms for cyan, blue, orange and red contours respectively. SiO contours are 5, 10, 20, and 40 in units of $\sigma$, with $\sigma$ = $38.5$, $49.8$, $38.8$, $49.1$ \mJybeamkms for cyan, blue, orange and red contours respectively. DCN spectra and fit (bottom right) of the high-mass PSC candidate of the G353.41 region.}

\end{figure*}

%%%%%%%%%%%%%%%%%%%%%%% W43-MM1 %%%%%%%%%%%%%%%%%%%%%%%%%%%%%%%%%%%%%
\begin{figure*}
    \label{appendix:W43-MM1_MPSC_fig}
    \centering
        \begin{minipage}[c]{0.49\textwidth}
            \centering
            \includegraphics[width=\textwidth]{Appendix/Appendix_figures/W43-MM1/W43-MM1_Source_6.pdf}
        \end{minipage}
        \begin{minipage}[c]{0.49\textwidth}
            \centering
            \begin{minipage}[c]{\textwidth}
                \centering
                \includegraphics[width=0.9\textwidth]{Appendix/Appendix_figures/W43-MM1/W43-MM1_Contours_Source_6.png}
            \end{minipage}
            \vfill
            \begin{minipage}[c]{\textwidth}
                \centering
                \includegraphics[width=0.7\textwidth]{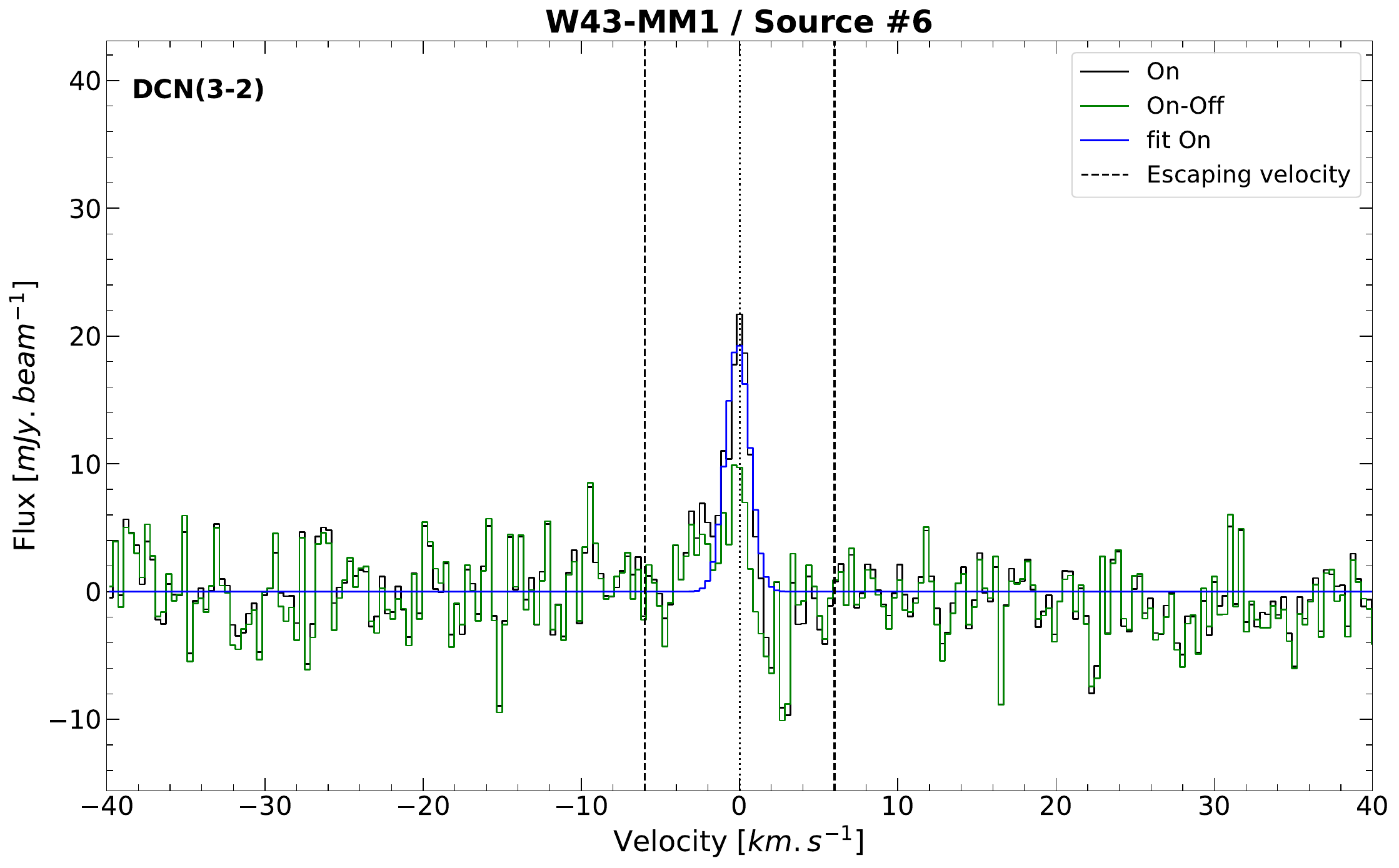}
            \end{minipage}
        \end{minipage}
    
    \vspace{0.2cm}
    
        \centering
        \begin{minipage}[c]{0.49\textwidth}
            \centering
            \includegraphics[width=\textwidth]{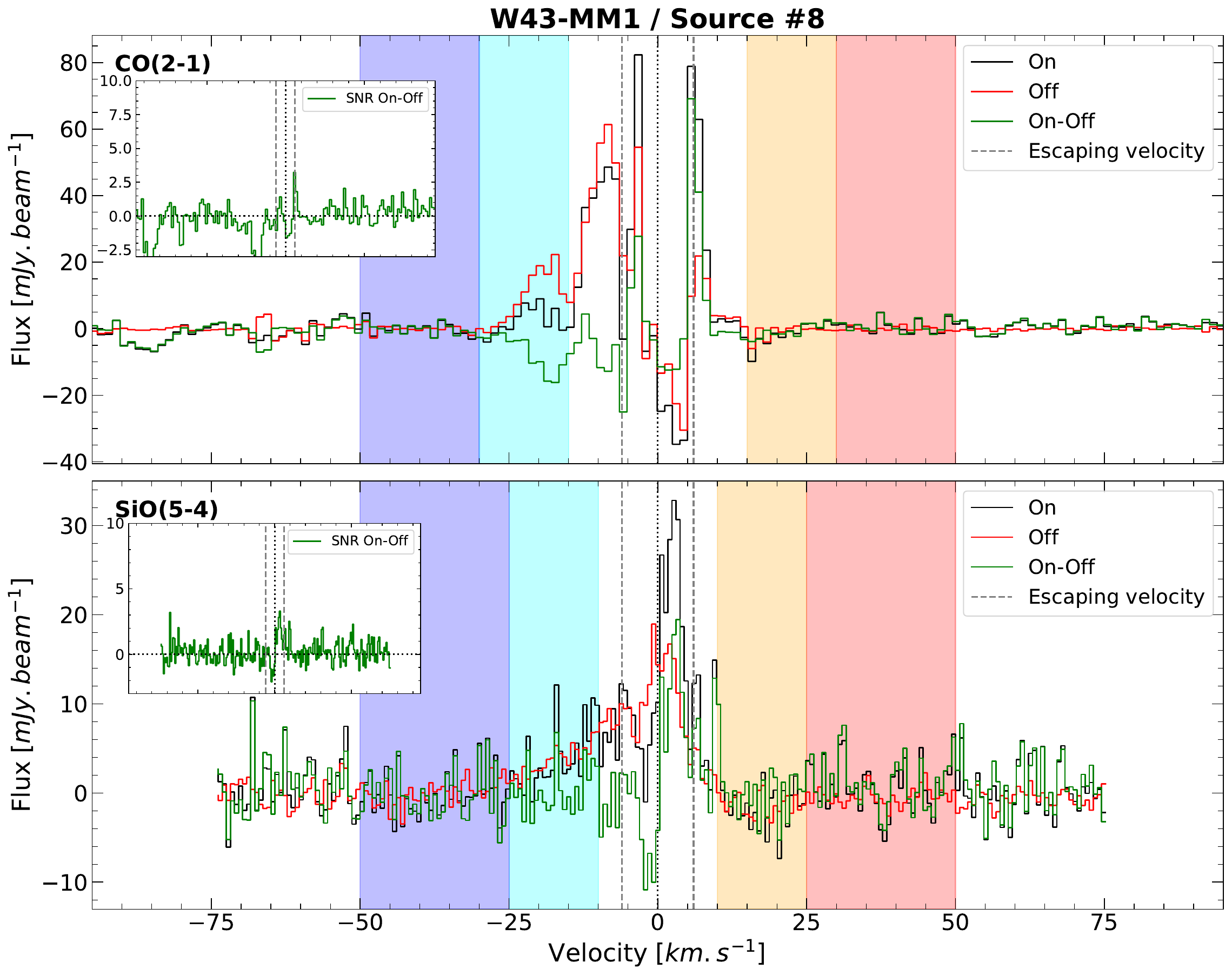}
        \end{minipage}
        \begin{minipage}[c]{0.49\textwidth}
            \centering
            \begin{minipage}[c]{\textwidth}
                \centering
                \includegraphics[width=0.9\textwidth]{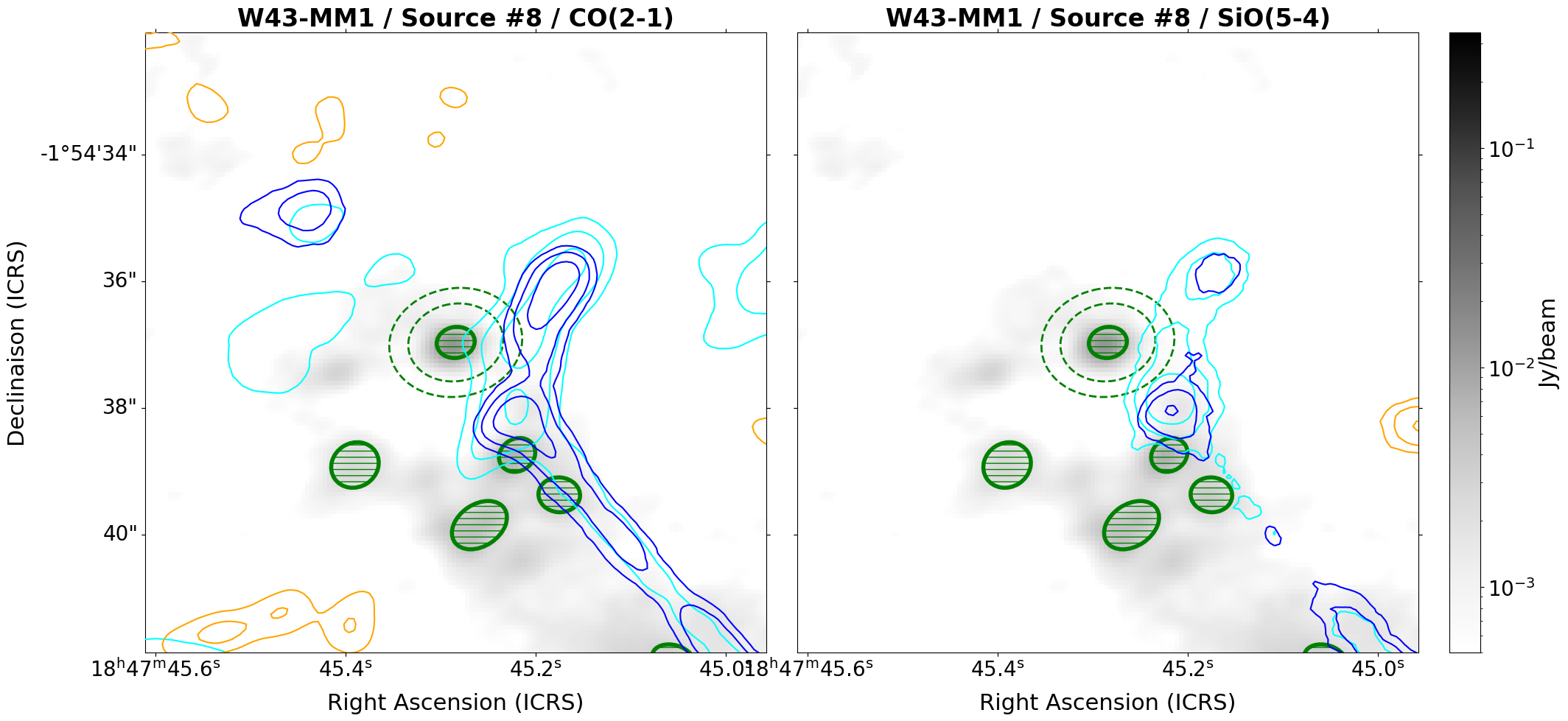}
            \end{minipage}
            \vfill
            \begin{minipage}[c]{\textwidth}
                \centering
                \includegraphics[width=0.7\textwidth]{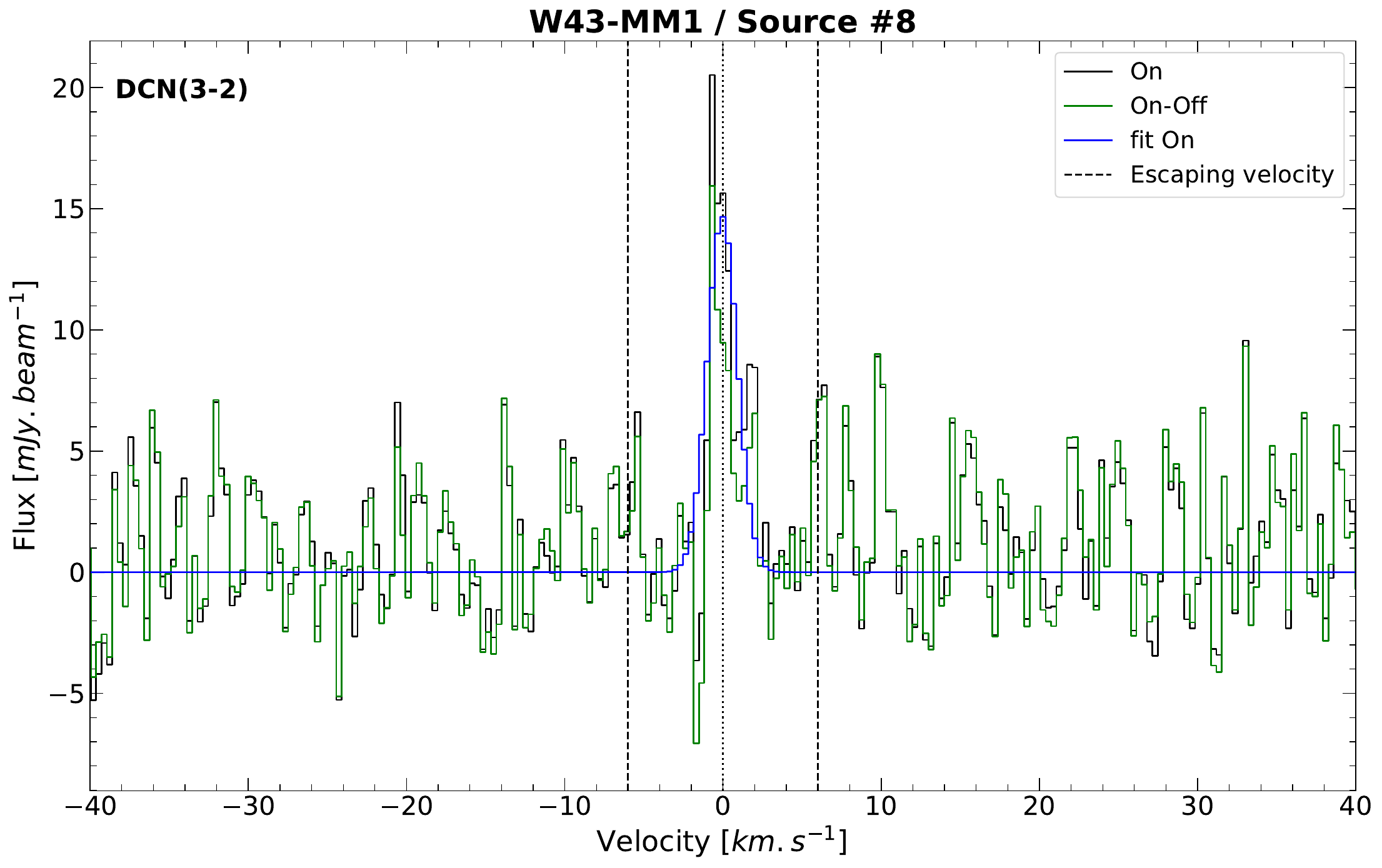}
            \end{minipage}
        \end{minipage}
    
    \vspace{0.2cm}
    
        \centering
        \begin{minipage}[c]{0.49\textwidth}
            \centering
            \includegraphics[width=\textwidth]{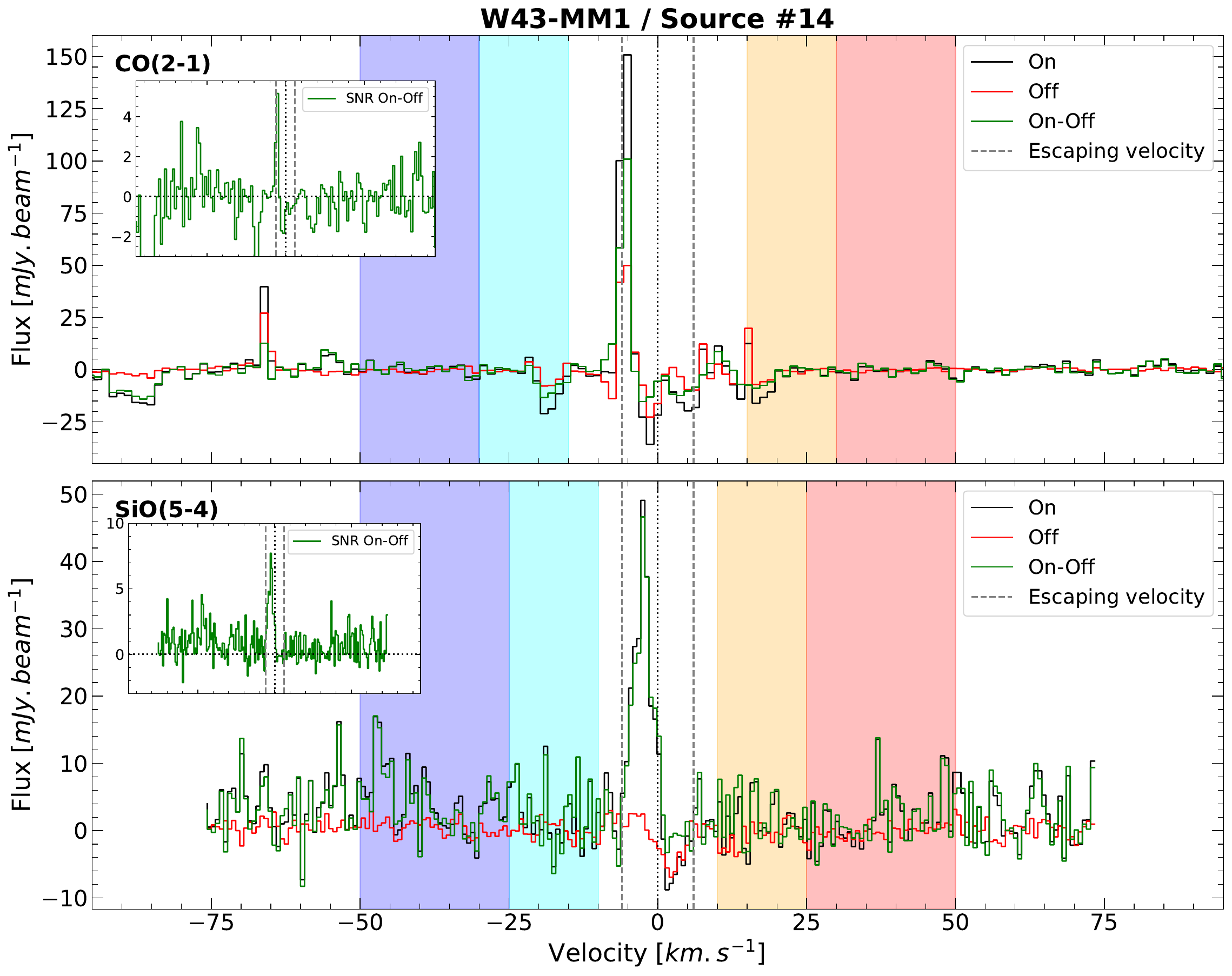}
        \end{minipage}
        \begin{minipage}[c]{0.49\textwidth}
            \centering
            \begin{minipage}[c]{\textwidth}
                \centering
                \includegraphics[width=0.9\textwidth]{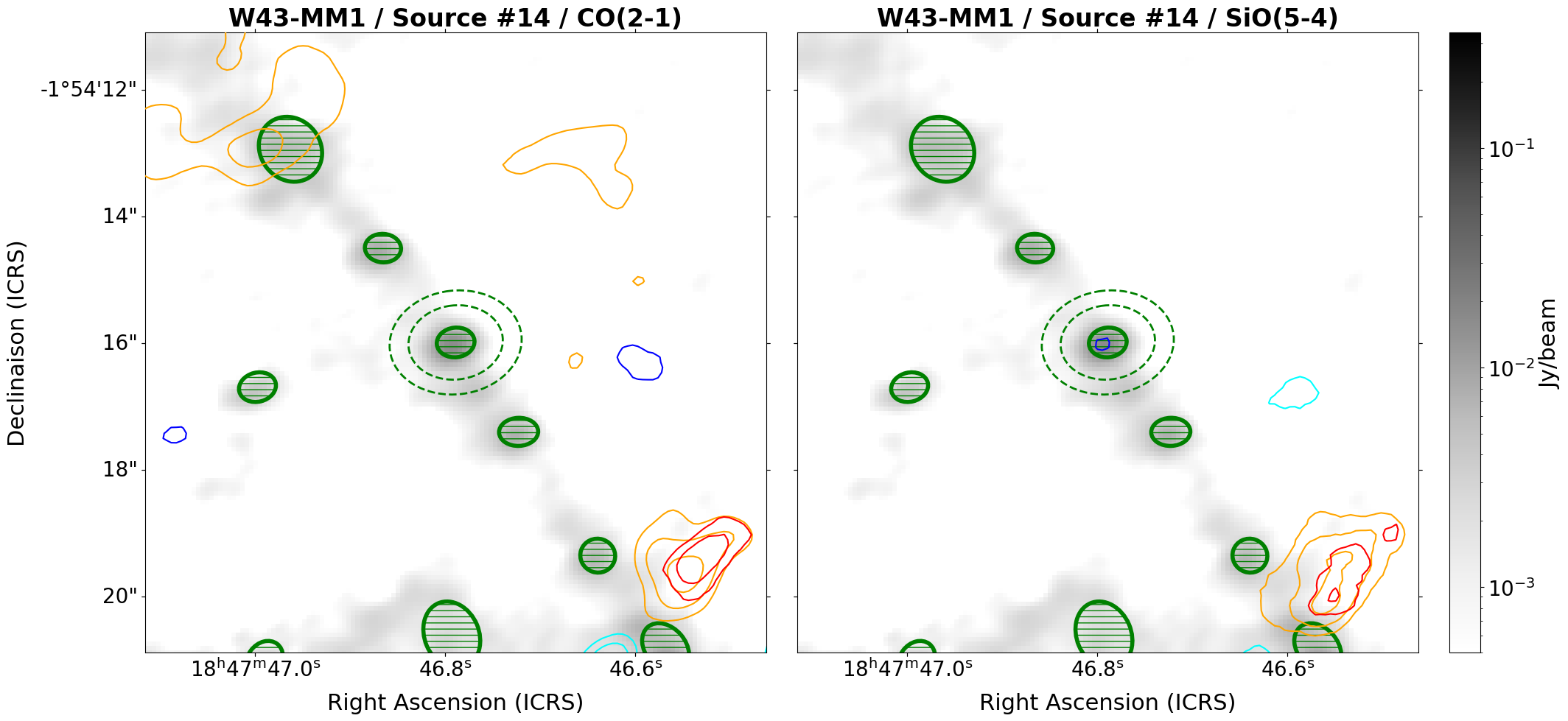}
            \end{minipage}
            \vfill
            \begin{minipage}[c]{\textwidth}
                \centering
                \includegraphics[width=0.7\textwidth]{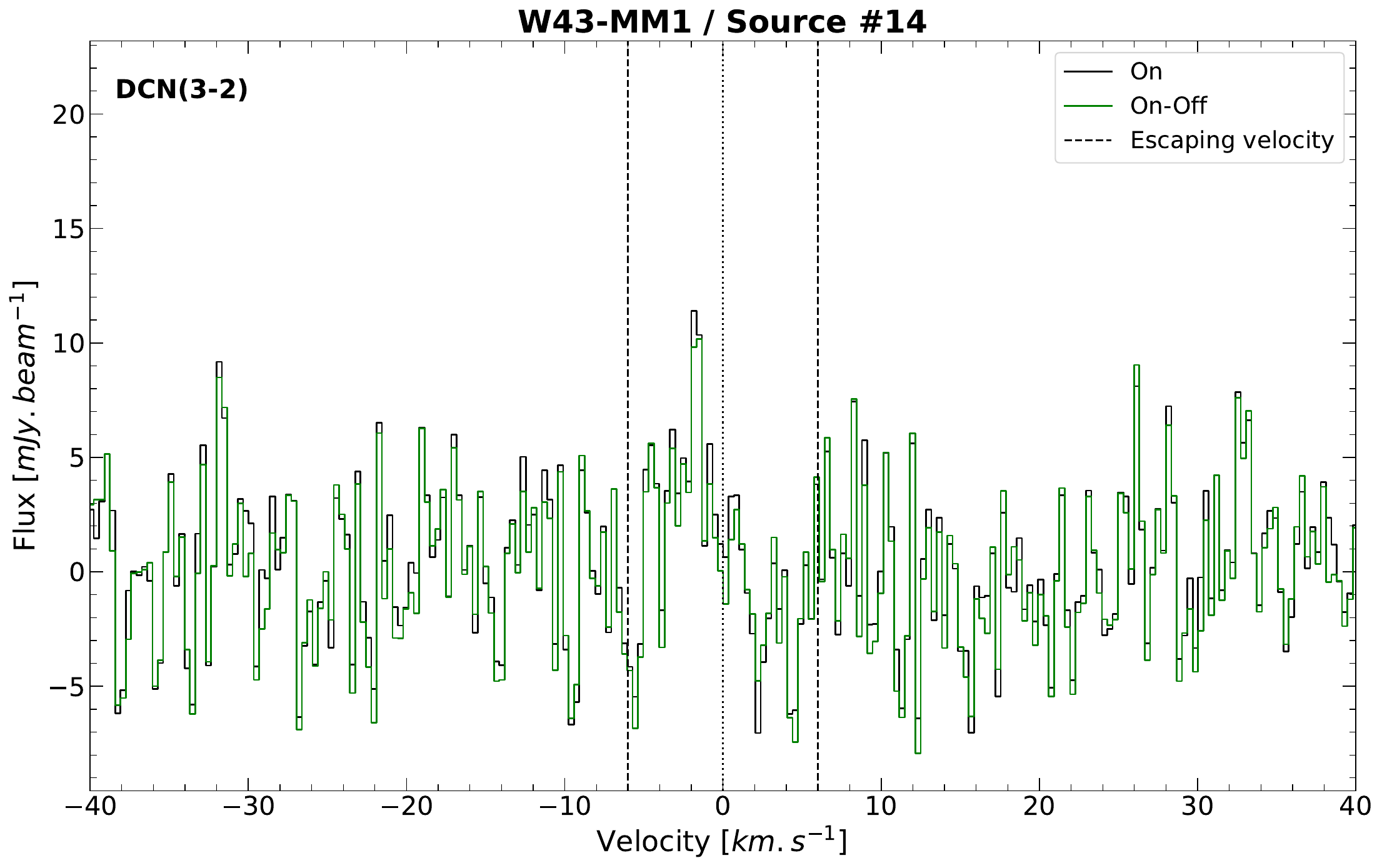}
            \end{minipage}
        \end{minipage}

   % \vskip -0.3cm
    \caption{CO and SiO spectra (left) and molecular outflow maps (top right) of the high-mass PSC candidates of the W43-MM1 region. CO contours are 10, 20, 40, and 80 in units of $\sigma$, with $\sigma$ = $32.6$, $18.6$, $49.5$, $15.3$ \mJybeamkms for cyan, blue, orange and red contours respectively. SiO contours are 10, 20, 40, and 80 in units of $\sigma$, with $\sigma$ = $6.4$, $7.4$, $7.3$, $7.2$ \mJybeamkms for cyan, blue, orange and red contours respectively. DCN spectra and fits (bottom right) of the high-mass PSC candidates of the W43-MM1 region.}

\end{figure*}

\begin{figure*}\ContinuedFloat

    \centering
        \begin{minipage}[c]{0.49\textwidth}
            \centering
            \includegraphics[width=\textwidth]{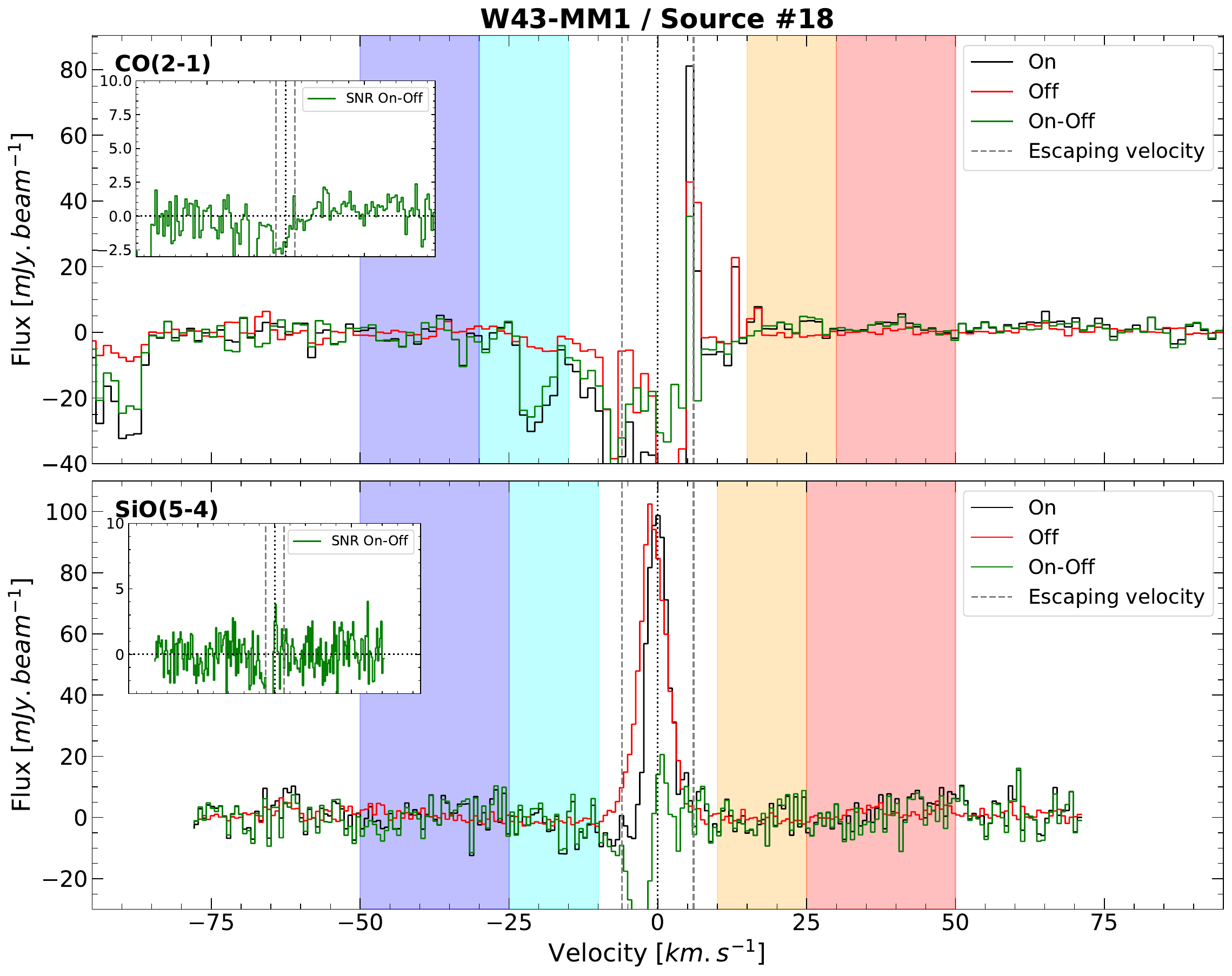}
        \end{minipage}
        \begin{minipage}[c]{0.49\textwidth}
            \centering
            \begin{minipage}[c]{\textwidth}
                \centering
                \includegraphics[width=0.9\textwidth]{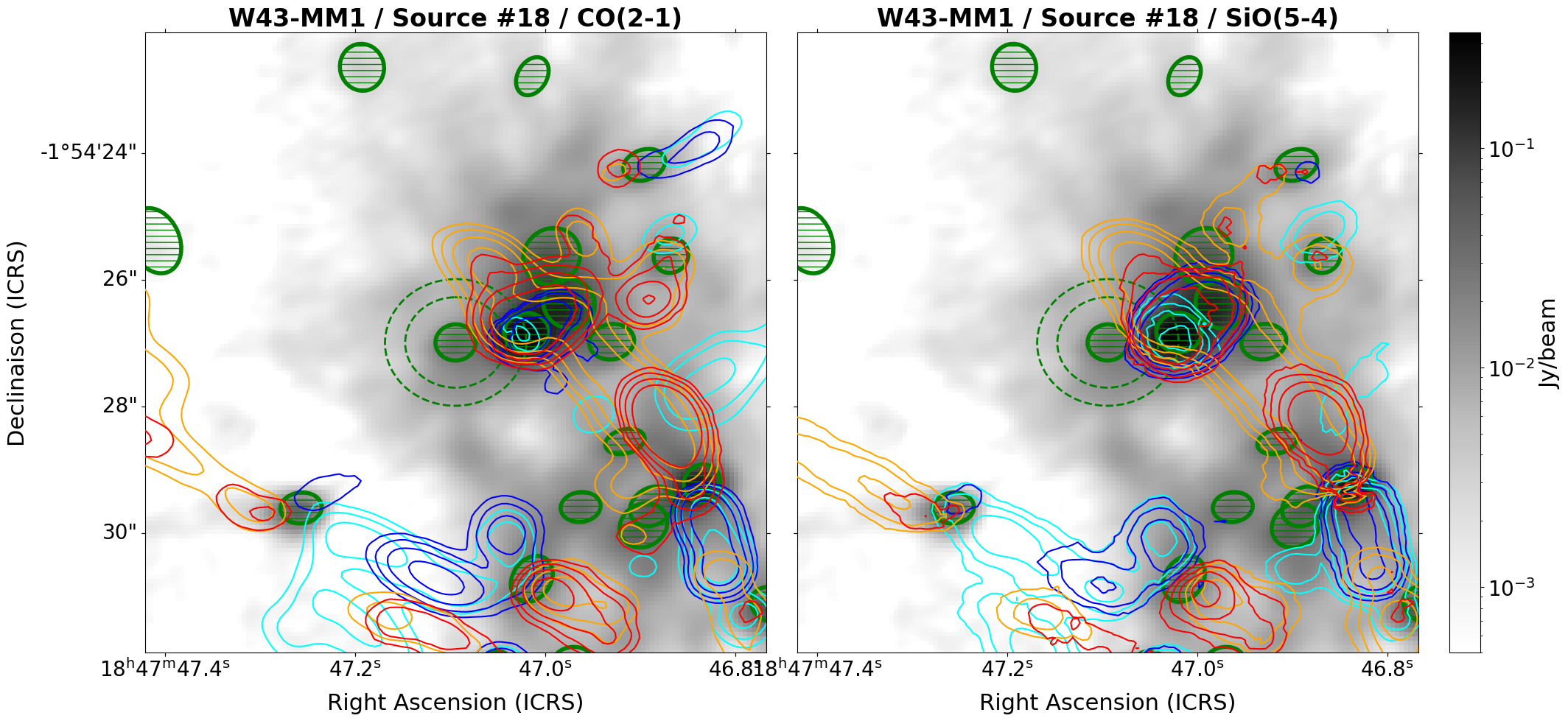}
            \end{minipage}
            \vfill
            \begin{minipage}[c]{\textwidth}
                \centering
                \includegraphics[width=0.7\textwidth]{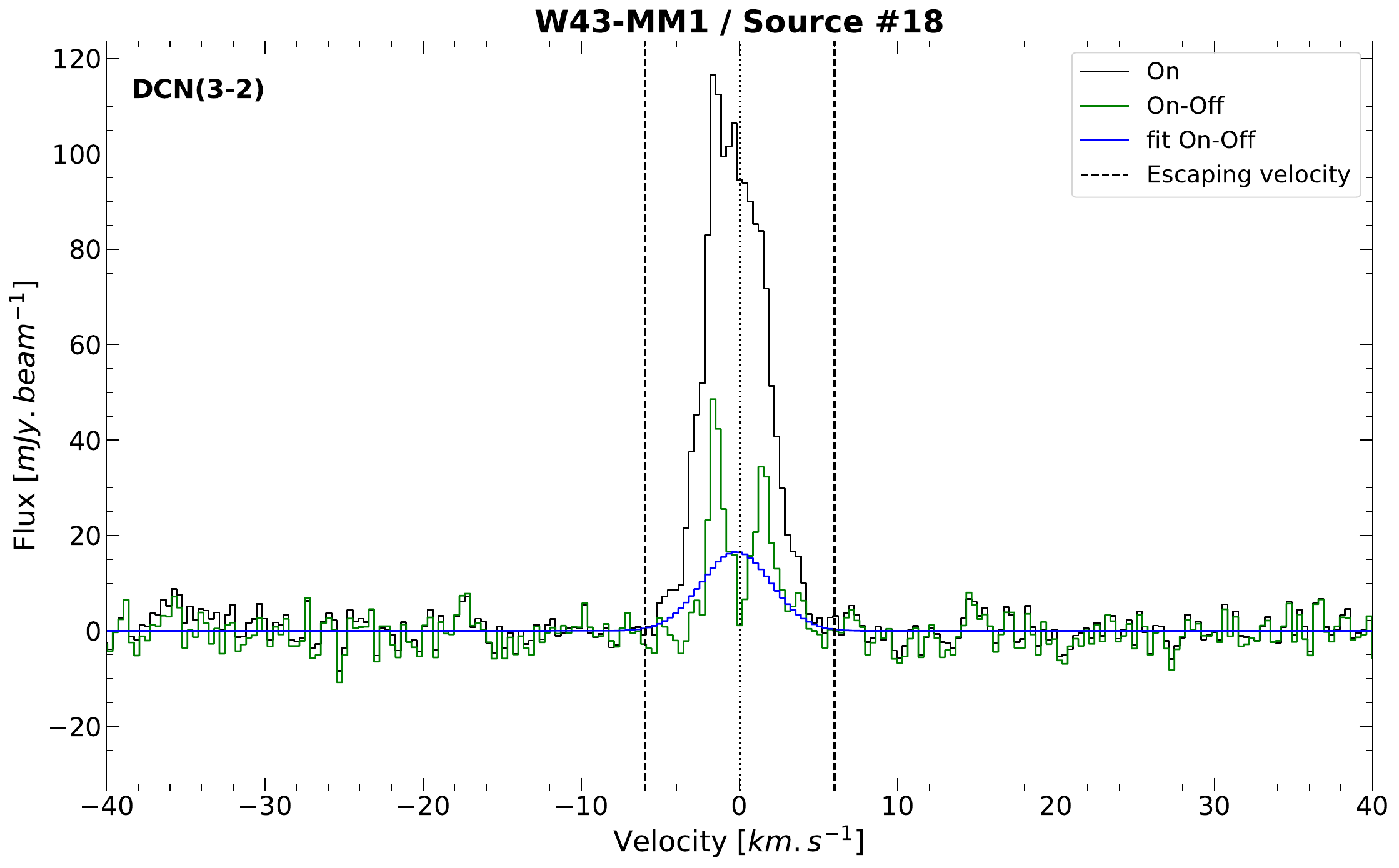}
            \end{minipage}
        \end{minipage}
        
    \vspace{0.2cm}
    
        \centering
        \begin{minipage}[c]{0.49\textwidth}
            \centering
            \includegraphics[width=\textwidth]{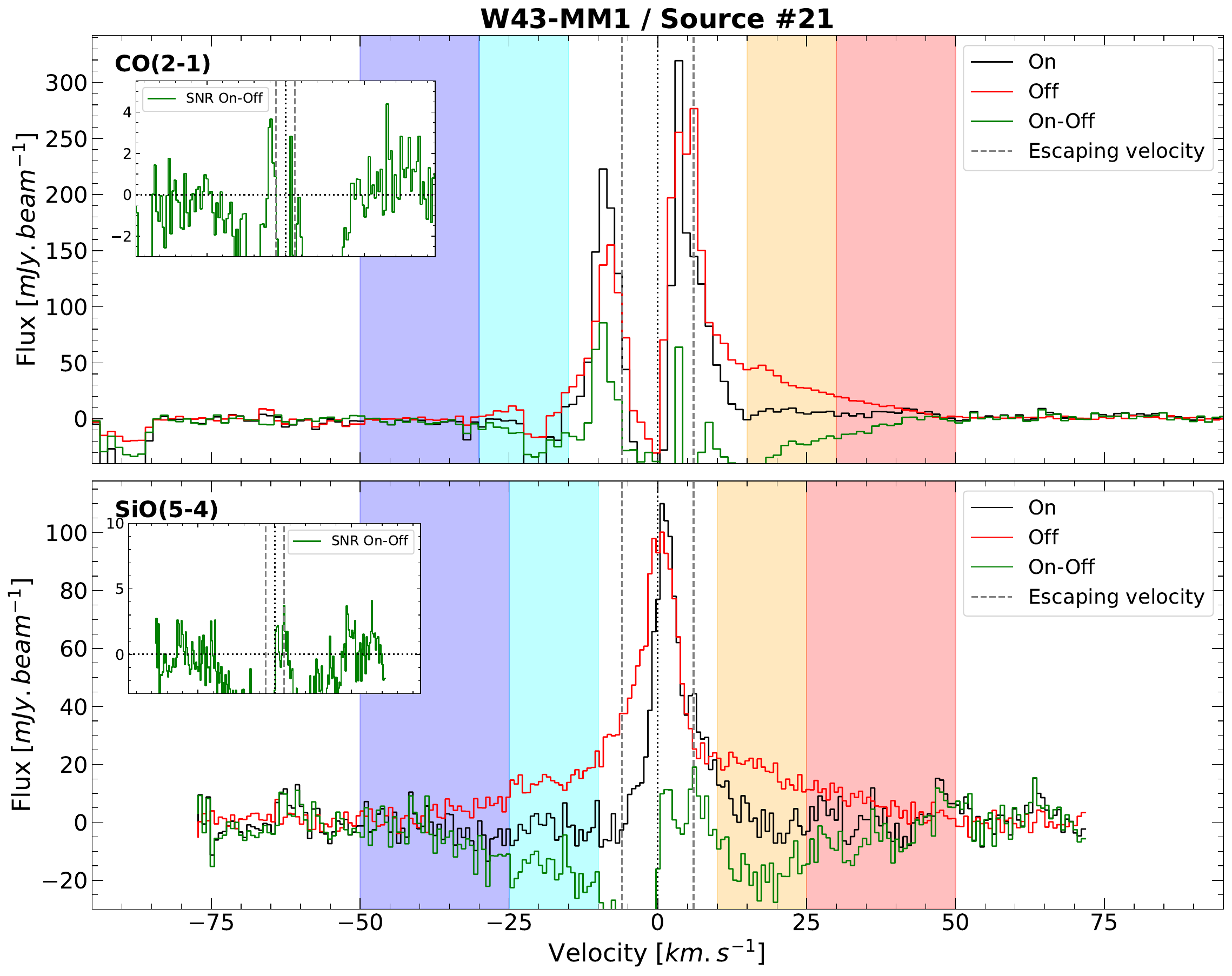}
        \end{minipage}
        \begin{minipage}[c]{0.49\textwidth}
            \centering
            \begin{minipage}[c]{\textwidth}
                \centering
                \includegraphics[width=0.9\textwidth]{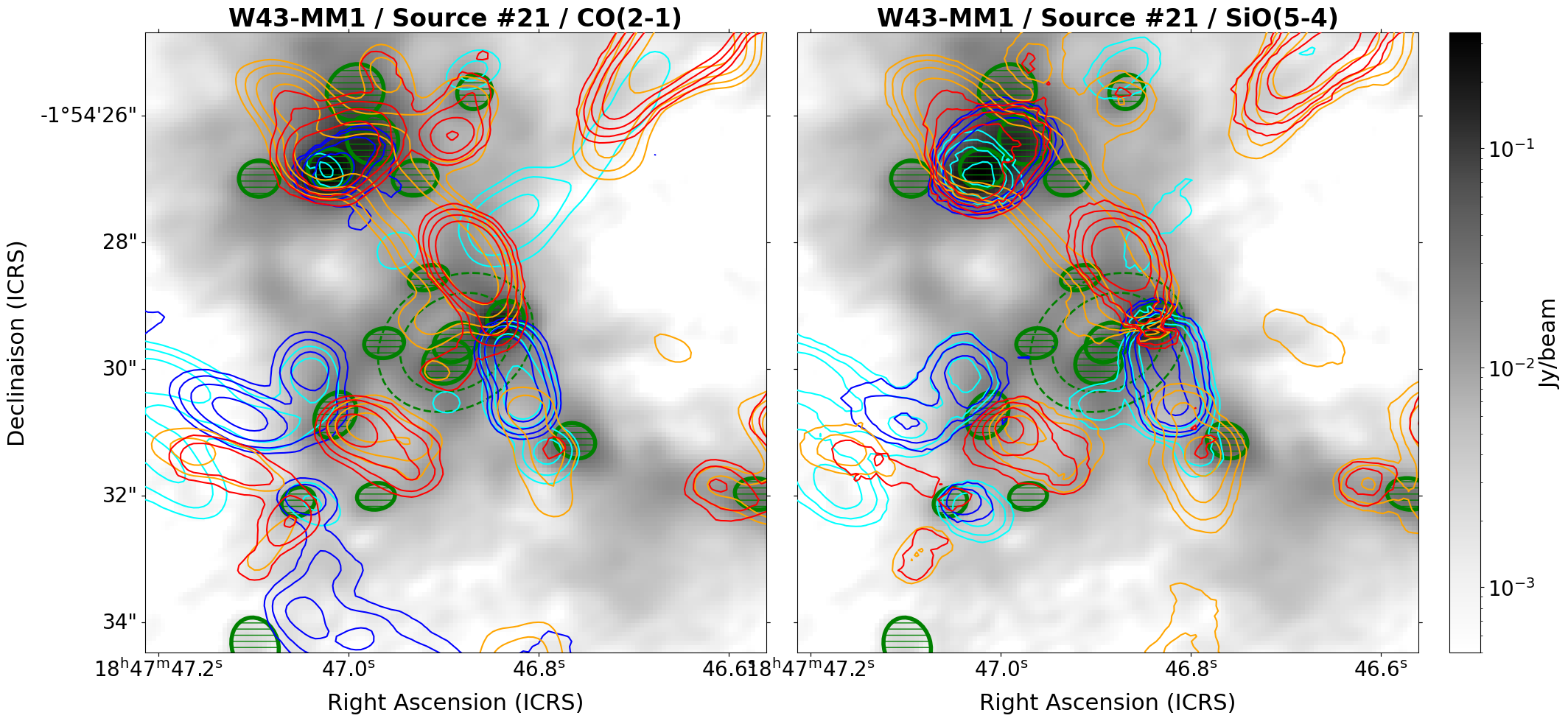}
            \end{minipage}
            \vfill
            \begin{minipage}[c]{\textwidth}
                \centering
                \includegraphics[width=0.7\textwidth]{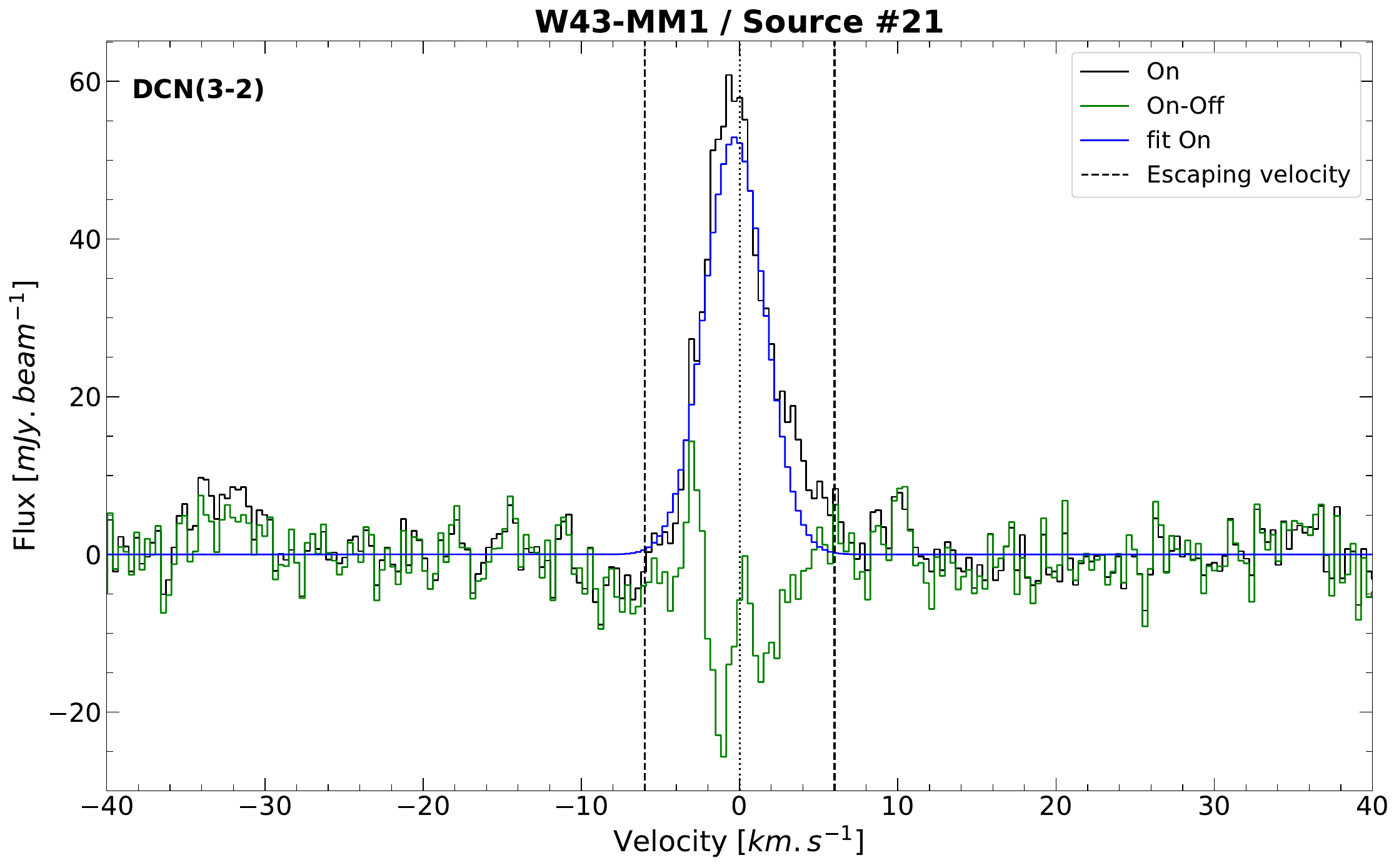}
            \end{minipage}
        \end{minipage}

    \vspace{0.2cm}

        \centering
        \begin{minipage}[c]{0.49\textwidth}
            \centering
            \includegraphics[width=\textwidth]{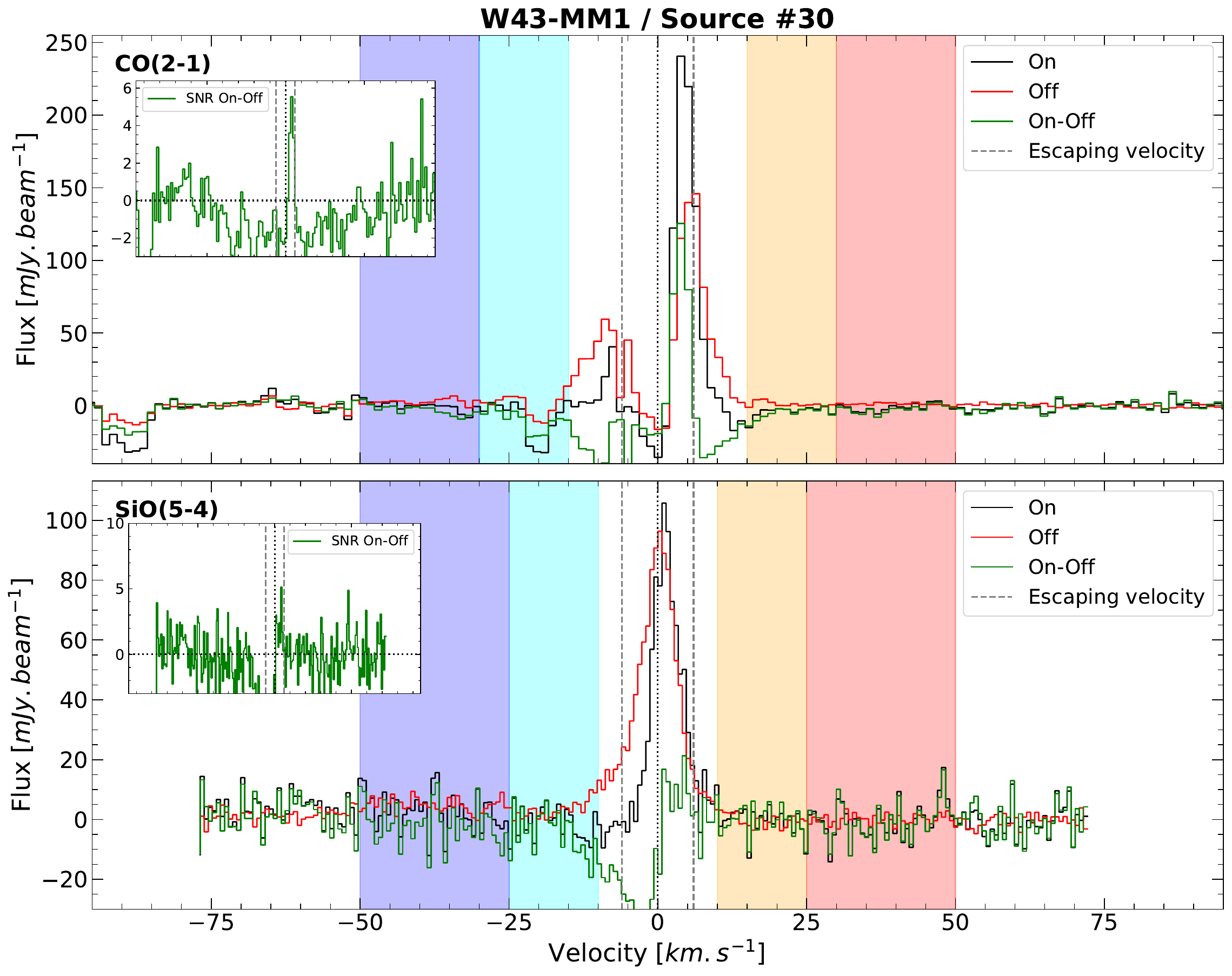}
        \end{minipage}
        \begin{minipage}[c]{0.49\textwidth}
            \centering
            \begin{minipage}[c]{\textwidth}
                \centering
                \includegraphics[width=0.9\textwidth]{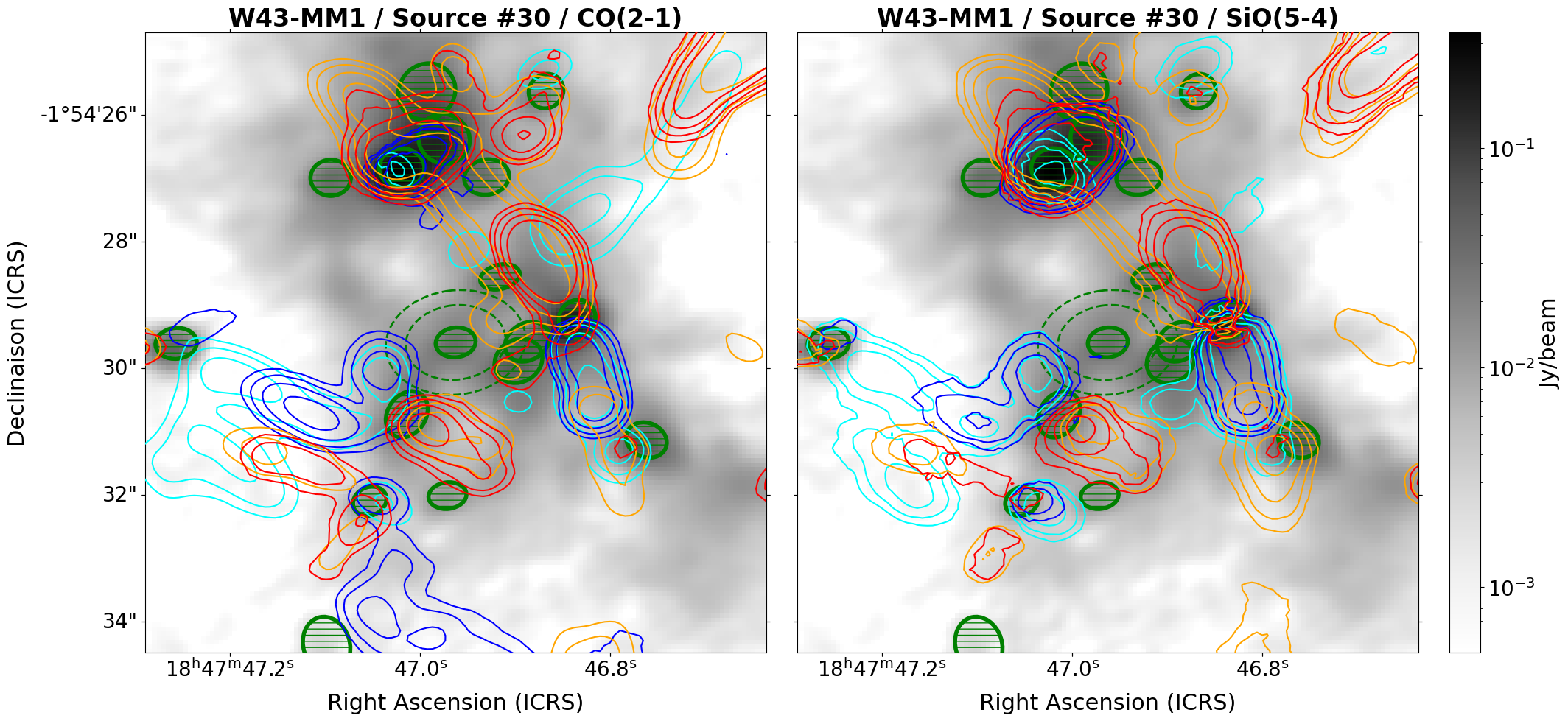}
            \end{minipage}
            \vfill
            \begin{minipage}[c]{\textwidth}
                \centering
                \includegraphics[width=0.7\textwidth]{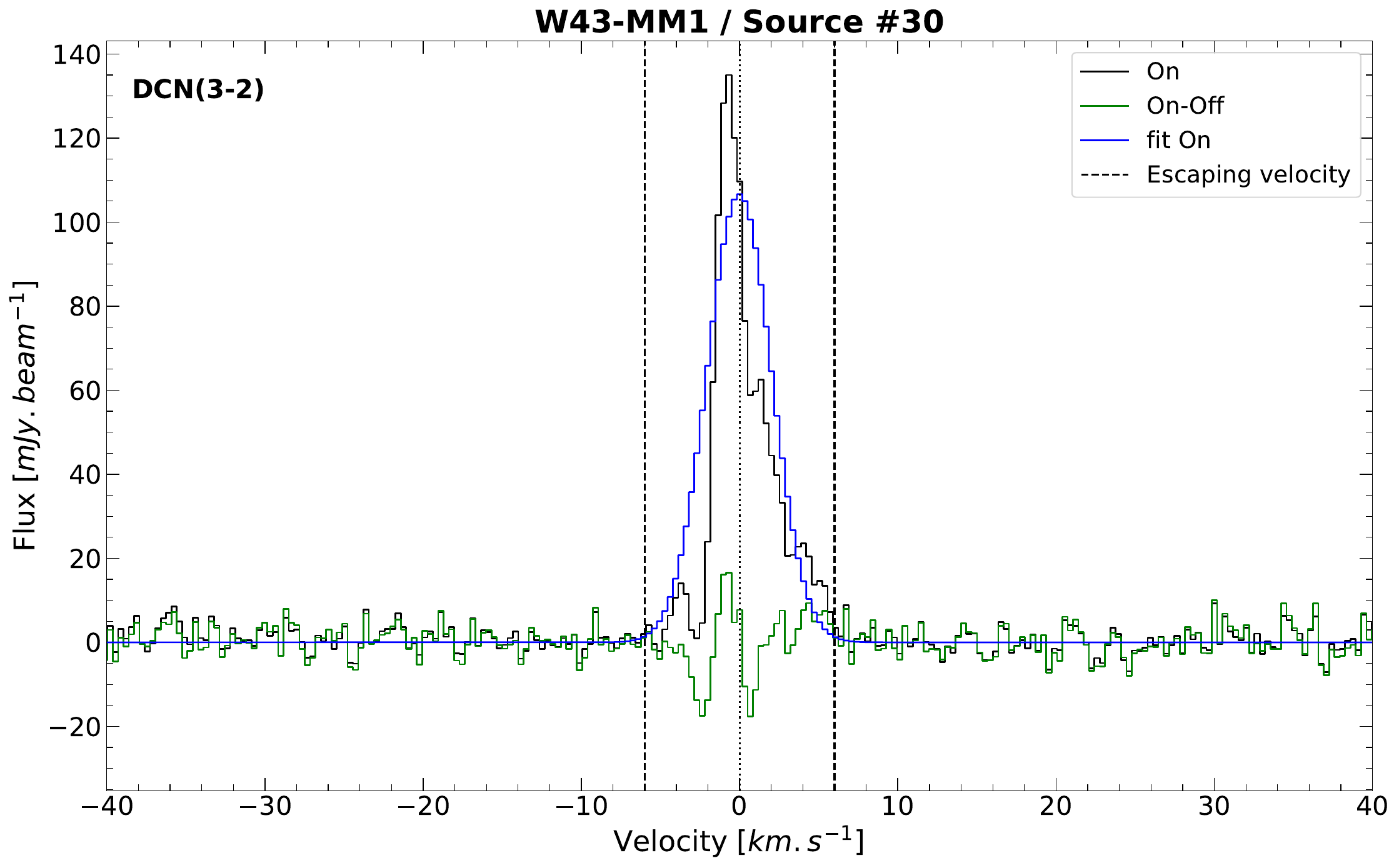}
            \end{minipage}
        \end{minipage}
        
    \caption{continued.}

\end{figure*}

\begin{figure*}\ContinuedFloat

    \centering
        \begin{minipage}[c]{0.49\textwidth}
            \centering
            \includegraphics[width=\textwidth]{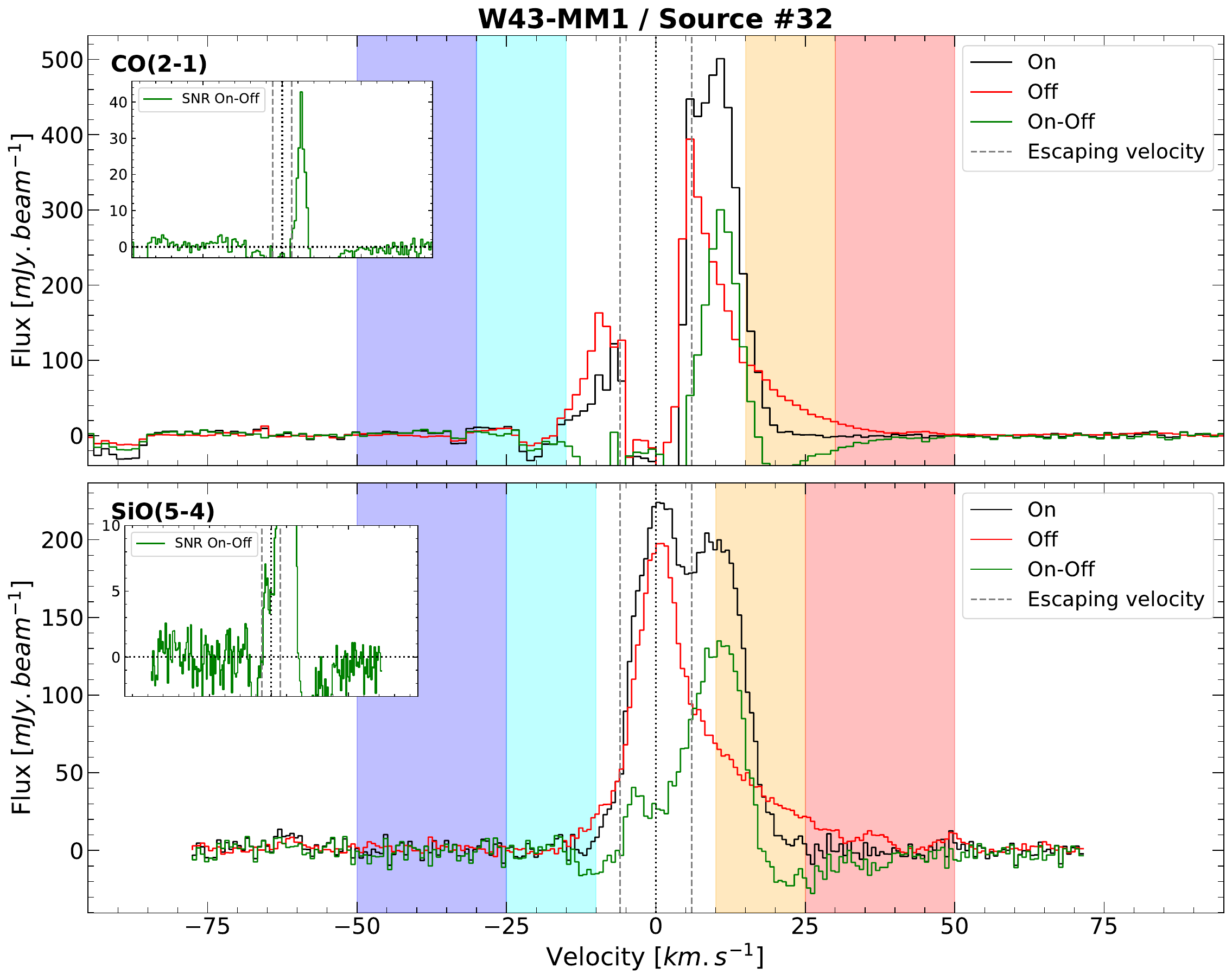}
        \end{minipage}
        \begin{minipage}[c]{0.49\textwidth}
            \centering
            \begin{minipage}[c]{\textwidth}
                \centering
                \includegraphics[width=0.9\textwidth]{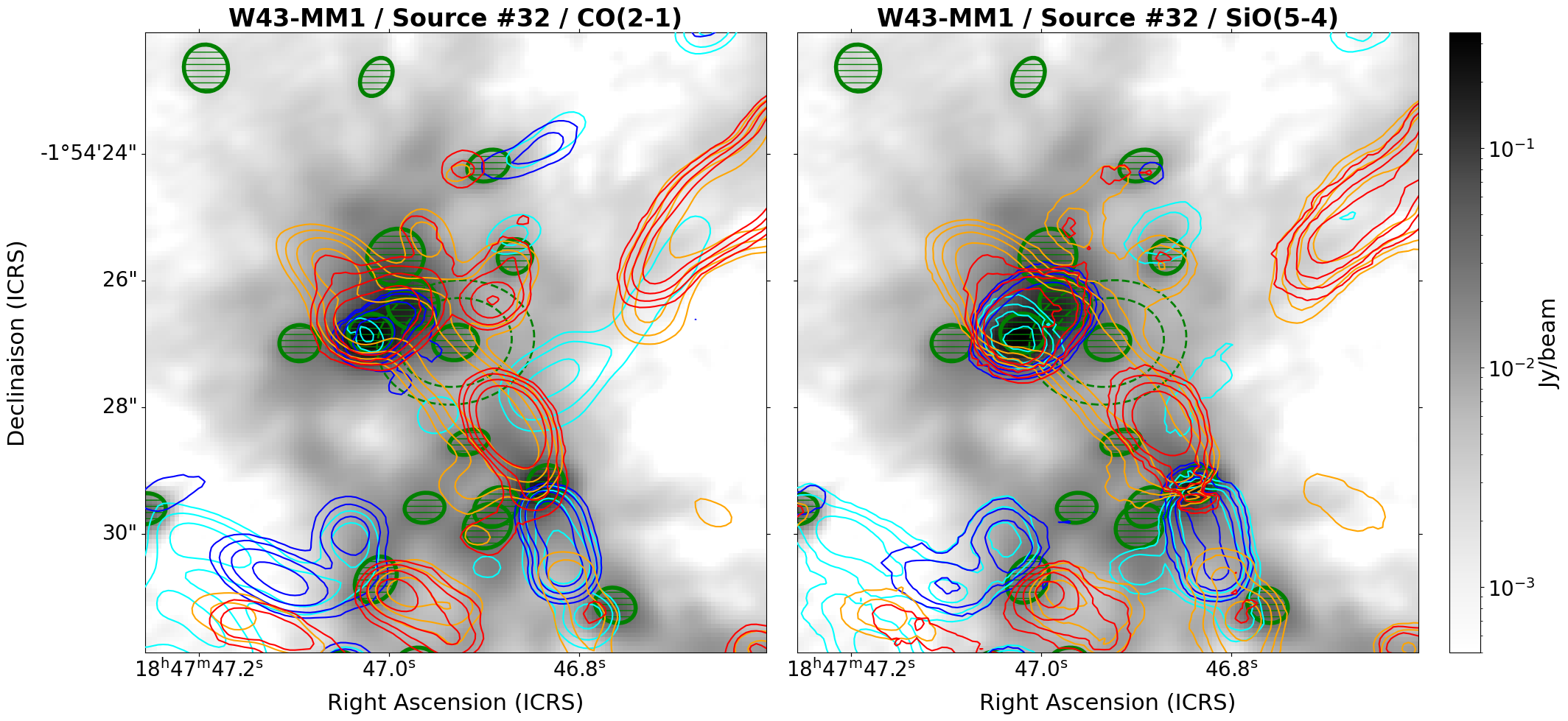}
            \end{minipage}
            \vfill
            \begin{minipage}[c]{\textwidth}
                \centering
                \includegraphics[width=0.7\textwidth]{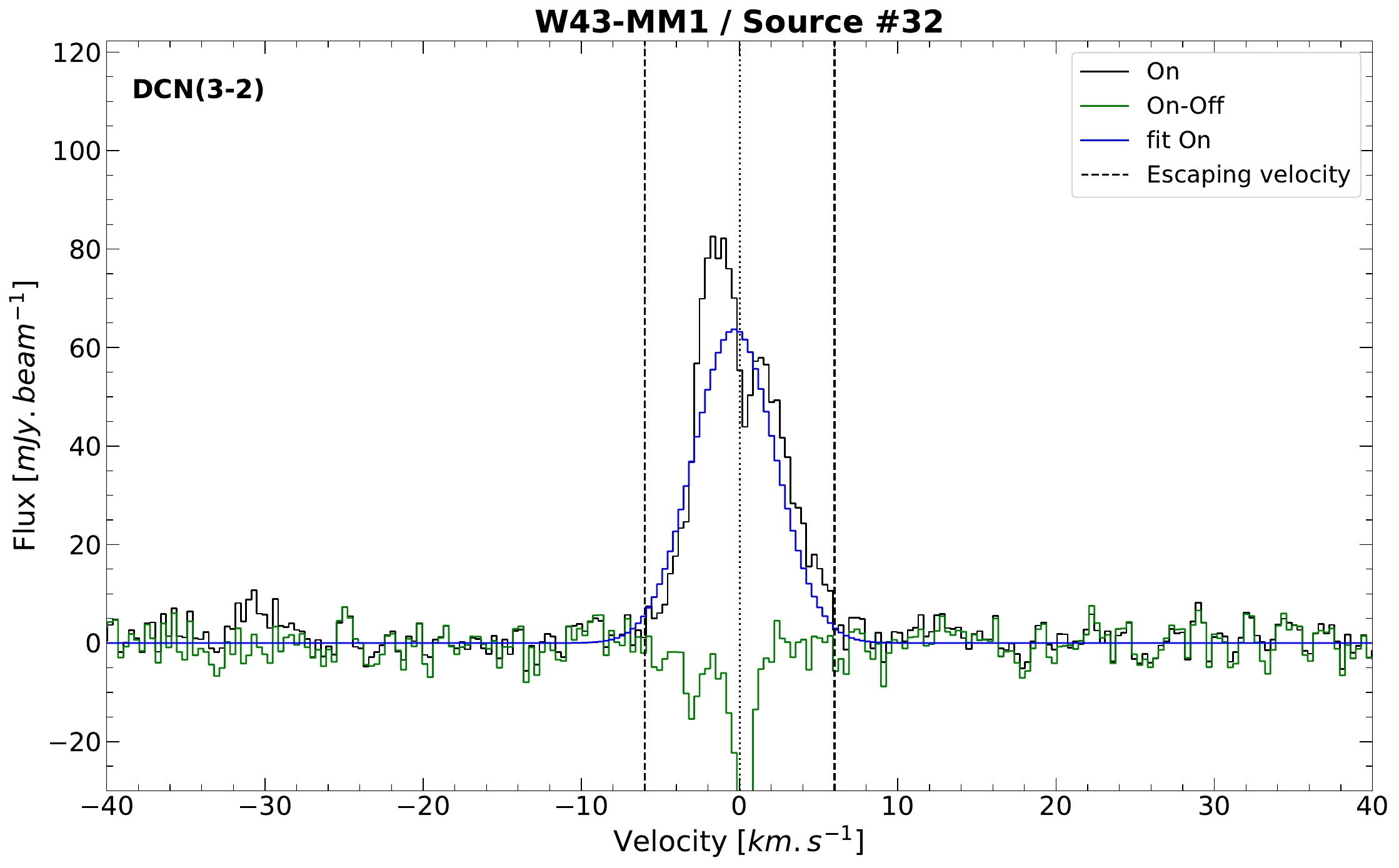}
            \end{minipage}
        \end{minipage}

    \caption{continued.}

\end{figure*}

%%%%%%%%%%%%%%%%%%%%%%% W43-MM2 %%%%%%%%%%%%%%%%%%%%%%%%%%%%%%%%%%%%%
\begin{figure*}
    \label{appendix:W43-MM2_MPSC_fig}
    \centering
        \begin{minipage}[c]{0.49\textwidth}
            \centering
            \includegraphics[width=\textwidth]{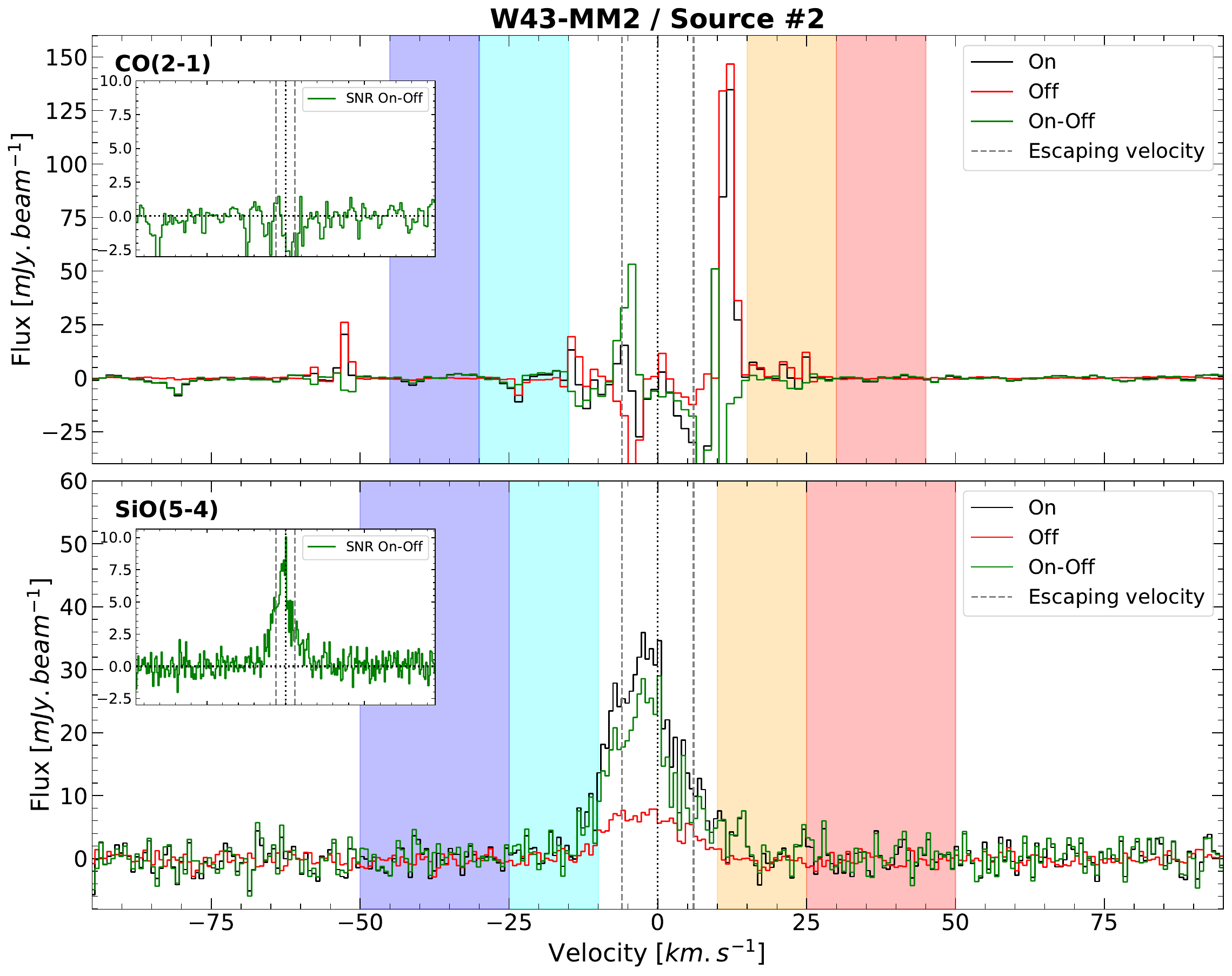}
        \end{minipage}
        \begin{minipage}[c]{0.49\textwidth}
            \centering
            \begin{minipage}[c]{\textwidth}
                \centering
                \includegraphics[width=0.9\textwidth]{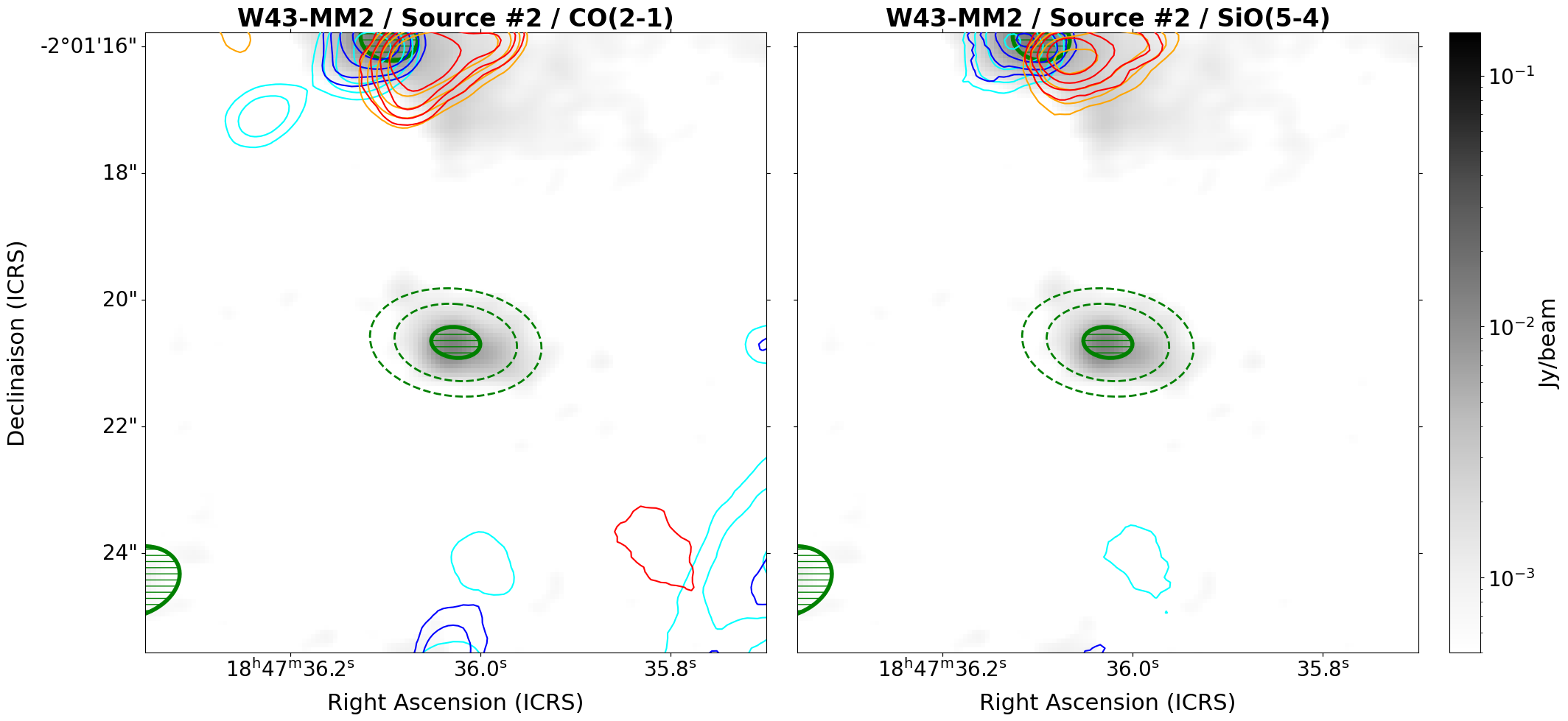}
            \end{minipage}
            \vfill
            \begin{minipage}[c]{\textwidth}
                \centering
                \includegraphics[width=0.7\textwidth]{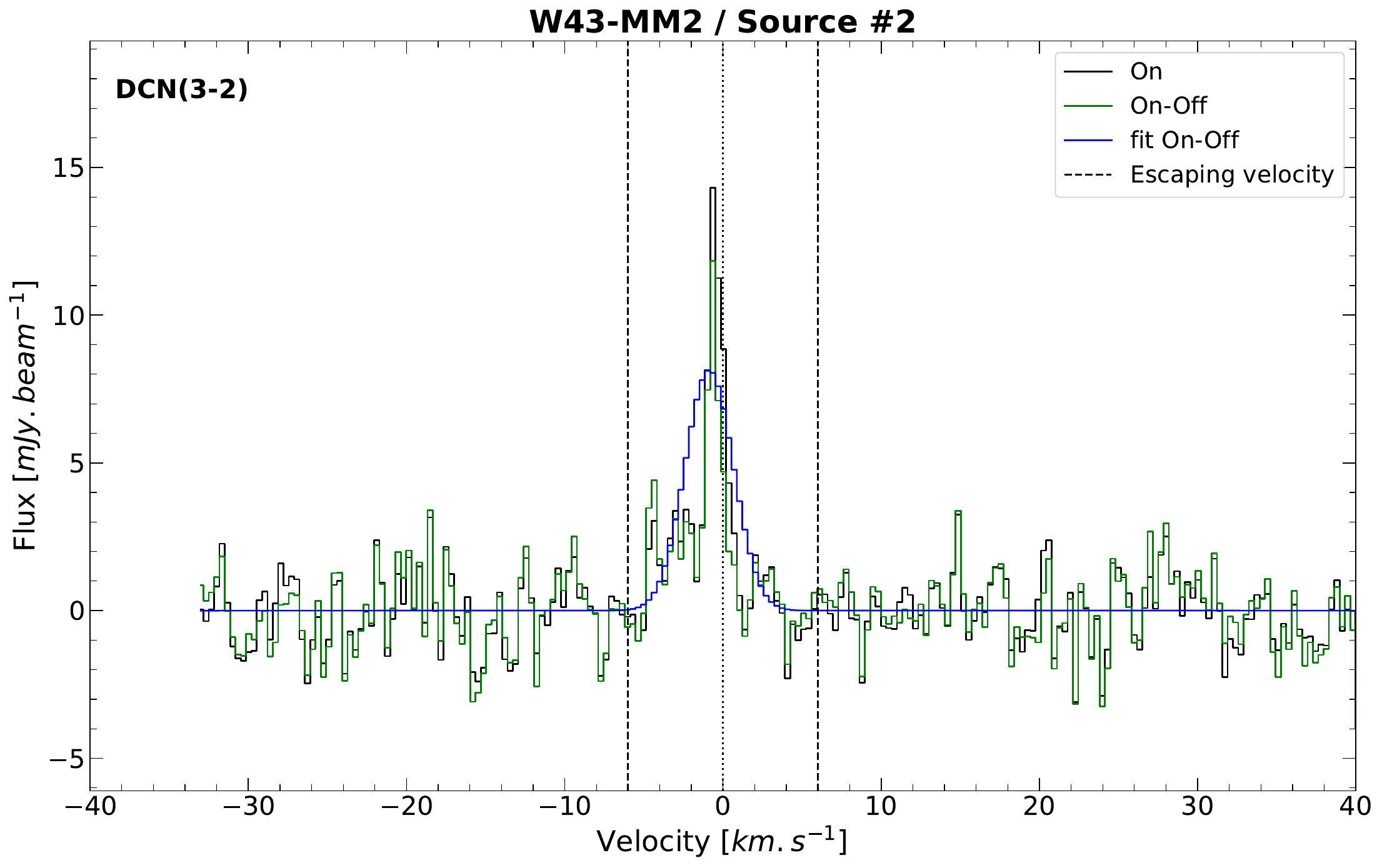}
            \end{minipage}
      \end{minipage}
    
    \vspace{0.2cm}
    
        \centering
                \begin{minipage}[c]{0.49\textwidth}
            \centering
            \includegraphics[width=\textwidth]{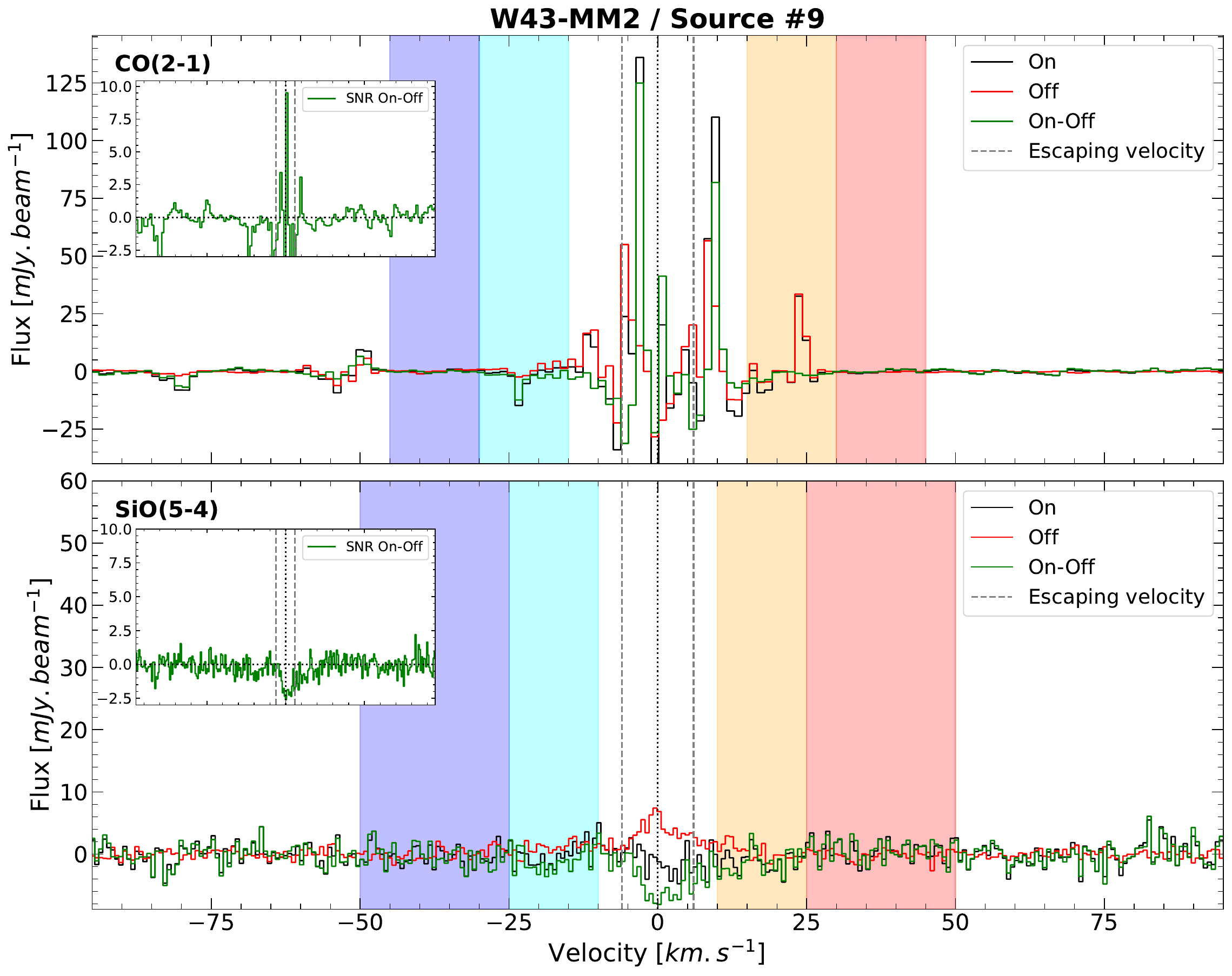}
        \end{minipage}
        \begin{minipage}[c]{0.49\textwidth}
            \centering
            \begin{minipage}[c]{\textwidth}
                \centering
                \includegraphics[width=0.9\textwidth]{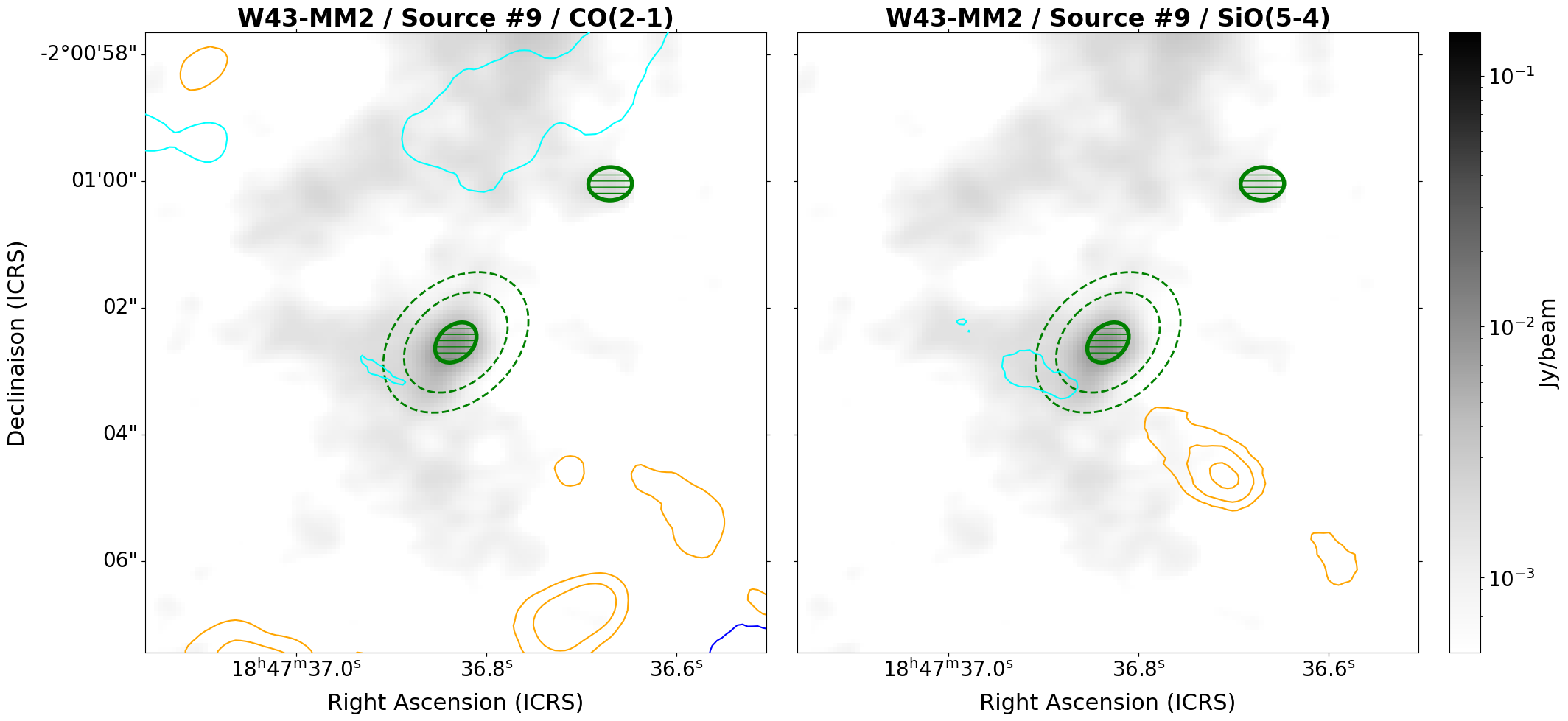}
            \end{minipage}
            \vfill
            \begin{minipage}[c]{\textwidth}
                \centering
                \includegraphics[width=0.7\textwidth]{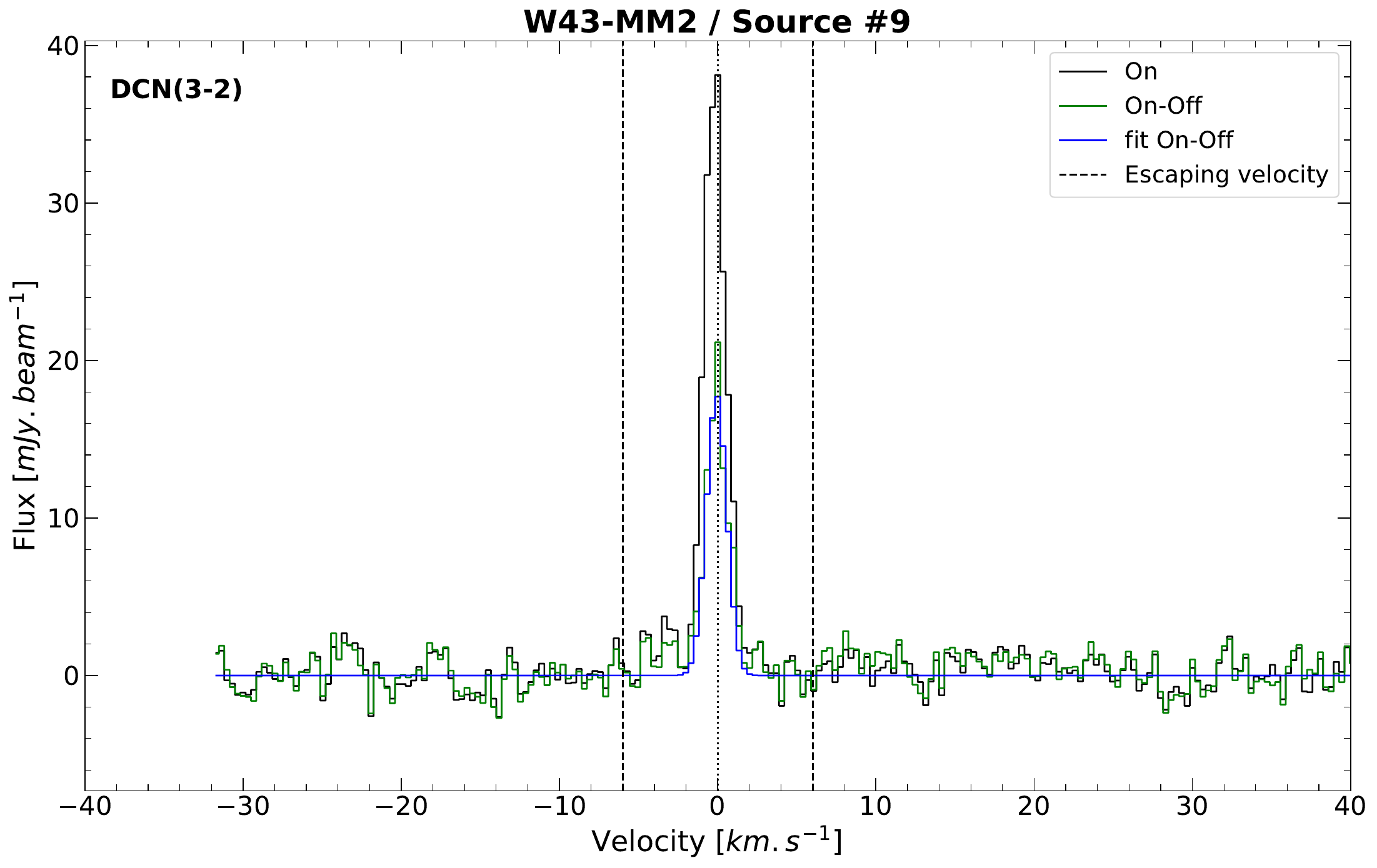}
            \end{minipage}
        \end{minipage}
    
    \vspace{0.2cm}
    
        \centering
        \begin{minipage}[c]{0.49\textwidth}
            \centering
            \includegraphics[width=\textwidth]{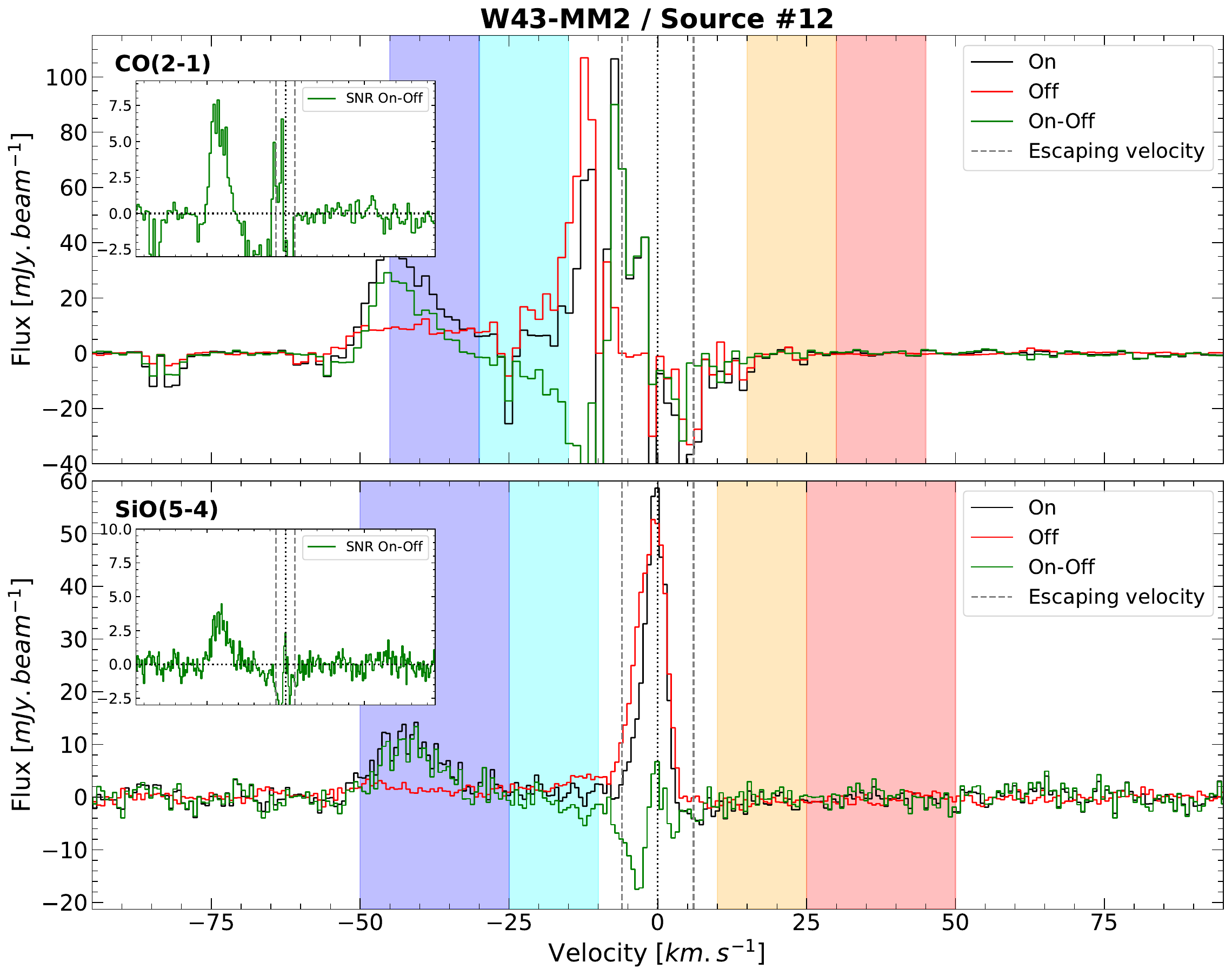}
        \end{minipage}
        \begin{minipage}[c]{0.49\textwidth}
            \centering
            \begin{minipage}[c]{\textwidth}
                \centering
                \includegraphics[width=0.9\textwidth]{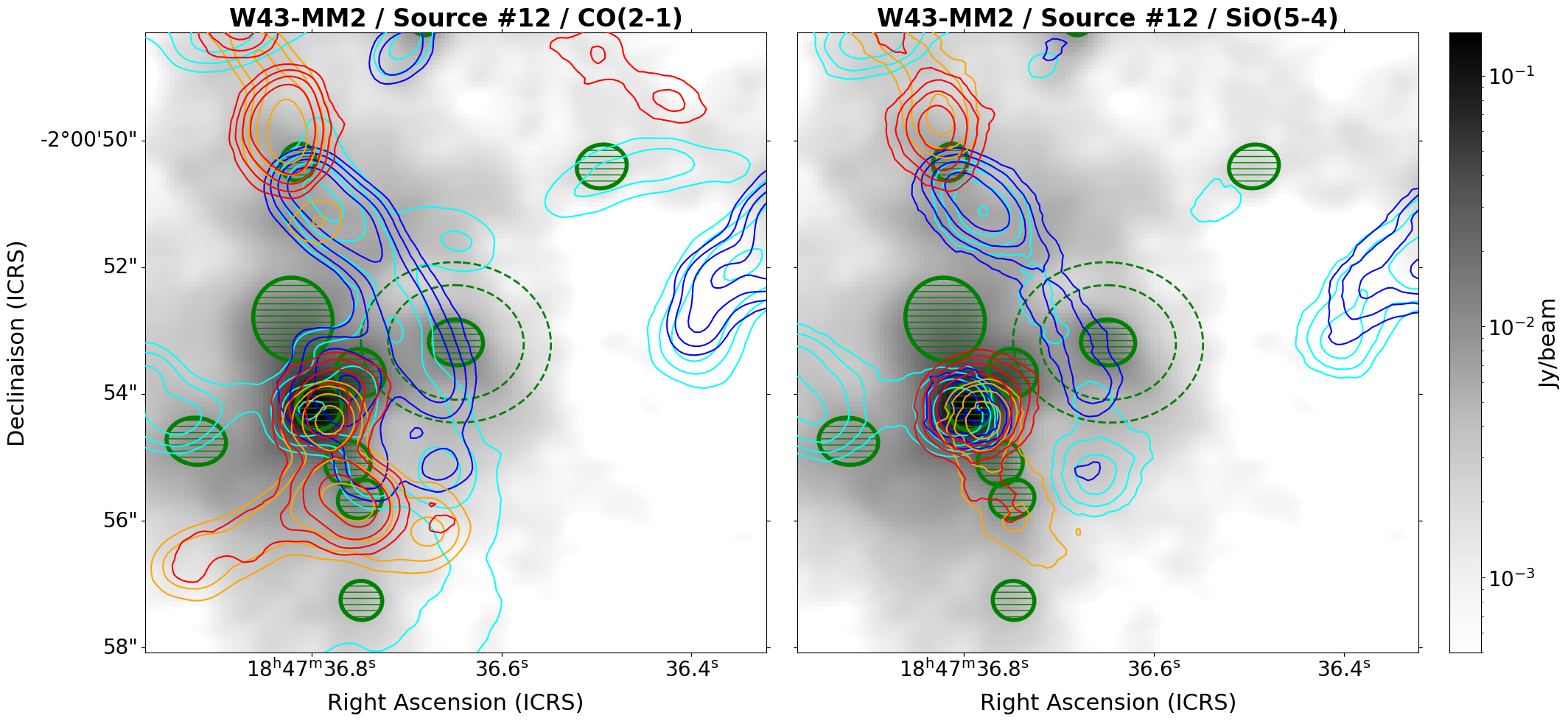}
            \end{minipage}
            \vfill
            \begin{minipage}[c]{\textwidth}
                \centering
                \includegraphics[width=0.7\textwidth]{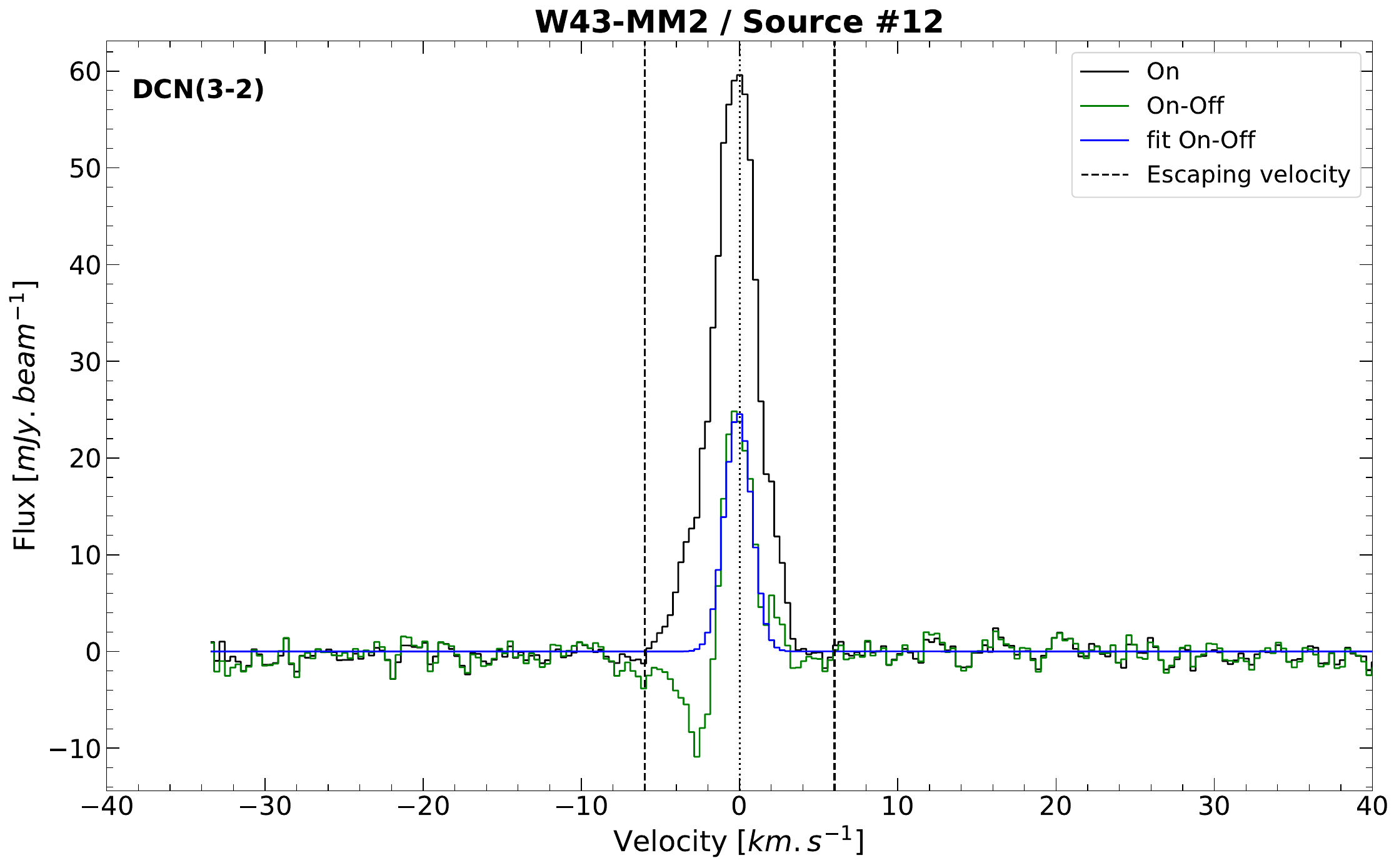}
            \end{minipage}
        \end{minipage}

   % \vskip -0.3cm
    \caption{CO and SiO spectra (left) and molecular outflow maps (top right) of the high-mass PSC candidates of the W43-MM2 region. CO contours are 10, 20, 40, and 80 in units of $\sigma$, with $\sigma$ = $18.9$, $11.0$, $18.1$, $6.4$ \mJybeamkms for cyan, blue, orange and red contours respectively. SiO contours are 10, 20, 40, and 80 in units of $\sigma$, with $\sigma$ = $4.4$, $5.7$, $4.5$, $5.7$ \mJybeamkms for cyan, blue, orange and red contours respectively. DCN spectra and fits (bottom right) of the high-mass PSC candidates of the W43-MM2 region.}

\end{figure*}

%%%%%%%%%%%%%%%%%%%%%%% W43-MM3 %%%%%%%%%%%%%%%%%%%%%%%%%%%%%%%%%%%%%
\begin{figure*}
    \label{appendix:W43-MM3_MPSC_fig}
    \centering
        \begin{minipage}[c]{0.49\textwidth}
            \centering
            \includegraphics[width=\textwidth]{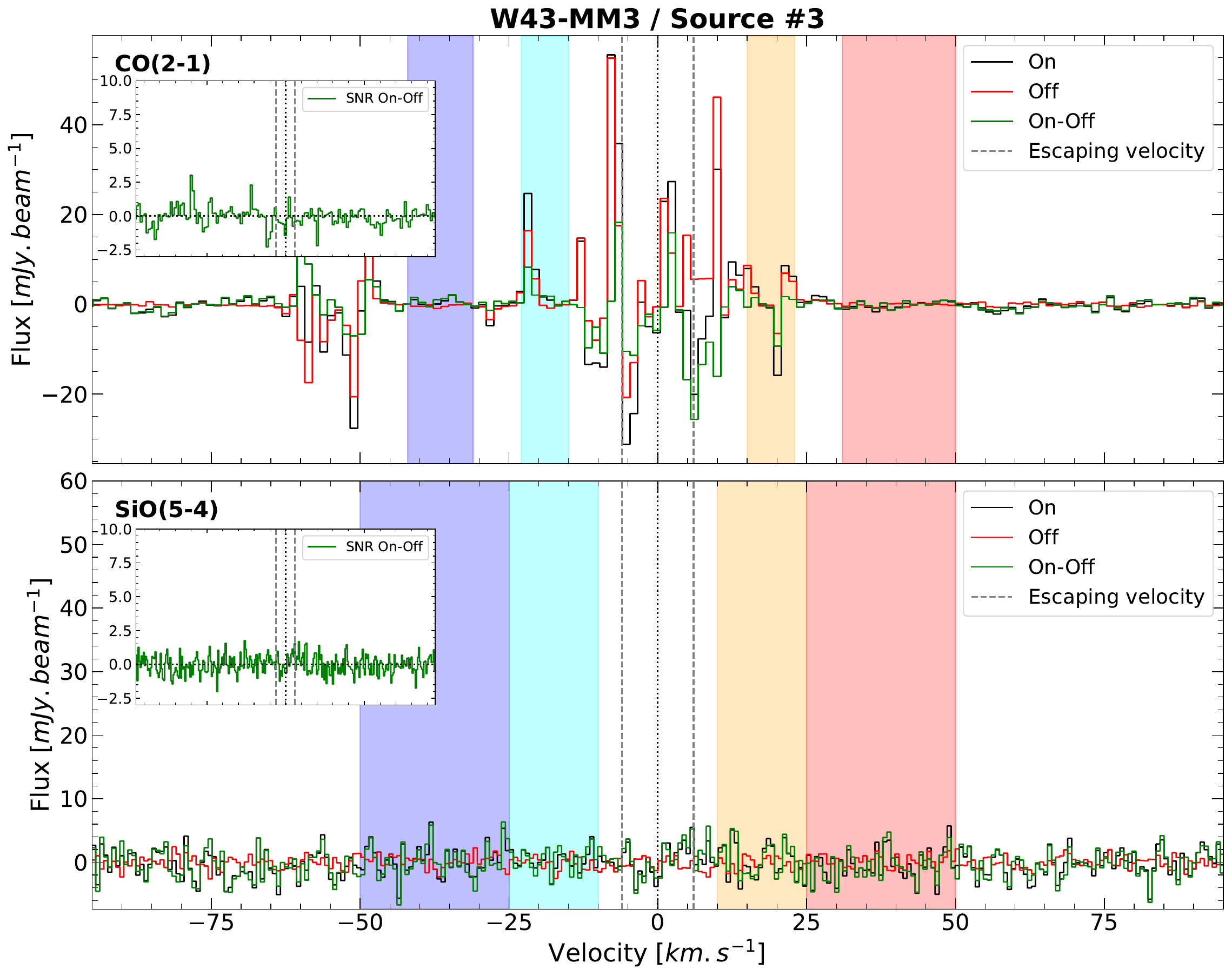}
        \end{minipage}
        \begin{minipage}[c]{0.49\textwidth}
            \centering
            \begin{minipage}[c]{\textwidth}
                \centering
                \includegraphics[width=0.9\textwidth]{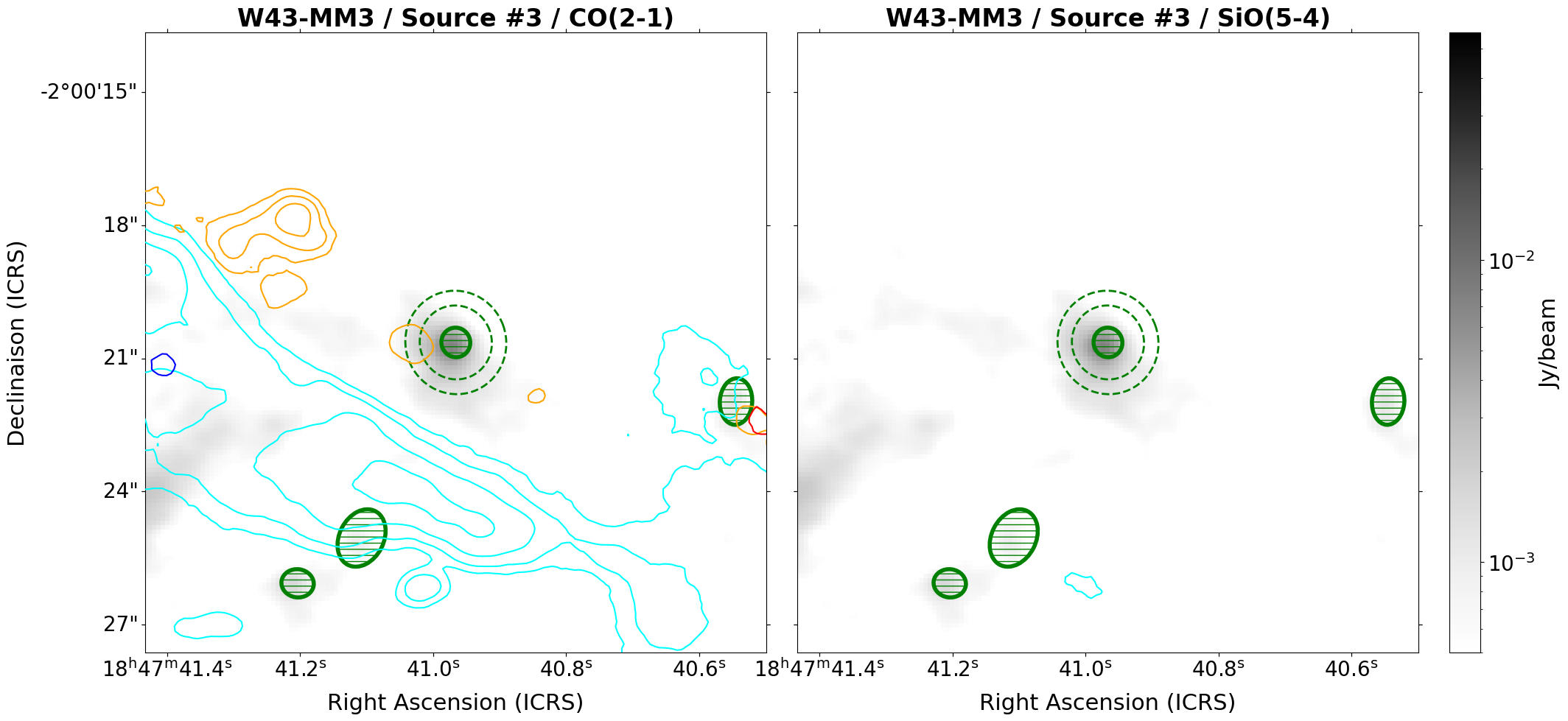}
            \end{minipage}
            \vfill
            \begin{minipage}[c]{\textwidth}
                \centering
                \includegraphics[width=0.7\textwidth]{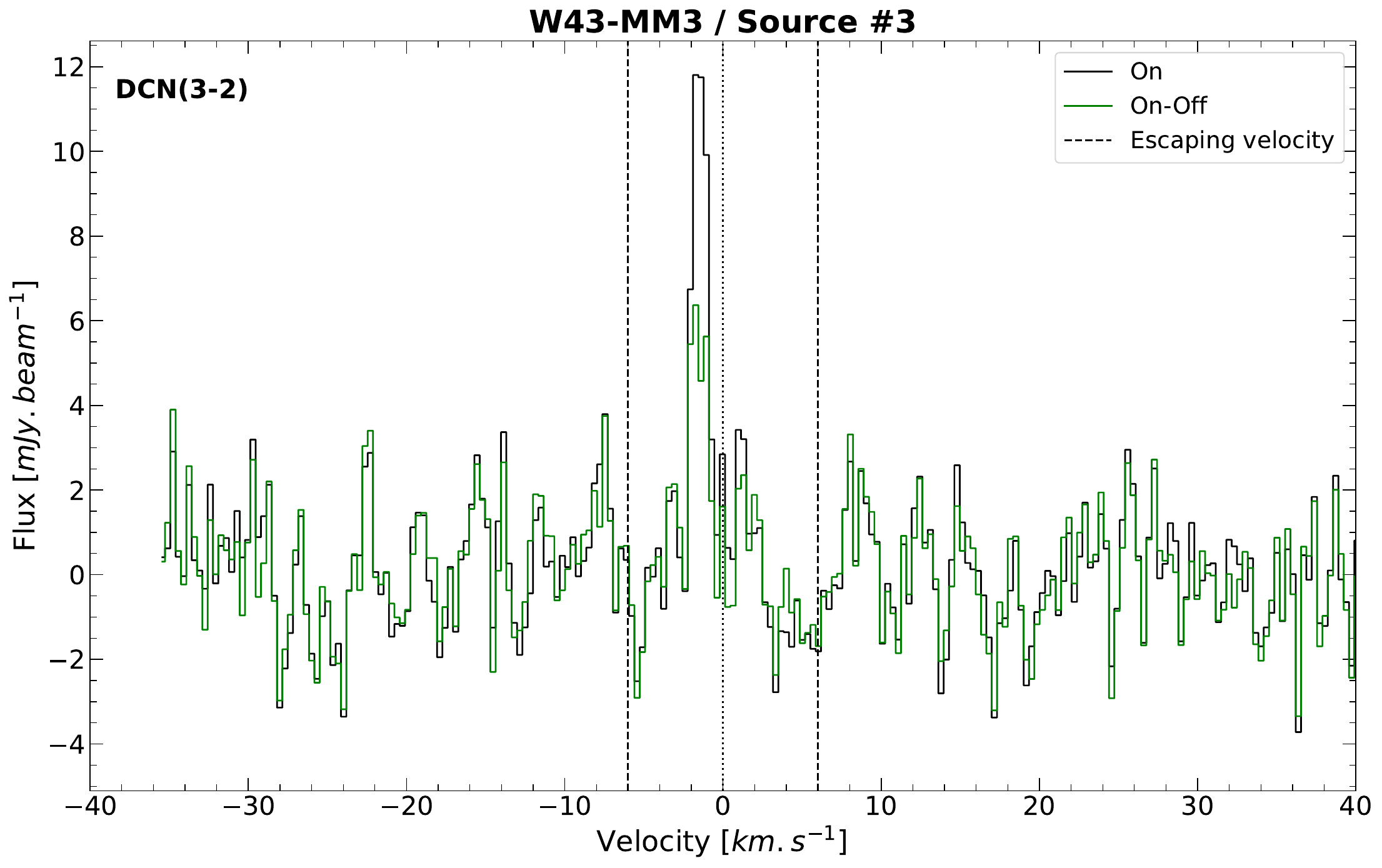}
            \end{minipage}
        \end{minipage}

   % \vskip -0.3cm
    \caption{CO and SiO spectra (left) and molecular outflow maps (top right) of the high-mass PSC candidate of the W43-MM3 region. CO contours are 5, 10, 20, and 40 in units of $\sigma$, with $\sigma$ = $14.3$, $12.6$, $10.8$, $9.0$ \mJybeamkms for cyan, blue, orange and red contours respectively. SiO contours are 5, 10, 20, and 40 in units of $\sigma$, with $\sigma$ = $5.7$, $7.2$, $5.5$, $7.2$ \mJybeamkms for cyan, blue, orange and red contours respectively. DCN spectra and fit (bottom right) of the high-mass PSC candidate of the W43-MM3 region.}

\end{figure*}

%%%%%%%%%%%%%%%%%%%%%%% W51-E %%%%%%%%%%%%%%%%%%%%%%%%%%%%%%%%%%%%%
\begin{figure*}
    \label{appendix:W51_E_MPSC_fig}
    \centering
        \begin{minipage}[c]{0.49\textwidth}
            \centering
            \includegraphics[width=\textwidth]{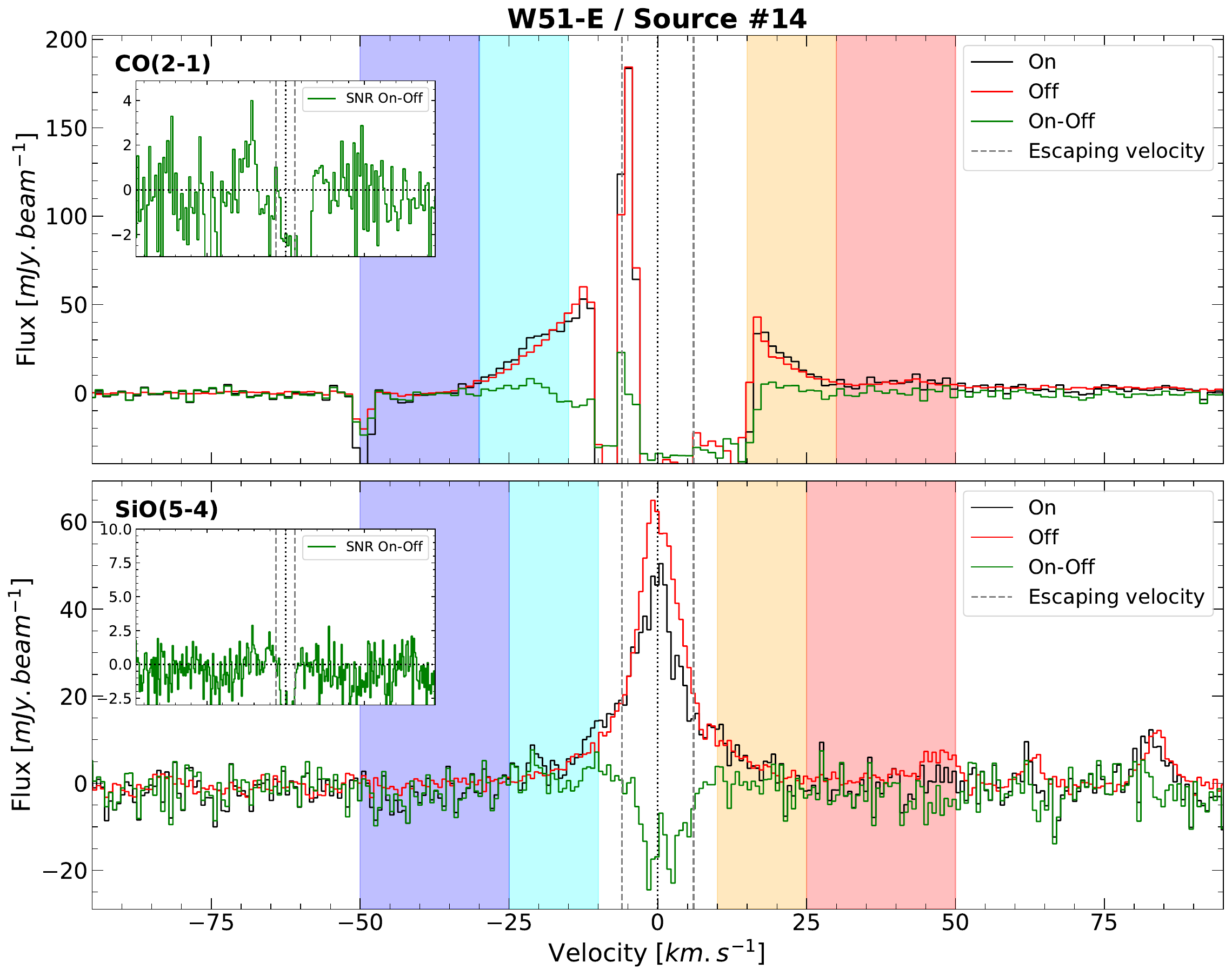}
        \end{minipage}
        \begin{minipage}[c]{0.49\textwidth}
            \centering
            \begin{minipage}[c]{\textwidth}
                \centering
                \includegraphics[width=0.9\textwidth]{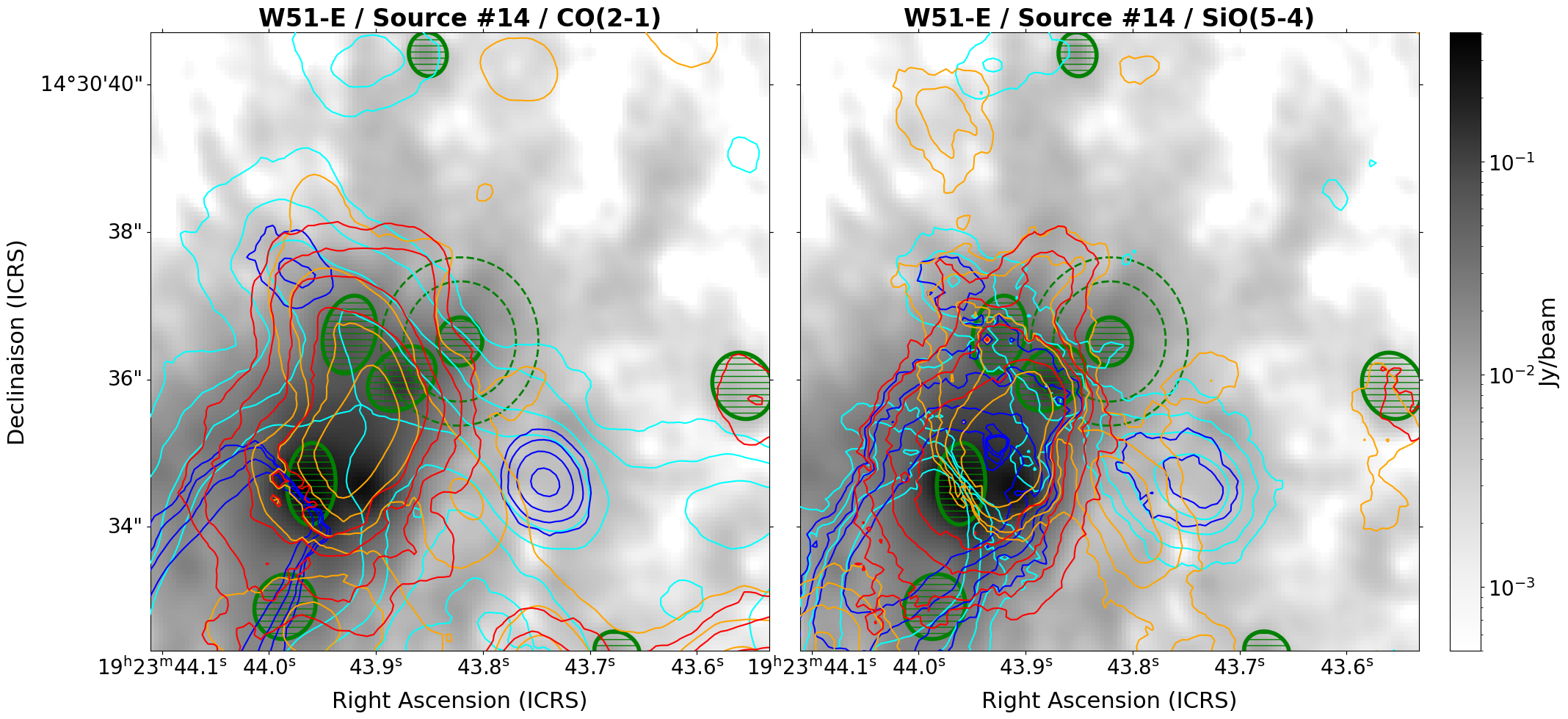}
            \end{minipage}
            \vfill
            \begin{minipage}[c]{\textwidth}
                \centering
                \includegraphics[width=0.7\textwidth]{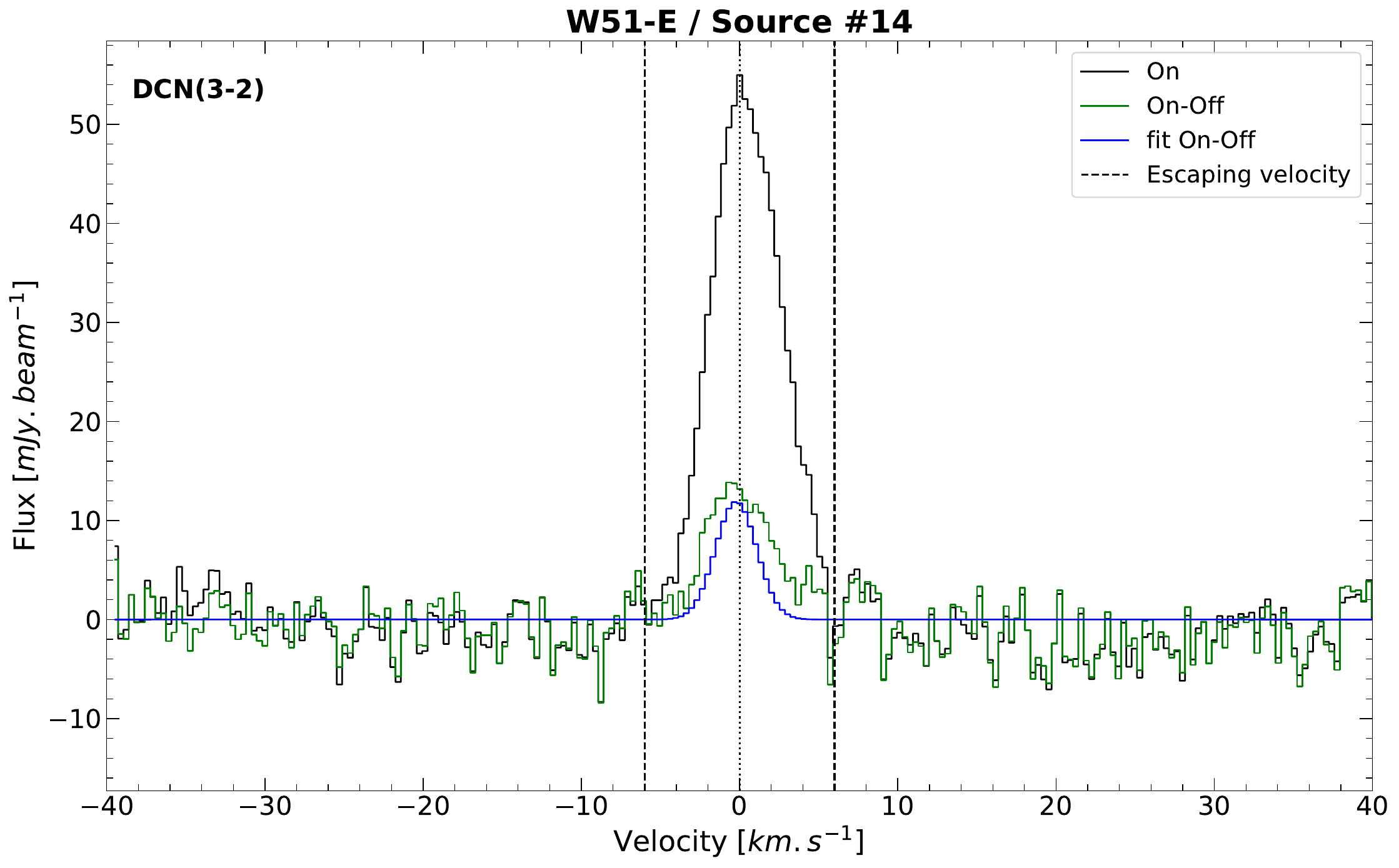}
            \end{minipage}
      \end{minipage}
    
    \vspace{0.2cm}
    
    \centering
        \begin{minipage}[c]{0.49\textwidth}
            \centering
            \includegraphics[width=\textwidth]{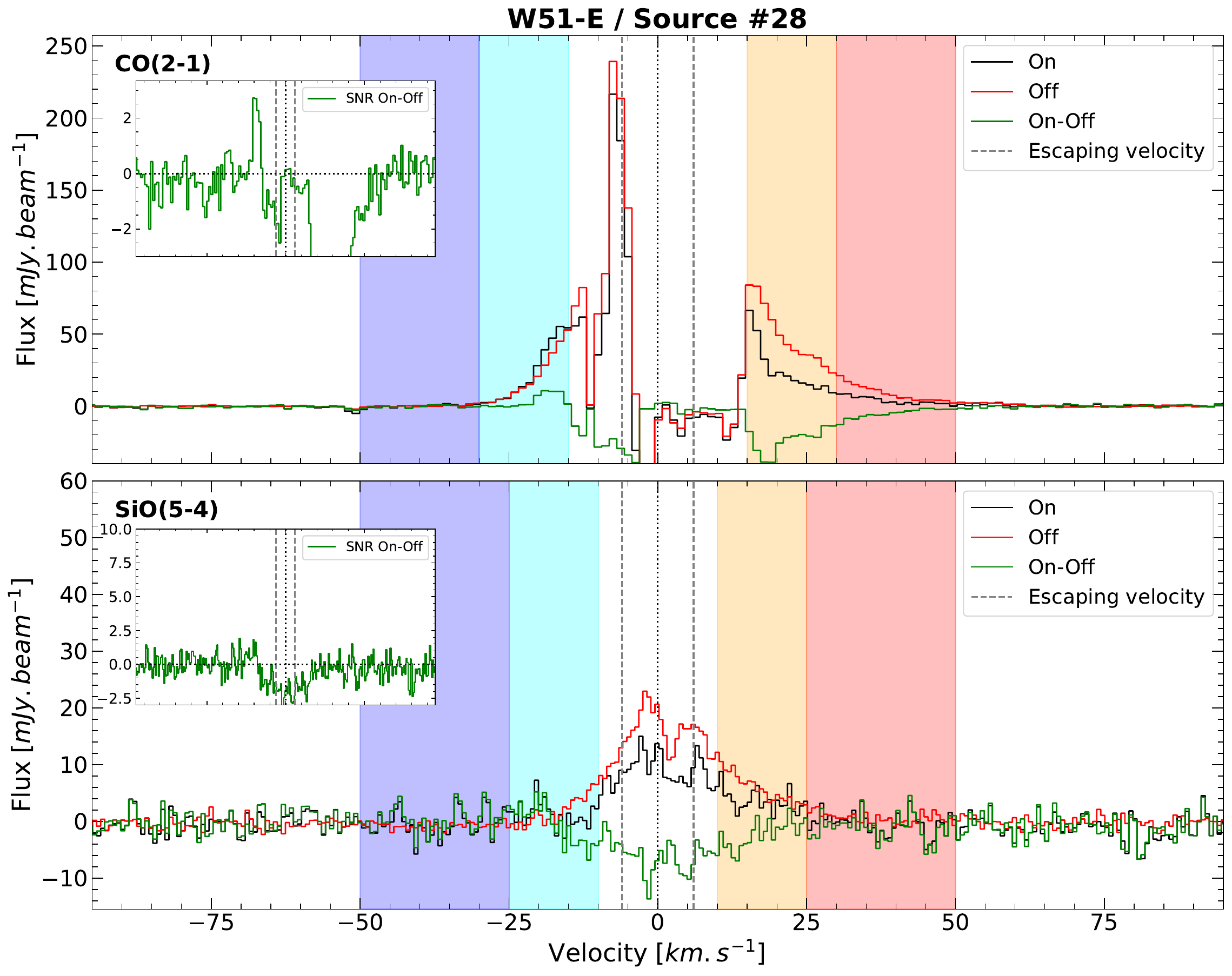}
        \end{minipage}
        \begin{minipage}[c]{0.49\textwidth}
            \centering
            \begin{minipage}[c]{\textwidth}
                \centering
                \includegraphics[width=0.9\textwidth]{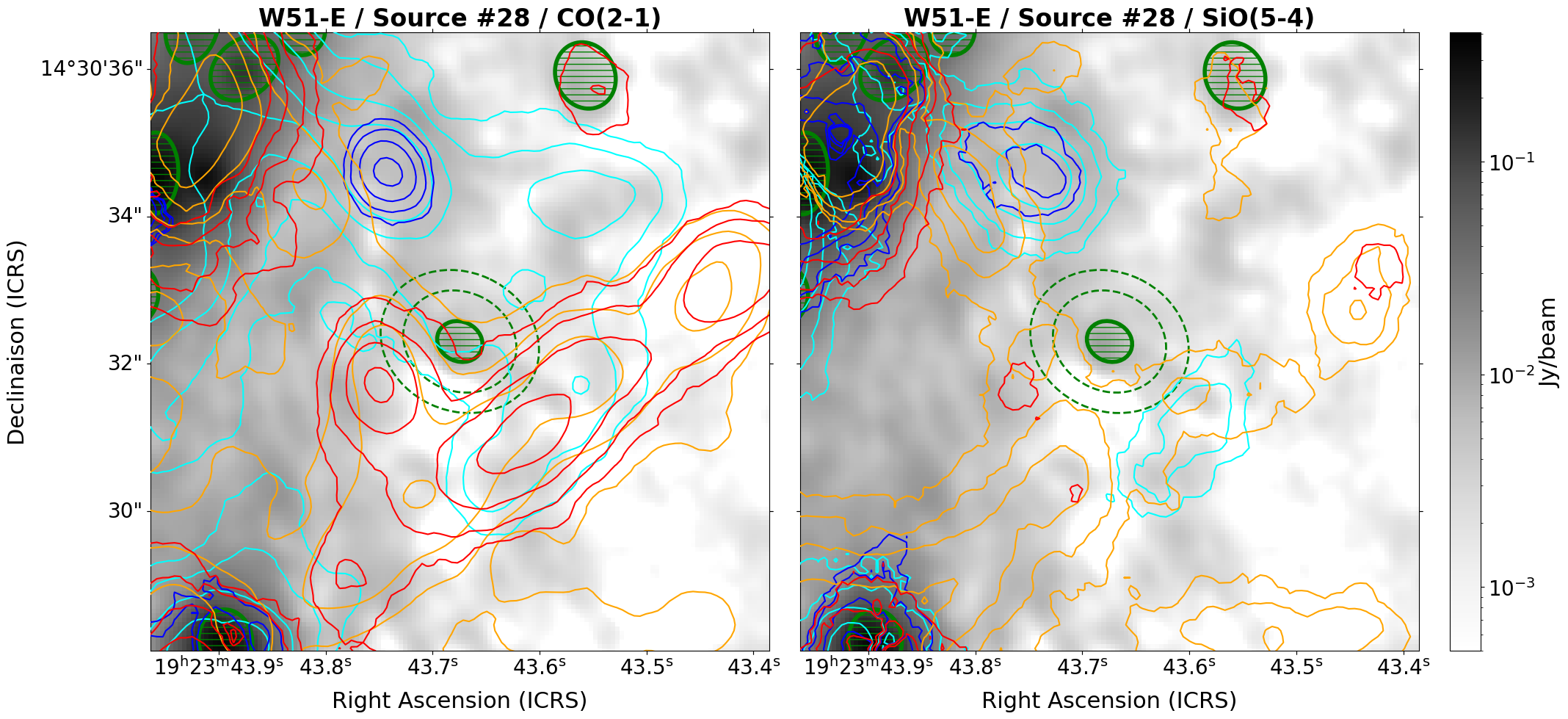}
            \end{minipage}
            \vfill
            \begin{minipage}[c]{\textwidth}
                \centering
                \includegraphics[width=0.7\textwidth]{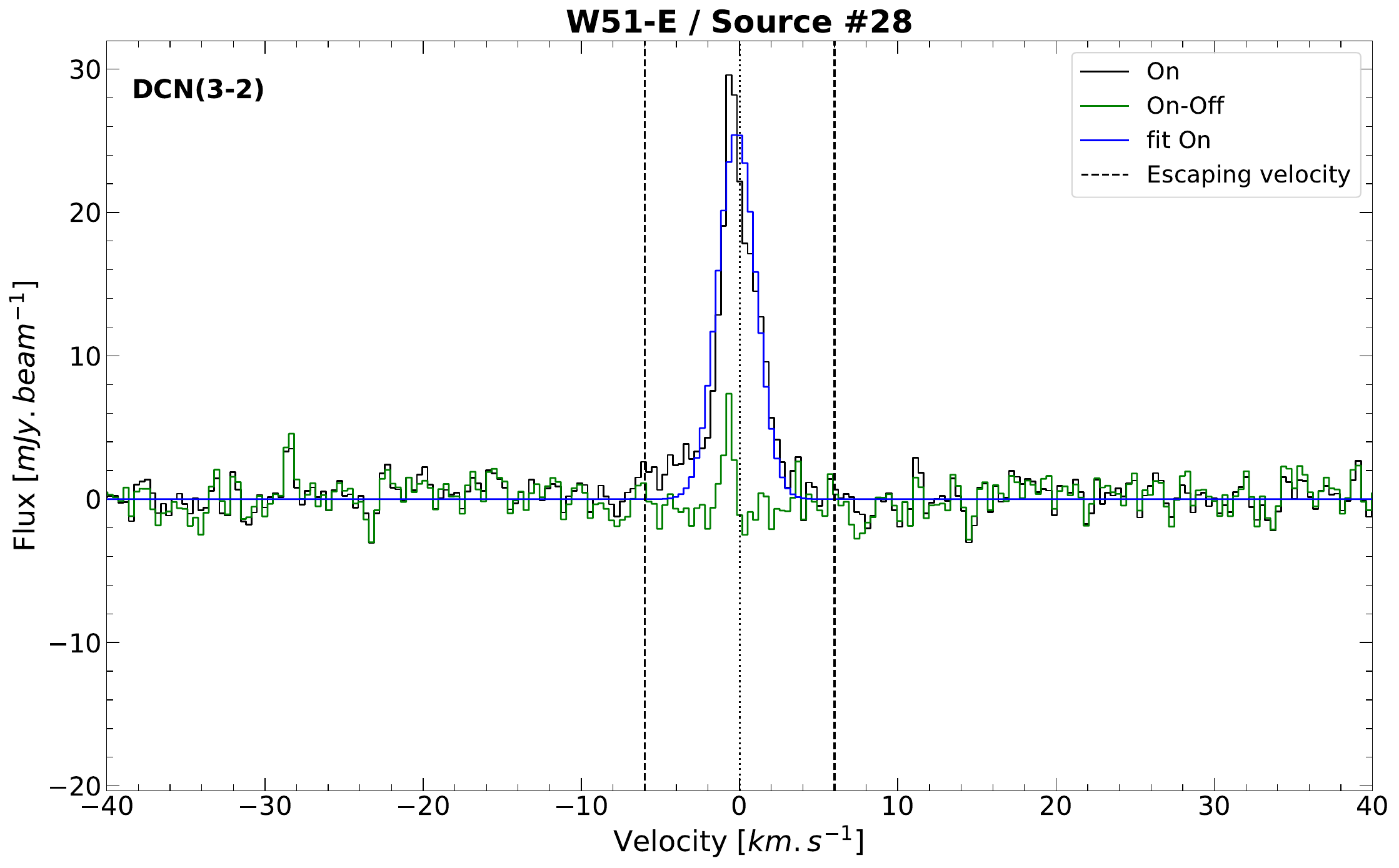}
            \end{minipage}
      \end{minipage}    
      
    \vspace{0.2cm}

        \centering
        \begin{minipage}[c]{0.49\textwidth}
            \centering
            \includegraphics[width=\textwidth]{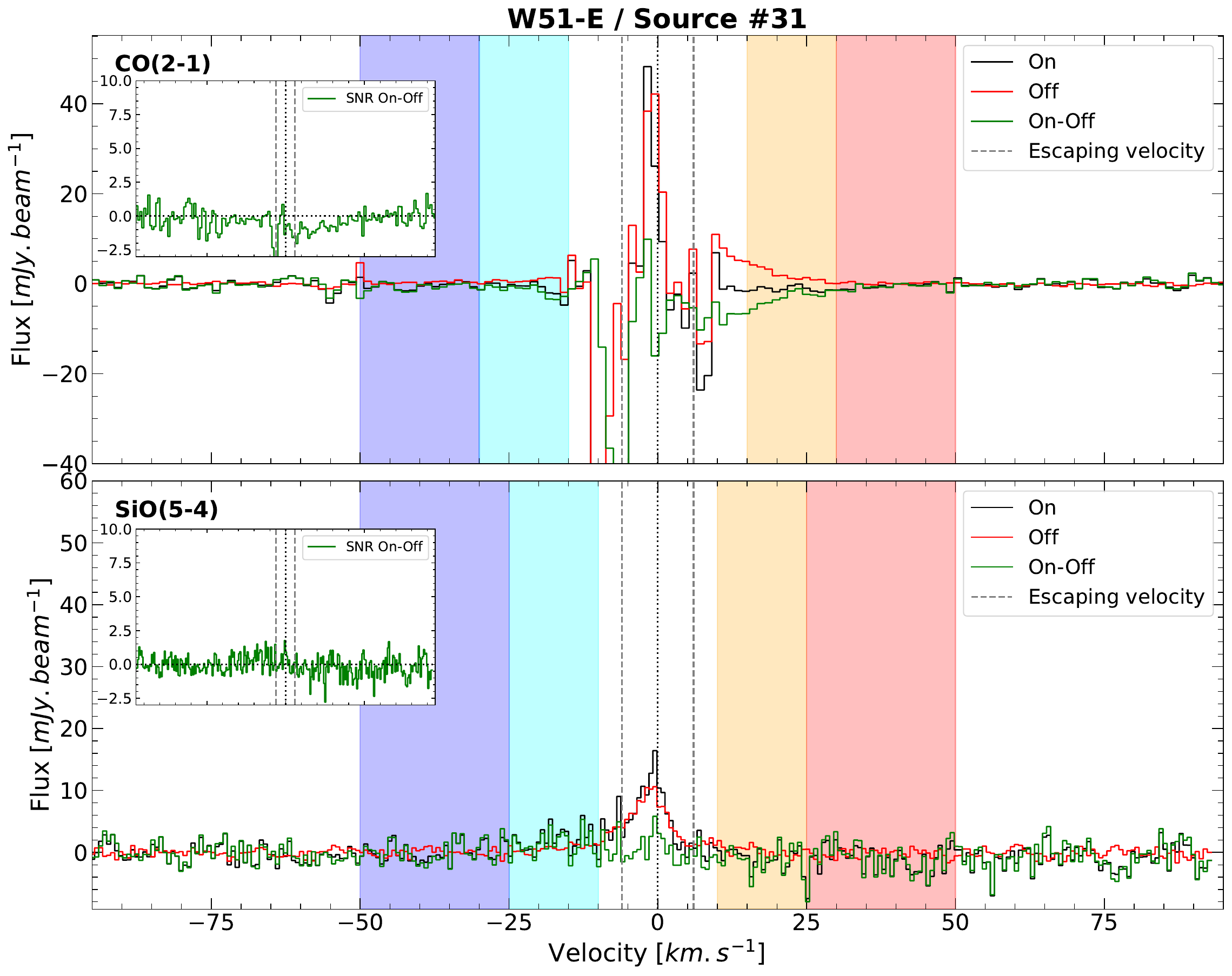}
        \end{minipage}
        \begin{minipage}[c]{0.49\textwidth}
            \centering
            \begin{minipage}[c]{\textwidth}
                \centering
                \includegraphics[width=0.9\textwidth]{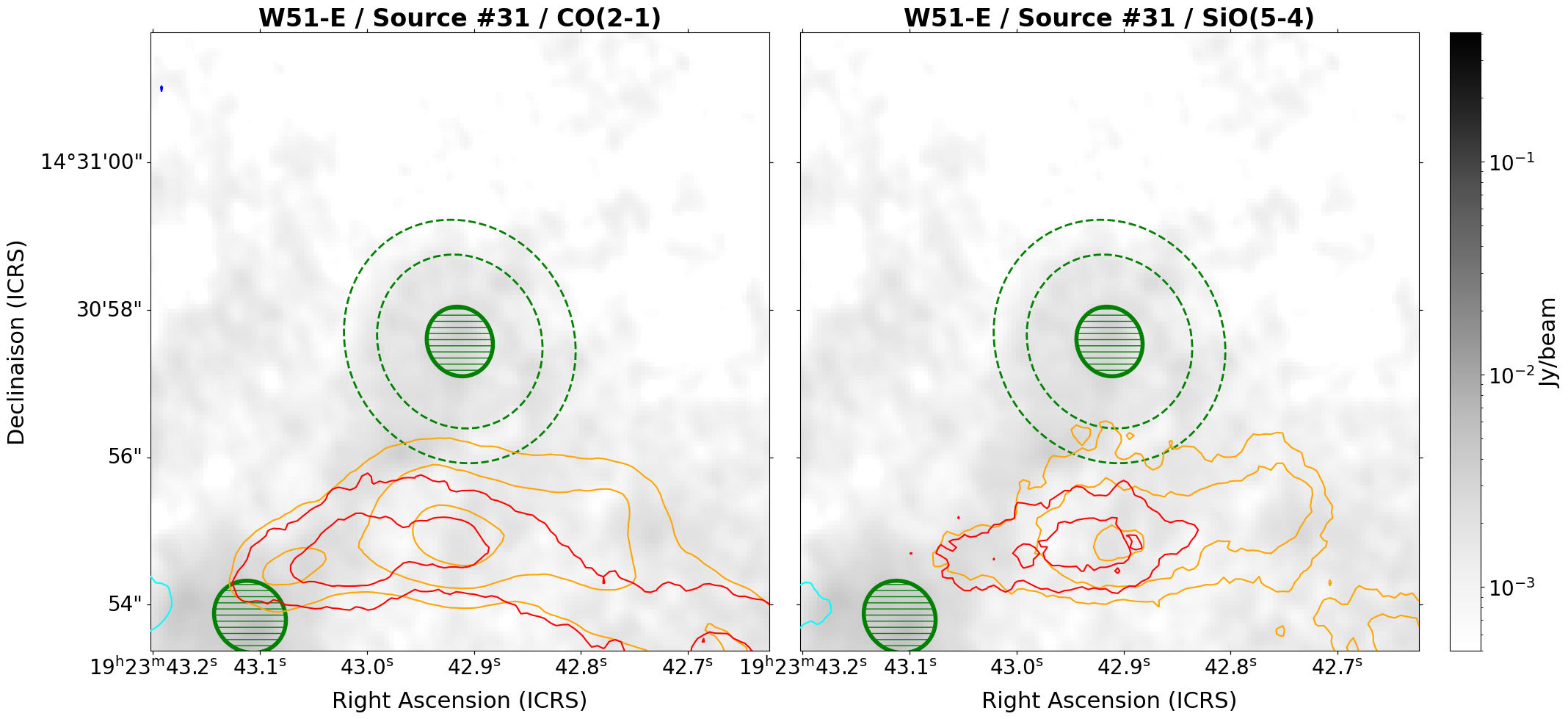}
            \end{minipage}
            \vfill
            \begin{minipage}[c]{\textwidth}
                \centering
                \includegraphics[width=0.7\textwidth]{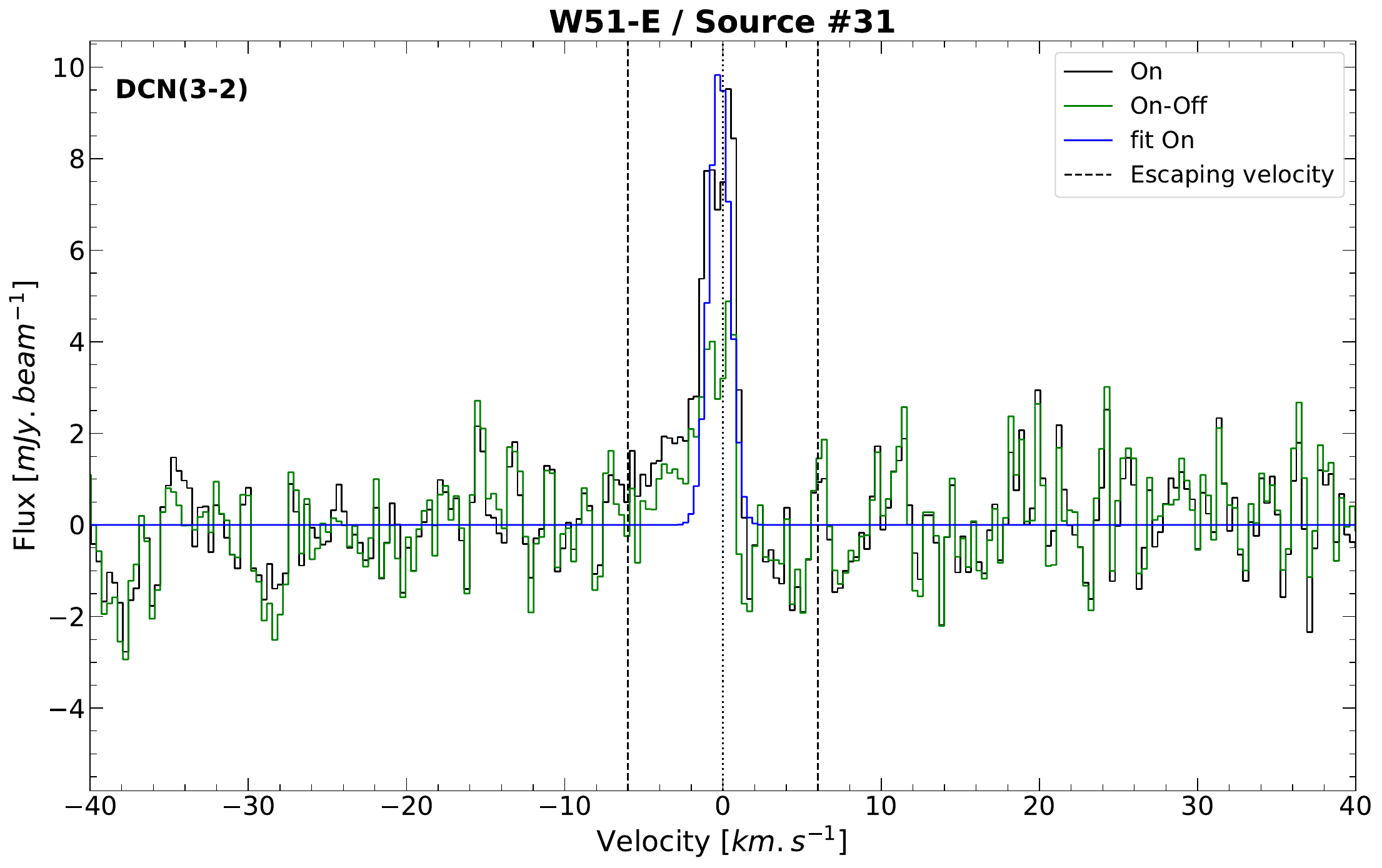}
            \end{minipage}
      \end{minipage}
    
    \vspace{0.2cm}

   % \vskip -0.3cm
    \caption{CO and SiO spectra (left) and molecular outflow maps (top right) of the high-mass PSC candidates of the W51-E region. CO contours are 10, 20, 40 and 80 in units of $\sigma$, with $\sigma$ = $13.3$, $7.5$, $25.7$, $10.2$ \mJybeamkms for cyan, blue, orange and red contours respectively. SiO contours are 10, 20, 40 and 80 in units of $\sigma$, with $\sigma$ = $4.7$, $6.1$, $4.6$, $6.0$ \mJybeamkms for cyan, blue, orange and red contours respectively. DCN spectra and fits (bottom right) of the high-mass PSC candidates of the W51-IRS2 region.}

\end{figure*}

%%%%%%%%%%%%%%%%%%%%%%% W51-IRS2 %%%%%%%%%%%%%%%%%%%%%%%%%%%%%%%%%%%%%
\begin{figure*}
    \label{appendix:W51-IRS2_MPSC_fig}
    \centering
        \begin{minipage}[c]{0.49\textwidth}
            \centering
            \includegraphics[width=\textwidth]{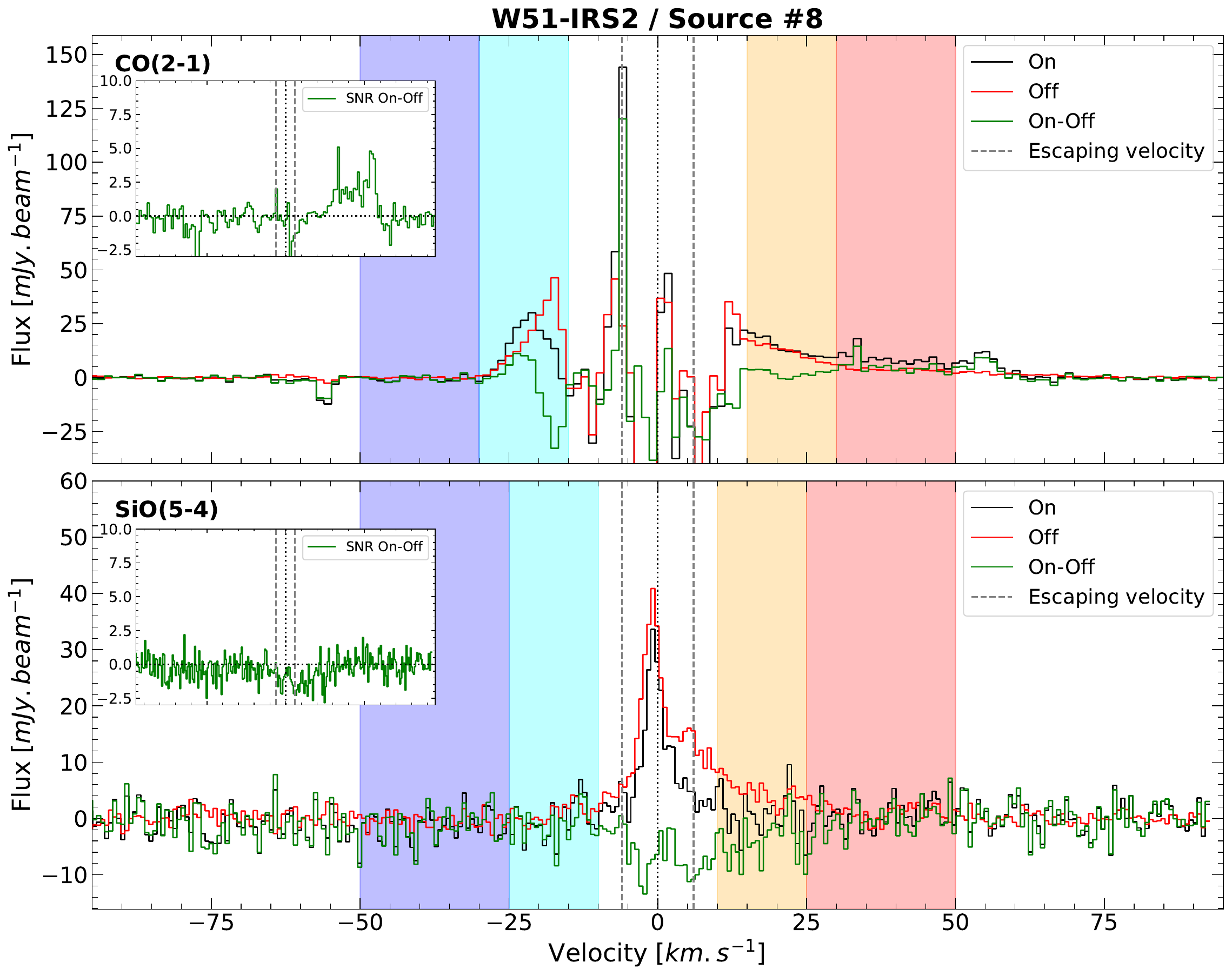}
        \end{minipage}
        \begin{minipage}[c]{0.49\textwidth}
            \centering
            \begin{minipage}[c]{\textwidth}
                \centering
                \includegraphics[width=0.9\textwidth]{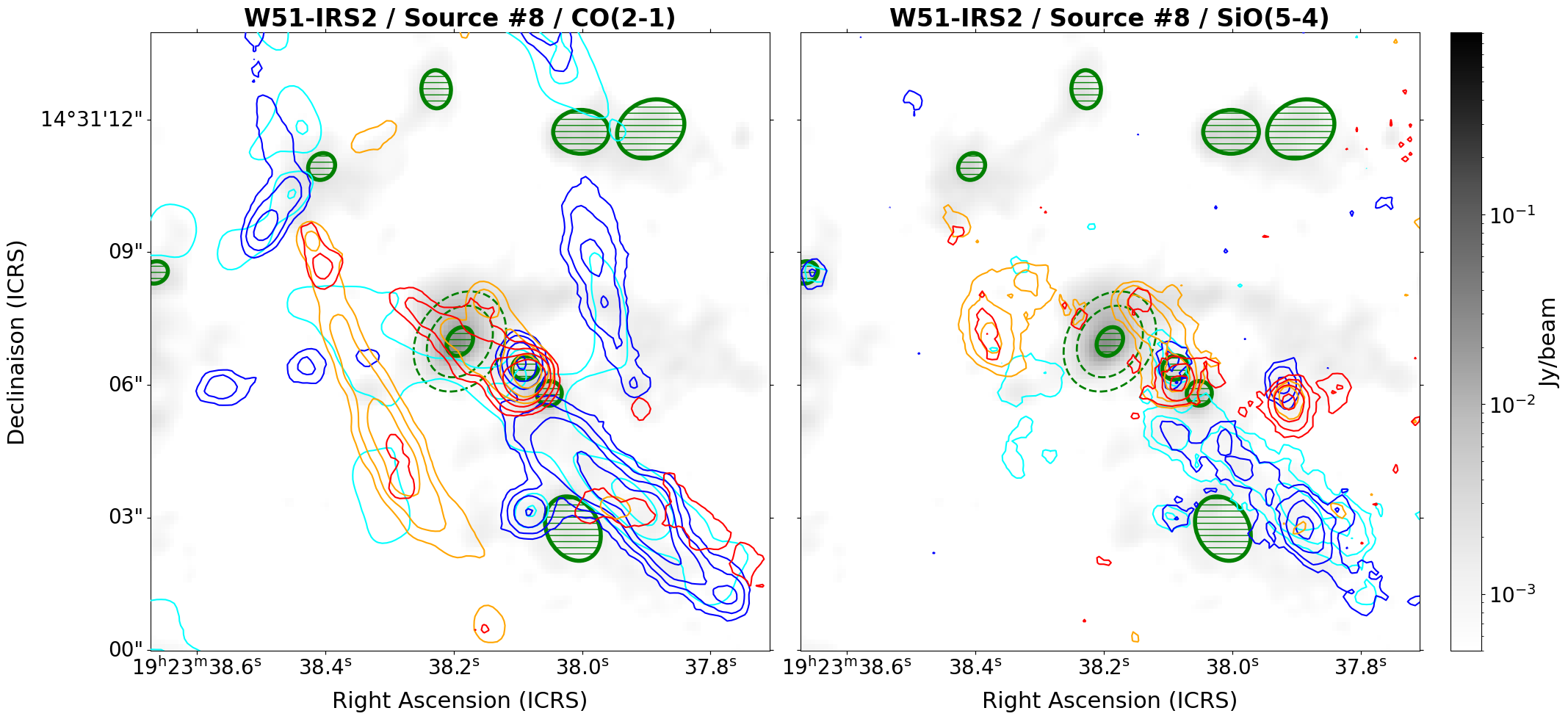}
            \end{minipage}
            \vfill
            \begin{minipage}[c]{\textwidth}
                \centering
                \includegraphics[width=0.7\textwidth]{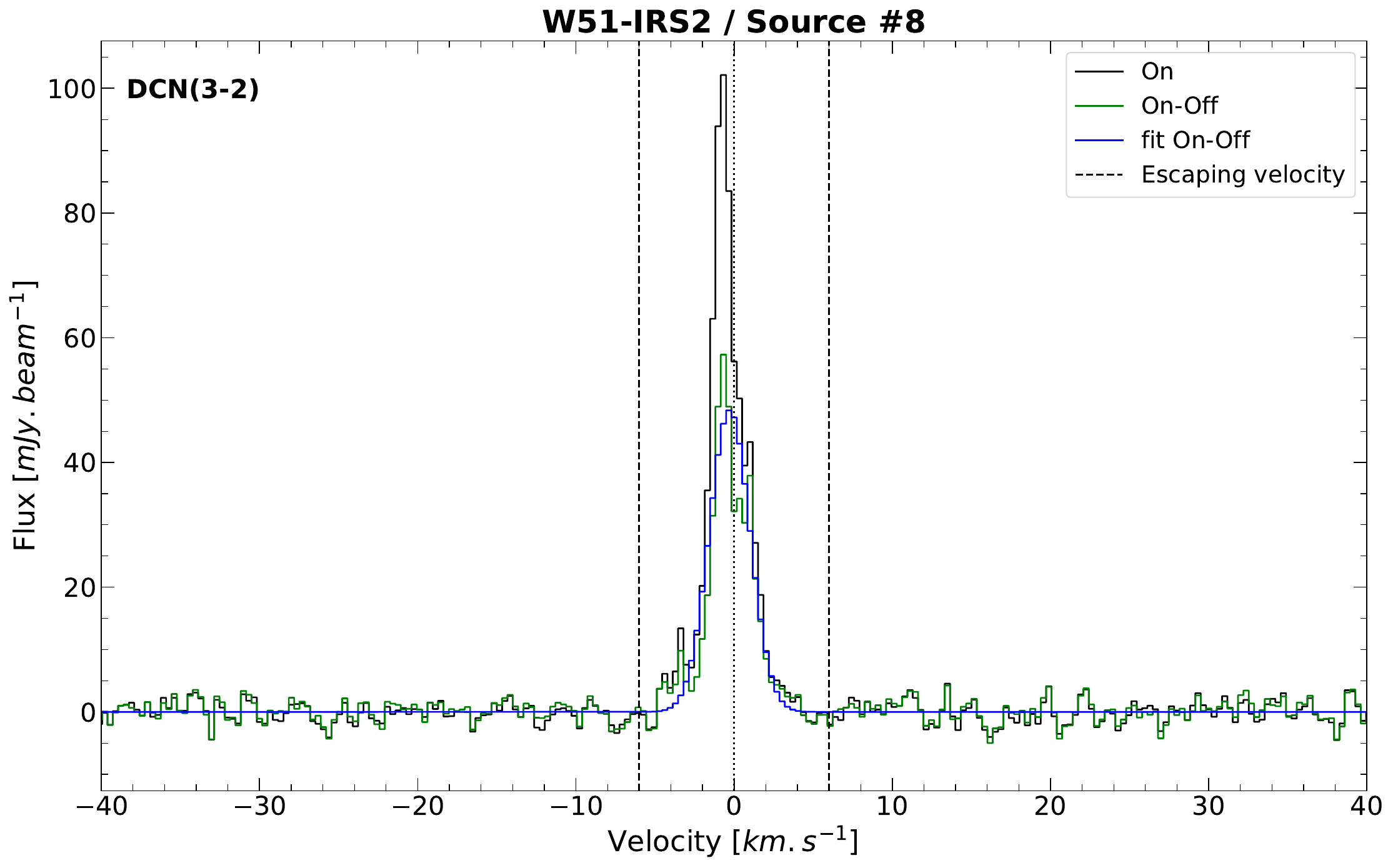}
            \end{minipage}
      \end{minipage}
    
    \vspace{0.2cm}
    
        \centering
        \begin{minipage}[c]{0.49\textwidth}
            \centering
            \includegraphics[width=\textwidth]{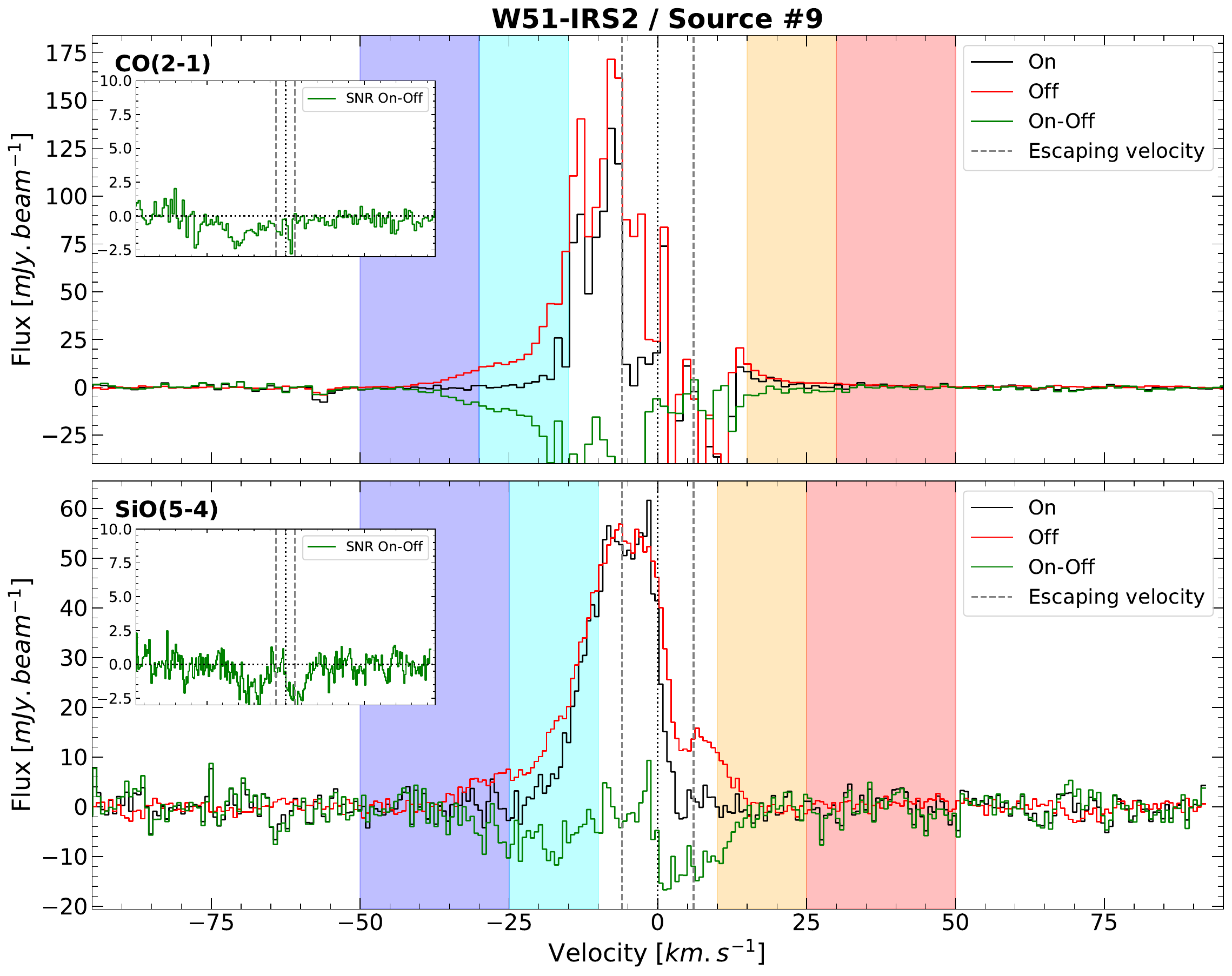}
        \end{minipage}
        \begin{minipage}[c]{0.49\textwidth}
            \centering
            \begin{minipage}[c]{\textwidth}
                \centering
                \includegraphics[width=0.9\textwidth]{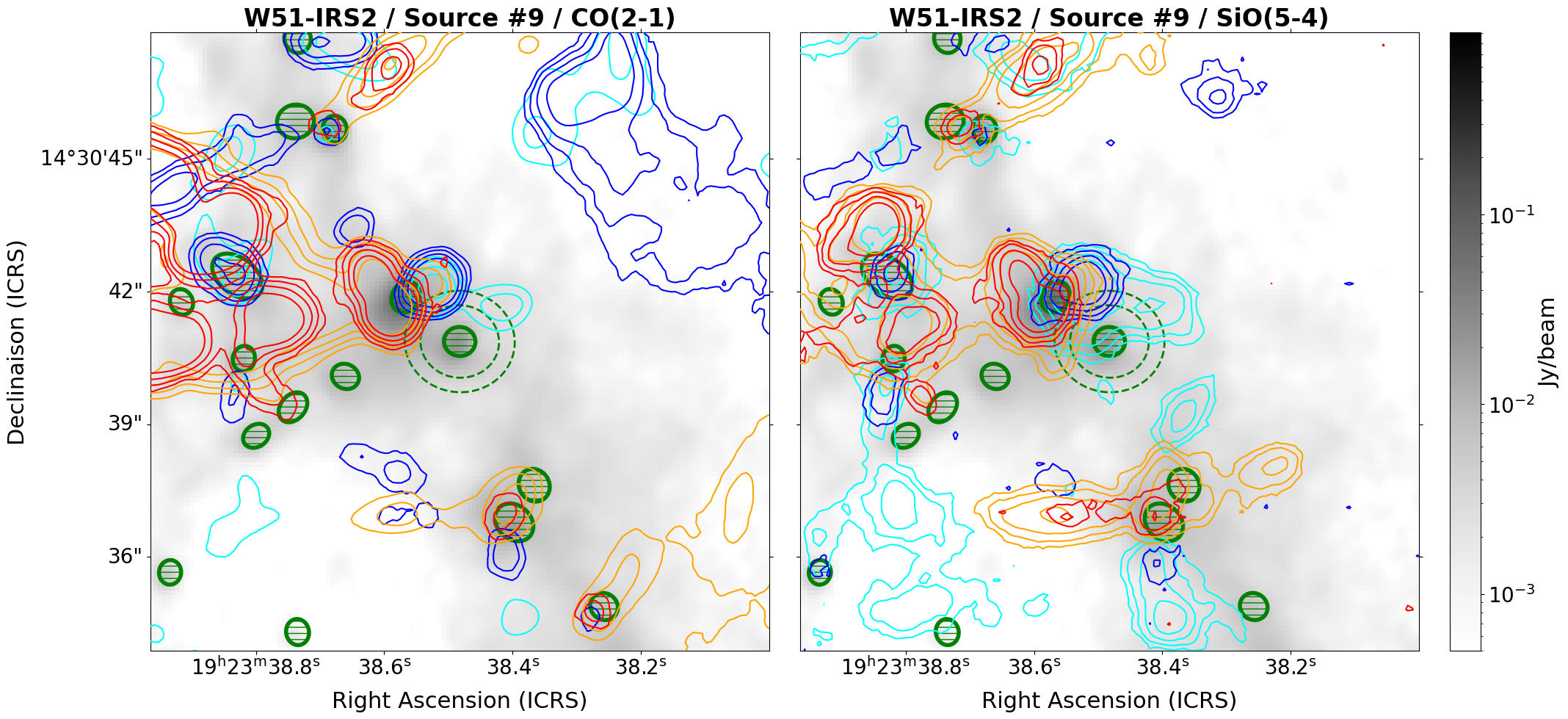}
            \end{minipage}
            \vfill
            \begin{minipage}[c]{\textwidth}
                \centering
                \includegraphics[width=0.7\textwidth]{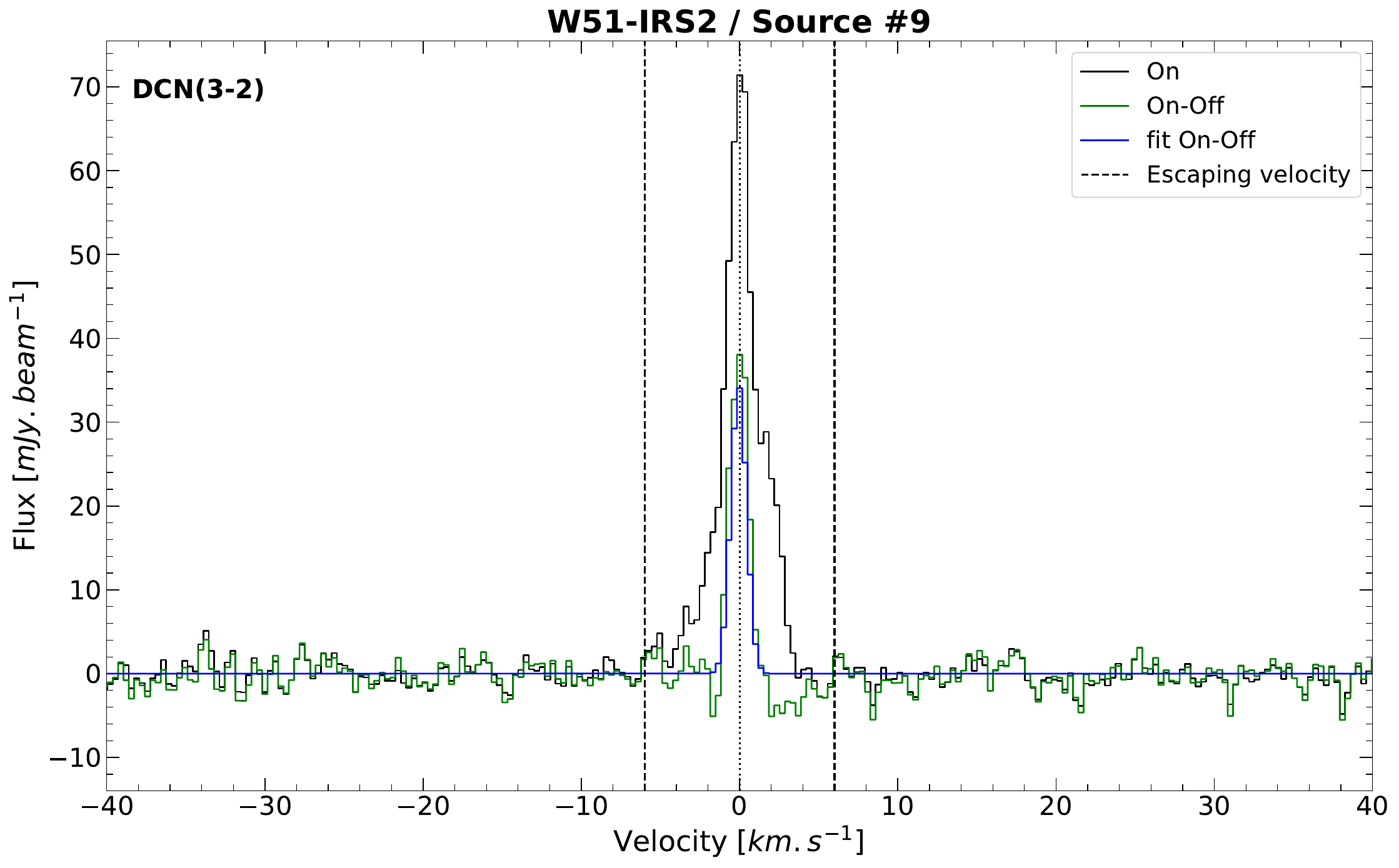}
            \end{minipage}
      \end{minipage}
    
    \vspace{0.2cm}
    
        \centering
        \begin{minipage}[c]{0.49\textwidth}
            \centering
            \includegraphics[width=\textwidth]{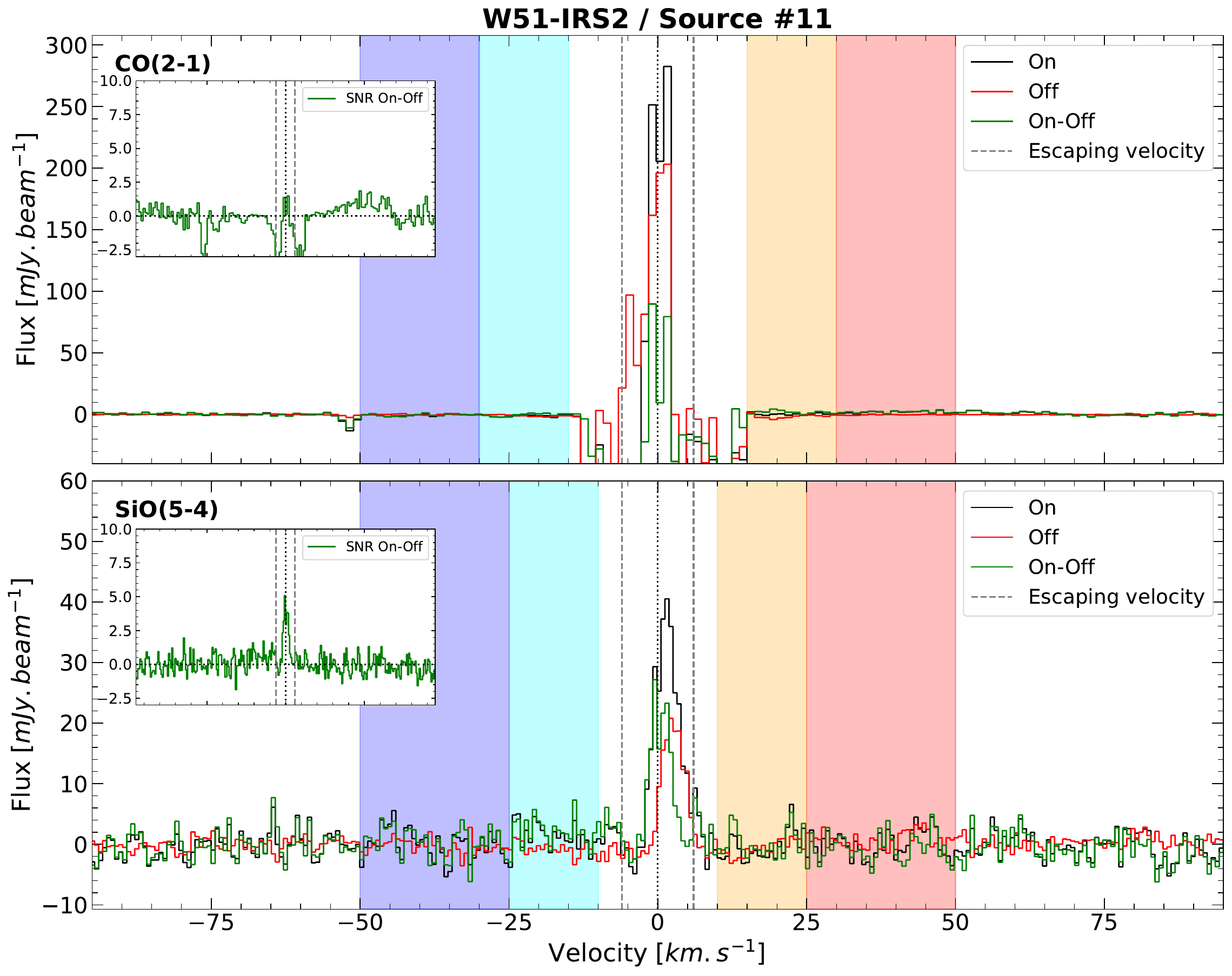}
        \end{minipage}
        \begin{minipage}[c]{0.49\textwidth}
            \centering
            \begin{minipage}[c]{\textwidth}
                \centering
                \includegraphics[width=0.9\textwidth]{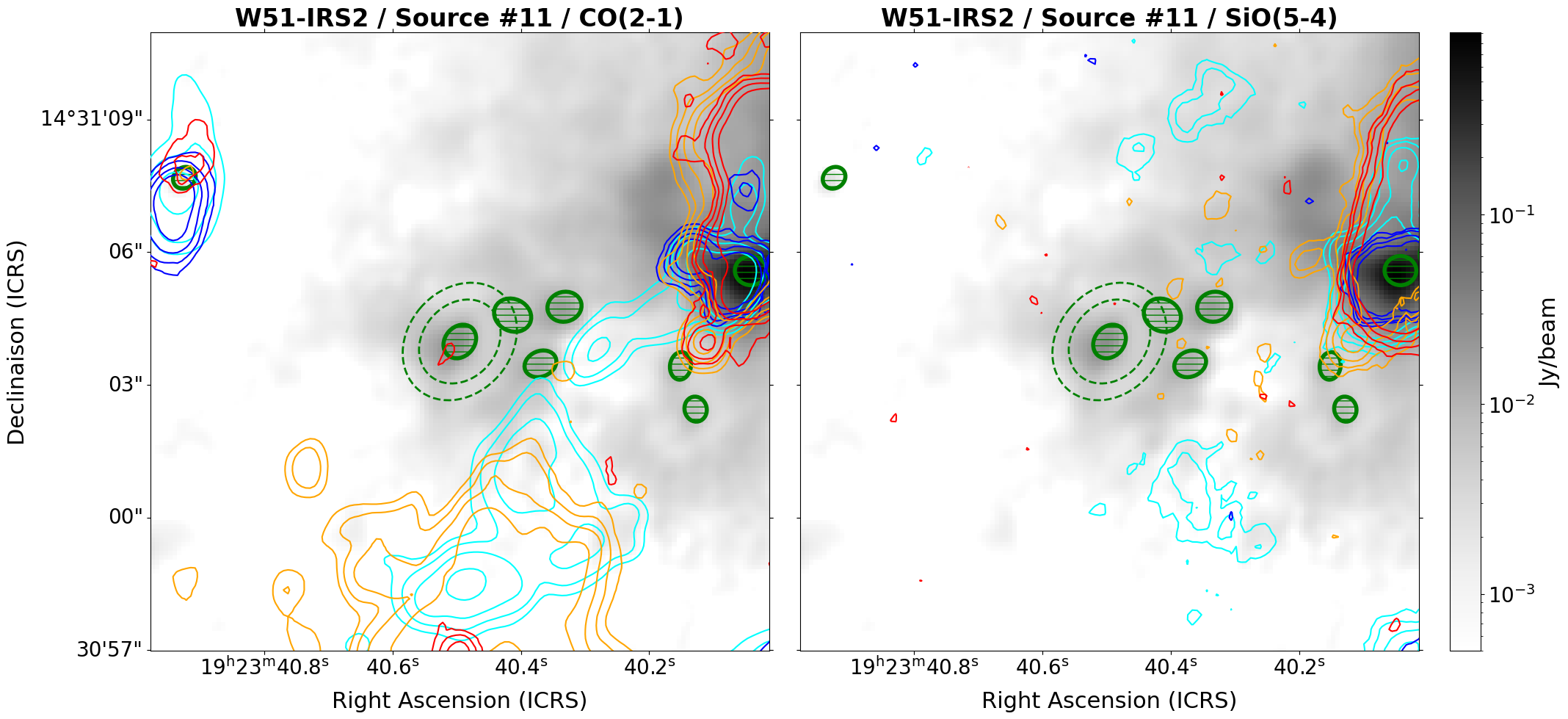}
            \end{minipage}
            \vfill
            \begin{minipage}[c]{\textwidth}
                \centering
                \includegraphics[width=0.7\textwidth]{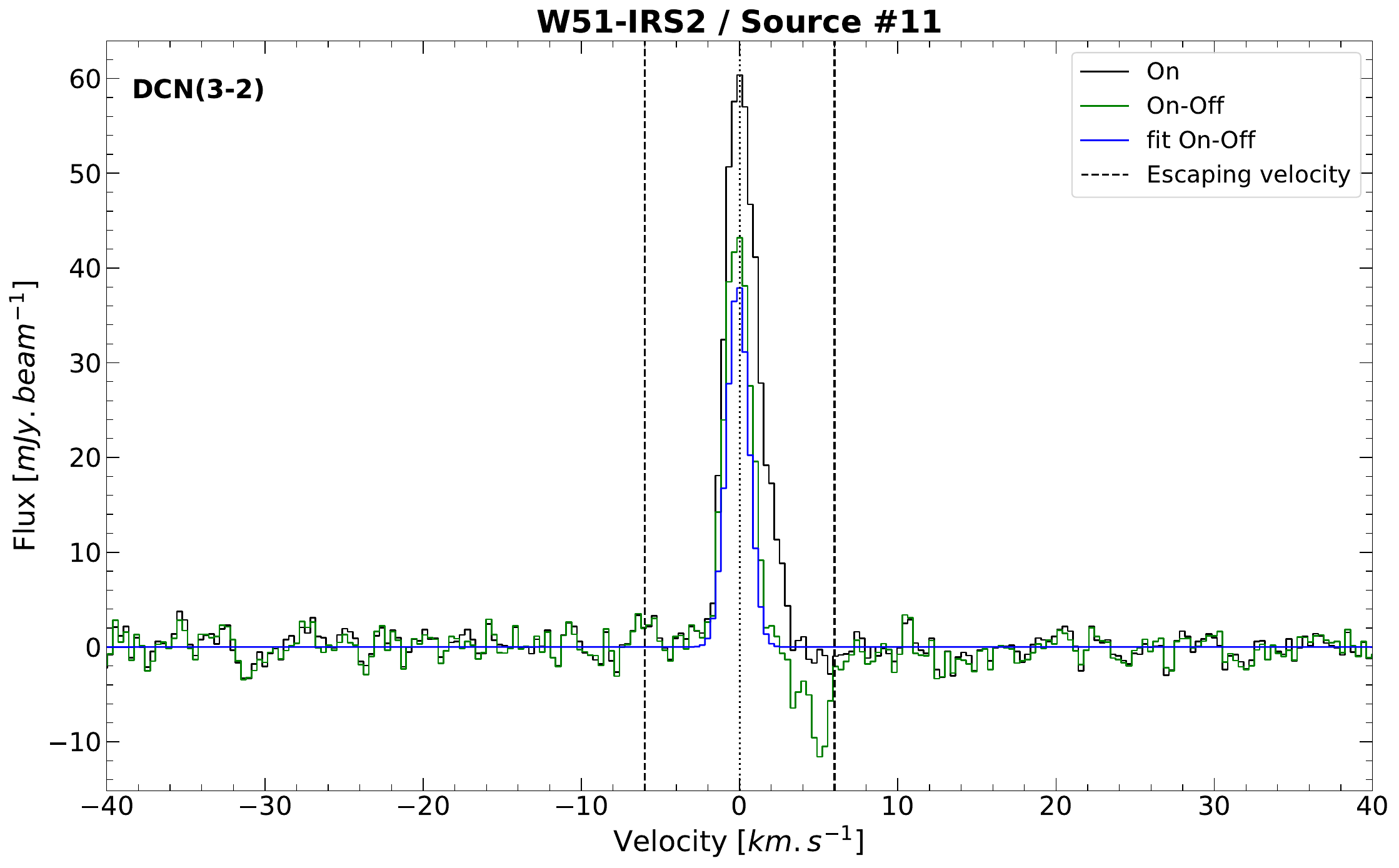}
            \end{minipage}
      \end{minipage}

   % \vskip -0.3cm
    \caption{CO and SiO spectra (left) and molecular outflow maps (top right) of the high-mass PSC candidates of the W51-IRS2 region. CO contours are 5, 10, 20, and 40 in units of $\sigma$, with $\sigma$ = $105.5$, $13.8$, $44.8$, $13.2$ \mJybeamkms for cyan, blue, orange and red contours respectively. SiO contours are 5, 10, 20, and 40 in units of $\sigma$, with $\sigma$ = $5.9$, $7.4$, $6.2$, $7.5$ \mJybeamkms for cyan, blue, orange and red contours respectively. DCN spectra and fits (bottom right) of the high-mass PSC candidates of the W51-IRS2 region.}

\end{figure*}

\begin{figure*}\ContinuedFloat

    \centering
        \begin{minipage}[c]{0.49\textwidth}
            \centering
            \includegraphics[width=\textwidth]{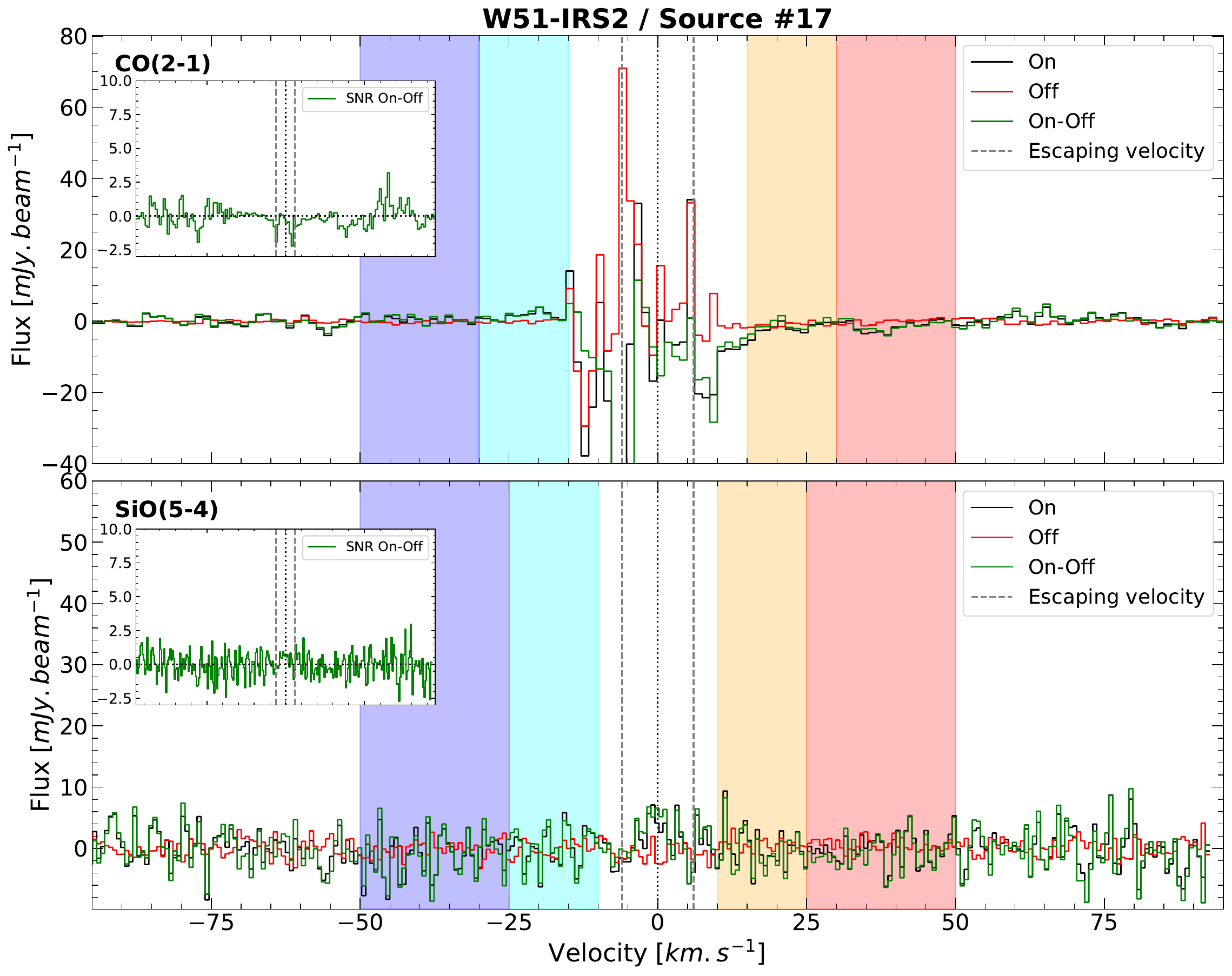}
        \end{minipage}
        \begin{minipage}[c]{0.49\textwidth}
            \centering
            \begin{minipage}[c]{\textwidth}
                \centering
                \includegraphics[width=0.9\textwidth]{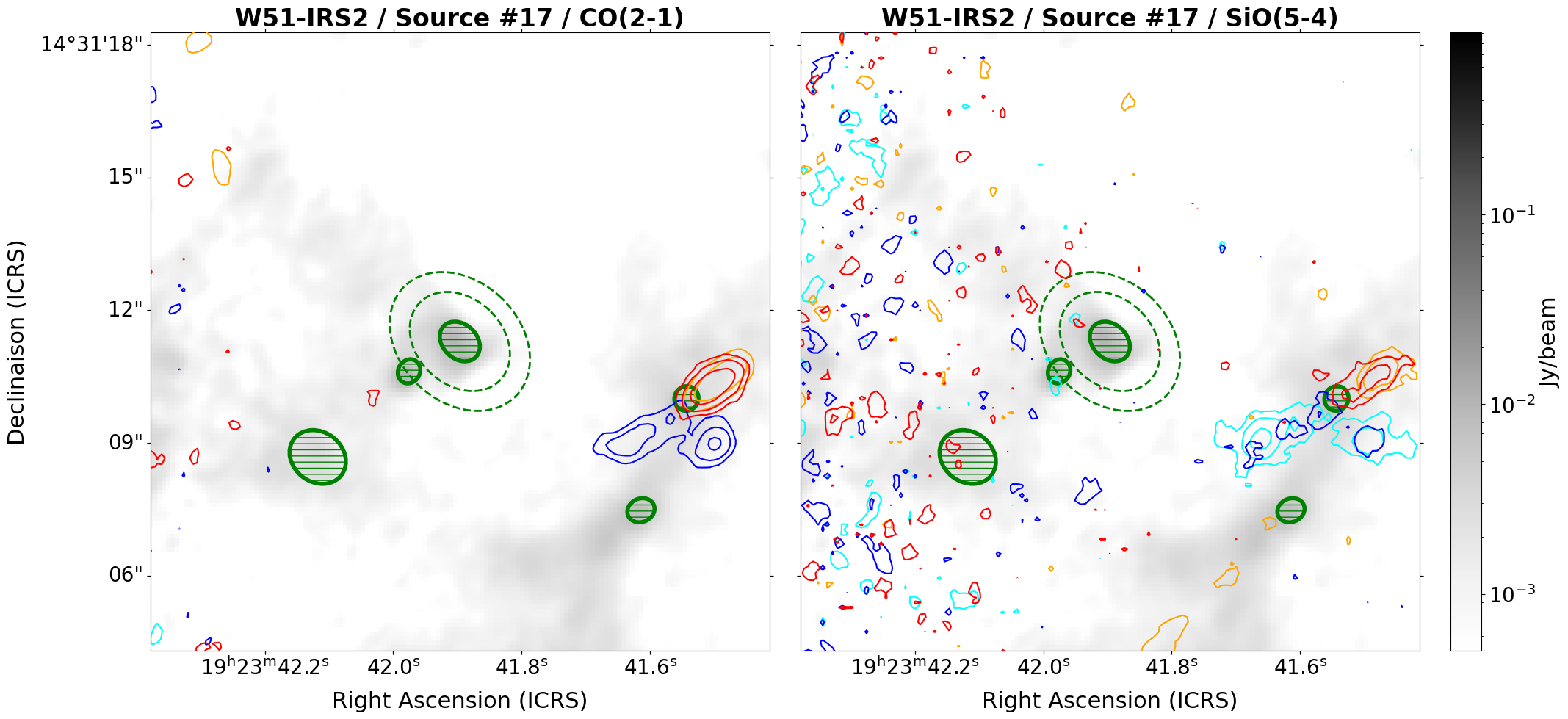}
            \end{minipage}
            \vfill
            \begin{minipage}[c]{\textwidth}
                \centering
                \includegraphics[width=0.7\textwidth]{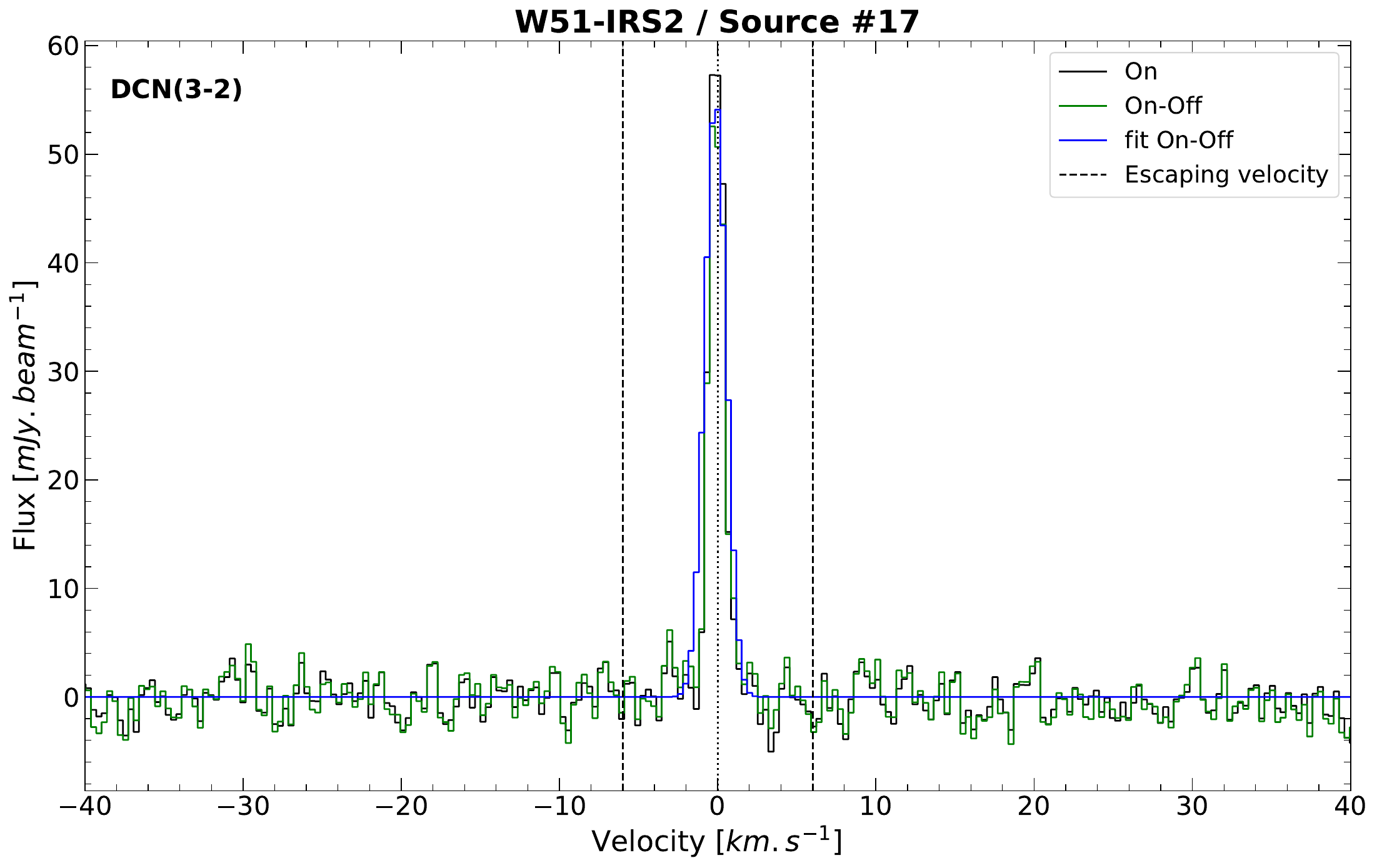}
            \end{minipage}
      \end{minipage}
    
    \vspace{0.2cm}
    
        \centering
        \begin{minipage}[c]{0.49\textwidth}
            \centering
            \includegraphics[width=\textwidth]{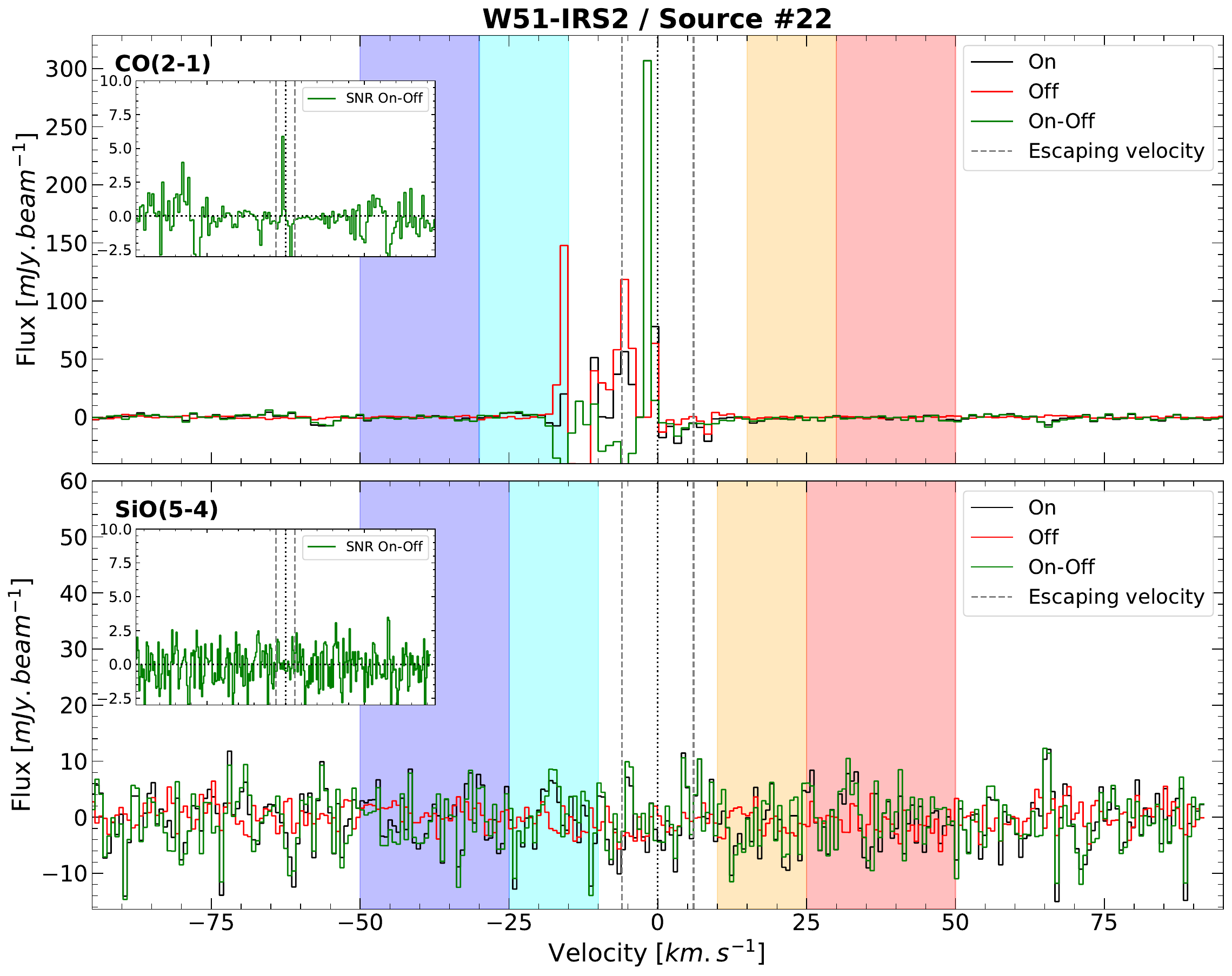}
        \end{minipage}
        \begin{minipage}[c]{0.49\textwidth}
            \centering
            \begin{minipage}[c]{\textwidth}
                \centering
                \includegraphics[width=0.9\textwidth]{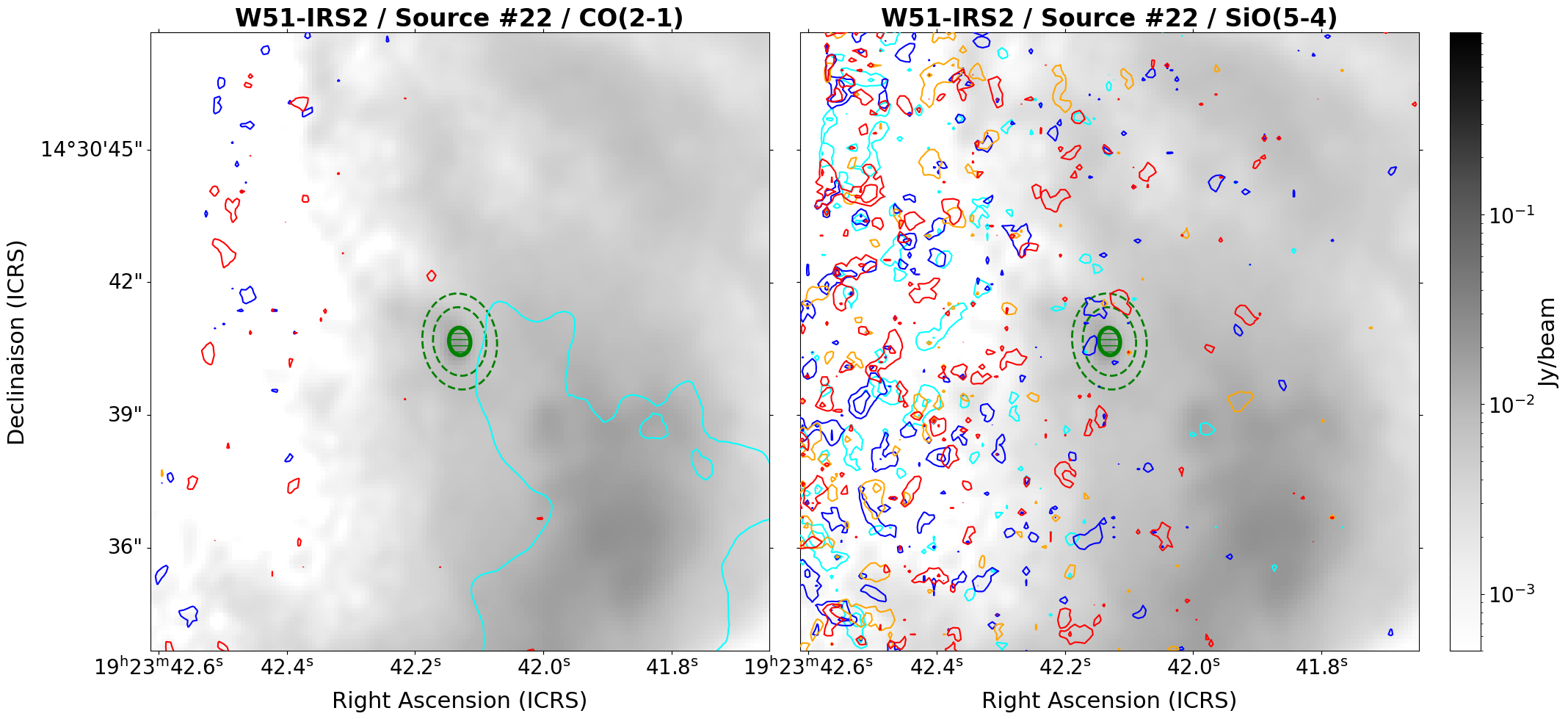}
            \end{minipage}
            \vfill
            \begin{minipage}[c]{\textwidth}
                \centering
                \includegraphics[width=0.7\textwidth]{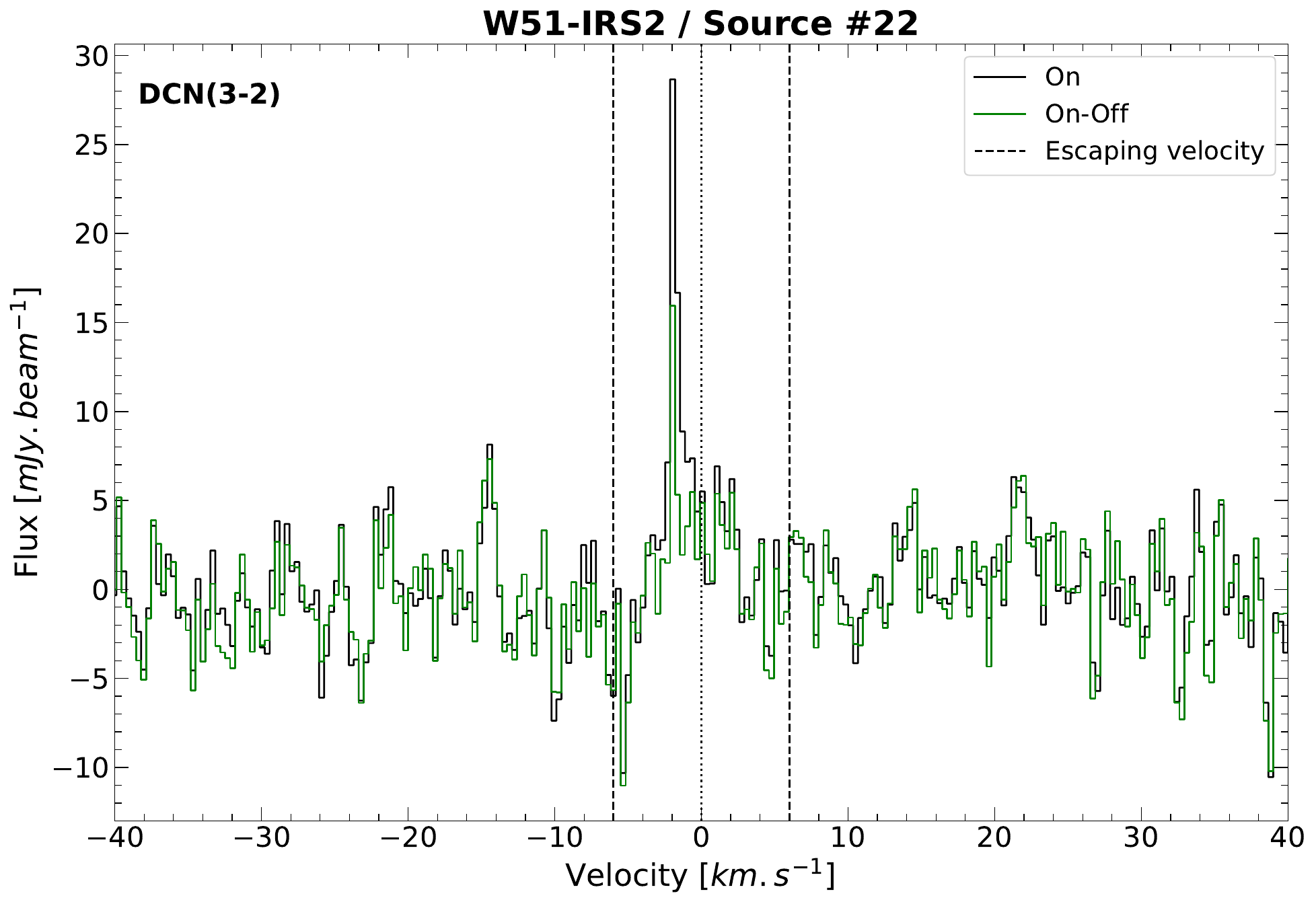}
            \end{minipage}
      \end{minipage}
    
    \vspace{0.2cm}
    
        \centering
        \begin{minipage}[c]{0.49\textwidth}
            \centering
            \includegraphics[width=\textwidth]{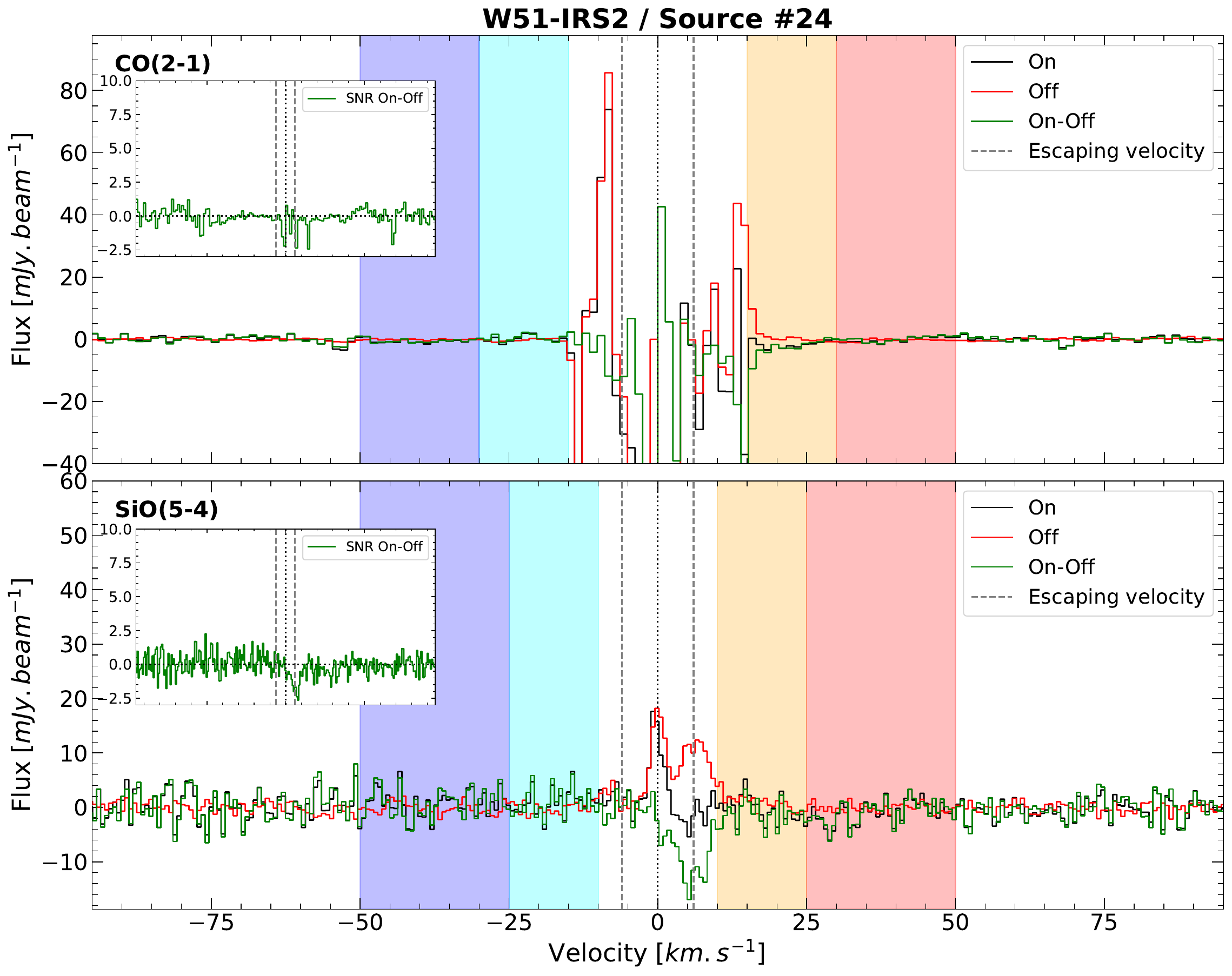}
        \end{minipage}
        \begin{minipage}[c]{0.49\textwidth}
            \centering
            \begin{minipage}[c]{\textwidth}
                \centering
                \includegraphics[width=0.9\textwidth]{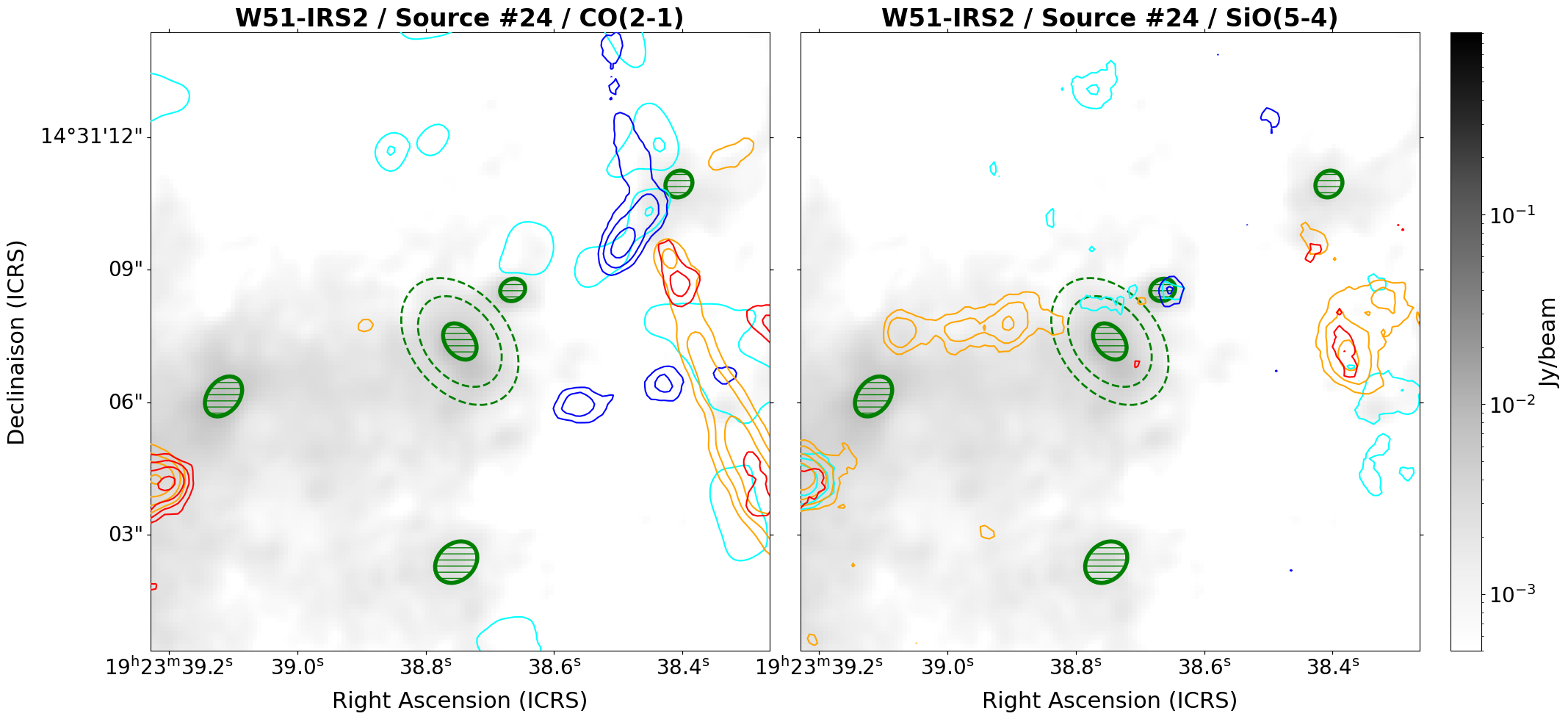}
            \end{minipage}
            \vfill
            \begin{minipage}[c]{\textwidth}
                \centering
                \includegraphics[width=0.7\textwidth]{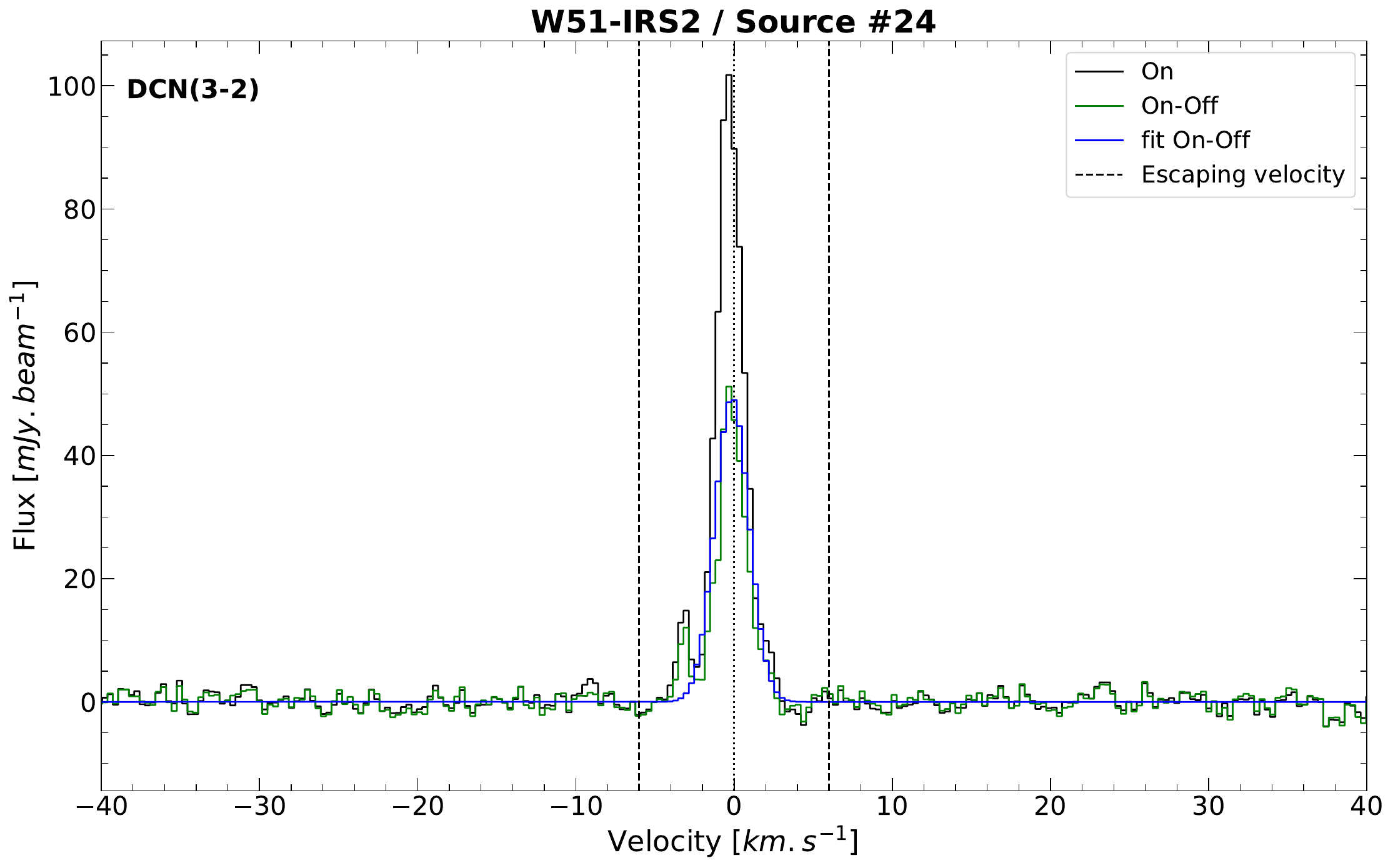}
            \end{minipage}
      \end{minipage}
    \caption{continued.}
\end{figure*}

\begin{figure*}\ContinuedFloat
    \centering
        \begin{minipage}[c]{0.49\textwidth}
            \centering
            \includegraphics[width=\textwidth]{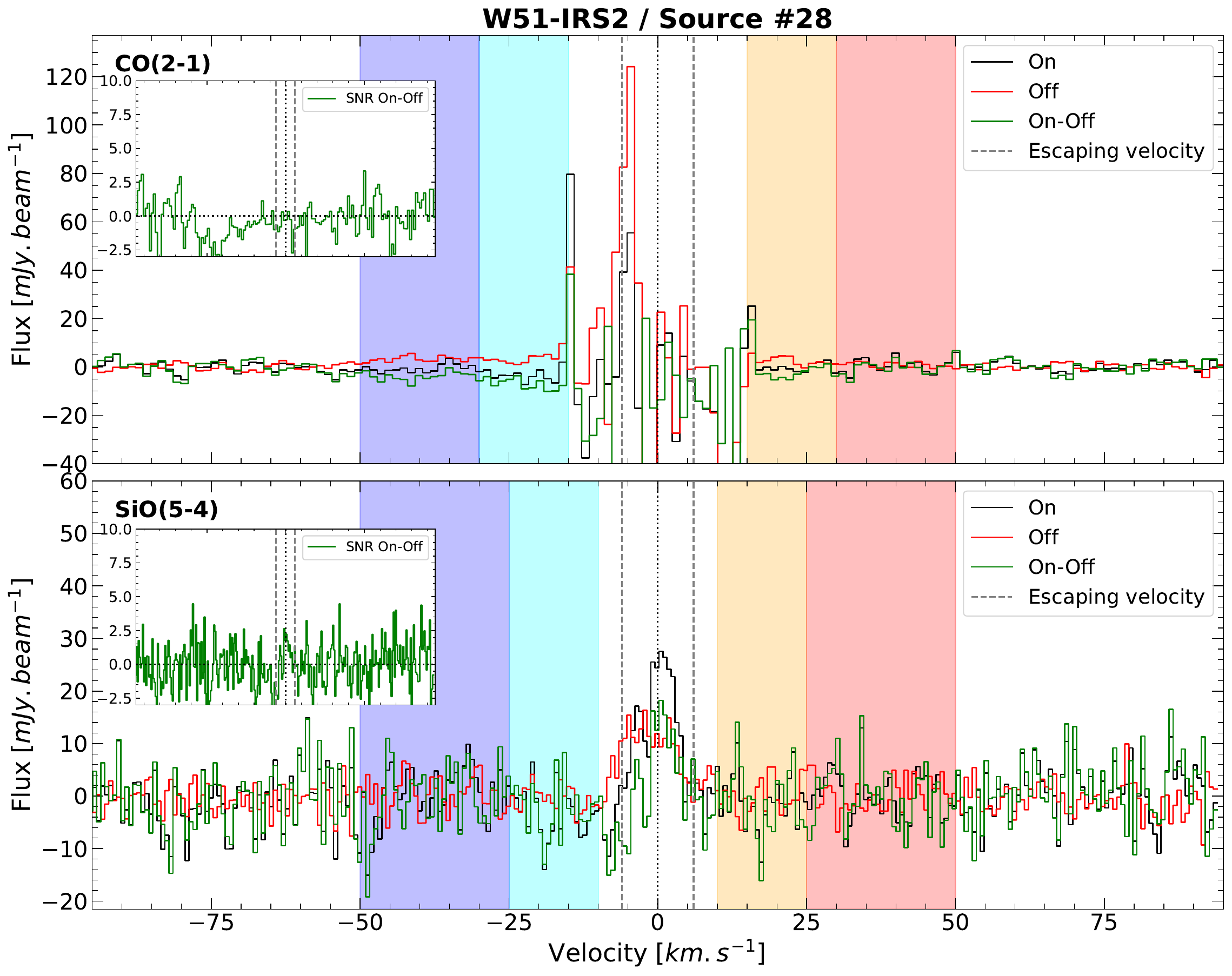}
        \end{minipage}
        \begin{minipage}[c]{0.49\textwidth}
            \centering
            \begin{minipage}[c]{\textwidth}
                \centering
                \includegraphics[width=0.9\textwidth]{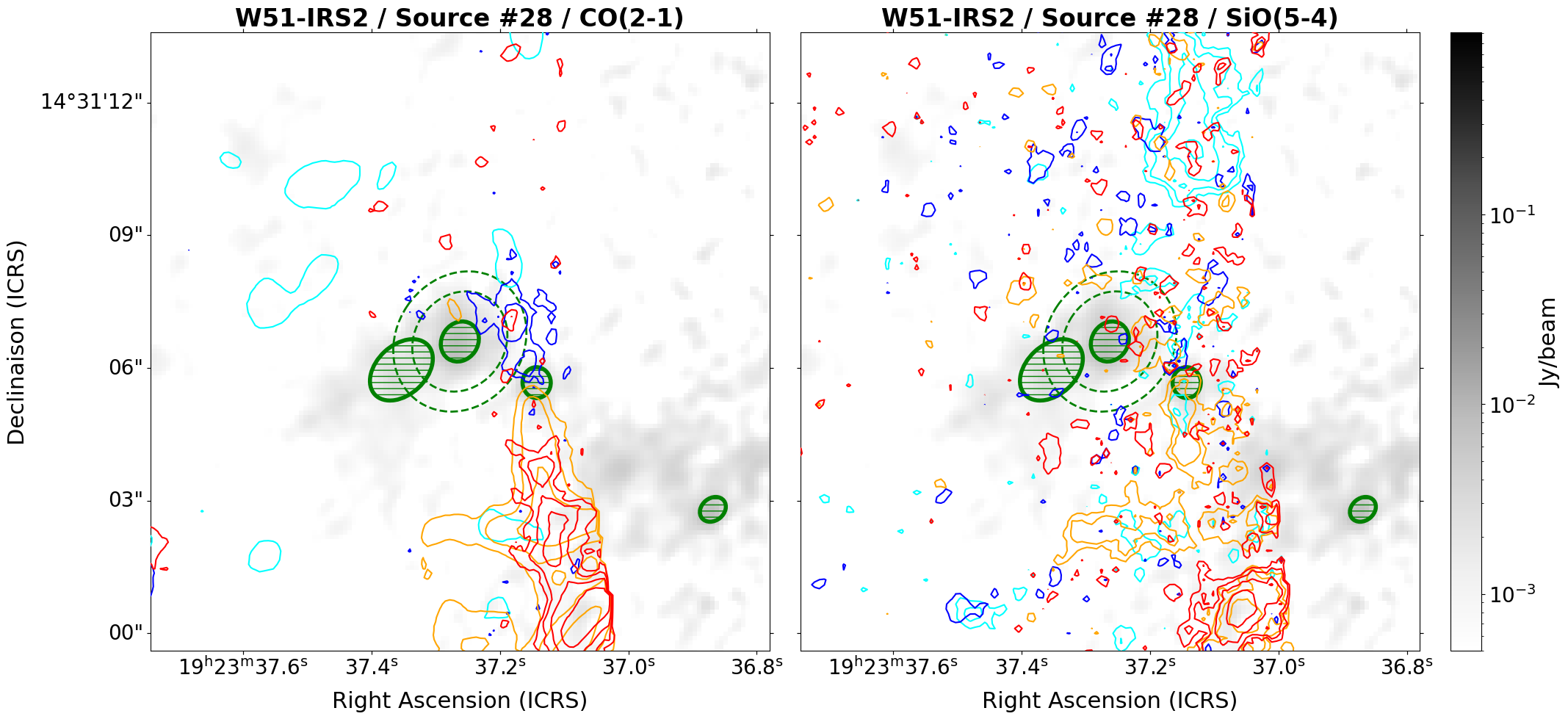}
            \end{minipage}
            \vfill
            \begin{minipage}[c]{\textwidth}
                \centering
                \includegraphics[width=0.7\textwidth]{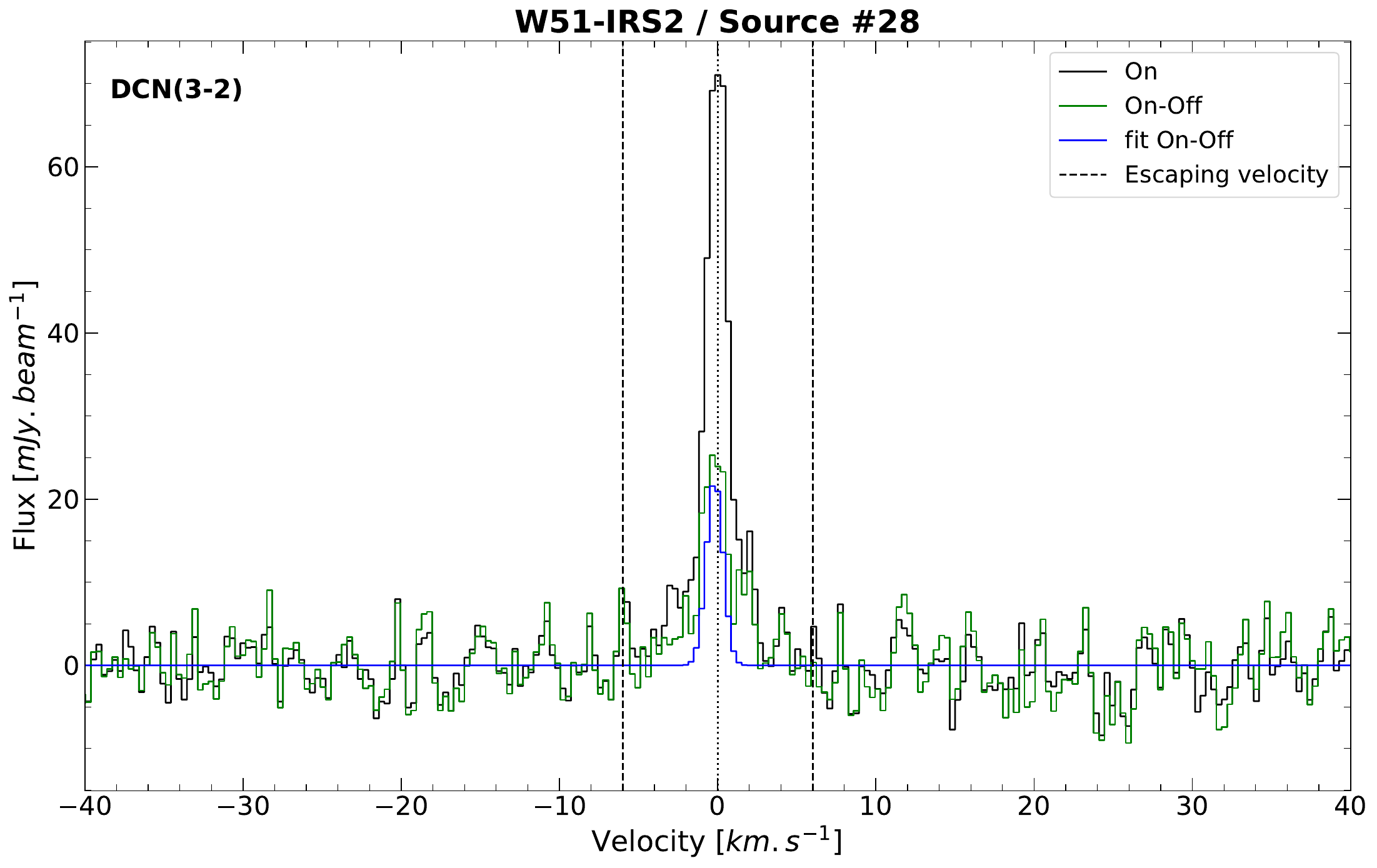}
            \end{minipage}
      \end{minipage}
    
    \vspace{0.2cm}
    
        \centering
        \begin{minipage}[c]{0.49\textwidth}
            \centering
            \includegraphics[width=\textwidth]{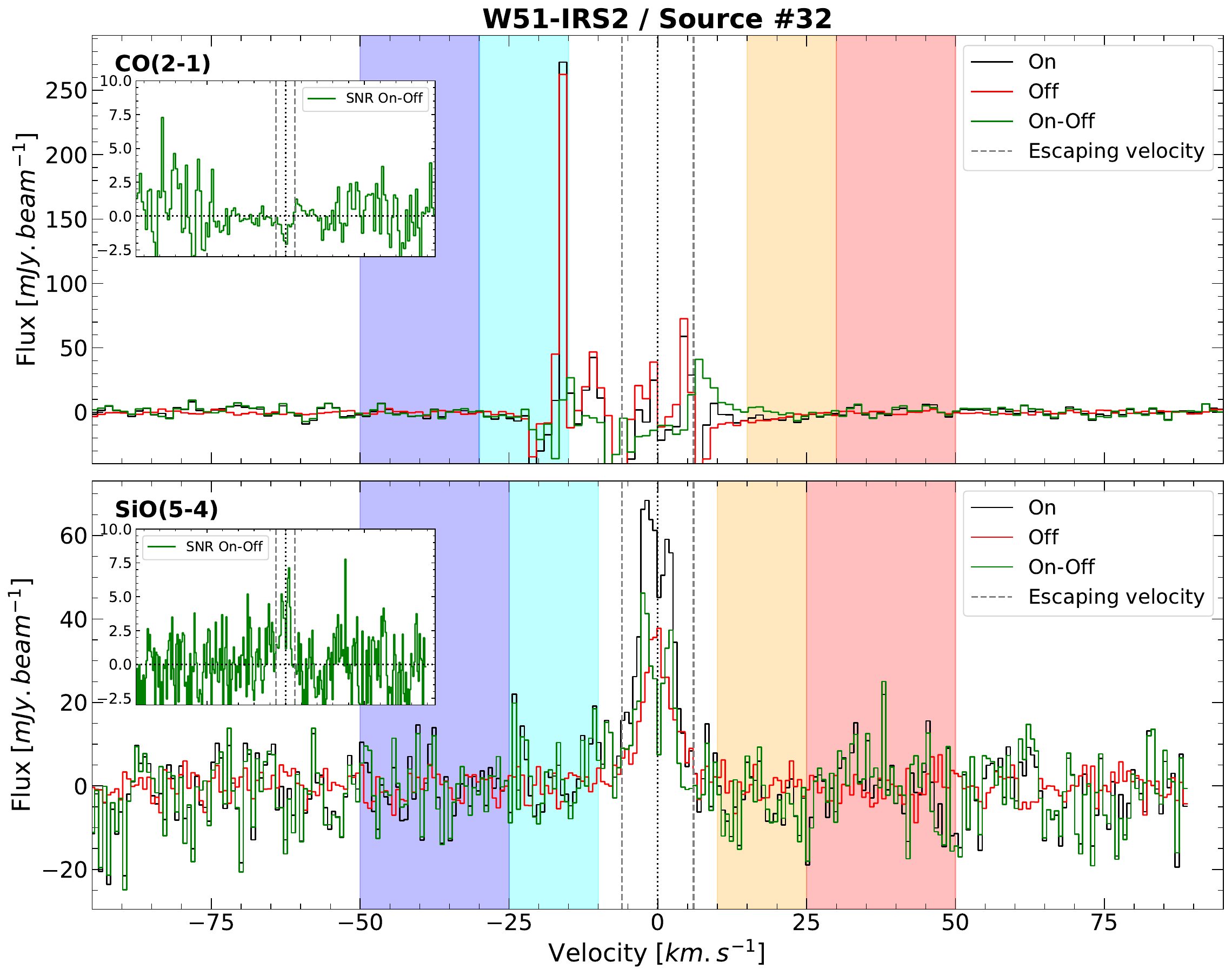}
        \end{minipage}
        \begin{minipage}[c]{0.49\textwidth}
            \centering
            \begin{minipage}[c]{\textwidth}
                \centering
                \includegraphics[width=0.9\textwidth]{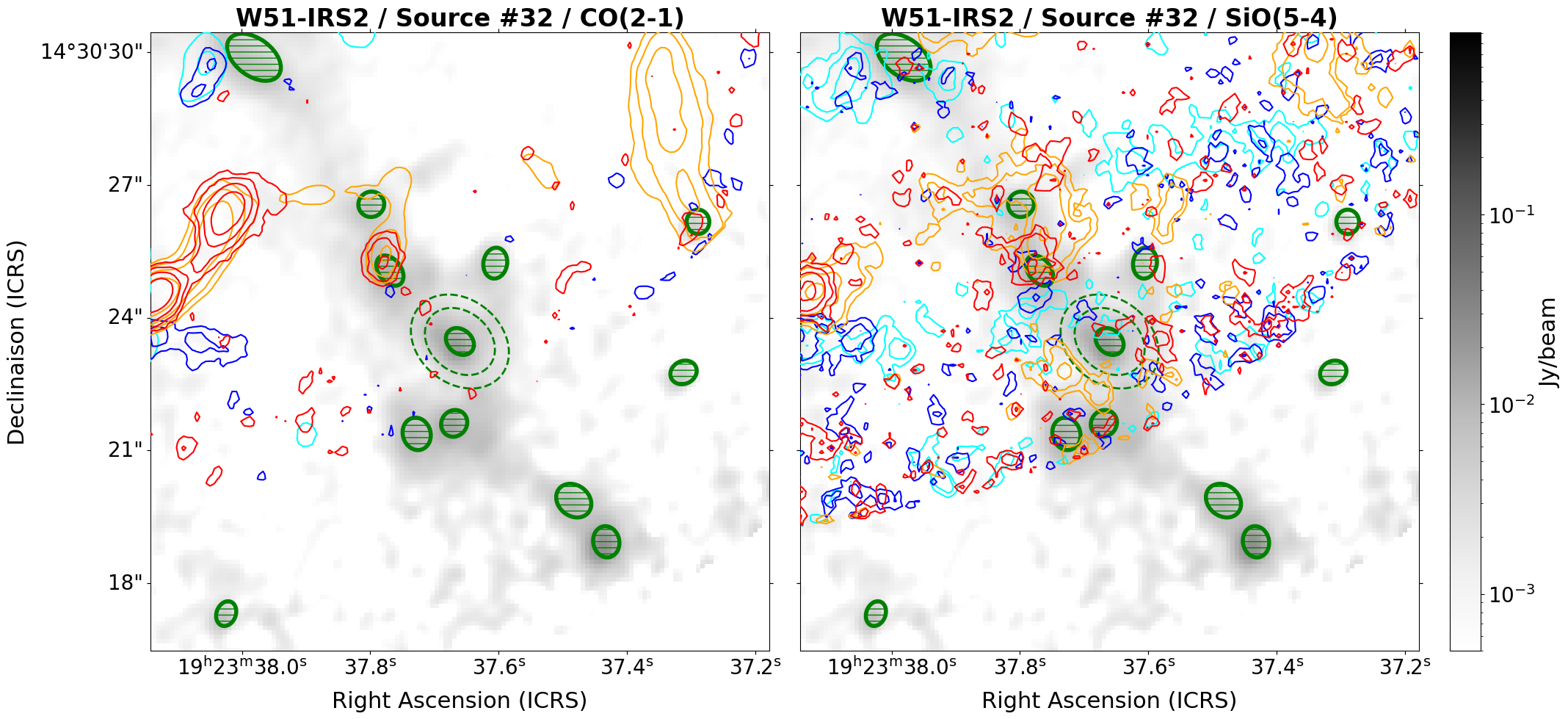}
            \end{minipage}
            \vfill
            \begin{minipage}[c]{\textwidth}
                \centering
                \includegraphics[width=0.7\textwidth]{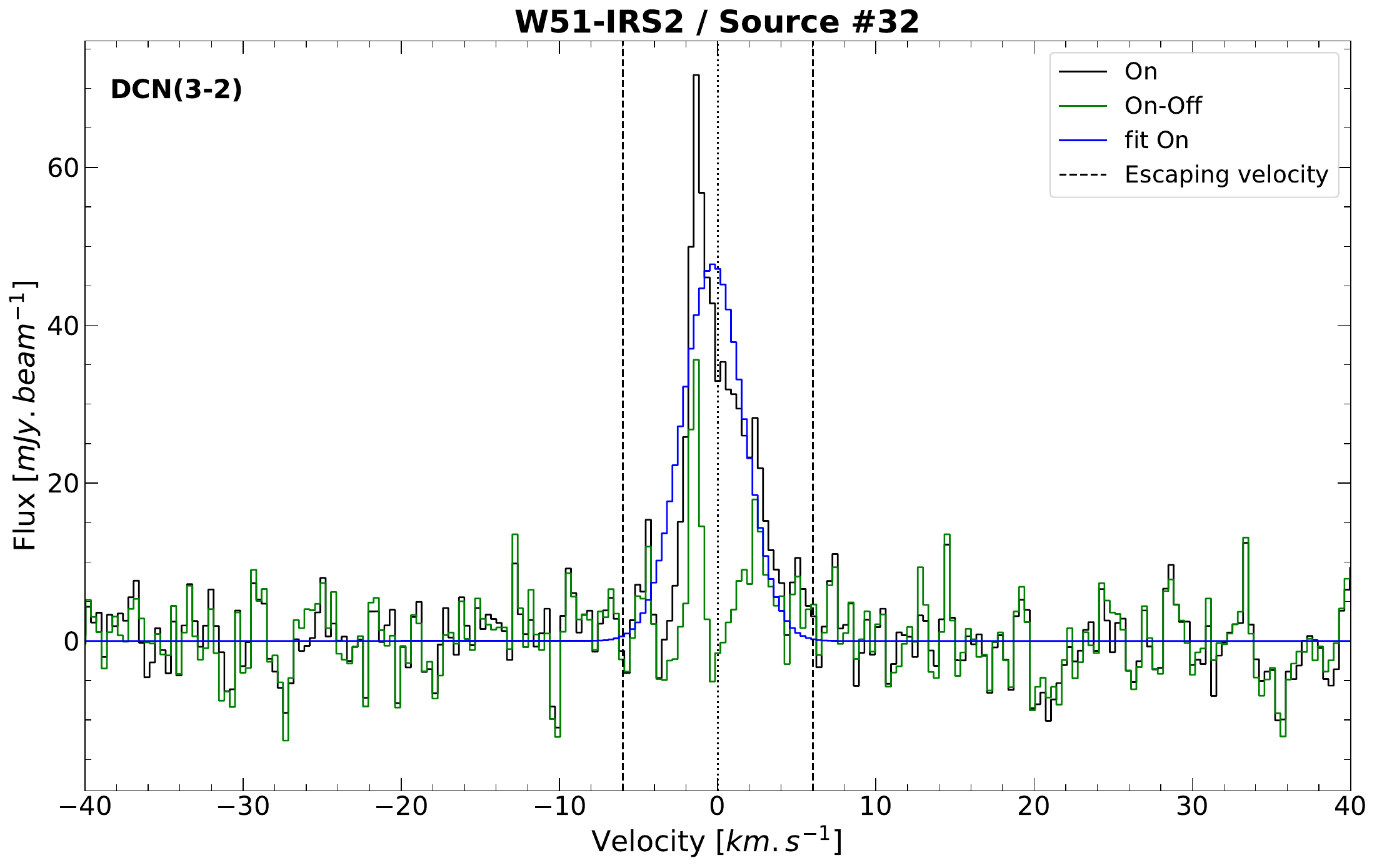}
            \end{minipage}
      \end{minipage}
    
    \vspace{0.2cm}

    \centering
        \begin{minipage}[c]{0.49\textwidth}
            \centering
            \includegraphics[width=\textwidth]{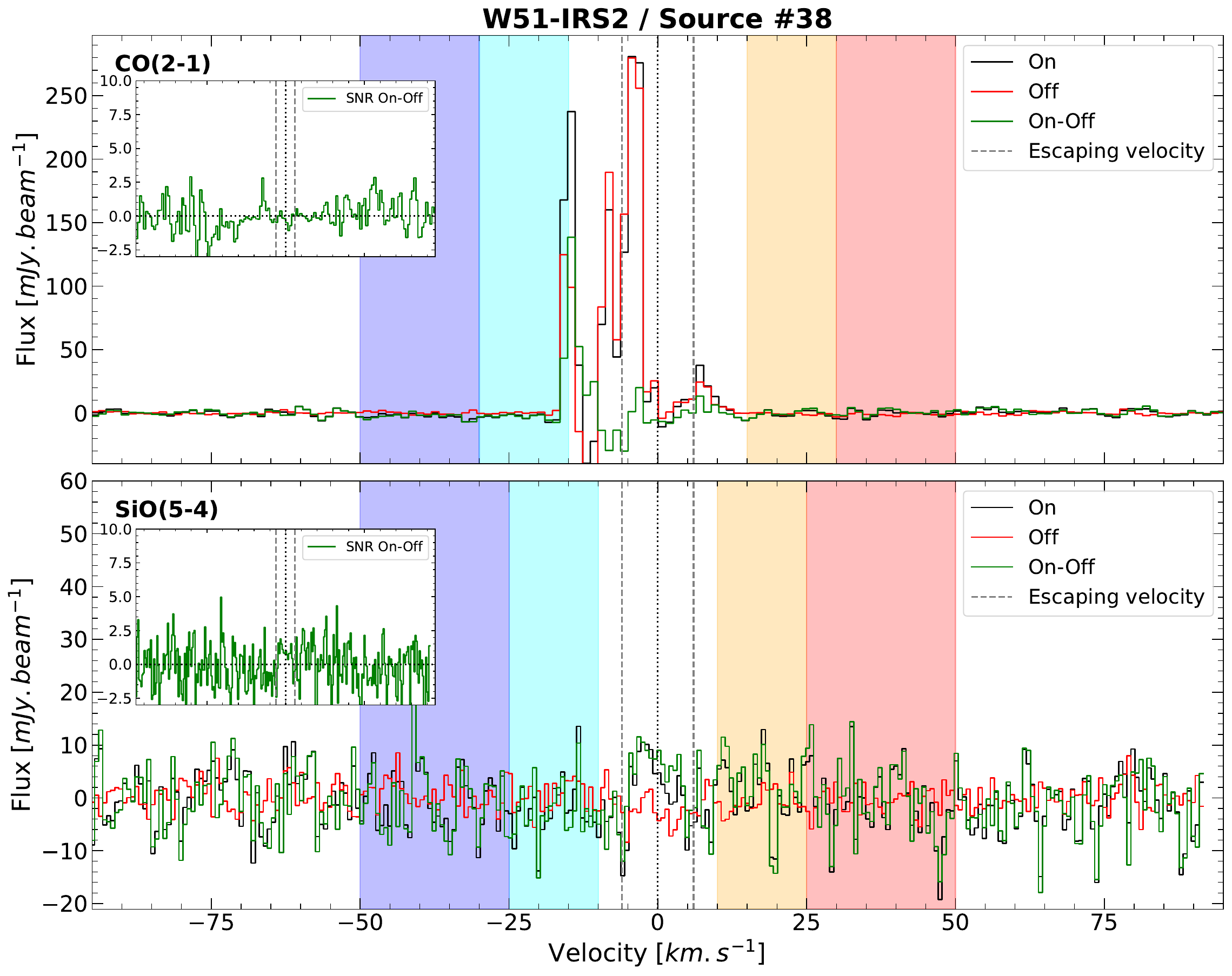}
        \end{minipage}
        \begin{minipage}[c]{0.49\textwidth}
            \centering
            \begin{minipage}[c]{\textwidth}
                \centering
                \includegraphics[width=0.9\textwidth]{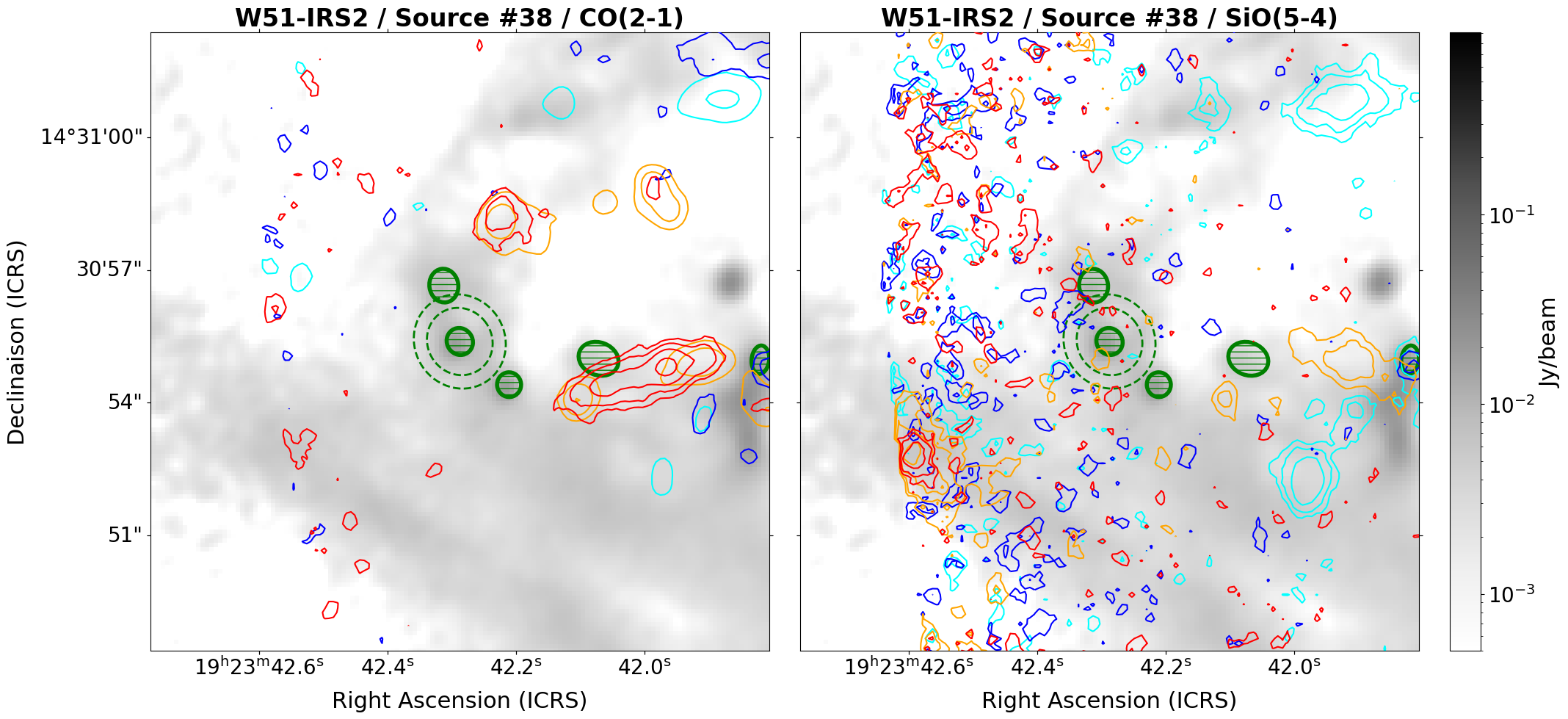}
            \end{minipage}
            \vfill
            \begin{minipage}[c]{\textwidth}
                \centering
                \includegraphics[width=0.7\textwidth]{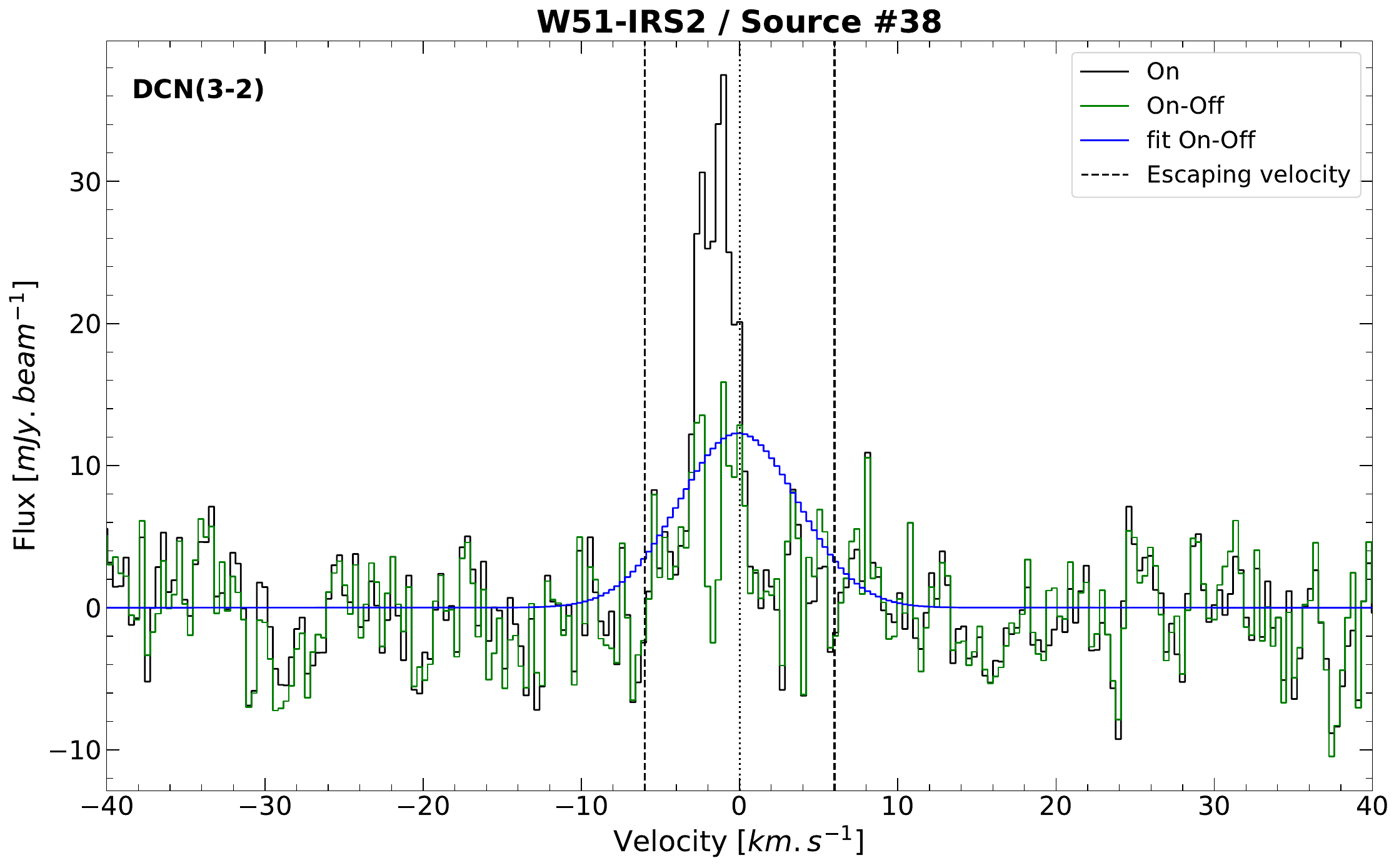}
            \end{minipage}
      \end{minipage}
    
    \caption{continued.}
\end{figure*}

\begin{figure*}\ContinuedFloat
    
    \centering
        \begin{minipage}[c]{0.49\textwidth}
            \centering
            \includegraphics[width=\textwidth]{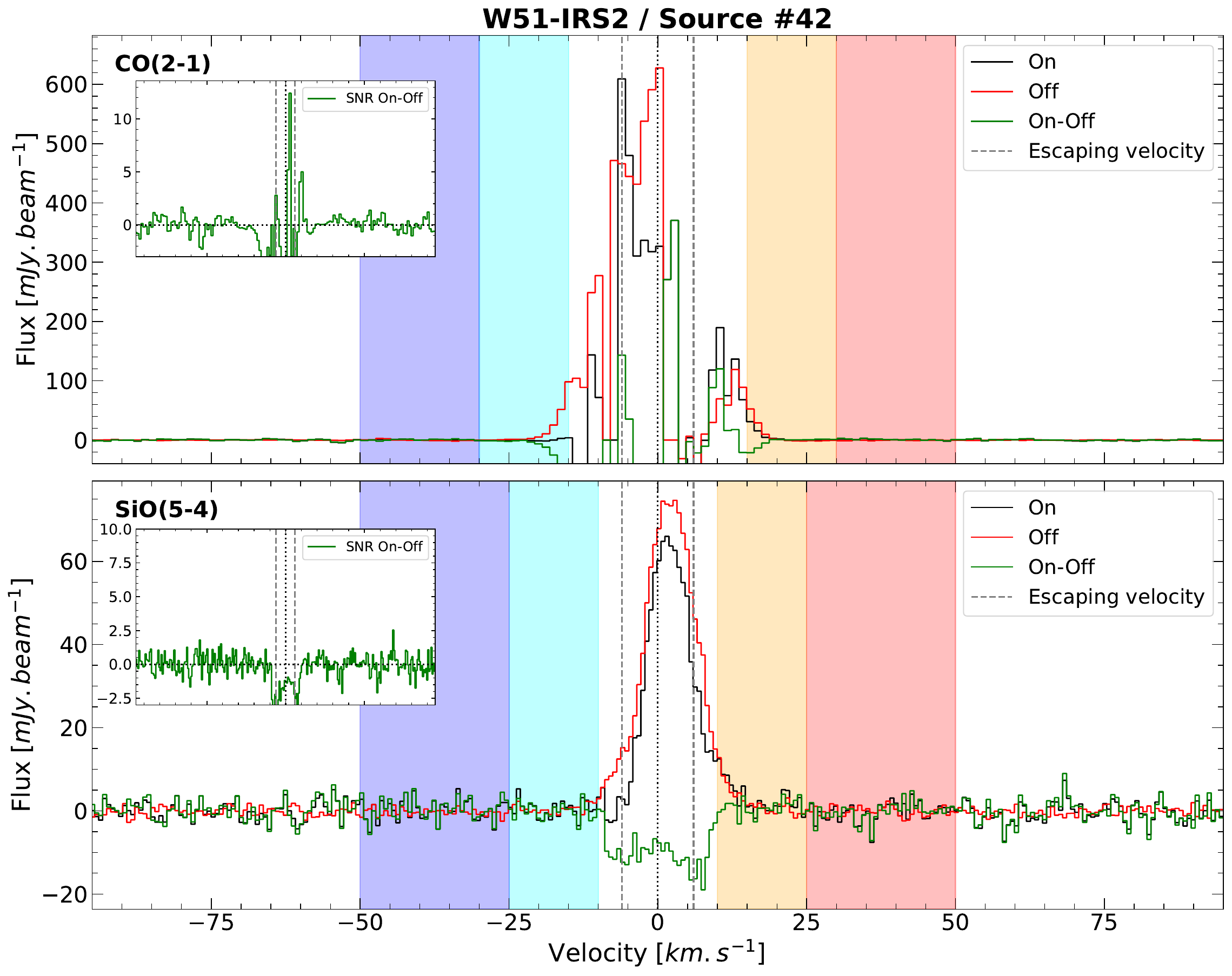}
        \end{minipage}
        \begin{minipage}[c]{0.49\textwidth}
            \centering
            \begin{minipage}[c]{\textwidth}
                \centering
                \includegraphics[width=0.9\textwidth]{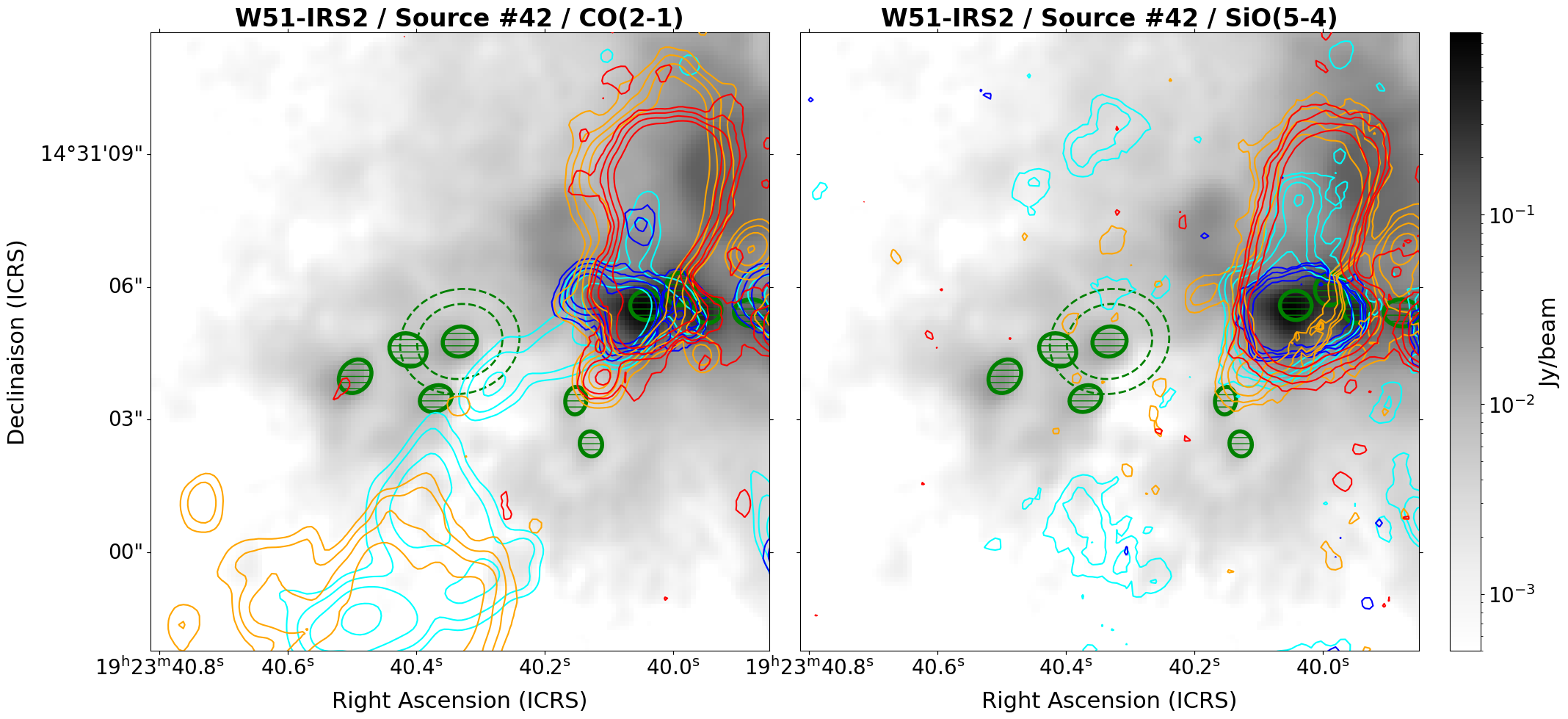}
            \end{minipage}
            \vfill
            \begin{minipage}[c]{\textwidth}
                \centering
                \includegraphics[width=0.7\textwidth]{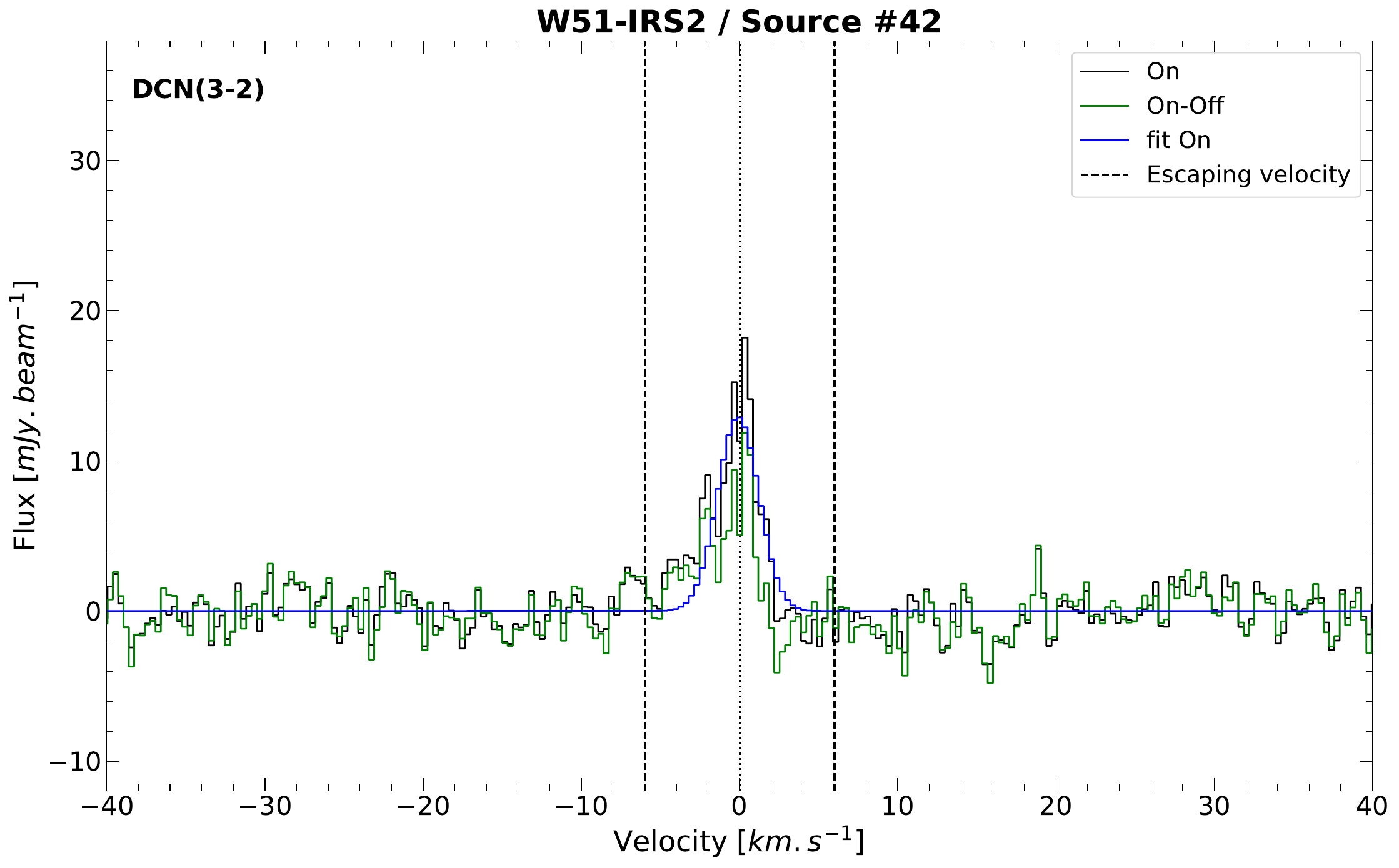}
            \end{minipage}
      \end{minipage}
    
    \caption{continued.}
\end{figure*}
%\input{Appendix/appendix_retrieved_MPSC}
% MPSC highlight in full field
\section{Location of the prestellar candidates in the ALMA-IMF regions} \label{appendix:MPSC_highlight_part}

Figure \ref{appendix:MPSC_highlight_figure} presents the fourteen ALMA-IMF fields with the prestellar cores candidates highlighted in blue with their number. The red and orange cores are all the high-mass and intermediate mass cores above  which have been studied and have been associated to an outflow (i.e. protostellar cores). The green cores remaining are all the cores which have not been classified.

\begin{figure*}
    \centering
    \includegraphics[width=0.49\textwidth]{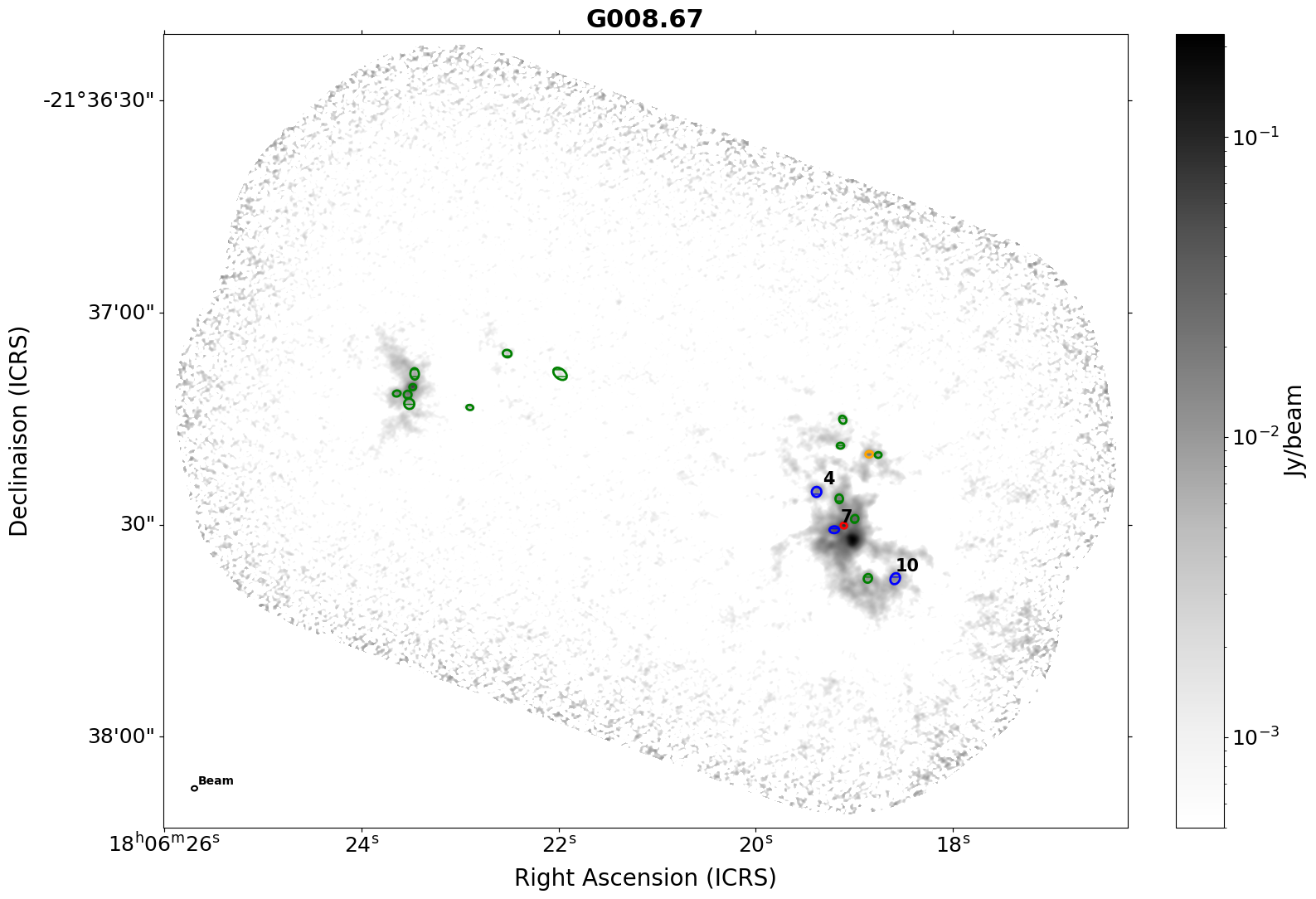}
    \includegraphics[width=0.49\textwidth]{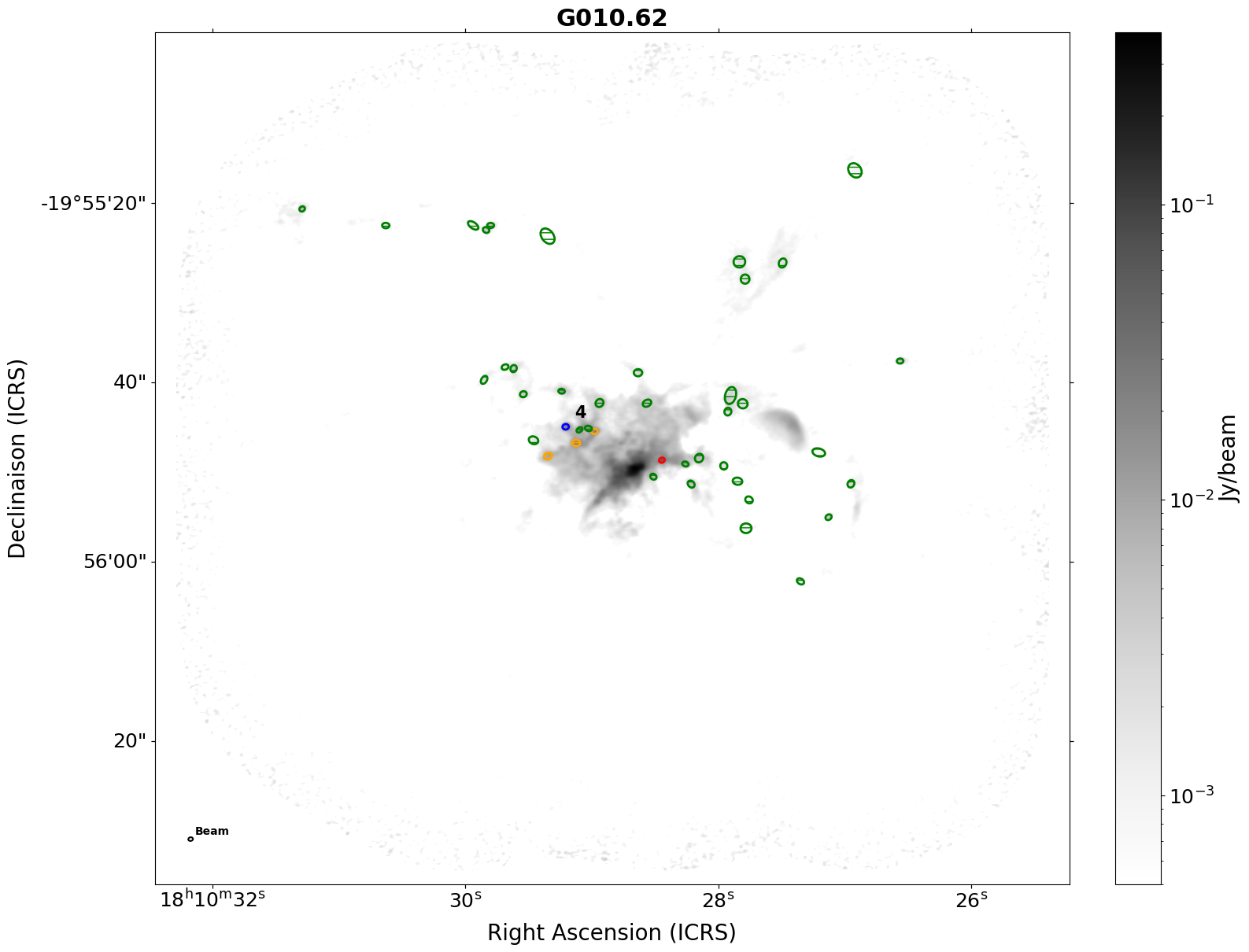}
    \includegraphics[width=0.49\textwidth]{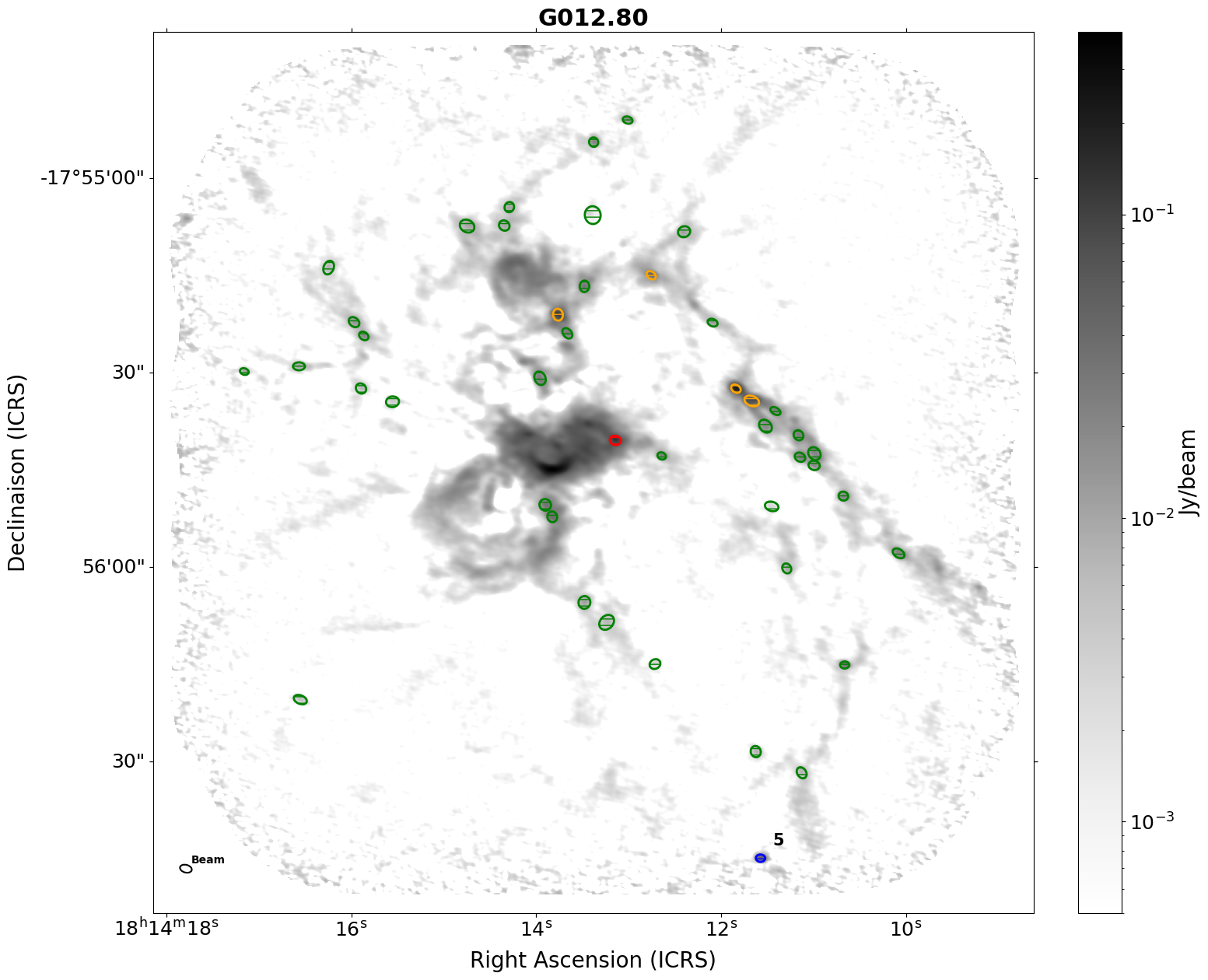}
    \includegraphics[width=0.49\textwidth]{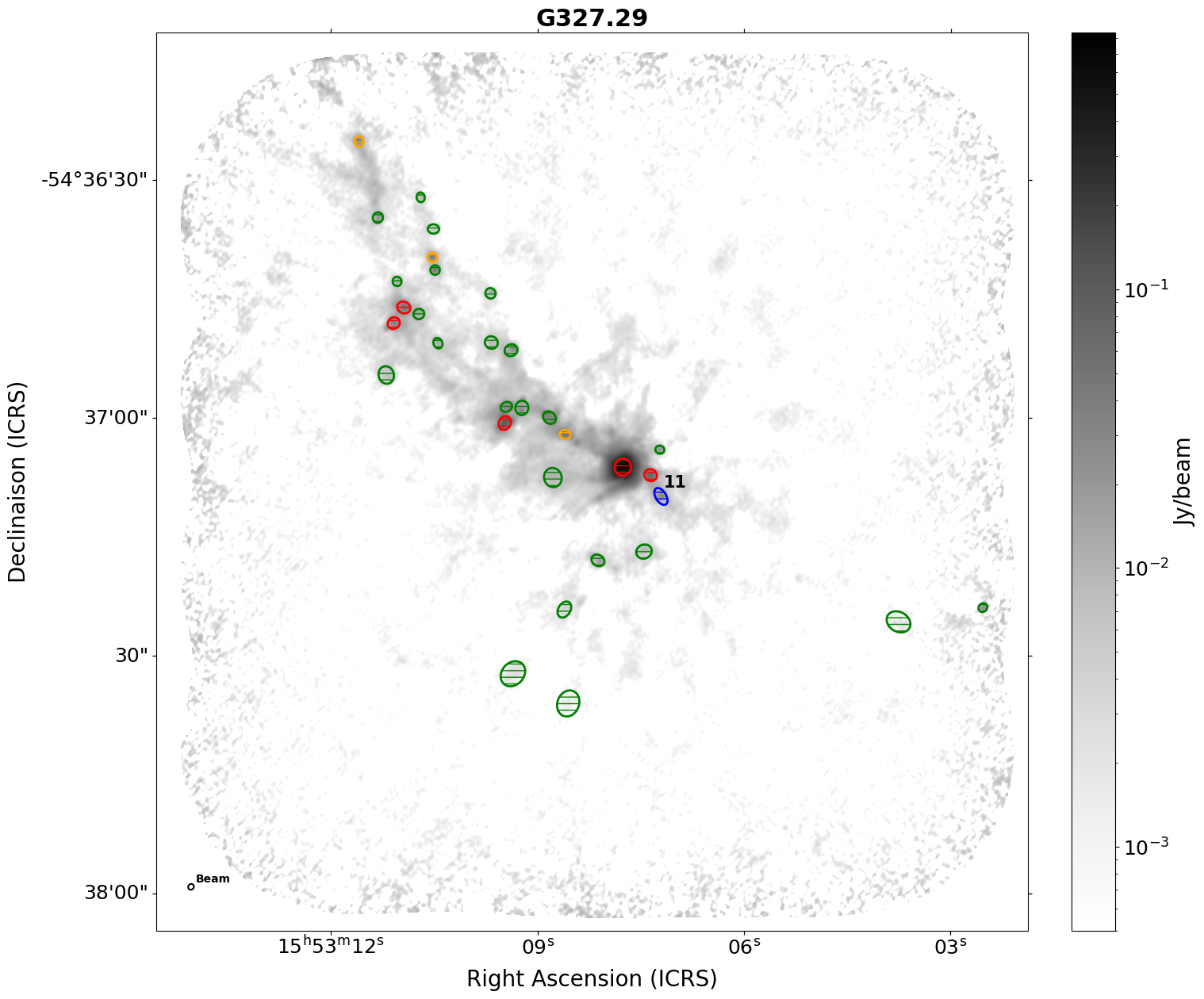}
    \includegraphics[width=0.49\textwidth]{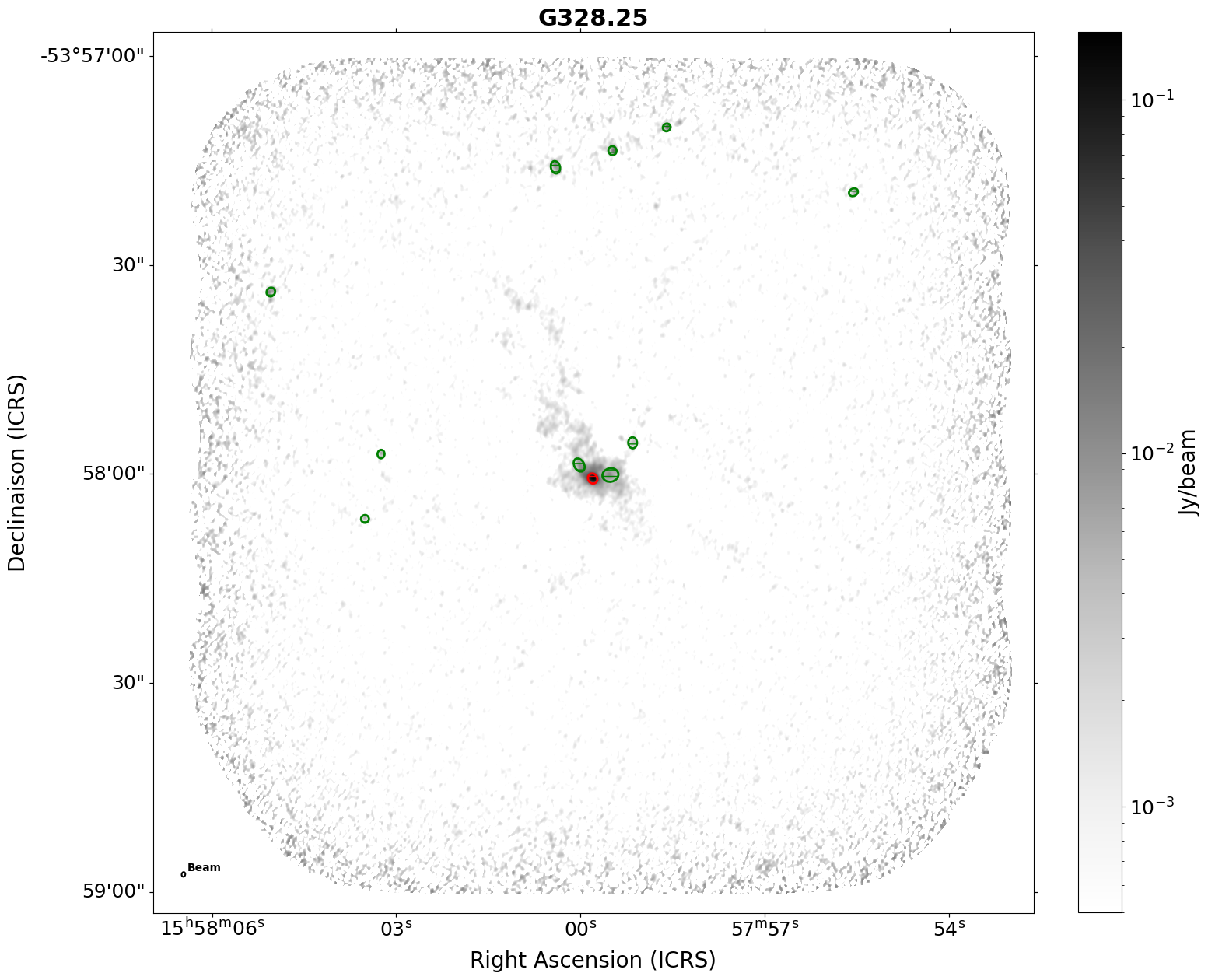}
    \includegraphics[width=0.49\textwidth]{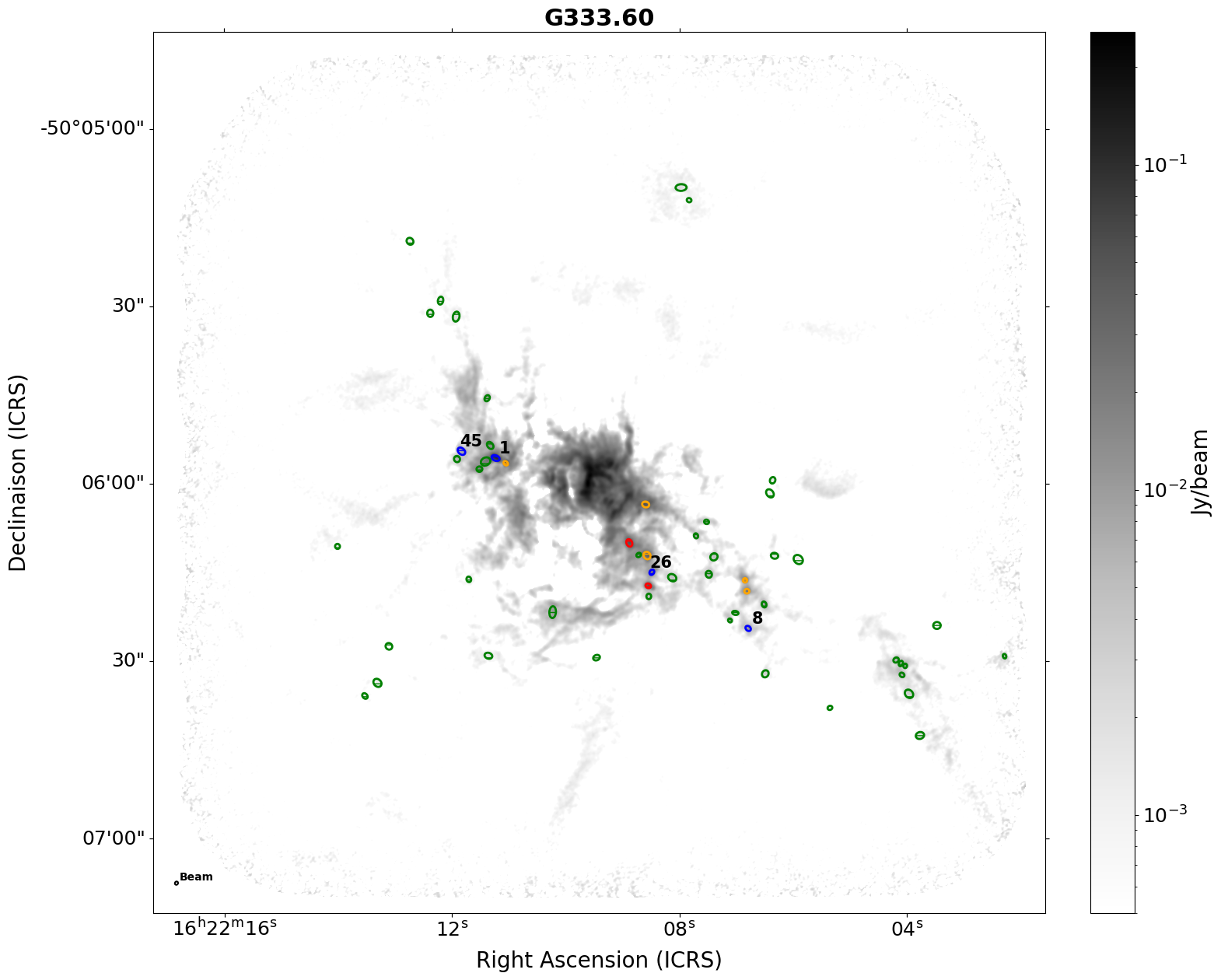}
    \caption{Locations of the high-mass PSC candidates in the ALMA-IMF regions. The candidates are the blue ellipses annotated with their number. The red cores are the high-mass protostellar cores (M\,$>$\,8\,M$_{\odot}$) and the orange ones are the intermediate mass protostellar cores (4\,M$_{\odot}<$\,M\,<\,8\,M$_{\odot}$. The other cores are the green ellipses which are non studied cores.}
    \label{appendix:MPSC_highlight_figure}
\end{figure*}

\begin{figure*}\ContinuedFloat
    \centering
    \includegraphics[width=0.49\textwidth]{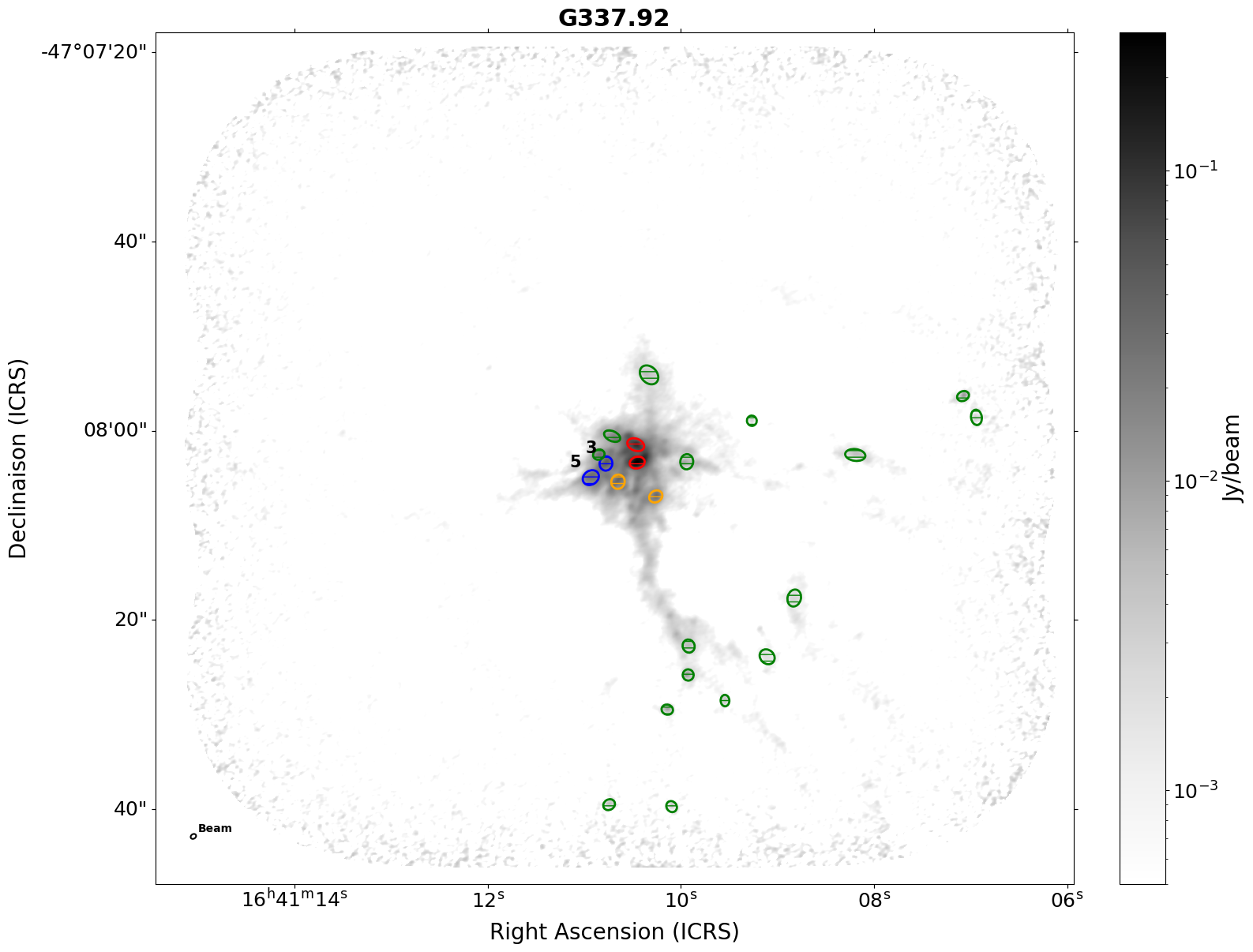}
    \includegraphics[width=0.49\textwidth]{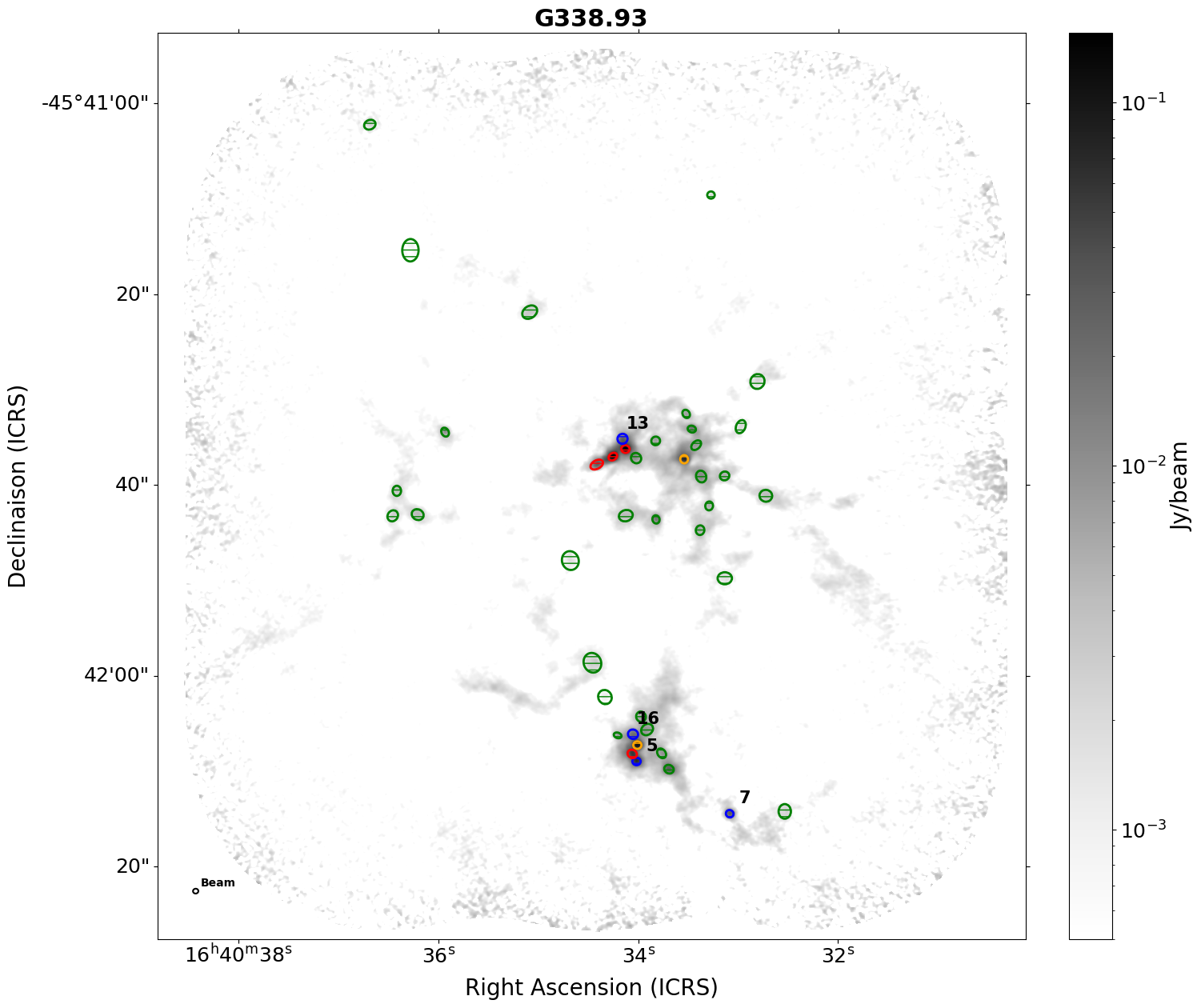}
    \includegraphics[width=0.49\textwidth]{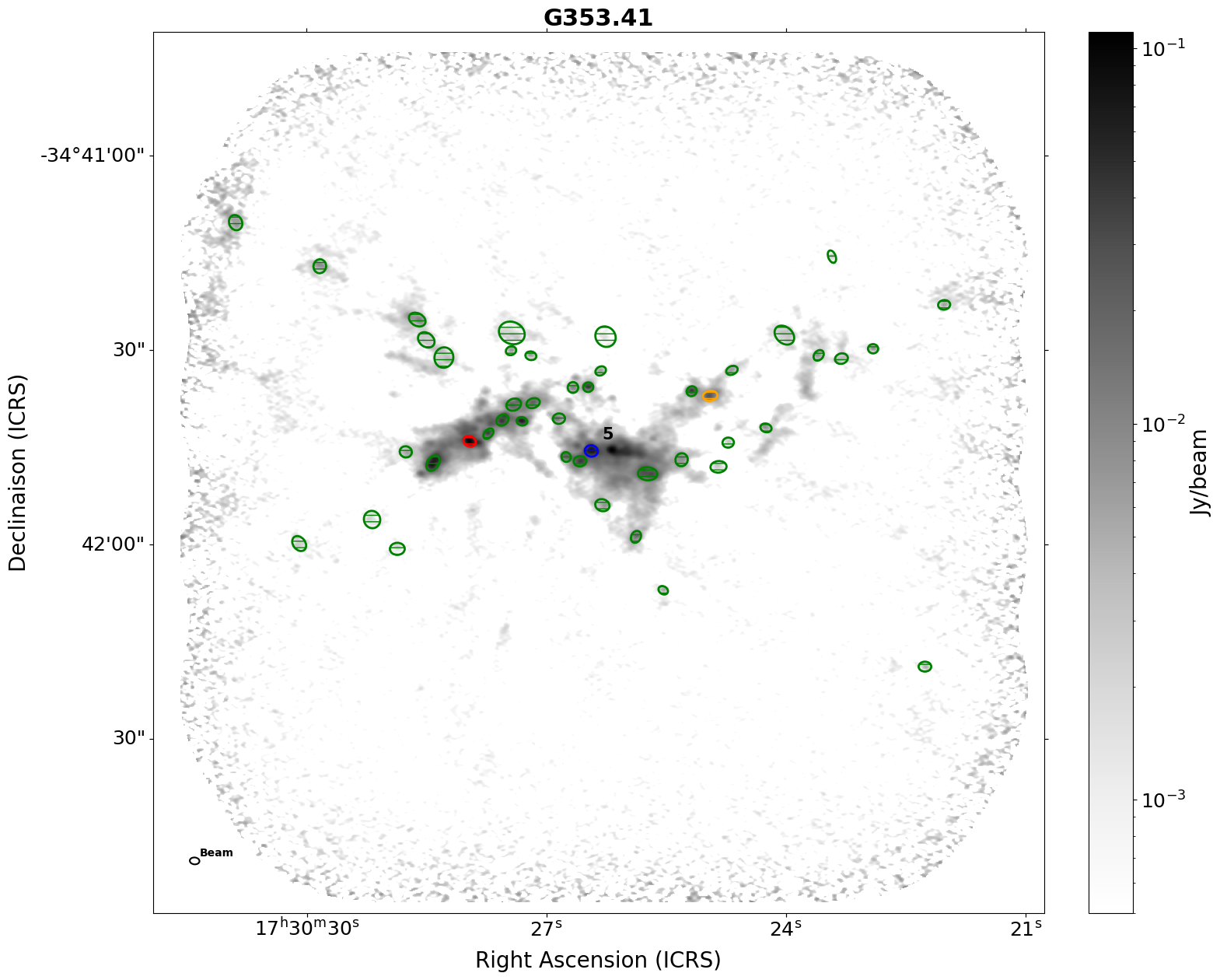}
    \includegraphics[width=0.49\textwidth]{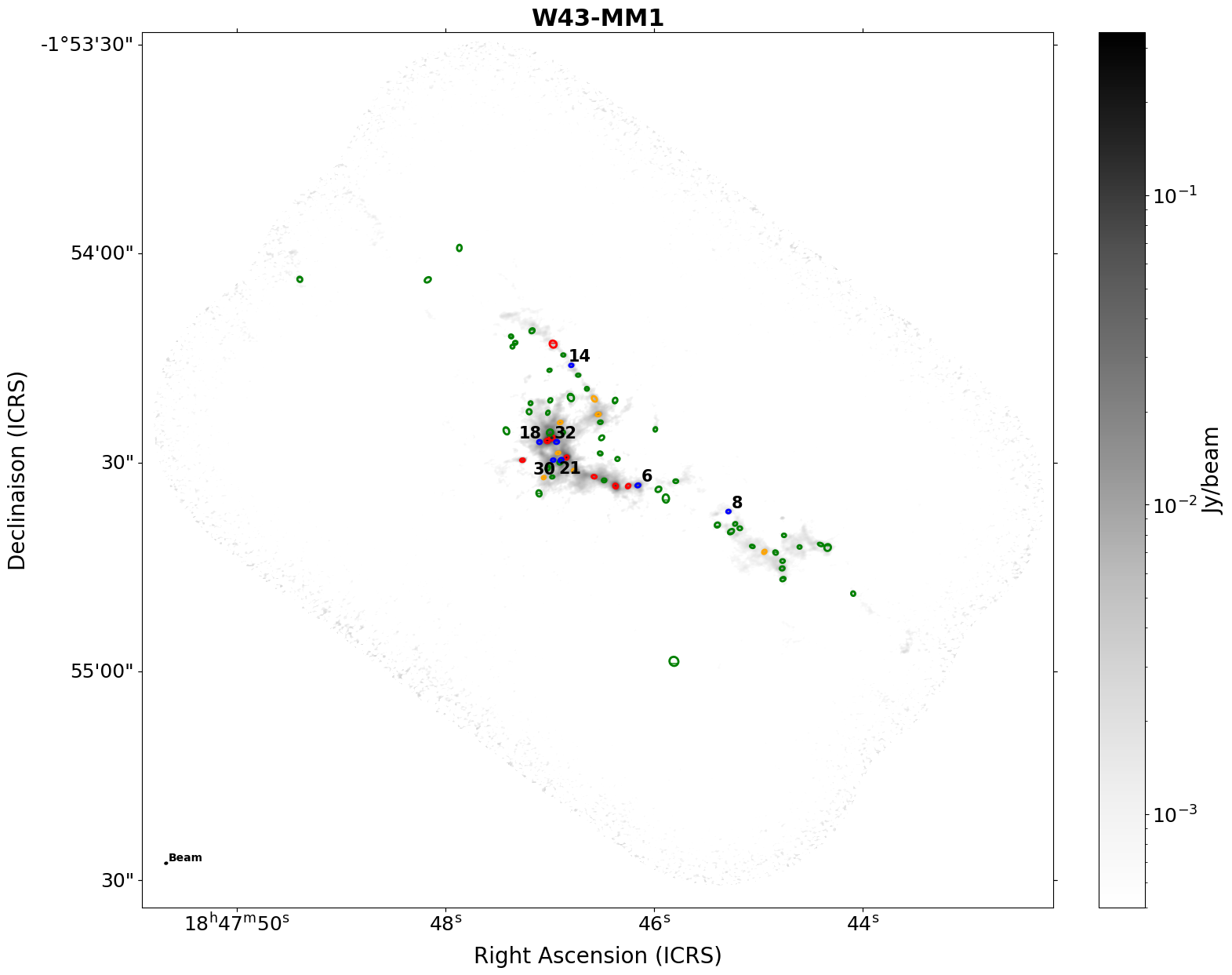}
    \includegraphics[width=0.49\textwidth]{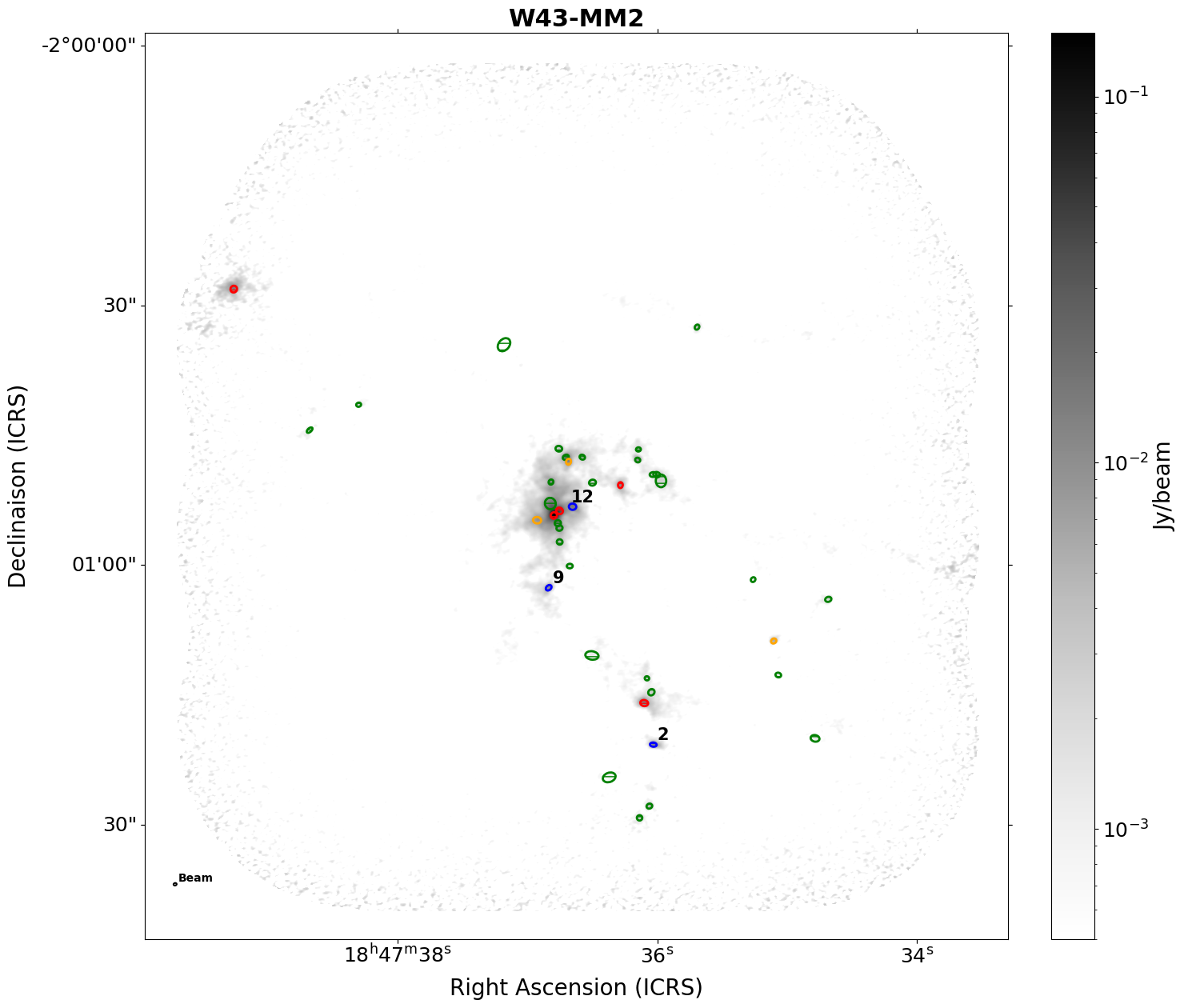}
    \includegraphics[width=0.49\textwidth]{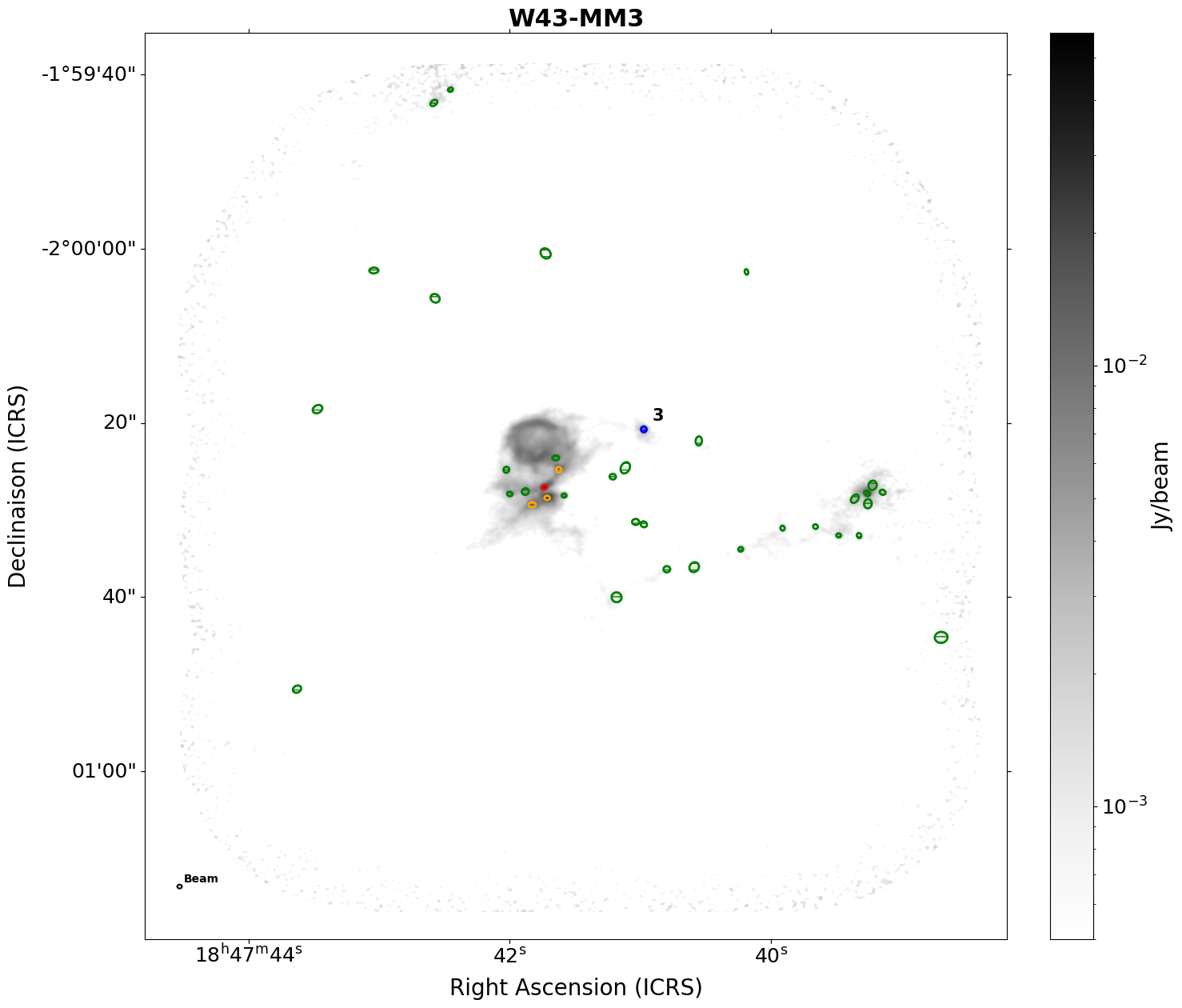}
    \caption{continued.}
\end{figure*}

\begin{figure*}\ContinuedFloat
    \centering
    \includegraphics[width=0.49\textwidth]{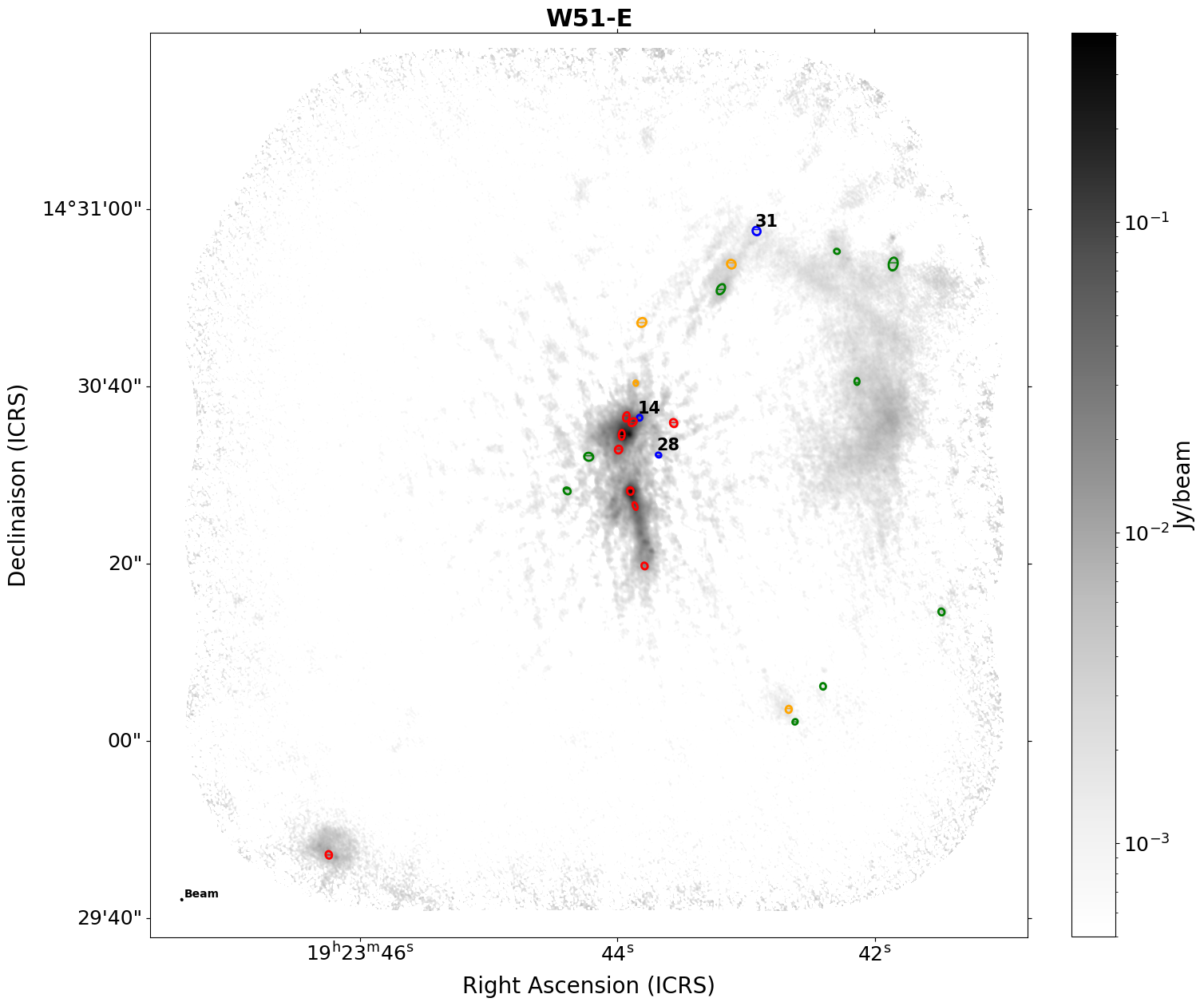}
    \includegraphics[width=0.49\textwidth]{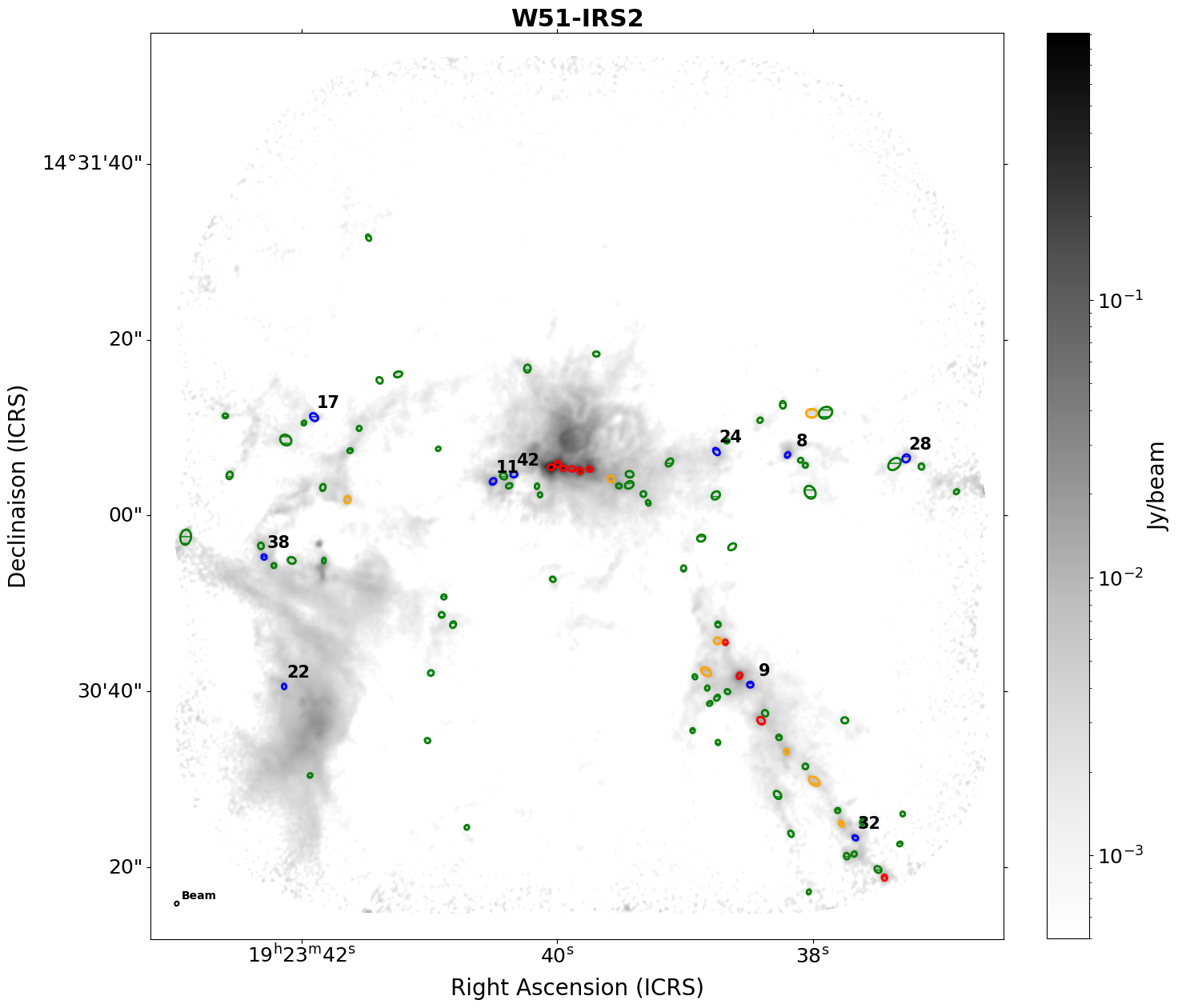}
    \caption{continued.}
\end{figure*}

\end{appendix}

\end{document}